\documentclass[12pt,a4paper,twoside,sort&compress]{book}
 \usepackage{slashed}
\usepackage{epsfig}
\usepackage{amsfonts}
\usepackage{amssymb}
\usepackage{amsmath}
\usepackage{setspace}
\usepackage{lscape}
\usepackage[latin1]{inputenc}
\usepackage{csquotes}
\usepackage{amsthm}
\usepackage{latexsym}
\usepackage{psfrag}
\usepackage{graphicx}
\usepackage{rotating}
\usepackage{graphics}
\usepackage{subfigure}
\usepackage{mathenv}
\usepackage{amsthm}
\usepackage{axodraw}
\def\r2{\sqrt 2}
\def\beq{\begin{equation}}
\def\eeq{\end{equation}}
\def\beqn{\begin{eqnarray}}
\def\eeqn{\end{eqnarray}}

\def\PL{{1-\gamma_5\over 2}}
\def\PR{{1+\gamma_5\over 2}}
\def\sinW2{\sin^2\theta_W}

\def\mz2{M_{z}^2}
\def\c2b{\cos 2\beta}

\def\mz{M_z}

\def\Fq2{F_{2}(q^2)}

\newcommand{\ra}{\rangle}

\newcommand{\be}{\begin{equation}}
\newcommand{\ee}{\end{equation}}
\newcommand{\ba}{\begin{eqnarray}}
\newcommand{\ea}{\end{eqnarray}}

\newcommand{\mchi}{m_{\chi}}
 \renewcommand{\mp}{M_p}
\def\Imag{\Im m\,}
\def\mg{m_{\tilde{G}}}
\def\msl{m_{\tilde{l}}}

\def\calO{{\cal{O}}}

\newcommand{\jxf}{J({\xf})}

\newcommand{\xf}{x_f}

\def\Imag{\Im m\,}
\def\mg{m_{\tilde{G}}}
\def\msl{m_{\tilde{l}}}

\def\calO{{\cal{O}}}

\def\gm{\gamma}

\def\mz{m_Z}

\def\mh{m_{H_1}}

\def\gm{\gamma}
\def\Re{{\rm Re}}

\def\prop#1{s-m_{#1}^2+i \Gamma_{#1}m_{#1}}
\def\slp#1{\widetilde{\ell}_{#1}}

\def\neu#1{\chi_{#1}^0}

\def\mslp#1{m_{\widetilde{\ell}_{#1}}}
\def\msli{m_{\widetilde{\ell}_{a}}}
\def\mslj{m_{\widetilde{\ell}_{b}}}
\def\mslk{m_{\widetilde{\ell}_{c}}}

\def\ml{m_{\ell}}

\def\mnmn{m_{\chi^0_i}m_{\chi^0_j}}

\def\Dijqp#1{D^{abij}_{#1}}

\def\neu#1{\chi_{#1}^0}
\def\neu#1{\chi_{#1}^0}

\def\gm{\gamma}
\def\Re{{\rm Re}}

\def\prop#1{s-m_{#1}^2+i \Gamma_{#1}m_{#1}}
\def\slp#1{\widetilde{\ell}_{#1}}

\def\slsl#1{\mslp{a}^{#1}\mslp{b}^{#1}}
\def\slpsl#1{\mslp{a}^{#1}+\mslp{b}^{#1}}
\def\slmsl#1{\mslp{a}^{#1}-\mslp{b}^{#1}}
\def\T{\mathcal{T}}
\def\F{\mathcal{F}}

\def\Fu{{\mathcal F}^u}
\def\Ft{{\mathcal F}^t}
\def\Tu{{\mathcal T}^u}
\def\Tt{{\mathcal T}^t}
\def\Y{\mathcal{Y}}
\def\mnq{m_{\chi_i^0}}
\def\mnp{m_{\chi_j^0}}

\def\N1{\widetilde N_1}

\def\stilde{\widetilde}

\newcommand{\newc}{\newcommand}
\newc{\sigmabar}{\overline\sigma}
\newc{\MSbar}{\overline{{\rm MS}}}
\newc{\DRbar}{\overline{{\rm DR}}}

\def\lam{\lambda}

\def\eps{\epsilon}
\def\stilde{\widetilde}
\def\ra{\rightarrow}
\def\newcdot{\kern.06em{\cdot}\kern.06em}

\catcode`@11
\def\seceqaa{\@addtoreset{equation}{section}
\def\theequation{A\arabic{equation}}}
\def\seceqbb{\@addtoreset{equation}{section}
\def\theequation{B\arabic{equation}}}
\def\seceqcc{\@addtoreset{equation}{section}
\def\theequation{C\arabic{equation}}}
\def\seceqdd{\@addtoreset{equation}{section}
\def\theequation{D\arabic{equation}}}
\catcode`@11
\def\beq{\begin{equation}}
\def\eeq{\end{equation}}
\def\bea{\begin{eqnarray}}
\def\eea{\end{eqnarray}}

\def\eps{\epsilon}

\def\mgss{M_{\tilde{g}}}
\def\mchi{M_{\tilde{\chi}}}

\def \mij {m_{ij}}
\def \mjk {m_{jk}}
\def \mik {m_{ik}}
\def \mm {m_{3/2}}
\def \mi {m_{\nu_i}}
\def \mj {m_{e_j}}
\def \mk {m_{e_k}}

\def \aa {m_{\tilde \nu_{iL}}}
\def \bb {m_{\tilde e_{jL}}}
\def \cc {m_{\tilde e_{kR}}}
\def \aaps {m_{\tilde \nu_{iL}}}
\def \bbps {m_{\tilde d_{jL}}}
\def \ccps {m_{\tilde d_{kR}}}
\def \aapt {m_{\tilde e_{iL}}}
\def \bbpt {m_{\tilde u_{jL}}}
\def \ccpt {m_{\tilde d_{kR}}}

\usepackage[latin1]{inputenc}
\usepackage{cite}
\usepackage{subfigure}
\usepackage[small,bf]{caption}
\usepackage{color}
\usepackage[Conny]{fncychap}
\ChNameVar{\LARGE\sffamily\bfseries} \ChNumVar{\Huge} \ChTitleVar{\Huge\sffamily\bfseries}
\ChRuleWidth{0.85pt} \ChNameUpperCase \ChTitleUpperCase 
\usepackage{fancyhdr}

\catcode`@=11
\def\seceqaa{\@addtoreset{equation}{section}
           \def\theequation{A\arabic{equation}}}
\def\seceqbb{\@addtoreset{equation}{section}
           \def\theequation{B\arabic{equation}}}
\def\seceqcc{\@addtoreset{equation}{section}
           \def\theequation{C\arabic{equation}}}
\def\seceqdd{\@addtoreset{equation}{section}
           \def\theequation{D\arabic{equation}}}
\def\seceqee{\@addtoreset{equation}{section}
           \def\theequation{E\arabic{equation}}}
\def\seceqff{\@addtoreset{equation}{section}
           \def\theequation{F\arabic{equation}}}
\def\seceqgg{\@addtoreset{equation}{section}
           \def\theequation{G\arabic{equation}}}
\def\seceqhh{\@addtoreset{equation}{section}
           \def\theequation{H\arabic{equation}}}
\def\seceqjj{\@addtoreset{equation}{section}
           \def\theequation{J\arabic{equation}}}
\def\seceqll{\@addtoreset{equation}{section}
           \def\theequation{L\arabic{equation}}}
\catcode`@=11
\def\sbar{\overline}
\def\stilde{\widetilde}

\def\conj{{{\rm c.c.}}}

\def\half{{1\over 2}}

\begin{document}
\frontmatter
\pagestyle{plain}
    \pagestyle{fancy}                       
    \fancyfoot{}                            
    \renewcommand{\chaptermark}[1]{         
      \markboth{\chaptername\ \thechapter.\ #1}{}} %
    \renewcommand{\sectionmark}[1]{         
      \markright{\thesection.\ #1}}         %
    \fancyhead[LE,RO]{\bfseries\thepage}    
    \fancyhead[RE]{\bfseries\leftmark}      
    \fancyhead[LO]{\bfseries\rightmark}     
    \renewcommand{\headrulewidth}{0.3pt}    
    \makeatletter
    \def\cleardoublepage{\clearpage\if@twoside \ifodd\c@page\else%
        \hbox{}%
        \thispagestyle{empty}
        \newpage%
        \if@twocolumn\hbox{}\newpage\fi\fi\fi}
    \makeatother
\addcontentsline{toc}{chapter}{Abstract}
\pagestyle{plain}

\vskip -0.9in 
\begin{center}
{\Large\bf Topics in Supergravity Phenomenology \footnote{Based on author's Ph.D. thesis written when at IIT Roorkee and defended on August 11, 2014.}}
\vskip 0.1in Mansi Dhuria \footnote{email: mansi@prl.res.in}
 \\
Theoretical Physics Division, Physical Research Laboratory, Ahmedabad - 380 009, India
\\
 \vskip 0.5 true in
\date{\today}
\end{center}
{\bf Abstract}: The first part of the review article is devoted to investigation of important phenomenological and particle-cosmology-related issues in the context of Type IIB string compactifications. 
After undertaking a brief review of (split) supersymmetry in the context of Beyond Standard Model Physics, we discuss the possibility of realizing ``$\mu$-split-like SUSY" scenario from a phenomenological model, which we show could be realizable locally as the large volume limit of a type IIB Swiss-Cheese Calabi-Yau orientifold involving a mobile space-time filling D3-brane localized at a nearly special Lagrangian three-cycle embedded in the "big" divisor (hence the local nature of the model's realization) and multiple fluxed stacks of  space-time filling D7-branes wrapping the same ``big" divisor. Naturally realizing split-SUSY scenario of N. Arkani-Hamed and S. Dimopoulos in our model, we show that the mass of one of the Higgs formed by a linear combination of two Higgs doublets (related to the D3-brane position moduli), can be produced to be of the order of 125 GeV whereas other Higgs as well as higgsino mass parameter to be very heavy- the ``$\mu$-split-like SUSY" scenario. The squarks'/sleptons' (the Wilson line moduli on D7-branes' world volume) masses also turn out to be very heavy.  Motivated by the fact that the gravitino appears as the Lightest Supersymmetric Particle (LSP) in our model, we explore the possibility of the gravitino as a viable cold dark matter candidate by showing its life time to be of the order of or greater than the age of the universe whereas lifetimes of decays of the co-NLSPs (the first generation squark/slepton and the lightest neutralino) to the LSP (the gravitino) turns out to be too short to disturb predictions of Big Bang Nucleosynthesis (BBN). Assuming non-thermal gravitino production mechanism,
we estimate the gravitino relic abundance to be around $0.1$ by evaluating the neutralino/slepton
annihilation cross sections and hence show that the former satisfies the requirement for a dark
matter candidate.  As another testing ground for split-SUSY scenarios, we estimate Electric Dipole Moment (EDM) of electron/neutron up to two-loops in our model. By explaining distinct ${\cal O}(1)$ CP violating phases associated with Wilson line moduli and position moduli, we show that it is possible to obtain dominant contribution of EDM of electron/neutron to be around $10^{-29}$ esu-cm at two-loop level by including heavy sfermions and a light Higgs, around $10^{-27}$ esu-cm for a two-loop Barr-Zee diagram involving W boson and SM interaction vertices, eEDM  to be around $10^{-33}$ esu-cm at one-loop by including heavy chargino and a light Higgs, and nEDM  to be around $10^{-32}$  esu-cm at one-loop by including  SM fermions and a light Higgs.  

In second part of the review article, we obtain local M-theory uplift of Type IIB background involving non-compact (resolved) warped deformed conifold and (N) space time filling D3-branes placed on the singularity of the conifold, (M) D5-branes wrapped around compact $S^2$, and ($N_{f}$) D7-branes wrapped around non-compact four-cycle via Strominger-Yau-Zaslow's mirror symmetry prescription \cite{SYZ3_Ts}. This work is mainly concerned with the investigation of hydro/thermodynamical aspects relevant to explain the behavior of thermal QCD with fundamental quarks, as well as to demonstrate the thermodynamical stability of the M-theory uplift. The uplift gives a black M3-brane solution whose near-horizon geometry near
$\theta_{1,2}=0,\pi$-branches, preserves $1/8$ SUSY. We propose a new MQGP limit corresponding to finite-string-coupling ($g_s \lesssim 1$)- large-t'Hooft-coupling regime of M-theory in addition to the one discussed in \cite{metrics}. Interestingly, we obtain $\eta/s=1/4 \pi$ for the uplift and the diffusion constant $ D \sim1/T$ for types IIB/IIA backgrounds in the both limits. The thermodynamical stability of M-theory uplift is checked by evaluating the D = 11 Euclideanized supergravity action upto $O(R^4,|G_4|^2)$  term in the two limits, and thereafter showing the positive sign of specific heat from the finite part of the action.

\clearpage
\begin{spacing}{1.34}
 \addcontentsline{toc}{chapter}{List of Figures}
 \listoffigures
 \end{spacing}
 \begin{spacing}{1.34}
\listoftables
 \addcontentsline{toc}{chapter}{List of Tables}
\end{spacing}
 \tableofcontents
 \clearpage
\mainmatter
\pagestyle{fancy}
\setcounter{chapter}{0}
\setcounter{section}{0}
\setcounter{subsection}{0}
\setcounter{tocdepth}{3}


\chapter{Introduction}
\vskip -0.5in
{\hskip1.4in{\it ``From time immemorial, man has desired to comprehend the complexity of nature in terms of as few elementary concepts as possible."}}

\hskip4.2in - Abdus Salam.

\graphicspath{{Chapter1/}{Chapter1/}}
\vskip -0.5in
\section{Overview}
Physics in the mid-20th century took off with  two important mathematical frameworks: (i) Quantum Field Theory \cite{qft} upon which microscopic description of the world is based, and (ii) General Relativity  \cite{gr} which successfully explains description of gravity at the large macroscopic scale. Both frameworks provide a good theoretical understanding to explain the empirical evidences, but at different energy scales. However, both quantum field theory as well as general relativity, on their own, lack any theoretical explanation which could describe what happened during the beginning of our universe.  This inability brought an interesting thought to the physicists's minds that all physics might be described by just one fundamental theory that unifies quantum field theory and General relativity. The idea of unification started out with the unification of electric and magnetic forces in an Electromagnetic theory proposed by James Clerk Maxwell in 1873. This continued to the unification of the Glashow-Salam-Weinberg electroweak theory and quantum chromodynamics (QCD) in 1970's, known as the Standard Model (SM) of particle physics. However, it turns out to be physically inconsistent when one wants to quantize gravity as a quantum field theory. At this point, the SM and gravity seem to be incompatible and unable to have the potential to answer a number of crucial questions like the microscopic details of black hole physics or the origin of the universe.
Nowadays,  prodigious underground accelerators and space crafts probe these two theories everyday to unravel the fundamental puzzles. Therefore it seems a necessity to reach at the level where physics might be able to give a consistent theoretical explanation to address the aforementioned issues. To do so, one should have a theory that combines the Standard Model and General Relativity. String theory is one prominent candidate to be such a unifying theory. It  provides a consistent unification of all fundamental forces of nature and a compelling UV completion of the supersymmetric field theories. Also, this ``theory of everything"  is of utmost importance to provide a wider and a more general view to one's thoughts of introducing notion of supersymmetry as well as  electroweak symmetry breaking in the extensions of Standard Model. The motivations for SUSY so far lie purely in the realm of particle physics, but there are more fundamental, although perhaps indirect, reasons to believe in SUSY and to understand the origin of dark matter and dark energy. On the other hand,  after years of efforts, the biggest and one of the most expensive experiments in the history of science (Large Hadron Collider) bought a revolutionary period in the era of fundamental physics with the discovery of the Higgs boson at LHC. The discovery has  opened the doors to Beyond Standard Model (BSM) Physics.

So,  one of the greatest challenges is to relate string theory to the observables in the low energy physics world, i.e.,  to reproduce all the characteristic features of the SM such as non-Abelian gauge group, Yukawa couplings, chiral fermions,  hierarchy between the electroweak scale $M_W$ and the Planck scale $M_P$, etc. This main task comes under the area of String Phenomenology.  
This fascinating area, due to its fundamental and mathematically structured character, 
 is able to provide a framework for computing all couplings of the
(MS)SM dynamically and give an explanation of the supersymmetry breaking at both low and high energy scales. The framework is generally based on two approaches:  (i) the top-down approach (global models), which starts from the fundamental theory and try to deduce from it all possible low energy observables (ii) bottom-up approach (local models), which try to construct consistent string models that incorporate as many SM features as possible. {\emph{The work presented in this review article belong to a local model-building approach of a string theory that is able to provide sufficient tools to reproduce characteristic features of Standard Model or the Minimal Supersymmetric Model without introducing any exotic particles.}}
\vskip -0.5in
  \section{Challenges of Standard Model}
We begin our discussion by reasoning out in brief the shortcomings of Standard Model. The SM of particle physics is believed to be a unified field theory combining electromagnetic, weak
and strong interactions \cite{SM1,SM2,SM3}. The particle content of the Standard Model is grouped into three categories: matter particles, force
carriers, and the Higgs particle. All the visible matter in the universe is described in terms of a limited number of fermionic constituents of matter: six quarks, three charged leptons and three light neutrinos.  It is the biggest triumph of particle physics to date, and has been tested very
extensively (predictions of the existence of the W and Z bosons, gluons, and the top and charm quarks before these particles were indirectly discovered experimentally). Nevertheless, through experimental and theoretical research, it is manifest that though the Standard Model precisely describes the phenomena within its domain, it is still incomplete.  Here, we list some of the problems of SM. 
\begin{itemize}
\item[--]
{\emph{Fine tuning in the Higgs sector}}:  At one-loop, the quantum corrections to the square of the Higgs mass scales as $m^{2}_H = m^{2}_{H}$(bare)$ + a^2 {\Lambda}^2$.  If $\Lambda_{SM}$ shoots up to the Planck scale, one has to do a fine-tuning of the around 1 in $10^{26}$ to get the Higgs mass of the order of 100 GeV as required by the electroweak theory. This is of course a very huge fine tuning problem. 
 \item[--]{\emph{Neutrino masses and mixings}}: Neutrinos are massless in the SM.  However,  observed
neutrino oscillation patterns proved that neutrinos oscillate from one flavor to another, implying small but non-zero masses. To account for these observations, it is important to implement a mechanism
to generate them.
\item[--] {\emph{Gravity}}: SM does not incorporate gravity.
\item[--] {\emph{Dark matter and dark energy}}: The existence of both have been confirmed by many independent observations such as gravitational lensing effects, CMB fluctuations, and more recently by Planck 2013. The Standard Model does not include any particle which can make up the dark
matter and also fails to give right contribution of dark energy of the universe.
\item[--]{\emph{Matter-anti-matter asymmetry}}:  Various considerations indicate that there is an imbalance of matter and anti-matter in our universe. Moreover, although SM has the means of
fulfilling the three Sakharov's conditions, it falls short to explain the baryon asymmetry of the universe.
\end{itemize}
In addition to aforementioned problems, there is a lack of natural explanation to understand the origin of the form of the scalar potential which exhibits electroweak symmetry breaking, a particular structure of Yukawa couplings, choices of gauge groups of SM and matter fields transforming under a particular gauge group. 
\section{Notion of Supersymmetry}
\subsection{Supersymmetry}
Supersymmetry \cite{martin,nilles,dienes4,utpal3} has been considered to be a proficient tool to anticipate new physics beyond the Standard Model  because of its ability to provide a solution to most of shortcomings of Standard Model  including gauge hierarchy problem, vacuum stability of Higgs boson from the point of view of particle physics, and to address fundamental puzzles such as providing astrophysical dark matter candidates,  dark energy and matter-antimatter asymmetry from the point of view of cosmology (see \cite{yajnik3,utpal4,dienes1}). The theory was discovered in the 70's  in the context of quantum field theory by establishing a relationship between elementary particles of different quantum nature. It  basically aims to provide a link between fermions and bosons by extending the Poincar\'{e} algebra to include spinorial generators that connect the fermionic degrees of freedom to the bosonic degrees of freedom and vice-versa \cite{wess+zumino,wess+zumino1,Wess_Bagger}.

Supersymmetry is accomplished by introducing a supercharge ${\cal Q}$ which is
an anticommuting spinor with the properties $ {\cal Q}|{\rm Boson}\rangle = {\rm fermion}, {\cal Q}|{\rm fermion}\rangle = {\rm Boson}.$
The generators satisfy super-poincar\'{e} algebra as follows:
 \begin{eqnarray}
 && \{{\cal Q}_{\alpha}, {\cal Q}^{\dagger}_{\dot \alpha}\} = 2 \sigma^{\mu}_{\alpha \dot\alpha} {\cal P}_{\mu}, 
\{{\cal Q}_{\alpha}, {\cal Q}_{\beta} \} =  \{{\cal Q}^{\dagger}_{\dot \alpha}, {\cal Q}^{\dagger}_{\dot \beta}\} = 0 \nonumber\\
 &&   [{\cal Q}_{\alpha}, {\cal P}^{\mu}] = [{\cal Q}^{\dagger}_{\dot \alpha}, P^{\mu}] = 0, 
 \end{eqnarray}
 where $\alpha$, $\beta$ are spinor indices, $\sigma^{\mu}$ are Pauli spin matrices and $P_{\mu}$ is generator of space-time translations. In a supersymmetric theory, single particle states are classified into irreducible representations of the
supersymmetry algebra that are defined as supermultiplets. Each supermultiplet carries fermionic 
 and bosonic states that are superpartners of each other and involve the same number of degrees of freedom. Particles
in a supermultiplet share the same quantum numbers and have the same mass.  In case of renormalizable ${\cal N}=1$ supersymmetric gauge theory, one considers a distinct copy of supersymmetry generators (${{\cal Q}, {\cal Q}^{\dagger}}$). The irreducible representation corresponding to two fermionic and bosonic degrees of freedom in a supermultiplet, when translated in terms of field theory language, give rise to combinations that include chiral/matter/scalar supermultiplet which contains a spin-1/2 Weyl fermion and a complex scalar, and gauge/vector supermultiplet which contains a massless spin-1 vector boson and its superpartner, a spin-1/2 Weyl fermion. In addition to this, if we also include quantum gravity, then we have the gravity supermultiplet which contains a spin-2 graviton and a spin-3/2 superpartner called the gravitino. Therefore, in generic ${\cal N}=1$ supersymmetric gauge theory, supermultiplets appear in three pairs: chiral multiplet $(\psi_\mu, \phi^{\mu})$ corresponding to fermion and its super partner scalar; gauge multiplet $(A^{\mu}, \lambda^{\mu})$ corresponding to neutral gauge boson and its superpartner gaugino; and the gravity multiplet $(g_{\mu \nu}, \psi^{\mu})$ corresponding to graviton and its superpartner gravitino.
 \paragraph {Gauge Hierarchy Problem:} \vskip -0.3in
 Historically, one of the major motivations which certainly demanded the need to introduce supersymmetry at TeV scale was to resolve gauge hierarchy problem \cite{godbole3,dienes2}.  As discussed above, the quantum (loop) corrections to the Higgs mass-squared, $m_H^2$ give a term which is quadratically divergent. For example, for a generic fermionic loop as given in Figure~1.1(a),
the quantum correction to the Higgs mass is given as:
\begin{equation}
\Delta m_H^2=-\frac{Y_f^2}{8\pi^2}[2\Lambda^2 + 6m_f^2 \ln(\Lambda/m_f)+...],
\label{quadf}
\end{equation}
where $Y_f$ is the Yukawa coupling  and $\Lambda$ is an ultraviolet cutoff which can be identified as the scale up to which SM remains valid, beyond which new physics appears. This contribution to the mass of the Higgs diverges quadratically with $\Lambda$. 
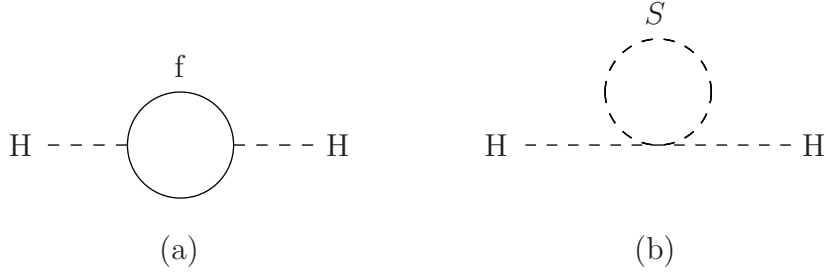
\begin{figure}
\begin{center}
\begin{picture}(145,105) (170,10)
 \DashLine(100,50)(130,50){4}
 \CArc(150,50)(20,0,180)
 \CArc(150,50)(20,180,0)
 \DashLine(170,50)(200,50){4}
   \Text(150,80)[]{{f}}
    \Text(90,50)[]{{H}}
       \Text(210,50)[]{{H}}
    \Text(150,10)[]{{(a)}}
   \DashLine(280,50)(380,50){4}
   \DashCArc(330,70)(20,180,0){4}
    \DashCArc(330,70)(20,0,180){4}
    \DashCArc(330,70)(20,180,0){4}
    \DashCArc(330,70)(20,0,180){4}
    \DashCArc(330,70)(20,180,0){4}
    \DashCArc(330,70)(20,0,180){4}
     \Text(270,50)[]{{H}}
     \Text(390,50)[]{{H}}
       \Text(330,100)[]{{$S$}}
        \Text(330,10)[]{{(b)}}
    \end{picture}
\end{center}
\vskip -0.1in
\caption{Quantum corrections to the mass-squared of the Higgs boson due to generic fermion (f) and scalar (S) in the loop.}
\label{fig:hierarchy}
 \end{figure}
Moreover, the one-loop diagram corresponding to scalar in the loop as shown in Figure~1.1(b), makes a
similarly divergent contribution given as:
\begin{equation}
\Delta m_H^2=\frac{\lambda_S}{16\pi^2}[\Lambda^2 - 2m_S^2 \ln(\Lambda/m_S)+...],
\label{quadS}
\end{equation}
\pagebreak
where $\lambda_S$ is the quartic coupling of the scalar $S$ to the Higgs boson. Comparing (\ref{quadf})
 and (\ref{quadS}), we realize that the quadratically-divergent terms  $\propto \Lambda^2$ would cancel if, corresponding to every fermion $f$ there exists a
scalar $S$ with quartic coupling $\lambda_S \; = \; 2 y_f^2 \,$. This is exactly the relationship imposed by supersymmetry! Therefore, it appears that
quadratic divergences disappear in supersymmetric field theories in a natural way. Moreover, the theory regulates quadratic divergence not just at the one-loop level
discussed above, but also at the multi-loop level.  
In addition to quadratic divergent piece, logarithimic correction to the physical Higgs mass squared is proportional  to scalar mass squared. Consequently, to obtain the Higgs mass of the order of TeV, scalar mass should also be with in ${\cal O}$(TeV) limit. {\emph{So the hierarchy problem implies that new physics beyond the Standard Model (in other words, the SUSY breaking scale) is likely to be around the TeV scale.}} 
\vskip -0.5in 
\subsection{Supersymmetric Lagrangian}
We first briefly review SUSY algebra \cite{martin,godbole3} to get a reasonable understanding of chiral and vector superfields.
SUSY algebra is realized as a translation in superspace $(x^{\mu}, \theta^{\alpha}, {\bar\theta}^{\dot\beta})$; $\theta^{\alpha}$ and ${\bar\theta}^{\dot\beta}$ represent Grassmannian variable. We can thus define a superfield as $\Phi(x^{\mu}, \theta^{\alpha},{\bar\theta}^{\dot\beta})$. The supercharges are represented as:
\begin{eqnarray}
&& {\cal Q}_{\alpha}= \frac{\partial}{\partial \theta^{\alpha}}- i(\sigma^{\mu})_{\alpha \dot\alpha}{\bar\theta^{\dot\alpha}}\partial_{\mu}, {\cal Q}_{\dot\beta}= \frac{\partial}{\partial \theta^{\dot\beta}}+ i {\theta^{\beta}}(\sigma^{\mu})_{\beta \dot\beta} \partial_{\mu},
\end{eqnarray}
and spinor derivatives are represented as:
 \begin{eqnarray}
D_{\alpha}={\partial}/{\partial \theta^{\alpha}}+ i{(\sigma^{\mu})}_{\alpha\dot\alpha}{{\bar\theta}^{\dot\alpha}{\partial_{\mu}}}, { \bar D}_{\dot\beta}=-{\partial}/{\partial{\bar \theta}^{\dot\beta}}-i{\theta}^{\beta}{(\sigma^{\mu})}_{\beta\dot\beta}{\partial_{\mu}}.
\end{eqnarray}
The chiral superfield is obtained by imposing condition ${\bar D}_{\dot\beta}{\Phi} = 0$. Since $\Phi$ depends on both ordinary and Grassmannian variable $\theta$, by expanding the same in power series in $\theta$,  one gets (for simplicity, we omit indices here):
\begin{eqnarray}
 \Phi(x,\theta)=\phi(x,\theta)+ \sqrt{2} \theta \psi(x,\theta)+\theta\theta F(x, \theta),
\end{eqnarray}
where  $\psi$ corresponds to Weyl fermion, $\phi$ represents two complex scalar fields  and F is an auxiliary field required to close the algebra. The supersymmetry transformations transforming different components of chiral superfields are given as:
\beqn
\label{gphitran}
&&
\delta \phi_i = \epsilon\psi_i, \delta \psi_{i\alpha} =
-i (\sigma^\mu \epsilon^\dagger)_{\alpha}\> \nabla_\mu\phi_i + \epsilon_\alpha F_i,
\nonumber\\
&&\delta F_i = -i \epsilon^\dagger \sigmabar^\mu \nabla_\mu \psi_i
\> + \> \sqrt{2} g (T^a \phi)_i\> \epsilon^\dagger \lambda^{\dagger a} .
\eeqn
where $ \epsilon^{\alpha}$  is an infinitesimal, anticommuting, two-component Weyl fermion object which parametrize the supersymmetry transformation. A vector superfield is real and obtained by imposing condition $V(x, \theta, {\bar\theta}) = V^{\dagger}(x, \theta, {\bar\theta})$. The component expansion of the vector superfield in Wess-Zumino gauge is simply given as:
\begin{eqnarray}
V^{a}({x,\theta,\bar\theta})= -\theta{\sigma}^{\mu}{\bar\theta}A^{a}_{\mu}(x)+i\theta{ \theta}{\bar \theta}{\bar\lambda}^{a}(x)- i {\bar \theta}{\bar \theta}{\theta}{\lambda}^{a}(x)+\frac{1}{2}\theta\bar\theta \bar\theta D^{a}(x).
\end{eqnarray}
The Supersymmetry transformations transforming different components of vector superfields are given as:
\beqn
\label{gphitran}
&&
\delta V_{\mu} = i \epsilon\sigma_{\mu}{\bar\lambda}+ i{\bar\alpha}{\sigma_{\mu}}{\lambda}, \delta \lambda =
\epsilon\sigma^{\mu\nu}(\partial_{\mu} A_{\nu}-\partial_{\nu}A_{\mu})+{\bar \alpha}D,
\nonumber\\
&&\delta D=\epsilon \sigma^{\mu}{\bar\lambda}+{\bar\epsilon}{\sigma^{\mu}}\partial_{\mu}{\lambda},
\eeqn
The auxiliary field can be eliminated by using the equation of motion, implying:
\begin{eqnarray}
&& F_i=\frac{\partial F}{\partial \phi_i}= W^{i}, D^{a}= -g_{a}({\phi^{*}_i}T^{a}\phi_i),
\end{eqnarray}
where W is holomorphic function of $\Phi$, known as superpotential.
The scalar potential is expressed in terms of auxiliary terms as follows:
\begin{eqnarray}
V=\sum_{i} {\frac{\partial W}{\partial {\phi_i}}}+\frac{1}{2}\Bigl(\sum_{i}\phi^{\dagger}_{i}T^{a}\phi_{i}\Bigl)^2.
\end{eqnarray}
By using an analytic form of superfields and Grassmannian algebra, the general form \\of Lagrangian is written in the following form:
\begin{eqnarray}
L= \int d^{4}{x} \left\{ d^{2}{\theta}d^{2}{\bar\theta}{\cal L}_{D}+\int d^{2}{\theta}{\cal L}_F +h.c.\right\},
\end{eqnarray}
where ${\cal L}_{D}$ is sum of vector superfields given by ${\cal L}_{D}= \phi^{\dagger}e^{g_{a}T^{a}V^{a}}\phi$,
and ${\cal L}_{F}$ is sum of scalar superfields given by ${\cal L}_{F}=\frac{1}{4}W^{a \alpha} W^{a}_{\alpha}+W(\Phi)$. The superfield $W^{a}$ corresponding to field strength is given by $W^a= -i{\lambda}^{a}+\theta D^{a}-{\sigma}^{\mu\nu}\theta F^{a}_{\mu\nu}-\theta\theta \sigma^{\mu}D_{\mu}{\lambda}^{a\dagger}$.
 \paragraph{Local supersymmetry:}\vskip -0.2in
  The global supersymmetry can be promoted to local supersymmetry \cite{local_nath1,localsusy,local_nath2} by making transformation parameter $\epsilon$ depend on space-time i.e $\delta \phi_i = \epsilon(x)\psi_i$. The requirement of  gauge invariance introduces  spin-3/2 fermion, known as gravitino and its scalar partner, the spin-2 massless graviton. So, by introducing space-time dependent parameter $\epsilon(x)$, the gravity appear naturally in the picture. This local minimal supersymmetric gauge theory is known as Supergravity (SUGRA). The resulting Lagrangian depends only two functions:  the K\"{a}hler function ${\cal G}$ and gauge kinetic function $f_{ab}$ given as:
 \begin{eqnarray}
{\cal G}={\kappa^{2}} K + ln \left[\kappa^{6}\left|W^{2}\right|\right], f_{ab}(\Phi_i)= \delta_{ab}( \frac{1}{g^{2}_a} - i \frac{\Theta_a}{8\pi^2}).
\end{eqnarray}
where $K$ is real function $K(\phi, \phi^{\dagger})$ called as K\"{a}hler potential, W is non-renormalizable superpotential, and both real and imaginary part of the gauge kinetic function ($f_{ab}$) appear as coefficients of  kinetic term corresponding to gauge field strength.
The scale of supersymmetry breaking is governed by gravitino mass $m_{\frac{3}{2}}=M_{P} e^{-{\langle \cal G \rangle}/2M^{2}_{P}}$.
  \subsection{Minimal Supersymmetric Standard Model (MSSM)}
We briefly review the minimal version of a supersymmetric Standard Model. This is the simplest extension of Standard Model formed by promoting each SM field to a superfield by considering an ordinary, non-extended supersymmetry \cite{martin,godbole3}. Each of the known fundamental particles is placed in either a chiral or gauge supermultiplet, and must have a superpartner with spin differing by 1/2 unit. The left-handed fermions in chiral supermultiplets transform differently under the gauge group than their right-handed parts.  The chiral and gauge supermultiplets of the MSSM are classified in Table~1.1. In addition to squarks and sleptons existing as superpartners of quarks and leptons, the MSSM contains two Higgs doublets and corresponding Higgsino doublets, as well gauginos as superpartners of the gauge bosons of the Standard Model \cite{mssmtable}.  
 \begin{table}[htbp]
 \label{table:mssm}
 \begin{center}
\vglue 0.6cm
\nopagebreak[4]
\renewcommand{\tabcolsep}{0.03cm}
\begin{tabular}{lllccc}
 \\ \hline \hline
Superfield & \ \ \ \ \ \ \ Bosons & \ \ \ \ \ \ \ Fermions &
$SU_c(3)$& $SU_L(2)$ & $U_Y(1)$ \\ \hline \hline Gauge  &&&&& \\
${\bf G^a}$   & gluon \ \ \ \ \ \ \ \ \ \ \ \ \ \ \  $g^a$ &
gluino$ \ \ \ \ \ \ \ \ \ \ \ \ \tilde{g}^a$ & 8 & 1 & 0 \\ ${\bf
V^k}$ & Weak \ \ \ $W^k$ \ $(W^\pm, Z)$ & wino, zino \
$\tilde{w}^k$ \ $(\tilde{w}^\pm, \tilde{z})$ & 1 & 3& 0 \\ ${\bf
V'}$   & Hypercharge  \ \ \ $B\ (\gamma)$ & bino \ \ \ \ \ \ \ \ \
\ \ $\tilde{b}(\tilde{\gamma })$ & 1 & 1& 0 \\ \hline Matter &&&&
\\ $\begin{array}{c} {\bf L_i} \\ {\bf E_i}\end{array}$ & sleptons
\ $\left\{
\begin{array}{l} \tilde{L}_i=(\tilde{\nu},\tilde e)_L \\ \tilde{E}_i =\tilde
e_R \end{array} \right. $ & leptons \ $\left\{ \begin{array}{l}
L_i=(\nu,e)_L
\\ E_i=e_R \end{array} \right.$ & $\begin{array}{l} 1 \\ 1 \end{array} $  &
$\begin{array}{l} 2 \\ 1 \end{array} $ & $\begin{array}{r} -1 \\ 2
\end{array} $ \\ $\begin{array}{c} {\bf Q_i} \\ {\bf U_i} \\ {\bf D_i}
\end{array}$ & squarks \ $\left\{ \begin{array}{l}
\tilde{Q}_i=(\tilde{u},\tilde d)_L \\ \tilde{U}_i =\tilde u_R \\
\tilde{D}_i =\tilde d_R\end{array}  \right. $ & quarks \ $\left\{
\begin{array}{l} Q_i=(u,d)_L \\ U_i=u_R^c \\ D_i=d_R^c \end{array}
\right.$ & $\begin{array}{l} 3
\\ 3^* \\ 3^* \end{array} $  & $\begin{array}{l} 2 \\ 1 \\ 1 \end{array} $ &
$\begin{array}{r} 1/3 \\ -4/3 \\ 2/3 \end{array} $ \\ \hline Higgs
&&&& \\ $\begin{array}{c} {\bf H_1} \\ {\bf H_2}\end{array}$ &
Higgses \ $\left\{
\begin{array}{l} H_1 \\ H_2 \end{array}  \right. $ & higgsinos \ $\left\{
 \begin{array}{l} \tilde{H}_1 \\ \tilde{H}_2 \end{array} \right.$ &
$\begin{array}{l} 1 \\ 1 \end{array} $  & $\begin{array}{l} 2 \\ 2
\end{array} $ &
$\begin{array}{r} -1 \\ 1
\end{array} $
 \\ \hline \hline
\end{tabular}
\caption{Particle Content of the Minimal Supersymmetric Standard Model.}
\end{center}\vspace{-0.5cm}
\end{table}
The superpotential for minimal supersymmetric model is as follows \cite{martin}:
\beq
W_{\rm MSSM} =
\sbar u {\bf y_u} Q H_u -
\sbar d {\bf y_d} Q H_d -
\sbar e {\bf y_e} L H_d +
\mu H_u H_d \> .
\label{MSSMsuperpot}
\eeq
${\rm y}_{\rm u,d,e}$ are related to the Yukawa couplings, $\sbar u, \sbar d, \sbar e, Q, L, H_{u,d}$ are chiral superfields, and $\mu$ corresponds to supersymmetric Higgs mass parameter. Since supersymmetry is a broken symmetry, the mass of missing superpartners must be higher than Standard Model particles otherwise it would have been observed at LHC long time ago. The SUSY can be broken by adding a set of soft terms in the MSSM Lagrangian though it lacks any theoretical motivation. In addition to that, there are models, including string-inspired models, in which supersymmetry can be broken in a hidden sector and then transmitted to visible sector via different type of interactions \cite{softsusy2,yajnik2}. However, the question of how supersymmetry is broken does not have a definitive answer yet. By avoiding the discussion about SUSY breaking mechanism(s), we write the general MSSM Lagrangian formed by including mass as well as interaction terms as follows  \cite{martin}:
\begin{eqnarray}
{\cal L}_{\rm soft}^{\rm MSSM} &=& -\half\left ( M_3 \stilde g\stilde g
+ M_2 \stilde W \stilde W + M_1 \stilde B\stilde B 
+\conj \right )
\nonumber
\\
&&
-\left ( \stilde {\sbar u} \,{\bf y_u}\, \stilde Q H_u
- \stilde {\sbar d} \,{\bf y_d}\, \stilde Q H_d
- \stilde {\sbar e} \,{\bf y_e}\, \stilde L H_d
+ \conj \right ) 
\nonumber
\\
&&
-\stilde Q^\dagger \, {\bf m^2_{Q}}\, \stilde Q
-\stilde L^\dagger \,{\bf m^2_{L}}\,\stilde L
-\stilde {\sbar u} \,{\bf m^2_{{\sbar u}}}\, {\stilde {\sbar u}}^\dagger
-\stilde {\sbar d} \,{\bf m^2_{{\sbar d}}} \, {\stilde {\sbar d}}^\dagger
-\stilde {\sbar e} \,{\bf m^2_{{\sbar e}}}\, {\stilde {\sbar e}}^\dagger
\nonumber \\
&&
- \, m_{H_u}^2 H_u^* H_u - m_{H_d}^2 H_d^* H_d
- \left ( b H_u H_d + \conj \right ) .
\label{MSSMsoft}
\end{eqnarray}
where $M_{1,2,3}$ correspond to mass terms of bino, wino and gluino, 
$m^{2}_{Q,L,u,d,e}$ are squark and slepton mass terms and finally $m^{2}_{H_u, H_d}$ correspond to mass terms for up and down type Higgs. In total, there appear to be 105 independent parameters in MSSM.   \paragraph {R-symmetry:} \vskip -0.3in
 Unlike SM, where the requirement of gauge invariance and renormalizability led to baryon and lepton number conservation automatically, the most general MSSM superpotential  consistent with 
the symmetries and renormalizability contains another set of terms which violate either baryon number  or  lepton number. The additional lepton-as well as baryon-number violating terms are given as  \cite{martin}:
\beqn
\label{WBviol}
W_{\Delta {\rm L} =1} &=&
{1\over 2} \lambda^{ijk} L_iL_j{\sbar e_k}
+ \lambda^{\prime ijk} L_i Q_j {\sbar d_k}
+ \mu^{\prime i} L_i H_u
 \nonumber \\
W_{\Delta {\rm B}= 1} &=& {1\over 2} \lambda^{\prime\prime ijk}
{\sbar u_i}{\sbar d_j}{\sbar d_k}.
\eeqn
These terms can lead to proton decay at unacceptable rates. To avoid this to happen, one needs to forbid these terms with the help of an underlying symmetry. There is a new global symmetry known as R-symmetry \cite{rsymmetry1,rsymmetry2,nilles3} which exists in  superspace and does not commute with supersymmetry. According to this, different components in the same supermultiplet have different R-charges. This basically transforms $\theta$ to $e^{i \alpha} \theta$, so $\theta$ has R-charge$=$1. Assuming that chiral super field  $\Phi$ also transforms with R-charge$=$ +1, scalar components $\phi$ have $R(\phi)=+1$ and fermions have $R(\psi)=0$. On the other hand, vector superfield is real having R-charge $= $0 . As a consequence of this, vector bosons $V_{\mu}$ have $R(V_{\mu})=0$ and gauginos have $R(\lambda)=1$. Now, invariance of Lagrangian requires ${\cal L}=\int d^{2}{\theta} W$ requires R-charge to be zero whereas $d^{2}{\theta}$ transforms with $R=-2$. If we go back to  the minimal superpotential  as given in equation (\ref{MSSMsuperpot}), the R-charges assigned to Higgs superfields and chiral superfields corresponding to SM particles are  $R(H, \bar H)= 1, R(Q, L, \bar U, \bar D, \bar E)=\frac{1}{2}$. Using this, one can check that terms given in the superpotential in equation (\ref{WBviol}) are not invariant under R-symmetry and therefore one can drop them out.
This continuous R-symmetry is somewhat problematic because gaugino Majorana masses get protected by R-symmetry. To  forbid these terms, even  a smaller symmetry like R-parity $R_{P} =(-1)^{3B +L+ 2S}$, is enough. According to this, all Standard Model particles achieve  +1 R-parity while all supersymmetric particles achieve -1 R-parity. 
 \subsection{Split/$\mu$-split Supersymmetry}
The phenomenological models invoke a particular supersymmetry breaking mechanism (along with supersymmetry breaking scale).  Most of these models mainly rely on  ${\cal O}$(TeV) supersymmetry breaking scale because of its ability to resolve serious gauge-hierarchy problem as mentioned above. Though it is possible to achieve gauge coupling unification and to obtain a good dark matter candidate using ${\cal O}$(TeV) scale-based models, yet the existence of a naturally large supersymmetric contribution to flavor changing neutral currents (FCNC), experimental value of electron dipole moment (EDM) for natural CP violating phase, and dimension-five proton decays are serious issues that can not be solved elegantly.  Moreover, the evidence \cite{lhc1,lhc2} obtained by the ATLAS and CMS experiments at CERN of a SM-like Higgs particle with mass $m_H \sim126$ GeV (for theoretical explanations on 126 GeV Higgs, see: \cite{nath3,godbole2}), also points towards some sort of (fine-tuned) SUSY in order to stabilize the vacuum at this value.

Motivated by obtaining an extremely small cosmological constant (see \cite{tye1}) and the string landscape scenario, the alternative to low-scale SUSY models was proposed by Arkani-Hamed and Dimopoulos (dubbed as `Split SUSY') in  \cite{HamidSplitSUSY} according to which SUSY is broken at an energy scale way beyond the collider search and could be even near the scale of grand unification (GUT). In this approach, the probability to sit in a vacuum corresponding to right amount of cosmological constant becomes higher (number of string vacua depends on different SUSY breaking scale)  in the `string theory landscape' though one has to assume fine-tuning in the Higgs sector. The particle spectrum consists of heavy scalar particles except one Higgs doublet which is finely tuned to be light, and light supersymmetric fermions (possibly gaugino and higgsino) in addition to light SM fermions. The scenario is emerging out to be quite interesting (though it abandons the primary reason for introducing supersymmetry) from the point of view of phenomenology because of the fact that heavy scalars mostly appearing as virtual particles in most of the particle decay studies, help to resolve many diverse issues of both particle physics and cosmology. With this, the model eliminates unrealistic features of MSSM while keeping essential features of the same (gauge coupling unification and possibility of a viable Cold Dark Matter (CDM) candidate) to be inherent. 
 \paragraph{$\mu$-split SUSY Model:} \vskip -0.3in
 A variant of  split SUSY scenario was proposed in \cite{mu split susy} with the motivation to resolve the famous $\mu$-problem.  Despite explaining many unresolved issues of phenomenology in the context of split SUSY, the notorius ${\mu}$-problem still remains unsolved according to which the stable vaccum that spontaneously breaks electroweak symmetry requires ${\mu}$ to be of the order of supersymmetry breaking scale. In case of split SUSY scenario one is assuming ${\mu}$ to be light while
 supersymmetry breaking scale to be very high. The other alternative to alleviate the $\mu$-problem has been discussed in \cite{musplitsusy} by further splitting the split SUSY by raising the $\mu$-parameter to a large value (which could be about the same as the sfermion mass or the SUSY breaking scale); this scenario is dubbed as ${\mu}$-split SUSY scenario.  In other words, at electroweak minima, radiative electroweak symmetry breaking condition, 
 \begin{equation}
 \frac{1}{2}M_z^2 = \frac{m_{H_2}^2- m_{H_1}^2{\tan^2{\beta}}}{1-{\tan^2{\beta}}}- \mu^2,
 \end{equation} can be satisfied naturally by assuming partial cancellation between high values of  ($m_{H_1}^2, m_{H_2}^2, \mu^2$) with some amount of fine-tuning. In addition to solving the $\mu$-problem, all the nice features of split supersymmetric model like gauge coupling unification, dark matter candidate, remain protected in this scenario.
     
To summarize, we briefly highlight the signatures as well as features of split SUSY below:
\begin{itemize}
\item[--] Gauge coupling unification still same as low-scale SUSY models \cite{habaokada,Kokorelis_1};
\item[--] A finely tuned Higgs of around ${\cal O}(100)$GeV  \cite{HamidSplitSUSY,Benakli};
\item[--] Heavy masses of squarks, sleptons and one of the Higgs doublets, heavy higgsino, and light gauginos \cite{Giudice_masses,sengupta1,sengupta2, musplitsusy};
\item[--] Long life time of gluino due to presence of heavy scalars existing as propagators in tree-level as well as one-loop decay diagrams of the same \cite{gambino_et_al,kCheung,Manuel_Toharia};
\item[--] 
Dimension-five proton decay get naturally suppressed due to ultra heavy scalar masses \cite{aspectssplitsusy};
\item[--]
Contribution to electron dipole moment (EDM) arising at two loop reach the order of current experimental limits ignoring the effect of CP violating phase as compared to the case of low energy supersymmetric models \cite{giudice_splitsusy, chang_splitsusy,godbole1};
\item[--]
Possibility of lightest neutralino/gravitino/axion as a potential Cold Dark Matter (CDM) candidate, depending on the model-building approach \cite{aspectssplitsusy, sudhirkumargupta, Fweng, masip, Provenza}.
 \end{itemize}
 {\emph{Our basic goal in the string particle cosmology part of the review article is to get a natural realization of $\mu$-split-like  supersymmetry scenario in the context of a phenomenological model embedded in a particular class of string compactification schemes.}}
\vskip -0.5in
 \section{String Phenomenology Background}
 The idea of supersymmetry breaking in MSSM/2HDM models lead to around a hundred
free parameters which introduce new CP violations phases, mass patterns, mixing angles etc., and a big flavor problem due to appearance of FCNC's. The parameters are fixed by hand so as to fit to the experimental results. In fact supersymmetry, though able to provide a candidate for CDM, does not generate the mass of the same naturally. The WIMP paradigm fixes the mass scale of O(TeV) in order to obtain the thermal cross-section to be able to produce relic abundance of the order 0.1. Similarly, there is nothing clear about the nature of scalar field as well as a particular mechanism to produce the inflation potential. As a consequence, both cosmology and particle physics point out the need  to embed (MS)SM into a more fundamental theory, a complete theory explaining all physical phenomena including constants, which can give a dynamical mechanism to explain apparently imaginative choices of these fundamental parameters. String theory \cite{Polchinski,gsw,becker} appears to be the most serious candidate for a fundamental theory which incorporates in principle the constituents of the SM and cosmology. 

Historically,  the concept of string theory came into picture in the late 1960's in an attempt to describe hadron forces and their spectrum with the understanding that specific particles correspond to specific oscillation modes (or quantum states) of the string.  This proposal emerged as a satisfying unified picture in that it postulates a single fundamental object (namely, the string) to describe the different observed hadrons. Though this area remained quite active in the initial stages, it was abandoned only after few years of its discovery because it encountered serious theoretical difficulties in describing the strong nuclear forces as a result of which QCD came along as a convincing theory of the strong interaction. However, several interesting breakthroughs were made in the mid-seventies. In 1971 Neveu, Schwarz and Ramond discovered the idea about how to supplement fermionic degrees of freedom in string theory and emphasized the requirement of world-sheet supersymmetry which led to the development of  space-time supersymmetry.  This space-time supersymmetry was eventually considered to be a generic feature of consistent string theories - hence the name superstring theory. In 1974, Scherk and Schwarz proposed that  problems of string theory could be turned into virtues if it were used for more ambitious purpose of construction of a quantum theory that unifies the description of gravity and other fundamental forces of nature, rather than a theory to describe hadrons and strong nuclear forces. The massless spin-2 particle in the string spectrum, was identified as the graviton and shown to interact at low energies with interaction strength given by the Planck scale. In principle, this fascinating theory has rich potential to provide a complete understanding of particle physics and of cosmology because of its fundamental and mathematically structured character. 

These strings (open or closed) are mathematical one-dimensional extended objects, characterized by a single parameter given by the string tension $\alpha'$, and the typical energy scale of the same given by 1/$\alpha'$. The interactions does not occur at a single point, but is smeared out
into an area already encoded in the world-sheet topology. At energies much below the string scale (corresponding to $\alpha' \rightarrow 0$), the string diagrams diminish to the usual field theoretic ones.
The theory is advantageous in that it manages to eliminate the strong divergences of graviton scattering amplitudes in field theory by replacing the notion of point particles by strings. 
 However, consistent formulation of a stable and
tachyon-free string theory demands a spacetime
supersymmetry and predict a ten-dimensional spacetime at weak coupling (six of which need to be compactified) instead of usual four-dimensional space-time.   
There are five  perturbative superstring
theories in ten-dimensional spacetime, termed Type IIA, Type IIB, Heterotic $E8 × E8$,
Heterotic SO(32) and Type I.  Beginning in the mid-nineties, Witten realized
that all these five different string theories shouldn't really be regarded as distinct theories. These might stem from a eleven-dimensional theory called M-theory \cite{Polchinski,gsw,becker} which has eleven-dimensional supergravity as its low energy
description.   
From a phenomenological point of view, the essential aim is to understand how the SM
or the MSSM may be obtained as a low energy limit of string theory; for nice review articles on string phenomenology, see: \cite{Antoniadis,dienes,ibanez}. The primary approach in this direction is to first reduce String/M-theory from ten(eleven) to four spacetime dimensions, the so called compactification; for nice papers and review articles on this topic, see: \cite{Giryavets+Kachru,grana,Kachru,Blumenhagen,shiu3}.
 To preserve Poincar\'{e} invariance in four dimension spacetime,
the ten (eleven)-dimensional metric is assumed to be a (possibly warped) product of the form
$\mathbb{M}^{9,1}= \mathbb{R}^{3,1} \times X$, where $X$ could be a very complicated Calabi-Yau(CY) manifold(orientifold)\cite{greene,Hubsch}. In reality, the number of choices for the Calabi-Yau manifolds is not known. The extra dimensions, when compactified on Calabi-Yau manifolds, are studied via the shape as well the size deformations of the internal space which give rise to set of massless neutral scalar fields known as moduli. For a typical compactification, the number of moduli are of the order of several hundreds.  In general,  the geometric moduli of a CY are classified into $h_{1,1}$ K\"{a}hler moduli and $h_{2,1}$ Complex Structure moduli. To obtain an effective 4D theory, it is of primary importance to stabilize the potential by giving these moduli vacuum expectation values (VEVs). This problem goes under the name of moduli stabilization. The cosmological moduli problem puts a constraint over the moduli masses according to which \cite{cosmod1,cosmod2}, $m_{mod} >10 \ TeV$. By far, the ${\cal N}=1$ compactification is more important because of its tendency to introduce the chiral spectrum of
massless fermions at low energies. Historically,  the compactification process has made a reasonable progress in heterotic  $E8 \times E8$ string theory because of its ability to produce ${\cal N}=1$ supersymmetry \cite{donagi}. However, it could not get a considerable attention because of subtle issues related to moduli stabilization, and presence of unfavourable exotic matter and anomalous extra $U(1)$'s while constructing Standard Model-like spectrum.  Amongst the four string theories, Type IIB string theory has received so much interest worldwide after the discovery of Dirichlet(D)-branes \cite{Dbranes}. In the presence of D-branes,  Type IIB/IIA theories have potential to evade no-go theorem of \cite{nogo}, and provide a new insight of generating non-Abelian gauge symmetries and chirality after compactifying on a Calabi-Yau orientifold. 

 Type IIB string compactifications appear to be, at present, the most promising way to link string theory with an effective low energy physics producing (MS)SM, due to the presence, within its framework, of reasonable solutions both to stabilize the moduli and to build local SM-like constructions. Type IIB superstring theories compactified on a Calabi-Yau three-fold results in ${\cal N} = 2$ supersymmetric theory in four dimensions \cite{greene}. The number of supersymmetry can be reduced from ${\cal N} = 2$ to ${\cal N} = 1$ by adding an appropriate set of D-branes. Nevertheless, the consistency requires the presence of appropriate orientifold planes (non-dynamical extended objects in string theory),  in contrast to D-branes, to cancel tadpoles originating from space-time filling D-branes. The essential aim in Type IIB string models is to obtain 4D effective ${\cal N}=1$ theory through orientifolding as well as dynamics of D-brane solutions by compactifying on a Calabi-Yau orientifold\footnote{One has to take care of the fact that  it is not the complete spectrum of Calabi-Yau harmonic forms that survives under orientifold symmetries. The orientifold action of Calabi-Yau as well as the states invariant under the same is described in detail in \cite{Jockers_thesis}.}.  The SM, or any of its possible generalisations, is believed to live on a stack of space-time filling D-branes, the so-called `brane-world scenario'.  In  Type IIB string theory one generally considers D1-, D3-, D5-, D7- and D9-branes as sources of RR $C_{2p+1}$ IIB forms. In addition to this, the non-trivial fluxes are turned on the internal geometry to  break ${\cal N} = 1$ supersymmetry spontaneously, and to stabilize the moduli fields except the K\"{a}hler moduli. However, the fluxes are tuned in such a way that they render the compactification manifold to be conformally Calabi-Yau \cite{grana}.  

From the point of view of cosmology/astrophysics, the realistic models obtained as an effective low energy description of Type IIB ${\cal N}=1$ supergravity should be such as to keep up with the primary requirement of getting de-Sitter (dS) vacua with a small cosmological constant. The first de-Sitter like vacuum in string theory framework  was obtained in KKLT compactification \cite{Kachru_et_al} which include additional non-perturbative effects (generated by an ED3-instanton or the gaugino condensation) in the superpotential. The non-perturbative term is responsible to violate no-scale structure and stabilize the K\"{a}hler (volume) modulus in a controlled manner if flux-dependent Gukov-Vafa-Witten superpotential is finely-tuned to be extremely small. The transition from the Anti de-Sitter (AdS) vacuum to supersymmetric de-Sitter (dS) vacuum is realized via a particular uplifting mechanism
such as by including an additional anti-D3-brane. Following a similar approach, another interesting class of compactification models have been proposed in \cite{lvs} in the context of Type IIB orientifold  with inclusion of perturbative $\alpha'^{3}$-corrections (of \cite{pertalpha}) in the K\"{a}hler potential. These perturbative corrections $\alpha'^{3}$-corrections as well as string-loop corrections provide a competing contribution to non-perturbative ones, thereby generating a possibility to give non-supersymmetric metastable AdS minima even without fine-tuning flux-dependent superpotential, but at exponentially large volume. This construction is referred to as LARGE Volume Scenarios (LVS). However, as in KKLT scenario, the de-Sitter vacuum in LVS models is also realized by including an additional ingredients (such as anti-D3-brane) used to uplift the vacuum from AdS to dS.  A consistent local compactification scheme, resulting in the de-Sitter metastable non-supersymmetric minima without introducing any additional uplifting term, has been provided in \cite{dSetal}  by including ``non-perturbative" $\alpha$'-corrections (coming from the world-sheet instantons) in the K\"{a}hler potential \cite{nonpert}. 

For addressing the phenomenological issues, one
considers a space-time filling D3-brane and multiple fluxed stacks of space-time filling D7-branes
wrapping a single four-cycle. In the context of intersecting brane world scenarios \cite{blumenhagen_ints,ibanez_ints,shiu1}, bifundamental leptons and quarks are obtained respectively from open strings stretched between U(2) and U(1) stacks, and U(3) and U(2) stacks of D7-branes; the adjoint gauge fields correspond to open strings starting and ending on the same D7-brane. With the local model-building approach of Large Volume Scenarios, the realistic constructions reproducing SM spectrum via D-branes require wrapping of D7-branes around blown-up cycle(s), similar to the technique used in models of branes at singularities (\cite{Verlinde,Aldazabal,conlon,trivedi3} and references therein). Here, one considers four stacks of different numbers of multiple D7-branes wrapping  a single divisor volume modulus (the "big" divisor $\Sigma_B$ in a Swiss-Cheese Calabi-Yau in\cite{D3_D7_Misra_Shukla}) but with different choices of
magnetic U(1) fluxes turned on, on the two-cycles which are non-trivial in the homology of divisor volume and not the ambient Swiss-cheese Calabi-Yau. By using the appropriate ${\cal N} = 1$ coordinates as obtained in \cite{Jockers_thesis} due to the presence of a single D3-brane and a single D7-brane wrapping the four-cycle (big divisor $\Sigma_B$ in a swiss-Cheese Calabi-Yau) along with D7-brane fluxes, the soft SUSY breaking parameters are calculated in \cite{Sparticles_Misra_Shukla}. The value of scalar masses so-obtained turns out to be quite high, thus indicating possibility of ``split SUSY-like scenario" in a ``large volume local D3/D7 model".  The split SUSY scenarios in the string landscape are getting considerable attention because of the absence of sign of squark and gluinos below 1 TeV at LHC. Recently, the possibility of a stable electroweak minima consistent with high scale SUSY breaking is explained in \cite{ibanez2} in the context of Type IIB orientifold/F-theory unified models. The signatures of split SUSY-like scenarios are also observed in the context of Type I, Type IIA string theoretic models (see \cite{type1string},\cite{Kokorelis}) as well as in orbifold GUT-based model \cite{utpal1}.{\emph{With this motivation, we construct a model, and show a possible local realization of large volume $D3/D7$ $\mu$-split-like  SUSY scenario which includes (non-)perturbative $\alpha^{\prime}$ corrections in the K\"{a}hler potential and non-perturbative instanton-corrections in the superpotential.}}
 \vskip -0.5in
\section{Plan of the Review Article}This review article has two parts. The first is concerned with the investigation of various phenomenological implications of
a ``local  Type IIB large volume D3/D7 $\mu$-split-like  supersymmetry model" to be able to produce order of
magnitude estimates which could be matched directly or indirectly with experiments. The second (smaller) part deals with the investigation of  thermodynamical and hydrodynamical aspects of local eleven-dimensional uplifts
of warped deformed conifold geometries relevant to the studies of thermal QCD-like theories.
The organization of the review article is as follows:

In {\bf {{Chapter 2}}}, we  describe the details of a phenomenological model which could possibly be obtained as a local  Type IIB Swiss-Cheese Calabi-Yau orientifold compactification in the large volume limit with a mobile space-time filling D3-brane restricted to a nearly special Lagrangian cycle in the Calabi-Yau and fluxed stacks of wrapped D7-branes and provides a natural realization of $\mu$-split-like  SUSY \cite{Dhuria+Misra_mu_Split_SUSY,gravitino_DM}. Further, we also argue that for the specific choice of values of VEV of position as well as Wilson line moduli and bulk moduli, there is a possibility to obtain a local meta-stable dS-like minimum corresponding to a positive definite potential. Using the modified  ${\cal N} = 1$ chiral coordinates due to presence of D3- and D7-branes, we construct the form of K\"{a}hler potential and superpotential relevant to generate mass scales of Standard Model as well as supersymmetric particles. By evaluating all possible effective Yukawa couplings in the context of ${\cal N} = 1$ gauged supergravity and subsequent Dirac mass terms,  we suggest the possibility of identification of fermionic superpartners of four Wilson line moduli with first generation leptons $e_L$ and $e_R$, and first generation quarks: $u/d_L$ and $u/d_R$.  The identification is used throughout to address other phenomenological implications (discussed in Chapters 2-4). Using an appropriate SUSY breaking mechanism, the mass scales of supersymetric as well as soft SUSY breaking parameters are obtained.  The high value of Wilson line moduli
masses (typically of the order of $10^{11}$- $10^{12}$ GeV respectively) corresponding to sleptons/squarks and D3-brane position moduli mass corresponding to both Higgs doublets  provide us with a concrete signature of split SUSY. To implement the split-SUSY proposal of N. Arkani-Hamed and S. Dimopoulos [2004], we calculate the mass of light Higgs formed by linear combination of two
Higgs doublets at electroweak scale. By assuming a small amount of fine-tuning, the eigenvalues of Higgs mass matrix translate into getting a light Higgs of order 125 GeV at EW scale while masses of another Higgs as well as higgsino mass parameter turn out to be very heavy even at EW scale, thus showing the possibility of realizing $\mu$- split SUSY scenario in the context of local  Type IIB LVS D3/D7 set-up. As an important benchmark of $\mu$-split-like  supersymmetry, we estimate life time of gluino using ${\cal N} = 1$ gauged supergravity action and estimate the same to be very high- thus providing another evidence of
$\mu$-split-like  SUSY. 
 
{\bf {{Chapter 3}}} involves some of the important issues related to particle cosmology in the context of string-inspired models. In particular,  we demonstrate the possibility of gravitino as a viable C(old) D(ark) M(atter) candidate respecting BBN constraints as well as reproducing a relic abundance of around 0.1 in the context of ${\cal N}=1$ gauged supergravity limit of $D3/D7$ $\mu$-split-like  SUSY model \cite{gravitino_DM}. The local $D3/D7$ $\mu$-split-like  SUSY model  provides gravitino as a L(ightest) S(upersymmetric) P(article) and lightest neutralino as well as sleptons and squarks as N(ext)-to-the-L(ightest) S(upersymmetric) P(article)'s.  With the assumption that sizable amount of gravitinos  are produced non-thermally by decay of NLSP,  life time of NLSP decay into the LSP should be such that it does not disturb the beautiful predictions of Big Bang nucleosynthesis (BBN). In this spirit, we carry out calculations corresponding to decay width of all important two- and three-body decay channels of neutralino as well as sleptons existing as co-NLSP's and gravitino existing as LSP in our model to explicitly show that (i)  the life time obtained in case of all (co)-NLSP decay channels is less than $10^2$ sec (onset of BBN era) in conformity with BBN constraints (ii) the life time for direct R-parity violating neutralino three-body decay into ordinary SM particles is longer than neutralino decays into the LSP which accentuate the fact that neutralino preferably decays into gravitino (LSP) only, thereby ensuring that the gravitino relic abundance is not diluted (iii) decay life time of gravitino turn out to be of the order of age of universe, thus showing possibility of being considered as a viable CDM candidate, (iv)  relic abundance of gravitino turns out to be around 0.1. For the last condition to get satisfied, we follow up the non-thermal production mechanism of gravitino according to which relic abundance of gravitino is given in terms of relic density of NLSP if co-NLSP's freeze out with appropriate thermal relic density before decaying
and then eventually decay into the gravitino. Since the freeze
out condition depends on thermal cross-section of such particles, we evaluate all
the important annihilation channels possible in case of neutralino annihilation and in the case
of slepton annihilation in the context of ${\cal N} =1$
gauged supergravity action. By satisfying the above requirements, we demonstrate the possibility of gravitino as a viable CDM candidate in our local $D3/D7$ $\mu$-split-like  SUSY model.

{\bf {{Chapter 4}}} encompasses the study related to obtaining order of magnitude estimates of electric dipole moment (EDM) of electron as well as neutron in the context of local $D3/D7$ $\mu$-split-like  SUSY model \cite{dhuria+misra_EDM}. By discussing the possible origin of generation of non-zero complex phases in above-mentioned
model, we emphasize the possibility of obtaining a sizable contribution of (electron/neutron)
EDM even at one-loop level due to presence of heavy supersymmetric fermions nearly isospectral
with heavy sfermions in addition to getting a significant value of (e/n) EDM at two-loop level. The
independent CP violating phases are generated from non-trivial distinct phase factors associated
with four Wilson line moduli (identified with first generation leptons and quarks and their
$SU(2)_L$-singlet cousins) as well as the D3-brane position moduli (identified with two Higgses)
and the same are sufficient to produce overall distinct phase factors corresponding to all possible
effective Yukawa's as well as effective gauge couplings that we have discussed in the context
of ${\cal N} = 1$ gauged supergravity action. However, the complex phases responsible to generate
non-zero EDM at one-loop level mainly appear from off-diagonal contribution of sfermion as
well as Higgs mass matrices at electroweak scale.  In our analysis, we obtain
dominant contributions of electron/neutron EDM around $d_{e}/e \sim {\cal O}(10^{-29})$ cm from two-loop
diagrams involving heavy sfermions and a light Higgs, and ${d_e}/e \sim {\cal O}(10^{-32})$ cm from one-loop
diagram involving heavy chargino and a light Higgs as propagators in the loop. The neutron
EDM get a dominant contribution of the order $d_{n}/e \sim {\cal O}(10^{-33})$ cm from one-loop diagram involving SM-like quarks and Higgs. Next, by conjecturing that the CP violating phases can appear from the linear combination of Higgs doublets obtained in the context of $\mu$-split-like  SUSY,
we also get an EDM of electron/neutron around ${\cal O}(10^{-27})$ esu-cm in case of two-loop diagram
involving $W^{\pm}$ bosons.

 {\bf {{Chapter 5}}} comprises study of few thermodynamical as well as hydrodynamical aspects relevant to the    thermal QCD by obtaining local M-theory uplift of Type IIB background involving non-compact (resolved) warped deformed conifold and (N) space time filling D3-branes placed on the singularity of the conifold, (M) D5-branes wrapped around compact $S^2$, and ($N_{f}$) D7-branes wrapped around non-compact four-cycle via the Ouyang embedding in the presence of a black-hole; Ouyang-Klebanov-Strassler(OKS)-Black-Hole(BH) background \cite{MQGP}. By satisfying the requirements of implementing SYZ mirror symmetry locally (as the resolved warped deformed conifold does not possess a `third' global killing isometry along  the `original' angular variable $\psi\in[0,4\pi]$ for implementing SYZ mirror symmetry) to obtain Type $IIA$ background (near $(\theta_{1,2},\psi)=(\langle\theta_{1,2}\rangle,\left\{0,2\pi,4\pi
\right\})$), we oxidize the so-obtained Type $IIA$ background to M-theory  and then argue that there exists a new `MQGP limit' ($\frac{g_sM^2}{N}<<1,g_sN>>1$ for finite $(M,g_s)$ more relevant to strongly-coupled QGP) not considered before. After obtaining so, we set up an approach to study the behaviour of hydrodynamical as well as thermodynamical quantities in both weak coupling but large t'Hooft coupling regime of M-theory ($g_s<<1$) as well as MQGP limit  accomplished by letting ($g_s \lesssim1$). For both limits we obtain a black M3-brane solution whose near-horizon geometry near the $\theta_{1,2} = 0$, branches, preserve $\frac{1}{8}$ supersymmetry. Then we investigate thermodynamical stability of solution in  Type IIB background as well as its M-theory uplift. The thermodynamical stability of the M-theory uplift is discussed by first evaluating finite 11D supergravity action and thereafter showing positivity of specific heat. On the other hand,  we evaluate hydrodynamical quantities such as shear viscosity-to-entropy ratio, and diffusion constant which naturally turns out to $\frac{1}{4 \pi}$ and $\frac{1}{2 \pi T}$ respectively in Type IIB,  Type IIA and its local M-theory uplift. 

Finally, {\bf {{Chapter 6}}} summarizes the main results and conclusions of
the work done in this review article along with interesting
future directions. An appendix is included.

\chapter{Phenomenological Model: Realization of $\mu$-split-like SUSY Scenario}
\vskip -0.5in
{\hskip1.4in{\it ``Do there exist many worlds, or is there but a single world? This is one of the most noble and exalted questions in the study of Nature."}}

\hskip3.0in - Albertus Magnus, c. 13th Century.

\graphicspath{{Chapter2/}{Chapter2/}}
\vskip -0.5in
\section{Introduction}
Over the past few years, producing realistic models satisfying both cosmological as well as phenomenological requirements from string compactifications, has proven to be a daunting challenge. To get the phenomenological implications of BSM models, they must implicate a particular SUSY breaking mechanism and hence scale of SUSY breaking. The recently proposed split SUSY model (based on high SUSY breaking scale) inspired by the need of fine-tuning the cosmological constant, is emerging to be quite interesting from the point of view of phenomenology because of the fact that heavy scalars mostly appearing as virtual particles in most of the particle decay studies, help to resolve many diverse issue of both particle physics and cosmology \cite{HamidSplitSUSY}. Therefore, it is quite interesting to realize split SUSY scenario within the string-theoretic framework. One of the interesting string compactification schemes, which eventually gives a metastable dS minima naturally, was presented in \cite{dSetal} by considering Type IIB compactified on the orientifold of a Swiss-Cheese Calabi-Yau in the LVS limit that includes non-(perturbative) corrections and non-perturbative instanton-corrections in superpotential. Further, the possibility of generating light fermion masses as well as heavy sfermion masses by including a space-time filling mobile $D3$-brane and stack(s) of (fluxed) $D7$-branes wrapping the ``Big" divisor in the aforementioned framework \cite{ferm_masses_MS}, has interestingly pointed towards the likelihood of split SUSY-like scenario in large volume Type IIB model. Motivated by this, we extend the model to include two-Wilson line moduli\cite{Dhuria+Misra_mu_Split_SUSY}, followed by a systematic four-Wilson line moduli model\cite{gravitino_DM}, and explicitly show a possible local realization of $\mu$-split-like SUSY scenario in the large volume limit by including D3/D7-branes.

Before elaborating upon the discussion of $\mu$-split-like  SUSY realization, in section {\bf 2}, we first provide details of a phenomenological model which could possibly be realized in a string-theoretic framework by considering, locally, the Swiss-Cheese Calabi-Yau type IIB string compactification in the large volume limit with a mobile space-time filling D3-brane restricted to a nearly special Lagrangian three-cycle in the Calabi-Yau and fluxed stacks of wrapped D7-branes. Then we summarize the construction of  four harmonic distribution one-forms localized along the mobile space-time filling $D3$-brane (restricted to the three-cycle). By calculating the intersection matrix valued in the Wilson line moduli sub-space $a_{I=1,...,h^{(0,1)}_-(\Sigma_B)}$, appearing in K\"{a}hler coordinate $T_B$, we write (the K\"{a}hler sector of the) K\"{a}hler potential that includes  four-Wilson line moduli $a_1, a_2, a_3, a_4$ and two-position moduli of a mobile space-time filling $D3$-brane restricted to the above-mentioned (nearly) SLag (corresponding to a local minimum). The estimate of the Dirac mass terms appearing in the ${\cal N}=1$ supergravity action of \cite{Wess_Bagger} calculated from superpotential and K\"{a}hler potential suggest that the fermionic superpartners of ${\cal A}_1$ and ${\cal A}_3$ correspond respectively to  the first generation leptons: $e_L$ and $e_R$, and the fermionic superpartners of ${\cal A}_2$ and ${\cal A}_4$ correspond respectively to the first generation quarks: $u_L$ and $u_R$.  We use this identification throughout  to discuss all phenomenological issues such as evaluation of soft SUSY breaking parameters as well as estimates of various three-point vertices in the context of gauged ${\cal N}=1$ supergravity action. Next, due to multiple D7-branes, we provide explicit bi-fundamental representations for the four-Wilson line moduli and two $D3$-brane position moduli (after having turned on appropriate two-form fluxes on the $D7$-brane world volume decomposing adjoint-valued matter fields into bi-fundamental matter fields) and speculate about the possible modification in the relevant ${\cal N}=1$ chiral coordinate in the presence of the same. It is also shown that the (effective) physical Yukawas change only by ${\cal O}(1)$ under an RG-flow from the string scale down to the EW scale. The discussion of phenomenological results  governing signatures of $\mu$-split-like  SUSY in this framework are  appended in sections {\bf 3}-{\bf 5}.  Section {\bf 3} has details of calculations of various soft SUSY breaking parameters.  In section {\bf 4}, we describe the fact that it is possible to obtain $125\ GeV$ mass for one of the Higgs doublets at EW scale. Section {\bf 5} discuss evaluation of neutralino/chargino mass matrix. In section {\bf 6}, we present a detailed study of gluino decay life time at tree-level as well as at one-loop level in the context of ${\cal N}=1$ gauged SUGRA. Finally we
summarize the chapter in section {\bf 7}.
 \vskip -0.5in
\section{The Model}
 Let us first briefly describe our model for which we present evidence that
it could be locally realized as a large volume $D3/D7$ Swiss-Cheese setup in
 \cite{gravitino_DM}. For an ${\cal N}=1$ compactification, we will take the phenomenological K\"{a}hler potential of our model to be:
\begin{eqnarray}
\label{K_pheno}
& & \hskip -0.5in K_{\rm Pheno} = - ln\left[-i(\tau-{\bar\tau})\right]
-ln\left(-i\int_{CY_3}\Omega\wedge{\bar\Omega}\right)\nonumber\\
& & \hskip -0.5in - 2\ ln\Biggl[a_B(\sigma_B + {\bar\sigma}_B - \gamma K_{\rm geom})^{\frac{3}{2}} - (\sum_{i}a_{S,i}(\sigma_{S,i} + {\bar\sigma}_{S,i} - \gamma K_{\rm geom}))^{\frac{3}{2}} + {\cal O}(1) {\cal V}\Biggr],
 \end{eqnarray}
where the divisor volumes $\sigma_\alpha$ are expressible in terms of ``K\"{a}hler" coordinates $T_\alpha, {\cal M}_{\cal I}$
\begin{eqnarray}
\label{sigma}
& & \sigma_\alpha\sim T_\alpha -\left[ i{\cal K}_{\alpha bc}c^b{\cal B}^c
+ i C^{{\cal M}_{\cal I}{\bar{\cal M}}_{\bar {\cal J}}}_\alpha({\cal V})Tr\left({\cal M}_{\cal I}{\cal M}^\dagger_{\bar {\cal J}}\right)\right],
\end{eqnarray}
$\alpha=\left(B,\{S,i\}\right)$ and ${\cal M}_{\cal I}$ being
  $SU(3_c)\times SU(2)_L$  bifundamental  matter field $a_{{\cal I}=2}$, $SU(3_c)\times U(1)_R$ bifundamental matter field
$a_{{\cal I}=4}$, $SU(2)_L\times U(1)_L$ bifundamental matter field $a_{{\cal I}=1}$,
$U(1)_L\times U(1)_R$ bifundamental matter field $a_{{\cal I}=3}$ along with
$SU(2)_L\times U(1)_L$ bifundamental $\tilde{z}_{1,2}$
 with the intersection matrix:
$ C^{a_I{\bar a}_{\bar J}}_\alpha\sim \delta^B_\alpha C^{I{\bar J}}_\alpha, C^{a_I\bar {\tilde{z}}_{\bar j}}_\alpha=0$, $\rho_{S,B}, {\cal G}^a = c^a - \tau b^a$ being complex axionic
fields ($\alpha,a$ running over the real dimensionality of a sub-space of the internal manifold's cohomology complex),  and the phenomenological  superpotential is given as under:
\begin{equation}
\label{W_pheno}
 W_{\rm Pheno}\sim\left(z_1^{18} + z_2^{18}\right)^{n^s}
e^{-n^s vol(\Sigma_S) - (\alpha_S z_1^2 + \beta_S z_2^2 + \gamma_S z_1z_2)},
\end{equation}
\noindent where the bi-fundamental $\tilde{z}_i$ in $K$ will be equivalent to the $z_{1,2}\in\mathbb C$ in $W$. It is
expected that ${\cal M}_{\cal I}, T_{S,B}, {\cal G}^a$ will constitute the ${\cal N}=1$ chiral coordinates. The
 intersection matrix elements $\kappa_{S/B ab}$ and the volume-dependent $C^{{\cal M}_{\cal I}{\bar{\cal M}}_{\bar {\cal  J}}}_\alpha({\cal V})$, are chosen in such a way that at a local (meta-stable) minimum:
 \begin{eqnarray}
 \label{extremum_i}
& &  \langle \sigma_S\rangle \sim \langle (T_S + {\bar T}_S)\rangle  - i C^{\tilde{z}_i\bar {\tilde{z}}_{\bar j}}({\cal V}) Tr\left(\langle\tilde{z}_i\rangle \langle \bar {\tilde{z}}_{\bar j}\rangle\right)\sim {\cal O}(1)
\nonumber\\
& &  \langle \sigma_B\rangle \sim \langle (T_B + {\bar T}_B)\rangle  - i C^{\tilde{z}_i\bar {\tilde{z}}_{\bar j}}({\cal V}) Tr\left(\langle\tilde{z}_i\rangle \langle\bar {\tilde{z}}_{\bar j}\rangle\right)
- i C^{a_I{\bar a}_{\bar J}}({\cal V}) Tr\left(\langle a_I\rangle \langle{\bar a}_{\bar J}\rangle\right)\sim e^{f \langle \sigma_S\rangle},\nonumber\\
& &
 \end{eqnarray}
 where $f$ is a fraction not too small as compared to 1, and the stabilized values of $T_\alpha$ around the meta-stable local minimum:
 \begin{equation}
 \label{extremum_ii}
 \langle\Re e T_S\rangle,\langle\Re e T_B\rangle\sim {\cal O}(1).
 \end{equation}
 In the context of ${\cal N}=1$ type IIB orientifolds, in (\ref{extremum_i}), $\alpha,a$ index respectively involutively even, odd sectors of $h^{1,1}(CY_3)$ under a holomorphic, isometric involution.
If the volume ${\cal V}$ of the internal manifold is large in string length units, one sees that one obtains a hierarchy between the stabilized values $\langle\Re e\tau_{S,B}\rangle$ but not $\langle\Re e T_{S,B}\rangle$.

We now argue how to realize (\ref{K_pheno}) - (\ref{extremum_ii}), locally, in string theory.
Consider type IIB compactified on the orientifold of a Swiss-Cheese Calabi-Yau
in the LVS limit that includes non-(perturbative) $\alpha^{\prime}$ corrections
 and non-perturbative instanton-corrections in superpotential. For phenomenological purposes, we will consider a space-time filling $D3$-brane
and multiple fluxed stacks of space-time filling $D7$-branes wrapping a single
four-cycle, the big divisor, with different choice of small
two-form fluxes turned on the different two-cycles homologously non-trivial from the point of view of this four-cycle's
Homology (for the purpose of decomposing initially adjoint-valued matter fields
to bi-fundamental matter fields, for generating the SM gauge groups and
to effect gauge-coupling unification at the string scale). Then  $z_{1,2}$ get identified with the $D3$-brane's position moduli,
$\tau$ is the axion-dilaton modulus and ${\cal G}^a$ are NS-NS and RR two-form axions complexified by the axion-dilaton modulus.

From Sen's orientifold-limit-of-F-theory point of view  corresponding to type IIB compactified on a Calabi-Yau three fold $Z$-orientifold with $O3/O7$ planes, one requires an elliptically fibered Calabi-Yau four-fold $X_4$ (with projection $\pi$) over a 3-fold $B_3(\equiv CY_3-$orientifold)  where $B_3$ can be taken to be  an $n$-twisted ${\mathbb {CP}}^1$-fibration over ${\mathbb {CP}}^2$ such that pull-back of the divisors in $CY_3$ automatically satisfy Witten's unit-arithmetic genus condition \cite{DDF}.  For $n=6$ \cite{DDF}, the Calabi-Yau three-fold $Z$ then turns out to be a unique Swiss-Cheese Calabi Yau  in ${ \mathbb{WCP}}^4_{[1,1,1,6,9]}[x_1:x_2:x_3:x_4:x_5]$ given by a smooth degree-18 hypersurface in ${ \mathbb{WCP}}^4[1,1,1,6,9]$; the exceptional divisor corresponding to resolution of a ${\mathbb  Z}_3$-singularity $x_1=x_2=x_3=0$, via the monomial-divisor map, is encoded as the $\phi x_1^6x_2^6x_3^6 (\phi$ being an involutively odd complex structure modulus) polynomial deformation in the defining hypersurface. Henceforth, we will be working in the non-singular coordinate patch $x_2=1$ in the large volume limit of the Swiss-Cheese Calabi-Yau.

The closed string moduli-dependent K\"{a}hler potential, includes perturbative (using \cite{BBHL}) and non-perturbative (using \cite{nonpert}) $\alpha^\prime$-corrections. Written out in (discrete subgroup of) $SL(2,{\bf Z})$(expected to survive orientifolding)-covariant form, the perturbative corrections are proportional to $\chi(CY_3)$ and non-perturbative $\alpha^\prime$ corrections are weighted by $\{n^0_\beta\}$, the genus-zero Gopakumar-Vafa invariants that count the number of genus-zero rational curves $\beta\in H_2^-(CY_3,{\bf Z})$. In fact, the closed string moduli-dependent contributions are dominated by the genus-zero Gopakumar-Vafa invariants which using Castelnuovo's theory of moduli spaces can be shown to be extremely large for compact projective varieties \cite{Klemm_GV} such as the one used.

Based on the study initiated in \cite{ferm_masses_MS,Dhuria+Misra_mu_Split_SUSY}, one of us (AM), inspired by the Donaldson's algorithm - see \cite{Donaldson_i} -  had obtained in \cite{review-iii}, an estimate of a nearly Ricci-flat Swiss Cheese metric for points finitely separated from $\Sigma_B$ based on the ansatz as given in \cite{gravitino_DM}. Using this ansatz, one sees that $h^{11}\sim h^{z_4^2{\bar z}_4^2}{\cal V}^{\frac{2}{3}}$.
Using (\ref{eq:K}), one can show that:
\begin{equation}
\label{eq:R11bar_i}
R_{z_i{\bar z}_j}\sim\frac{\sum_{n=0}^8a_n\left(h^{z_4^2{\bar z}_4^2}\right)^n {\cal V}^{\frac{n}{3}}}{\left(1+{\cal O}(1)h^{z_4^2{\bar z}_4^2} {\cal V}^{\frac{1}{3}}\right)^2{\cal V}^{\frac{1}{18}}\left(\sum_{n=0}^3b_n\left(h^{z_4^2{\bar z}_4^2}\right)^n {\cal V}^{\frac{n}{3}}\right)^2}.
\end{equation}
Solving numerically: $\sum_{n=0}^8a_n\left(h^{z_4^2{\bar z}_4^2}\right)^n {\cal V}^{\frac{n}{3}}=0$, as was assumed, one (of the eight values of) $h^{z_4^2{\bar z}_4^2}$, up to a trivial K\"{a}hler transformation, turns out to be ${\cal V}^{-\frac{1}{3}}, {\cal V}\sim 10^6$. Curiously, from GLSM-based analysis, we had seen in \cite{D3_D7_Misra_Shukla} that on $\Sigma_B(z_4=0)$, the argument of the logarithm received the most dominant contribution from the FI-parameter $r_2\sim{\cal V}^{\frac{1}{3}}$ which using this value of $h^{z_4^2{\bar z}_4^2}$, is in fact, precisely what one obtains even now. Using this value of $h^{z_4^2{\bar z}_4^2}$, one obtains: $R_{z_i{\bar z}_4},R_{z_4{\bar z}_4}\sim10^{-1}$. Further, as has been assumed that the metric components $g_{z_{1,2}{\bar z}_4}$ are negligible as compared to $g_{z_i{\bar z}_j}$ - this was used in \cite{ferm_masses_MS} in showing the completeness of the basis spanning $H^{1,1}_-$ for a large volume holomorphic isometric involution restricted to (\ref{eq:near_slag_i}),
 $z_1\rightarrow-z_1,z_{2,3}\rightarrow z_{2,3}$, which is now born out explicitly, wherein the latter turn out to about 10$\%$ of the former. One does not need the aforementioned restriction on the geometric Swiss-Cheese metric for large volume involutions restricted to (\ref{eq:near_slag_i}), of the following type. In the $x_2\neq0$-coordinate patch, the defining degree-18 hypersurface in ${\mathbb {WCP}}^4_{1,1,1,6,9}[x_1:x_2:x_3:x_4:x_5]$ can be written as: $1 + z_1^{18} + z_2^{18} + z_3^3 - \psi z_1z_2z_3z_4 - \phi z_1^6z_2^6 = 0, z_1=\frac{x_1}{x_2}, z_2=\frac{x_3}{x_2}, z_3=\frac{x_4}{x_1^6}, z_4=\frac{x_5}{x_1^9}$. The same can be rewritten as: $\left(i z_4\right)^2 - \psi z_1z_2z_3z_4 =  z_3^3 + \phi z_1^6z_2^6 - z_1^{18} - z_2^{18}$. This is can therefore be thought of as the following Weierstrass variety:
 $\begin{array}{rrc}T^2(z_3,z_4(z_3)) & \rightarrow&{\mathbb {WCP}}^4_{1,1,1,6,9}(z_1,z_2,z_3)\\
 & & \downarrow\pi \\
 & & {\mathbb CP}^2(z_1,z_2)\\
  \end{array}$. Now, assuming that the complex structure modulus $\psi$ has been stabilized to an infinitesimal value so that one can disregard the polynomial deformation proportional to $\psi$, and defining $\chi_{1,2}\equiv z_{1,2}^6$, this elliptic fibration structure can thought also of as:
  $\begin{array}{rrc}T^2(z_3,z_4(z_3)) & \rightarrow&{\mathbb {WCP}}^4_{1,1,1,6,9}(\chi_1,\chi_2,z_3)\\
 & & \downarrow\pi \\
 \left({\rm Complex\ curve}\right){\cal C}(\chi_1)&\longrightarrow & {\mathbb {CP}}^2(\chi_1,\chi_2)\\
 & & \downarrow\pi^\prime\\
 & & {\mathbb {CP}}^1(\chi_2)
  \end{array}$, corresponding to $\left(i z_4\right)^2\approx z_3^3 + \phi \chi_1\chi_2 - \chi_1^3 - \chi_2^3$. Alternatively, one can also think of the Weierstrass variety as:
  $\begin{array}{rrc}T^2(-\chi_1,z_4(-\chi_1)) & \rightarrow&{\mathbb {WCP}}^4_{1,1,1,6,9}(\chi_1,\chi_2,z_3)\\
 & & \downarrow\pi \\
 \left({\rm Complex\ curve}\right){\cal C}(-z_3)&\longrightarrow & {\mathbb {CP}}^2(-z_3,\chi_2)\\
 & & \downarrow\pi^\prime\\
 & & {\mathbb {CP}}^1(\chi_2)
  \end{array}$, corresponding to $\left(i z_4\right)^2\approx\left(-\chi_1\right)^3 + \phi \chi_1\chi_2 + z_3^3 - \chi_2^3$. Therefore, near (\ref{eq:near_slag_i}), one can define a large volume holomorphic involution: $\chi_1\leftrightarrow-z_3, \phi\rightarrow-\phi$(the fact that complex structure modulus $\phi$ is involutively odd, is also used in (\ref{eq:sigmas_Ts})). We will assume that in the coordinate patch
(but not globally):
$|z_1|\equiv {\cal V}^{\frac{1}{36}},\ |z_2|\equiv{\cal V}^{\frac{1}{36}},\ |z_3|\equiv{\cal V}^{\frac{1}{6}}$, the Calabi-Yau is diffeomorphic to the Swiss-Cheese $\mathbb {{WCP}}^4_{1,1,1,6,9}[18]$. The
defining hypersurface for the same is: $x_1^{18}+x_2^{18}+x_3^{18}+x_4^3+x_5^2
-18\psi\prod_{i=1}^5x_i-3\phi (x_1x_2x_3)^6=0$. This can be thought of as the following
hypersurface in an ambient complex four-fold: $P(x_1,...,x_5;\xi)=0$ after resolution
of the $\mathbb Z_3$-singularity\cite{Rummel_et_al} [the $x_4$ and $x_5$ have been switched
relative to \cite{DDF}; $n=6$ $\mathbb {CP}^1$-fibration over $\mathbb{CP}^2$ with
projective coordinates $x_{1,2,3},x_4,x_5$ of \cite{DDF} is equivalent to $n=-6$ with projective coordinates
$x_{1,2,3,},x_5,x_4$ - see \cite{Denef_Les_Houches}] with the toric data for the same given by:
$$\begin{array}{c|cccccc}
& x_1 & x_2 & x_3 & x_4 & x_5 & \xi \\ \hline
Q^1 & 1 & 1 & 1 & 6 & 0 & 9 \\
Q^2 & 0 & 0 & 0 & 1 & 1 & 2 \\
\end{array}.$$
In the coordinate patch: $x_2\neq0$(implying one is away
from the $\mathbb Z_3$-singular $(0,0,0,x_4,x_5)$ in
$\mathbb{WCP}^4_{1,1,1,6,9}[18]$), $\xi\neq0$, one sees that the following are
the gauge-invariant coordinates: $z_1=\frac{x_1}{x_2}, z_2=\frac{x_3}{x_2},
z_3=\frac{x_4^2}{x_2^3\xi}, z_4=\frac{x_5^2x_2^9}{\xi}.$
We henceforth assume
the Calabi-Yau hypersurface to be written in this coordinate patch as:
$z_1^{18} + z_2^{18} + {\cal P}(z_{1,2,3,4};\psi,\phi)=0$.
The divisor $\left\{x_5=0\right\}\cap\left\{P(x_{1,2,3,4,5};\xi)=0\right\}$
is rigid with $h^{0,0}=1$  (See \cite{Rummel_et_al})  satisfying the Witten's
unit-arithmetic genus condition, and that the Calabi-Yau volume can
be written as
${\rm vol}(CY_3)=\frac{\tau_4^{\frac{3}{2}}}{18} -
\frac{\sqrt{2}\tau_5^{\frac{3}{2}}}{9},$  implying that the `small divisor'
$\Sigma_s$ is $\left\{x_5=0\right\}\cap\left\{z_1^{18} + z_2^{18} + {\cal P}(z_{1,2,3},z_4=0;\psi,\phi)=0\right\}$
and the big divisor $\Sigma_B$ is  $\left\{x_4=0\right\}\cap\left\{z_1^{18} + z_2^{18} + {\cal P}(z_{1,2,4},z_3=0;\psi,\phi)=0\right\}.$ Alternatively, using
the toric data of \cite{candelas_et_al_11169}:
$$\begin{array}{c|cccccc}
& x_1 & x_2 & x_3 & x_4 & x_5 & \xi \\ \hline
Q^1 & 1 & 1 & 1 & 0 & 0 & -3 \\
Q^2 & 0 & 0 & 0 & -2 & -3 & -1 \\
\end{array},$$ one can verify that
$\left\{\xi=0\right\}\cap\left\{P^\prime(x_{1,2,3,4,5};\xi)=0
\right\}$ is the rigid blow-up mode with $h^{0,0}=1$ (which can be easily
verified using cohomCalg\footnote{We thank P.Shukla for verifying the same.})
and one can define gauge-invariant coordinates in the $x_2\neq0,x_4\neq0$
coordinate-patch: $z_1=\frac{x_1}{x_2}, z_2=\frac{x_3}{x_2},
z_3=\frac{(x_5x_1)^2}{x_4^3}, z_4=\frac{(x_6x_1^3)^2}{x_4}.$ \\
Interestingly, around the  three-cycle
\begin{equation}
\label{eq:near_slag_i}
C_3:|z_1|\sim {\cal V}^{\frac{1}{36}},\ |z_2|\sim{\cal V}^{\frac{1}{36}},\ |z_3|\sim{\cal V}^{\frac{1}{6}},
\end{equation}
the Calabi-Yau can be thought of,
locally, as a complex three-fold ${\cal M}_3$ which is a $T^3$(swept out by ($arg z_1$,$arg z_2$, $arg z_3$)-fibration over a large base $(|z_1|,|z_2|,|z_3|)$; precisely apt for application of mirror symmetry as three T-dualities a la S(trominger) Y(au) Z(aslow)), $C_3$ is almost a s(pecial) Lag(rangian) sub-manifold because it satisfies the requirement that  $f^*J\approx0,\ \Re e\left(f^*e^{i\theta}\Omega\right)\biggr|_{\theta=\frac{\pi}{2}}\approx {\rm vol}(C_3),\ \Im m\left(f^* e^{i\theta}\Omega\right)\biggr|_{\theta=\frac{\pi}{2}}\approx0$ where $f:C_3\rightarrow CY_3$). Let us see if these requirements hold up. Using the geometric K\"{a}hler potential  as given in \cite{gravitino_DM}, the geometric metric near $C_3$ is estimated to be:
\begin{equation}
\label{eq:geom met}
g_{i{\bar j}}\sim\frac{{\cal O}(10^{-1})}{2}\left(
\begin{array}{lll}
 \frac{1}{ \sqrt[18]{V}} & \frac{1}{ \sqrt[18]{V}} & \frac{1}{ V^{7/36}} \\
 \frac{1}{ \sqrt[18]{V}} & \frac{1}{ \sqrt[18]{V}} & \frac{1}{ V^{7/36}} \\
 \frac{1}{ V^{7/36}} & \frac{1}{ V^{7/36}} & \frac{1}{ \sqrt[3]{V}}
\end{array}
\right),
\end{equation}
The same was necessary in
explicitly demonstrating $h^{1,1}_-\neq0$ in \cite{ferm_masses_MS} for the large volume involution: $z_1\rightarrow -z_1, z_{2,3}\rightarrow z_{2,3}$. \\
$(i)$ Using (\ref{eq:geom met}), one sees that:
\begin{eqnarray}
\label{eq:J_pull back}
& & f^*(ds_6^2)\sim0.05\left[\sum_{i,j=1}^2{\cal V}^{-\frac{1}{18}}\left(dz_id{\bar z}_{\bar j}\right)+{\cal V}^{-\frac{1}{3}}|dz_3|^2\right]\Biggr|_{C_3}\nonumber\\
& & \sim 0.05\left[\sum_{i,j=1}^2d (arg z_i)d (arg{\bar z}_{\bar j}) + d (arg z_3) d (arg {\bar z}_3)\right].
\end{eqnarray}
This implies $f^*J\sim0.05$, which is small.\\
$(ii)$ In the $x_2\neq0$- coordinate patch with $z_1=\frac{x_1}{x_2}, z_2=\frac{x_3}{x_2}, z_3=\frac{x_4}{x_2^6}, z_4=\frac{x_5}{x_2^6}$, by the Griffiths residue formula, one obtains: $\Omega=\frac{dz_1\wedge dz_2\wedge dz_4}{\frac{\partial P}{\partial z_3}}=\frac{dz_1\wedge dz_2\wedge dz_4}{3z_3^2-\psi z_1z_2z_4}$. The coordinate $z_4$ can be solved for using:
\begin{equation}
\label{eq:x4_i}
z_4=\frac{\psi z_1z_2z_3\pm\sqrt{\psi^2(z_1z_2z_3)^2-4(1+z_1^{18}+z_2^{18}+z_3^3-\phi (z_1z_2)^6)}}{2}\sim \psi {\cal V}^{\frac{2}{9}}\pm i{\cal V}^{\frac{1}{4}},
\end{equation}
which implies that the $T^3$ will never degenerate as the roots will never coincide:
\begin{eqnarray}
\label{eq:x4_ii}
& & dz_4\sim dz_3\left(\frac{\psi z_1z_2}{2} \pm\frac{\left(2\psi^2z_1^2z_2^2z_3 - 12 z_3^2\right)}{4\sqrt{\psi^2(z_1z_2z_3)^2-4(1+z_1^{18}+z_2^{18}+z_3^3-\phi (z_1z_2)^6)}}\right) \nonumber\\
&& \hskip 0.35in+ (...)dz_1 + (...)dz_2 \nonumber\\
&& \hskip 0.2in  \sim\left(\frac{\psi z_1z_2}{2} \pm
  1.5i\frac{z_3^2}{\sqrt{z_1^{18}+z_2^{18}+z_3^3}}\right)dz_3 +
(...)dz_1 + (...)dz_2, \nonumber\\
&&  \hskip 0.35in {\rm near}\ (\ref{eq:near_slag_i}),
|\psi|<<1,|\phi|<<1.
\end{eqnarray}
 Therefore, restricted to (\ref{eq:near_slag_i}) and substituting (\ref{eq:x4_i}) and (\ref{eq:x4_ii}) in the aforementioned expression of $\Omega$, one obtains:
\begin{eqnarray}
\label{eq:Omega}
& & \Omega\sim\frac{\left(\psi {\cal V}^{\frac{1}{18}}\pm 1.5i{\cal V}^{\frac{1}{12}}\right)}{3{\cal V}^{\frac{1}{3}}-\psi {\cal V}^{\frac{1}{18}}\left[\psi {\cal V}^{\frac{2}{9}}\pm i{\cal V}^{\frac{1}{4}}\right]}dz_1\wedge dz_2\wedge dz_3 \sim  \left(\frac{\psi{\cal V}^{-\frac{5}{18}}}{3}\pm 0.5i {\cal
    V}^{-\frac{1}{4}}\right)
dz_1\wedge dz_2\wedge dz_3.\nonumber\\
\end{eqnarray}
Hence, if one could estimate  the pull-back of the nowhere vanishing
holomorphic three-form as:
\begin{eqnarray}
\label{eq:f*Omega}
& & f^*\Omega\sim\left(\frac{\psi}{3}{\cal
    V}^{-\frac{5}{18}+\frac{1}{18}+\frac{1}{6}}+
0.5i{\cal V}^{-\frac{1}{4}+\frac{1}{18}+\frac{1}{6}}\right)d(arg
z_1)\wedge d(arg z_2)\wedge d(arg z_3)\Biggr|_{{\cal V}\sim10^6}
\nonumber\\
& & \sim\left(0.2\psi\pm0.3i\right)d(arg z_1)\wedge d(arg z_2)\wedge d(arg z_3),
\end{eqnarray}
and:
\begin{eqnarray}
\label{eq:vol C_3}
& &  {\rm vol}(C_3)\sim\sqrt{f^*g}f^*(dz_1\wedge dz_2\wedge dz_3)\sim\sqrt{(0.05)^3}{\cal V}^{\frac{2}{9}}d(arg
z_1)\wedge d(arg z_2)\wedge d(arg z_3)\Biggr|_{{\cal V}\sim10^6}\nonumber\\
& & \sim0.2d(arg z_1)\wedge d(arg z_2)\wedge d(arg z_3),
\end{eqnarray}
then relative to a phase $e^{-i\theta},\theta=\frac{\pi}{2}$,
$\Im m\left(f^*e^{-i\theta}\Omega\right)\sim 0.2\psi d(arg z_1)\wedge
d(arg z_2)\wedge d(arg z_3)$, which for $|\psi|<<1$, is close to
zero; (equivalently)$\Re\left(f^*e^{-i\theta}\Omega\right)\sim$
vol($C_3)$. This implies that $C_3$ is nearly/almost a special Lagrangian
sub-manifold
\cite{Kachru+McGreevy_slag}.

The ${\cal N}=1$ chiral co-ordinates  with the inclusion of mobile $D3$-brane
position moduli $z_{1,2}$ (which we identify with the $\Sigma_B$ coordinates) and mulitple matrix-valued $D7$-branes Wilson line moduli
${a_I}$ were guessed at in  \cite{gravitino_DM}. 
 the ${\cal N}=1$ chiral coordinates: $T_{\alpha=B,S},{\cal G}^{a=1,2},\zeta^A,z^{i=1,2}$
(See \cite{Jockers_thesis,D3_D7_Misra_Shukla} for definitions of all) - in particular:
\begin{eqnarray}
\label{eq:sigmas_Ts}
& & \sigma_\alpha\sim T_\alpha -\left( i\kappa_{\alpha bc}c^b{\cal B}^c + \kappa_\alpha + \frac{i}{(\tau - {\bar\tau})}\kappa_{\alpha bc}{\cal G}^b({\cal G}^c
- {\bar {\cal G}}^c) \right.\nonumber\\
 & & \left.+ i\delta^B_\alpha\kappa_4^2\mu_7l^2C_\alpha^{I{\bar J}}a_I{\bar a_{\bar J}} + \frac{3i}{4}\delta^B_\alpha\tau Q_{\tilde{f}} + i\mu_3l^2(\omega_\alpha)_{i{\bar j}} z^i\bigl({\bar z}^{\bar j}-\frac{i}{2}{\bar z}^{\tilde{a}}({\bar{\cal P}}_{\tilde{a}})^{\bar j}_lz^l\bigr)\right),
\end{eqnarray}
where (i) $\kappa_4$ is related to four-dimensional Newton's constant, $\mu_3$ and $\mu_7$ are $D3$ and $D7$-brane tensions; (ii) $\kappa_{\alpha ab}$'s are triple intersection integers of the CY orientifold; (iii)
 $c^a$ and $b^a$ are coefficients of RR and NS-NS two forms expanded in odd basis of $H^{(1,1)}_{{\bar\partial},-}(CY)$; (iv) $C^{I{\bar J}}_\alpha=\int_{\Sigma^B}i^*\omega_\alpha\wedge A^I\wedge A^{\bar J}$, $\omega_\alpha\in H^{(1,1)}_{{\bar\partial},+}(CY_3)$ and $A^I$ forming a basis for $H^{(0,1)}_{{\bar\partial},-}(\Sigma^B)$ - immersion map is defined as: $i:\Sigma^B\hookrightarrow CY_3$, $a_I$ is defined via a Kaluza-Klein reduction of the $U(1)$ gauge field (one-form) $A(x,y)=A_\mu(x)dx^\mu P_-(y)+a_I(x)A^I(y)+{\bar a}_{\bar J}(x){\bar A}^{\bar J}(y)$, where $P_-(y)=1$ if $y\in\Sigma^B$ and -1 if $y\in\sigma(\Sigma^B)$; (v)
 $z^{\tilde{a}}, \tilde{a}=1,...,h^{2,1}_-(CY_3),$ are $D=4$ complex structure deformations of the CY orientifold, $\left({\cal P}_{\tilde{a}}\right)^i_{\bar j}\equiv\frac{1}{||\Omega||^2}{\bar\Omega}^{ikl}\left(\chi_{\tilde{a}}\right)_{kl{\bar j}}$, i.e.,
${\cal P}:TCY_3^{(1,0)}\longrightarrow TCY_3^{(0,1)}$ via the transformation:
$z^i\stackrel{\rm c.s.\ deform}{\longrightarrow}z^i+\frac{i}{2}z^{\tilde{a}}\left({\cal P}_{\tilde{a}}\right)^i_{\bar j}{\bar z}^{\bar j}$, $z^i$ are scalar fields corresponding to geometric fluctuations of $D3$-brane inside the Calabi-Yau and defined via: $z(x)=z^i(x)\partial_i + {\bar z}^{\bar i}({\bar x}){\bar\partial}_{\bar i}$, and (vi)
 $Q_{\tilde{f}}\equiv l^2\int_{\Sigma^B}\tilde{f}\wedge\tilde{f}$, where $\tilde{f}\in\tilde{H}^2_-(\Sigma^B)\equiv{\rm coker}\left(H^2_-(CY_3)\stackrel{i^*}{\rightarrow}H^2_-(\Sigma^B)\right)$.

We estimate the intersection matrices $C_{I\bar{J}}^{B}$ in an appendix of \cite{gravitino_DM} by constructing harmonic one forms. Following the previous constructions in \cite{D3_D7_Misra_Shukla,ferm_masses_MS}, the harmonic distribution one-forms can be constructed by integrating: $dA_I=\left(P_{\Sigma_B}(z_{1,2,3})\right)^Idz_1\wedge dz_2$, with $I=1,2(done),3,4$ where
\begin{equation}
\label{eq:A I}
A_I\sim \delta\left(|z_3|-{\cal V}^{\frac{1}{6}}\right)\delta\left(|z_1|-{\cal V}^{\frac{1}{36}}\right)\delta\left(|z_2|-{\cal V}^{\frac{1}{36}}\right)\left[\omega_I(z_1,z_2)dz_1 + \tilde{\omega}_I(z_1,z_2)dz_2\right].
\end{equation}
 As the defining hypersurface of the Swiss-Cheese Calabi-Yau in $x_2\neq0$-coordinate patch
will be $z_1^{18} + z_2^{18} + ...$ which near $C_3$ (implying that the other two coordinates will scale like
${\cal V}^{\frac{1}{6}}, {\cal V}^{\frac{1}{6}} - {\cal V}^{\frac{1}{4}}$) receives the most dominant contributions from the monomials $z_1^{18}$ and
$z_2^{18}$ it is sufficient to consider ${\cal P}_{\Sigma_S}\Bigr|_{D3|_{{\rm near}\ C_3\hookrightarrow\Sigma_B}},{\cal P}_{\Sigma_B}|_{{\rm near}\ C_3\hookrightarrow\Sigma_B}\sim z_1^{18} + z_2^{18}$ with the understanding $|{\cal P}(z_{1,2,3},z_4=0;\phi,\psi)|_{C_3},|{\cal P}(z_{1,2,4},z_3;\phi,\psi)|_{C_3}<|z_1^{18}+z_2^{18}|$. One sees that $A_I$ is harmonic only on $\Sigma_B$ and not at any other generic locus in the Calabi-Yau manifold; (\ref{eq:A I}) shows that $A_I$ are distribution one-forms on $\Sigma_B$ localized along the $D3$-brane which is localized on the three-cycle $C_3$ of (\ref{eq:near_slag_i}). Writing $A_I(z_1,z_2)=\omega_I(z_1,z_2)dz_1+\tilde{\omega}_I(z_1,z_2)dz_2$,\footnote{Intuitively, these distribution one-forms could be thought of as
the holomorphic square-root of a Poincare dual of a four-cycle.} where $\omega(-z_1,z_2)=\omega(z_1,z_2)$, $\tilde{\omega}(-z_1,z_2)=-\tilde{\omega}(z_1,z_2)$ and $\partial_1\tilde{\omega}=-\partial_2\omega$, one obtains:
 \begin{eqnarray}
\label{eq:A_1234}
& & A_1|_{C_3}\sim - z_1^{18}z_2^{19}dz_1 + z_1^{19}z_2^{18}dz_2, A_2|_{C_3}\sim - z_1^{18}z_2dz_1 + z_2^{18}z_1 dz_2,\nonumber\\
& & A_3|_{C_3}\sim -z_1^{18}z_2^{37}dz_1 -z_2^{18}z_1^{37}dz_1, A_4|_{C_3}\sim -z_1^{36}z_2^{37}dz_1 + z_2^{36}z_1^{37}dz_2.
\end{eqnarray}
Utilizing above, one can calculate the intersection numbers $C_{I \bar J}$ corresponding to set of four-Wilson line moduli. The resultant contribution of various $C_{I \bar J}$'s are given in an appendix of \cite{gravitino_DM}.

Let's now look at the term quadratic in the $D3$-brane position moduli in $T_{\alpha=B,S}$ which is given by: $\left(\omega_\alpha\right)_{i{\bar j}}z^i\left({\bar z}^{\bar j} - \frac{i}{2}\left({\cal P}_{\tilde{a}}\right)^{\bar j}_{\ l}{\bar z}^{\tilde{a}}z^l\right)$, where $\left({\cal P}_{\tilde{a}}\right)^{\bar j}_{\ l}=\frac{\Omega^{{\bar j}{\bar k}{\bar m}}\left(\chi_{\tilde{a}}\right)_{{\bar k}{\bar m}l}}{||\Omega||^2}$\\
 $\sim\frac{g^{j_1{\bar j}}g^{k_1{\bar k}}g^{m_1{\bar m}}\Omega_{j_1k_1m_1}\left(\chi_{\tilde{a}}\right)_{{\bar k}{\bar m}l}}{g^{i_2{\bar i}_2}g^{j_3{\bar j}_3}g^{k_2{\bar k}_2}\Omega_{i_2j_2k_2}{\bar\Omega}_{{\bar i}_2{\bar j}_2{\bar k}_2}}$.
As $\omega_\alpha$ forms a basis of $H^{1,1}_+(CY_3)$ and $\chi_{\tilde{a}}$ forms a basis of $H^{2,1}_-(CY_3)$, this does not therefore depend on the choice of the divisor $\alpha$.
By Griffith's residue formula, $\Omega=\Omega_{124}dz_1\wedge dz_2\wedge dz_4$ where $\Omega_{124}=\frac{1}{\frac{\partial P(z_1,z_2,z_3,z_4)}{\partial z_3}}=\frac{1}{3z_3^2-\psi z_1z_2z_4}$. From $P(z_1,z_2,z_3,z_4)=0$, one obtains:
\begin{eqnarray}
\label{eq:z_3}
& &  z_3\sim \frac{\psi z_1z_2z_4}{\left(-\left(z_1^{18}+z_2^{18}+z_4^2-\phi z_1^6z_2^6\right)+\sqrt{\left(\psi z_1z_2z_4\right)^3+\left(z_1^{18}+z_2^{18}+z_4^2-\phi z_1^6z_2^6\right)^2}\right)^{\frac{1}{3}}}\nonumber\\
& & \hskip -0.15in + \left(-\left(z_1^{18}+z_2^{18}+z_4^2-\phi z_1^6z_2^6\right)+\sqrt{\left(\psi z_1z_2z_4\right)^3+\left(z_1^{18}+z_2^{18}+z_4^2-\phi z_1^6z_2^6\right)^2}\right)^{\frac{1}{3}},
\end{eqnarray}
which for $z_{1,2}\sim {\cal V}^{\frac{1}{36}}, z_4\sim {\cal V}^{\frac{1}{6}}$ yields $z_3\sim {\cal V}^{\frac{1}{6}}$. Hence, $\Omega_{124}\sim{\cal V}^{-\frac{1}{3}}$.
The polynomial deformation coefficient $\psi$ is a complex structure deformation modulus and given that it must be of the type $z^{\tilde{a}}$, i.e., odd under the holomorphic isometric involution $\sigma:z_1\rightarrow -z_1,z_{2,4}\rightarrow z_{2,4}$, $\sigma:\psi\rightarrow-\psi$. So, given that $\sigma^*\Omega=-\Omega$, hence $\sigma:\Omega_{124}\rightarrow\Omega_{124}$. Now, $\frac{1}{2}\left({\cal P}_{\tilde{a}}\right)^{\bar j}_{\ l}\sim\frac{\left(\chi_{\tilde{a}}\right)_{{\bar 4}{\bar 2}1}}{{\bar\Omega}_{{\bar 1}{\bar 2}{\bar 4}}}\sim\frac{{\cal V}^{\frac{1}{3}}}{2}\Biggr|_{{\cal V}\stackrel{<}{\sim}10^6}\sim{\cal O}(10)$. Assuming the complex structure moduli $z^{\tilde{a}}$ are stabilized at value ${\cal O}(10)z^{\tilde{a}}\sim{\cal O}(1)$ for ${\cal V}~{\lesssim}10^6$, one sees that contribution quadratic in $z^iz^j$ goes like $(\omega_\alpha)_{i{\bar j}}(z^i{\bar z}^{\bar j})$. $\omega_{B,S}$ are Poincare-duals of $\Sigma_{B,S}$ respectively. Hence, $\omega_{B,S}=\delta (P_{\Sigma_{B,S}})dP_{\Sigma_{B,S}}\wedge \delta({\bar P}_{\Sigma_{B,S}})d{\bar P}_{\Sigma_{B,S}}$.
Therefore one can argue (\cite{gravitino_DM,D3_D7_Misra_Shukla}) that near
$
|z_{1,2}|\equiv{\cal V}^{\frac{1}{36}}M_P,|z_3|\equiv{\cal V}^{\frac{1}{6}}M_P, |a_1|\equiv{\cal V}^{-\frac{2}{9}}M_P,
|a_2|\equiv{\cal V}^{-\frac{1}{3}}M_P, |a_3|\equiv{\cal V}^{-\frac{13}{18}}M_P,|a_4|\equiv{\cal V}^{-\frac{11}{9}}M_P;
\zeta^{A=1,...,h^{0,2}_-(\Sigma_B|_{C_3})}\equiv0$ (implying rigidity of the non-rigid $\Sigma_B$); ${\cal G}^a\sim\frac{\pi}{{\cal O}(1)k^a(\sim{\cal O}(10))}M_P,
$
 one obtains a local meta-stable dS-like minimum corresponding to the positive definitive potential
$e^KG^{T_S{\bar T}_S}|D_{T_S}W|^2$ \footnote{The CCB minima \cite{casas} are usually avoided if, in terms of our local setup, the square of the trilinear ${\cal A}$ couplings have an upper bound given in terms of a linear combination of squark mass squared and the mass squared of the Higgs doublets (the mass squared of the $D3$-brane position moduli + Higgs mass parameter squared). Writing the above as:   $ |{\cal A}|^2 \leq  O(1) m_{\tilde{q}}^2 + {\cal O}(1) m_{Z_i}^2 + {\cal O}(1) \mu^2$, we have verified that for the Calabi-Yau volume(which is a free parameter in our model) ${\cal V}\sim 1/{\cal O}(1) \times 10^5$ [in string length units], the inequality is satisfied at the string scale and the inequality is approximately saturated at the EW scale. Of course, this inequality is usually obtained by assuming that all sparticles have acquired the same vev at the CCB minimun, which is not true for us. Hence this was only to get an idea about the CCB issue. More relevant to our setup is the fact that the local $F$-term potential (the $D$-term is 2-form-flux-suppressed in the dilute flux approximation that we work with) with the mobile space-time filling $D3$-brane restricted to a nearly special Lagrangian 3-cycle, corresponds to the norm-squared of the K\"{a}hler vector obtained by the K\"{a}hler-covariant derivative of the ED3-instanton superpotential w.r.t. to the complexified big/small divisor modulus at the string/EW scale and is hence positive semi-definite. Thus, there is no `UFB' problem. Further, at the string and EW scale, the CCB minimum corresponding to non-zero vevs for the $D3$-brane position moduli identified with two Higgses and Wilson line moduli, identified with the sleptons and squarks, is approximately degenerate with SM-like vacua with non-zero vevs only for the Higgs.  Further, the $SU(3)_c$-violating vertices in our model relevant to, e.g., gluino decays, (N)LSP decays, etc., despite the non-zero vevs for sleptons and squarks, are highly volume-suppressed as compared to the $SU(3)_c$-preserving vertices.}stabilizing
${\rm vol}\left(\Sigma_B\right) = \Re e ({\sigma}_B)\sim{\cal V}^{\frac{2}{3}}, {\rm vol}\left(\Sigma_S\right) = \Re e{\sigma}_S\sim{\cal V}^{\frac{1}{18}}$ such that $\Re e T_S\sim{\cal V}^{\frac{1}{18}}$.\\
In the dilute flux approximation, the gauge couplings corresponding to the gauge theories living on stacks of $D7$ branes wrapping the ``big" divisor $\Sigma_B$  will given by:
\begin{eqnarray}
\label{eq:1overgsquared}
 \frac{1}{g_{j{=SU(3)\ {\rm or}\ SU(2)}}^2}&  = & \Re(T_{B}) + ln\left(\left.P\left(\Sigma_S\right)\right|_{D3|_{\Sigma_B}}\right) + ln\left(\left.{\bar P}\left(\Sigma_S\right)\right|_{D3|_{\Sigma_B}}\right)\nonumber\\
 & & + {\cal O}\left({\rm U(1)-Flux}_j^2\right).
\end{eqnarray}
Near the aforementioned stabilized values of the open string moduli,
\begin{eqnarray}
\label{eq:Casquared}
& & \left|C_{1{\bar 1}}|a_1|^2\right| \sim {\cal V}^{\frac{2}{3}},\ \left|C_{1{\bar 2}}\left(a_1{\bar a}_2 + h.c.\right)\right| \sim {\cal V}^{\frac{1}{18}},\ \left|C_{1{\bar 3}}\left(a_1{\bar a}_3 + h.c.\right)\right| \sim {\cal V}^{\frac{2}{3}},\nonumber\\
& & \left|C_{1{\bar 4}}\left(a_1{\bar a}_4 + h.c.\right)\right| \sim {\cal V}^{\frac{2}{3}},\
\left|C_{2{\bar 2}}|a_2|^2\right| \sim {\cal V}^{-\frac{5}{9}},\
\left|C_{2{\bar 3}}\left(a_2{\bar a}_3 + h.c.\right)\right| \sim {\cal V}^{\frac{1}{18}},\nonumber\\
& & \left|C_{2{\bar 4}}\left(a_2{\bar a}_4 + h.c.\right)\right| \sim {\cal V}^{\frac{1}{18}},\
\left|C_{3{\bar 3}}|a_3|^2\right| \sim {\cal V}^{\frac{2}{3}},\
\left|C_{3{\bar 4}}\left(a_3{\bar a}_4 + h.c.\right)\right| \sim {\cal V}^{\frac{2}{3}},\nonumber\\
& & \left|C_{4{\bar 4}}|a_4|^2\right| \sim {\cal V}^{\frac{2}{3}},
\end{eqnarray}
we see that there is the possibility that 
\begin{eqnarray}
Vol(\Sigma_B)+C_{I{\bar J}}a_I{\bar a}_{\bar J} + h.c.\sim {\cal V}^{\frac{1}{18}},
\end{eqnarray}
and hence $\frac{1}{g_{j{=SU(3)\ {\rm or}\ SU(2)}}^2} \sim {\cal V}^{\frac{1}{18}}\sim {\cal O}(1)$,
   as according to equation (\ref{eq:1overgsquared}). 

The K\"{a}hler potential relevant to all the calculations in chapters {\bf 2-4} (without being careful about ${\cal O}(1)$ constant factors) is given as under \footnote{One should include the perturbative $(\alpha^\prime)^3$ correction $\sum_{(m,n)\mathbb{Z}^2/(0,0)}\frac{\chi(CY_3)}{|m + n \tau|^3}$ and the perturbative $(\alpha^\prime)^2$ correction $\sqrt{\sigma_B + {\bar\sigma}_B}$ \cite{Rummel_et_al} as shifts in the argument of the logarithm. Further, due to the presence of the mobile $D3$-brane, one needs to include a $-\gamma K_{\rm geom}, \gamma\sim\frac{1}{\cal V}$ so that one considers $a_B(\tau_B + {\bar\tau}_B - \gamma K_{\rm geom})^{\frac{3}{2}} - a_S(\tau_S + {\bar\tau}_S - \gamma K_{\rm geom}) + \sum_\beta n^0_\beta(...)$. However, as was verified in \cite{gravitino_DM} using the Donaldson's algorithm, $\gamma K_{\rm geom}\sim\frac{ln {\cal V}}{\cal V}<<\langle\sigma_S\rangle(\sim {\cal V}^{\frac{1}{18}}),\langle\sigma_B\rangle(\sim{\cal V}^{\frac{2}{3}})$; it is hence justified to drop $\gamma K_{\rm geom}$ from the K\"{a}hler potential.}:
 \begin{eqnarray}
\label{eq:K}
& &  K\sim-2 ln\left(a_B\Biggl[\frac{T_B+{\bar T}_B}{M_P} - \mu_3(2\pi\alpha^\prime)^2
\frac{\left\{|z_1|^2 + |z_2|^2 + z_1{\bar z}_2 + z_2{\bar z}_1\right\}}{M_P^2}
 +{\cal V}^{\frac{10}{9}}\frac{|a_1|^2}{M_P^2}\right.\nonumber\\
 & &+{\cal V}^{\frac{11}{18}}
 \frac{\left(a_1{\bar a}_2+h.c.\right)}{M_P^2} +{\cal V}^{\frac{1}{9}}\frac{|a_2|^2}{M_P^2} + {\cal V}^{\frac{29}{18}}
 \frac{\left(a_1{\bar a}_3+h.c.\right)}{M_P^2}+ {\cal V}^{\frac{10}{9}}
 \frac{\left(a_2{\bar a}_3+h.c.\right)}{M_P^2} \nonumber \\
 & & + {\cal V}^{\frac{19}{9}}\frac{|a_3|^2}{M_P^2} + {\cal V}^{\frac{19}{9}}\frac{\left(a_1{\bar a}_4 + a_4{\bar a}_1\right)}{M_P^2} +  {\cal V}^{\frac{29}{18}}\frac{\left(a_2{\bar a}_4 + a_4{\bar a}_2\right)}{M_P^2}+ {\cal V}^{\frac{47}{18}}\frac{\left(a_3{\bar a}_4 + a_4{\bar a}_3\right)}{M_P^2}\nonumber \\
 & & + {\cal V}^{\frac{28}{9}}\frac{|a_4|^2)}{M_P^2}
\Biggr]^{3/2}- \nonumber\\
& & \left. a_S\left(\frac{{T_S+{\bar T}_S}}{M_P}-\mu_3(2\pi\alpha^\prime)^2 \frac{\left\{|z_1|^2 + |z_2|^2 + z_1{\bar z}_2 + z_2{\bar z}_1\right\}}{M_P^2}\right)^{3/2} +\sum n^0_\beta(...)\right),
\end{eqnarray}
\vskip -0.09in
\noindent 
and ED3 generated non-perturbative superpotential  is given by \cite{nonpert,Ganor1_2}:
\begin{eqnarray}
\label{eq:W}
\begin{array}{r}
  W\sim\left({\cal P}_{\Sigma_S}\Bigr|_{D3|_{{\rm near}\ C_3\hookrightarrow\Sigma_B}}\sim z_1^{18} + z_2^{18}\right)^{n^s}\sum_{m_a}e^{i\tau\frac{m^2}{2}}e^{in^s G^am_a} e^{i n^s T_s},
\end{array}
\end{eqnarray}
\vskip -0.09in
\noindent which is like (\ref{W_pheno}) assuming $G^a, \tau$ has been stabilized. The  genus-zero Gopakumar-Vafa invariants (which for projective varieties are very large) prefix the $h^{1,1}_-$-valued real axions $b^a,c^a$. In general there are no known globally defined involutions, valid for all Calabi-Yau volumes, for which $h^{1,1}_-(CY_3)\neq0, h^{0,1}_-(\Sigma_B
)\neq0$.
However, as mentioned earlier, in the spirit of the involutive mirror symmetry
implemented a la SYZ prescription in terms of a triple of T dualities along a
local $T^3$ in the large volume limit, we argued in \cite{ferm_masses_MS},
e.g., $z_1\rightarrow-z_1$ would, restricted to $C_3$, generate non-zero
$h^{1,1}_-\left(\begin{array}{c} T^3(arg z_{1,2,3})\rightarrow{\cal M}_3(z_{1,2,3})\\
\hskip1in\downarrow \\
\hskip 0.3in M_3(|z_1|,|z_2|,|z_3|) \end{array}  \right)$. An example of holomorphic involutions near $C_3$  not requiring a
large Calabi-Yau volume has been discussed in \cite{gravitino_DM}. However, even if $h^{1,1}_-=0$, one can self-consistently stabilize $c^a,b^a$ to zero and $\sigma_s,\sigma_b$ to ${\cal V}^{\frac{1}{18}}, {\cal V}^{\frac{2}{3}}$ such that the K\"{a}hler potential continues to be stabilized at $- 2 ln {\cal V}$.

The evaluation of ``physical"/normalized Yukawa couplings, soft SUSY breaking parameters and various 3-point vertices needs the matrix generated from the mixed double derivative of the K\"{a}hler potential to be a diagonalized matrix. After diagonalization  the corresponding eigenvectors of the same are given by:
\begin{eqnarray}
\label{eq:eigenvectors}
  &&{\cal A}_4\sim a_4 + {\cal V}^{-\frac{3}{5}}a_3 + {\cal V}^{-\frac{6}{5}} a_1+{\cal V}^{-\frac{9}{5}} a_2 + {\cal V}^{-2}\left(z_1+z_2\right),\nonumber\\
&&{\cal A}_3\sim -a_3 + {\cal V}^{-\frac{3}{5}}a_4 - {\cal V}^{-\frac{3}{5}}a_1 - {\cal V}^{-\frac{7}{5}}a_2 + {\cal V}^{-\frac{8}{5}}\left(z_1+z_2\right),\nonumber\\
&&  {\cal A}_1\sim a_1 - {\cal V}^{-\frac{3}{5}}a_3 + {\cal V}^{-1}a_2 - {\cal V}^{-\frac{6}{5}}a_4+ {\cal V}^{-\frac{6}{5}}\left(z_1+z_2\right),\nonumber\\
&&{\cal A}_2\sim - a_2 - {\cal V}^{-1}a_1 + {\cal V}^{-\frac{7}{5}}a_3 - {\cal V}^{-\frac{3}{5}}\left(z_1+z_2\right),\nonumber\\
&&{\cal Z}_2\sim - \frac{\left(z_1+z_2\right)}{\sqrt 2}  - {\cal V}^{-\frac{6}{5}}a_1 + {\cal V}^{-\frac{3}{5}}a_2 + {\cal V}^{-\frac{8}{5}}a_3+ {\cal V}^{-2}a_4,\nonumber\\
&&{\cal Z}_1\sim \frac{\left(z_1-z_2\right)}{\sqrt 2}  - {\cal V}^{-\frac{6}{5}}a_1 + {\cal V}^{-\frac{3}{5}}a_2 + {\cal V}^{-\frac{8}{5}}a_3+ {\cal V}^{-2}a_4.
\end{eqnarray}
and the numerical eigenvalues are estimated to be:
\begin{eqnarray}
\label{eq:eigenvals}
 && K_{{\cal Z}_1{\cal Z}_1} \sim10^{-5},K_{{\cal Z}_2{\cal Z}_2}\sim 10^{-3}, K_{{\cal A}_1{\cal A}_1}\sim 10^4,\nonumber\\
 && K_{{\cal A}_2{\cal A}_2}\sim10^{-2}, K_{{\cal A}_3{\cal A}_3}\sim 10^7, K_{{\cal A}_4{\cal A}_4}\sim 10^{12}.
\end{eqnarray}
The system of equations (\ref{eq:eigenvectors}) can be solved to yield:
 \begin{eqnarray}
\label{eq:a_I+z_i}
& & z_1\sim \frac{\left({\cal Z}_2-{\cal Z}_1\right)}{\sqrt 2} - {\cal V}^{-\frac{6}{5}}{\cal A}_1 - {\cal V}^{-\frac{3}{5}}{\cal A}_2 + {\cal V}^{-\frac{8}{5}}{\cal A}_3 + {\cal V}^{-2}{\cal A}_4,\nonumber\\
& & z_2\sim -\frac{\left({\cal Z}_2+ {\cal Z}_1\right)}{\sqrt 2} - {\cal V}^{-\frac{6}{5}}{\cal A}_1 - {\cal V}^{-\frac{3}{5}}{\cal A}_2 + {\cal V}^{-\frac{8}{5}}{\cal A}_3 + {\cal V}^{-2}{\cal A}_4,\nonumber\\
& & a_1\sim {\cal V}^{-\frac{6}{5}}{\cal Z}_1 + {\cal V}^{-\frac{7}{5}}{\cal Z}_2 - {\cal A}_1 - \frac{{\cal A}_2}{\cal V} + {\cal V}^{-\frac{3}{5}}{\cal A}_3 + {\cal V}^{-\frac{6}{5}}{\cal A}_4,\nonumber\\
& & a_2\sim {\cal V}^{-\frac{3}{5}}{\cal Z}_1 + \frac{{\cal Z}_2}{\cal V} - \frac{{\cal A}_1}{\cal V}
+ {\cal A}_2 - {\cal V}^{-\frac{7}{5}}{\cal A}_3 + {\cal V}^{-\frac{9}{5}}{\cal A}_4,\nonumber\\
& & a_3\sim {\cal V}^{-\frac{7}{5}}{\cal Z}_1 + {\cal V}^{-\frac{9}{5}}{\cal Z}_2 - {\cal V}^{-\frac{3}{5}}{\cal A}_1 - {\cal V}^{-\frac{7}{5}}{\cal A}_2 - {\cal A}_3 + {\cal V}^{-\frac{3}{5}}{\cal A}_4,\nonumber\\
& & a_4\sim \frac{{\cal Z}_1}{{\cal V}^2} + {\cal V}^{-\frac{11}{5}}{\cal Z}_2 + \frac{{\cal A}_1}{{\cal V}} - {\cal V}^{-\frac{9}{5}}{\cal A}_2 + {\cal V}^{-\frac{3}{5}}{\cal A}_3 + {\cal A}_4.
\end{eqnarray}
 We will now consider four stacks of $D7$-branes: a stack of 3, a stack of 2 and two stacks of 1. The matter fields: L-quarks and their superpartners will be valued in the bifundamentals $(3,{\bar 2})$ under $SU(3)_c\times SU(2)_L$; the L-leptons and their superpartners will be valued in the bifundamentals
$(2,\bar{-1})$ of $SU(2)_L\times U(1)_Y$.  Before the  fluxes (\ref{eq:flux})  are turned on,
the Wilson line moduli are valued in the adjoint of $U(7)$. With the following choice of fluxes:
{\footnotesize{
 \begin{equation}
\label{eq:flux}
F = \left(\begin{array}{ccccccc}f_1 & 0 & 0 & 0 & 0 & 0&0\\
0 & f_1 & 0 & 0 & 0 & 0&0\\
0 & 0 & f_1 & 0 & 0 & 0&0\\
0 & 0 & 0 & f_2 & 0 & 0&0\\
0 & 0 & 0 & 0 & f_2 & 0&0\\
0 & 0 & 0 & 0 & 0 & f_3&0\\
0 & 0 & 0 & 0 & 0 & 0 & f_4
\end{array}\right),
\end{equation}}}
the $U(7)$ is broken down to $U(3)\times U(2)\times U(1)\times U(1)$. So, the bifundamental Wilson line super-moduli ${\cal A}_I$  will be represented as:
\begin{equation}
\label{eq:Q L_I}
{\cal A}_I=\sum a_I^{ab}e_{ab} + \sum\theta \tilde{a}_I^{ab}e_{ab},\ \left[(a,b)=1,...,7\right],
\end{equation}
where $\left(e_{ab}\right)_{ij}=\delta_{ai}\delta_{bj}$. Hence,
{\footnotesize{
  \begin{equation}
 \label{eq:e_L}
{\cal A}_1 =
 \left(\begin{array}{ccccccc}
 0 & 0 & 0 & 0 & 0 & 0 & 0\\
 0 & 0 & 0 & 0 & 0 & 0 & 0\\
 0 & 0 & 0 & 0 & 0 & 0 & 0\\
 0 & 0 & 0 & 0 & 0 & \tilde{\nu}_e + \theta\nu_e   & 0\\
 0 & 0 & 0 & 0 & 0 & \tilde{e} + \theta e & 0\\
 0 & 0 & 0 & \bar{\tilde{\nu}}_e + {\bar\theta}{\bar\nu}_e  &  \bar{\tilde{e}} + {\bar\theta}{\bar e} & 0 & 0\\
 0 & 0 & 0 & 0 & 0 & 0 & 0
 \end{array} \right),
\end{equation}
  \begin{equation}
\label{eq:Q L}
{\cal A}_2 =
\left(\begin{array}{ccccccc}
0 & 0 & 0 & \tilde{u}  + \theta u & 0 & 0 & 0\\
0 & 0 & 0 & 0 & 0 & 0 & 0\\
0 & 0 & 0 & 0 & 0 & 0 & 0\\
\bar{\tilde{u}}  + {\bar\theta}{u}^\dagger &  0 & 0 & 0 & 0 & 0 & 0\\
0 &  0 & 0 & 0 & 0 & 0 & 0\\
0 & 0 & 0 & 0 & 0 & 0 & 0\\
0 & 0 & 0 & 0 & 0 & 0 & 0
\end{array}\right),
\end{equation}
and
\begin{equation}
\label{eq:e_R}
{\cal A}_3 =
\left(\begin{array}{ccccccc}
0 & 0 & 0 & 0 & 0 & 0 & 0\\
0 & 0 & 0 & 0 & 0 & 0 & 0\\
0 & 0 & 0 & 0 & 0 & 0 & 0\\
 0 &  0 & 0 & 0 & 0 & 0 & 0\\
0 &  0 & 0 & 0 & 0 & 0 & 0\\
0 & 0 & 0 & 0 & 0 & 0 & \tilde{e}_R + \theta e_R \\
0 & 0 & 0 & 0 & 0 & \bar{\tilde{e}}_R + {\bar\theta}{e}_R^\dagger & 0
\end{array}\right),
\end{equation}
\begin{equation}
 \label{eq:f_R}
 {\cal A}_4 =
\left(\begin{array}{ccccccc}
 0 & 0 & 0 & 0 & 0 & 0 & \tilde{u}_R + \theta u_R\\
 0 & 0 & 0 & 0 & 0 & 0 & 0\\
 0 & 0 & 0 & 0 & 0 & 0 & 0\\
 0 & 0 & 0 & 0 & 0 & 0 & 0 \\
 0 & 0 & 0 & 0 & 0 & 0 \\
 0 & 0 & 0 & 0 &  0 & 0 & 0 \\
 \bar{\tilde{u}}_R + {\bar\theta}u_R^\dagger & 0 & 0 & 0 & 0 & 0 & 0
 \end{array} \right).
\end{equation}
}}
Unlike the ${\cal A}_I$  which correspond to matter fields
corresponding to open strings stretched between two stacks of $D7$
branes (with different two-form fluxes turned on their world
volumes), the Higgses would arise as a geometric moduli corresponding
to the fluctuations in the position of the mobile space-time filling
$D3$-brane. Now, we will see if one can construct appropriate  $a_I$ (bi-fundamental)
and $z_i$(mimicking bi-fundamental fields) such that
(\ref{eq:eigenvectors}), (\ref{eq:e_L}) - (\ref{eq:f_R}), as well as:
{\footnotesize{
\begin{equation}
\label{eq:H_u}
{\cal Z}_1=\left(\begin{array}{ccccccc}
 0 & 0 & 0 & 0 & 0 & 0 & 0\\
 0 & 0 & 0 & 0 & 0 & 0 & 0\\
 0 & 0 & 0 & 0 & 0 & 0 & 0\\
 0 & 0 & 0 & 0 & 0 & H_u + \theta\tilde{H}_u   & 0\\
 0 & 0 & 0 & 0 & 0 & 0 & 0\\
 0 & 0 & 0 & \bar{H}_u + {\bar\theta}\tilde{H}_u^\dagger  &  0 & 0 & 0\\
 0 & 0 & 0 & 0 & 0 & 0 & 0
 \end{array} \right),
\end{equation}
and
\begin{equation}
\label{eq:H_d}
{\cal Z}_2=\left(\begin{array}{ccccccc}
 0 & 0 & 0 & 0 & 0 & 0 & 0\\
 0 & 0 & 0 & 0 & 0 & 0 & 0\\
 0 & 0 & 0 & 0 & 0 & 0 & 0\\
 0 & 0 & 0 & 0 & 0 & 0   & 0\\
 0 & 0 & 0 & 0 & 0 & H_d + \theta\tilde{H}_d & 0\\
 0 & 0 & 0 &0  &  \bar{H}_d + {\bar\theta}\tilde{H}_d^\dagger & 0 & 0\\
 0 & 0 & 0 & 0 & 0 & 0 & 0
 \end{array} \right),
\end{equation}
}}
 are satisfied. It was shown in \cite{gravitino_DM}  the following solution:
 {\footnotesize{
\begin{eqnarray*}
\label{eq:sols_a_I}
& &    a_1 = \left(\begin{array}{ccccccc}
0 & 0 & 0 & \xi_1^{14}{\cal V}^{-\frac{7}{5}} \tilde{u}_L & 0 & 0 &
\xi_1^{17}{\cal V}^{-\frac{11}{5}} \tilde{u}_R \\
0 & 0 & 0 & 0 & 0 & 0 & 0 \\
0 & 0 & 0 & 0 & 0 & 0 & 0 \\
\xi_1^{14}{\cal V}^{-\frac{7}{5}} {\bar{\tilde{u}}}_L & 0 & 0 & 0 & 0 &
\xi_1^{46}\left(\frac{\tilde{e}_L}{2} + {\cal V}^{-\frac{8}{5}}H_u\right) & 0 \\
0 & 0 & 0 & 0 & 0 & \xi_1^{56}{\cal V}^{-\frac{8}{5}} H_d & 0 \\
0 & 0 & 0 &{\bar\xi}_1^{46}\left( \frac{{\bar{\tilde{e}}}_L}{2} + {\cal V}^{-\frac{8}{5}}{\bar H}_u\right) &
{\bar\xi}_1^{56}{\cal V}^{-\frac{8}{5}} {\bar H}_d & 0 & -\xi_1^{67} {\cal V}^{-\frac{4}{5}} e_R
\\
{\bar\xi}_1^{17}{\cal V}^{-\frac{11}{5}} {\bar{\tilde{u}}}_R & 0 & 0 & 0 & 0 &
-{\bar\xi}_1^{67}  {\cal V}^{-\frac{4}{5}} {\bar{\tilde{e}}}_R
\end{array}\right),\nonumber\\
& &  \nonumber\\
&&  \nonumber\\
&&  \nonumber\\
& &  a_2 = \left(\begin{array}{ccccccc}
0 & 0 & 0 & 0.3\xi_2^{14} \tilde{u}_L & 0 & 0 &
\xi_2^{17}{\cal V}^{-\frac{13}{5}} \tilde{u}_R \\
0 & 0 & 0 & 0 & 0 & 0 & 0 \\
0 & 0 & 0 & 0 & 0 & 0 & 0 \\
0.3\xi_2^{14} {\bar{\tilde{u}}}_L & 0 & 0 & 0 & 0 &
\xi_2^{46}\left(-{\cal V}^{-\frac{7}{5}}\tilde{e}_L + {\cal V}^{-\frac{4}{5}}H_u\right) & 0 \\
0 & 0 & 0 & 0 & 0 & 5\xi_2^{56}{\cal V}^{-\frac{6}{5}} H_d & 0 \\
0 & 0 & 0 &{\bar\xi}_2^{46}\left( -{\cal V}^{-\frac{7}{5}}{\bar{\tilde{e}}}_L +
  {\cal V}^{-\frac{4}{5}}{\bar H}_u\right) &
5{\bar\xi}_2^{56}{\cal V}^{-\frac{6}{5}} {\bar H}_d & 0 & -\xi_2^{67} {\cal V}^{-\frac{9}{5}} e_R
\\
{\bar\xi}_2^{17}{\cal V}^{-\frac{13}{5}} {\bar{\tilde{u}}}_R & 0 & 0 & 0 & 0 &
-{\bar\xi}_2^{67}  {\cal V}^{-\frac{9}{5}} {\bar{\tilde{e}}}_R
\end{array}\right),\nonumber\\
\end{eqnarray*}
\begin{eqnarray*}
& & a_3 = \left(\begin{array}{ccccccc}
0 & 0 & 0 & \xi_3^{14}\frac{\tilde{u}_L}{{\cal V}} & 0 & 0 &
-6\xi_3^{17} \frac{\tilde{u}_R}{{\cal V}} \\
0 & 0 & 0 & 0 & 0 & 0 & 0 \\
0 & 0 & 0 & 0 & 0 & 0 & 0 \\
{\bar\xi}_3^{14}\frac{{\bar{\tilde{u}}}_L}{{\cal V}} & 0 & 0 & 0 & 0 &
\xi_3^{46}\left(-{\cal V}^{-\frac{4}{5}}\tilde{e}_L + {\cal V}^{-\frac{8}{5}}H_u\right) & 0 \\
0 & 0 & 0 & 0 & 0 & \xi_3^{56}{\cal V}^{-\frac{11}{5}} H_d & 0 \\
0 & 0 & 0 &{\bar\xi}_3^{46}\left(-{\cal V}^{-\frac{4}{5}}\tilde{e}_L + {\cal V}^{-\frac{8}{5}}H_u\right) &
5{\bar\xi}_3^{56}{\cal V}^{-\frac{11}{5}} {\bar H}_d & 0 & 0.3\xi_3^{67} e_R
\\
{\bar\xi}_3^{17}{\cal V}^{-\frac{13}{5}} {\bar{\tilde{u}}}_R & 0 & 0 & 0 & 0 &
0.3{\bar\xi}_3^{67}  {\bar{\tilde{e}}}_R
\end{array}\right),\nonumber\\
\end{eqnarray*}
\begin{eqnarray}
\label{eq:sols_a_I_ii}
& &  a_4 = \left(\begin{array}{ccccccc}
0 & 0 & 0 & 0.08\xi_4^{14}\tilde{u}_L & 0 & 0 &
0.2\xi_4^{17} \tilde{u}_R \\
0 & 0 & 0 & 0 & 0 & 0 & 0 \\
0 & 0 & 0 & 0 & 0 & 0 & 0 \\
0.08{\bar\xi}_4^{14}{\bar{\tilde{u}}}_L & 0 & 0 & 0 & 0 &
\xi_4^{46}{\cal V}^{-\frac{7}{5}}\tilde{e}_L & 0 \\
0 & 0 & 0 & 0 & 0 & \xi_4^{56}{\cal V}^{-\frac{14}{5}} H_d & 0 \\
0 & 0 & 0 &{\bar\xi}_4^{46}{\cal V}^{-\frac{7}{5}}{\bar{\tilde{e}}}_L &
{\bar\xi}_4^{56}{\cal V}^{-\frac{14}{5}} {\bar H}_d & 0 & -7\xi_4^{67} e_R
\\
0.2{\bar\xi}_4^{17} {\bar{\tilde{u}}}_R & 0 & 0 & 0 & 0 &
-7{\bar\xi}_4^{67}  {\bar{\tilde{e}}}_R
\end{array}\right).\nonumber\\
\end{eqnarray}
\begin{eqnarray*}
z_1=\left(\begin{array}{ccccccc}
0 & 0 & 0 & \alpha_1^{14}\frac{\tilde{u}_L}{{\cal V}} & 0 & 0 &
5\alpha_1^{17}{\cal V}^{-\frac{14}{5}} \tilde{u}_R \\
0 & 0 & 0 & 0 & 0 & 0 & 0 \\
0 & 0 & 0 & 0 & 0 & 0 & 0 \\
{\bar\alpha}_1^{14}\frac{{\bar{\tilde{u}}}_L}{{\cal V}} & 0 & 0 & 0 & 0 &
\alpha_1^{46}\left({\cal V}^{-\frac{9}{5}}\tilde{e}_L - \frac{H_u}{\sqrt{2}}\right) & 0 \\
0 & 0 & 0 & 0 & 0 & \alpha_1^{56} \frac{H_d}{\sqrt{2}} & 0 \\
0 & 0 & 0 &{\bar\alpha}_1^{46}\left({\cal V}^{-\frac{9}{5}}{\bar {\tilde e}}_L -
  \frac{{\bar H}_u}{\sqrt{2}}\right) &
{\bar\alpha}_1^{56} \frac{{\bar H}_d}{\sqrt{2}} & 0 & \alpha_1^{67} {\cal
  V}^{-\frac{11}{5}}e_R
\\
5{\bar\alpha}_1^{17}{\cal V}^{-\frac{14}{5}} {\bar{\tilde{u}}}_R & 0 & 0 & 0 & 0 &
{\bar\alpha}_1^{67} {\cal
  V}^{-\frac{11}{5}}  {\bar{\tilde{e}}}_R
\end{array}\right),\nonumber\\
\end{eqnarray*}
\begin{eqnarray}
\label{eq:sols z_i}
& & z_2=\left(\begin{array}{ccccccc}
0 & 0 & 0 & \alpha_2^{14}{\cal V}^{-\frac{4}{5}}{\tilde{u}_L} & 0 & 0 &
5\alpha_2^{17}{\cal V}^{-\frac{13}{5}} \tilde{u}_R \\
0 & 0 & 0 & 0 & 0 & 0 & 0 \\
0 & 0 & 0 & 0 & 0 & 0 & 0 \\
{\bar\alpha}_2^{14}{\cal V}^{-\frac{4}{5}}{\bar{u}_L} & 0 & 0 & 0 & 0 &
\alpha_2^{46}\left(6{\cal V}^{-\frac{8}{5}}\tilde{e}_L - \frac{H_u}{\sqrt{2}}\right) & 0 \\
0 & 0 & 0 & 0 & 0 & -\alpha_2^{56} \frac{H_d}{\sqrt{2}} & 0 \\
0 & 0 & 0 &{\bar\alpha}_2^{46}\left({\cal V}^{-\frac{8}{5}}\bar {\tilde e}_L -
  \frac{{\bar H}_u}{\sqrt{2}}\right) &
-{\bar\alpha}_2^{56} \frac{{\bar H}_d}{\sqrt{2}} & 0 &
\alpha_2^{67}\frac{{\tilde e_R}}{{\cal V}^{-2}}
\\
5{\bar\alpha}_2^{17}{\cal V}^{-\frac{13}{5}} {\bar{\tilde{u}}}_R & 0 & 0 & 0 & 0 &
{\bar\alpha}_2^{67}\frac{\bar {\tilde e_R}}{{\cal V}^{-2}}
\end{array}\right).\nonumber\\
& &
\end{eqnarray}}}
In (\ref{eq:sols_a_I_ii}) and (\ref{eq:sols z_i}), $\xi_{I,i}^{ab}$ and $\alpha_{I,i}^{ab}, 
1\leq a\leq 6, 1\leq b\leq7$ are ${\cal O}(1)$ numbers. The intersection matrix was worked out in \cite{gravitino_DM}. Assuming
that the complex structure moduli
$z^{\tilde{a}=1,...,h^{2,1}_-(CY_3)}$ are stabilized at very small
values, which is in fact already assumed in writing
(\ref{eq:sigmas_Ts}) which has been written upon inclusion of terms up
to linear in the complex structure moduli, let us define a modified
intersection matrix in the $a_I-z_i$ moduli space:
\begin{eqnarray}
\label{eq:CIJbar mod}
& & {\cal C}^{{\cal I}{\cal J}}=C^{I{\bar J}},\ {\cal I}=I, {\cal
  J}={\bar J};\nonumber\\
  &&   {\cal C}^{{\cal I}{\cal J}}
=\mu_3\left(2\pi\alpha^\prime\right)^2\left(\omega_\alpha
\right)^{i{\bar j}},\ {\cal I}=i, {\cal
  J}={\bar j};\nonumber\\
& &  {\cal C}^{{\cal I}{\cal J}}=0,\ {\cal I}=I, {\cal
  J}={\bar j},\ {\rm etc.}.
\end{eqnarray}
Now, from (\ref{eq:sols_a_I}) and (\ref{eq:sols z_i}), one can show
that for $|z_i|\sim0.8{\cal V}^{\frac{1}{36}},\ {\cal V}\sim10^5$:
\begin{eqnarray}
\label{eq:tr-1}
& & {\cal C}^{{\cal I}\bar{\cal J}}Tr({\cal M}_{\cal I} {\cal M}_{\cal
  J}^\dagger)\sim
C^{a_1\bar{a}_1}|\tilde{e}_L|^2 + C^{a_2\bar{a}_2}|\tilde{u}_L|^2 +
C^{a_3\bar{a}_3}|\tilde{e}_R|^2
+ C^{a_4\bar{a}_4}|\tilde{u}_R|^2 \nonumber\\
&& +\mu_3\left(2\pi\alpha^\prime\right)^2|H_u|^2,
\end{eqnarray}
where ${\cal M}_{\cal I}\equiv a_I, z_i$. 

Now, using (\ref{eq:eigenvectors}), one can show that in the large volume limit:
 ${\cal C}^{{\cal A}_I\bar{\cal A}_{\bar J}}\sim C^{a_I\bar{a}_{\bar J}}$, 
 ${\cal C}^{{\cal Z}_i\bar{\cal Z}_{\bar j}}\sim {\cal C}^{z_i\bar{z}_j}$. Clubbing together the Wilson line moduli and the $D3$-brane position moduli into a single vector: ${\cal M}_{\Lambda}\equiv {\cal A}_I,{\cal Z}_i$, then one sees from (\ref{eq:e_L}) - (\ref{eq:H_d}), and (\ref{eq:tr-1}): 
 \begin{equation}
 {\cal C}^{\Lambda\bar{\Sigma}}Tr\left({\cal M}_\Lambda{\cal M}_{\Sigma}^\dagger\right)\sim {\cal C}^{{\cal I}\bar{\cal J}}Tr({\cal M}_{\cal I} {\cal M}_{\cal
  J}^\dagger).
  \end{equation}
  In the large volume and rigid limit of $\Sigma_B(\zeta^A=0$ which corresponds to a local minimum), perhaps  ${\cal C}^{\Lambda\bar{\Sigma}}Tr\left({\cal M}_\Lambda{\cal M}_{\Sigma}^\dagger\right)$
   is invariant under moduli transformations in the $({\cal A}_I,{\cal Z}_i)/(a_I,z_i)$-subspace of the open-string moduli space, which would imply that $C^{I{\bar J}}a_I{\bar a}_{\bar J} + \mu_3\left(\alpha^\prime\right)^2\left(\omega_B\right)_{i{\bar j}}z^i{\bar z}^{\bar j}$ for multiple  $D7$-branes, in a basis that diagonalizes $g_{{\cal M}_{\cal I}\bar{\cal M}_{\bar J}}$ at stabilized values of the open string moduli,  is replaced by (\ref{eq:tr-1}).
   
   The effective Yukawa couplings can be calculated using 
 \begin{equation}
 \hat{Y}^{\rm eff}_{C_i C_j C_k}\equiv\frac{e^{\frac{K}{2}}Y^{\rm eff}_{C_i C_j C_k}}{\sqrt{K_{C_i{\bar C}_i}K_{C_j{\bar C}_j}K_{C_k{\bar C}_k}}}.
 \end{equation}, $C_i$ being an open string modulus which for us is $\delta{\cal Z}_{1,2},\delta{\cal A}_{1,2,3,4}$, where, e.g., $Y^{\rm eff}_{{\cal Z}_i{\cal A}_{1/2}{\cal A}_{3/4}}$ is the ${\cal O}({\cal Z}_i)$-coefficient in the following mass term in the ${\cal N}=1$ gauged supergravity action of \cite{Wess_Bagger}: $e^{\frac{K}{2}}{\cal D}_{{\bar{\cal A}}_{1/2}}D_{{\bar{\cal A}}_{3/4}}{\bar W}{\bar\chi}^{{\cal A}_{1/3}}\chi^{{\cal A}_{2/4}}$. Using (\ref{eq:a_I+z_i}), and:
\begin{eqnarray}
\label{eq:expKover2_calD_DW}
& & e^{\frac{K}{2}} {\cal D}_{z_1} D_{z_1}W = {\cal V}^{-\frac{43}{72}},\  e^{\frac{K}{2}} {\cal D}_{z_1} D_{a_1}W = {\cal V}^{-\frac{89}{72}},\
 e^{\frac{K}{2}} {\cal D}_{z_1} D_{a_2}W = {\cal V}^{-\frac{125}{72}},\nonumber\\
& & e^{\frac{K}{2}} D_{z_1} D_{a_3}W = {\cal V}^{-\frac{25}{36}},\ e^{\frac{K}{2}} D_{z_1} D_{a_4}W = {\cal V}^{-\frac{17}{72}},\
e^{\frac{K}{2}} D_{a_1} D_{a_1}W = {\cal V}^{-\frac{71}{72}},\nonumber\\
& & e^{\frac{K}{2}} {\cal D}_{a_1} D_{a_2}W = {\cal V}^{-\frac{107}{72}},\  e^{\frac{K}{2}} {\cal D}_{a_1} D_{a_3}W = {\cal V}^{-\frac{35}{72}},\
e^{\frac{K}{2}} {\cal D}_{a_1} D_{a_4}W = {\cal V}^{\frac{1}{72}},\nonumber\\
& & e^{\frac{K}{2}} {\cal D}_{a_2} D_{a_2}W = {\cal V}^{-\frac{1}{72}},\ e^{\frac{K}{2}} {\cal D}_{a_2} D_{a_3}W = {\cal V}^{-\frac{3}{4}},\
e^{\frac{K}{2}} {\cal D}_{a_2} D_{a_4}W = {\cal V}^{-\frac{5}{9}},\nonumber\\
& & e^{\frac{K}{2}} {\cal D}_{a_3} D_{a_3}W = {\cal V}^{\frac{1}{72}},\ e^{\frac{K}{2}} {\cal D}_{a_3} D_{a_4}W = {\cal V}^{\frac{37}{72}},\  e^{\frac{K}{2}} {\cal D}_{a_4} D_{a_4}W = {\cal V}^{\frac{73}{72}},
\end{eqnarray}
 one can verify that
$e^{\frac{K}{2}}{\cal D}_{{\cal A}_I}D_{{\cal A}_J}\bar{W}\sim e^{\frac{K}{2}}{\cal D}_{a_I}D_{a_J}\bar{W}$. Now, $\left(e^{\frac{K}{2}}{\cal D}_{\bar{\lambda}}D_{\bar{\Sigma}}\bar{W}\right)\bar{\chi}_L^{{\lambda}}\chi_R^{\Sigma}=\left(e^{\frac{K}{2}}{\cal D}_{\bar{\cal I}}D_{\bar{\cal J}}\bar{W}\right)\bar{\chi}_L^{{\cal I}}\chi_R^{\cal J}$, where $\lambda/\Sigma$ index ${\cal A}_I, {\cal Z}_J$ and ${\cal I}/{\cal J}$ index $a_I,z_i$. What is interesting is that using (\ref{eq:eigenvectors}) and (\ref{eq:a_I+z_i}) e.g.,
\begin{eqnarray}
\label{eq:eq_mass_terms_diag_non-diag_basis}
& & \left(e^{\frac{K}{2}}{\cal D}_{\bar{\cal A}_{1,2}}D_{\bar{\cal A}_{3,4}}\bar{W}\right)\bar{\chi}_L^{{\cal A}_{1,2}}\chi_R^{{\cal A}_{3,4}}\sim\left(e^{\frac{K}{2}}{\cal D}_{\bar{a}_{1,2}}D_{\bar{a}_{3,4}}\bar{W}\right)\bar{\chi}_L^{{\cal A}_{1,2}}\chi_R^{{\cal A}_{3,4}}=\nonumber\\
&& {\hskip -0.3in} \left(\frac{\partial\bar{\cal A}_{1,2}}{\partial\bar{\cal M}_{\cal I}}\right)\left(\frac{\partial\bar{\cal A}_{3,4}}{\partial\bar{\cal M}_{\bar J}}\right) \left(e^{\frac{K}{2}}{\cal D}_{\bar{a}_{1,2}}D_{\bar{a}_{3,4}}\bar{W}\right)\bar{\chi}_L^{{\cal M}_{\cal I}}\chi_R^{{\cal M}_{\cal J}}  \sim \left(e^{\frac{K}{2}}{\cal D}_{\bar{a}_{1,2}}D_{\bar{a}_{3,4}}\bar{W}\right)\bar{\chi}_L^{{a}_{1,2}}\chi_R^{a_{3,4}}.
\end{eqnarray}
This implies that the mass terms, in the large volume limit, are invariant under diagonalization of the open string moduli, and \\
 $\bullet$\begin{equation}
\label{eq:Yhatz1a1a2} \frac{{\cal O}({\cal Z}_i)\ {\rm term\ in}\
e^{\frac{K}{2}}{\cal D}_{{\cal A}_1}D_{{\cal A}_3}W}{\sqrt{K_{{\cal Z}_i\bar{\cal Z}_i}K_{{\cal A}_1\bar{\cal A}_1}K_{{\cal A}_3\bar{\cal A}_3}}}\equiv
\hat{Y}^{\rm eff}_{{\cal Z}_i{\cal A}_1{\cal A}_3}\sim 10^{-3}\times {\cal V}^{-\frac{4}{9}},
\end{equation}
which implies that for ${\cal V}\sim 10^5, \langle {\cal Z}_i\rangle\sim 246 GeV$,
 $\langle {\cal Z}_i\rangle \hat{Y}_{{\cal Z}_1{\cal A}_1{\cal A}_3}\sim MeV$ - about the mass of the electron!\\
$\bullet$
\begin{equation}
\label{eq:Yhatz1a1a2}\frac{{\cal O}({\cal Z}_i)\ {\rm term\ in}\
e^{\frac{K}{2}}{\cal D}_{{\cal A}_2}D_{{\cal A}_4}W}{\sqrt{K_{{\cal Z}_i\bar{\cal Z}_i}K_{{\cal A}_2\bar{\cal A}_2}K_{{\cal A}_4\bar{\cal A}_4}}}\equiv
\hat{Y}^{\rm eff}_{{\cal Z}_i{\cal A}_2{\cal A}_4}\sim 10^{-\frac{5}{2}}\times {\cal V}^{-\frac{4}{9}},
\end{equation}
which implies that for ${\cal V}\sim 10^5,\langle {\cal Z}_i\rangle\sim 246 GeV$,
 $\langle z_i\rangle \hat{Y}_{{\cal Z}_i{\cal A}_1{\cal A}_2}\sim10MeV$ - close to the mass of the up quark!
\\
$~~~$There is an implicit assumption that the VEV of the $D3$-brane position moduli, identified with the neutral components of two Higgs doublets, can RG-flow down to $246 GeV$ - that the same is possible was shown in \cite{ferm_masses_MS}. The RG-flow of the effective physical Yukawas are expected to change by ${\cal O}(1)$ under an RG flow from the string scale down to the EW scale. This can be motivated by looking at RG-flows of the physical Yukawas $\hat{Y}_{\Lambda\Sigma\Delta}$ in MSSM-like models:
\begin{equation}
\label{eq:Yhat_RG-1}
 \frac{d\hat{Y}_{\Lambda\Sigma\Delta}}{dt} = \gamma_{\Lambda}^\kappa\hat{Y}_{\kappa\Sigma\Delta}
+ \gamma_{\Sigma}^{\kappa}\hat{Y}_{\Lambda\kappa\Delta} + \gamma_{\Delta}^\kappa\hat{Y}_{\Lambda\Sigma\kappa},
\end{equation}
 where the anomalous dimension matrix $\gamma_\Lambda^\kappa$, at one loop, is defined as: \\
 $\gamma_\Lambda^\kappa = \frac{1}{32\pi^2}\left(\hat{Y}_{\Lambda\Psi\Upsilon}\hat{Y}^*_{\kappa\Psi\Upsilon} - 2 \sum_{(a)}g_{(a)}^2C_{(a)}(\Phi_\Lambda)\delta_{{\cal I}}^{\kappa}\right)$, $(a)$ indexes the three gauge groups and $\Lambda$, etc. index
the diagonalized basis fields ${\cal A}_I,{\cal Z}_i$. Using
(\ref{eq:W}), one can show that at $M_s$(string scale):
\begin{eqnarray}
\label{eq:Yhats}
& & \hat{Y}_{{\cal Z}_i{\cal Z}_j{\cal Z}_k}\sim{\cal
  V}^{\frac{1}{8}},\ \hat{Y}_{{\cal Z}_i{\cal Z}_i{\cal
    A}_1}\sim{\cal
  V}^{-\frac{79}{40}},\
\hat{Y}_{{\cal Z}_i{\cal Z}_i{\cal A}_2}\sim{\cal
  V}^{-\frac{31}{40}},\ \hat{Y}_{{\cal Z}_i{\cal Z}_i{\cal A}_4}\sim{\cal
  V}^{-\frac{143}{40}},\nonumber\\
& &\hat{Y}_{{\cal Z}_i{\cal Z}_i{\cal A}_3}\sim{\cal
  V}^{-\frac{107}{40}},\ \hat{Y}_{{\cal Z}_i{\cal A}_1{\cal
    A}_1}\sim{\cal
  V}^{-\frac{163}{40}},\  \hat{Y}_{{\cal Z}_i{\cal A}_1{\cal
    A}_2}\sim{\cal
  V}^{-\frac{23}{8}},\ \hat{Y}_{{\cal Z}_i{\cal A}_1{\cal
    A}_3}\sim{\cal
  V}^{-\frac{191}{40}},\nonumber\\
& & \hat{Y}_{{\cal Z}_i{\cal A}_1{\cal
    A}_4}\sim{\cal
  V}^{-\frac{227}{40}},\  \hat{Y}_{{\cal Z}_i{\cal A}_2{\cal A}_2}\sim{\cal
  V}^{-\frac{67}{40}},\ \hat{Y}_{{\cal Z}_i{\cal A}_2{\cal A}_3}\sim{\cal
  V}^{-\frac{143}{40}},\  \hat{Y}_{{\cal Z}_i{\cal A}_2{\cal A}_4}\sim{\cal
  V}^{-\frac{179}{40}}, \nonumber\\
& & \  \hat{Y}_{{\cal Z}_i{\cal A}_3{\cal A}_3}\sim{\cal
  V}^{-\frac{219}{40}},\ \hat{Y}_{{\cal Z}_i{\cal A}_3{\cal
    A}_4}\sim{\cal
  V}^{-\frac{51}{8}},\  \hat{Y}_{{\cal Z}_i{\cal A}_4{\cal
    A}_4}\sim{\cal
  V}^{-\frac{55}{8}},\ \hat{Y}_{{\cal A}_1{\cal A}_1{\cal
    A}_1}\sim{\cal V}^{-\frac{247}{40}},\nonumber\\
& & \hat{Y}_{{\cal A}_1{\cal A}_1{\cal A}_2}
\sim{\cal V}^{-\frac{199}{40}},\  \hat{Y}_{{\cal A}_1{\cal A}_1{\cal A}_3}
\sim{\cal V}^{-\frac{55}{8}},\  \hat{Y}_{{\cal A}_1{\cal A}_1{\cal A}_4}
\sim{\cal V}^{-\frac{311}{40}},\ \hat{Y}_{{\cal A}_1{\cal A}_2{\cal A}_2}\sim
\sim{\cal V}^{-\frac{151}{40}},\nonumber\\
& & \ \hat{Y}_{{\cal A}_1{\cal A}_2{\cal A}_3}
\sim{\cal V}^{-\frac{227}{40}},\ \hat{Y}_{{\cal A}_1{\cal A}_2{\cal A}_4}
\sim{\cal V}^{-\frac{263}{40}},\ \hat{Y}_{{\cal A}_2{\cal A}_4{\cal A}_5}
\sim{\cal V}^{-\frac{339}{40}},\   \hat{Y}_{{\cal A}_1{\cal A}_4{\cal A}_4}
\sim{\cal V}^{-\frac{75}{8}},\nonumber\\
& &\ \hat{Y}_{{\cal A}_2{\cal A}_2{\cal
    A}_2}\sim{\cal V}^{-\frac{103}{40}},\ \hat{Y}_{{\cal A}_2{\cal A}_2{\cal A}_3}
\sim{\cal V}^{-\frac{179}{40}},\  \hat{Y}_{{\cal A}_2{\cal A}_2{\cal A}_4}
\sim{\cal V}^{-\frac{43}{8}},\  \hat{Y}_{{\cal A}_2{\cal A}_3{\cal A}_3}
\sim{\cal V}^{-\frac{299}{60}},\nonumber\\
& &  \hat{Y}_{{\cal A}_2{\cal A}_3{\cal A}_4}
\sim{\cal V}^{-\frac{263}{40}},\  \hat{Y}_{{\cal A}_4{\cal A}_3{\cal A}_3}
\sim{\cal V}^{-\frac{367}{40}},\  \hat{Y}_{{\cal A}_1{\cal A}_3{\cal A}_3}
\sim{\cal V}^{-\frac{283}{40}},\ \hat{Y}_{{\cal A}_2{\cal A}_4{\cal A}_4}
\sim{\cal V}^{-\frac{327}{40}},\nonumber\\
& &   \hat{Y}_{{\cal A}_3{\cal A}_4{\cal A}_4}
\sim{\cal V}^{-\frac{403}{40}},\  \hat{Y}_{{\cal A}_3{\cal A}_3{\cal A}_3}
\sim{\cal V}^{-\frac{331}{40}},\ \hat{Y}_{{\cal A}_4{\cal A}_4{\cal A}_4}
\sim{\cal V}^{-\frac{439}{40}}.
\end{eqnarray}
From (\ref{eq:Yhats}), one sees that $\hat{Y}_{{\cal Z}_i{\cal
    Z}_i{\cal Z}_i}(M_s)\sim{\cal V}^{\frac{1}{8}}$ is the most
dominant physical Yukawa at the string scale. Hence,
\begin{equation}
\label{eq:anom_dimen-2}
\gamma_\Lambda^\Gamma\hat{Y}_{\Gamma\Sigma\Delta}\sim\frac{1}{32\pi^2}\hat{Y}_{\Lambda{\cal
  Z}_i{\cal Z}_i}\hat{Y}^*_{{\cal Z}_i{\cal Z}_i{\cal
  Z}_i}\hat{Y}_{{\cal Z}_i\Sigma\Delta}
-\frac{2\sum_{(a)}g_{(a)}^2C_{(a)}(\Phi_\Gamma)\delta_{\Lambda}^{\Gamma}\hat{Y}_{\Gamma\Sigma\Delta}}{32\pi^2}.
\end{equation}
Now, the first term in (\ref{eq:anom_dimen-2}) is volume suppressed as
compared to the second term at $M_s$; let us assume that the same will
be true up to the EW scale. Using the one-loop solution for
$\frac{g_{(a)}^2(t)}{16\pi^2}
=\frac{ \frac{\beta_{(a)}}{b_{(a)}}}{1 + \beta_{(a)}t}$ in
(\ref{eq:anom_dimen-2}) and therefore (\ref{eq:Yhat_RG-1}), one
obtains:
\begin{equation}
\label{eq:Yhat_sol_1}
\frac{d ln\hat{Y}_{\Lambda\Sigma\Delta}}{dt}\sim -2\left(\sum_{(a)}
\frac{C_{(a)}(\Lambda)\frac{\beta_{(a)}}{b_{(a)}}}{1 + \beta_{(a)}t}
+ \sum_{(a)}
\frac{C_{(a)}(\Sigma)\frac{\beta_{(a)}}{b_{(a)}}}{1 + \beta_{(a)}t} +
\sum_{(a)}
\frac{C_{(a)}(\Delta)\frac{\beta_{(a)}}{b_{(a)}}}{1 + \beta_{(a)}t}\right),
\end{equation}
whose solution yields:
\begin{equation}
\label{eq:Yhat_sol-2}
\hat{Y}_{\Lambda\Sigma\Delta}(t)\sim\hat{Y}_{\Lambda\Sigma\Delta}(M_s)
\prod_{(a)=1}^3\left(1 + \beta_{(a)}t\right)^{\frac{-2\left(C_{(a)}(\Lambda)
    +C_{(a)}(\Sigma) + C_{(a)}(\Delta)\right) }{b_{(a)}}}.
\end{equation}
The solution (\ref{eq:Yhat_sol-2}) justifies the assumption that all
$\hat{Y}_{\Lambda\Sigma\Delta}$s change only by ${\cal O}(1)$ as one
RG-flows down from the string to the EW scale. A similar argument for the RG-evolution of $\hat{Y}^{\rm eff}_{{\cal Z}_i{\cal A}_I{\cal A}_J}$s would proceed as follows. We will for definiteness and due to relevance to the preceding discussion on lepton and quark masses, consider $\frac{d\hat{Y}^{\rm eff}_{{\cal Z}_1{\cal A}_1{\cal A}_3}}{dt}$ and $\frac{d\hat{Y}^{\rm eff}_{{\cal Z}_1{\cal A}_2{\cal A}_4}}{dt}$. Now, 
\begin{equation}
 \frac{d\hat{Y}^{\rm eff}_{{\cal Z}_1{\cal A}_1{\cal A}_3}}{dt} = \gamma_{{\cal Z}_1}^\Lambda\hat{Y}^{\rm eff}_{\Lambda{\cal A}_1{\cal A}_3} + \gamma_{{\cal A}_1}^\Lambda\hat{Y}^{\rm eff}_{{\cal Z}_1\Lambda{\cal A}_3}
+ \gamma_{{\cal A}_3}^\Lambda\hat{Y}^{\rm eff}_{{\cal Z}_1{\cal A}_1\Lambda},
\end{equation}
 where $ \gamma_{{\cal Z}_1}^\Lambda\ni\hat{Y}_{{\cal Z}_1\Upsilon\Sigma}\hat{Y}^*_{\Lambda\Upsilon\Sigma}\sim
\hat{Y}_{{\cal Z}_1{\cal Z}_1{\cal Z}_1}\hat{Y}^*_{\Lambda{\cal Z}_1{\cal Z}_1}$, {etc.}, implying:
\begin{eqnarray}
\label{eq:Yeffzia1a3-III}
& & \frac{d\hat{Y}^{\rm eff}_{{\cal Z}_1{\cal A}_1{\cal A}_3}}{dt}\ni \hat{Y}_{{\cal Z}_1{\cal Z}_1{\cal Z}_1}\hat{Y}^*_{{\cal Z}_1{\cal Z}_1{\cal Z}_1}\hat{Y}^{\rm eff}_{{\cal Z}_1{\cal A}_1{\cal A}_3}
+ \hat{Y}_{{\cal Z}_1{\cal Z}_1{\cal Z}_1}\hat{Y}^*_{{\cal A}_1{\cal Z}_1{\cal Z}_1}\hat{Y}^{\rm eff}_{{\cal A}_1{\cal A}_1{\cal A}_3}\nonumber\\
& &
+ \hat{Y}_{{\cal Z}_1{\cal Z}_1{\cal Z}_1}\hat{Y}^*_{{\cal A}_2{\cal Z}_1{\cal Z}_1}\hat{Y}^{\rm eff}_{{\cal A}_2{\cal A}_1{\cal A}_3} \hat{Y}_{{\cal Z}_1{\cal Z}_1{\cal Z}_1}\hat{Y}^*_{{\cal A}_3{\cal Z}_1{\cal Z}_1}\hat{Y}^{\rm eff}_{{\cal A}_3{\cal A}_1{\cal A}_3}\hat{Y}_{{\cal Z}_1{\cal Z}_1{\cal Z}_1}\hat{Y}^*_{{\cal A}_4{\cal Z}_1{\cal Z}_1}\hat{Y}^{\rm eff}_{{\cal A}_4{\cal A}_1{\cal A}_3},
\end{eqnarray}
and a similar equation for $\frac{d\hat{Y}^{\rm eff}_{{\cal Z}_1{\cal A}_2{\cal A}_4}}{dt}$.
Using (\ref{eq:Yhats}) and:
\begin{eqnarray}
\label{eq:yeff_I or i a1a3 and a2a4}
& & e^{\frac{K}{2}}{\cal D}_{a_{1/2}}D_{a_{3/4}}W\ni {\cal V}^{-\frac{4}{9}}\left(z_i-{\cal V}^{\frac{1}{36}}\right) + {\cal V}^{\frac{2}{3}}\left(a_1-{\cal V}^{-\frac{2}{9}}\right)
+ {\cal V}^{-\frac{5}{6}}\left(a_2-{\cal V}^{-\frac{1}{3}}\right)\nonumber\\
& &  {\hskip 1.1in}+
{\cal V}^{\frac{1}{6}}\left(a_3-{\cal V}^{-\frac{13}{18}}\right)
+ {\cal V}^{\frac{7}{6}}\left(a_4-{\cal V}^{-\frac{11}{9}}\right),
\end{eqnarray}
and (\ref{eq:Yeffzia1a3-III}), one sees that at $M_s$:
\begin{eqnarray}
\label{eq:Yeffzia1a3+a2a4-IV}
& & \gamma_{{\cal Z}_1}^\Lambda\hat{Y}^{\rm eff}_{\Lambda{\cal A}_1{\cal A}_3} + \gamma_{{\cal A}_1}^\Lambda\hat{Y}^{\rm eff}_{{\cal Z}_1\Lambda{\cal A}_3}
+ \gamma_{{\cal A}_3}^\Lambda\hat{Y}^{\rm eff}_{{\cal Z}_1{\cal A}_1\Lambda}\ni \frac{{\cal V}^{\frac{1}{8}+\frac{1}{8} - \frac{4}{9}}}{\sqrt{K_{{\cal Z}_1\bar{\cal Z}_1}K_{{\cal A}_1\bar{\cal A}_1}K_{{\cal A}_3\bar{\cal A}_3}}}\Biggr|_{{\cal V}\sim10^5}\sim{\cal V}^{-\frac{143}{180}},\nonumber\\
& &  \gamma_{{\cal Z}_1}^\Lambda\hat{Y}^{\rm eff}_{\Lambda{\cal A}_2{\cal A}_4} + \gamma_{{\cal A}_2}^\Lambda\hat{Y}^{\rm eff}_{{\cal Z}_1\Lambda{\cal A}_4}
+ \gamma_{{\cal A}_4}^\Lambda\hat{Y}^{\rm eff}_{{\cal Z}_1{\cal A}_2\Lambda}\ni \frac{{\cal V}^{\frac{1}{8}+ \frac{1}{8} - \frac{4}{9}}}{\sqrt{K_{{\cal Z}_1\bar{\cal Z}_1}K_{{\cal A}_2\bar{\cal A}_2}K_{{\cal A}_4\bar{\cal A}_4}}}\Biggr|_{{\cal V}\sim10^5}\sim{\cal V}^{-\frac{25}{36}}.
\end{eqnarray}
Hence, at the string scale, the anomalous dimension matrix contribution is sub-dominant as compared to the gauge coupling-dependent contribution, and by a similar assumption (justified by the solution below)/reasoning:
\begin{equation}
\label{eq:Yeffhat_sol}
\hat{Y}^{\rm eff}_{\Lambda\Sigma\Delta}(t)\sim\hat{Y}^{\rm eff}_{\Lambda\Sigma\Delta}(M_s)
\prod_{(a)=1}^3\left(1 + \beta_{(a)}t\right)^{\frac{-2\left(C_{(a)}(\Lambda)
    +C_{(a)}(\Sigma) + C_{(a)}(\Delta)\right) }{b_{(a)}}}.
\end{equation}
This suggests that possibly, the fermionic superpartners of ${\cal A}_1$ and ${\cal A}_3$ correspond respectively to  $e_L$ and $e_R$ and the fermionic superpartners of ${\cal A}_2$ and ${\cal A}_4$ correspond respectively to the first generation $u_L$ and $u_R$, and ${\cal Z}_I (I=1,2)$ correspond to two Higgs doublets.\\
$~~~$For the purposes of evaluation of various interaction vertices needed to evaluate decay width of gluino in section {\bf 6}, N(LSP) decay channels as well as EDM of electron/neutron in the forthcoming chapters, we will be using the following terms (written out in four-component notation or their two-component analogs and utilizing/generalizing results of \cite{Jockers_thesis}) in the ${\cal N}=1$ gauged supergravity action of Wess and Bagger \cite{Wess_Bagger} with the understanding that $m_{{\rm moduli/modulini}}<<m_{\rm KK}\Bigl(\sim\frac{M_s}{{\cal V}^{\frac{1}{6}}}\Bigr|_{{\cal V}\sim10^{5}/10^{6}} $ $\sim10^{14}GeV\Bigr)$, $M_s=\frac{M_P}{\sqrt{{\cal V}}}\Biggr|_{{\cal V}\sim10^{5/6}}\sim10^{15}GeV$, and that for multiple $D7$-branes, the non-abelian gauged isometry group\footnote{As explained in  \cite{Jockers_thesis}, one of the two Pecci-Quinn/shift symmetries along the RR two-form axions $c^a$ and the four-form axion $\rho_B$ gets gauged due to the dualization of the Green-Schwarz term $\int_{{\bf R}^{1,3}}dD^{(2)}_B\wedge A$ coming from the KK reduction of the Chern-Simons term on $\Sigma_B\cup\sigma(\Sigma_B)$ - $D^{(2)}_B$ being an RR two-form axion. In the presence of fluxes (\ref{eq:flux}) for multiple $D7$-brane fluxes, the aforementioned Green-Schwarz is expected to be modified to $Tr\left(Q_B\int_{{\bf R}^{1,3}}dD^{(2)}_B\wedge A\right)$, which after dualization in turn modifies the covariant derivative of $T_B$ and hence the killing isometry.}, corresponding to the killing vector $6i\kappa_4^2\mu_7\left(2\pi\alpha^\prime\right)Q_B\partial_{T_B}, \\Q_B=\left(2\pi\alpha^\prime\right)\int_{\Sigma_B}i^*\omega_B\wedge P_-\tilde{f}$ arising due to the elimination of of the two-form axions $D_B^{(2)}$ in favor of the zero-form axions
$\rho_B$ under the KK-reduction of the ten-dimensional four-form axion \cite{Jockers_thesis} (which results in a modification of the covariant derivative of $T_B$ by an additive shift given by $6i\kappa_4^2\mu_7\left(2\pi\alpha^\prime\right)Tr(Q_B A_\mu)$)  can be identified with the SM group (i.e. $A_\mu$ is the SM-like adjoint-valued gauge field \cite{Wess_Bagger}):
\begin{eqnarray}
\label{eq:WB_gSUGRA N=1}
& & {\cal L} = g_{YM}g_{T_B {\bar{\cal J}}}Tr\left(X^{T_B}{\bar\chi}^{\bar{\cal J}}_L\lambda_{\tilde{g},\ R}\right)\nonumber\\
&& +ig_{{\cal I}\bar{\cal J}}Tr\Biggl({\bar\chi}^{\bar{\cal I}}_L\Bigl[\slashed{\partial}\chi^{\cal I}_L+\Gamma^i_{Mj}\slashed{\partial} a^M\chi^{\cal J}_L \frac{1}{4}\left(\partial_{a_M}K\slashed{\partial} a_M - {\rm c.c.}\right)\chi^{\cal I}_L\Bigr] \Biggr)\nonumber\\
&&
 +\frac{e^{\frac{K}{2}}}{2}\left({\cal D}_{\bar{\cal I}}D_{\cal J}\bar{W}\right)Tr\left(\chi^{\cal I}_L\chi^{\cal J}_R\right) + g_{T_B\bar{T}_B}Tr\left[\left(\partial_\mu T_B - A_\mu X^{T_B}\right)\left(\partial^\mu T_B - A^\mu X^{T_B}\right)^\dagger\right]  \nonumber\\
&&
+ g_{T_B{\cal J}}Tr\left(X^{T_B}A_\mu\bar{\chi}^{\cal J}_L\gamma^\nu\gamma^\mu\psi_{\nu,\ R}\right) + \bar{\psi}_{L,\ \mu}\sigma^{\rho\lambda}\gamma^\mu\lambda_{\tilde{g},\ L}F_{\rho\lambda} + \bar{\psi}_{L,\ \mu}\sigma^{\rho\lambda}\gamma^\mu\lambda_{\tilde{g},\ L}W^+_\rho W^-_\lambda\nonumber\\
 & & + Tr\left[\bar{\lambda}_{\tilde{g},\ L}\slashed{A}\left(6\kappa_4^2\mu_7(2\pi\alpha^\prime)Q_BK + \frac{12\kappa_4^2\mu_7(2\pi\alpha^\prime)Q_Bv^B}{\cal V}\right) \lambda_{\tilde{g},\ L}\right]\nonumber\\
 & & + \frac{e^K G^{T_B\bar{T}_B}}{\kappa_4^2}6i\kappa_4^2(2\pi\alpha^\prime)Tr\left[Q_BA^\mu\partial_\mu
 \left(\kappa_4^2\mu_7(2\pi\alpha^\prime)^2C^{I\bar{J}}a_I\bar{a}_{\bar J}\right)\right] -\frac{f_{ab}}{4}F^a_{\mu\nu}F^{b\ mu\nu}  \nonumber\\
 & &+ \frac{1}{8}f_{ab}\epsilon^{\mu\nu\rho\lambda}F^a_{\mu\nu}F^b_{\rho\lambda} - \frac{i\sqrt{2}}{4} g \partial_{i/I}f_{ab} Tr\left(\frac{12\kappa_4^2\mu_7(2\pi\alpha^\prime)Q_B^av^B}{\cal V}{\bar\lambda}_{\tilde{g},L}^b\chi^{i/I}_R\right) + {\rm h.c.}\nonumber\\
 & & - \frac{\sqrt{2}}{4}\partial_{i/I}f_{ab}Tr\left({\bar\lambda}_{\tilde{g},R}^a\sigma^{\mu\nu}\chi^{i/I}_L\right)F_{\mu\nu}^b + {\rm h.c.}.
\end{eqnarray}
To evaluate the contribution of various 3-point interaction vertices in context of ${\cal N}=1$ gauged supergravity defined above, one needs to evaluate the full moduli space metric in fluctuations linear in different kind of moduli corresponding to different SM particles. The expansion of moduli space metric, its inverse and derivatives w.r.t to each moduli in terms of fluctuations linear in $z_i\rightarrow z_1 + {\cal V}^\frac{1}{36}{M_P}, a_1\rightarrow a_1 + {\cal V}^{-\frac{2}{9}}{M_P},a_2\rightarrow a_2 + {\cal V}^-{\frac{1}{3}}{M_P}, a_3\rightarrow a_3 + {\cal V}^{-\frac{13}{18}}{M_P},a_3\rightarrow a_3 + {\cal V}^{-\frac{11}{9}}{M_P}$ are given in an appendix  of \cite{gravitino_DM}. 
\vskip -0.5in
\section{Computation of Soft Terms}
We briefly describe evaluation of various soft supersymmetric as well as supersymmetry breaking parameters in the model involving four-Wilson line moduli. The various soft terms are calculated by power series expansion of  superpotential as well as K\"{a}hler potential,
\begin{eqnarray}
\label{eq:KWexp}
&& W= {\hat W}(\Phi) + \mu(\Phi) {\cal Z}_I {\cal Z}_{J}+ \frac{1}{6} Y_{IJK}(\Phi){\cal M}^{I}{\cal M}^{J} {\cal M}^{K}+...,\nonumber\\
 && K = {\hat K}(\Phi,\bar\Phi) + K_{I {\bar J}}(\Phi,\bar\Phi){\cal M}^{I}{\cal M}^{\bar J} + Z(\Phi,\bar\Phi){\cal M}^{I}{\cal M}^{\bar J}+....,
 \end{eqnarray}
where ${\cal M}^{I}=({\cal Z}^{I},{\cal A}^{I}).$
The soft SUSY breaking parameters are calculated by expanding ${\cal N}=1$ supergravity potential, $V= e^{K}\left(K^{I \bar J}D_{I}W D_{\bar J}{\bar W}- 3 \left|W\right|^2\right)$, in the powers of matter fields ${\cal M}^{I}$, after expanding superpotential as well as K\"{a}hler potential as according to equation (\ref{eq:KWexp}).
In gravity-mediated supersymmetry breaking, SUSY gets spontaneously broken in bulk sector by giving a vacuum expectation value to auxiliary F-terms. Hence, to begin with, one needs to evaluate the bulk $F$-terms which in turn entails evaluating the bulk metric. Writing the K\"{a}hler sector of the K\"{a}hler potential as:
 \begin{eqnarray}
&& K\sim-2 ln\left[\left(\sigma_B + \bar{\sigma}_B - \gamma K_{\rm geom}\right)^{\frac{3}{2}} - \left(\sigma_S + \bar{\sigma}_S - \gamma K_{rm geom}\right)^{\frac{3}{2}}\right.\nonumber\\
  && \left. + \sum_{\beta\in H_2^-(CY_3)}n^0_\beta\sum_{(n,m)} cos\left( i n k\cdot(G - \bar{G})g_s + m k\cdot (G + \bar{G}) \right)\right],
  \end{eqnarray} 
  and working near $sin \left( i n k\cdot(G - \bar{G})g_s + m k\cdot (G + \bar{G}) \right)=0$ - corresponding to a local minimum - generates the following components of the bulk metric: 
 \begin{equation}G_{m\bar{n}}\sim
\left(\begin{array}{cccc} {\cal V}^{-\frac{37}{36}} & {\cal V}^{-\frac{59}{36}} & 0 & 0\\
{\cal V}^{-\frac{59}{36}} & {\cal V}^{-\frac{4}{3}} & 0 & 0 \\
0 & 0 & {\cal O}(1) & {\cal O}(1) \\
0 & 0 & {\cal O}(1) & {\cal O}(1)
\end{array}\right),
\end{equation}
which therefore produces the following inverse: 
\begin{equation}
G^{m\bar{n}}\sim\left(\begin{array}{cccc} {\cal V}^{\frac{37}{36}} & {\cal V}^{\frac{13}{18}} & 0 & 0 \\
{\cal V}^{\frac{13}{18}} & {\cal V}^{\frac{4}{3}} & 0 & 0 \\
0 & 0 & {\cal O}(1) & {\cal O}(1) \\
0 & 0 & {\cal O}(1) & {\cal O}(1)
\end{array}\right).
\end{equation}
After spontaneous SUSY breaking, the gravitino mass is given by:
\begin{equation}
m_{3/2}= e^{K}\left|W\right|^2\sim {\cal V}^{-\frac{n^s}{2}-1}M_P.
\end{equation}
 Given that bulk $F$-terms are defined as \cite{softsusy2}: $F^m=e^{\frac{K}{2}}G^{m\bar{n}}D_{\bar{n}}\bar{W}$, one obtains:
\begin{equation}
\label{eq:bulk_F}
F^{\sigma_S}\sim{\cal V}^{-\frac{n^s}{2} + \frac{1}{36}}M_P^2,\ F^{\sigma_B}\sim{\cal V}^{-\frac{n^s}{2} - \frac{5}{18}}M_P^2,\ F^{G^a}\sim{\cal V}^{-\frac{n^s}{2}-1}M_P^2,
\end{equation}
The gaugino mass is given as:
\begin{equation}
\label{eq:gaugino_mass}
m_{\tilde{g}}=\frac{F^m\partial_m T_B}{Re T_B}\lesssim{\cal V}^{\frac{2}{3}}m_{3/2}.
\end{equation}
The analytic form of scalar masses obtained via spontaneous symmetry breaking is given as \cite{softsusy2}: 
$m^{2}_{I}= (m^{2}_{\frac{3}{2}} +V_{0}) - F^{\bar m} F^{n} {\partial_{\bar m}}{\partial_n}\log K_{I {\bar I}}$.
These were calculated in \cite{gravitino_DM} to yield
\begin{equation}
\label{eq:mass_Zi}
m_{{\cal Z}_i}\sim {\cal V}^{\frac{59}{72}}m_{3/2}, m_{{\cal A}_1}\sim\sqrt{{\cal V}}m_{3/2},
\end{equation}
implying a universality in the open moduli masses.
 
Further in \cite{gravitino_DM}, we showed  that the universality in the trilinear $A$-couplings \cite{softsusy2},
\begin{equation}
\label{ew:A}
A_{{\cal I}{\cal J}{\cal K}}=F^m\left(\partial_mK + \partial_m ln Y_{{\cal I}{\cal J}{\cal K}}
+ \partial_m ln\left(K_{{\cal I}\bar{\cal I}}K_{{\cal J}\bar{\cal J}}K_{{\cal K}\bar{\cal K}}\right)\right),
\end{equation}
that was seen in the case of the $D3$-brane position moduli and a single Wilson line modulus in \cite{D3_D7_Misra_Shukla}, is preserved even for the current four-Wilson-line moduli setup and

\begin{equation}
\label{eq:A-univ}
A_{{\cal I}{\cal J}{\cal K}}\sim{\cal V}^{\frac{37}{36}}m_{3/2}\sim\hat{\mu}_{{\cal Z}_1{\cal Z}_2}.
\end{equation}
The physical higgsino mass parameter $\hat{\mu}_{{\cal Z}_1{\cal Z}_2}$ turned out to be given by:
\begin{equation}
\label{eq:muhat_Z1Z2}
\hat{\mu}_{{\cal Z}_1{\cal Z}_2}=\frac{e^{\frac{K}{2}}\mu_{{\cal Z}_1{\cal Z}_2}}{\sqrt{K_{{\cal Z}_1\bar{\cal Z}_1}K_{{\cal Z}_2\bar{\cal Z}_2}}}\sim{\cal V}^{\frac{19}{18}}m_{3/2}.
\end{equation}
 Further,
\begin{eqnarray}
\label{eq:muhatB_Z1Z2}
& & \left(\hat{\mu}B\right)_{{\cal Z}_1{\cal Z}_2}=\frac{e^{-i arg(W)+\frac{K}{2}}}{\sqrt{K_{{\cal Z}_1\bar{\cal Z}_1}K_{{\cal Z}_2\bar{\cal Z}_2}}}F^m\left(\partial_mK \mu_{{\cal Z}_1{\cal Z}_2} + \partial_m\mu_{{\cal Z}_1{\cal Z}_2} - \mu_{{\cal Z}_1{\cal Z}_2}\partial_m ln\left(K_{{\cal Z}_1\bar{\cal Z}_1}K_{{\cal Z}_2\bar{\cal Z}_2}\right)\right)\nonumber\\
& &  \sim\hat{\mu}_{{\cal Z}_1{\cal Z}_2}\left(F^m\partial_mK+F^{\sigma_S} - F^m\partial_m ln\left(K_{{\cal Z}_1\bar{\cal Z}_1}K_{{\cal Z}_2\bar{\cal Z}_2}\right)\right) \sim{\cal V}^{\frac{19}{18}+\frac{37}{36}}m^2_{3/2}\sim\hat{\mu}^2_{{\cal Z}_1{\cal Z}_2},
\end{eqnarray}
an observation which will be very useful in obtaining a light Higgs of mass $125 \ GeV$.
\section{Realizing Large Volume $\mu$-split-like  SUSY}
In the previous section, we calculated masses of position moduli which we have identified with Higgs doublets as according to the argument given below equation (\ref{eq:Yeffhat_sol}) at string scale. In order to obtain the possibility of getting a light Higgs at electroweak scale using the fine-tuning argument of \cite{HamidSplitSUSY}, in this section, we first analyse the contribution of masses of Higgs doublets obtained after SUSY breaking at electroweak scale.  The Higgs doublets are defined as:
\beqn
\label{massmat}
&& H_{1}=D_{h_{11}} H_{u} +D_{h_{12}}H_{d}, H_{2}=D_{h_{21}}H_{u} +D_{h_{22}} H_{d}.
\eeqn
where $D_h=\left(\begin{array}{ccccccc}
\cos \frac{\theta_h}{2} & -\sin \frac{\theta_h}{2} e^{-i\phi_{h}}\\
\sin \frac{\theta_h}{2} e^{i\phi_{h}} & \cos \frac{\theta_h}{2}
 \end{array} \right),
D_{h}^\dagger M_{h}^2 D_h={\rm diag}(M^{2}_{H_1},M^{2}_{H_2})$, 
and $\tan \theta_h=
\frac{2|M_{h_{21}}^2|}{M_{h_{11}}^2-M_{h_{22}}^2}$ for a particular range of ${{-\pi}\over {2}} \leq  \theta_h \leq {{\pi}\over {2}}$.

The Higgs masses after soft supersymmetry breaking is given by $(m_{Z_i}^{2}+\hat{\mu}_{Z_i}^{2})^{1/2}$ (where $m_{Z_i}$'s correspond to  mobile $D3$- Brane position moduli masses (to be identified with soft Higgs scalar mass parameter)) and the Higgsino mass is given by  $\hat{\mu}_{Z_i}$. However, due to lack of universality in moduli masses but universality in trilinear $A_{ijk}$ couplings, we use solution of RG flow equation for moduli masses as given in \cite{Nath+Arnowitt}. Using the same,
\begin{equation}
\label{eq:heavy_H_I}
m_{{\cal Z}_{1}}^{2}(t)=m_{o}^{2}(1+\delta_1)+m_{1/2}^{2} g(t)+\frac{3}{5}S_0p,
\end{equation}
where  
\begin{equation}
S_0=Tr(Ym^2)=m_{{\cal Z}_2}^2-m_{{\cal Z}_1}^2+\sum_{i=1}^{n_g}(m_{\tilde q_L}^2-2
m_{\tilde u_R}^2 +m_{\tilde d_R}^2 - m_{\tilde l_L}^2 + m_{\tilde e_R}^2),
\end{equation} where all the masses are at the string scale and
 $n_g$ is the number of generations. $p$ is defined by
$p=\frac{5}{66}[1-(\frac{\tilde\alpha_1(t)}
{\tilde\alpha_1(M_s)})]$
where  $\tilde\alpha_1\equiv g_1^2/(4\pi)^2$ and  $g_1$ is the
$U(1)_Y$ gauge coupling constant. Further,
\begin{equation}
\label{eq:light_H_I}
 m_{{\cal Z}_{2}}^{2}(t) = m_0^2\Delta_{{\cal Z}_2} + m_{1/2}^{2}e(t) + A_{o}m_{1/2}f(t) + m_{o}^{2}h(t) -k(t)A_{o}^{2} - \frac{3}{5}S_0p,
\end{equation}
where $\Delta_{{\cal Z}_2}$ is given by $\Delta_{{\cal Z}_2}=\frac{(D_0-1)}{2}(\delta_2+\delta_3+\delta_4)+ \delta_2;  D_0=1-6 {\cal Y}_t \frac{F(t)}{E(t)}$.
 Here ${\cal Y}_t\equiv\hat{Y}_t^2(M_s)/(4\pi)^2$ where $\hat{Y}_t(M_s)$ is the physical top Yukawa coupling at the string scale which following \cite{Ibanez_et_al} will be set to 0.08, and $E(t)=(1+\beta_3t)^{\frac{16}{3b_3}}(1+\beta_2t)^{\frac{3}{b_2}}
(1+\beta_1t)^{\frac{13}{9b_1}}$.
   $\beta_i\equiv\alpha_i(M_s)b_i/4\pi$
($\alpha_1=(5/3)\alpha_Y$), $b_i$ are
the one loop
beta function coefficients defined  by $(b_1,b_2,b_3)=
(33/5,1,-3)$,
and $F(t)=\int_0^t E(t)dt$.

In the dilute flux approximation,
$g_1^2(M_S)=g_2^2(M_S)=g_3^2(M_S)$. To ensure $E(t)\in\bf{R}$, the $SU(3)$-valued $1+\beta_3t >0$, which for $t=57$ implies that $g_3^2(M_s)<\frac{(4\pi)^2}{3\times57}\sim{\cal O}(1)$. Hence,  $g_3^2(M_s)=0.4$ is what we will be using (see \cite{Dhuria+Misra_mu_Split_SUSY}). From (\ref{eq:heavy_H_I}) and (\ref{eq:light_H_I}), the results of scalar masses as given in section {\bf 3}, and the values of t-dependent functions given in an appendix of \cite{Dhuria+Misra_mu_Split_SUSY}, one sees that:
\begin{eqnarray}
\label{eq:heavy_H_II}
 && m_{{\cal Z}_1}^2(M_{EW})\sim m_{{\cal Z}_1}^2(M_s)+{(0.39)}{\cal V}^{\frac{4}{3}}m_{3/2}^2+\frac{1}{22}\times\frac{19\pi}{100}\times S_0,\\
\label{eq:light_H_II}
&&m_{{\cal Z}_2}^2(M_{EW})\sim  {(0.32)}{\cal V}^{\frac{4}{3}}m_{3/2}^2+{(-0.03)}n^s\hat{\mu}_{{\cal Z}_1{\cal Z}_2}{\cal V}^{\frac{2}{3}}m_{3/2}\nonumber\\
&&+{(0.96)} m_0^2-{(0.01)}(n^s)^2\hat{\mu}_{{\cal Z}_1{\cal Z}_2} -\frac{19\pi}{2200}\times S_0,
\end{eqnarray}
where we used $A_{{\cal Z}_i{\cal Z}_i{\cal Z}_i}\sim n^s\hat{\mu}_{{\cal Z}_1{\cal Z}_2}$ (\ref{eq:A-univ}).
The solution for RG flow equation for $\hat{\mu}^2$ to one loop order is given by \cite{Nath+Arnowitt}:
\begin{eqnarray}
\label{eq:muhat_I}
& &{\hskip -0.4in} \hat{\mu}^2_{{\cal Z}_i{\cal Z}_i}=-\left[m_0^2 C_1+A_0^2 C_2 +m_{\frac{1}{2}}^2C_3+m_{\frac{1}{2}}
A_0C_4-\frac{1}{2}M_Z^2 +\frac{19\pi}{2200}\left(\frac{tan^2\beta+1}{tan^2\beta-1}\right)S_0\right],
\end{eqnarray}
  wherein $C_1=\frac{1}{tan^2\beta-1}(1-\frac{3 D_0-1}{2}tan^2\beta) + 
\frac{1}{tan^2\beta-1}\Bigl(\delta_1-\delta_2tan^2\beta - \frac{ D_0-1}{2}(\delta_2 +
\delta_3+\delta_4)tan^2\beta\Bigr)$, $C_2=-\frac{tan^2\beta}{tan^2\beta-1}k(t),  C_3=-\frac{1}{tan^2\beta-1}\left(g(t)- tan^2\beta e(t)\right),~C_4=-\frac{tan^2\beta}{tan^2\beta-1}f(t)$, and the functions $e(t),f(t),g(t),k(t)$ are defined
in an appendix of \cite{Dhuria+Misra_mu_Split_SUSY}.  In the large $tan\beta$ (but less than 50)-limit, one sees that:
\begin{eqnarray}
\label{eq:muhat_II}
&& \hat{\mu}_{{\cal Z}_1{\cal Z}_2}^2\sim-\Bigl[-m_0^2-{(0.01)}(n^s)^2\hat{\mu}_{{\cal Z}_1{\cal Z}_2}^2+{(0.32)}{\cal V}^{\frac{4}{3}}m_{3/2}^2-1/2 M_{EW}^2\nonumber\\
&&+{(0.03)}{\cal V}^{\frac{2}{3}}n^s\hat{\mu}_{{\cal Z}_1{\cal Z}_2}m_{3/2} +\frac{19\pi}{2200}S_0\Bigr].
\end{eqnarray}
From (\ref{eq:light_H_II}) and (\ref{eq:muhat_II}), one therefore sees that the mass-squared of one of the two Higgs doublets, $m_{H_d}^2$, at the $EW$ scale is given by:
\begin{equation}
\label{eq:muhat_III}
  m^2_{H_d}=m^2_{{\cal Z}_2}+\hat{\mu}^2_{{\cal Z}_i{\cal Z}_i}= 2m^2_{0}-{(0.06)}{\cal V}^{\frac{2}{3}}n^s\hat{\mu}_{{\cal Z}_1{\cal Z}_2}m_{3/2}+\frac{1}{2}M_{EW}^2-\frac{19\pi}{1100}S_0.
\end{equation}
From \cite{D3_D7_Misra_Shukla}, we notice, ${\cal V}^{\frac{2}{3}}\hat{\mu}_{{\cal Z}_1{\cal Z}_2}m_{3/2}\sim m_0^2$, using which in (\ref{eq:muhat_III}), one sees that for an ${\cal O}(1)\ n^s$,
\begin{equation}
\label{eq:muhat_IV}
m^2_{H_d}(M_{EW})\sim  1.94 m_0^2 + \frac{1}{2}M_{EW}^2-\frac{19\pi}{1100}S_0.
\end{equation}
We have assumed at $ m^2_{{\cal Z}_2}(M_s)\sim m^2_0$ (implying $\delta_2=0$ but $ \delta_{1,3,4}\neq0$). 
  Further,
\begin{eqnarray}
\label{eq:heavy_Higgs}
& & m^2_{H_u}(M_{EW})=\left(m^2_{{\cal Z}_1}+\hat{\mu}_{{\cal Z}_1{\cal Z}_2}^2\right)(M_{EW})
 \sim m^2_{{\cal Z}_1}(M_s)(1+ \delta_1)+\frac{1}{2}M_{EW}^2 + m_0^2 \nonumber\\
  & & {\hskip 0.8in} -{(0.03)}{\cal V}^{\frac{2}{3}}n^s\hat{\mu}_{{\cal Z}_1{\cal Z}_2}m_{3/2}+ {(0.01)}(n^s)^2\hat{\mu}_{{\cal Z}_1{\cal Z}_2}^2  \nonumber\\
  & & {\hskip 0.8in} \sim (1.97+ \delta_1) m^2_0 +\frac{1}{2}M_{EW}^2 + {(0.01)}(n^s)^2\hat{\mu}_{{\cal Z}_1{\cal Z}_2}^2.
\end{eqnarray}
 By assuming $\left(\hat{\mu}B\right)_{{\cal Z}_1{\cal Z}_2}\sim\hat{\mu}^2_{{\cal Z}_1{\cal Z}_2}$ - see (\ref{eq:muhatB_Z1Z2}) -  to be valid at the string and EW scales, the Higgs mass matrix at the $EW$-scale can thus be expressed as:
\begin{eqnarray}
\label{eq:Higss_mass_matrix}
& & \left(\begin{array}{cc}
m^2_{H_u} & \hat{\mu}B\\
\hat{\mu}B & m^2_{H_d}\end{array}\right) \sim\left(\begin{array}{cc}
m^2_{H_u} & \xi\hat{\mu}^2\\
\xi\hat{\mu}^2 & m^2_{H_d}
\end{array}\right),
\end{eqnarray}
$\xi$ being an appropriately chosen ${\cal O}(1)$ constant - see (\ref{eq:evs_3}). The eigenvalues are given by $ \frac{1}{2}\biggl(m^2_{H_u}+m^2_{H_d}\pm\sqrt{\left(m^2_{H_u}-m^2_{H_d}\right)^2+4\xi^2\hat{\mu}^4}\biggr).$
Using equation (\ref{eq:muhat_II}), for $\delta_1 ={\cal O}(0.1)$ and ${\cal O}(1)\ n^s$, we have
\vskip -0.4in
\begin{eqnarray}
\label{eq:evs_1}
& & m^2_{H_u}+m^2_{H_d}\sim   m_0^2-0.06S_0+...  \, m^2_{H_u}-m^2_{H_d}\sim  m_0^2+0.06S_0+... \ ,\nonumber\\
& & \hat{\mu}_{{\cal H}_u{\cal H}_d}^2\sim   m_0^2 -0.03S_0+... 
\end{eqnarray}
\vskip -0.2in
Utilizing above, one sees that the eigenvalues are:
\begin{eqnarray}
\label{eq:evs_2}
& {\hskip -0.4in}m^2_{H_{1,2}}= m_0^2 -0.06S_0+.... &\pm\sqrt{\left(  m_0^2 +0.06S_0+...\right)^2+4\xi^2\left(m_0^2- 0.03S_0\right)^2}. 
\end{eqnarray}
 Considering 
 \begin{equation}
 \label{eq:evs_3}
 S_0\sim (-4.23) m_0^2, {\rm and}~~ \xi^2 \sim  {\frac{1}{5}} + \frac{1}{16}\frac{m^2_{EW}}{m_0^2},
 \end{equation}
  translates into getting  a light Higgs (corresponding to the negative sign of the square root) of the order $125 \ GeV$ and both heavy Higgs (corresponding to the positive sign of the square root) and higgsino mass parameter $\hat\mu$ of the order ${\cal V}^{\frac{59}{72}}m_{\frac{3}{2}}$.  This shows the possibility of realizing  ``$\mu$-split-like  SUSY" scenario in the context of LVS phenomenology named as ``Large Volume $\mu$-split-like  SUSY" scenario.
  \vskip -0.5in
 \section{Neutralino/Chargino Mass Matrix}
 In this section we work out the eigenalues and eigenvectors of the neutralino (chargino) mass matrix  
 formed by mixing neutral (charged) gaugino and neutral (charged) higgsino after electroweak symmetry breaking. In ${\cal N}=1$ gauged supergravity, the interaction vertex corresponding to Higgs-gaugino-higgsino term is given by ${\cal L}= g_{YM}g_{T^B{\cal Z}^i}X^B\tilde{H}^i\lambda^{i} + \partial_{{\cal Z}_i}T_B D^{B}\tilde{H}^i\lambda^{i} $ where $\lambda^{i}$ corresponds to gaugino (such as the Bino/Wino). Expanding the same in the fluctuations linear in $Z_i$, we have
\begin{equation}
\label{eq:gYM_ReT_II}
g_{YM}g_{T^B{\cal Z}^i}X^{B}= {\tilde f} {\cal V}^{-2}\frac{{\cal Z}_i}{M_P}, (\partial_{{\cal Z}_i}T_B) D^{B}\sim   {\tilde f} {\cal V}^{-\frac{4}{3}}\frac{{\cal Z}_i}{M_P}
\end{equation}
After giving a VEV to ${\cal Z}_i$, the interaction vertex corresponding to mixing between gaugino and higgsino will be given as:
\begin{equation}
\label{eq:gYM_ReT_II}
C^{\tilde \lambda^{+}-{{\tilde H}^{-}_{d}}}/C^{\tilde \lambda^{-}-{{\tilde H}^{+}_{u}}}/C^{\tilde \lambda^{0}-{{\tilde H}^{0}_{u}}/{\tilde H}^{0}_{d}}=\frac{ {\tilde f} {\cal V}^{-\frac{4}{3}}}{\sqrt{K_{{\cal Z}_i {\cal Z}_i}K_{{\cal Z}_i {\cal Z}_i}}}\sim   {\tilde f} {\cal V}^{-\frac{1}{3}}\frac{v}{M_P},{\rm where}~v=246~GeV.
\end{equation}
Using value of  $M_{{\tilde \lambda}^{0}} = {\cal V}^{\frac{2}{3}} m_{\frac{3}{2}}$, $M_{{{\tilde H}^{0}}_u/{{\tilde H}^{0}}_d}= {\cal V}^{\frac{59}{72}}m_{\frac{3}{2}}$ and $m_{\frac{3}{2}}= {\cal V}^{-2}M_P$  at electroweak scale,  yields a neutralino mass matrix\footnote{We keep both neutral gaugino (Wino as well as Bino) at the same footing, hence consider $3\times 3$ neutralino mass matrix.}:
\begin{eqnarray}
\label{eq:neut_mass}
& & \left(\begin{array}{ccc}
  {\cal V}^{-\frac{4}{3}}  &  \left(\frac{v}{M_P}\right) {\tilde f}{\cal V}^{-\frac{1}{3}} &  \left(\frac{v}{M_P}\right) {\tilde f}{\cal V}^{-\frac{1}{3}}\\
- \left(\frac{v}{M_P}\right) {\tilde f}{\cal V}^{-\frac{1}{3}}& 0 & {\cal V}^{-\frac{85}{72}}\\
- \left(\frac{v}{M_P}\right) {\tilde f}{\cal V}^{-\frac{1}{3}}& {\cal V}^{-\frac{85}{72}} & 0
\end{array}\right)M_P,
\end{eqnarray}
with eigenvalues:
\begin{eqnarray}
\label{eq:neut evalues}
& & \Biggl\{\frac{1}{V^{85/72}}, \frac{  {\cal V}^{4/3}- {\cal V}^{85/72}+\sqrt{  {\cal V}^{8/3}-2  {\cal V}^{181/72}+ {\cal V}^{85/36}+4 {\tilde f}^2  v^2 {\cal V}^{109/36}}}{2
   V^{157/72}},\nonumber\\
& &  \frac{ {\cal V}^{4/3}+ {\cal V}^{85/72}+\sqrt{  {\cal V}^{8/3}-2   {\cal V}^{181/72}+ {\cal V}^{85/36}+4 {\tilde f}^2  v^2 {\cal V}^{109/36}}}{2
  {\cal V}^{157/72}}\Biggr\}M_P
\end{eqnarray}
and normalized eigenvectors:
\begin{eqnarray}
\label{eq:neutralinos_i}
& & \tilde{\chi}_1^0\sim\frac{-{\tilde H}^{0}_{u}+{\tilde H}^{0}_{d}}{\sqrt{2}},~ {\rm and}~m_{{\tilde \chi}^{0}_{1}}\sim {\cal V}^{\frac{59}{72}}m_{\frac{3}{2}},\nonumber\\
& & \tilde{\chi}_2^0\sim \left(\frac{v}{  M_P}{\tilde f}{\cal V}^{\frac{5}{6}}\right){\tilde \lambda}^0+\frac{{\tilde H}^{0}_{u}+{\tilde H}^{0}_{d}}{\sqrt{2}}~ {\rm and}~m_{{\tilde \chi}^{0}_{2}}\sim {\cal V}^{\frac{59}{72}}m_{\frac{3}{2}}, \nonumber\\
& & \tilde{\chi}_3^0\sim-{\tilde \lambda}^0+\left(\frac{v}{M_P}{\tilde f}{\cal V}^{\frac{5}{6}}\right)\left({\tilde H}^{0}_{u}+{\tilde H}^{0}_{d}\right)~ {\rm and}~m_{{\tilde \chi}^{0}_{3}}\sim {\cal V}^{\frac{2}{3}}m_{\frac{3}{2}}.
\end{eqnarray}
For  higgsino-doublets
${\tilde H_u}=\left({\tilde H}^{0}_u , {\tilde H}^{+}_u \right), {\tilde H}_d=\left({\tilde H}^{-}_d, {\tilde H}^{0}_d \right)$,
the chargino mass matrix is given as:
\begin{equation}
 M_{{\tilde \chi}^{-}}=\left(\begin{array}{ccccccc}
  M^{2}_{{\tilde H}^{-}_{d}} & C^{\tilde \lambda^{-}-{{\tilde H}^{-}_{d}}}\\
C^{\tilde \lambda^{-}-{{\tilde H}^{-}_{d}}} &  M^{2}_{{\tilde \lambda}^{-}}
 \end{array} \right),  M_{{\tilde \chi}^{+}}=\left(\begin{array}{ccccccc}
  M^{2}_{{\tilde H}^{+}_{u}} & C^{\tilde \lambda^{+}-{{\tilde H}^{+}_{u}}}\\
C^{\tilde \lambda^{+}-{{\tilde H}^{+}_{u}}} &  M^{2}_{{\tilde \lambda}^{+}}
 \end{array} \right)
\end{equation}
Incorporating value of  $M_{{\tilde \lambda}^{+}}=M_{{\tilde \lambda}^{-}}= {\cal V}^{\frac{2}{3}} m_{\frac{3}{2}}$, $M_{{{\tilde H}^{-}}_{d}}=  M_{{{\tilde H}^{+}}_{u}}= {\cal V}^{\frac{59}{72}}m_{\frac{3}{2}}$ and $m_{\frac{3}{2}}= {\cal V}^{-2}M_P$  at electroweak scale, we have
\begin{equation}
 M_{{\tilde \chi}^{\pm}}=\left(\begin{array}{ccccccc}
  {\cal V}^{-\frac{4}{3}} & \left(\frac{v}{M_P}\right) {\tilde f}{\cal V}^{-\frac{1}{3}}\\
\left( \frac{v}{M_P}\right){\tilde f}{\cal V}^{-\frac{1}{3}}  &V^{-\frac{85}{72}}
 \end{array} \right)M_P,
\end{equation}
giving eigenvalues:
\begin{eqnarray}
\label{eq:charginoevalues}
& & \Biggl\{\frac{  {\cal V}^{4/3}+ {\cal V}^{85/72}-\sqrt{  {\cal V}^{8/3}-2  {\cal V}^{181/72}+ {\cal V}^{85/36}+4 {\tilde f}^2  v^2 {\cal V}^{109/36}}}{2
   V^{157/72}},\nonumber\\
& & \frac{ {\cal V}^{4/3}+ {\cal V}^{85/72}+\sqrt{  {\cal V}^{8/3}-2   {\cal V}^{181/72}+ {\cal V}^{85/36}+4 {\tilde f}^2  v^2 {\cal V}^{109/36}}}{2
  {\cal V}^{157/72}}\Biggr\}M_P,  \end{eqnarray}
and normalized eigenvectors:
\begin{eqnarray}
\label{eq:charginoI}
&& {\tilde \chi}^{+}_{1}= -{\tilde H}^{+}_{u}+\left(\frac{v}{M_P}{\tilde f}{\cal V}^{\frac{5}{6}}\right){\tilde \lambda}^{+}_{i},{\tilde \chi}^{-}_{1}=  -{\tilde H}^{-}_{d}+\left(\frac{v}{M_P}{\tilde f}{\cal V}^{\frac{5}{6}}\right){\tilde \lambda}^{-}_{i},~ {\rm and}~m_{{\tilde \chi}^{\pm}_{1}}\sim {\cal V}^{\frac{59}{72}}m_{\frac{3}{2}}\nonumber\\
&&  {\tilde \chi}^{+}_{2}= {\tilde \lambda}^{+}_{i} +\left(\frac{v}{M_P}{\tilde f}{\cal V}^{\frac{5}{6}}\right){\tilde H}^{+}_{u},{\tilde \chi}^{-}_{2}= {\tilde \lambda}^{-}_{i} +\left(\frac{v}{M_P}{\tilde f}{\cal V}^{\frac{5}{6}}\right) {\tilde H}^{-}_{d},~ {\rm and}~m_{{\tilde \chi}^{\pm}_{2}}\sim {\cal V}^{\frac{2}{3}}m_{\frac{3}{2}}.
\end{eqnarray}
The results of mass scales of all SM as well as superpartners are summarized in Table~2.1.
 \begin{table}[htbp]
\centering
\begin{tabular}{|l|l|} \hline
Quark mass & $ M_{q}\sim O(10) MeV $\\
Lepton mass & $ M_{l}\sim {\cal O}(1) MeV $\\  \hline
Gravitino mass &  $ m_{\frac{3}{2}}\sim{\cal V}^{-\frac{n^s}{2} - 1} M_{P}; n_s=2$ \\
Gaugino mass & $ M_{\tilde g}\sim  {\cal V}^{\frac{2}{3}}m_{\frac{3}{2}}$\\
(Lightest) Neutralino/Chargino & $ M_{{\chi^0_3}/{\chi^{\pm}_3}}\sim {\cal V}^{\frac{2}{3}}m_{\frac{3}{2}}$\\  
mass & \\
\hline
$D3$-brane position moduli  & $ m_{{\cal Z}_i}\sim {\cal V}^{\frac{59}{72}}m_{\frac{3}{2}}$ \\
(Higgs) mass & \\
Wilson line moduli & $ m_{\tilde{\cal A}_I}\sim {\cal V}^{\frac{1}{2}}m_{\frac{3}{2}}$\\
(sfermion mass ) & ${I =1,2,3,4}$\\ \hline
A-terms & $A_{pqr}\sim n^s{\cal V}^{\frac{37}{36}}m_{\frac{3}{2}}$\\
& $\{p,q,r\} \in \{{{\tilde{\cal A}_I}},{{\cal Z}_i}\}$\\ \hline
Physical $\mu$-terms & $\hat{\mu}_{{\cal Z}_i{\cal Z}_j}$ \\
 (Higgsino mass) & $\sim{\cal V}^{\frac{37}{36}}m_{\frac{3}{2}}$ \\ \hline
Physical $\hat{\mu}B$-terms & $\left(\hat{\mu}B\right)_{{\cal Z}_1{\cal Z}_2}\sim{\cal V}^{\frac{37}{18}}m_{\frac{3}{2}}^2$\\ \hline
\end{tabular}
\caption{Mass scales of first generation of SM as well supersymmetric, and soft SUSY breaking parameters.}
\label{table:mass scales}
\end{table}
\vskip -0.5in
\section{Gluino Decays}
One of the other striking features of $\mu$-split-like  SUSY scenario is based on longevity of gluinos
since the squarks which mediate its decay, are ultra-heavy. In this spirit, we estimate the gluino decay life time at tree-level as well as one-loop level in the context of ${\cal N}=1$ gauged supergravity limit of  our ``local large volume $D3/D7$ model". Since squarks are heavier than gluino, two-body decays of gluino into squarks and quarks are kinematically forbidden at tree-level.  

 \underline{${\bf \tilde{g}\rightarrow q \ {\bar q} \ \chi^{0}_n}$}
 
We first discuss gluino three-body decays that involve the process like $\tilde{g}\rightarrow q{\bar q}\chi^{0}_n$ - $\tilde{g}$ being a gaugino, $q/{\bar q}$ being quark/anti-quark and $\chi_n$ being a neutralino. More specifically, e.g., the gluino decays into an anti-quark and an off-shell squark and the off-shell squark decays into a quark and neutralino. The eigenvalues as well as eigenvectors of neutralino are evaluated in section {\bf 5}. For gluino-decay studies, it is $\tilde{\chi}_3^0$ - largely a gaugino $\lambda^0$ with a small admixture of the higgsinos - which will be relevant. For squark-quark-neutralino vertex, we will work out squark-quark-gaugino vertex and also the squark-quark-higgsino vertex, and then add these contributions.
\begin{figure}
 \begin{center}
\begin{picture}(300,170)(50,0)
\Text(90,130)[]{$\tilde{g}$}
\Line(60,120)(110,120)
\Gluon(60,120)(110,120){5}{4}
\ArrowLine(110,120)(140,150)
\Text(147,150)[]{${q}_I$}
\DashArrowLine (130,90)(110,120){4}
\Text(107,100)[]{$\tilde{q}_i$}
\ArrowLine(160,120)(130,90)
\Text(167,120)[]{${\bar q_J}$}
\ArrowLine(130,90)(160,60)
\Text(167,65)[]{$\tilde{\chi}_3^0$}
\Text(110,30)[]{(a)}
\Text(250,130)[]{$\tilde{g}$}
\Line(210,120)(260,120)
\Gluon(210,120)(260,120){5}{4}
\ArrowLine(290,150)(260,120)
\Text(297,150)[]{$\bar{q}_J$}
\DashArrowLine(260,120)(280,90){4}{}
\Text(257,100)[]{$\tilde{q}_j$}
\ArrowLine(280,90)(310,120)
\Text(320,120)[]{${q}_I$}
\ArrowLine(310,60)(280,90)
\Text(317,65)[]{$\tilde{\chi}_3^0$}
\Text(280,30)[]{(b)}
\end{picture}
\end{center}
\vskip-0.5in
\caption{Three-body gluino decay diagrams.}
\label{fig:treegdec}
\end{figure}
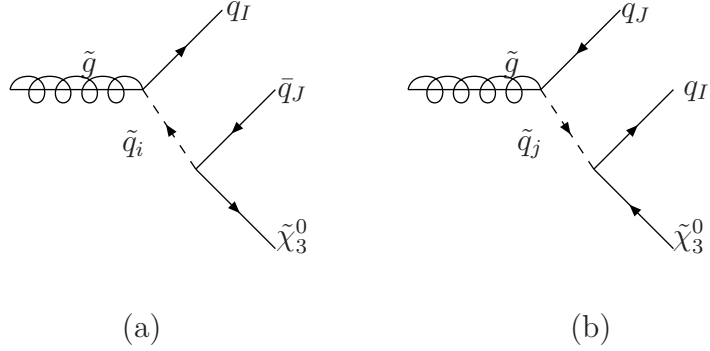
To calculate the contribution of various interaction vertices shown in Figure~2.1 in the context of ${\cal N}=1$ gauged supergravity, one needs to consider the following terms of gauged supergravity action  of Wess and Bagger \cite{Wess_Bagger}.
\begin{eqnarray}
\label{eq:fermion+fermion+sfermion}
& & {\hskip -0.3in}{\cal L} = g_{YM}g_{\alpha {\bar J}}X^\alpha{\bar\chi}^{\bar J}\tilde{\lambda^{0}}+ig_{i{\bar J}}{\bar\chi}^{\bar J}\left[{\bar\sigma}\cdot\partial\chi^i+\Gamma^i_{Lk}{\bar\sigma}\cdot\partial a^L\chi^k
+\frac{1}{4}\left(\partial_{a_L}K{\bar\sigma}\cdot a_L - {\rm c.c.}\right)\chi^i\right] \nonumber\\
& & {\hskip -0.3in}+\frac{e^{\frac{K}{2}}}{2}\left({\cal D}_iD_JW\right)\chi^i\chi^J+{\rm h.c.},
\end{eqnarray}
where $X^\alpha$ corresponds to the components of a killing isometry vector. From here, one notes that $X^\alpha=-6i\kappa_4^2\mu_7Q_\alpha$, where $\alpha=B, Q_\alpha=2\pi\alpha^\prime\int_{T_B}i^*\omega_\alpha\wedge P_-\tilde{f}$ where $P_-$ is a harmonic zero-form on $\Sigma_B$ taking value +1 on $\Sigma_B$ and $-1$ on $\sigma(\Sigma_B)$ - $\sigma$ being a holomorphic isometric involution as part of the Calabi-Yau orientifold - and $\tilde{f}\in\tilde{H}^2_-(\Sigma^B)\equiv{\rm coker}\left(H^2_-(CY_3)\stackrel{i^*}{\rightarrow}H^2_-(\Sigma^B)\right)$. Also,
\begin{eqnarray}
\label{eq:DdW_def}
{\cal D}_iD_jW=\partial_i\partial_jW + \left(\partial_i\partial_jK\right)W+\partial_iKD_JW +
\partial_JKD_iW - \left(\partial_iK\partial_jK\right)W - \Gamma^k_{ij}D_kW. \nonumber\\
\end{eqnarray}
The gaugino-quark-squark vertex is given as:
 $g_{YM}g_{{ J}B^{*}}X^{{*}B}{\bar\chi}^{\bar J}\lambda_{\tilde{g}}+ + \partial_{a_2}T_B D^{B}{\bar\chi}^{\bar a_1}\lambda^{0}$. Utilizing (\ref{eq:K}) and $X^{B}=-6i\kappa_4^2\mu_7Q_{B}, \kappa_4^2\mu_7\sim \frac{1}{\cal V}, g_{YM}\sim{\cal V}^{-\frac{1}{36}}, \footnote{A small note on dimensional analysis: $\kappa_4^2\mu_7\left(2\pi\alpha^\prime\right)Q_Bg_{\sigma^B{\bar a}_{\bar j}}{\bar\chi}^{\bar j}\lambda$ has dimensions $M_P^4$ - utilizing $\kappa_4^2\mu_7\sim{\cal V}^{-1}\left(2\pi\alpha^\prime\right)^{-3}$, one sees that $Q_B$ has dimensions of $\left(2\pi\alpha^\prime\right)^2M_P$. Using the definition of $Q_B$, we will estimate $Q_B$ by
${\cal V}^{\frac{1}{3}}\left(2\pi\alpha^\prime\right)^2$, where the integral of $i^*\omega_B\wedge P_-\tilde{f}$ over $\Sigma_B$ is approximated by integrals of $i^*\omega_B$ and $P_-\tilde{f}$ over two-cycles non-trivial in the cohomology of $\Sigma_B$ and estimated/parametrized as ${\cal V}^{\frac{1}{3}}\left(2\pi\alpha^\prime\right)$ and ${\tilde f}$ respectively.}Q_B\sim{\cal V}^{\frac{1}{3}}(2\pi\alpha^\prime)^2\tilde{f},$
 \begin{equation}
\label{eq:gYMgBIbar_zi_aI}
g_{YM}g_{B {\bar a}_2}\rightarrow-{\cal V}^{-\frac{5}{4}}(a_2-{\cal V}^{-\frac{1}{3}}), \partial_{a_2}T_B \rightarrow {\cal V}^{\frac{1}{9}} (a_2-{\cal V}^{-\frac{1}{3}}).  \nonumber\\
\end{equation}
 The dominant contribution to the physical gaugino-quark-squark vertex is proportional to :
\begin{equation}
\label{eq:gaugino-quark-squark1}
C^{\tilde{\lambda^{0}}{u_L}\tilde {u_L}}\sim \frac{ {\cal V}^{-\frac{11}{9}} {\tilde f}}{{\left(\sqrt{\hat{K}_{{\cal A}_2{\bar {\cal A}}_2}}\right)}{\left(\sqrt{\hat{K}_{{\cal A}_2{\bar {\cal A}}_2}}\right)}}\sim \tilde{f}\left({\cal V}^{-\frac{4}{5}}\right).
\end{equation}
The contribution of gluino-quark/antiquark-squark vertex is same as gaugino(neural)-quark/antiquark-squark, and therefore proportional to:
\begin{eqnarray}
\label{eq:G}
 G^{{q/{\bar q}}_{{\cal A}_2}}_{\tilde{q}_{{\cal A}_2}}= C^{\tilde{\lambda^{0}}{u_L}\tilde {u_L}}\sim \tilde{f}{\cal V}^{-\frac{4}{5}}{\tilde q}\chi^{q/ \bar q}\lambda_{\tilde{g}},\lambda_{\tilde{g}}~ {\rm corresponds~ to ~gluino}.
\end{eqnarray}
The neutralino-quark-squark vertex of Figure~2.1 is given by considering the contribution of
$$\frac{e^{\frac{K}{2}}}{2}\left({\cal D}_ID_{\bar A_2}W\right)\chi^{{\cal Z}_I}{\chi^{c^{{\cal {A}}_2}}} +ig_{{\bar {\cal Z}_I}{\bar {A_2}}}{\bar\chi}^{{\cal Z}_I}\left[{\bar\sigma}\cdot\partial+\Gamma^{{\cal A}_2}_{{\cal A}_2{\bar A_2}}{\bar\sigma}\cdot\partial {\cal A}_2{\chi^{c^{{\cal {A}}_2}}}
+\frac{1}{4}\left(\partial_{{\cal A}_2}K{\bar\sigma}\cdot {\cal A}_2 - {\rm c.c.}\right){\chi^{c^{{\cal {A}}_2}}} \right]$$
 $\chi^I$ is $SU(2)_L$ higgsino, $\chi^{A_2}$ corresponds to $SU(2)_L$ quark and ${{\cal A}_2}$ corresponds to left-handed squark,
and  $g_{{\bar I}{\bar {\cal A}_2}}=0$. In terms of undiagonalized basis, consider ${\cal D}_iD_{\bar a_2}W= \left(\partial_i\partial_{\bar a_2}K\right)W+\partial_iKD_{\bar a_2}W +
\partial_{\bar a_2}KD_iW - \left(\partial_iK\partial_{\bar a_2}K\right)W,$ where $a_2, z_i$ correspond to undiagonalized moduli fields.
 
Considering  $a_2\rightarrow a_2+{\cal V}^{-\frac{1}{3}}{M_P}$, using equations (\ref{eq:K}), (\ref{eq:W}) and picking up the component linear in  $z_1$ as well as fluctuations of $a_2$, one gets:
\begin{eqnarray}
\label{eq:DKDW32}
& & {\hskip -0.4in} e^{\frac{K}{2}}\left(\left(\partial_i\partial_{\bar a_2}K\right)W+\partial_iKD_{\bar a_2}W +
\partial_{\bar a_2}KD_iW - \left(\partial_iK\partial_{\bar a_2}K\right)W\right)  \sim ({\cal V}^{-\frac{20}{9}})z_i \delta a_2.
\end{eqnarray}
Based on the discussion below (\ref{eq:expKover2_calD_DW}),  $e^{\frac{K}{2}}{\cal D}_i D_{{\cal A}_2}W \chi^{{\cal Z}_i}{\chi^{c^{{\cal {A}}_2}}}\sim {\cal V}^{-\frac{20}{9}}{\cal Z}_i\delta{\cal A}_2 \chi^{{\cal Z}_i}{\chi^{c^{{\cal {A}}_2}}},$
 and the contribution of physical higgsino-quark-squark vertex will be given as :
\begin{eqnarray}
\label{eq:Higgsino-quark-squark1}
& &  C^{{\tilde H^{c}}_L {u_L} \tilde {u_L}}\sim\frac{{\cal V}^{-\frac{20}{9}}\langle{\cal Z}_i\rangle}{{\sqrt{\hat{K}^{2}_{{\cal Z}_i{\bar{\cal Z}}_i}{\hat{K}_{{\cal A}_2{\bar {\cal A}}_2}}{\hat{K}_{{\cal A}_2{\bar {\cal A}}_2}}}}} \sim {\cal V}^{-\frac{4}{5}}.
\end{eqnarray}
Following equation (\ref{eq:neutralinos_i}), the neutralino-quark squark vertex will be given by:
\begin{eqnarray}
\label{eq:X}
 {\hskip -0.2in}X^{{q/{\bar q}}}_{\tilde{q}}= C^{{\chi^{0}_3} {u_L}\tilde {u_L}}\sim C^{\tilde{\lambda^{0}}{u_L}\tilde {u_L}}+ {\tilde f}{\cal V}^{\frac{5}{6}}\left(\frac{v}{ M_{P}}\right) C^{H^{c}_L {u_L} \tilde {u_L}} \sim \tilde{f}{\cal V}^{-\frac{4}{5}}{{\chi^{0}_3} {u_L}\tilde {u_L}}.
\end{eqnarray}
Using RG analysis of coefficients of the effective dimension-six gluino decay operators as given in
\cite{gambino_et_al}, it was shown in \cite{Dhuria+Misra_mu_Split_SUSY} that these coefficients at the EW scale are of the same order as that at the squark mass scale.

The analytical formulae to calculate decay width for three-body tree-level gluino decay channel as given in \cite{Manuel_Toharia} is:
\begin{eqnarray}
\label{eq:neutralinowidth}
&&\Gamma(\tilde{g}\to\chi_{\rm n}^{o}q_{{}_I} \bar{q}_{{}_J} )
=  {g_s^2\over256 \pi^3 \mgss^3 } \sum_{i,j} \int ds_{13} ds_{23}\ {1\over2}
{\cal R}e\Big(A_{ij}(s_{23}) + B_{ij}(s_{13}) \nonumber\\
&&{\hskip 1.3in} -  2 \eps_{n} \eps_{{g}}\
C_{ij}(s_{23},s_{13})\Big), 
\end{eqnarray}
where the integrand is the square of the spin-averaged total amplitude
and $i,j=1,2,..,6$ are the indices of the squarks mediating the decay. The limits of integration in (\ref{eq:neutralinowidth}) are given in \cite{Dhuria+Misra_mu_Split_SUSY}.
The $A_{ij}$ terms in (\ref{eq:neutralinowidth}) represent the contributions from the gluino decay channel involving gluino$\rightarrow$squark+quark and squark$\rightarrow$neutralino+anti-quark, whereas
 the $B_{ij}$ terms come from channel gluino $\rightarrow$squark+anti-quark and squark$\rightarrow$neutralino+quark. The same are defined in \cite{Dhuria+Misra_mu_Split_SUSY}.
Utilizing the results as given in (\ref{eq:G}) and (\ref{eq:X}),

{\small $A_{ij}\Bigl(Tr\left[G^{{\bar q}}_{\tilde{q}}G^{{\bar q}}_{\tilde{q}}\ ^\dagger\right]
Tr\left[X^{q}_{\tilde{q}}X^{q}_{\tilde{q}}\ ^\dagger\right]\Bigr)\sim \tilde{f^4}{\cal V}^{-\frac{16}{5}}$},
{\small $B_{ij}\Bigl(Tr\left[G^{q_{a_1}}_{\tilde{q}_{a_1}}G^{q_{a_1}}_{\tilde{q}_{a_1}}\ ^\dagger\right]
Tr\left[X^{{\bar q}_{a_1}}_{\tilde{q}_{a_1}}X^{{\bar q}_{a_1}}_{\tilde{q}_{a_1}}\ ^\dagger\right]\Bigr)$ $\sim \tilde{f^4}{\cal V}^{-\frac{16}{5}}$}, 
{\small $C\Bigl(Tr\left[G^{{\bar q}}_{\tilde{q}_{a_1}}G^{q}_{\tilde{q}}\ ^\dagger X^{q}_{\tilde{q}}X^{{\bar q}}_{\tilde{q}}\ ^\dagger\right]\Bigr)\sim \tilde{f^4}{\cal V}^{-\frac{16}{5}}$}. 

Substituting these values for various vertex elements and solving equation (\ref{eq:neutralinowidth}) by using the limits of integration as given in \cite{Dhuria+Misra_mu_Split_SUSY},  the dominating contribution of decay width  is:
\begin{eqnarray}
 \label{eq: Decay width 1}
 \hspace{-.5cm}\Gamma(\tilde{g}\to\chi_{\rm n}^{o}q_{{}_I} \bar{q}_{{}_J} )
&\sim&{g_s^2  \over256 \pi^3  {\cal V}^2 m_{\frac{3}{2}}^3 }\left[
{\tilde{f}^4  {\cal V}^{-\frac{16}{5}} m_{\frac{3}{2}}^4 }{\cal V}^{\frac{10}{3}}+ \tilde{f}^4{\cal V}^{-\frac{16}{5}}{\cal V}^{\frac{10}{3}} m_{\frac{3}{2}}^4 +\tilde{f}^4{\cal V}^{-\frac{16}{5}} {\cal V}^{\frac{10}{3}} m_{\frac{3}{2}}^4\right] \nonumber\\
& & {\hskip -1.1in} \sim { g_s^2\over256 \pi^3  {\cal V}^2 m_{\frac{3}{2}}^3 }\left( \tilde{f}^4{\cal V}^{-\frac{16}{5}}{\cal V}^{\frac{10}{3}} m_{\frac{3}{2}}^4  \right)\sim O(10^{-4})\tilde{f}^4{\cal V}^{-\frac{19}{5}}{M_P} \ \sim O(10^{-5})\tilde{f}^4 GeV.
 \end{eqnarray}
 \paragraph{Upper bound on dilute flux $\tilde{f}$:}
  In case of ${\cal N}=1$ gauged supergravity, the modified D-term scalar potential in presence of background $D7$ fluxes is defined as \cite{jockersetal}:
   \begin{equation}
   \label{eq:Dterm}
V_D = {\frac{108\kappa_4^2\mu_7}{{{\cal{K}}^2}{ReT_B}}}({{\cal{K}}_{Pa}}{\cal{B}}^a-{\cal{Q}_{B}}{\emph{v}}^{B})^2.
\end{equation}
The first term of $V_D$ can be minimized for ${\cal{B}}^a= 0$ and second term has extra contribution coming from additional D7-brane fluxes.
Now, the corresponding F-term scalar potential in LVS limit has been calculated in \cite{D3_D7_Misra_Shukla} as:
\begin{equation}
\label{eq:Fterm}
V_f\sim e^K G^{\sigma^{{\alpha}}{\bar{\sigma}^{\bar{\alpha}}}}D_{\sigma^{\alpha}}W^\alpha{\bar{D}_{\bar{\sigma}^{\bar{\alpha}}}}\bar{W}\sim{\cal{V}}^{19/18}m_{3/2}^2
\sim{\cal{V}}^{-3} M^{2}_{P}.
\end{equation}
 In case of dilute flux approximation $V_D << V_f$. Therefore, from (\ref{eq:Dterm}) and (\ref{eq:Fterm}),
 \begin{equation}
 {\frac{108\kappa_4^2\mu_7}{{{\cal{K}}^2}{ReT_B}}}({\cal{Q}_{B}}{\emph{v}}^{B})^2 << {\cal{V}}^{-3} M^{2}_{P},
 \end{equation}
 where $Q_B\sim{\cal V}^{\frac{1}{3}}{\tilde f}\left(2\pi\alpha^\prime\right)^2M_P$, $ \kappa_4^2\mu_7\sim\frac{1}{\cal{V}}, {\cal{K}}=1/6Y$(volume of physical Calabi Yau), ${ReT_B}\sim {\cal{V}}^{1/18}$
 (volume of ``Big" Divisor) and ${\cal{V}}^{B}\sim{\cal{V}}^{1/3}$(internal volume of 2-cycle),
 solving this: $  {\tilde f}^2{\cal V}^{-\frac{31}{18}}<<{\cal{V}}^{-3}$, i.e ${\tilde f}<< {\cal V}^{-\frac{23}{36}}\equiv10^{4}$ for ${\cal V}=10^5$.
  
By considering ${{\tilde f}^2}\sim{10}^{-10}$, the decay width of gluino i.e equation (\ref{eq: Decay width 1}) becomes:
$ \Gamma(\tilde{g}\to\chi_{\rm n}^{o}q_{{}_I} \bar{q}_{{}_J} )\sim O(10^{-5}){\tilde f}^4 < {\cal O}(10^{-25})GeV$.
 Further, life time of gluino is given as:
   \begin{eqnarray}
   \tau &=&\frac{\hbar}{\Gamma}\sim\frac{10^{-34} Jsec}{10^{-5}{\tilde f}^4 GeV}\sim\frac{10^{-19}}{{\tilde f}^4} > 10 \ sec.
   \end{eqnarray}
   
$\underline{\bf{\tilde{g}\rightarrow\tilde{\chi}_3^0 \ g}}$

\begin{figure}
\begin{center}
\begin{picture}(500,170)(50,0)
\Text(140,135)[]{$\tilde{g}$}
\Line(110,120)(160,120)
\Gluon(110,120)(160,120){5}{4}
\ArrowLine (160,120)(190,150)
\Text(180,152)[]{$q_I$}
\DashArrowLine (190,90)(160,120){4}
\Text(170,100)[]{$\tilde{q}_R$}
\DashArrowLine (190,150)(190,90){4}
\Gluon(190,90)(230,90){5}{4}
\Text(240,90)[]{$g_\mu$}
\ArrowLine(190,150)(230,150)
\Text(240,150)[]{$\tilde{\chi}_3^0$}
\Text(190,60)[]{(a)}
\Line(310,120)(360,120)
\Gluon(310,120)(360,120){5}{4}
\Text(340,135)[]{$\tilde{g}$}
\DashArrowLine (360,120)(390,150){4}
\Text(380,152)[]{$\tilde{q}_R$}
\ArrowLine (390,90)(360,120)
\Text(370,100)[]{$q_I$}
\ArrowLine (390,150)(390,90)
\Gluon(390,90)(430,90){5}{4}
\Text(440,90)[]{$g_\mu$}
\ArrowLine(390,150)(430,150)
\Text(440,150)[]{$\tilde{\chi}_3^0$}
\Text(390,60)[]{(b)}
\end{picture}
\end{center}
\vskip -0.7in
\caption {Diagrams contributing to one-loop two-body gluino decay.}
\label{fig:loopgdc}
\end{figure}
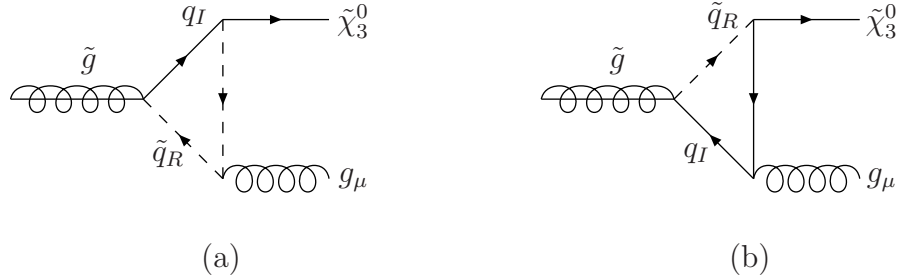
Relevant to Figs.~2.2(a) and 2.2(b), the gluino-quark-squark vertex and the neutralino-quark-squark vertex will again be given by equations (\ref{eq:G}) and (\ref{eq:X}). The quark-quark-gluon vertex relevant to Figure~2.2, from \cite{Wess_Bagger} is given by:
  \begin{eqnarray}
  \label{eq:qqgl 1}
& &  g_{I{\bar J}}{\bar\chi}^{\bar J}{\bar\sigma}\cdot A\ {\rm Im}\left(X^BK + i D^B\right)\chi^I,\nonumber\\
& & \sim g_{YM}g_{{\cal A}_2 {\bar {\cal A}_2}}{\bar\chi}^{\bar {{\cal A}_2}}{\bar\sigma}\cdot A\left\{6\kappa_4^2\mu_72\pi\alpha^\prime Q_BK+\frac{12\kappa_4^2\mu_72\pi\alpha^\prime Q_Bv^B}{\cal V}\right\},
  \end{eqnarray}
   $\chi^{{\cal A}_2}$ and ${\bar\chi}^{\bar {{\cal A}_2}}$ correspond to quarks and antiquarks.
 $X=X^B\partial_B=-12i\pi\alpha^\prime\kappa_4^2\mu_7Q_B\partial_{T_B}$  corresponds to the killing isometry vector and $D$ term generated is given as: $D^B=\frac{4\pi\alpha^\prime\kappa_4^2\mu_7Q_Bv^B}{\cal V}$.
Utilizing  value of $Q_B\sim{\cal V}^{\frac{1}{3}}f\left(2\pi\alpha^\prime\right)^2M_P$, $g_{{\cal A}_2{\bar {\cal A}}_{\bar 2}}\sim{\cal V}^{-\frac{5}{9}}, v^B\sim{\cal V}^{\frac{1}{3}}, Q_B\sim{\cal V}^{\frac{1}{3}}(2\pi\alpha^\prime)^2\tilde{f}$,  it has been shown in chapter {\bf 4} that quark-quark-gluon vertex will be of ${\cal O}(1)$.

The gauge kinetic term for squark-squark-gluon vertex, relevant to Figure~2.2(b), will be given by $ {\frac{1}{\kappa_4^2{\cal V}^2}}G^{T_B{\bar T}_B}\tilde{\bigtriangledown}_\mu T_B\tilde{\bigtriangledown}^\mu {\bar T}_{\bar B}$.  Using the value of $G^{T_B{\bar T}_B}(EW)= {\cal V}^{\frac{7}{3}}$ (for detailed explanation, see chapter {\bf 4}), the given term generates the required squark-squark-gluon vertex as follows:
\begin{eqnarray}
\label{eq:sq sq gl}
& &  {\frac{6i\kappa_4^2\mu_72\pi\alpha^\prime Q_BG^{\sigma_B{\bar\sigma}_B}}{\kappa_4^2{\cal V}^2}}A^\mu\partial_\mu\left(\kappa_4^2\mu_7(2\pi\alpha^\prime)^2C_{2{\bar 2}}{\cal A}_2{\bar {\cal A}}_{\bar 2}\right)\nonumber\\
&& \xrightarrow[{\small \kappa_4^2\mu_7(2\pi\alpha^\prime)^2C_{2{\bar 2}}\sim{\cal V}^{\frac{1}{9}}}]{\small G^{\sigma_B{\bar\sigma}_B}\sim{\cal V}^{\frac{7}{3}},}\frac{{\cal V}^{\frac{7}{9}}\epsilon\cdot\left(2k-(p_{\tilde{\chi}_3^0}+p_{\tilde{g}})\right)}{\left(\sqrt{\hat{K}_{{\cal A}_2{\bar {\cal A}}_2}}\right)^2} \sim O({10}^{2})\tilde{f}{\cal V}^{\frac{7}{9}}\left[ 2\epsilon\cdot k-\epsilon\cdot\left(p_{\tilde{\chi}_3^0}+p_{\tilde{g}}\right)\right]\nonumber\\
&& \sim \tilde{f}{\cal V}^{\frac{53}{45}}, {\rm for~{\cal V}\sim {10}^5}.
\end{eqnarray}
In \cite{Dhuria+Misra_mu_Split_SUSY}, it has been discussed that behavior of Wilson coefficients corresponding to  two-body gluino decay does not change much upon RG evolution to EW scale.  Using the vertices calculated above relevant to Figs.~2.2(a) and 2.2(b), and the Feynman rules of \cite{2comp}, one obtains for the scattering amplitude:
\begin{eqnarray}
\label{eq:M_I}
& & {\cal M}\sim\tilde{f}^3M_P \int\frac{d^4k}{\left(2\pi\right)^4} \times{\cal V}^{-\frac{4}{5}}\left(\frac{i{\bar\sigma}\cdot k}{k^2-m_q^2+i\epsilon}\right)\left({\cal V}^{-\frac{4}{5}}
\right)\left(\frac{i}{\left[\left(k-p_{\tilde{G}}\right)^2-m^2_{\tilde{q}}+i\epsilon\right]}
\right)\nonumber\\
& & \times\left({\cal V}^{\frac{53}{45}}\right)\Bigl(\frac{i}
{\left[\left(k-p_{\tilde{g}}\right)^2-m^2_{\tilde{q}}+i\epsilon\right]}\Bigr)\nonumber\\
& & + \tilde{f}^2M_P \int\frac{d^4k}{\left(2\pi\right)^4} \times {\cal V}^{-\frac{4}{5}}\Biggl(\frac{i}
{\left[\left(k+p_{\tilde{\chi}_3^0}\right)^2-m^2_{\tilde{q}}+i\epsilon\right]}\Biggr)\left({\cal V}^{-\frac{4}{5}}
\right)\left(\frac{i{\bar\sigma}\cdot k}{k^2-m_q^2+i\epsilon}\right)\nonumber\\
& & \times\Biggl(\frac{i{\bar\sigma}\cdot\left(k-p_{g_\mu}\right)}{\left[
\left(k-p_{g_\mu}\right)^2-m^2_q+i\epsilon\right]}\Biggr).
\end{eqnarray}
Using the 1-loop integrals of \cite{Pass+Velt}:
\begin{eqnarray}
\label{eq:passvelt}
& & \frac{1}{i}\int\frac{d^4k}{\left(2\pi\right)^4}\frac{\left(k_\mu,\ k_\mu k_\nu\right)}{\left(k^2-m_1^2+i\epsilon\right)
\left[\left(k+p_1\right)^2-m_2^2+i\epsilon\right]\left[\left(k+p_1+p_2\right)^2-m_3^2+i\epsilon\right]}\nonumber\\
& & =4\pi^2\Biggl[p_{1\mu}C_{11}+p_{2\mu}C_{12}, p_{1\mu}p_{1\nu}C_{21}+p_{2\mu}p_{2\nu}C_{22}+\left(p_{1\mu}p_{2\nu}+p_{1\nu}p_{2\mu}\right)C_{23}+\eta_{\mu\nu}C_{24}\Biggr];
\nonumber\\
& & (a)~m_1=m_q, m_2=m_3=m_{\tilde{q}};\ p_1=-p_{\tilde{\chi}_3^0}, p_2=-p_{g_\mu};\nonumber\\
& & (b)~m_1=m_3=m_q, m_2=m_{\tilde{q}};\ p_1=p_{\tilde{\chi}_3^0}, p_2=-p_{\tilde{g}}.
\end{eqnarray}
\vskip -0.4in
The one-loop three-point functions $C_{ij}$'s relevant for cases (a) and (b) have been calculated in \cite{Dhuria+Misra_mu_Split_SUSY} and are given as under:
\begin{eqnarray}
\label{eq: three point functions}
& & C^{(a)}_{24}= O(10^6), C^{(b)}_{24}=  O(10^6),\nonumber\\
&& C^{(a)}_0= \frac{1}{4{\pi}^2}\times 10^{-21}\sim O(1)\times {10}^{-23}~{GeV}^{-2},\nonumber\\
&&C^{(b)}_{11}=  O(10) \times {10}^{-16} {GeV}^{-2}, C^{(b)}_{12}=  O(10)\times {10}^{-16}~{GeV}^{-2},\nonumber\\
&&C^{(b)}_{21}= C^{(b)}_{22}= C^{(b)}_{23}\sim O(1) \times {10}^{-10}~{GeV}^{-2}, C^{(b)}_0=  \sim O(1)\times {10}^{-23}~{GeV}^{-2}. \nonumber\\
  \end{eqnarray}
Now, equation (\ref{eq:M_I}) can be evaluated to yield:
\begin{eqnarray}
\label{eq:M_II}
& & {\bar u}(p_{\tilde{\chi}_3^0})\Biggl(\tilde{f}^3{\cal V}^{-0.4}\Bigl[\left\{{\bar\sigma}\cdot p_{\tilde{\chi}_3^0}C^{(a)}_{11}+{\bar\sigma}\cdot p_{g_\mu}C^{(a)}_{12}\right\}(2\epsilon\cdot p_{\tilde{\chi}_3^0}) \left\{{\bar\sigma}\cdot p_{\tilde{\chi}_3^0}\epsilon\cdot p_{\tilde{\chi}_3^0}C^{(a)}_{21}\right.+\nonumber\\
&& \left.+{\bar\sigma}\cdot p_{g_\mu}\epsilon\cdot p_{\tilde{\chi}_3^0}C^{(a)}_{23}
+{\bar\sigma}\cdot\epsilon C^{(a)}_{24}\right\}\Bigr]\nonumber\\
& & + \tilde{f}^2{\cal V}^{-1.6}\Bigl[-\left\{{\bar\sigma}\cdot p_{\tilde{\chi}_3^0}C^{(b)}_{11}
+{\bar\sigma}\cdot p_{\tilde{g}}C^{(b)}_{12}\right\}{\bar\sigma}\cdot\epsilon{\bar\sigma}\cdot p_{g_\mu}+
{\bar\sigma}\cdot p_{\tilde{\chi}_3^0}{\bar\sigma}\cdot\epsilon {\bar\sigma}\cdot p_{\tilde{\chi}_3^0}C^{(b)}_{21} \nonumber\\
& & + {\bar\sigma}\cdot p_{\tilde{g}}{\bar\sigma}\cdot\epsilon {\bar\sigma}\cdot p_{\tilde{g}}C^{(b)}_{22}  -\biggl({\bar\sigma}\cdot p_{\tilde{\chi}_3^0}{\bar\sigma}\cdot\epsilon{\bar\sigma}\cdot p_{\tilde{g}} + {\bar\sigma}\cdot p_{\tilde{g}}{\bar\sigma}\cdot\epsilon {\bar\sigma}\cdot p_{\tilde{\chi}_3^0}\biggr)C^{(b)}_{23}+ \nonumber\\
&&{\bar\sigma}_\mu{\bar\sigma}\cdot\epsilon{\bar\sigma}^\mu C^{(b)}_{24}\Bigr]\Biggr)
u(p_{\tilde{g}}),
\end{eqnarray}
which equivalently could be rewritten as:
\begin{equation}
\label{eq:M_III}
 {\bar u}(p_{\tilde{\chi}_3^0})\Biggl[{\bar\sigma}\cdot{\cal A}+{\bar\sigma}\cdot p_{\tilde{\chi}_3^0}{\bar\sigma}\cdot\epsilon{\bar\sigma}\cdot{\cal B}_1 + {\bar\sigma}\cdot p_{g_\mu}
{\bar\sigma}\cdot\epsilon{\bar\sigma}\cdot{\cal B}_2 + D_3{\bar\sigma}_\mu{\bar\sigma}\cdot\epsilon{\bar\sigma}^\mu
C^{(b)}_{24} \Biggr]u(p_{\tilde{g}}),
\end{equation}
where
\vskip -0.5in
\begin{eqnarray}
\label{eq:AB1B2}
& & {\cal A}^\mu\equiv \tilde{f}^3{\cal V}^{-0.4}\left[p_{\tilde{\chi}_3^0}^\mu\epsilon\cdot p_{\tilde{\chi}_3^0}\left(2C^{(a)}_{11} + C^{(a)}_{21}\right) + p_{g_\mu}^\mu\epsilon\cdot p_{\tilde{\chi}_3^0}\left(C^{(a)}_{12} + C^{(a)}_{23}\right) + \epsilon^\mu C^{(a)}_{24}\right],\nonumber\\
& & {\cal B}_1^\mu\equiv  \tilde{f}^2{\cal V}^{-1.6}\left[-p_{g_\mu}^\mu\left(C^{(b)}_{11} + C^{(b)}_{12} + C^{(b)}_{23}-C^{(b)}_{22}\right) + p_{\tilde{\chi}_3^0}^\mu\left(C^{(b)}_{21} + C^{(b)}_{22} - 2C^{(b)}_{23}\right)\right],\nonumber\\
& & {\cal B}_2^\mu\equiv \tilde{f}^2{\cal V}^{-1.6}\left[p_{g_\mu}^\mu\left(C^{(b)}_{12}+C^{(b)}_{22}\right) + p_{\tilde{\chi}_3^0}^\mu\left( C^{(b)}_{22} - C^{(b)}_{23}\right)\right], {D_3}\equiv \tilde{f}^2{\cal V}^{-1.6}.
\end{eqnarray}
Replacing ${\bar u}(p_{\tilde{\chi}_3^0}){\bar\sigma}\cdot p_{\tilde{\chi}_3^0}$ by $m_{\tilde{\chi}_3^0}{\bar u}(p_{\tilde{\chi}_3^0})$ and ${\bar\sigma}\cdot p_{\tilde{g}}u(p_{\tilde{g}})$ by $m_{\tilde {g}}$, and using $\epsilon\cdot p_{\tilde{\chi}_3^0}=0$, (\ref{eq:M_III}) can be simplified to:
\begin{eqnarray}
\label{eq:M_IV}
&&{\cal M}\sim {\bar u}(p_{\tilde{\chi}_3^0})\left[A{\bar\sigma}\cdot\epsilon + B_1{\bar\sigma}\cdot\epsilon
{\bar\sigma}\cdot p_{\tilde{\chi}_3^0} + B_2{\bar\sigma}\cdot p_{\tilde{g}}{\bar\sigma}\cdot\epsilon +
D_1{\bar\sigma}\cdot p_{\tilde{g}}{\bar\sigma}\cdot\epsilon{\bar\sigma}\cdot p_{\tilde{\chi}_3^0}\right. \nonumber\\
&&\left. + {D_3} {\bar\sigma}_\mu{\bar\sigma}\cdot\epsilon{\bar\sigma}^\mu C^{(b)}_{24}\right]u(p_{\tilde{g}}),
\end{eqnarray}
where,
\vskip -0.5in
\begin{eqnarray}
\label{eq:AB1B2D1 defs}
&&A\equiv {\cal V}^{-0.4} \tilde{f}^3 C^{(a)}_{24} - m_{\tilde{g}}m_{\tilde{\chi}_3^0} {\cal V}^{-1.6} \tilde{f}^2\Bigl\{
 C^{(b)}_{11} + 2 C^{(b)}_{12} + C^{(b)}_{23}-C^{(b)}_{22}\Bigr\},\nonumber\\
&& B_1\equiv {\cal V}^{-1.6}\tilde{f}^2\left(C^{(b)}_{11} + 2 C^{(b)}_{12} +  C^{(b)}_{21}\right)m_{\tilde{\chi}_3^0}, B_2\equiv {\cal V}^{-1.6}\tilde{f}^2\left(C^{(b)}_{12}+C^{(b)}_{22}\right)m_{\tilde{g}},\nonumber \\
&& D_1\equiv {\cal V}^{-1.6}\tilde{f}^2\left(- C^{(b)}_{12}- C^{(b)}_{23}\right).
\end{eqnarray}
 Utilizing values of C's calculated in (\ref{eq: three point functions}),
 $   A \sim {\cal O}(10^{5})\tilde{f}^2, B_1\sim {\cal O}({10^{-7}})\tilde{f}^2{GeV}^{-2}, B_2\sim {\cal O}({10^{-7}})\tilde{f}^2{GeV}^{-2}$, $D_1\sim {\cal O}({10^{-18}})\tilde{f}^2{GeV}^{-2};{D_3}\sim {\cal V}^{-1.6}\tilde{f}^2$.
\begin{eqnarray}
\label{eq:spinavGamma}
& & {\rm and}~~  \sum_{\tilde{g}\ {\rm and}\ \tilde{\chi}_3^0\ {\rm spins}}|{\cal M}|^2
\sim Tr\Biggl(\sigma\cdot p_{\tilde{\chi}_3^0}\left[A{\bar\sigma}\cdot\epsilon + B_1{\bar\sigma}\cdot\epsilon
{\bar\sigma}\cdot p_{\tilde{\chi}_3^0} + B_2{\bar\sigma}\cdot p_{\tilde{g}}{\bar\sigma}\cdot\epsilon \right. \nonumber\\
& & \left. +
D_1{\bar\sigma}\cdot p_{\tilde{g}}{\bar\sigma}\cdot\epsilon{\bar\sigma}\cdot p_{\tilde{\chi}_3^0}  + D_3 {\bar\sigma}_\mu{\bar\sigma}\cdot\epsilon{\bar\sigma}^\mu C^{(b)}_{24}\right]\nonumber\\
& & \times\sigma\cdot p_{\tilde{g}}\left[A{\bar\sigma}\cdot\epsilon + B_1{\bar\sigma}\cdot\epsilon
{\bar\sigma}\cdot p_{\tilde{\chi}_3^0} + B_2{\bar\sigma}\cdot p_{\tilde{g}}{\bar\sigma}\cdot\epsilon +
D_1{\bar\sigma}\cdot p_{\tilde{g}}{\bar\sigma}\cdot\epsilon{\bar\sigma}\cdot p_{\tilde{\chi}_3^0} \right.\nonumber\\
&&\left. + D_3 {\bar\sigma}_\mu{\bar\sigma}\cdot\epsilon{\bar\sigma}^\mu C^{(b)}_{24}\right]^\dagger\Biggr),
\end{eqnarray}
\vskip -0.3in
which at  $p^0_{\tilde{\chi}_3^0}=\sqrt{m_{\tilde{\chi}_3^0}^2c^4+\rho^2},
p^1_{\tilde{\chi}_3^0}=p^2_{\tilde{\chi}_3^0}=p^3_{\tilde{\chi}_3^0}=\frac{\rho}{\sqrt{3}}
=\frac{1}{\sqrt{3}}\frac{c\left(m_{\tilde{g}}^2 - m_{\tilde{\chi}_3^0}^2\right)}{2m_{\tilde{g}}}$,
yields:
\begin{eqnarray}
\label{eq:trace}
& &  \frac{1}{256}{m_{\tilde{g}}}^2 \biggl[6 {m^{2}_{\tilde{g}}}({B_1}+{D_1} {m_{\tilde{g}}})^2+\left\{8
   {A_1}+16 D_3{C_{24}}+{m_{\tilde{g}}} \left(\left(5+\sqrt{3}\right) {B_1}+8 {B_2}\right. \right. \nonumber\\
   && \left. \left. +\left(5+\sqrt{3}\right) {D_1}
   {m_{\tilde{g}}}\right)\right\}^2\biggr],
   \end{eqnarray}
in the rest frame of the gluino.\\
Incorporating results of (\ref{eq:AB1B2D1 defs}) in equation (\ref{eq:trace}), one gets, 
\begin{equation}
 \sum_{\tilde{g}\ {\rm and}\ \tilde{\chi}_3^0\ {\rm spins}}|{\cal M}|^2 \sim 0.2 {D^{2}_1} m_{\tilde{g}}^6. 
 \end{equation}
 Now, using standard two-body decay results (See \cite{Griffiths_particle}), the decay width $\Gamma$ is given by the following expression: $
\Gamma=\frac{\sum_{\tilde{g}\ {\rm and}\ \tilde{\chi}_3^0\ {\rm spins}}|{\cal M}|^2\left(m_{\tilde{g}}^2-m_{\tilde{\chi}_3^0}^2\right)}{16\pi\hbar m_{\tilde{g}}^3}.$
Using value of $D_1\sim O({10^{-18}}){\tilde f}^2 \ {GeV}^{-2}$ as given above, $m_{\tilde{g}}\sim {\cal V}^{-\frac{4}{3}}{M_P} \ GeV$, and
$m^{2}_{\tilde{\chi_3^0}}\sim m^{2}_{\tilde{g}}+ \frac{m_{\tilde g} {\tilde f}^2  v^2 {\cal V}^{\frac{2}{3}}}{M_{P}}$ ${GeV}^{2}$, two-body decay width is given as:
\begin{equation}
\label{eq:Gamma1}
\Gamma=  \frac{(0.2) {D_1^2} m_{\tilde {g}}^6} {16\pi m_{\tilde{g}}^3} \sim \frac{O({10}^{-2}) \tilde{f}^4.{10^{-36}}.{m_{\tilde {g}}}^6 (2\frac{m_{\tilde g} {\tilde f}^2  v^2 {\cal V}^{\frac{2}{3}}}{ M_{P}})} {m_{\tilde{g}}^3}\sim{10}^{-4} \tilde{f}^4 GeV.
\end{equation}
Considering $\tilde{f} < {10}^{-5}$ as calculated above, $\Gamma < {10}^{-24}$ GeV.  Life time of gluino is given as:
   \begin{eqnarray}
   \tau &=&\frac{\hbar}{\Gamma}\sim\frac{10^{-34} Jsec}{10^{-4}{\tilde f}^4 GeV}\sim\frac{10^{-20}}{{\tilde f}^6} sec > {10}^{10}sec.
 \end{eqnarray}
\underline{${\bf \tilde{g}\rightarrow\tilde{G} \ q \ {\bar q}}$}\\
We now consider the three-body decay of the gluino into goldstino and a quark and an anti-quark: $\tilde{g}\rightarrow\tilde{G}+q+{\bar q}$. The Feynman diagrams corresponding to this particular decay are shown in Figure~2.3.
\begin{figure}
\begin{center}
\begin{picture}(250,180)(50,0)
\Text(90,130)[]{$\tilde{g}$}
\Line(60,120)(110,120)
\Gluon(60,120)(110,120){5}{4}
\Vertex(110,120){3.5}
\ArrowLine(110,120)(140,150)
\Text(149,150)[]{${q}_{a_2}$}
\DashArrowLine (130,90)(110,120){4}
\Text(107,100)[]{$\tilde{q}_{a_2}$}
\ArrowLine(160,120)(130,90)
\Text(167,120)[]{${\bar{q}_{a_2}}$}
\Line(130,90)(160,60)
\Line(128,86)(158,56)
\Vertex(129,88){3.5}
\Text(167,65)[]{${\tilde G}$}
\Text(110,30)[]{(a)}
\Text(250,130)[]{$\tilde{g}$}
\Line(210,120)(260,120)
\Gluon(210,120)(260,120){5}{4}
\ArrowLine (290,150)(260,120)
\Vertex(260,120){3.5}
\Text(299,150)[]{$\bar{q}_{a_2}$}
\DashArrowLine(260,120)(280,90){4}{}
\Text(257,100)[]{$\tilde{q}_{a_4}$}
\ArrowLine(280,90)(310,120)
\Text(320,120)[]{$q_{a_2}$}
\Line(280,90)(310,60)
\Line(278,86)(308,56)
\Vertex(279,88){3.5}
\Text(317,65)[]{${\tilde G}$}
\Text(280,30)[]{(b)}
\end{picture}
\vskip-0.3in
\caption{Three-body gluino decay into goldstino.}
\end{center}
\label{fig:gludecgoldstno}
\end{figure}
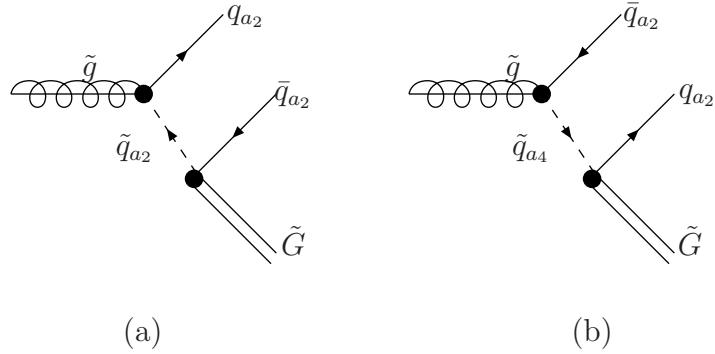
The gluino-(anti-)quark-squark vertex will again be given by (\ref{eq:G}). The gravitino-quark-squark vertex would come from a term of type  ${\bar\psi}_\mu
\tilde{q}_Lq_L$, which in ${\cal N}=1$ gauged supergravity action of \cite{Wess_Bagger} is given by:
\begin{equation}
\label{eq:gravitino-q-sq}
-g_{I{\bar J}}\left(\partial_\mu {\bar {\cal A}}^{\bar J}\right)\chi^I\sigma^\nu{\bar\sigma}_\mu\psi_\nu - \frac{i}{2}e^{\frac{K}{2}}\left(D_IW\right)\chi^I\sigma^\mu{\bar\psi}_\mu + {\rm h.c.}.
\end{equation}
From \cite{gravitinomodexp}, the gravitino field can be decomposed into the spin-$\frac{1}{2}$ goldstino field $\tilde{G}$
via: 
$\psi_\nu=\rho_\nu + \sigma_\nu\tilde{G},\ \tilde{G}=-\frac{1}{3}\sigma^\mu\psi_\mu$,
$\rho_\nu$ being a spin-$\frac{3}{2}$ field.
Hence, the goldstino-content of (\ref{eq:gravitino-q-sq}), using $\sigma^\nu{\bar\sigma}^\mu\sigma_\nu=-2{\bar\sigma}^\nu$,
is given by: $2g_{I{\bar J}}\chi^I\left(\partial_\mu {\bar {\cal A}}^{\bar J}\right){\bar\sigma}^\mu\tilde{G} + \frac{3i}{2}e^{\frac{K}{2}}\left(D_IW\right)\chi^I\tilde{G} + {\rm h.c.}$. Utilizing
\begin{eqnarray}
\label{eq:Goldstino-s-sq_II}
& & g_{{\cal A}_2{\bar {\cal A}}_{2}}\sim{\cal V}^{-\frac{5}{9}}~ {\rm and}~ e^{\frac{K}{2}}D_{{\cal A}_2}W\Biggr|_{{\cal A}_2\rightarrow {\cal A}_{2} +{\cal V}^{-\frac{1}{3}}}\sim {\cal V}^{-\frac{37}{18}}{\cal A}_{2},
\end{eqnarray}
one obtains:
\begin{eqnarray}
\label{eq:Goldstino-s-sq_I}
& & {\hskip -0.5in}2g_{{\cal A}_{2}{\bar {\cal A}}_{2}}\chi^I\left({\bar\sigma}\cdot\frac{p_{q_{{\cal A}_2}}+ p_{\tilde{G}}}{M_P}\right){\bar\sigma}^\mu\tilde{G} + \frac{3i}{2}e^{\frac{K}{2}}\left(D_{{\cal A}_2}W\right)\chi^I\tilde{G} + {\rm h.c.}\nonumber\\
& & {\hskip -0.5in}\sim  {\cal V}^{-\frac{5}{9}}\chi^I\left(\frac{m_{G}}{M_P}\right){{\cal A}_2}\tilde{G} + {\cal V}^{-\frac{5}{9}}\chi^I\left(\frac{m_{q_{{\cal A}_2}}}{M_P}\right){{\cal A}_2}\tilde{G}+ {\cal V}^{-\frac{37}{18}}{\cal A}_2\chi^I\tilde{G} \sim {\cal V}^{-\frac{37}{18}}{\cal A}_2\chi^I\tilde{G},
\end{eqnarray}
for $m_{q_{{\cal A}_2}}\sim O(10)MeV, m_G \sim 0$ and ${\cal V}\sim 10^5.$

The physical goldstino-quark-squark vertex will be given as:
\begin{equation}
\tilde{G}^{q_{{\cal A}_2}}_{\tilde{q}_{{\cal A}_2}}\sim \frac{{\cal V}^{-\frac{1}{18}}{\cal A}_2\chi^I\tilde{G}}{\left(\sqrt{\hat{K}_{{\cal A}_2{\bar {\cal A}}_2}}\sqrt{\hat{K}_{{\cal A}_2{\bar {\cal A}}_2}}\right)}
\sim O({10}^{2}){\cal V}^{-\frac{37}{18}}\sim {\cal V}^{-\frac{15}{9}} ~{\rm for}~ {\cal V}\sim {10}^5.
 \end{equation}
 For this particular case:-

{\small $A_{ij}\Bigl(Tr\left[G^{{\bar q}_{{\cal A}_2}}_{\tilde{q}_{{\cal A}_2}}G^{{\bar q}_{{\cal A}_2}}_{\tilde{q}_{{\cal A}_2}}\ ^\dagger\right]
Tr\left[\tilde{G}^{q_{{\cal A}_2}}_{\tilde{q}_{{\cal A}_2}}\tilde{G}^{q_{{\cal A}_2}}_{\tilde{q}_{{\cal A}_2}}\ ^\dagger\right]\Bigr)\sim \tilde{f}^2{\cal V}^{-\frac{8}{5}}.{\cal V}^{-\frac{10}{3}}\sim \tilde{f}^2{\cal V}^{-5}$},

 {\small $B_{ij}\Bigl(Tr\left[G^{q_{{\cal A}_2}}_{\tilde{q}_{{\cal A}_2}}G^{q_{{\cal A}_2}}_{\tilde{q}_{{\cal A}_2}}\ ^\dagger\right]
Tr\left[\tilde{G}^{{\bar q}_{{\cal A}_2}}_{\tilde{q}_{{\cal A}_2}}\tilde{G}^{{\bar q}_{{\cal A}_2}}_{\tilde{q}_{{\cal A}_2}}\ ^\dagger\right]\Bigr)\sim \tilde{f}^2{\cal V}^{-\frac{8}{5}}.{\cal V}^{-\frac{10}{3}}\sim \tilde{f}^2{\cal V}^{-5}$},

{\small $C\Bigl(Tr\left[G^{{\bar q}_{{\cal A}_2}}_{\tilde{q}_{{\cal A}_2}}G^{q_{{\cal A}_2}}_{\tilde{q}_{{\cal A}_2}}\ ^\dagger \tilde{G}^{q_{{\cal A}_2}}_{\tilde{q}_{{\cal A}_2}}\tilde{G}^{{\bar q}_{{\cal A}_2}}_{\tilde{q}_{{\cal A}_2}}\ ^\dagger\right]\Bigr)\sim \tilde{f}^2{\cal V}^{-\frac{8}{5}}.{\cal V}^{-\frac{10}{3}}\sim \tilde{f}^2{\cal V}^{-5}$}.
\vskip 0.05in
Limits of integration as given in \cite{Dhuria+Misra_mu_Split_SUSY} are:
\begin{eqnarray}
&& {\hskip -0.2in} s_{23 \max }= \left({\cal V}^{\frac{2}{3}}m_{\frac{3}{2}}-m_q\right){}^2,\ s_{23 \min }= m_q^2, s_{13 \max }= {\cal V}^{\frac{4}{3}}m_{\frac{3}{2}}^2-s_{23}, s_{13 \min }= 0.
\end{eqnarray}
where $m_{\tilde{g}}= {\cal V}^{\frac{2}{3}}m_{\frac{3}{2}} \sim {10}^{11} GeV, m_{\tilde{G}}= 0$.
Utilizing the values of vertex elements calculated above and from (\ref{eq:neutralinowidth}), we calculate decay width for gluino as:
\begin{eqnarray}
 \label{eq: Decay width goldstino}
\Gamma(\tilde{g}\to\chi_{\rm n}^{o}q_{{}_I} \bar{q}_{{}_J} )
&\sim&{g_s^2 \over256 \pi^3 {\cal V}^2 m_{\frac{3}{2}}^3 }\left[-
\tilde{f}^2{\cal V}^{-5} m_{\frac{3}{2}}^4 {\cal V}^{\frac{8}{3}}+  \tilde{f}^2{\cal V}^{-5}{\cal V}^{\frac{8}{3}} m_{\frac{3}{2}}^4 - \tilde{f}^2{\cal V}^{-5} {\cal V}^{\frac{8}{3}}m_{\frac{3}{2}}^4\right] \nonumber\\
& & \sim { g_s^2 \over256 \pi^3 {\cal V}^2 m_{\frac{3}{2}}^3 }( \tilde{f}^2 {\cal V}^{-5}{\cal V}^{\frac{8}{3}} m_{\frac{3}{2}}^4 ) \sim O({10}^{-4}){\cal V}^{-\frac{13}{3}}m_{\frac{3}{2}}\tilde{f}^2 \ GeV  \nonumber\\
& &< 10^{-17}\tilde{f}^2 GeV < O(10^{-25}) GeV.
 \end{eqnarray}
 \vskip -0.3in
The life time of gluino is given as:
   \begin{eqnarray}
   \tau &=&\frac{\hbar}{\Gamma}\sim\frac{10^{-34} Jsec}{10^{-17}f^2 GeV}\sim\frac{10^{-7}}{f^2}> 10^{3}sec.
   \end{eqnarray}
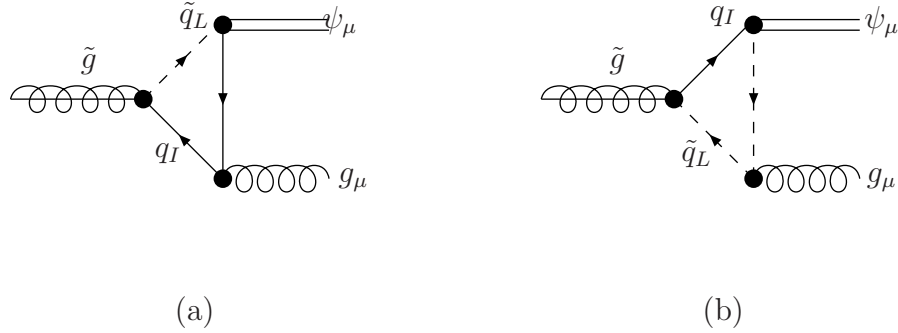
\begin{figure}
\begin{center}
\begin{picture}(1000,170)(50,0)
\Text(140,135)[]{$\tilde{g}$}
\Line(110,120)(160,120)
\Vertex(160,120){3.5}
\Gluon(110,120)(160,120){5}{4}
\DashArrowLine (160,120)(190,150){4}
\Text(180,152)[]{$\tilde{q}_L$}
\ArrowLine(190,90)(160,120)
\Text(170,100)[]{${q_I}$}
\ArrowLine(190,150)(190,90)
\Vertex(190,90){3.5}
\Gluon(190,90)(230,90){5}{4}
\Text(240,90)[]{$g_\mu$}
\Vertex(190,148){3.5}
\Line(190,150)(230,150)
\Line(190,146)(230,146)
\Text(235,150)[]{$\psi_\mu$}
\Text(180,40)[]{(a)}
\Line(310,120)(360,120)
\Vertex(360,120){3.5}
\Gluon(310,120)(360,120){5}{4}
\Text(340,135)[]{$\tilde{g}$}
\ArrowLine (360,120)(390,150)
\Text(380,152)[]{$q_I$}
\DashArrowLine (390,90)(360,120){4}
\Text(370,100)[]{$\tilde{q}_L$}
\DashArrowLine (390,150)(390,90){4}
\Vertex(390,90){3.5}
\Gluon(390,90)(430,90){5}{4}
\Text(440,90)[]{$g_\mu$}
\Vertex(390,148){3.5}
\Line(390,150)(430,150)
\Line(390,146)(430,146)
\Text(440,150)[]{$\psi_\mu$}
\Text(380,40)[]{(b)}
\end{picture}
 \end{center}
 \vskip -0.7in
\caption{One-loop gluino decay into goldstino and gluon.}
\label{fig:treedecgold}
\end{figure}
   \underline{${\bf \tilde{g}\rightarrow g_\mu \ \tilde{G}}$}
   
The matrix element for two-body decay of the gluino into a goldstino and a gluon (see Figure~2.4) is given as:
\begin{eqnarray}
\label{eq:MGold_I}
& & {\cal M}\sim {\tilde{f}}^2\int\frac{d^4k}{\left(2\pi\right)^4} \times{\cal V}^{-\frac{4}{5}}\left(\frac{i{\bar\sigma}\cdot k}{k^2-m_q^2+i\epsilon}\right)\left({\cal V}^{-\frac{15}{9}}\right)\left(\frac{i}{\left[\left(k-p_{\tilde{G}}\right)^2-m^2_{\tilde{q}}+i\epsilon\right]}
\right)\nonumber\\
& & \times\left({\cal V}^{\frac{53}{45}}\epsilon\cdot\left(2k-p_{\tilde{G}}-p_{\tilde{g}}\right)\right)\left(\frac{i}
{\left[\left(k-p_{\tilde{g}}\right)^2-m^2_{\tilde{q}}+i\epsilon\right]}\right) +\nonumber\\
& &{\tilde{f}}\int\frac{d^4k}{\left(2\pi\right)^4} \times{\cal V}^{-\frac{4}{5}}\left(\frac{i}
{\left[\left(k+p_{\tilde{G}}\right)^2-m^2_{\tilde{q}}+i\epsilon\right]}\right)\left({\cal V}^{-\frac{15}{9}}\right)\left( {\bar\sigma}\cdot\epsilon\right)\nonumber\\
& & \times\left(\frac{i{\bar\sigma}\cdot\left(k-p_{g_\mu}\right)}{\left[
\left(k-p_{g_\mu}\right)^2-m^2_q+i\epsilon\right]}\right).
\end{eqnarray}
As discussed in \cite{Dhuria+Misra_mu_Split_SUSY}, the Wilson coefficients corresponding to gluino-goldstino- gluon coupling do not change much upon RG evolution to EW scale. Going through an analysis similar to the ${\tilde g} \rightarrow \tilde{\chi}_3^0  g$, it was shown in \cite{gravitino_DM}:
\begin{eqnarray}
\label{eq:spinavGammaGold}
& &  \sum_{\tilde{g}\ {\rm and}\ \tilde{G}\ {\rm spins}}|{\cal M}|^2
\sim   Tr\biggl(\sigma\cdot p_{\tilde{G}}\Bigl[A_2{\bar\sigma}\cdot\epsilon + B_3{\bar\sigma}\cdot p_{\tilde{g}}{\bar\sigma}\cdot\epsilon + D_2{\bar\sigma}\cdot p_{\tilde{g}}{\bar\sigma}\cdot\epsilon{\bar\sigma}\cdot p_{\tilde{G}} \nonumber\\
& & + D_4{\bar\sigma}_\mu{\bar\sigma}\cdot\epsilon{\bar\sigma}^\mu C^{(b)}_{24}\Bigr] \nonumber\\
&& {\hskip -0.2in}  \times\sigma\cdot p_{\tilde{g}}\left[A_2{\bar\sigma}\cdot\epsilon + B_3{\bar\sigma}\cdot p_{\tilde{g}}{\bar\sigma}\cdot\epsilon + D_2{\bar\sigma}\cdot p_{\tilde{g}}{\bar\sigma}\cdot\epsilon{\bar\sigma}\cdot p_{\tilde{G}} + D_4{\bar\sigma}_\mu{\bar\sigma}\cdot\epsilon{\bar\sigma}^\mu C^{(b)}_{24}\right]^\dagger\biggr),
\end{eqnarray}
where  $A_2\equiv  {10}^{-6} {\tilde f}^2 $, $B_3\equiv ({10}^{-23} {\tilde f}) {GeV}^{-1}$, $D_2\equiv  ({10}^{-34} {\tilde f}) {GeV}^{-2}$, $D_4 \equiv {\cal V}^{-2.4} {\tilde f}.$
At $p^0_{\tilde{G}}=m_{\tilde{g}}/2,p^1_{\tilde{G}}=p^2_{\tilde{G}}=p^3_{\tilde{G}}=\frac{m_{\tilde{g}}}{2\sqrt{3}}$, the matrix amplitude squared becomes:
\begin{equation}
\label{eq:spinavgammaGold II}
 {m^{2}_{\tilde{g}}}\left[{D_2}^2 {m^{4}_{\tilde{g}}}+\frac{1}{6} \left(6 {A_2}+12{D_4}
   {C_{24}^{(b)}}+{m_{\tilde{g}}} \left(6 {B_3}+\left(3+\sqrt{3}\right) {D_2} {m_{\tilde{g}}}\right)\right)^2\right]\sim  \tilde{f}^4 {A^{2}_2} {m^{2}_{\tilde{g}}}.
\end{equation}
So, using results from \cite{Griffiths_particle}, the decay width comes out to be equal to:
\begin{equation}
\label{eq:GammaGold}
\Gamma=\frac{ \sum_{\tilde{g}\ {\rm and}\ \tilde{G}\ {\rm spins}}|{\cal M}|^2}{16\pi\hbar m_{\tilde{g}}}\sim O({10}^{-1}) {A^{2}_2} {m_{\tilde{g}}} \sim {10}^{-2}\tilde{f}^4 \ GeV.
\end{equation}
Considering value of $\tilde{f}^2 \sim{10}^{-10}$, $\Gamma < {10}^{-22}$ GeV.  Life time of gluino is given as:
   \begin{eqnarray}
   \tau &=&\frac{\hbar}{\Gamma}\sim\frac{10^{-34} Jsec}{10^{-2}{\tilde f}^4 \ GeV}\sim\frac{10^{-21}}{{\tilde f}^4} sec > {10^{-1}}sec.
   \end{eqnarray}
\section{Results and Discussion}
We have studied in detail the possibility of generating $\mu$-split-like  SUSY scenario in the context of local type IIB Swiss-cheese orientifold (involving isometric holomorphic involution) compactifications in the L(arge) V(olume) S(cenarios). After giving details of the geometric framework, we showed the possibility of obtaining a local positive semi-definite potential for specific choice of values of VEV of position as well as Wilson line moduli and bulk moduli.  Using the modified  ${\cal N} = 1$ chiral co-ordinates due to presence of D3- and D7-branes, we constructed the form of K\"{a}hler potential and superpotential relevant to describe the phenomenological results. By evaluating all possible effective Yukawa couplings in the context of ${\cal N} = 1$ gauged supergravity, and thereafter showing that effective Yukawas change only by ${\cal O}(1)$ under RG flow evolution from string scale down to electroweak scale, we realized that Dirac mass term corresponding to fermionic superpartners of two (${\cal A}_1$ and ${\cal A}_3$) of the aforementioned four-Wilson line moduli, matched very well with the order of Dirac mass of the electron, and the Dirac mass term corresponding to fermionic superpartners of the remaining two (${\cal A}_2$ and ${\cal A}_4$) of the aforementioned four-Wilson line moduli, matched very well with the order of Dirac masses of the first generation SM-like quarks. Therefore, the fermionic superpartners of ${\cal A}_1$ and ${\cal A}_3$ could be identified, respectively with $e_L$ and $e_R$, and the fermionic superpartners of ${\cal A}_2$ and ${\cal A}_4$ could be identified, respectively with the first generation quarks: $u/d_L$ and $u/d_R$. After evaluating soft SUSY breaking parameters, Wilson line moduli masses (identified with sfermions) turned out to be very heavy at string scale.  Based on the arguments given in \cite{Sparticles_Misra_Shukla}, according to which there was an ${\cal O}(1)$ change in the value of Wilson line modulus mass under an RG flow from string scale down to EW scale, we assumed that squark/slepton masses will remain heavy at EW scale too. The highlights of this chapter are:
\begin{enumerate}
\item
By evaluating Higgs mass matrix at electroweak scale, we showed the possibility of obtaining one light Higgs of order 125 GeV at EW scale while masses of another Higgs as well as higgsino mass parameter to be very large; the primary indication of $\mu$-split-like  SUSY.
\item
By evaluating relevant interaction vertices using ${\cal N}=1$ gauged supergravity action, we estimated life time of gluino to be very high- thus providing another concrete signature of $\mu$-split-like  SUSY.
\end{enumerate}

 
\chapter{Particle Cosmology in the Context of Type IIB Local D3/D7 $\mu$-split-like SUSY Model}
\vskip -0.5in
{\hskip1.4in{\it ``The full cosmos consists of the physical stuff and consciousness. Take away consciousness and it's only dust; add consciousness and you get things, ideas, and time."}}

\hskip3.0in - Neal Stephenson, Anathem.

\graphicspath{{Chapter3/}{Chapter3/}}
\vskip -0.5in
\section{Introduction}
It is challenging to provide a suitable choice of vacuum (local minimum) of string theory to infer various cosmological and phenomenological issues, and other fundamental physics. One of the most exciting unresolved issues in particle physics and cosmology is the nature of Dark Matter (DM) in the Universe. Over the years, astronomical and cosmological observations have put significant constraints on its expected properties. Theoretical status of DM generally emerges in the context of theories beyond Standard Model. It is well known that supersymmetric models with conserved R-parity contain a stable neutralino which has been assumed to be a good candidate for cold dark matter. However, in models coupled to gravity and even in other scenarios of supersymmetry breaking mediation including gauge mediation, the gravitino (the supersymmetric partner of graviton) stands out as probably the most natural and attractive candidate for DM. The long-lived gravitinos were generated in the early universe. In the standard big-bang cosmology, they were in thermal equilibrium and then because of their weak gravitational interactions, frozen out while they were relativistic. In this case their abundance might overclose the universe known as ``cosmological gravitino problem", the resolution of which is quite natural because  abundance get diluted as universe experiences through inflationary phase. Therefore, cosmologically relevant gravitino abundance is then recreated in the reheating phase after inflation by inelastic $2\rightarrow 2$ scattering  and $1\rightarrow 2$ decay processes of particles from the thermal bath where abundance goes linear in the reheating temperature $T_R$. However, in string/M theory-inspired models, production of DM particles significantly alters because of the presence of moduli because decay of modulus increases the entropy by several amount, thereby decreasing the overall relic abundance of gravitino. Therefore, sizable amount of gravitinos can be produced by   (non-thermal) production of gravitinos LSP formed by decays of moduli and  can even dominate the thermal production of gravitinos in the early plasma, discussed in \cite{Nakamura,Bobby_Acharya,dutta_et_al}. Even in particle physics models, sufficient amount of relic density of LSP can be produced by decays of  N(ext-to) L(ightest) S(upersymmetric) P(article) and has been studied in literature \cite{Fweng}.

In addition to getting a light Higgs and high life time of gluino, the mass scales of superpartners evaluated in the context of local type IIB large volume big divisor $D3/D7$ framework provide us with gravitino as the L(ightest) S(upersymmetric) P(article) and sleptons/squarks as N(ext-to) L(ightest)~ S(upers
ymmetric) P(article)s with  (Bino/Wino-type) gaugino/neutralino, in the dilute-flux approximation,  only differing in their masses  by an ${\cal O}(1)$ factor. This helps in shedding a light on the possibility of gravitino as a potential DM candidate. However, the more explicit realization of the same requires life time of the LSP to be around and preferably more than the age of the universe and life times of co-NLSP's to be small enough to disturb the beautiful predictions of Big Bang Nucleosynthesis (BBN). With this motivation, in section {\bf 2} of this chapter, we show  that  the presence of very suppressed non-zero R-parity violating couplings and high squark masses helps to reduce the decay width and hence lifetime of the gravitino LSP becomes very long (of the order greater than age of the Universe) satisfying the requirement of potential dark matter candidate. In {\bf 3.1} and {\bf 3.3}, we study decay width and life time of two-body and hadronic three-body decays of NLSP's (gauginos) and co-NLSP's (sleptons/squarks), life times of which  ensure that energy released from both gauge boson and hadronic decays do not disturb the the predictions arising from BBN and cosmic microwave background, etc. In {\bf 3.2}, we consider R-parity violating decays of (Wino/Bino-dominated) neutralino into ordinary particles (even R-parity), high life time of which, as compared to life time for decay of neutralino into LSP, ensures that relic abundance of gravitino is not diluted. Going ahead, in order to be able to perform a reliable comparison between theoretical predictions and improving measurements of the relic abundance from underground DM searches, in section {\bf 4}, we evaluate the relic abundance of gravitino with the assumption that almost all of the dark matter particles are produced non-thermally via decays of sleptons/neutralino existing as co-NLSP's in our case. Finally we summarize our results in section {\bf 5}.

\vskip -0.5in
 \section{Gravitino Decays}
Decays of the gravitino are, in general, driven by any of the trilinear R-violating couplings which  necessarily involve squark or slepton as propagators. In MSSM, stability of LSP is governed by conserved R-parity however same is not a restrictive condition in `` $\mu$-split-like  SUSY''.  Some of the R-parity-violating interactions are not necessarily negligible  and the consideration of LSP to be a viable dark matter candidate needs the contribution of the same to be evaluated. Here, the large squark masses in $\mu$-split-like  SUSY helps to suppress the decay width. In BSM/2HDM models, one considers R-parity violating coupling generated from R-parity violating superpotential. Here also, we consider the similar R-parity violating couplings, but evaluated in the context of effective ${\cal N}=1$ gauged supergravity action.
\subsection{Two-Body Gravitino Decays}
We discuss the decays of the gravitino into neutrino and gauge bosons as well as the light Higgs and neutrinos.
\begin{figure}
   \begin{center}
    \begin{picture}(135,137) (15,-18)
   \Line(100,50)(115,24)
   \DashArrowLine(125,41)(115,24){4}
   \Vertex(115,24){2.5}
   \Text(136,19)[]{{{$q$}}}
   \LongArrow(126,24)(135,8)
   \ArrowLine(115,24)(130,-2)
   \Photon(100,50)(115,24){5}{2}
   \Text(102,84)[]{{{$k$}}}
   \LongArrow(103,73)(112,88)
   \Photon(100,50)(130,102){5}{4}
   \Text(30,50)[]{{{$\psi_{\mu}$}}}
   \Text(70,65)[]{{{$p$}}}
   \LongArrow(61,59)(79,59)
   \Line(40,52)(100,52)\Line(40,48)(100,48)
   \Vertex(100,50){2.5}
   \Text(130,50)[]{{{$\left\langle\tilde{\nu}\right\rangle$}}}
   \Text(135,-10)[]{{{$\nu$}}}
   \Text(135,111)[]{{{$\gamma$}}}
   \Text(97,30)[]{{{$\tilde{\lambda}^0$}}}
  \end{picture}
  \caption{Two-body gravitino decay: $\psi_\mu\rightarrow\nu+\gamma$.}
  \end{center}
  \end{figure}
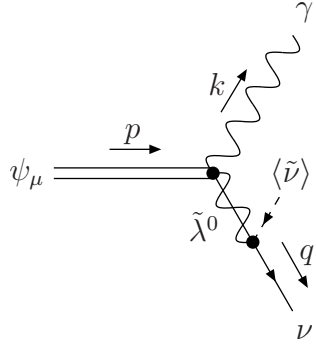
In ${\cal N}=1$ SUGRA the gaugino-$\langle$sneutrino$\rangle$-neutrino vertex is given by:
\begin{equation}
\label{eq:gaugino-sneu-neu}
{\cal O}(a_1)-{\rm term\ in}\ \frac{g_{YM}g_{T^B{\bar a}_1}X^{T^B}}{\left(\sqrt{K_{{\cal A}_1{\bar{\cal A}}_1}}\right)^2}
\sim\frac{{\cal V}^{-\frac{11}{9}}}{10^4}.
\end{equation}
Similarly the gravitino-gauge-boson-$\langle\tilde{\nu}\rangle-\nu$-vertex in the ${\cal N}=1$ SUGRA Lagrangian is given by the ${\cal O}(a_1)$-term in $g_{YM}g_{a_1{\bar T}^{B}}X^{T^B}Z^0/A_\mu{\bar\nu}\gamma^\nu\gamma^\mu\psi_\nu$, which is: $\frac{g_{YM}g_{a_1{\bar T}^B}X^{T^B}}{\left(\sqrt{K_{{\cal A}_1{\bar{\cal A}}_1}}\right)^2}\sim\frac{{\cal V}^{-\frac{11}{9}}}{10^4}$.
Using \cite{Grefe_i}, one sees that the lifetime of the decay $\psi_\mu\rightarrow\gamma+\nu_e$ is given by:
\begin{eqnarray}
\label{eq:psi-nu+phot}
& & \Gamma\left(\psi_\mu\rightarrow\gamma+\nu_e\right)=\frac{1}{64\pi}\left(\frac{\langle {\cal A}_1\rangle}{\langle {\cal Z}_i\rangle}\right)^2\frac{m^3_{3/2}}{M_P^2}\left|U_{\tilde{\gamma}\tilde{Z}}\right|^2\left(\frac{V^{-\frac{11}{9}}}{10^4}\right)^2
\nonumber\\
& & \sim\frac{1}{64\pi}\left(\frac{\langle{\cal V}^{-\frac{2}{9}}\rangle}{{\cal V}^{\frac{1}{36}}}\right)^2\frac{m^2_{3/2}}{M_P^2}
m_Z^2\left(\frac{M_{\lambda_1}-M_{\lambda_2}}{M_{\lambda_1}M_{\lambda_2}}\right)^2sin^2\theta_W cos^2\theta_W
\left(\frac{V^{-\frac{11}{9}}}{10^4}\right)^2.
\end{eqnarray}
We have a gaugino mass degeneracy (up to ${\cal O}(1)$ factors) implying: $M_{\lambda_1}\sim M_{\lambda_2}$. One hence gets a very small decay width and an extremely enhanced lifetime.
   \begin{figure}
   \begin{center}
 \parbox{6.3cm}{
  \begin{picture}(135,140) (15,-18)
   \Line(100,50)(115,24)
   \DashArrowLine(125,41)(115,24){4}
   \Vertex(115,24){2.5}
   \Text(136,19)[]{{{$q$}}}
   \LongArrow(126,24)(135,8)
   \ArrowLine(115,24)(130,-2)
   \Photon(100,50)(115,24){5}{2}
   \Text(102,84)[]{{{$k$}}}
   \LongArrow(103,73)(112,88)
   \Photon(100,50)(130,102){5}{4}
   \Text(30,50)[]{{{$\psi_{\mu}$}}}
   \Text(70,65)[]{{{$p$}}}
   \LongArrow(61,59)(79,59)
   \Line(40,52)(100,52)\Line(40,48)(100,48)
   \Vertex(100,50){2.5}
   \Text(130,50)[]{{{$\left\langle\tilde{\nu}\right\rangle$}}}
   \Text(135,-10)[]{{{$\nu$}}}
   \Text(135,111)[]{{{$Z^0$}}}
   \Text(97,30)[]{{{$\tilde{\chi}^0$}}}
  \end{picture}
 }\;+\;\parbox{6.3cm}{
  \begin{picture}(155,140) (15,-18)
   \Text(30,50)[]{{{$\psi_{\mu}$}}}
   \Text(70,65)[]{{{$p$}}}
   \LongArrow(61,59)(79,59)
   \Line(40,52)(100,52)\Line(40,48)(100,48)
   \Vertex(100,50){2.5}
   \Text(135,-10)[]{{{$\nu_{\tau}$}}}
   \DashArrowLine(130,50)(100,50){4}
   \Text(142,50)[]{{{$\left\langle\tilde{\nu}_{\tau}\right\rangle$}}}
   \Text(128,32)[]{{{$q$}}}
   \LongArrow(118,36)(127,21)
   \ArrowLine(100,50)(130,-2)
   \Text(102,84)[]{{{$k$}}}
   \LongArrow(103,73)(112,88)
   \Photon(100,50)(130,102){5}{4}
   \Text(135,111)[]{{{$Z^0$}}}
  \end{picture}
 }
 \caption{Two-body gravitino decay: $\psi_\mu\rightarrow Z^0+\nu$.}
 \end{center}
 \end{figure}
Again, using (\ref{eq:gaugino-sneu-neu}) and \cite{Grefe_i}, the decay width for $\psi_\mu\rightarrow Z+\nu$ is given by:
\begin{eqnarray}
\label{eq:gravitino_Z+nu}
& & \Gamma\left(\psi_\mu\rightarrow Z+\nu\right)\sim \frac{1}{64\pi}\left(\frac{\langle {\cal A}_1\rangle}{\langle {\cal Z}_i\rangle}\right)^2\frac{m^3_{3/2}}{M_P^2}\left(1-\frac{M^2_Z}{m^2_{3/2}}\right)^2\left(\frac{V^{-\frac{11}{9}}}{10^4}\right)^2
\nonumber\\
& & \times\Biggl[\left(\frac{M_Z}{M_{\lambda}}\right)^2\times{\cal O}(1) + {\cal O}(1)\left|1 + sin\beta cos\beta \left(\frac{M^2_Z}{M_\lambda\hat{\mu}_{{\cal Z}_1{\cal Z}_2}}\right)\right|^2 + \left(\frac{M_Z}{m_{3/2}}\right)\left(\frac{m_Z}{M_\lambda}\right)\nonumber\\
&& \left(1 + sin\beta cos\beta \left(\frac{M^2_Z}{M_\lambda\hat{\mu}_{{\cal Z}_1{\cal Z}_2}}\right)\right)\times{\cal O}(1)\Biggr] \sim\frac{{\cal V}^{-\frac{1}{2}-6-\frac{22}{9}}}{64\pi\times 10^{8}}\times {\cal O}(1)M_P.
\end{eqnarray}
Using  $M_Z\sim 90 GeV$; $m_{ \lambda}\sim {\cal V}^{\frac{2}{3}} m_{\frac{3}{2}} $ and  $\hat{\mu}_{{\cal Z}_1{\cal Z}_2}\sim {\cal V}m_{\frac{3}{2}}$ from eqs. (\ref{eq:gaugino_mass}) and (\ref{eq:muhat_Z1Z2}),
\begin{eqnarray}
\label{eq:gravitino_Z+nu1}
& & \Gamma\sim\frac{{\cal V}^{-\frac{1}{2}-6-\frac{22}{9}}}{64\pi\times 10^{8}}\times {\cal O}(1)M_P.
\end{eqnarray}
 This, for ${\cal V}\sim10^5$ yields $\tau\sim 10^{22}s$.
 \begin{figure}
 \begin{center}
 \parbox{4.5cm}{
  \begin{picture}(135,137) (15,-18)
   \Text(30,50)[]{{{$\psi_{\mu}$}}}
   \SetWidth{0.5}
   \Text(70,65)[]{{{$p$}}}
   \LongArrow(61,59)(79,59)
   \Line(40,52)(100,52)\Line(40,48)(100,48)
   \Vertex(100,50){2.5}
   \Text(135,-10)[]{{{$\nu$}}}
   \Text(128,32)[]{{{$q$}}}
   \LongArrow(118,36)(127,21)
   \ArrowLine(100,50)(130,-2)
   \DashArrowLine(115,76)(100,50){4}
   \Text(109,96)[]{{{$k$}}}
   \LongArrow(110,85)(119,101)
   \DashLine(115,76)(130,102){4}
   \Line(108,74)(122,78)\Line(113,83)(117,69)
   \Text(135,111)[]{{{$h$}}}
   \Text(95,67)[]{{{$\tilde{\nu}^*$}}}
  \end{picture}
 }\;+\;\parbox{4.5cm}{
  \begin{picture}(135,137) (15,-18)
   \Line(100,50)(115,24)
   \DashArrowLine(125,41)(115,24){4}
   \Vertex(115,24){2.5}
   \Text(136,19)[]{{{$q$}}}
   \LongArrow(126,24)(135,8)
   \ArrowLine(115,24)(130,-2)
   \Photon(100,50)(115,24){5}{2}
   \Text(102,84)[]{{{$k$}}}
   \LongArrow(103,73)(112,88)
   \DashLine(100,50)(130,102){4}
   \Text(30,50)[]{{{$\psi_{\mu}$}}}
   \Text(70,65)[]{{{$p$}}}
   \LongArrow(61,59)(79,59)
   \Line(40,52)(100,52)\Line(40,48)(100,48)
   \Vertex(100,50){2.5}
   \Text(130,50)[]{{{$\left\langle\tilde{\nu}\right\rangle$}}}
   \Text(135,-10)[]{{{$\nu$}}}
   \Text(135,111)[]{{{$h$}}}
   \Text(97,30)[]{{{$\tilde{\lambda}^0$}}}
  \end{picture}
 }
 \vskip -0.1in
\caption{Two-body gravitino decay: $\psi_\mu\rightarrow h + \nu$.}
\end{center}
\end{figure}
The mass-like insertion is given by $m^2_{\tilde{\nu}h}$, which is the coefficient of $\tilde{\nu}h$ in the ${\cal N}=1$ SUGRA action. To estimate this, we note that near $|z_1|\sim|z_2|\sim0.7{\cal V}^{\frac{1}{36}},|a_1|\sim{\cal V}^{-\frac{2}{9}},|a_2|\sim{\cal V}^{-\frac{1}{9}},|a_3|\sim{\cal V}^{-\frac{13}{18}},|a_4|\sim{\cal V}^{-\frac{11}{9}}$, the potential can be approximated by
\begin{eqnarray}
\label{eq:Potential}
& & V\sim e^KG^{T_S{\bar T}_S}|D_{T_S}W|^2 \sim\frac{z^{72}e^{-2 n^s (vol(\Sigma_S) + \mu_3 z^2)}\sqrt{{\cal V}^{\frac{1}{18}}+\mu_3z^2}}{\Lambda},
\end{eqnarray}
where ${\Lambda}={\cal V}+\left({\cal V}^{\frac{2}{3}}+(\alpha_{12}{\cal V}^{\frac{5}{18}} + \alpha_{13}{\cal V}^{\frac{8}{9}} + \alpha_{14}{\cal V}^{\frac{8}{9}})a_1 + (\alpha_{11}{\cal V}^{\frac{10}{9}}a_1 + \alpha_{12}{\cal V}^{\frac{5}{18}}  + \alpha_{13}{\cal V}^{\frac{8}{9}} \right. \\
 \left. + \alpha_{14}{\cal V}^{\frac{8}{9}})|a_1| + \mu_3z^2\right)^{\frac{3}{2}} - \left({\cal V}^{\frac{1}{18}} + \mu_3 z^2\right)^{\frac{3}{2}}$.

The coefficient of the ${\cal O}\left((z - 0.6{\cal V}^{\frac{1}{36}})(a_1 - {\cal V}^{-\frac{2}{9}})\right)/\sqrt{K_{Z_1{\bar Z}_1}K_{{\cal A}_1{\bar{\cal A}}_1}}$ in
(\ref{eq:Potential}) turns out to be:
\begin{equation}
\label{eq:msnuhsquared}
m^2_{\tilde{\nu}h}\sim{\cal V}^{\frac{23}{18}}m^2_{3/2}\times10^{-16}.
\end{equation}
The gravitino-gaugino-Higgs vertex
in the ${\cal N}=1$ gauged SUGRA action will be given by the term : $
g_{YM}D^{T^B}{\bar\psi}_\mu\gamma^\mu\lambda=0$, on shell gravitino. Hence, the second diagram involving a gaugino NLSP, does not contribute. However, if the gaugino is replaced by a neutralino, then the higgsino-component of this neutralino will contribute via the term: $
g_{{\cal Z}_1{\bar Z}_1}\partial_\nu{\cal Z}_1{\bar\psi}_{\mu R}\gamma^\nu\gamma^\mu\tilde{H}^0_L$
in ${\cal N}=1$ SUGRA action. This yields a vertex:$
\tilde{f}\frac{g_{{\cal Z}_1{\bar{\cal Z}}_1}\gamma_\mu\slashed{k}}{\left(\sqrt{K_{{\cal Z}_1{\bar{\cal Z}}_1}}\right)^2}\sim \tilde{f}\gamma_\mu\slashed{k}.$
Further, the Higgsino-$\langle\tilde{\nu}_e\rangle$-$\nu_e$ vertex will be given by ${\cal O}(a_1-z_i)$ term in
$e^{\frac{K}{2}}\left({\cal D}_{{\bar a}_1}D_{z_i}{\bar W}\right){\bar\nu}_{e\ L}\tilde{H}^0_R$, which after assigning $\langle z_i\rangle$ a value of $\sim{\cal V}^{\frac{1}{36}}$ yields:
\begin{equation}
\label{eq:vev z_i-vev nu_e-Higgsino-nu_e}
\tilde{f}{\cal V}^{-\frac{61}{36}}\langle\nu_e\rangle.
\end{equation}
Hence, using (\ref{eq:vev z_i-vev nu_e-Higgsino-nu_e}) and \cite{Grefe_i}:
\begin{eqnarray}
\label{eq:Gamma_gravtoh+nu_e}
& & \Gamma\left(\psi_\mu\rightarrow h+\nu_e\right)\sim\frac{1}{384\pi}\left(\frac{m^3_{3/2}}{M_P^2}\right)\left|\frac{m^2_{\tilde{\nu}_eh}}{m_h^2-m^2_{\tilde{\nu}_e}}
+sin\beta cos\beta \frac{m_Z^2}{M_{\tilde{g}}\hat{\mu}_{{\cal Z}_1{\bar{\cal Z}}_2}}\frac{\langle\nu_e\rangle}{\langle {\cal Z}_i\rangle}\tilde{f}^2{\cal V}^{-\frac{61}{36}}\right|^2\nonumber\\
& & \sim 10^{-3}{\cal V}^{-6+\frac{5}{9}}\times10^{-32}M_P\Biggr|_{{\cal V}\sim10^5}\sim10^{-60}M_P,
\end{eqnarray}
which yields a lifetime of around $10^{17}s$.
\vskip -0.5in
\subsection{Three-Body R-parity Violating Gravitino Decays}
In this subsection, we will be considering gravitino decays involving three types of R-parity violating vertices that appear in R-parity violating superpotential \cite{Gmoreau}: $$ W_{\slashed{R}_p} = {\lambda}_{ijk} L_iL_j E^c_k +
{\lambda}^{\prime}_{ijk} L_i Q_j D^c_k+   {\lambda}^{\prime \prime}_{ijk}U_i^cD_j^cD_k^c + \mu_i H L_i. $$
\begin{figure}[t!]
\label{decaydiag1}
\begin{center}
\begin{picture}(150,97)(100,100)
\Vertex(150,158){3.5}
\Line(100,160)(150,160)
\Line(100,156)(150,156)
\ArrowLine(150,160)(190,190)
\DashArrowLine (190,130)(150,158)5
\Vertex(190,130){3.5}
\ArrowLine(190,130)(230,160)
\ArrowLine(230,100)(190,130)
\Text(95,168)[]{$\psi_\mu$}
\Text(240,165)[]{$e_{jL}$}
\Text(240,100)[]{$e_{kR}$}
\Text(196,197)[]{${\nu_{iL}}$}
\put(155,132){${\stilde {\nu}}_{iL}$}
\put(165,80){(a)}
\end{picture}
\hspace{0.9cm}
\begin{picture}(150,97)(100,100)
\Line(100,160)(150,160)
\Line(100,156)(150,156)
\Vertex(150,158){3.5}
\ArrowLine(150,158)(190,190)
\DashArrowLine (190,130)(150,158)5
\Vertex(190,130){3.5}
\ArrowLine(190,130)(230,160)
\ArrowLine(230,100)(190,130)
\Text(95,168)[]{$\psi_\mu$}
\Text(240,165)[]{${\nu}_{iL}$}
\Text(240,100)[]{$e_{kR}$}
\Text(196,197)[]{${e_{jL}}$}
\put(155,132){${\stilde e}_{jL}$}
\put(165,80){(b)}
\end{picture}
\end{center}
\vspace{0.8cm}
\begin{picture}(150,97)(100,100)
\Line(250,160)(300,160)
\Line(250,156)(300,156)
\Vertex(300,158){3.5}
\ArrowLine(340,190)(300,160)
\DashArrowLine (300,158)(340,130)5
\Vertex(340,130){3.5}
\ArrowLine(380,160)(340,130)
\ArrowLine(340,130)(380,100)
\Text(245,168)[]{$\psi_\mu$}
\Text(390,165)[]{${\bar\nu}^c_{iR}$}
\Text(390,100)[]{$e_{jL}$}
\Text(346,197)[]{${e_{kL}}$}
\put(305,132){${\stilde e}_{kR}$}
\put(315,80){(c)}
\end{picture}
\phantom{xxx}
\label{fig:gravdeca}
\vskip 0.3in
\caption{Three-body gravitino decays involving $\slashed{R}_p\ \lambda_{ijk}$ coupling.}
\end{figure}
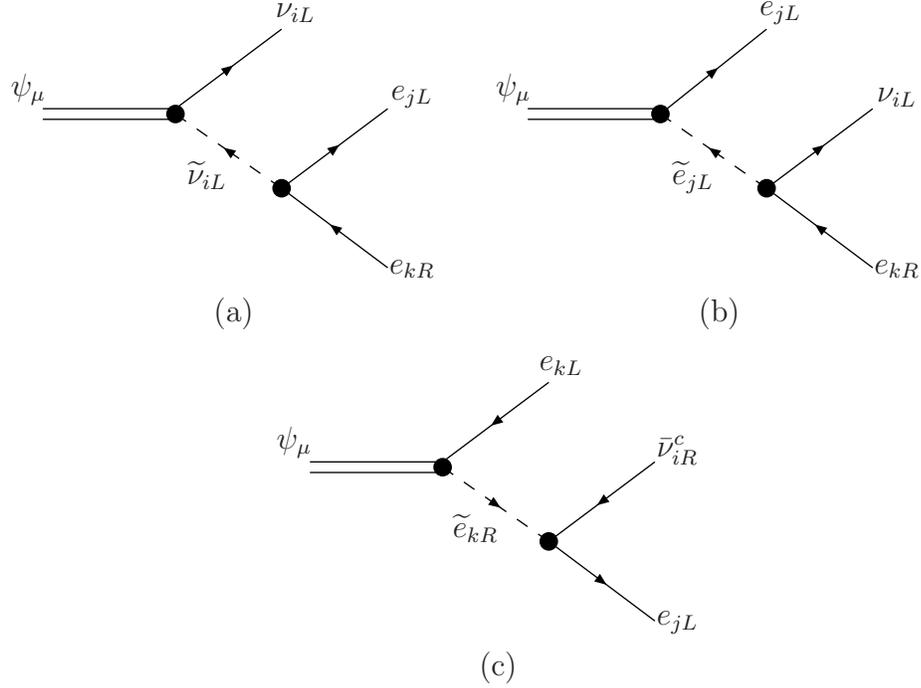
\underline{\bf {Decays involving $\lambda_{ijk}$ coupling}}

The ${\tilde \nu_{iL}}-{e_{jL}}-{e_{kR}}$ vertex corresponding to Figure~3.4(a) is given by considering the contribution of the term: $
{\cal L }= \frac{e^{\frac{K}{2}}}{2}\left({\cal D}_{{\cal A}_1} D_{{\cal A}_3}W\right){\bar \chi}^{{\cal A}_1}\chi^{{\cal A}_3}$.
We first expand $ {\cal D}_{a_1}D_{a_3}W$ in the fluctuations linear in $a_1\rightarrow a_1+{\cal V}^{-\frac{2}{9}}{M_P}$ by utilizing equations (\ref{eq:K}) and (\ref{eq:W}). On simplifying and using an argument discussed in equation (\ref{eq:eq_mass_terms_diag_non-diag_basis}) according to which $\frac{e^{\frac{K}{2}}}{2} {\cal D}_{a_1}D_{a_3}W \chi^{{\cal A}_1}\chi^{{\cal A}_3} \sim e^{\frac{K}{2}}{\cal D}_{{\cal A}_1}D_{{\cal A}_3}W\chi^{{\cal A}_1}\chi^{{\cal A}_3}$, we have $e^{\frac{K}{2}}{\cal D}_{{\cal A}_1}D_{{\cal A}_3}W\chi^{{\cal A}_1}\chi^{{\cal A}_3} \sim {\cal V}^{-\frac{19}{72}}\delta{{\cal A}_1}$. Using this, the physical ${\tilde \nu_{iL}}-{e_{jL}}-{e_{kR}}$ vertex will be given as:
\begin{eqnarray}
\label{eq:2a1}
C^{{\tilde \nu_{iL}} {e_{jL}} {e_{kR}}} \sim \frac{{\cal V}^{-\frac{19}{72}}}{{\sqrt{\hat{K}_{{\cal A}_1{\bar{\cal A}}_1}{\hat{K}_{{\cal A}_1{\bar{\cal A}}_1}}{\hat{K}_{{\cal A}_3{\bar {\cal A}}_3}}}}}\sim \frac{{\cal V}^{-\frac{19}{72}}}{\sqrt {10^{15}}}\sim {\cal V}^{-\frac{7}{4}}, {\rm for}~{\cal V}\sim {10^5}.
\end{eqnarray}
The physical ${\tilde e_{jL}}-{\nu_{iL}}-{e_{kR}}$ vertex of Figure~3.4(b) is same as the vertex ${\tilde \nu_{iL}}-{e_{jL}}-{e_{kR}}$, and therefore,
\begin{equation}
\label{eq:2b1}
C^{{\tilde e_{jL}} {\nu_{iL}}{e_{kR}}}\sim C^{{\tilde \nu_{iL}} {e_{jL}} {e_{kR}}} \sim {\cal V}^{-\frac{7}{4}}.
\end{equation}
The ${\tilde e_{kR}}-{\bar\nu^c_{iR}}-{e_{jL}}$ vertex corresponding to Figure~3.4(c) is given by considering the contribution of: $\frac{e^{\frac{K}{2}}}{2}\left({\cal D}_{{\cal {\bar A}}_1} D_{{\cal A}_1}W\right)\chi^{ {\cal A}^{c}_1}\chi^{{\cal A}_1}$. By expanding ${\cal D}_{a_1}D_{\bar {a_1}}W $ in the fluctuations linear in $a_3\rightarrow a_3+{\cal V}^{-\frac{13}{18}}{M_P}$, on simplifying,  we have $\frac{e^{\frac{K}{2}}}{2} {\cal D}_{a_1}D_{\bar {a_1}}W \equiv  e^{\frac{K}{2}}{\cal D}_{{\cal {\bar A}}_1}D_{{\cal A}_1}W\sim {\cal V}^{-\frac{1}{3}}\delta{a_3},$
implying that the physical ${\tilde e_{kR}}-{\bar\nu^c_{iR}}-{e_{jL}}$ vertex will be given as:
\begin{eqnarray}
\label{eq:2c1}
C^{{\tilde e_{kR}}{\bar\nu^c_{iL}}{e_{jL}}} \sim \frac{{\cal V}^{-\frac{1}{3}}}{{\sqrt{\hat{K}_{{\cal A}_1{\bar{\cal A}}_1}{\hat{K}_{{\cal A}_1{\bar{\cal A}}_1}}{\hat{K}_{{\cal A}_3{\bar {\cal A}}_3}}}}}\sim \frac{{\cal V}^{-\frac{1}{3}}}{\sqrt {10^{15}}}\sim {\cal V}^{-\frac{7}{4}}, {\rm for}~{\cal V}\sim {10^5}.
\end{eqnarray}
Now, the matrix amplitude for all three Feynman diagrams corresponding to Figure~3.4 will be given as: $
|M|^2= \vert {M_a +M_b+M_c} \vert^2$. The analytical results for the full matrix  amplitude summed over spins in terms of  pure and cross terms are given in \cite{Gmoreau}. Utilizing their results, we will estimate matrix  amplitude for  both pure and cross terms to calculate decay width for the process $\psi_\mu{\to} \nu_i e_j \bar e_k$ in our framework. Strictly speaking, we will be neglecting fermion mass as compared to gravitino and squark masses.
 Introducing kinematic variables
 \begin{equation}
 2 p(\nu_i) \cdot p(e_j)
=
(1 - z_{e_k}) \mm^2 ,
\
 2 p(e_j) \cdot p(e_k)
=
(1 - z_{\nu_i}) \mm^2 ,
\ 2 p(\nu_i) \cdot p(e_k)  = (1 - z_{e_j}) \mm^2.
\end{equation}
In view of above, we can express the following definitions in terms of kinematic variables as given below:
\begin{eqnarray}
\label{eq:mii}
m^2_{ij} &= & (p(\nu_i)+p(e_j))^2 \sim  2p(\nu_i).p(e_j)= (1-z_{e_k})\mm^2 \nonumber\\
m^2_{jk} & = & (p(e_j)+p(e_k))^2 \sim 2p(e_j).p(e_k)=(1-z_{\nu_i})\mm^2 \nonumber\\
m^2_{ik} & = & (p(\nu_i)+p(e_k))^2 \sim  2p(\nu_i).p(e_k)=(1-z_{e_j})\mm^2.
\end{eqnarray}
Utilizing the form of expressions given in an appendix of \cite{Gmoreau}, and simplifying using (\ref{eq:mii}), we have
\begin{eqnarray}
\vert M_a \vert^2 &=& {1 \over 3} {( C^{{\tilde \nu_{iL}} {e_{jL}} {e_{kR}}})^2\over M^2_{pl}
(m^2_{jk}-\aa^2)^2}
(\mm^2-\mjk^2+\mi^2) (\mjk^2-\mj^2-\mk^2) \cr &&
\bigg ( {(\mm^2+\mjk^2-\mi^2)^2 \over 4 \mm^2}-\mjk^2 \bigg )
\nonumber\\
 &=& {1 \over 3} {( C^{{\tilde \nu_{iL}} {e_{jL}} {e_{kR}}})^2\over M^2_{pl}
((1-z_{e_i})\mm^2-\aa^2)^2}
[z_{\nu_i}(1-z_{\nu_i}) (\frac{z^2_{\nu_i}}{4})],\nonumber\\
& & \sim {1 \over 3} {(C^{{\tilde \nu_{iL}} {e_{jL}} {e_{kR}}})^2\over M^2_{pl}
\aa^4}
[z_{\nu_i}(1-z_{\nu_i}) (\frac{z^2_{\nu_i}}{4})].
\label{amp1a}
\end{eqnarray}
Here we have neglected gravitino mass as compared to sfermion mass.
With  same steps of similar procedure, we have
\begin{eqnarray}
\vert M_{b} \vert^2 &=& {1 \over 3} {(C^{{\tilde e_L} {\nu_L}{e_R}})^2 \over M^2_{pl}
(m^2_{ik}-\bb^2)^2}
(\mm^2-\mik^2+\mj^2) (\mik^2-\mi^2-\mk^2)  \cr &&
{\hskip -0.6in} \bigg ( {(\mm^2+\mik^2-\mj^2)^2 \over 4 \mm^2}-\mik^2 \bigg ) \sim  {1 \over 3} {(C^{{\tilde e_L} {\nu_L}{e_R}})^2 \over M^2_{pl}
(\bb^2)^2}[z_{e_j}(1-z_{e_j}) (\frac{z^2_{e_j}}{4} )], \\
\label{amp2}
\nonumber\\
 \vert M_{c} \vert^2 &=& {1 \over 3} {(C^{{\tilde e_R}{\bar\nu^c_R}{e_L}})^2 \over M^2_{pl}
(m^2_{ij}-\cc^2)^2}
(\mm^2-\mij^2+\mk^2) (\mij^2-\mi^2-\mj^2) \cr &&
{\hskip -0.6in} \bigg ( {(\mm^2+\mij^2-\mk^2)^2 \over 4 \mm^2}-\mij^2 \bigg ) \sim {1 \over 3}{(C^{{\tilde e_R}{\bar\nu^c_R}{e_L}})^2 \over M^2_{pl}
(\cc^2)^2}[z_{e_k}(1-z_{e_k}) (\frac{z^2_{e_k}}{4})], \\
\label{amp3}
\nonumber\\
2 Re(M_a M^\dagger_{b})&=&{1 \over 3} {( C^{{\tilde \nu_{iL}} {e_{jL}} {e_{kR}}}.C^{{\tilde e_L} {\nu_L}{e_R}})
\over M^2_{pl} (m^2_{jk}-\aa^2) (m^2_{ik}-\bb^2)} \bigg [
(\mik^2 \mjk^2 - \mm^2 \mk^2 - \mi^2 \mj^2) \cr && {\hskip -0.4in} \bigg (
(\mm^2+\mk^2-\mi^2-\mj^2)
-{1 \over 2 \mm^2}(\mm^2+\mjk^2-\mi^2) \cr && {\hskip -0.4in}  (\mm^2+\mik^2-\mj^2) \bigg )
+ {1 \over 2} (\mij^2 - \mi^2 - \mj^2) (\mjk^2 - \mj^2 - \mk^2) \cr && {\hskip -0.4in} 
(\mik^2 - \mi^2 - \mk^2) - {\mi^2 \over 2} (\mjk^2 - \mj^2 - \mk^2)^2
- {\mj^2 \over 2} (\mik^2 - \mi^2 - \mk^2)^2 \cr && {\hskip -0.4in} 
- {\mk^2 \over 2} (\mij^2 - \mi^2 - \mj^2)^2 + 2 \mi^2 \mj^2 \mk^2 \bigg ]\nonumber\\
& & {\hskip -1.0in}\sim  {2 \over 3} {( C^{{\tilde \nu_{iL}} {e_{jL}} {e_{kR}}}.C^{{\tilde e_L} {\nu_L}{e_R}})
\over M^2_{pl} \aa^2 \bb^2} (1-z_{\nu_i})(1-z_{e_j})(-1-z_{e_k}+2 z_{\nu_i}+2 z_{e_j}-z_{\nu_i}.z_{e_k}),
\label{amp12}
\end{eqnarray}
\begin{eqnarray}
2 Re(M_{b} M^\dagger_{c}) &=& {1 \over 3} {(C^{{\tilde e_L} {\nu_L}{e_R}}.C^{{\tilde e_R}{\bar\nu^c_R}{e_L}})
\over M^2_{pl} (m^2_{ik}-\bb^2) (m^2_{ij}-\cc^2)} \bigg [
(\mij^2 \mik^2 - \mm^2 \mi^2 - \mj^2 \mk^2) \cr && \bigg (
(\mm^2+\mi^2-\mj^2-\mk^2)
-{1 \over 2 \mm^2}(\mm^2+\mik^2-\mj^2) \cr && (\mm^2+\mij^2-\mk^2) \bigg )
+ {1 \over 2} (\mij^2 - \mi^2 - \mj^2) (\mjk^2 - \mj^2 - \mk^2) \cr &&
(\mik^2 - \mi^2 - \mk^2) - {\mi^2 \over 2} (\mjk^2 - \mj^2 - \mk^2)^2
- {\mj^2 \over 2} (\mik^2 - \mi^2 - \mk^2)^2 \cr &&
- {\mk^2 \over 2} (\mij^2 - \mi^2 - \mj^2)^2 + 2 \mi^2 \mj^2 \mk^2 \bigg ]\nonumber\\
& & {\hskip -0.7in}\sim  {2 \over 3} {(C^{{\tilde e_L} {\nu_L}{e_R}}.C^{{\tilde e_R}{\bar\nu^c_R}{e_L}})
\over M^2_{pl} \aa^2 \bb^2} (1-z_{e_j})(1-z_{e_k})(-1-z_{\nu_i}+2 z_{e_j}+2 z_{e_k}-z_{e_j}.z_{e_k}),\\
\label{amp23}
\nonumber\\
2 Re(M_a M^\dagger_{c}) &=& {1 \over 3} {( C^{{\tilde \nu_{iL}} {e_{jL}} {e_{kR}}}.C^{{\tilde e_R}{\bar\nu^c_R}{e_L}})
\over M^2_{pl} (m^2_{jk}-\aa^2) (m^2_{ij}-\cc^2)} \bigg [
(\mij^2 \mjk^2 - \mm^2 \mj^2 - \mi^2 \mk^2) \cr && \bigg (
(\mm^2+\mj^2-\mi^2-\mk^2)
-{1 \over 2 \mm^2}(\mm^2+\mjk^2-\mi^2) \cr && (\mm^2+\mij^2-\mk^2) \bigg )
+ {1 \over 2} (\mij^2 - \mi^2 - \mj^2) (\mjk^2 - \mj^2 - \mk^2) \cr &&
(\mik^2 - \mi^2 - \mk^2) - {\mi^2 \over 2} (\mjk^2 - \mj^2 - \mk^2)^2
- {\mj^2 \over 2} (\mik^2 - \mi^2 - \mk^2)^2 \cr &&
- {\mk^2 \over 2} (\mij^2 - \mi^2 - \mj^2)^2 + 2 \mi^2 \mj^2 \mk^2 \bigg ]\nonumber\\
& & {\hskip -1.0in}\sim  {2 \over 3} {( C^{{\tilde \nu_{iL}} {e_{jL}} {e_{kR}}}.C^{{\tilde e_R}{\bar\nu^c_R}{e_L}})
\over M^2_{pl} \aa^2 \bb^2} (1-z_{\nu_i})(1-z_{e_k})(-1-z_{e_j}+2 z_{\nu_i}+2 z_{e_k}-z_{\nu_i}.z_{e_k}).
\label{amp13}
\end{eqnarray}
Utilizing the results from (\ref{amp1a})- (\ref{amp13}), one gets the following form:
\begin{eqnarray}
\label{finampfig1}
& & |M|^2= {1 \over 3} {( C^{{\tilde \nu_{iL}} {e_{jL}} {e_{kR}}})^2\over M^2_{pl}
\aa^4}
[z_{\nu_i}(1-z_{\nu_i}) (\frac{z^2_{\nu_i}}{4})]+ {1 \over 3} {(C^{{\tilde e_L} {\nu_L}{e_R}})^2 \over M^2_{pl}
(\bb^2)^2}[z_{e_j}(1-z_{e_j}) (\frac{z^2_{e_j}}{4} )]\nonumber\\
& & + {1 \over 3}{(C^{{\tilde e_R}{\bar\nu^c_R}{e_L}})^2 \over M^2_{pl}
(\cc^2)^2}[z_{e_k}(1-z_{e_k}) (\frac{z^2_{e_k}}{4})] \nonumber\\
& & + {2 \over 3} {( C^{{\tilde \nu_{iL}} {e_{jL}} {e_{kR}}}\cdot C^{{\tilde e_L} {\nu_L}{e_R}})
\over M^2_{pl} \aa^2 \bb^2} (1-z_{\nu_i})(1-z_{e_j})(-1-z_{e_k}+2 z_{\nu_i}+2 z_{e_j}-z_{\nu_i}.z_{e_k})\nonumber\\
& & +  {2 \over 3} {(C^{{\tilde e_L} {\nu_L}{e_R}}\cdot C^{{\tilde e_R}{\bar\nu^c_R}{e_L}})
\over M^2_{pl} \aa^2 \bb^2} (1-z_{e_j})(1-z_{e_k})(-1-z_{\nu_i}+2 z_{e_j}+2 z_{e_k}-z_{e_j}.z_{e_k})+
\nonumber\\
& & {2 \over 3} {( C^{{\tilde \nu_{iL}} {e_{jL}} {e_{kR}}}\cdot C^{{\tilde e_R}{\bar\nu^c_R}{e_L}}).
\over M^2_{pl} \aa^2 \bb^2} (1-z_{\nu_i})(1-z_{e_k})(-1-z_{e_j}+2 z_{\nu_i}+2 z_{e_k}-z_{\nu_i}.z_{e_k}).
\end{eqnarray}
The differential decay rate follows:
\begin{eqnarray}
\frac{d^2 \Gamma}{dz_{e_j} d z_{e_k}}
 &=&
\frac{N_c \mm}{2^8 \pi^3}
\biggl (\frac{1}{2} \sum_{\rm spins} |{\cal M}|^2 \biggr )
\end{eqnarray}
Putting the result of $|M|^2$ in above,
\begin{eqnarray}
& & \frac{d^2 \Gamma}{dz_{e_j} d z_{e_k}}\sim
\frac{N_c \mm}{2^9 \pi^3}{1 \over 3} [{( C^{{\tilde \nu_{iL}} {e_{jL}} {e_{kR}}})^2\over M^2_{pl}
\aa^4}
[z_{\nu_i}(1-z_{\nu_i}) (\frac{z^2_{\nu_i}}{4})]+ {1 \over 3} {(C^{{\tilde e_L} {\nu_L}{e_R}})^2 \over M^2_{pl}
(\bb^2)^2}[z_{e_j}(1-z_{e_j}) (\frac{z^2_{e_j}}{4} )]\nonumber\\
& & + {1 \over 3}{(C^{{\tilde e_R}{\bar\nu^c_R}{e_L}})^2 \over M^2_{pl}
(\cc^2)^2}[z_{e_k}(1-z_{e_k}) (\frac{z^2_{e_k}}{4})] \nonumber\\
& & + {2 \over 3} {( C^{{\tilde \nu_{iL}} {e_{jL}} {e_{kR}}}\cdot C^{{\tilde e_L} {\nu_L}{e_R}})
\over M^2_{pl} \aa^2 \bb^2} (1-z_{\nu_i})(1-z_{e_j})(-1-z_{e_k}+2 z_{\nu_i}+2 z_{e_j}-z_{\nu_i}.z_{e_k})\nonumber\\
& & +  {2 \over 3} {(C^{{\tilde e_L} {\nu_L}{e_R}}\cdot C^{{\tilde e_R}{\bar\nu^c_R}{e_L}})
\over M^2_{pl} \aa^2 \bb^2} (1-z_{e_j})(1-z_{e_k})(-1-z_{\nu_i}+2 z_{e_j}+2 z_{e_k}-z_{e_j}.z_{e_k})+
\nonumber\\
& & {2 \over 3} {( C^{{\tilde \nu_{iL}} {e_{jL}} {e_{kR}}}\cdot C^{{\tilde e_R}{\bar\nu^c_R}{e_L}}).
\over M^2_{pl} \aa^2 \bb^2} (1-z_{\nu_i})(1-z_{e_k})(-1-z_{e_j}+2 z_{\nu_i}+2 z_{e_k}-z_{\nu_i}.z_{e_k})].
\end{eqnarray}
Using  $0 < z_{e_j} < 1, 1 - z_j <  z_{e_k}  <1$,
and  the numerical estimates of masses, ${\mm}\sim {\cal V}^{-2}M_P, \aa^2= \bb^2= \cc^2 \sim m^{2}_{{\cal A}_1}\sim{\cal V} \mm$  as given in equation (\ref{eq:mass_Zi}), after integrating, decay width reduces to
\begin{eqnarray}
& & \Gamma\sim
\frac{N_c \mm^7}{(2^9.3.120) \pi^3. M^2_{pl}.{\cal V}^2 \mm^4} \biggl[( C^{{\tilde \nu_{iL}} {e_{jL}} {e_{kR}}})^2
+  (C^{{\tilde e_L} {\nu_L}{e_R}})^2  + (C^{{\tilde e_R}{\bar\nu^c_R}{e_L}})^2 \nonumber\\
& & {\hskip -0.25in}\ + {3 \over 4}\bigg(( C^{{\tilde \nu_{iL}} {e_{jL}} {e_{kR}}}\cdot C^{{\tilde e_L} {\nu_L}{e_R}})
 + (C^{{\tilde e_L} {\nu_L}{e_R}}\cdot C^{{\tilde e_R}{\bar\nu^c_R}{e_L}})+ ( C^{{\tilde \nu_{iL}} {e_{jL}} {e_{kR}}}\cdot C^{{\tilde e_R}{\bar\nu^c_R}{e_L}})\bigg)\biggr].
\end{eqnarray}
Utilizing the set of results given in eqs. (\ref{eq:2a1}) - (\ref{eq:2c1}), decay width simplifies to
\begin{eqnarray}
& &  \Gamma\sim
\frac{N_c \mm^7 (
  {\cal V}^{-\frac{7}{2}})}{(2^9.3.120) \pi^3. M^2_{pl}.{\cal V}^2 \mm^4}  \sim  \frac{1}{10^6} \frac{{\cal V}^{-\frac{11}{2}}.\mm^3}{M^2_{pl}}\sim 10^{-45.5} GeV ; {\rm for~{{\cal V}\sim10^5}}.
\end{eqnarray}
Life time  will be given as
\begin{eqnarray}
 \tau &=&\frac{\hbar}{\Gamma}\sim\frac{10^{-34} Jsec}{10^{-45.5} GeV}\sim O(10^{21}) sec.
 \end{eqnarray}
\underline{\bf {Decays involving $\lambda'_{ijk}$ coupling.}}

\begin{figure}[t!]
\label{decaydiag1}
\begin{center}
\begin{picture}(150,87)(100,100)
\Line(100,160)(150,160)
\Line(100,156)(150,156)
\Vertex(150,158){3.5}
\ArrowLine(150,158)(190,190)
\DashArrowLine (190,130)(150,158)5
\Vertex(190,130){3.5}
\ArrowLine(190,130)(230,160)
\ArrowLine(230,100)(190,130)
\Text(95,168)[]{$\psi_\mu$}
\Text(240,165)[]{$d_{jL}$}
\Text(240,100)[]{$d_{kR}$}
\Text(196,197)[]{${\nu_{iL}}$}
\put(155,132){${\stilde \nu}_{iL}$}
\put(165,80){(a)}
\end{picture}
\hspace{0.9cm}
\begin{picture}(150,97)(100,100)
\Line(100,160)(150,160)
\Line(100,156)(150,156)
\Vertex(150,158){3.5}
\ArrowLine(150,158)(190,190)
\DashArrowLine (190,130)(150,158)5
\ArrowLine(190,130)(230,160)
\ArrowLine(230,100)(190,130)
\Vertex(190,130){3.5}
\Text(95,168)[]{$\psi_\mu$}
\Text(240,165)[]{${\nu}_{iL}$}
\Text(240,100)[]{$d_{kR}$}
\Text(196,197)[]{${d_{jL}}$}
\put(155,132){${\stilde d}_{jL}$}
\put(165,80){(b)}
\end{picture}
\end{center}
\vspace{0.7cm}
\begin{picture}(150,97)(100,100)
\Line(250,160)(300,160)
\Line(250,156)(300,156)
\Vertex(300,158){3.5}
\ArrowLine(340,190)(300,158)
\Vertex(340,130){3.5}
\DashArrowLine (300,158)(340,130)5
\ArrowLine(380,160)(340,130)
\ArrowLine(340,130)(380,100)
\Text(245,168)[]{$\psi_\mu$}
\Text(390,165)[]{${\bar\nu}^c_{iR}$}
\Text(390,100)[]{$d_{jL}$}
\Text(346,197)[]{${d_{kR}}$}
\put(305,132){${\stilde d}_{kR}$}
\put(315,80){(c)}
\end{picture}
\phantom{xxx}
\vskip 0.3in
\caption{Three-body gravitino decays involving $\slashed{R}_p\ \lambda^\prime_{ijk}$ coupling.}
\end{figure}
 The ${\tilde \nu_L}-{d_{jL}}-{d_{kR}}$  vertex corresponding to Figure~3.5(a) in the context of ${\cal N}=1$ gauged supergravity action is given by the term: $\frac{e^{\frac{K}{2}}}{2}\left({\cal D}_{{\bar {\cal A}}_2}D_{{\cal A}_4}W\right)\chi^{{\cal A}^{c}_2}\chi^{{\cal A}_4}$. By expanding ${\cal D}_{\bar{a_2}}D_{a_4}W$ in the fluctuations linear in $a_1$ by using equations (\ref{eq:K}) and (\ref{eq:W}), we have $e^{\frac{K}{2}}{\cal D}_{\bar {a_2}}D_{a_4}W \equiv e^{\frac{K}{2}}{\cal D}_{{\cal {\bar A}}_2}D_{{\cal A}_4}W\sim{\cal V}^{-\frac{1}{3}}\delta{\cal A}_1$.
Therefore,  the contribution of physical ${\tilde \nu_L}-{d_{jL}}-{d_{kR}}$ vertex is given by
\begin{eqnarray}
\label{eq:2A2}
C^{ d^{c}_L \tilde {l_L} u_L}\sim \frac{{\cal V}^{-\frac{1}{3}}}{{\sqrt{\hat{K}_{{\cal A}_2 {\cal A}^{c}_2}{\hat{K}_{{\cal A}_1{\bar {\cal A}}_1}}{\hat{K}_{{\cal A}_4{\bar {\cal A}}_4}}}}}\sim (10)^{-\frac{14}{2}}{\cal V}^{-\frac{1}{3}} \sim {\cal V}^{-\frac{5}{3}} {\rm for}~ {\cal V}\sim {10}^5.
\end{eqnarray}
The  ${\tilde d_{jL}}-{\nu_{iL}}-{d_{kR}}$ vertex corresponding to Figure~3.5(b) is given by considering the contribution of $\frac{e^{\frac{K}{2}}}{2}\left({\cal D}_{\bar{\cal A}_1}D_{{\cal A}_2}W\right){\chi^{c^{{\cal {A}}_1}}}\chi^{{\cal A}_2}+h.c$ in ${\cal N}=1$ gauged SUGRA. By expanding ${\cal D}_{\bar a_1}D_{a_2}W$ in the fluctuations linear in  $a_4\rightarrow a_4+{\cal V}^{-\frac{11}{9}}$  with the help of eqs. (\ref{eq:K}) and (\ref{eq:W}), one has $e^{\frac{K}{2}}{\cal D}_{\bar {a_1}}D_{a_2}W  \equiv \frac{e^{\frac{K}{2}}}{2}\left({\cal D}_{\bar{\cal A}_1}D_{{\cal A}_2}W\right) \sim {\cal V}^{-\frac{1}{3}}\delta {\cal A}_4 $,  and the physical ${\tilde d_{jL}}-{\nu_{iL}}-{d_{kR}}$ vertex is given by:
\begin{eqnarray}
\label{eq:2b2}
& & {\hskip-0.7in} C^{l_L \tilde {d_R} u_L}\sim\frac{{\cal V}^{-\frac{1}{3}}}{{\sqrt{\hat{K}_{{\cal A}_1{\bar{\cal A}}_1}{\hat{K}_{{\cal A}_2{\bar {\cal A}}_2}}{\hat{K}_{{\cal A}_4{\bar {\cal A}}_4}}}}}  \sim  {\cal V}^{-\frac{5}{3}} {\rm for}~ {\cal V}\sim {10}^5.
\end{eqnarray}
The ${\tilde d_{kR}}-{\bar\nu^c_{iR}}-{d_{jL}}$ vertex of Figure~3.5(c) is given by considering $\frac{e^{\frac{K}{2}}}{2}\left({\cal D}_{{\cal {\bar A}}_1} D_{{\cal A}_4}W\right){\chi^{c^{{\cal {A}}_1}}}\chi^{{\cal A}_4}$ in ${\cal N}=1$ gauged SUGRA. By picking up the component of ${\cal D}_{a_1} D_{a_4}W$ in the fluctuations linear in $a_2$ by using eqs.~(\ref{eq:K}) and (\ref{eq:W}), $e^{\frac{K}{2}}{\cal D}_{\bar {a_1}}D_{a_4}W \equiv e^{\frac{K}{2}}{\cal D}_{{\cal {\bar A}}_1}D_{{\cal A}_4}W \sim {\cal V}^{-\frac{1}{3}}\delta{\cal A}_2  $.
The physical ${\tilde d_{kR}}-{\bar\nu^c_{iR}}-{d_{jL}}$ vertex is given by:
\begin{eqnarray}
\label{eq:2C2}
C^{ d^{c}_L \tilde {l_L} l_L}\sim \frac{{\cal V}^{-\frac{1}{3}}}{{\sqrt{\hat{K}_{{\cal A}_2{\bar{\cal A}}_2}{\hat{K}_{{\cal A}_1{\bar {\cal A}}_1}}{\hat{K}_{{\cal A}_4{\bar {\cal A}}_4}}}}}\sim (10)^{-\frac{14}{2}}{\cal V}^{-\frac{1}{3}} \sim {\cal V}^{-\frac{5}{3}} {\rm for}~ {\cal V}\sim {10}^5.
\end{eqnarray}
The analytical result of matrix amplitude for all three Feynman diagrams corresponding to Figure~3.5 is same as calculated in Figure~3.4 by replacing  $\aa \to \aaps \ (\aapt)$, $\bb \to \bbps \ (\bbpt)$ and
$\cc \to \ccps \ (\ccpt)$ and $C^{{\tilde \nu_{iL}} {e_{jL}} {e_{kR}}} \to C^{{\tilde \nu_{iL}} {d_{jL}} {d_{kR}}}$, $C^{{\tilde e_{jL}} {\nu_{iL}}{e_{kR}}}$ $ \to C^{{\tilde d_{jL}} {\nu_{iL}} {d_{kR}}}$.
 $C^{{\tilde e_{kR}}{\bar\nu^c_{iR}}{e_{jL}}} \to C^{{\tilde d_{kR}}{\bar\nu^c_{iR}}{d_{jL}}}$. Doing so and  using same numerical estimates of masses ${\mm}\sim {\cal V}^{-2}M_P, \aaps^2= \bbps^2= \ccps^2 \sim {\cal V}^{\frac{1}{2}}\mm$ in our model, after  integrating, decay width of gravitino corresponding to Figure~3.5 reduces to
\begin{eqnarray}
& & \Gamma\sim
\frac{N_c \mm^7}{(2^9.3.120) \pi^3. M^2_{pl}.{\cal V}^2 \mm^4} \biggr[( C^{{\tilde \nu_{iL}} {d_{jL}} {d_{kR}}})^2
+  (C^{{\tilde d_{jL}} {\nu_{iL}} {d_{kR}}})^2  + (C^{{\tilde d_{kR}}{\bar\nu^c_{iL}}{d_{jL}}})^2 + \nonumber\\
& &  {\hskip -0.2in} {3 \over 4}\biggl(C^{{\tilde \nu_{iL}} {d_{jL}} {d_{kR}}} \cdot C^{{\tilde d_{jL}} {\nu_{iL}} {d_{kR}}}
 + C^{{\tilde d_L} {\nu_L} {d_{kR}}}\cdot C^{{\tilde d_{kR}}{\bar\nu^c_{iL}}{d_{jL}}}+ C^{{\tilde \nu_{iL}} {d_{jL}} {d_{kR}}} \cdot C^{{\tilde d_{kR}}{\bar\nu^c_{iR}}{d_{jL}}}\biggr)\biggr].
\end{eqnarray}
Utilizing the set of results given in eqs.~(\ref{eq:2A2}) - (\ref{eq:2C2}), decay width simplifies to
\begin{eqnarray}
& & \Gamma\sim
\frac{N_c \mm^7 \  (
  {\cal V}^{-\frac{10}{3}}) }{(2^9.3.120) \pi^3. M^2_{pl}.{\cal V}^2 \mm^4} \sim  \frac{1}{10^6} \frac{{\cal V}^{-\frac{16}{3}}.\mm^3}{M^2_{pl}}\sim 10^{-44} GeV ; {\rm for~{{\cal V}\sim10^5}}.
\end{eqnarray}
Life time  will be given as:
\begin{eqnarray}
 \tau &=&\frac{\hbar}{\Gamma}\sim\frac{10^{-34} Jsec}{10^{-44} GeV}\sim O(10^{20}) sec.
 \end{eqnarray}
\underline{\bf {Decays involving $\lambda''_{ijk}$ coupling}}

\begin{figure}[t!]
\label{decaydiag1}
\begin{center}
\begin{picture}(150,97)(100,100)
\Line(100,160)(150,160)
\Line(100,156)(150,156)
\Vertex(150,158){3.5}
\ArrowLine(150,160)(190,190)
\DashArrowLine (190,130)(150,160)5
\ArrowLine(190,130)(230,160)
\ArrowLine(230,100)(190,130)
\Text(95,168)[]{$\psi_\mu$}
\Text(240,165)[]{$d_{jL}$}
\Text(240,100)[]{$d^{c}_{kL}$}
\Text(196,197)[]{${u_{iR}}$}
\put(155,132){${\stilde u}_{iR}$}
\put(165,83){(a)}
\end{picture}
\hspace{0.2cm}
\begin{picture}(150,97)(100,100)
\Vertex(150,158){3.5}
\Line(100,160)(150,160)
\Line(100,156)(150,156)
\ArrowLine(150,158)(190,190)
\DashArrowLine (190,130)(150,158)5
\Vertex(190,130){3.5}
\ArrowLine(190,130)(230,160)
\ArrowLine(230,100)(190,130)
\Text(95,168)[]{$\psi_\mu$}
\Text(240,165)[]{${u}_{iL}$}
\Text(240,100)[]{$d^{c}_{kL}$}
\Text(196,197)[]{${d_{jR}}$}
\put(155,132){${\stilde d}_{jR}$}
\put(165,83){(b)}
\end{picture}
\end{center}
\vskip 0.1in
\caption{Three-body gravitino decays involving $\slashed{R}_p\ \lambda^{\prime\prime}_{ijk}$ coupling.}
\end{figure}
The ${\tilde u_{iR}}-{d_{jR}}-{d^c_{kL}}$ vertex corresponding to Figure~3.6(a) is given by considering term $
\frac{e^{\frac{K}{2}}}{2}\left({\cal D}_{{\cal {\bar A}}_4} D_{{\cal A}_4}W\right)\chi^{ {\cal A}^{c}_4}\chi^{{\cal A}_4}$ of ${\cal N}=1$ gauged SUGRA. By expanding ${\cal D}_{a_4}D_{\bar {a_4}}W$ in the fluctuations linear in $a_4\rightarrow a_4+{\cal V}^{-\frac{11}{9}}{M_P}$ by utilizing equations (\ref{eq:K}) and (\ref{eq:W}), on simplifying, one obtains: $e^{\frac{K}{2}}{\cal D}_{{\cal {\bar A}}_4}D_{{\cal A}_4}W \equiv \frac{e^{\frac{K}{2}}}{2} {\cal D}_{a_4}D_{\bar {a_4}}W \sim {\cal V}^{\frac{13}{6}}\delta{{\cal A}_4}$, implying that the contribution of physical ${\tilde u_{iR}}-{d_{jR}}-{d^c_{kL}}$ vertex is given as under:
\begin{eqnarray}
\label{eq:4a2}
  C^{{\tilde u_{iR}} {d_{jR}} {d^c_{kL}}} \sim \frac{{\cal V}^{\frac{13}{6}}}{{\sqrt{\hat{K}_{{\cal A}_4{\bar{\cal A}}_4}{\hat{K}_{{\cal A}_4{\bar{\cal A}}_4}}{\hat{K}_{{\cal A}_4{\bar {\cal A}}_4}}}}}\sim \frac{{\cal V}^{\frac{13}{6}}}{\sqrt {10^{36}}}\sim {\cal V}^{-\frac{43}{30}}; {\rm for}~{\cal V}\sim {10^5}.
\end{eqnarray}
Similarly, the ${\tilde d_{jR}}-{d^c_{kL}}-{u_{iR}}$ vertex of  Figure~3.6(b) is same as ${\tilde u_{iR}}-{d_{jR}}-{d^c_{kL}}$ vertex corresponding to Figure~3.6(a) and is given as under:
\begin{eqnarray}
\label{eq:4b2}
C^{{\tilde d_{jR}} {d^c_{kL}} {u_{iR}}} \sim C^{{\tilde u_{iR}} {d_{jL}} {d^c_{kL}}} \sim {\cal V}^{-\frac{43}{30}}, {\rm for}~{\cal V}\sim {10^5}.
\end{eqnarray}
 Again, introducing  kinematic variable
$
 2 p(u_i) \cdot p(d_j)
=
(1 - z_{d_k}) \mm^2 ,\
 2 p(d_j) \cdot p(d_k)
=
(1 - z_{u_i}) \mm^2 ,
\
 2 p(u_i) \cdot p(d_k)=(1 - z_{d_j}) \mm^2
$, and
\begin{eqnarray}
\label{eq:mij}
m^2_{ij} &= & (p(u_i)+p(d_j))^2 \sim  2p(u_i).p(d_j)= (1-z_{d_k})\mm^2 \nonumber\\
m^2_{jk} & = & (p(d_j)+p(d_k))^2 \sim 2p(d_j).p(d_k)=(1-z_{u_i})\mm^2 \nonumber\\
m^2_{ik} & = & (p(u_i)+p(d_k))^2 \sim  2p(u_i).p(d_k)=(1-z_{d_j})\mm^2,
\end{eqnarray}
Again utilizing the form of expressions given in an appendix of \cite{Gmoreau} for ${\lambda^{\prime \prime}_{ijk}}$ coupling, and evaluating the squared amplitude summed over the spins for the gravitino decay reaction $\psi_\mu \stackrel{\lambda^{\prime \prime}_{ijk}}{\to} u_i d_j d_k$ (see \cite{gravitino_DM} for details), decay width is given as:
\begin{eqnarray}
& &    \Gamma\sim
\frac{N_c \mm^7}{(2^9.3.120) \pi^3. M^2_{pl}.{\cal V}^2 \mm^4} \biggr[( C^{{\tilde u_{iR}} {d_{jR}} {d^c_{kR}}})^2
+  (C^{{\tilde d_{jR}} {u_{iR}}{d^c_{kR}}})^2  + \nonumber\\
&&{3 \over 4}\bigg( C^{{\tilde u_{iR}} {d_{jR}} {d^c_{kR}}}\cdot C^{{\tilde d_{jR}} {u_{iR}}{d^c_{kR}}})
 \bigg)\biggr].
\end{eqnarray}
Utilizing the set of results given in eqs.~(\ref{eq:4a2}) - (\ref{eq:4b2}), decay width simplifies to
\begin{eqnarray}
& & {\hskip -0.5in} \Gamma\sim
\frac{N_c \mm^7 (
{\cal V}^{-\frac{43}{15}} )}{(2^9.3.120) \pi^3. M^2_{pl}.{\cal V}^2 \mm^4}   \sim  \frac{1}{10^6} \frac{{\cal V}^{-\frac{73}{15}}.\mm^3}{M^2_{pl}}\sim 10^{-42} GeV ; {\rm for~{{\cal V}\sim 10^5}},
\end{eqnarray}
and therefore, the life time will be given as:
\begin{eqnarray}
 \tau &=&\frac{\hbar}{\Gamma}\sim\frac{10^{-34} Jsec}{10^{-42} GeV}\sim O(10^{18}) sec.
 \end{eqnarray}
\section{NLSP Decays}
With in the approach of considering non-thermal production of gravitino (LSP), all of its relic density is produced by non-thermal decays of NLSP decaying into LSP and therefore, it is important to determine whether life time of these decays do not affect the standard abundance of D, 3He, 4He, and 7Li, and hence predictions of Big Bang Nucleosynthesis (BBN). The BBN predicts the universal abundances of D, 3He, 4He, and 7Li, essentially fixed by an average lifetime  $\tau \sim 180 sec$. In our study, we  are considering both radiative as well as hadronic decay modes of NLSP's. Therefore, if life time of NLSP's are more than $ 10^2$ sec, the high energy photons emitted via radiative decays might destroy the abundance of light elements by inducing photo-dissociation of same, hence reducing the standard baryon-to-photon ratio and resulting in baryon-poor universe . Similarly, during the hadronic decay, the released hadronic energy can produce mesons/charged pions causing interconversions between background proton and neutron (p / n conversion) that alters the neutron-to proton ratio, resulting in the change of 4He abundance and hence destroying this success of standard BBN. The bounds on primordial abundance of late decaying particle, taking into account hadronic decay mode of the same, have been discussed in detail in \cite{kohri_BBN}. 

In view of the above discussion, we explicitly calculate the
decay widths and hence life times for possible NLSP candidates. The mass spectrum given in Table~2.1 suggests that sleptons/squarks and (Bino/Wino-type)gaugino/(Bino/Wino-type)neutralino exists as N(ext-to) L(ightest) S(upersymmetric) P(article)s in the dilute-flux approximation. As discussed in section {\bf 5} of chapter {{\bf 2}}, the diagonalization of neutralino mass matrix corresponds to smallest eigenvalue of neutralino of the order ${\cal V}^{\frac{2}{3}}m_{3/2}$ (similar as gaugino mass) given by the following eigenvector:
\begin{equation}
\label{eq:chi3}
\tilde{\chi}_3^0\sim-{\tilde \lambda}^0+(\tilde{f}{\cal V}^{\frac{5}{6}})\frac{v}{M_P}\left(\tilde{H}^{0}_u+\tilde{H}^{0}_d\right) + h.c,
\end{equation}
where $\tilde{H}^{0}_{u,d}$ are the higgsinos. We first evaluate the lifetimes corresponding to two- and three-body decays of Wino/Bino-type gauginos/neutralino to gravitino (LSP) followed by the lifetime corresponding to two- and three-body decays of the slepton to gravitino (LSP) using ${\cal N}=1$ gauged SUGRA. Further, there is a possibility for neutralino to directly decay into ordinary quarks via R-parity violating three-body decays. Due to the presence of two competing kind of decays, results of life time of both will decide, whether decay of neutralino into ordinary quarks effects the abundance of gravitino or not. To verify the same, we have also evaluated the three-body decay diagrams of neutralino into ordinary SM particles in the context of  ${\cal N}=1$ gauged SUGRA.
\subsection{Gaugino Decays}
In this subsection, we discuss two- and three-body decay modes of the Wino/Bino-type gauginos.
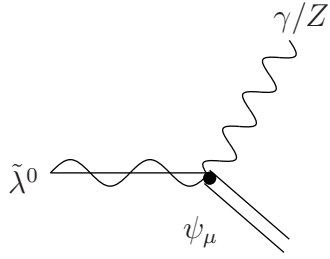
\begin{figure}
\begin{center}
    \begin{picture}(135,137) (15,-18)
   \Line(100,53)(130,27)
   \Line(98,48)(128,22)
   \Photon(40,52)(100,52){5}{2}
   \Photon(100,50)(130,102){5}{4}
   \Text(30,50)[]{{{$\tilde{\lambda}^0$}}}
   \Line(40,52)(100,52)
   \Vertex(100,50){2.5}
   \Text(135,111)[]{{{$\gamma/Z$}}}
   \Text(97,30)[]{{{$\psi_\mu$}}}
  \end{picture}
   \end{center}
   \vskip-0.6in
   \caption{Two-body gaugino-decay diagram.}
   \end{figure}
The relevant gravitino-gaugino-photon/$Z$-boson vertex in Figure~3.7  is obtained in the ${\cal N}=1$ SUGRA action of \cite{Wess_Bagger} from $\frac{1}{M_P}{\bar\psi}_\rho\sigma^{\mu\nu}\gamma^\rho\lambda_L^{(a)}F^{(a)}_{\mu\nu}, (a)$ corresponding to the three gauge groups. The decay width for $\tilde{B}\rightarrow\psi_\mu+\gamma$ is given by (See \cite{Hasenkamp}):
\begin{eqnarray}
\label{eq:Bino_grav+ph}
& & {\hskip -0.4in} \Gamma\left(\tilde{B}\rightarrow\psi_\mu+\gamma\right)=\frac{cos^2\theta_W}{48\pi M_P^2}\frac{m^5_{\tilde{B}}}{m^2_{3/2}}
\left(1-\frac{m^2_{3/2}}{m_{\tilde{B}}^2}\right)^3\left(1+3\frac{m^2_{3/2}}{m_{\tilde{B}}^2}\right)\sim\frac{m^5_{\tilde{B}}}{m^2_{3/2}M_P^2}.
\end{eqnarray}
For $m_{\tilde{B}}\sim m_{\tilde{g}}\sim{\cal V}^{\frac{2}{3}}m_{\frac{3}{2}}$ as given in equation (\ref{eq:gaugino_mass}) and $m_{3/2}\sim{\cal V}^{-2}M_P$ for $n^s=2$,  above decay process implies a lifetime of Bino ($\tilde{B}$) of around $10^{-30}s$.

Similarly, the decay width for $\tilde{B}\rightarrow\psi_\mu+Z$ is given by (See \cite{Hasenkamp}):
\begin{eqnarray}
\label{eq:Bino_grav+Z}
& & \Gamma\left(\tilde{B}\rightarrow\psi_\mu+Z\right)=\frac{cos^2\theta_W}{48\pi M_P^2}\frac{m^5_{\tilde{B}}}{m^2_{3/2}}
\sqrt{1-2\left(\frac{m_{\psi_\mu}^2}{m_{\tilde{B}}^2} + \frac{m^2_Z}{m^2_{\tilde{B}}}\right)+\left(\frac{m^2_{\psi_\mu}}{m_{\tilde{B}}}-\frac{m^2_Z}{m^2_{\tilde{B}}}\right)^2}\nonumber\\
& & \times\biggl[
\left(1-\frac{m^2_{3/2}}{m_{\tilde{B}}^2}\right)^2\left(1+3\frac{m^2_{3/2}}{m_{\tilde{B}}^2}\right)-\frac{m_Z^2}{m_{\tilde{B}}^2}
\Bigl\{3+\frac{m^3_{3/2}}{m^3_{\tilde{B}}}\left(\frac{m_{3/2}}{m_{\tilde{B}}}-12\right)-\nonumber\\
&& \frac{x_Z^2}{x_{\tilde{B}}^2}\left(
3-\frac{m^2_{3/2}}{m^2_{\tilde{B}}}-\frac{m^2_Z}{m^2_{\tilde{B}}}\right)\Bigr\}\biggr]\ \sim\frac{m^5_{\tilde{B}}}{m^2_{3/2}M_P^2}\sim{\cal V}^{-\frac{8}{3}}M_P.
\end{eqnarray}
So, (\ref{eq:Bino_grav+Z}) implies a lifetime of around $10^{-30}s$.

\begin{figure}
\begin{center}
\begin{picture}(150,97)(100,100)
\Line(100,160)(150,160)
\Photon(100,160)(150,160){5}{2}
\Vertex(150,160){4.6}
\Line(150,162)(190,190)
\Line(152,157)(192,185)
\Photon (150,160)(190,130){5}{4}
\Vertex(190,130){4}
\ArrowLine(230,160)(190,130)
\ArrowLine(190,130)(230,100)
\Text(95,168)[]{$\tilde{B}$}
\Text(240,165)[]{${\bar q}$}
\Text(240,100)[]{$q$}
\Text(196,197)[]{$\psi_\mu$}
\put(155,132){$\gamma/Z$}
\end{picture}
\end{center}
\vskip -0.2in
\caption{Three-body gaugino-decay diagram.}
\end{figure}
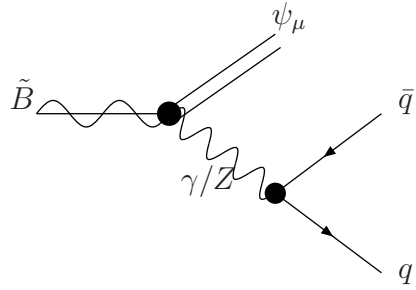
The gauge boson-quark-anti-quark vertex in the ${\cal N}=1$ gauged SUGRA will be accompanied by $\frac{g_{{\cal A}_2{\bar {\cal A}}_2}g_{YM}\left(X^{T^B}K+iD^{T^B}\right)}{\left(\sqrt{K_{{\cal A}_2{\bar{\cal A}}_2}}\right)^2}$ and value of the same has been shown to be same as its SM value i.e ${\cal O}(1)$  in chapter {{\bf 4}}. Using further the results of \cite{Hasenkamp} and gaugino mass ${\tilde \lambda^0}\sim {\cal V}^{\frac{2}{3}}m_{\frac{3}{2}}$ from equation (\ref{eq:gaugino_mass}), one obtains the following result for the decay width ${\tilde \lambda^0}\rightarrow\psi_\mu+u+{\bar u}$ mediated by $\gamma/Z$ boson:
\begin{eqnarray}
\label{eq:gaugino_grav+u+ubar}
& & \Gamma\left({\tilde \lambda^0}\rightarrow\psi_\mu+u+{\bar u}\right)\sim\frac{g^2_{YM} m^5_{\tilde {\tilde \lambda^0}}}{32\left(2\pi\right)^3m^2_{3/2}M_P^2}\sim10^{-17}M_P \sim {\cal O}(10),
\end{eqnarray}
which yields a lifetime of around $10^{-25}s$.

We will now discuss three-body gaugino decays into gravitino mediated by squarks. To avoid channel overlap, for simplicity, we will assume that the gaugino decays mediated by vector bosons and those mediated by squarks, involve different gauginos.
Utilizing the conditions $\overline{\psi^{(+)}_\mu}\gamma^\mu=\gamma^\mu\psi_\mu^{(-)}=0$, the amplitudes for the above two diagrams can be written as:
\begin{eqnarray}
\label{eq:gaugino_to_grav_i}
& & M_{(a)}\sim2p_{\tilde{q}}^\mu\frac{1}{M_P}\left(\overline{\psi^{(+)}_\mu}(p_\psi)
v(p_{\bar{q}})\right)\left(\frac{i}{p_{\tilde{q}}^2-m_{\tilde{q}}^2 + i\epsilon}\right)
\left(\bar{u}(p_q)V_{\lambda-q-\tilde{q}}\lambda^{(+)}_{\tilde{g}}\right)\nonumber\\
& & M_{(b)}\sim2p_{\tilde{q}}^\mu\frac{1}{M_P}\left(\bar{u}(p_q){\psi^{(-)}_\mu}(p_\psi)
v(p_{\bar{q}})\right)\left(\frac{i}{p_{\tilde{q}}^2-m_{\tilde{q}}^2 + i\epsilon}\right)
\left(\overline{\lambda^{(-)}_{\tilde{g}}} V_{\lambda-q-\tilde{q}}v(p_{\bar{q}})\right),
\end{eqnarray}
$V_{\lambda-q-\tilde{q}}\sim\tilde{f}{\cal V}^{-\frac{4}{5}}$ being the gaugino-quark-squark vertex.
From (\ref{eq:gaugino_to_grav_i}), one obtains the following helicities averaged sum:
\begin{eqnarray}
\label{eq:gaugino_to_grav_ii}
& & \sum_{s_\psi=\pm\frac{3}{2},\pm\frac{1}{2},s_q,s_{\bar{q}},s_{\lambda{g}}=\pm\frac{1}{2}}
M_{a)} M_{(a)}^\dagger\sim|V_{\lambda-q-\tilde{q}}|^2\frac{p^\mu_{\tilde{q}}p^\nu_{\tilde{q}}}{M_P^2}
Tr\left[{\cal P}^{(+)}_{\mu\nu}\left(-\slashed{p}_{\bar{q}} + m_{\bar{q}}\right)\right]\nonumber\\
&& Tr\left[\left(\slashed{p}_{\lambda_{\tilde{g}}} + m_{\lambda_{\tilde{g}}}\right)\left(\slashed{p}_q + m_q\right)\right], \\
& & \sum_{s_\psi=\pm\frac{3}{2},\pm\frac{1}{2},s_q,s_{\bar{q}},s_{\lambda{g}}=\pm\frac{1}{2}}
M_{b)} M_{(b)}^\dagger\sim|V_{\lambda-q-\tilde{q}}|^2\frac{p^\mu_{\tilde{q}}p^\nu_{\tilde{q}}}{M_P^2}
Tr\left[{\cal P}^{(-)}_{\mu\nu}\left(\slashed{p}_{{q}} + m_{{q}}\right)\right]\nonumber\\
&&Tr\left[\left(- \slashed{p}_{\lambda_{\tilde{g}}} + m_{\lambda_{\tilde{g}}}\right)\left(- \slashed{p}_{\bar{q}} + m_q\right)\right], \end{eqnarray}
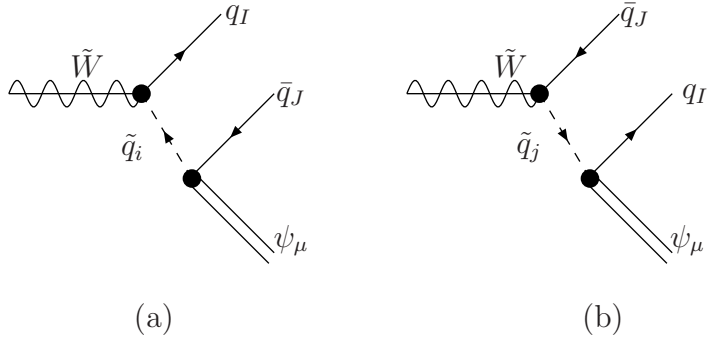
\begin{figure}
 \begin{center}
\begin{picture}(350,200)(50,-50)
\Text(90,132)[]{$\tilde{W}$}
\Line(60,120)(110,120)
\Photon(60,120)(110,120){5}{4}
\Vertex(110,120){3.5}
\ArrowLine(110,120)(140,150)
\Text(147,150)[]{${q}_I$}
\DashArrowLine(129,88)(110,120){4}
\Text(107,100)[]{$\tilde{q}_i$}
\ArrowLine(160,120)(130,90)
\Text(167,120)[]{${\bar q_J}$}
\Vertex(129,88){3.5}
\Line(130,90)(160,60)
\Line(128,86)(158,56)
\Text(167,65)[]{$\psi_\mu$}
\Text(115,35)[]{(a)}
\Text(250,132)[]{$\tilde{W}$}
\Line(210,120)(260,120)
\Photon(210,120)(260,120){5}{4}
\Vertex(260,120){3.5}
\ArrowLine(290,150)(260,120)
\Text(297,150)[]{$\bar{q}_J$}
\DashArrowLine(260,120)(279,88){4}{}
\Text(257,100)[]{$\tilde{q}_j$}
\ArrowLine(280,90)(310,120)
\Text(320,120)[]{${q}_I$}
\Vertex(279,88){3.5}
\Line(280,90)(310,60)
\Line(278,86)(308,56)
\Text(317,65)[]{$\psi_\mu$}
\Text(285,35)[]{(b)}
\end{picture}
\end{center}
\vskip-1.2in
\caption{Three-body gaugino decays into the gravitino.}
\end{figure}
and utilizing $\bar{u}(p_q){\psi^{(-)}_\mu}(p_\psi)=-\overline{\psi^{(+)}_\mu}(p_\psi)
v(p_{\bar{q}}), \overline{\lambda^{(-)}_{\tilde{g}}} v(p_{\bar{q}})=-\bar{u}(p_q)\lambda^{(+)}_{\tilde{g}} $ to rewrite $M_{(b)}$ and therefore obtain:
\begin{eqnarray}
\label{eq:gaugino_to_grav_iii}
&&\sum_{s_\psi=\pm\frac{3}{2},\pm\frac{1}{2},s_q,s_{\bar{q}},s_{\lambda{g}}=\pm\frac{1}{2}}\Re M_{(a)}M_{(b)}^\dagger\sim |V_{\lambda-q-\tilde{q}}|^2\frac{p^\mu_{\tilde{q}}p^\nu_{\tilde{q}}}{M_P^2}
Tr\left[{\cal P}^{(+)}_{\mu\nu}\left(-\slashed{p}_{\bar{q}} + m_{\bar{q}}\right)\right]\nonumber\\
&&Tr\left[\left(\slashed{p}_{\lambda_{\tilde{g}}} + m_{\lambda_{\tilde{g}}}\right)\left(\slashed{p}_q + m_q\right)\right].
\end{eqnarray}
In (\ref{eq:gaugino_to_grav_i}) - (\ref{eq:gaugino_to_grav_iii}), the positive and negative energy gravitino projectors are given by:
\begin{eqnarray}
\label{eq:P_pm_grav}
&&{\cal P}^{(\pm)}\equiv -\left(\slashed{p}_{3/2} \pm m_{3/2}\right)\Bigl[\Bigl(\eta_{\mu\nu} - \frac{p_{3/2,\mu}p_{3/2,\nu}}{m_{3/2}^2}\Bigr) - \frac{1}{3}\Bigl(\eta_{\mu\sigma} - \frac{p_{3/2,\mu}p_{3/2,\sigma}}{m_{3/2}^2}\Bigr) \nonumber\\
&& \Bigl(\eta_{\nu\lambda} - \frac{p_{3/2,\lambda}p_{3/2,\lambda}}{m_{3/2}^2}\Bigr)\gamma^\sigma\gamma^\lambda\Bigr].
\end{eqnarray}
A typical term that one would need to calculate is:
\vskip -0.4in
\begin{eqnarray}
\label{eq:gaugino_to_grav_iv}
& & p^\mu_{\tilde{q}} p^\nu_{\tilde{q}} Tr\left[{\cal P}^{(\pm)}_{\mu\nu}\left(\eta\slashed{p}_{\eta} + m_{\bar{q}}\right)\right],\ \eta=\pm\ {\rm corresponding\ to}\ p_{\eta}=p_q/p_{\bar{q}}\nonumber\\
& & \sim 4\Bigl\{\Bigl(m^2_{\tilde{q}} - \frac{\left(p_{\tilde{q}}\cdot p_{3/2}\right)^2}{m^2_{3/2}}\Bigr)\Bigl(\eta p_{3/2}\cdot p_\eta\pm m_{3/2}m_\eta\Bigr) \nonumber\\
 & & - \Bigl(p^{\tilde{q}}_\mu - \frac{p_{\tilde{q}}\cdot p_{3/2} p_\mu^{3/2}}{m^2_{3/2}}\Bigr)\Bigl(p_{\tilde{q}}^\mu - \frac{p_{\tilde{q}}\cdot p_{3/2} p^\mu_{3/2}}{m^2_{3/2}}\Bigr)\Bigl(\frac{\eta}{3}p_\eta\cdot p_{3/2} \pm m_{3/2}m_\eta\Bigr)\Bigr\}.
\end{eqnarray}
Using results of \cite{Manuel_Toharia}, to get an estimate of the decay width, one sees that:
\begin{eqnarray}
\label{eq:gaugino_to_grav_v}
& &  \Gamma\left(\tilde{W}\rightarrow q + \bar{q} + \psi_\mu\right) \sim
{\rm Max}\Biggl[\frac{1}{m_{\lambda_{\tilde{g}}}^3}\int_{s_{23}=m^2_{3/2}}^{m^2_{\lambda_{\tilde{g}}}}ds_{23}
\int_{s_{13}=\frac{m_{3/2}^2m^2_{\lambda_{\tilde{g}}}}{s_{23}}}^{m_{3/2}^2 + m_{\lambda_{\tilde{g}}}^2 - s_{23}}  ds_{13} \nonumber\\
&&
\frac{(\ref{eq:gaugino_to_grav_iv})|V_{\lambda-\tilde{q}-q}|^2\times
\left(p_{\lambda_{\tilde{g}}}\cdot p_q + m_{\lambda_{\tilde{g}}}m_q\right)}{(s_{23}^2 - m_{\tilde{q}}^2)^2, (s_{13}^2 - m_{\tilde{q}}^2)^2,  (s_{23}^2 - m_{\tilde{q}}^2)(s_{13}^2 - m_{\tilde{q}}^2)}\Biggr|_{{\cal V}\sim10^5, m_{q,\bar{q}}=0}\Biggr]\nonumber\\
& & \sim {\rm Max}\left(10^{-32},10^{-30}, 10^{-32}\right)M_P = 10^{-28}M_P,
\end{eqnarray}
implying that the corresponding lifetime would be $10^{-15}s$.

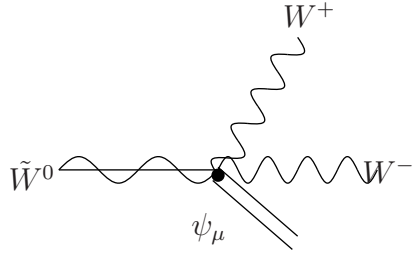
\begin{figure}
\begin{center}
    \begin{picture}(135,137) (15,-18)
   \Line(100,53)(130,27)
   \Line(98,48)(128,22)
   \Photon(40,52)(100,52){5}{2}
   \Photon(100,50)(130,102){5}{4}
   \Text(30,50)[]{{{$\tilde{W}^0$}}}
   \Line(40,52)(100,52)
   \Photon(100,52)(160,52){5}{4}
   \Text(165,52)[]{$W^-$}
   \Vertex(100,50){2.5}
   \Text(135,111)[]{{{$W^+$}}}
   \Text(97,30)[]{{{$\psi_\mu$}}}
  \end{picture}
   \end{center}
   \vskip-0.6in
   \caption{Contact-vertex three-body decay diagram.}
   \end{figure}
In the approximation, $\frac{m_W}{m_{\tilde{W}}}\approx0,\frac{m_{3/2}}{m_{\tilde{W}}}\approx0$ (valid in the limit of massses $m_{\tilde{W}}\sim{\cal V}^{\frac{2}{3}}m_{\frac{3}{2}}$ and $m_{3/2}\sim{\cal V}^{-2}M_P$ as given in equation (\ref{eq:gaugino_mass})), the decay
width for $\tilde{W}^0\rightarrow\psi_\mu+W^++W^-$ is given by (See \cite{Hasenkamp}):
\begin{eqnarray}
\label{eq:Wino_grav+W+_W-}
& & \Gamma\left(\tilde{W}^0\rightarrow\psi_\mu+W^++W^-\right)\approx\frac{g^2_{YM}m_{\tilde{W}}^9}{34560\left(2\pi\right)^3m^4_WM^{2}_P}
\sim\frac{{\cal V}^{-8}M^{5}_P}{M^{4}_W},
\end{eqnarray}
which implies a lifetime of around $10^{-66}s$. Treating the Wino/Bino gauginos as the co-NLSPs, we hence conclude that the two- and three-body gaugino decays into the gravitino (LSP), respect the BBN bounds. If one were to calculate similar neutralino decays into the gravitino, then for the higgsino contribution to the neutralino-gauge-gravitino vertex, one will have the following suppression factor:
$\left(\tilde{f} \frac{v}{M_P}{\cal V}^{\frac{5}{6}}\right) \times {\cal O}(z_i)$ term in $g_{\rm YM}
\frac{g_{T_B \bar{z}_i}}{\sqrt{K_{{\cal Z}_i\bar{\cal Z}_i}}} \times \frac{\langle z_i\rangle}{M_P}$, which is $\tilde{f}^2 ( \frac{v}{M_P}) {\cal V}^{\frac{5}{6} - \frac{5}{3}} \times 10^{\frac{5}{2}}.$ For ${\cal V}\sim10^5$, this is approximately, $10^{-26}$. The corresponding gaugino-gauge-gravitino vertex will, without worrying about the gamma matrices, be of ${\cal O}\left(\frac{p_Z}{M_P}\right)$. Now, even at low $Z$-momenta, this is around $10^{-26}$ for $M_Z\sim10^2 GeV$. Hence, the neutralino and gaugino decay (e.g., into gravitino and Z boson) widths, will approximately be the same.
\subsection{ R-Parity Violating Neutralino Decays}
\begin{figure}[t!]
\label{decaydiag1}
\begin{center}
\begin{picture}(150,100)(-100,100)
\ArrowLine(-100,160)(-50,160)
\ArrowLine(-50,160)(-10,190)
\DashArrowLine(-10,130)(-50,160)5
\ArrowLine(-10,130)(30,160)
\ArrowLine(-10,130)(30,100)
\Text(-80,174)[]{${\chi^{0}_3}\,(p_i)$}
\Text(43,170)[]{${u_L} \,(k_{u})$}
\Text(55,100)[]{${d^{c}_L}\,(k_{d})$}
\Text(-3,197)[]{${l_L} \, (k_{l})$}
\put(-50,130){$\stilde {l}_L$}
\put(-45,80){(a)}
\end{picture}
\hspace{0.9cm}
\begin{picture}(150,100)(0,-25)
\ArrowLine(0,35)(50,35)
\ArrowLine(50,35)(90,65)
\DashArrowLine(50,35)(90,5)5
\ArrowLine(90,5)(130,35)
\ArrowLine(90,5)(130,-25)
\Text(20,49)[]{${\chi^{0}_3}\,(p_i)$}
\Text(140,44)[]{${l_L}\,(k_{l})$}
\Text(155,-21)[]{${u_L} \,(k_{u})$}
\Text(97,72)[]{${d}^c_L\,(k_{d})$}
\put(50,5){$\stilde {d}_R$}
\put(55,-45){(b)}
\end{picture}

\end{center}
\vspace{0.6cm}
\begin{center}
\begin{picture}(150,97)(100,100)
\ArrowLine(100,160)(150,160)
\ArrowLine(150,160)(190,190)
\DashArrowLine(190,130)(150,160)5
\ArrowLine(190,130)(230,160)
\ArrowLine(190,130)(230,100)
\Text(120,174)[]{${\chi^{0}_3} \,(p_i)$}
\Text(240,170)[]{$l_L\,(k_{l})$}
\Text(255,100)[]{$d^{c}_L\,(k_{d})$}
\Text(196,197)[]{$ u_L \,(k_{u})$}
\put(155,132){$\stilde {u}_L$}
\put(165,80){(c)}
\end{picture}
\phantom{xxx}
\end{center}
\vskip 0.3in
\caption{Feynman diagrams for the R-parity
violating decays of Neutralino ${\chi^{0}_3}$.}
\end{figure}
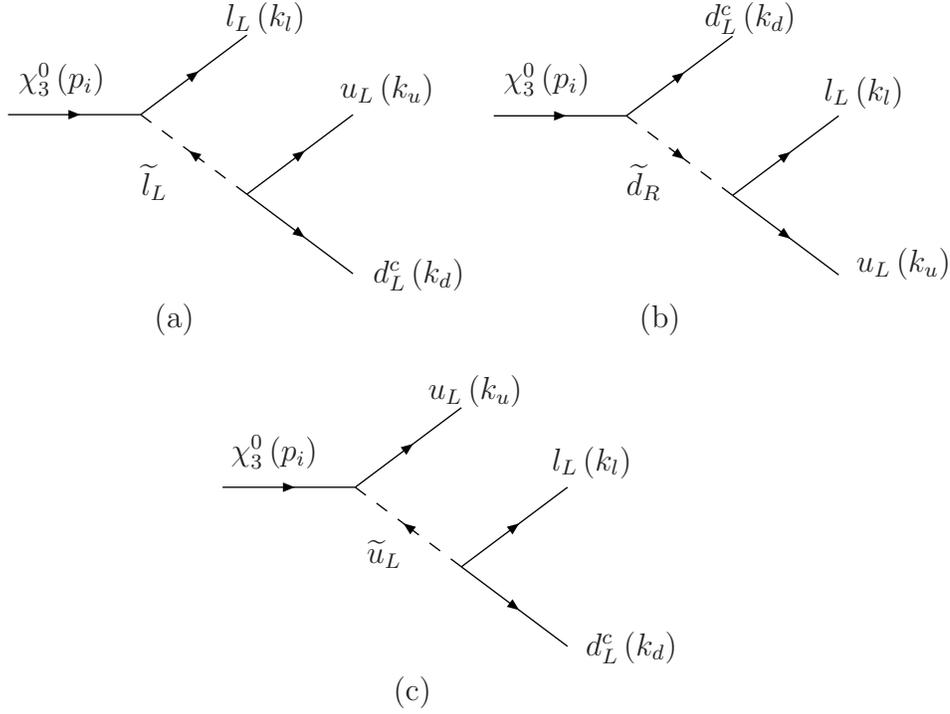
In this section, we  evaluate  the decay width of  tree-level diagrams of neutralino mediated by sleptons/squarks, and also involving R-parity violating vertices as shown in Figure~3.11. The life time calculation requires the evaluation of various matrix amplitudes corresponding to various decay channels. In particular, to evaluate the contribution of neutralino-squark-quark/neutralino-slepton-lepton, we will explicitly  work out squark-quark-gaugino vertex and squark-quark-higgsino vertex, and then add these contributions as according to equation (\ref{eq:chi3}). To calculate the contribution of various interaction vertices shown in Fig.~3.11 in the context of ${\cal N}=1$ gauged SUGRA, one needs to consider the following terms of ${\cal N}=1$ gauged SUGRA \cite{Wess_Bagger}.
\begin{eqnarray}
\label{eq:fermion+fermion+sfermion}
& & {\hskip -0.3in}{\cal L} = g_{YM}g_{\alpha {\bar J}}X^\alpha{\bar\chi}^{\bar J}\lambda_{\tilde{g}} +ig_{i{\bar J}}{\bar\chi}^{\bar I}\left[{\bar\sigma}\cdot\partial\chi^i+\Gamma^i_{Lj}{\bar\sigma}\cdot\partial a^L\chi^j
+\frac{1}{4}\left(\partial_{a_L}K{\bar\sigma}\cdot a_L - {\rm c.c.}\right)\chi^i\right] \nonumber\\
& & {\hskip -0.3in}+\frac{e^{\frac{K}{2}}}{2}\left({\cal D}_iD_JW\right)\chi^i\chi^J+{\rm h.c.},
\end{eqnarray}
$X^\alpha$ is defined in text below equation (\ref{eq:fermion+fermion+sfermion}).
 As discussed above, the neutralino-lepton-slepton  vertex corresponding to Figure~3.11(a), is given by considering the contribution of gaugino-slepton-lepton vertex with a small admixture of higgsino-lepton-slepton vertex as given in (\ref{eq:chi3}). Since neutralino is of Majorana nature, in two-component notation, the contribution of  higgsino-lepton-slepton vertex is given as:  
 \begin{eqnarray}
&& \frac{e^{\frac{K}{2}}}{2}\left({\cal D}_{{\cal Z}_1}D_{\bar {\cal A}_1}W\right)\chi^{{\cal Z}_i}{\chi^{c^{{{\cal A}}_1}}} +ig_{{\bar I}{\bar {A_1}}}{\bar\chi}^{Z_i}\biggl[{\bar\sigma}\cdot\partial{{\chi^{c^{{{\cal A}}_1}}}}+\Gamma^{{\cal A}_1}_{{\cal A}_1{\bar A_1}}{\bar\sigma}\cdot\partial {\cal A}_1{\chi^{c^{A_1}}} \nonumber\\
 &&
+\frac{1}{4}\left(\partial_{{\cal A}_3}K{\bar\sigma}\cdot {\cal A}_1 - {\rm c.c.}\right){\chi^{c^{{\cal {A}}_1}}}\biggr].
\end{eqnarray}
 $\chi^{\cal {Z}}$ is $SU(2)_L$ higgsino, $\chi^{{\cal A}_1}$ corresponds to $SU(2)_L$ electron and ${\tilde {\cal A}_1}$ corresponds to left-handed squark and $g_{{\bar I}{\bar {\cal A}_1}}=0$. Strictly speaking, $SU(2)$ EW symmetry gets spontaneously broken for higgsino-lepton-slepton vertex, however the effective Lagrangian respects $SU(2)$ symmetry. Therefore to calculate the contribution of same, basic idea is to generate a term of the type $l_L \tilde{l}_L{\tilde H^{c}}_L H_L$ wherein $\chi^{l_L}$ and $H_L$ are respectively the $SU(2)_L$ quark and Higgs doublets, $\tilde{l}_L$ is also an $SU(2)_L$ doublet and ${\tilde H^{c}}_L$ is $SU(2)_L$  higgsino doublet . After spontaneous breaking of the EW symmetry when $H^0$ in $H_L$ acquires a non-zero vev $\langle H^0\rangle$, this term generates: $\langle H^0\rangle {\tilde H^{c}}_L{l_L}\tilde{l}_L$. \\
 ~~{\hskip 0.1in}Now, in terms of undiagonalized basis, we first consider ${\cal D}_iD_{\bar a_1}W$ where $a_1, z_i$ correspond to undiagonalized moduli fields. Using equations (\ref{eq:K}) and (\ref{eq:W}), picking up the component linear in $z_1$ as well as linear in fluctuation $ (a_1-{\cal V}^{-\frac{2}{9}}{M_P})$, and
further using an argument $ e^{\frac{K}{2}}{\cal D}_i D_{\bar {{\cal A}_1}}W \sim {\cal O}(1) e^{\frac{K}{2}}{\cal D}_i D_{{\bar a}_1}W$,  we have $e^{\frac{K}{2}}{\cal D}_i D_{{\cal A}_1}W \sim {\cal V}^{-\frac{31}{18}}{{\cal Z}_i}\delta{\cal A}_1$. The contribution of physical  higgsino-lepton-slepton vertex after giving VEV to ${\cal Z}_I$  will be given as :
\begin{eqnarray}
\label{eq:Higgsino-lepton-slepton}
& &  C^{{\tilde H^{c}}_L {l_L} \tilde {l_L}}\sim\frac{{\cal V}^{-\frac{31}{18}}\langle{\cal Z}_i\rangle}{{\sqrt{\hat{K}^{2}_{{\cal Z}_i{\bar{\cal Z}}_i}{\hat{K}_{{\cal A}_1{\bar {\cal A}}_1}}{\hat{K}_{{\cal A}_1{\bar {\cal A}}_1}}}}} \sim {\cal V}^{-\frac{3}{2}}.
\end{eqnarray}
The gaugino-lepton-slepton vertex  in ${\cal N}=1$ gauged supergravity action is given by 
\begin{equation}
{\cal L}= g_{YM}g_{{ J}\bar{T}_B}X^{{*}B}{\bar\chi}^{\bar J}\lambda_{\tilde{g}}+ \partial_{a_1}T_B D^{B}{\bar\chi}^{\bar a_1}\lambda_{\tilde{g}}.
 \end{equation}
 Utilizing (\ref{eq:K}), $X^{B}=-6i\kappa_4^2\mu_7Q_{T_B}, \kappa_4^2\mu_7\sim \frac{1}{\cal V},  Q_{T_B}\sim{\cal V}^{\frac{1}{3}}(2\pi\alpha^\prime)^2\tilde{f}$,
and $g_{YM}g_{{T_B} {\bar {\cal A}}_1}\sim  {\cal V}^{-\frac{2}{9}} ({\cal A}_1-{\cal V}^{-\frac{2}{9}}M_P),$ 
Using $T_{B}=Vol(\sigma_B)- C_{I{\bar J}}a_I{\bar a}_{\bar J} + h.c.$, (intersection matrices $C_{I{\bar J}}$ are given in an appendix of \cite{gravitino_DM}),  $\partial_{{\cal A}_1} T_B\rightarrow {\cal V}^{\frac{10}{9}} ({\cal A}_1-{\cal V}^{-\frac{2}{9}})$ and
the dominant contribution to the physical gaugino-lepton-slepton vertex is proportional to :
\begin{equation}
\label{eq:gaugino-lepton-slepton}
C^{\lambda_{\tilde{g}} {l_L}\tilde {l_L}}\sim \frac{g_{YM}g_{{T_B} {\bar {\cal A}_1}}X^{T_B}\sim {\cal V}^{-\frac{2}{9}} {\tilde f}}{{\left(\sqrt{\hat{K}_{{\cal A}_1{\bar {\cal A}}_1}}\right)}{\left(\sqrt{\hat{K}_{{\cal A}_1{\bar {\cal A}}_1}}\right)}}\sim  \tilde{f} {\cal V}^{-1}.
\end{equation}
By adding the contribution of (\ref{eq:Higgsino-lepton-slepton}) and (\ref{eq:gaugino-lepton-slepton}) as according to equation (\ref{eq:chi3}), the physical neutralino-lepton-slepton vertex is given as:
\begin{eqnarray}
\label{eq:neutralino-lepton-slepton}
& & {\hskip -0.5in} C^{\chi^0_3 {l_L}\tilde {l_L}}:   \tilde{f} {\cal V}^{-1}  + \tilde{f} {\frac{v}{M_P}}\left({\cal V}^{-\frac{2}{3}}\right)  \sim \tilde{f} \left({\cal V}^{-1}\right)~{\rm for}~ {\cal V}\sim {10}^5.
\end{eqnarray}
Now, the  ${\tilde l_L}-{u_L}-{d^{c}_L}$  vertex corresponding to Figure~3.11(a) and the ${\tilde \nu_L}-{d_{jL}}-{d_{kR}}$ vertex as calculated in section {\bf 2} get identified with same set of moduli space superfields and given by similar interaction vertex. Therefore, the contribution of physical ${\tilde l_L}-{u_L}-{d^{c}_L}$  vertex is same as the ${\tilde \nu_L}-{d_{jL}}-{d_{kR}}$ vertex as given in (\ref{eq:2A2}).
\begin{eqnarray}
\label{eq:3a2}
C^{{\tilde l_L}{u_L}{d^{c}_L}}\sim {\cal V}^{-\frac{5}{3}}, {\rm for}~{\cal V}\sim {10^5}
\end{eqnarray}
Similar to the above calculation, the contribution of  higgsino-quark-squark relevant to the neutralino-quark-squark vertex of Figure~3.11(b) is given by: $\frac{e^{\frac{K}{2}}}{2}\left({\cal D}_iD_{{\cal A}_4}W\right){\bar \chi}^ {i}\chi^{{\cal A}_4} +ig_{{\bar i }{{\cal A}_4}}{\bar\chi}^{I}\left[{\bar\sigma}\cdot\partial\chi^{ {{\cal A}_4}}+\Gamma^{{{\cal A}_4}}_{{\cal A}_4{{\cal A}_4}}{\bar\sigma}\cdot\partial {{\cal A}_4}{\chi}^{{{\cal A}_4}}
+\frac{1}{4}\left(\partial_{{\cal A}_4}K{\bar\sigma}\cdot {\cal A}_4 - {\rm c.c.}\right)\chi^{{{\cal A}_4}}\right].$
Working in diagonalized basis, $g_{{\bar i }{{\cal A}_4}}$=0.
By expanding ${\cal D}_iD_{a_4}W$ in the fluctuations linear in $a_4\rightarrow a_4+{\cal V}^{-\frac{11}{9}}{M_P}$ by using equations (\ref{eq:K}) and (\ref{eq:W}), one obtains: $e^{\frac{K}{2}}{\cal D}_iD_{{\cal A}_4}W \equiv e^{\frac{K}{2}}{\cal D}_iD_{a_4}W \sim  {\cal V}^{\frac{71}{72}}\delta {\cal A}_4$, and the physical higgsino-quark-squark takes the form as below :
\begin{eqnarray}
\label{eq:Higgsino-quark-squark}
& & {\hskip -0.6in}C^{H^{c}_L \bar d^{c}_L \tilde {d_R}} \sim\frac{{\cal V}^{\frac{71}{72}}}{{\sqrt{\hat{K}_{{\cal Z}_i{\bar{\cal Z}}_i}{\hat{K}_{{\cal A}_4{\bar {\cal A}}_4}}{\hat{K}_{{\cal A}_4{\bar {\cal A}}_4}}}}}  \sim \left[{(10)^{-\frac{19}{2}}}{\cal V}^{\frac{71}{72}}\right]\sim {\cal V}^{-\frac{11}{12}}.
\end{eqnarray}
The interaction vertex corresponding to gaugino-quark-squark vertex of Figure~3.11(b) is given by $g_{YM}g_{{{{\bar {\cal A}}_4}}{\bar{T}_B}}X^{{*}B}{\chi^{c^{{\cal {A}}_4}}}\lambda_{\tilde{g}}$. For a K\"{a}hler moduli space, $g_{{{{\bar {\cal A}}_4}}{\bar{T}_B}}=0$ i.e the  gaugino-quark-squark vertex does not contribute to this particular vertex. In view of this, the physical neutralino-quark-squark vertex  of Figure~3.11(b) as according to equation (\ref{eq:chi3}) is given as:
\begin{eqnarray}
\label{eq:neutralino-quark-squark}
& & C^{{\chi^0}_3\bar d^{c}_L \tilde {d_R}}:   \left(\tilde{f}{\frac{v}{M_P}}\right) {\cal V}^{-\frac{1}{12}}.\end{eqnarray}
The  ${\tilde u_L}-{l_L}-{d^{c}_L}$ vertex corresponding to Figure~3.11(b) and ${\tilde d_{jL}}-{\nu_{iL}}-{d_{kR}}$vertex calculated in section {\bf 2} get identified with same set of moduli space superfields and given by similar interaction vertex. Therefore, the contribution of physical ${\tilde u_L}-{l_L}-{d^{c}_L}$ vertex is same as given in (\ref{eq:2b2}). i.e,
\begin{eqnarray}
\label{eq:lepton-squark-quark1}
C^{{\tilde u_L}{l_L}{d^{c}_L}}\sim C^{{\tilde d_{jL}} {\nu_{iL}} {d_{kR}}} \sim {\cal V}^{-\frac{5}{3}}, {\rm for}~{\cal V}\sim {10^5}
\end{eqnarray}
The neutralino-quark-squark vertex of Figure~3.11(c)  is given as considering the contribution of
$\frac{e^{\frac{K}{2}}}{2}\left({\cal D}_iD_{\bar A_2}W\right)\chi^{{\cal Z}_i}{\chi^{c^{{\cal {A}}_2}}} +ig_{{\bar {\cal Z}_i}{\bar {A_2}}}{\bar\chi}^{{\cal Z}_i}\Bigl[{\bar\sigma}\cdot\partial+\Gamma^{{\cal A}_2}_{{\cal A}_2{\bar A_2}}{\bar\sigma}\cdot\partial {\cal A}_2{\chi^{c^{{\cal {A}}_2}}}
+\frac{1}{4}\bigl(\partial_{{\cal A}_2}K{\bar\sigma}\cdot {\cal A}_2$ $ - {\rm c.c.}\bigr){\chi^{c^{{\cal {A}}_2}}} \Bigr]$,
 $\chi^I$ is $SU(2)_L$ higgsino, $\chi^{A_2}$ corresponds to $SU(2)_L$ quark and ${{\cal A}_2}$ corresponds to left-handed squark.
and  $g_{{\bar I}{\bar {\cal A}_2}}=0$. By expanding ${\cal D}_i D_{\bar {a_2}}W$ in the fluctuations
 linear in  $z_1$ and then $(a_2- {\cal V}^{-\frac{1}{3}}{M_P})$ by using equations (\ref{eq:K}) and (\ref{eq:W}), $e^{\frac{K}{2}}{\cal D}_i D_{{\cal A}_2}W \equiv e^{\frac{K}{2}}{\cal D}_i D_{\bar {a_2}}W \sim {\cal V}^{-\frac{20}{9}}{\cal Z}_i\delta{\cal A}_2 $, where ${\cal A}_2$ correspond to left- handed squark. The contribution of physical higgsino-quark-squark vertex will be given as :
\begin{eqnarray}
\label{eq:Higgsino-quark-squark1}
& &  C^{{\tilde H^{c}}_L {u_L} \tilde {u_L}}\sim\frac{{\cal V}^{-\frac{20}{9}}\langle{\cal Z}_i\rangle}{{\sqrt{\hat{K}^{2}_{{\cal Z}_i{\bar{\cal Z}}_i}{\hat{K}_{{\cal A}_2{\bar {\cal A}}_2}}{\hat{K}_{{\cal A}_2{\bar {\cal A}}_2}}}}} \sim {\cal V}^{-\frac{4}{5}}.
\end{eqnarray}
 As discussed for gaugino lepton-slepton vertex of Figure~3.11(a), the gaugino- quark-squark vertex corresponding to Figure 3.11(c) in ${\cal N}=1$ gauged SUGRA is given by $g_{YM}g_{{ J}B^{*}}X^{{*}B}{\bar\chi}^{\bar J}\lambda_{\tilde{g}}+ + \partial_{a_2}T_B D^{B}{\bar\chi}^{\bar a_1}\lambda^{0}$. Utilizing equation (\ref{eq:K}) and value of $X^{B}=-6i\kappa_4^2\mu_7Q_{B}$ as mentioned earlier, this time, we get:
$g_{YM}g_{B {\bar a}_2}\rightarrow-{\cal V}^{-\frac{5}{4}}(a_2-{\cal V}^{-\frac{1}{3}})$, $\partial_{a_2}T_B \rightarrow {\cal V}^{\frac{1}{9}} (a_2-{\cal V}^{-\frac{1}{3}}) $. Hence,  the dominant contribution to the physical gaugino-quark-squark vertex is proportional to :
\begin{equation}
\label{eq:gaugino-quark-squark1}
C^{\lambda_{\tilde{g}} {u_L}\tilde {u_L}}\sim \frac{ {\cal V}^{-\frac{11}{9}} {\tilde f}}{{\left(\sqrt{\hat{K}_{{\cal A}_2{\bar {\cal A}}_2}}\right)}{\left(\sqrt{\hat{K}_{{\cal A}_2{\bar {\cal A}}_2}}\right)}}\sim \tilde{f}\left({\cal V}^{-\frac{4}{5}}\right).
\end{equation}
Following equation (\ref{eq:chi3}), the neutralino-quark squark vertex will be given by :
\begin{eqnarray}
\label{eq:neutralino-quark-squark1}
\hskip -1in C^{{\chi^{0}_3} {u_L}\tilde {u_L}}\sim C^{\lambda_{\tilde{g}} {u_L}\tilde {u_L}}+ {\tilde f}{\cal V}^{\frac{5}{6}}\frac{v}{M_P} C^{H^{c}_L {u_L} \tilde {u_L}} \sim \tilde{f}{\cal V}^{-\frac{4}{5}}.
\end{eqnarray}
 Finally, the  ${\tilde d_R}-{l_L}-{u_L}$ vertex corresponding to Figure~3.11(c) and ${\tilde d_{kR}}-{\bar\nu^c_{iR}}-{d_{jL}}$  vertex vertex calculated in section {\bf 2} get identified with same set of moduli space superfields and given by similar interaction vertex . Hence,  contribution of physical ${\tilde d_R}-{l_L}-{u_L}$ vertex of Figure~3.11(c) is same as given in (\ref{eq:2C2}).
\begin{eqnarray}
\label{eq:lepton-squark-quark2}
C^{{\tilde d_R}{l_L}{u_L}} \sim {\cal V}^{-\frac{5}{3}}, {\rm for}~{\cal V}\sim {10^5}.
\end{eqnarray}
Now, working in two component notation, in order to calculate the decay width, we will be using the decay width formula as given by H.Dreiner et al. in \cite{2comp}. Because of Majorana nature of neutralino, we are considering both right-handed as well as left-handed incoming neutralino though the dominant contribution occurs in case of vertices corresponding to right-handed neutralino. Henceforth, we will be using incoming right-handed neutralino in the matrix amplitude calculation.

Guided by their notations, right-handed incoming and left-handed outgoing are denoted by wave function $ y^\dagger_i \equiv
 y^\dagger (\boldsymbol{\vec p}_i ,\lam_i)$, $ x^\dagger_l \equiv  x^\dagger
(\boldsymbol{\vec p}_l ,\lam_l)$, $ x^\dagger_u \equiv  x^\dagger
(\boldsymbol{\vec p}_u ,\lam_u)$, and $ x^\dagger_d \equiv  x^\dagger
(\boldsymbol{\vec p}_d ,\lam_d)$, respectively . Using the numerical estimate of vertices as calculated in set of equations (\ref{eq:neutralino-lepton-slepton})-(\ref{eq:lepton-squark-quark2}) for all three Feynman diagrams given in Figure~3.11, the corresponding contributions to the decay amplitude is:
\begin{eqnarray}
& & i{\cal M}_1 =
{\tilde{f}{\cal V}^{-1}{\cal V}^{-\frac{5}{3}}}
\left (
\frac{i}{(p_i - k_l)^2 - m_{\tilde l_L}^2}
\right )   y^\dagger_i  x^\dagger_l  x^\dagger_u  x^\dagger_d\
\\
& &  i{\cal M}_2 =
{\tilde f}{\cal V}^{-\frac{1}{12}}\frac{v}{M_P}{\cal V}^{-\frac{5}{3}}\left (
\frac{ i}{(p_i - k_d)^2 -  m_{\tilde d_R}^2}
\right )\nonumber
 y^\dagger_i  x^\dagger_d  x^\dagger_l  x^\dagger_u\
\\
 & &  i{\cal M}_3 =
{\tilde f}{\cal V}^{-\frac{4}{5}}{\cal V}^{-\frac{5}{3}} \left (
\frac{ i}{(p_i - k_u)^2 -  m_{\tilde u_L}^2}
\right ) y^\dagger_i  x^\dagger_u  x^\dagger_d  x^\dagger_l.
\end{eqnarray}
Neglecting all of the final state fermion masses, one can express kinematic variables in the form as: $z_l\equiv  2 p_i \newcdot k_l/m_{\chi^0_3}^2
= 2 E_l/m_{\chi^0_3},  z_d \equiv  2 p_i \newcdot k_d/m_{\chi^0_3}^2
= 2 E_d/m_{\chi^0_3}, z_u \equiv  2 p_i \newcdot k_u/m_{\chi^0_3}^2
= 2 E_u/m_{\chi^0_3}$,
and the total matrix amplitude can be rewritten as:
\begin{eqnarray}
{\cal M} =
c_1  y^\dagger_i  x^\dagger_l  x^\dagger_u  x^\dagger_d
+
c_2  y^\dagger_i  x^\dagger_d  x^\dagger_l  x^\dagger_u
+
c_3  y^\dagger_i  x^\dagger_u  x^\dagger_d  x^\dagger_l,
\end{eqnarray}
For $m_{\chi^0_3}\sim {\cal V}^{\frac{2}{3}}m_{\frac{3}{2}}$ and $m_{l_L}= m_{\tilde d_R} =m_{\tilde u_L}\sim {\cal V}^{\frac{1}{2}}m_{\frac{3}{2}}$ as given in eqs. (\ref{eq:gaugino_mass}) and (\ref{eq:mass_Zi}), we get:
\begin{eqnarray}
&&{\hskip -0.02in} c_1 \equiv {\tilde{f}{\cal V}^{-\frac{3}{2}}{\cal V}^{-\frac{5}{3}}} /[
m_{\tilde l_L}^2 - m_{\chi^0_3}^2 (1 - z_l)] = -{\tilde{f}{\cal V}^{-1}{\cal V}^{-\frac{5}{3}}} /{m^2_{\frac{3}{2}}}[{\cal V}- {\cal V}^{\frac{4}{3}}
 (1 - z_l)] \nonumber\\
&& c_2 \equiv {\tilde f}{\cal V}^{-\frac{1}{12}}{\frac{v}{M_P}}{\cal V}^{-\frac{5}{3}}/[
m_{\tilde d_R}^2 - m_{\chi^0_3}^2 (1 - z_d)]= -{\tilde f}{\cal V}^{-\frac{1}{12}}{\frac{v}{M_P}}{\cal V}^{-\frac{5}{3}}/{m^2_{\frac{3}{2}}}[{\cal V}- {\cal V}^{\frac{4}{3}} (1 - z_d)] \nonumber
\\
&&{\hskip -0.02in} c_3 \equiv {\tilde f}{\cal V}^{-\frac{4}{5}}{\cal V}^{-\frac{5}{3}}/[
m_{\tilde u_L}^2 - m_{\chi^0_3}^2 (1 - z_u)] =  -{\tilde f}{\cal V}^{-\frac{4}{5}}{\cal V}^{-\frac{5}{3}}/{m^2_{\frac{3}{2}}}[{\cal V}- {\cal V}^{\frac{4}{3}} (1 - z_u)].
\label{eq:RPVNdecaycthree}
\end{eqnarray}
As explained in \cite{2comp}, by applying Fierz identity, one can reduce the number of terms, i.e
${\cal M} =
(c_1 - c_3)  y^\dagger_i  x^\dagger_l  x^\dagger_u  x^\dagger_d
+
(c_2 - c_3)  y^\dagger_i  x^\dagger_d  x^\dagger_l  x^\dagger_u
$, and $
|{\cal M}|^2 = |c_1 - c_3|^2  y^\dagger_i  x^\dagger_l x_\mu y_i
                x^\dagger_u  x^\dagger_d x_d x_u
+ |c_2 - c_3|^2  y^\dagger_i  x^\dagger_d x_d y_i
                 x^\dagger_l  x^\dagger_u x_u x_l
- 2 {\rm Re}[(c_1 - c_3) (c_2^* - c_3^*)
 y^\dagger_i  x^\dagger_l x_l x_u  x^\dagger_u  x^\dagger_d x_d y_i]$.
 Summing over the fermion spins by using identities,
$ \sum_s x_\alpha({\boldsymbol{\vec p}},s)
x^\dagger_{\dot{\beta}}({\boldsymbol{\vec p}},s) =
+ p\newcdot\sigma_{\alpha\dot{\beta}}
\,,
\sum_s x^{\dagger\dot{\alpha}}({\boldsymbol{\vec p}},s)
 x^{\beta}({\boldsymbol{\vec p}},s) =
+ p\newcdot\sigmabar^{\dot{\alpha}\beta}
\,,
 \sum_s y^{\dagger\dot{\alpha}}({\boldsymbol{\vec p}},s)
y^{\beta}({\boldsymbol{\vec p}},s)
= + p \newcdot \sigmabar^{\dot{\alpha}\beta},
 \sum_s y_\alpha({\boldsymbol{\vec p}},s)
y^\dagger_{\dot{\beta}}({\boldsymbol{\vec p}},s) =
+ p \newcdot \sigma_{\alpha\dot{\beta}}
$, one obtains:
\begin{eqnarray}
\sum_{\rm spins} |{\cal M}|^2 &=&
|c_1 - c_3|^2 {\rm Tr}[k_l \newcdot \sigmabar p_i \newcdot \sigma]
              {\rm Tr}[k_d \newcdot \sigmabar k_u \newcdot \sigma]
+
|c_2 - c_3|^2 {\rm Tr}[k_d \newcdot \sigmabar p_i \newcdot \sigma]
              {\rm Tr}[k_u \newcdot \sigmabar k_l \newcdot \sigma]
\nonumber \\ &&
- 2 {\rm Re} \bigl [(c_1 - c_3) (c_2^* - c_3^*)
{\rm Tr}[ k_l \newcdot \sigmabar k_u \newcdot \sigma k_d \newcdot \sigmabar
          p_i \newcdot \sigma] \bigr ]\,.
\end{eqnarray}
Applying the trace formulae and using $
+ 2 k_l \newcdot k_d
=
(1 - z_u) m_{\chi^0_3}^2 ,
 + 2 k_l \newcdot k_u
=
(1 - z_d) m_{\chi^0_3}^2 ,
+ 2 k_d \newcdot k_u = (1 - z_l) m_{\chi^0_3}^2
$, one gets:
\begin{eqnarray}
\label{eq:sumspins}
\sum_{\rm spins} |{\cal M}|^2
&=& 4 |c_1 - c_3|^2 p_i \newcdot k_l \, k_d \newcdot k_u
  + 4 |c_2 - c_3|^2 p_i \newcdot k_d \, k_l \newcdot k_u
\nonumber \\ &&
 {\hskip -0.2in} -4  {\rm Re} \bigl [(c_1 - c_3) (c_2^* - c_3^*)]
  (k_l \newcdot k_u\, p_i \newcdot k_d
  + p_i \newcdot k_l \,k_d \newcdot k_u
  - k_l \newcdot k_d \, p_i \newcdot k_u)
\nonumber \\[3pt]
&&{\hskip -0.9in} = m_{\chi^0_3}^4 \Bigl [
    |c_1|^2 z_l (1 - z_l)
    + |c_2|^2 z_d (1 - z_d)
    + |c_3|^2 z_u (1 - z_u)
    - 2{\rm Re}[c_1 c_2^*] (1 - z_l)(1 - z_d)
    \nonumber \\ &&
    {\hskip -0.8in} - 2{\rm Re}[c_1 c_3^*] (1 - z_l)(1 - z_u)
    - 2{\rm Re}[c_2 c_3^*] (1 - z_d)(1 - z_u)
    \Bigr ]\,.
\end{eqnarray}
The differential decay rate follows: $
\frac{d^2 \Gamma}{dz_l d z_d}
=
\frac{N_c m_{\chi^0_3}}{2^8 \pi^3}
\biggl (\frac{1}{2} \sum_{\rm spins} |{\cal M}|^2 \biggr)$,
where  $N_c = 3$  and the
kinematic limits are $
0 < z_l < 1, 1 - z_l <  z_d  <1.$
Adopting the technique as used in \cite{2comp}, the total decay width for the above  Feynman diagrams is given as:
\begin{eqnarray}
\label{eq:gamma}
\Gamma = \frac{N_c m_{\chi^0_3}^5}{2^{8}\cdot3 \pi^3}
\left ( |c_1^{\prime *}|^2 + |c_2^{\prime *}|^2 + |c_3^{\prime *}|^2 - {\rm Re}[
c_1^{\prime *} c_2^{\prime *} + c_1^{\prime *} c_3^{\prime *}
+ c_2^{\prime *} c_3^{\prime *}] \right ),
\end{eqnarray}
where $c_i^{\prime *}, i=1,2,3$ are obtained from  $c_i$'s  by integrating $z_{l},z_{d}$ in the aforementioned limits. Therefore, on simplifying, the result comes out to be given by:
\begin{eqnarray}
& & {\hskip-0.4in}\Gamma \sim \frac{3. {  {\cal V}^{\frac{10}{3}} m^5_{\frac{3}{2}}}{\tilde f}^2. {\cal V}^{-\frac{74}{15}}}{2^{8}\cdot3. \pi^3 {\cal V}^{\frac{8}{3}}m^4_{\frac{3}{2}}} \sim {10}^{-4}{\cal V}^{\frac{2}{3}-{\frac{74}{15}}}m_{\frac{3}{2}}{\tilde f}^2  \sim {10}^{-18}{\tilde f}^2.
 \end{eqnarray}
 In dilute flux approximation, considering ${\tilde f}^2 < {10}^{-8}$, life time of neutralino:
\begin{eqnarray}
   \tau &=&\frac{\hbar}{\Gamma}\sim\frac{10^{-34} Jsec}{{10}^{-18}{\tilde f}^2}\sim\frac{10^{-34} Jsec}{10^{-26}GeV} > O(10) sec.
   \end{eqnarray}
\subsection{Slepton/Squark Decays}
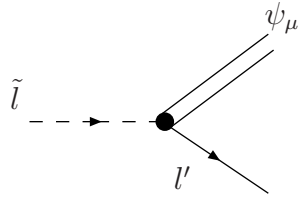
\begin{figure}
\begin{center}
\begin{picture}(150,97)(100,100)
\DashArrowLine(100,157)(150,157)5
\Line(150,160)(190,190)
\Line(152,154)(192,184)
\Vertex(151,157){3.5}
\ArrowLine(150,157)(190,130)
\Text(95,168)[]{${\tilde{l}}$}
\Text(196,197)[]{${\psi_\mu}$}
\put(155,132){${l^\prime}$}
%
\end{picture}
\end{center}
\vskip -0.3in
\caption{Two-Body Slepton/Squark Decay.}
\end{figure}
The decay width for $\tilde{l}/\tilde{q}\rightarrow l/q+\psi_\mu$ (as shown in Figure~3.12) is given by:
\begin{equation}
\label{eq:sfermiontofermion+grav}
\Gamma\left(\tilde{l}/\tilde{q}\rightarrow l/q+\psi_\mu\right)\sim\frac{m_{\tilde{l}/\tilde{q}}^5}{m_{3/2}^2M_P^2}\sim{\cal V}^{-\frac{7}{2}}M_P,
\end{equation}
implying a lifetime of around $10^{-25.5}s$, satisfying the BBN constraints.

As explained in \cite{Feng_et_al_sleptondecay}, the following set of effective operators are relevant to three-body slepton decays:
\begin{eqnarray}
\label{eq:Ois}
{\cal O}_1 &=& \overline{l'_h}\ p_{\tilde{l}}\cdot\tilde{G}^c\
p_{l'}\cdot\epsilon^*, {\cal O}_2 = \overline{l'_h}\ p_l\cdot\tilde{G}^c \
p_{\tilde{l}}\cdot\epsilon^*,{\cal O}_3 =\overline{l'_h}\ p_{\tilde{G}}\cdot\tilde{G}^c
\ p_{\tilde{l}}\cdot\epsilon^*, \nonumber\\
{\cal O}_4 &=&i\ \overline{l'_h}\ \gamma\cdot\epsilon^*\
p_V\cdot \tilde{G}^c,
O_5 =\overline{l'_h}\ \gamma\cdot p_V\
\gamma\cdot\epsilon^*\ p_{\tilde{l}} \cdot \tilde{G}^c,
{\cal O}_6 =i\ \overline{l'_h}\ \gamma\cdot p_V
\ \gamma\cdot\epsilon^* \ p_V \cdot \tilde{G}^c \nonumber\\
{\cal O}_7 &=& i\ \overline{l'_h}\ \gamma\cdot p_{\tilde{G}}\
\gamma\cdot\epsilon^* \ p_V \cdot \tilde{G}^c,
{\cal O}_8 = i \ \overline{l'_h}\ \epsilon^*\cdot\tilde{G}^c,
{\cal O}_9 = i \ \overline{l'_h}\  \gamma\cdot p_V\
\epsilon^*\cdot \tilde{G}^c \nonumber\\
{\cal O}_{10} &=& i \ \overline{l'_h}\ \gamma\cdot p_{\tilde{G}}
\ \gamma\cdot p_V \epsilon^*\cdot\tilde{G}^c\ ,
\end{eqnarray}
where
$\epsilon_\mu$ is the polarization of gauge boson $V$.  Notice that
for an on-shell gravitino in the final state, ${\cal O}_3=0$ by using
the gravitino equation of motion. Using the following volume-suppression factors in various vertices in Figs.~3.13(a)-(d) above:
\begin{eqnarray}
\label{eq:vertices}
& & \tilde{l}-\tilde{G}-l\ {\rm vertex}: \frac{g_{{\cal A}_1{\bar{\cal A}}_1}}{K_{{\cal A}_1{\bar{\cal A}}_1}}=1; l-l^\prime-V\ {\rm vertex}: \frac{g_{{\cal A}_1{\bar{\cal A}}_1}\tilde{f}{\cal V}^{-\frac{2}{3}} ln{\cal V}}{K_{{\cal A}_1{\bar{\cal A}}_1}};\nonumber\\
& & \tilde{l}-\tilde{l}-V\ {\rm vertex}: \frac{\left(\frac{\kappa_4^2\mu_7Q^BG^{T^B{\bar T}^B}}{{\cal V}}\right)\kappa_4^2\mu_7C_{a_1{\bar a}_1}}{\kappa_4^2K_{{\cal A}_1{\bar{\cal A}}_1}}\sim\frac{\tilde{f}{\cal V}^{-2}}{10^4};  \tilde{l}-l^\prime-\tilde{G}\ {\rm vertex}: \frac{g_{{\cal A}_1{\bar{\cal A}}_1}}{K_{{\cal A}_1{\bar{\cal A}}_1}}=1,\nonumber\\
\end{eqnarray}
and the operators of (\ref{eq:Ois}), the matrix elements for three-body
slepton decay [diagrams (a)- (d)] are:
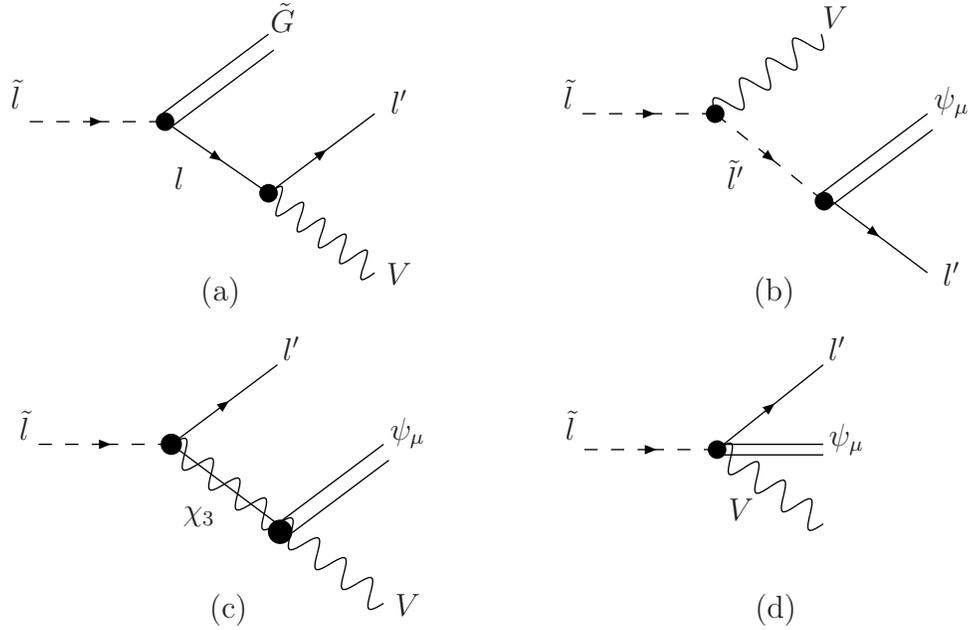
\begin{figure}[t!]
\label{decaydiag1}
\begin{center}
\begin{picture}(150,97)(100,100)
\DashArrowLine(100,157)(150,157)5
\Vertex(151,157){3.5}
\Line(150,160)(190,190)
\Line(152,154)(192,184)
\ArrowLine (151,157)(190,130)
\ArrowLine(190,130)(230,160)
\Vertex(190,130){3.5}
\Photon(190,130)(230,100){5}{5}
\Text(95,168)[]{${\tilde{l}}$}
\Text(240,165)[]{$l^\prime$}
\Text(240,100)[]{$V$}
\Text(196,197)[]{${\tilde{G}}$}
\put(155,132){${l}$}
\put(165,90){(a)}
\end{picture}
\hspace{1.8cm}
\begin{picture}(150,97)(100,100)
\DashArrowLine(100,160)(150,160)5
\Vertex(150,160){3.5}
\Photon(150,160)(190,190){5}{4}
\DashArrowLine (150,160)(191,127)5
\Line(190,130)(230,160)
\Line(192,124)(232,154)
\Vertex(191,127){3.5}
\ArrowLine(190,130)(230,100)
\Text(95,168)[]{$\tilde{l} $}
\Text(240,165)[]{$\psi_\mu$}
\Text(240,100)[]{$l^\prime$}
\Text(196,197)[]{$V$}
\put(155,132){$\tilde{l}^\prime$}
\put(165,90){(b)}
\end{picture}
\end{center}
\vspace{0.5cm}
\begin{picture}(150,97)(60,100)
\DashArrowLine(100,160)(150,160)5
\ArrowLine(150,160)(190,190)
\Line (150,160)(190,130)
\Vertex(150,160){4}
\Photon(150,160)(190,130){5}{4}
\Vertex(191,127){4.6}
\Line(190,130)(230,160)
\Line(192,124)(232,154)
\Photon(190,130)(230,100){5}{4}
\Text(95,168)[]{${\tilde{l}}$}
\Text(240,165)[]{$\psi_\mu$}
\Text(240,100)[]{$V$}
\Text(196,197)[]{$l^\prime$}
\put(155,132){$\chi_3$}
\put(165,95){(c)}
\end{picture}
\phantom{xxx}
%
\begin{picture}(150,97)(30,100)
\DashArrowLine(100,158)(150,158)5
\ArrowLine(150,158)(190,190)
\Photon(150,158)(190,130){5}{4}
\Vertex(150,158){3.5}
\Line(150,160)(190,160)
\Line(150,156)(190,156)
\Text(95,168)[]{${\tilde{l}}$}
\Text(196,197)[]{$l^\prime$}
\put(193,160){$\psi_\mu$}
\put(155,132){$V$}
\put(165,95){(d)}
\end{picture}
\caption{Three-body slepton decays.}
\end{figure}
\begin{eqnarray}
\label{eq:Ms}
{\cal {M}}_V^a &\sim& \frac{\tilde{f}{\cal V}^{-\frac{11}{18}}}{M_P}
\frac{1}{s_{23}} \left(2{\cal O}_1-{\cal O}_5 \right),
{\cal M}_V^b \sim \frac{\tilde{f}{\cal V}^{-2}}{10^4M_P}
\frac{1}{m^2_{\tilde{G}}+m_V^2-s_{13}-s_{23}}
\left({\cal O}_2+{\cal O}_3 \right)\nonumber \\
{\cal {M}}_V^c &\sim& \sum_i \frac{10\tilde{f}{\cal V}^{-\frac{3}{2}}}{M_P} \frac{1}{s_{13}-s_{\chi_3}}
\left[m_{\chi_3} \left({\cal O}_4-{\cal O}_9 \right)
- 4 \left({\cal O}_6+{\cal O}_7 - {\cal O}_{10} \right)
+ m_V^2{\cal O}_8 \right]
\nonumber\\
{\cal {M}}_V^d &\sim& i \frac{\tilde{f}{\cal V}^{-\frac{11}{9}}}{10^4M_P}{\cal O}_8 \ ,
\end{eqnarray}
where $\chi_3$ denotes the lightest neutralino in our set up, and
$ s_{12}=(p_{\tilde{G}}+p_{l'})^2, s_{13}=(p_{\tilde{G}}+p_V)^2, s_{23}=(p_{l'}+p_V)^2$. Notice that $s_{12}+s_{13}+s_{23}=\msl^2+\mg^2+m_Z^2$ and the differential decay width is
$d\Gamma=\frac{1}{(2\pi)^3}\frac{1}{32m^3_{\tilde{l}}}
 |{\cal M}|^2 dm^2_{13}dm^2_{23}$, where ${\cal M}={\cal M}_V^a+{\cal M}_V^b+{\cal M}_V^c+{\cal M}_V^d$.
The sum of the matrix elements can be written as:
\begin{eqnarray}
{\cal{M}}(\tilde{l}_h\to l' V \tilde{G})
= {\cal M}_V^a+{\cal M}_V^b+{\cal M}_V^c + {\cal M}_V^d
= \sum_{i=1\ldots 10} {\cal{M}}_i \calO_i \ ,
\end{eqnarray}
where the ${\cal{M}}_i$ can be read off from
Eqs.~(\ref{eq:Ms}) above as:
\begin{eqnarray}
\label{eq:Mis}
& & {\cal M}_{1,5}\sim\frac{\tilde{f}{\cal V}^{-\frac{11}{18}}}{M_P s_{23}};\ {\cal M}_{2,3}\sim\frac{\left(\frac{\tilde{f}{\cal V}^{-2}}{10^4}\right)}{M_P(m_{3/2}^2 - s_{13} - s_{23})}; {\cal M}_{4,9}\sim\frac{m_{\chi}\left(10\tilde{f}{\cal V}^{-\frac{3}{2}}\right)}{M_P(s_{13}-m_{\chi_3}^2)};\nonumber\\
& &
{\cal M}_{6,7,10}\sim\frac{M_{4,9}}{m_{\chi_3}}; {\cal M}_8\sim\frac{1}{M_P}\left(\frac{m_V^2\left(10\tilde{f}{\cal V}^{-\frac{3}{2}}\right)}{s_{13}-m_{\chi_3}^2} + \frac{\tilde{f}{\cal V}^{-\frac{11}{9}}}{10^4}\right).
\end{eqnarray}
The squared matrix
element is
\begin{eqnarray}
\left| {\cal{M}}(\tilde{l}_h \to l' V \tilde{G}) \right|^2
&=& \sum_{i=1\ldots 10} |{\cal{M}}_i|^2 \calO_{i,i} + \sum_{i,j=1\ldots 10}^{i<j} \Re({\cal{M}}_i {\cal{M}}_j^*)
\calO_{i,j}^{\text re} \nonumber \\
&&
+ \sum_{i,j=1\ldots 10}^{i<j} \Imag({\cal{M}}_i {\cal{M}}_j^*)
\calO_{i,j}^{\text im} \ .
\label{eq:square}
\end{eqnarray}
All the ${\cal{M}}_i$ are real except for ${\cal M}_8$, which has both
real and imaginary components.  The only non-zero contributions to the
last term in Eq.~(\ref{eq:square}) come from $\Imag({\cal{M}}_i
{\cal{M}}_8^*)$ ($i<8$) and $\Imag({\cal{M}}_8 {\cal{M}}_j^*)$
($j>8$). The resultant contribution of the same are evaluated in an appendix of \cite{gravitino_DM}. Utilizing that,
\begin{equation}
\label{eq:Gamma_sltograv+lep+V}
\Gamma\left(\tilde{l}\rightarrow l^\prime+\tilde{G}+V\right)\sim
 \int_{m^2_{3/2}}^{2m^2_{\tilde{l}}}ds_{13}\int_{m_V^2}^{m^2_{\tilde{l}}}ds_{23}\frac{\Re\left({\cal M}_4{\cal M}_8^*\right){\cal O}^{\rm re}_{4,8}}{m^3_{\tilde{l}}}\sim10^{-15}M_P,
\end{equation}
implying that the associated life-time of this decay is about $10^{-28}$s, which respects the BBN constraints.
The summarized results of life time of various (N)LSP candidates are given in Table~3.1.
\begin{table}[htbp]
\centering
\begin{tabular}{|l|l|}
\hline
{\bf Decay channels}   &  {\bf Average life time}  \\ \hline
  Gaugino decays:  $\tilde{B}\rightarrow\psi_\mu Z/{\gamma} $ & $ 10^{-30}s $\\
  $\tilde B\xrightarrow[]{Z}\psi_\mu u {\bar u} $  & $ 10^{-13}s $\\
  $\tilde {W}\xrightarrow[]{\tilde q}\psi_\mu u {\bar u} $  & $ 10^{-25}s $ \\

$\tilde{W}^0\rightarrow\psi_\mu W^+ W^- $ & $10^{-66}s $ \\ \hline
Gluino decays: $ \tilde{g}\rightarrow \chi_{\rm n}^{o}q_{{}_I} \bar{q}_{{}_J} $ & ${10}^{1} s$\\
$\tilde{g}\rightarrow\tilde{\chi}_3^0 g $ & ${10}^{10}s $ \\
 $\tilde{g}\rightarrow \psi_{\mu} q_{{}_I} \bar{q}_{{}_J}$ & $ {10}^{3}s $\\
$ \tilde{g}\rightarrow \psi_{\mu} g $ & $ {10}^{-1}s $\\ \hline
RPV Neutralino decay: $\chi^0_3\rightarrow u {\bar d} e^{-}$ & $ 10 s $\\ \hline
Slepton decays: $\tilde{l}\rightarrow l^\prime \tilde{G} V $ & $ 10^{-28}s $\\
$ \tilde{l}/\tilde{q}\rightarrow l/q \psi_\mu $  & $ 10^{-25.5}s $\\ \hline
Gravitino decays: $\psi_\mu\rightarrow h \nu_e $ & $ 10^{17}s $ \\
$ \psi_\mu\rightarrow \nu\gamma,\nu Z  $ & $10^{22}s$ \\
$\psi_\mu \rightarrow L_iL_j E^c_k$ & $ 10^{21}s $ \\
$\psi_\mu \rightarrow L_i Q_j D^c_k$ & $ 10^{20}s $ \\
$\psi_\mu \rightarrow U_i^c D_j^c D_k^c$ & $ 10^{18}s $ \\  \hline
\end{tabular}
\label{table:decay_lifetime}
\caption{Life time estimates of various N(LSP) decay channels.}
\end{table}
\vskip -0.5in
\section{Relic Abundance of Gravitino}
As discussed in section {\bf 2}, explicit calculations yielding the gravitino-decay life time of the order greater than  age of the universe, justify considering gravitino as a viable dark matter candidate in our model. Keeping in mind the fact that relic density of dark matter particle should be within the limits provided by recent WMAP observations and other direct and indirect experiments, this section is devoted to calculating the relic abundance of the gravitino as a dark matter particle. The constraints on the cosmological gravitino abundance have been discussed in \cite{KHLOPOV_LINDE}. Assuming that re-heating temperature will be  low enough to produce an appropriate amount of dark matter in case of heavy gravitino, we focus on the case of gravitinos produced only in the decays of co-NLSP's and show that the same are produced in sufficient numbers to constitute all of non-baryonic dark matter.  The mass scales and  life time estimates of sleptons and neutralino discussed in section {\bf 3} manifestly indicate the same to be valid co-NLSPs which freeze out with appropriate thermal relic density before decaying and then eventually decay into the gravitino. Therefore, gravitino then inherits much of the relic density of the neutralino/slepton. Therefore, in the following subsections, we evaluate relic density of neutralino $\chi^{0}_3$ and leptons.
\subsection{Neutralino Density Calculations}
The number density of CDM candidate from the early universe depends sensitively on the annihilation cross section of such particles. Amongst the various approaches used to calculate the thermal cross section given in the literature, we rely on the partial-wave expansion approach used in \cite{Takeshi_Leszek} to calculate the annihilation cross section for each possible process. Since local $D3/D7$ $\mu$-split-like  SUSY model as discussed in chapter {\bf 2} includes neutral light Higgs boson, heavy Higgs boson and superpartners of other neutral particles, we consider possible annihilation channels which proceed via these particles only. The important channels that we will be discussing are: ${{\chi}^0_3 {\chi}^0_3\rightarrow hh}$, ${{\chi}^0_3 {\chi}^0_3\rightarrow ZZ}$, ${{\chi}^0_3 {\chi}^0_3\rightarrow ff}$.

The general procedure which is being followed in partial wave expansion approach, is given as follows.
As given in \cite{Keith.a.olive}, in QFT, the general cross-section is given as:
\begin{equation}
\sigma v_{\rm M\o l}= \frac{1}{4E_1E_2}\int dLIPS {|\cal M|}^2
\end{equation}
\begin{equation}
{\rm where,}~dLIPS = (2\pi)^4 \delta^4(p_1 +p_2-\sum {p_j}) \prod \int {\frac{d^3{p_1}}{(2\pi)^3 2{\pi}_0}}.
\end{equation}
The thermally-averaged product of the neutralino pair-annihilation
cross section and their relative velocity $\langle \sigma v_{M \o l} \rangle$ is given as \cite{Takeshi_Leszek,Keith.a.olive}:
\begin{equation}
\label{eq:thermalcrosc}
\langle \sigma v_{\rm M\o l} \rangle(T)=
\frac{\int d^3p_1 d^3p_2\, \sigma v_{\rm
M\o l}\, e^{-E_1/T} e^{-E_2/T}} {\int d^3p_1 d^3p_2\,  e^{-E_1/T} e^{-E_2/T} },
\end{equation}
where $p_1= (E_1, {\bf p}_1)$ and  $p_2= (E_2, {\bf p}_2)$
are the 4-momenta of the two colliding particles, and $T$ is the temperature of the bath.
Given the complexity of the general cross section, it is difficult to solve it analytically.
Alternatively, one finds a way to get the solution by expressing $\langle \sigma v_{M\o l}\rangle$ in terms of $x=\frac{T}{M}$. For this, one defines: %
\begin{equation}
w(s)=\frac{1}{4}\int dLIPS {|\cal M|}^2=E_1E_2 \sigma v_{\rm M\o l}.
\end{equation}
Incorporating the value of $\sigma v_{\rm M\o l}$ in equation (\ref{eq:thermalcrosc}),
\begin{equation}
\label{eq:thermalcross}
\langle \sigma v_{\rm M\o l} \rangle(T)=
\frac{\int d^3p_1 d^3p_2\, w(s) e^{-E_1/T} e^{-E_2/T}} {\int d^3p_1 d^3p_2\, E_1E_2 e^{-E_1/T} e^{-E_2/T} }.
\end{equation}
By defining change of variables discussed in \cite{Keith.a.olive} i.e by expressing momentum and energy in terms of x, carrying over the integration in terms of x,  $\langle\sigma v_{\rm M\o l}\rangle$ takes the form given below:
\begin{equation}
\label{eq:thermalcross}
\langle\sigma v_{\rm M\o l}\rangle=\frac{1}{m_\chi^2}
\left[w-\frac{3}{2}\left(2w-w'\right) x +
{\mathcal O} (x^2)\right]_{s=4m_\chi^2} \equiv a + bx + {\mathcal
O}(x^2).
\end{equation}
The coefficients $a$ and $b$ summed over all possible final states $f_1 f_2$ are defined in \cite{Takeshi_Leszek}, and given as:
\begin{eqnarray}
\label{eq:abf1f2}
& &   a=
\sum_{f_1 f_2}c \, \theta\left(4\mchi^2-(m_{f_1}+m_{f_2})^2 \right)\,
v_{f_1 f_2}\,\widetilde{a}_{f_1 f_2}, \nonumber\\
& &   b=
\sum_{f_1 f_2}c \, \theta\left(4\mchi^2-(m_{f_1}+m_{f_2})^2 \right)\,
v_{f_1 f_2} \Bigg\{ \widetilde{b}_{f_1 f_2}+  \widetilde{a}_{f_1 f_2}
\Bigl[ -3+\frac{3}{4}v_{f_1 f_2}^{-2} \Bigr(
      \frac{m_{f_1}^{2}+m_{f_2}^{2}}{2 \mchi^{2}}
   + \nonumber\\
&&\frac{(m_{f_1}^{2}-m_{f_2}^{2})^{2}}{8 \mchi^{4}}\Bigr)\Bigr]
                     \Bigg\}. \nonumber
\end{eqnarray}
The analytic expressions of {\bf a} and {\bf b}  for the s, t and u channels of various possible annihilation processes are given in \cite{Takeshi_Leszek}. Utilizing the same, we calculate the numerical estimates of {\bf a} and {\bf b} for all kinematically possible annihilation processes in our model. The idea is to first calculate the required vertices corresponding to different annihilation processes in the context of ${\cal N}=1$ gauged supergravity action and then use their estimates to estimate a and b coefficients. Following the same formalism as used in section {\bf 3}, utilizing the ${\cal N}=1$ gauged supergravity action of Wess and Bagger \cite{Wess_Bagger}, we first obtain the numerical estimates of required vertices.

\underline{\bf {Higgsino-higgsino-Higgs vertex}}
\begin{eqnarray}
&&{\cal L}=\frac{e^{\frac{K}{2}}}{2}\left({\cal D}_{{\cal Z}_1} D_{{\cal Z}_1}W\right) \chi^{{\cal Z}_1}_L \bar\chi^{{\cal Z}_i}_R +ig_{{\cal Z}_1{\bar {{\cal Z}_i}}}{\bar\chi}^{\bar {{\cal Z}_1}}_L \Bigl[{\gamma}\cdot\partial\chi^{{\cal Z}_i}_L+\Gamma^{{\cal Z}_i}_{{\cal Z}_i{{\cal Z}_i}}{\gamma}\cdot\partial {{\cal Z}_i}\chi^{{\cal Z}_i}_L
\nonumber\\
&& +\frac{1}{4}\left(\partial_{{\cal Z}_i}K{\gamma}\cdot {\cal Z}_i - {\rm c.c.}\right)\chi^{{\cal Z}_i}_L\Bigr],\end{eqnarray}
where $\chi^{{\cal Z}_i}_L/\bar\chi^{{\cal Z}_i}_R $ corresponds to left-/right-handed components of the higgsino.
Since $SU(2)_L$ symmetry gets spontaneously broken for higgsino-higgsino-Higgs vertex, therefore the idea is to  first expand aforementioned term  in the fluctuations quadratic in $z_i$ such that one of the two $z_i$'s acquires a VEV.
Utilizing $z_i\rightarrow z_i +{\cal V}^{\frac{1}{36}}{M_P}$ and thereafter solving with the help of equations (\ref{eq:K}) and (\ref{eq:W}),  one has $e^{\frac{K}{2}}{\cal D}_{{\cal {\bar Z}}_i}D_{{\cal Z}_i}W \equiv \frac{e^{\frac{K}{2}}}{2} {\cal D}_{z_i}D_{\bar {z_i}}W \sim {\cal V}^{-\frac{16}{9}}\langle{z_i}\rangle \delta z_i $, and
 \begin{equation}
 \label{eq:massterm}
 e^{\frac{K}{2}}{\cal D}_{{\cal {\bar Z}}_i}D_{{\cal Z}_i}W \chi^{{{\cal Z}_i}}_L \bar\chi^{{\cal Z}_i}_R \sim  \left({\cal V}^{-\frac{16}{9}}\langle{\cal Z}_i\rangle\right)\delta{\cal Z}_i \chi^{{{\cal Z}_i}}_L\bar\chi^{{\cal Z}_i}_R\sim \left({\cal V}^{-\frac{7}{4}} \right)\delta{\cal Z}_i \chi^{{{\cal Z}_i}}_L\bar\chi^{{\cal Z}_i}_R.
 \end{equation}
Using $\chi^{{\cal Z}_i} \sim {\cal V}^{\frac{59}{72}}m_{3/2}$ for the higgsino mass at EW scale (see below (eq.~\ref{eq:evs_3})) and $m_{3/2}={\cal V}^{-2}{M_P}$,  one obtains:
\begin{eqnarray}
\label{eq:kinetic}
& &  g_{{\cal Z}_i{\bar {{\cal Z}_i}}}{\bar\chi}^{\bar {{\cal Z}_i}}_L {\gamma}\cdot\partial\chi^{{\cal Z}_i}_L \rightarrow \frac{{\cal V}^{-\frac{37}{36}}\langle{\cal Z}_i\rangle }{M_P}\delta{\cal Z}_i{\bar\chi}^{\bar {{\cal Z}_i}}_L{\gamma}\cdot p_{\chi^{{\cal Z}_i}}\chi^{{\cal Z}_i}_L\sim {\cal V}^{-2}{\bar\chi}^{\bar {{\cal Z}_i}}_L\delta{{\cal Z}_i}\chi^{{\cal Z}_i}_L; \nonumber\\
& &  g_{{\cal Z}_i{\bar {{\cal Z}_i}}}{\bar\chi}^{\bar {{\cal Z}_i}}_L\Gamma^{{\cal Z}_i}_{{\cal Z}_i{{\cal Z}_i}}{\gamma}\cdot\partial {{\cal Z}_i}\chi^{{\cal Z}_i}_L  \rightarrow \frac{{\cal V}^{-\frac{25}{36}}\langle{\cal Z}_i\rangle }{M_P}\delta{\cal Z}_i{\bar\chi}^{\bar {{\cal Z}_i}}_L{\gamma}\cdot (p_{\chi^{{\cal Z}_i}}+p_{\bar \chi^{{\cal Z}_i}})\chi^{{\cal Z}_i}_L \sim {\cal V}^{-\frac{5}{3}}{\bar\chi}^{\bar {{\cal Z}_i}}_L\delta{{\cal Z}_i}\chi^{{\cal Z}_i}_L ; \nonumber\\
& &  g_{{\cal Z}_i{\bar {{\cal Z}_i}}}{\bar\chi}^{\bar {{\cal Z}_i}}_L\frac{1}{4}\left(\partial_{{\cal Z}_i}K{\gamma}\cdot {\cal Z}_i - {\rm c.c.}\right)\chi^{{\cal Z}_i}_L  \rightarrow \frac{{\cal V}^{-\frac{4}{3}}\langle{\cal Z}_i\rangle }{M_P}\delta{\cal Z}_i{\bar\chi}^{\bar {{\cal Z}_i}}_L{\gamma}\cdot (p_{\chi^{{\cal Z}_i}}+p_{\bar \chi^{{\cal Z}_i}})\chi^{{\cal Z}_i}_L\nonumber\\
& & \sim {\cal V}^{-\frac{83}{36}}{\bar\chi}^{\bar {{\cal Z}_i}}_L\delta{{\cal Z}_i}\chi^{{\cal Z}_i}_L.
\end{eqnarray}
Incorporating results of  (\ref{eq:massterm}) and (\ref{eq:kinetic}), the physical Higgsino-Higgsino- Higgs vertex will be given as:
\begin{eqnarray}
\label{eq:CHHh}
& & C^{\tilde {H}^0 \tilde {H}^0 H_i}= \frac{1}{{\sqrt{(\hat{K}_{{\cal Z}_i{\bar{\cal Z}}_i})^4}}}\left[({\cal V}^{-\frac{7}{4}})+ ({\cal V}^{-2}+{\cal V}^{-\frac{5}{3}} +{\cal V}^{-\frac{83}{36}})\right]\sim {\cal V}^{\frac{1}{4}}.
\end{eqnarray}
Here, $\tilde{H}^0 \sim \frac{\tilde{H}^{0}_u+\tilde{H}^{0}_d}{\sqrt{2}}$ corresponds to physical higgsino\footnote{Working in the sublocus, where position moduli ${z_1}$ and ${z_2}$ are considered to be equivalent, for notational simplification, we will write higgsino superfield  $\chi^{\frac{1}{\sqrt{2}}({{\cal Z}_1+ {\cal Z}_2})}\sim\chi^{{\cal Z}_i}$.}. The physical light Higgs is defined as $H_1=  \frac{{H}^{0}_{u}-{H}^{0}_{d}}{\sqrt{2}}$,and heavy Higgs as $H_2=  \frac{{H}^{0}_{u}+{H}^{0}_{d}}{\sqrt{2}}$. Henceforth, we will be using these notations in the calculations.

\underline{\bf{Gaugino-higgsino-Higgs vertex}}
\begin{equation}
{\cal L}=g_{YM}g_{B {\bar {\cal Z}_i}}X^B \bar \chi^{{\cal Z}_i}_L \lambda_L+ \partial_{{\cal Z}_i}T_B D^{B}\tilde{H}^i\lambda^{i}.
\end{equation}
 Expanding the same in the fluctuations linear in ${\delta Z}_i$, we have $
g_{YM}g_{T^B{\cal Z}^i}X^{B}= {\tilde f} {\cal V}^{-2}\delta{{\cal Z}_i} , (\partial_{{\cal Z}_i}T_B) D^{B}\sim   {\tilde f} {\cal V}^{-\frac{4}{3}}\delta{{\cal Z}_i}$, and the physical gaugino-higgsino-Higgs vertex works out to yield:
\begin{eqnarray}
\label{eq:CgHh}
& & C^{H_i{\tilde H}^0\lambda_L}=  \frac{{\cal V}^{-\frac{4}{3}}\tilde{f}}{{\left(\sqrt{\hat{K}_{{\cal Z}_i{\bar {\cal Z}}_i}}\right)^2}}\sim \tilde{f}\left({10}^{5}{\cal V}^{-\frac{4}{3}}\right) \sim \tilde{f}{\cal V}^{-\frac{1}{3}}.
\end{eqnarray}

\underline{\bf{Gaugino-gaugino-Higgs vertex}}
\begin{eqnarray}
& & {\cal L}= i{\bar {\lambda_L}}{\gamma}^m {\frac{1}{4}(K_{{\cal Z}_i}{\partial}_m {{\cal Z}_i}- c.c.)}{\lambda}_{L},
\end{eqnarray} where  ${\lambda}_{L}$ corresponds to gaugino. Here also, the aforementioned vertex  does not preserve $SU(2)_L$ symmetry - one has to obtain the term linear in $\langle z_i\rangle$. In terms of undiagonalized basis,
$\partial_{z_i}K\sim{\cal V}^{-\frac{2}{3}}\langle z_i\rangle$, and using
$\partial_{{\cal Z}_i}K\sim  {\cal O}(1) \partial_{z_i}K$, we have:
$\partial_{{\cal Z}_i}K\sim {\cal V}^{-\frac{2}{3}}\langle{\cal Z}_i\rangle$, incorporating the same
\begin{eqnarray}
\label{eq:Cggh1}
 & & C^{H_i{\bar {\lambda_L}}{\lambda}_{L}}:  \frac{{\cal V}^{-\frac{2}{3}}\langle{\cal Z}_i\rangle{\bar {\lambda_L}}\frac{\slashed{\partial}{{\cal Z}_i}}{M_P}{\lambda_L}}{{\sqrt{(\hat{K}_{{\cal Z}_1{\bar {\cal Z}}_1}})^2}}\sim {10}^{5} {\cal V}^{-\frac{23}{36}}h {\bar {\lambda_L}}\frac{\slashed{p}_h}{M_P}{\lambda_L}\sim {\cal V}^{-\frac{23}{36}}{h}{\bar {\lambda_L}}\frac{{\gamma}\cdot({p_{{\bar {\lambda_L}}} + p_{{\lambda}_{L}}})}{M_P}{\lambda}_{L} \nonumber\\
 & & \sim{10}^{5} {\cal V}^{-\frac{23}{36}}h{\bar {\lambda_L}}\frac{m_{\tilde g}}{M_P}{\lambda}_{L}\sim {\cal V}^{-\frac{35}{36}}h{\bar {\lambda_L}}{\lambda}_{L}
 \end{eqnarray}
 
\underline{\bf{Higgs-Higgs-Higgs vertex:}} The value of effective Higgs triple-interaction vertex in SM  is given by $ C^{H_i H_i H_i} \sim \frac{M^{2}_{H}}{v}$, v is electroweak VEV.

\underline{\bf{ Z boson-Higgs-Z boson vertex:}}
The effective vertex has been calculated in \cite{dhuria+misra_EDM} by using gauge kinetic term $Re(T_B) F^2$ and the value of the same has been shown to be same as given by Standard Model i.e $C^{ZZH_i} \sim \frac{M^{2}_{Z}}{v}$.

\underline{\bf{Higgsino-higgsino-Z boson vertex}}
\begin{eqnarray}
  & &  {\cal L}= g_{{\cal Z}^I{\cal Z}^J}{\bar\chi}^{{\cal Z}^I}\slashed{Z}\ {\rm Im}\left(X^BK + i D^B\right)\chi^{{\cal Z}^I},
\end{eqnarray}
$\chi^{z_1}$ corresponds to higgsino. Using values of $X^B$ and $D^B$ as mentioned earlier,  $g_{{\cal Z}_1{\bar {\cal Z}}_{\bar 1}}\sim{\cal V}^{-\frac{2}{3}}$ as from (\ref{eq:K}), yields value of physical higgsino-higgsino- Z boson vertex given by:
\begin{eqnarray}
 \label{eq:CHHZ}
& & C^{{\tilde H}^0 {\tilde H}^0 Z}\sim \frac{\left({\cal V}^{-\frac{23}{18}}\right)\tilde{f}}{\Bigl(\sqrt{\hat{K}_{{\cal Z}_1{\bar {\cal Z}}_1}}\Bigr)^2\sim O({10}^{-5})} \sim \left(O({10}^{5})\tilde{f}{\cal V}^{-\frac{23}{18}}\right)\sim \tilde{f}{\cal V}^{-\frac{5}{18}}.
 \end{eqnarray}
\underline{{\bf Gaugino-gaugino-Z boson vertex}}
\begin{eqnarray}
  & &  {\cal L}= {\bar \lambda^L}\slashed{Z}\ {\rm Im}\left(X^BK + i D^B\right){\lambda^L},\nonumber\\
& & \sim {\bar \lambda^L}{\gamma}\cdot A\left\{6\kappa_4^2\mu_72\pi\alpha^\prime Q_BK+\frac{12\kappa_4^2\mu_72\pi\alpha^\prime Q_Bv^B}{\cal V}\right\}{\lambda^L}
  \end{eqnarray}
   ${\lambda^L}$ corresponds to gaugino.
Again, using values of  $ v^B\sim{\cal V}^{\frac{1}{3}}$ and $Q_B\sim{\cal V}^{\frac{1}{3}}(2\pi\alpha^\prime)^2\tilde{f}$, yields value of physical gaugino-gaugino-Z boson vertex as:
\begin{eqnarray}
 \label{eq:CggZ}
& & C^{{\tilde \lambda^0}{\tilde \lambda^0} Z}\sim \left({\cal V}^{-\frac{11}{18}}\right)\tilde{f} \sim \left(\tilde{f}{\cal V}^{-\frac{11}{18}}\right).
 \end{eqnarray}
 \underline{{\bf Higgs-Fermion-Fermion Interaction}}
\begin{equation}
\label{eq:HFF}
{\cal L}\ni ig_{{\cal A}_I{\bar {{\cal A}_I}}}{\bar\chi}^{\bar {{\cal A}_I}}_L \left[\slashed{\partial}\chi^{{\cal A}_I}_L+\Gamma^{{\cal A}_I}_{{\cal Z}_1{{\cal A}_I}}\slashed{\partial} {{\cal Z}_i}\chi^{{\cal A}_I}_L
+\frac{1}{4}\left(\partial_{{\cal Z}_i}K\slashed{\partial} {\cal Z}_i - {\rm c.c.}\right)\chi^{{\cal A}_I}_L\right]
\end{equation}
$\chi^{{\cal A}_I}_L,I=1,2 $ corresponding to first  generation of L-handed leptons and quarks.
\\
\underline{Higgs-lepton-lepton interaction:} utilizing ${\cal Z}_1= \delta {\cal Z}_1 +{\cal V}^{\frac{1}{36}}{M_P}$,  $m_{\chi^{{\cal A}_1}_L} \sim {\cal O}(1) MeV $ and $m_{3/2}={\cal V}^{-2}{M_P}$, $g_{{\cal A} _1{\bar {{\cal A} _1}}} \ra \frac{{\cal V}^{-\frac{2}{9}}\langle z_i\rangle\delta z_i}{M_P^2}, \Gamma^{{\cal A} _1}_{{\cal Z}_i{{\cal A} _1}}\ra \frac{{\cal V}^{-\frac{2}{3}}{\langle z_i\rangle}}{M_P}, \partial_{{\cal Z}_i}K\ra\frac{{\cal V}^{-\frac{2}{3}}{\langle z_i\rangle}}{M_P}$, one gets:
\begin{eqnarray}
\label{eq:kinetic1}
& & g_{{\cal A}_1{\bar {{\cal A}_1}}}{\bar\chi}^{\bar {{\cal A}_1}}_L \slashed{\partial}\chi^{{\cal A}_1}_L \sim  \ra \frac{{\cal V}^{-\frac{2}{9}}\langle{\cal Z}_I\rangle}{M_P}\delta{\cal Z}_I{\bar\chi}^{\bar {{\cal A}_1}}_L\slashed{p}_{\chi^{{\cal A}_1}}\chi^{{\cal A}_1}_L\sim {\cal V}^{-\frac{7}{36}}{\bar\chi}^{\bar {{\cal A}_1}}_L\delta{{\cal Z}_i}\chi^{{\cal A}_1}_L; \nonumber\\
& & {\hskip -0.2in} g_{{\cal A}_1{\bar {{\cal A}_1}}}{\bar\chi}^{\bar {{\cal A}_1}}_L\Gamma^{{\cal A}_1}_{{\cal Z}_i{{\cal A}_1}}\slashed{\partial} {{\cal Z}_i}\chi^{{\cal A}_1}_L   \ra \frac{{\cal V}^{-\frac{2}{9}}\langle{\cal Z}_I\rangle }{M_P}\delta{\cal Z}_I{\bar\chi}^{\bar {{\cal A}_1}}_L (\slashed{p}_{\chi^{{\cal A}_1}} + \slashed{p}_{\bar \chi^{{\cal A}_1}})\chi^{{\cal A}_1}_L  \sim {\cal V}^{-\frac{7}{36}}{\bar\chi}^{\bar {{\cal A}_1}}_L\delta{{\cal Z}_i}\chi^{{\cal A}_1}_L; \nonumber\\
& & g_{{\cal A}_1{\bar {{\cal A}_1}}}{\bar\chi}^{\bar {{\cal A}_1}}_L\frac{1}{4}\left(\partial_{{\cal Z}_i}K{\gamma}\cdot {\cal Z}_i - {\rm c.c.}\right)\chi^{{\cal A}_1}_L    \ra \frac{{\cal V}^{-\frac{2}{9}}\langle{\cal Z}_i\rangle }{M_P}\delta{\cal Z}_i{\bar\chi}^{\bar {{\cal A}_1}}_L (\slashed{p}_{\chi^{{\cal A}_1}} + \slashed{p}_{\bar \chi^{{\cal A}_1}})\chi^{{\cal A}_1}_L\nonumber\\
& & \sim {\cal V}^{-\frac{7}{36}}{\bar\chi}^{\bar {{\cal A}_1}}_L\delta{{\cal Z}_I}\chi^{{\cal A}_1}_L.
\end{eqnarray}
Incorporating results of (\ref{eq:kinetic1}) in (\ref{eq:HFF}), the physical Higgs-lepton-lepton vertex is given as:
\begin{eqnarray}
\label{eq:llH}
& & C^{{{\cal A}_1}_L \bar {{\cal A}_1}_L H_i}= \frac{1}{{\sqrt{(\hat{K}_{{\cal Z}_i{\bar{\cal Z}}_i})^2.(\hat{K}_{{\cal A}_1{\bar{\cal A}}_1})^2}}}\left[{\cal V}^{-\frac{7}{36}}\right] \sim 10^{1} \left[ {\cal V}^{-\frac{7}{36}}\right]\sim {\cal O}(1).
\end{eqnarray}
\underline{Higgs-quark-quark interaction:} utilizing ${\cal Z}_1 = \delta {\cal Z}_1 +{\cal V}^{\frac{1}{36}}{M_P}$,   $m_{\chi^{{\cal A}_2}_L} \sim
{\cal O}(5) MeV $ and $m_{3/2}={\cal V}^{-2}{M_P}$, $g_{{\cal A}_2{\bar {{\cal A}_2}}} \ra \frac{{\cal V}^{-\frac{11}{9}}{{\cal Z}_i}\langle {\cal Z}_i\rangle}{M_P^2}, \Gamma^{{\cal A}_2}_{{\cal Z}_i{{\cal A}_2}}\ra \frac{{\cal V}^{-\frac{2}{3}}{\langle {\cal Z}_i\rangle}}{M_P}, \partial_{{\cal Z}_i}K\ra\frac{{\cal V}^{-\frac{2}{3}}{\langle z_i\rangle}}{M_P} $, one gets:
\begin{eqnarray}
\label{eq:kinetic2}
& &   g_{{\cal A}_2{\bar {{\cal A}_2}}}{\bar\chi}^{\bar {{\cal A}_2}}_L \slashed{\partial}\chi^{{\cal A}_2}_L \sim   \ra \frac{{\cal V}^{-\frac{11}{9}}\langle{\cal Z}_i\rangle }{M_P}\delta{\cal Z}_i{\bar\chi}^{\bar {{\cal A}_2}}_L\slashed{p}_{\chi^{{\cal A}_2}}\chi^{{\cal A}_2}_L\sim {\cal V}^{-\frac{43}{36}}{\bar\chi}^{\bar {{\cal A}_2}}_L\delta{{\cal Z}_i}\chi^{{\cal A}_2}_L; \nonumber\\
& & g_{{\cal A}_2{\bar {{\cal A}_2}}}{\bar\chi}^{\bar {{\cal A}_2}}_L\Gamma^{{\cal A}_2}_{{\cal Z}_i{{\cal A}_2}}\slashed{\partial} {{\cal Z}_i}\chi^{{\cal A}_2}_L \sim \ra \frac{{\cal V}^{-\frac{11}{9}}\langle{\cal Z}_I\rangle }{M_P}\delta{\cal Z}_I{\bar\chi}^{\bar {{\cal A}_2}}_L (\slashed{p}_{\chi^{{\cal A}_2}} + \slashed{p}_{\bar \chi^{{\cal A}_2}})\chi^{{\cal A}_2}_L \sim {\cal V}^{-\frac{43}{36}}{\bar\chi}^{\bar {{\cal A}_2}}_L\delta{{\cal Z}_i}\chi^{{\cal A}_2}_L; \nonumber\\
& & g_{{\cal A}_2{\bar {{\cal A}_2}}}{\bar\chi}^{\bar {{\cal A}_2}}_L.\frac{1}{4}\left(\partial_{{\cal Z}_i}K{\gamma}\cdot {\cal Z}_i - {\rm c.c.}\right)\chi^{{\cal A}_2}_L  \ra \frac{{\cal V}^{-\frac{11}{9}}\langle{\cal Z}_i\rangle }{M_P}\delta{\cal Z}_i{\bar\chi}^{\bar {{\cal A}_2}}_L (\slashed{p}_{\chi^{{\cal A}_2}} + \slashed{p}_{\bar \chi^{{\cal A}_2}})\chi^{{\cal A}_2}_L\nonumber\\
& & \sim  {\cal V}^{-\frac{43}{36}} {\bar\chi}^{\bar {{\cal A}_2}}_L\delta{{\cal Z}_i}\chi^{{\cal A}_2}_L.
\end{eqnarray}
\vskip -0.2in
{\hskip -0.2in} Incorporating results of  (\ref{eq:kinetic2}) in (\ref{eq:HFF}), the physical Higgs-quark-quark vertex is:\begin{eqnarray}
\label{eq:qqH}
& & C^{{{\cal A}_2}_L \bar {{\cal A}_2}_L H_i}= \frac{1}{{\sqrt{(\hat{K}_{{\cal Z}_i{\bar{\cal Z}}_i})^2.(\hat{K}_{{\cal A}_2{\bar{\cal A}}_2})^2}}}\left[{\cal V}^{-\frac{43}{36}}\right] \sim 10^{-7} \left[ {\cal V}^{-\frac{43}{36}} \right]\sim  {\cal V}^{-\frac{13}{5}} .
\end{eqnarray}
\begin{equation}
\label{Cffhi}
{\hskip -1.2in} {\rm We~would~estimate:}~C^{f \bar f  H_i}= {\rm Max}\left[C^{{{\cal A}_1}_L \bar {{\cal A}_1}_L H_i}, C^{{{\cal A}_2}_L \bar {{\cal A}_2}_L h}\right] \sim {\cal O}(1).
\end{equation}
\underline{{\bf Fermion-fermion-Z boson vertex}}
\begin{eqnarray}
  & &  {\hskip -0.2in} {\cal L}\ni g_{{\cal A}_I{\cal A}_I}{\bar\chi}^{{\cal A}_I}{\gamma}\cdot A\ {\rm Im}\left(X^BK + i D^B\right)\chi^{{\cal A}_I}, I={1,2}.
  \end{eqnarray}
$\chi^{{\cal A}_{1,2}}$  correspond to first  generation of leptons and quarks. Using values of $X^B$ and $D^B$
as mentioned earlier, $g_{{\cal A}_1{\bar {\cal A}}_{\bar 1}}\sim{\cal V}^{\frac{4}{9}}$, $g_{{\cal A}_2{\bar {\cal A}}_{ 2}}\sim{\cal V}^{-\frac{5}{9}}$ using equation (\ref{eq:K}), the contribution of physical fermion-fermion- Z boson vertex  $C^{f \bar f Z}\sim {\cal O}(1)$(see chapter {\bf 4}).

Using the physical eigenstates of neutralino mass matrix (eq. \ref{eq:neutralinos_i}), the interaction vertices involving neutralino's and Higgs will be given as:  
\begin{eqnarray}
\label{eq:chichih}
& & C^{\chi^0_3 \chi^0_1 H_i}=  C^{{\tilde \lambda^0} \tilde{H}^0 H_i}+  (\tilde{f}{\cal V}^{\frac{5}{6}}\frac{v}{M_P})C^{\tilde{H}^0 \tilde{H}^0 H_i},\nonumber\\
& & C^{\chi^0_3 \chi^0_2 H_i}=  (\tilde{f}{\cal V}^{\frac{5}{6}}\frac{v}{M_P})C^{{\tilde \lambda^0} {\tilde \lambda^0} H_i}+ (\tilde{f}^2{\cal V}^{\frac{5}{3}}\frac{v^2}{m^{2}_{pl}}) C^{{\tilde \lambda^0} \tilde{H}^0 H_i}+  (\tilde{f}{\cal V}^{\frac{5}{6}}\frac{v}{M_P})C^{\tilde{H}^0 \tilde{H}^0 H_i},\nonumber\\
& &  C^{\chi^0_3 \chi^0_3 h}=  C^{{\tilde \lambda^0} {\tilde \lambda^0} h}+  (\tilde{f}{\cal V}^{\frac{5}{6}}\frac{v}{M_P})  C^{{\tilde \lambda^0} \tilde{H}^0 H_i}+ (\tilde{f}^2{\cal V}^{\frac{5}{3}}\frac{v^2}{m^{2}_{pl}})  C^{\tilde{H}^0 \tilde{H}^0 H_i}.
\end{eqnarray}
 Now, using the set of results given in equation no (\ref{eq:CHHh}), (\ref{eq:CgHh}) and (\ref{eq:Cggh1}), the contribution of vertices appearing in equation (\ref{eq:chichih}) are as follows:
\begin{eqnarray}
\label{eq:Nchichih}
& & C^{\chi^0_3 \chi^0_1 H_i}= {\tilde f}{\cal V}^{-\frac{1}{3}}, C^{\chi^0_3 \chi^0_2 H_i}\sim  \tilde{f} {\cal V}^{\frac{13}{12}}\frac{v}{M_P}, C^{\chi^0_3 \chi^0_3 H_i}\sim {\cal V}^{-\frac{35}{36}}.
\end{eqnarray}
Similar, the interactions involving neutralino's and Z boson will be given by:
\begin{eqnarray}
\label{eq:chichiZ}
& & C^{\chi^0_3 \chi^0_1 Z}=   (\tilde{f}{\cal V}^{\frac{5}{6}}\frac{v}{M_P}) C^{\tilde{H}^0 \tilde{H}^0 Z}, C^{\chi^0_3 \chi^0_2 Z}=  (\tilde{f}{\cal V}^{\frac{5}{6}}\frac{v}{M_P}) ( C^{{\tilde \lambda^0} {\tilde \lambda^0} Z}+ C^{\tilde{H}^0 \tilde{H}^0 Z} ),\nonumber\\
& &  C^{\chi^0_3 \chi^0_3 Z}=  C^{{\tilde \lambda^0} {\tilde \lambda^0} Z}+  (\tilde{f}^2{\cal V}^{\frac{5}{3}}\frac{v^2}{m^{2}_{pl}})  C^{\tilde{H}^0 \tilde{H}^0 Z}
\end{eqnarray}
 Using set of results given in eqs. (\ref{eq:CHHZ}) and (\ref{eq:CggZ}), the contribution of vertices appearing in equation (\ref{eq:NchichiZ}) are as follows:
 \begin{eqnarray}
\label{eq:NchichiZ}
& & C^{\chi^0_3 \chi^0_1 Z}=   \tilde{f^2}{\cal V}^{\frac{5}{9}}\frac{v}{M_P},  C^{\chi^0_3 \chi^0_2 Z} \sim \tilde{f^2}{\cal V}^{\frac{5}{9}}\frac{v}{M_P}, C^{\chi^0_3 \chi^0_3 Z} \sim \tilde{f} {\cal V}^{-\frac{11}{18}}.
\end{eqnarray}
 Since now we have got the estimates of interaction vertices required to calculate partial wave coefficients for  ${{\chi}^0_3 {\chi}^0_3\rightarrow H_1 H_1}$, ${{\chi}^0_3 {\chi}^0_3\rightarrow ZZ}$ and ${{\chi}^0_3 {\chi}^0_3\rightarrow ff}$ annihilation processes,  we are in a position to  estimate the values of partial wave {\bf a} and {\bf b} coefficients sequentially just by using the form of analytical results provided in \cite{Takeshi_Leszek}.
 
{$~~\bf{{\chi}^0_3 {\chi}^0_3\rightarrow H_1 H_1}$}

In this case, we consider annihilation diagrams mediated via $s$--channel Higgs exchange and t-channel ${\chi}^{0}_i$ exchange. Using the following analytic expressions of $\widetilde{a}_{H_1 H_1}$ and $\widetilde{b}_{H_1 H_1}$ as given in \cite{Takeshi_Leszek}, we calculate the numerical values of the same.
 
 \underline{s-channel Higgs--boson($H_1,H_2$) exchange}
   \begin{eqnarray}
   \label{eq:ahh}
 &&  \widetilde{a}_{H_1 H_1}^{(H_1,H_2)}= 0,  \widetilde{b}_{H_1 H_1}^{(H_1,H_2)}=  \frac{3}{64\,\pi} \left |
     \sum_{r=H_1,H_2} \frac{C^{H_1 H_1r}\: C^{\chi^0_3\chi^0_3 r}}
       {4\, \mchi^{2}-m_{r}^{2}+i\, \Gamma_{r}\, m_{r}}
                 \right |^{2} .
    \end{eqnarray}
Utilizing the value of mass $m_{H_1} =125 GeV$ and $ m_{H_2}\sim {\cal V}^{\frac{59}{72}} m_{\frac{3}{2}}$ as calculated in chapter {\bf 2}, $m_{\chi^0_3}\sim {\cal V}^{\frac{2}{3}}m_{\frac{2}{3}}\sim {\cal V}^{-\frac{4}{3}}M_P$
and  $C^{H_1 H_1H_1}\sim 10^2 GeV, C^{\chi^0_3 \chi^0_3 H_1}\sim  C^{\chi^0_3 \chi^0_3 H_2}\sim  {\cal V}^{-\frac{35}{36}} $ from above and assuming $\Gamma_{H_1,H_2} < m_{H_1,H_2}$, after simplifying, we have
\begin{eqnarray}
\label{bhhh}
& & \widetilde{b}_{H_1 H_1}^{(H_1,H_2)} \sim \frac{3}{64\,\pi}\times O(10)^{-51}{GeV}^{-2}~{\rm for}~ {\cal V}\sim {10}^5.
\end{eqnarray}
 \underline{t-channel {neutralino\ ($\chi_{i}^{0}$) exchange:}}
    \begin{eqnarray}
  &&  \widetilde{a}_{H_1 H_1}^{(\chi^0)}= 0, \widetilde{b}_{H_1 H_1}^{(\chi^0)}= \frac{1}{16\,\pi}
      \sum_{i,j=1}^{3}  (C^{\chi_{i}^0 \chi^0_3 H_1})^2
        (C^{\chi_{j}^0 \chi^0_3, H_1 * })^2
         \frac{1}{\Delta_{hi}^{2}\,\Delta_{hj}^{2} } \nonumber
\\
       & & \times \Big[4 \,m_{\chi^0_3}^{2}\,(m_{\chi^0_3}^{2}-{m^{2}_{H_1}} )^{2}
           +4\,m_{\chi^0_3}\,(m_{\chi^0_3}^{2}-{m^{2}_{H_1}})\,
           (m_{\chi^0_3}+m_{\chi^0_i})\,\Delta_{hi} \nonumber \\
       & &\hspace{0.3in} +3\,(m_{\chi^0_3}+m_{\chi^0_i})\,(m_{\chi^0_3}+m_{\chi^0_j})\,
             \Delta_{hi}\,\Delta_{hj}\Big],
     \end{eqnarray}
where $\Delta_{hi}\equiv\,{m^{2}_{H_1}}-m_{\chi^0_3}^{2}-m_{\chi^0_{i}}^{2}$.
Utilizing the values of masses given above,
$\Delta_{h1}\equiv\,{m^{2}_{H_1}}-m^{2}_{\chi^0_3}-m^{2}_{\chi^0_{1}} \sim {\cal V}^2{m^2_\frac{3}{2}}$,
$\Delta_{h2}\equiv\,{m^{2}_{H_1}}-m^{2}_{\chi^0_3}-m^{2}_{\chi^0_{2}} \sim {\cal V}^2{m^2_\frac{3}{2}}$,
$\Delta_{h3}\equiv\,{m^{2}_{H_1}}-m^{2}_{\chi^0_3}-m^{2}_{\chi^0_{3}} \sim {\cal V}^{\frac{4}{3}}{m^2_\frac{3}{2}}$,
 and as shown in \cite{gravitino_DM},
  \begin{eqnarray}
  \label{bchihh}
{\hskip -0.5in} \widetilde{b}_{H_1H_1}^{(\chi^0)} \sim \frac{1}{16\,\pi} \times O(10)^{-42}{GeV}^{-2} ~{\rm for}~ {\cal V}\sim  {10}^5~{\rm and }~{\tilde f}\sim {10}^{-4}.
     \end{eqnarray}
 \underline{{Higgs ($H_1,H_2$)--neutralino\ ($\chi_{i}^{0}$) interference term:}}
   \begin{eqnarray}
  &&  \widetilde{a}_{H_1 H_1}^{(H_1,H_2-\chi^0)}=0, \  \nonumber\\
  && \widetilde{b}_{H_1H_1}^{(H_1,H_2-\chi^0)}= \frac{1}{16\,\pi}
      \sum_{i=1}^{3} Re \left[\sum_{r=H_1,H_2} \left(\frac{C^{H_1H_1r}\:
       C_{S}^{\chi\chi r}}{4\, m_{\chi^0_3}^{2}-m_{r}^{2}
            +i\, \Gamma_{r}\, m_{r}}\right)^{*}\, C_{S}^{\chi^0_{i} \chi\, {H_1}}
       C_{S}^{\chi^0_{i} \chi\, {H_1}} \right] \nonumber \\
       & & \times \frac{[2\,m_{\chi^0_3}\,(m_{\chi^0_3}^{2}-{m^2_{H_1}})
         +3\,(m_{\chi^0_3}+m_{\chi^0_3}i)\,\Delta_{hi}]}{\Delta_{hi}^{2}}.
     \end{eqnarray}
After simplifying as shown in \cite{gravitino_DM}, we get
      \begin{eqnarray}
       \label{bhchihh}
     \widetilde{b}_{H_1H_1}^{({H_1},{H_2}-\chi^0)}\sim \frac{1}{16\,\pi} \times O(10)^{-26}{GeV}^{-2} ~{\rm for}~ {\cal V}\sim {10}^5.
     \end{eqnarray}
Utilizing results of equations (\ref{bhhh}), (\ref{bchihh}), (\ref{bhchihh}),
\begin{eqnarray}
\label{sumbhh}
& & \widetilde{a}_{H_1H_1} =0, \widetilde{b}_{H_1H_1} = \widetilde{b}_{H_1H_1}^{(H_1,H_2)}
+\widetilde{b}_{H_1H_1}^{(\chi^{0})}
+\widetilde{b}_{H_1H_1}^{(H_1,H_2-\chi^{0})} \sim  O(10)^{-26}{GeV}^{-2}
\end{eqnarray}
 
 {$~~ \bf{{\chi}^0_3 {\chi}^0_3\rightarrow Z Z}$}
 
 In this case also, we consider annihilation diagrams mediated via $s$--channel Higgs exchange and t-channel ${\chi}^{0}_i$ exchange. Using the following analytic expressions of $\widetilde{a}_{Z Z}$ and $\widetilde{b}_{Z Z}$ as given in \cite{Takeshi_Leszek}, we obtain the numerical values of the same.

\underline{Higgs--boson $(H_1,H_2)$ exchange:}
   \begin{eqnarray}
   && \widetilde{a}_{ZZ}^{(H_1,H_2)} = 0, \widetilde{b}_{ZZ}^{(H_1,H_2)}= \frac{3}{64\,\pi}
      \left |\sum_{r=H_1,H_2} \frac{C^{ZZr}\:
       C^{\chi^0_3 \chi^0_3 r}}{ s-m_{r}^{2}+i\, \Gamma_{r}\, m_{r}} \right |^{2}
               \frac{3\,{m^{4}_Z}-4\,{m^{2}_Z}\,{m_{\chi^0_3}}^2+4\,{m^{4}_{\chi^0_3}}}{{m^{4}_Z}}.\nonumber\\
   \end{eqnarray}
 Using values of $m_{H_1} =125 GeV, m_{H_2}\sim {\cal V}^{-\frac{85}{72}} M_P$, $m_{\chi^0_3}\sim {\cal V}^{-\frac{4}{3}}M_P, m_{Z}\sim 90 GeV $; $C^{ZZH_i} \sim {\cal O}(10^2), C^{\chi^0_3 \chi^0_3 H_1}\sim   {\cal V}^{-\frac{35}{36}} $ from above; and further assuming $\Gamma_{H_{1,2}} < m_{H_{1,2}}$ in our case, one gets:
 \begin{eqnarray}
\label{eq:bhZZ}
& & \widetilde{b}_{ZZ}^{(H_1,H_2)} \sim  {\cal O}(10)^{-16}{GeV}^{-2}~{\rm for}~ {\cal V}\sim {10}^5.
\end{eqnarray}
\underline{Neutralino\ ($\chi_{i}^{0}$) exchange:}
   \begin{eqnarray}
   \label{eq:a_ZZchi0}
    \widetilde{a}_{ZZ}^{(\chi^0)} & = & \frac{1}{4\,\pi}\sum_{i,j=1}^{3}
    |C^{\chi^0_i \chi^0_3Z}|^2 |C^{\chi^0_j \chi^0_3Z}|^2
   \frac{({m^{2}_{\chi^0_3}}-{m^{2}_Z})}{\Delta_{Zi}\Delta_{Zj}},
\end{eqnarray}
where $\Delta_{Zi}\equiv\,{m^{2}_Z}-{m^{2}_{\chi^0_3}}-{m^{2}_{\chi^0_i}}$.
For the given neutralino mass eigenstates,
$\Delta_{Z1}\equiv\,{m^{2}_Z}-{m^{2}_{\chi^0_3}}-{m^{2}_{\chi^0_1}}\sim -{\cal V}^{2}m^2_{\frac{3}{2}}$, $\Delta_{Z2}\equiv\,{m^{2}_Z}-{m^{2}_{\chi^0_3}}-{m^{2}_{\chi^0_2}}\sim - {\cal V}^{2}m^2_{\frac{3}{2}}$, $\Delta_{Z3}\equiv\,{m^{2}_Z}-{m^{2}_{\chi^0_3}}-{m^{2}_{\chi^0_3}}\sim -2{\cal V}^{\frac{4}{3}}m^2_{\frac{3}{2}}$.
Using these values and values of interaction vertices as given in (\ref{eq:NchichiZ}), (\ref{eq:a_ZZchi0}) reduces to
\begin{eqnarray}
\label{eq:achiZZ}
  &&  \widetilde{a}_{ZZ}^{(\chi^0)} \sim \frac{1}{4\,\pi}{m^2_{\chi^0_3}}.\frac{{\tilde f}^4{\cal V}^{-\frac{22}{9}}}{{\cal V}^{\frac{8}{3}}m^4_{\frac{3}{2}}} \sim  {\cal O}(10)^{-54} {GeV}^{-2}.
\end{eqnarray}
As from \cite{Takeshi_Leszek}, the analytical expression of $\widetilde{b}_{ZZ}^{(\chi^0)}$ is defined as following:
\begin{eqnarray}
\label{eq:bZZ}
    \widetilde{b}_{ZZ}^{(\chi^0)} & = & \frac{1}{16\,\pi}
       \sum_{i,j=1}^{3} \frac{1}{{m_Z}^4\Delta_{Zi}^3 \Delta_{Zj}^3}  |C^{\chi^0_i \chi^0_3Z}|^2 |C^{\chi^0_j \chi^0_3Z}|^2 \, \nonumber\\
    &&   \times\Big[D_{ij}^{(1)} \Delta_{Zi}^2
       +D_{ij}^{(2)}\,\Delta_{Zi} \Delta_{Zj}
       +D_{ij}^{(3)}\,\Delta_{Zi}^2 \Delta_{Zj}
       +D_{ij}^{(4)}\,\Delta_{Zi}^2 \Delta_{Zj}^2\Big], 
          \end{eqnarray}
  where $D_{ij}$ are defined as well as evaluated in \cite{gravitino_DM}. Incorporating results of the same, values of $\Delta_{Z1,Z2,Z3}$ and interaction vertices as given in equation (\ref{eq:NchichiZ}) in equation (\ref{eq:bZZ}), the same reduces to:
\begin{eqnarray}
\label{eq:bchiZZ}
    \widetilde{b}_{ZZ}^{(\chi^0)}\sim  {\cal O}(10)^{-14} {GeV}^{-2}.
    \end{eqnarray}
\underline{Higgs $(H_1,H_2)$--neutralino\ ($\chi_{i}^{0}$) interference term:}
   \begin{eqnarray}
  & &   \widetilde{a}_{ZZ}^{(H_1,H_2-\chi^{0})} =  0, \nonumber
\\
  & &  \widetilde{b}_{ZZ}^{(H_1,H_2-\chi^{0})} =  \frac{1}{16\,\pi}
      \sum_{i=1}^{3} Re \left[\left(\sum_{r=H_1,H_2} \frac{C^{ZZr}\:
       C_{S}^{\chi^0_3 \chi^0_3 r}}
         { 4{m_{\chi^0_3}}^2-m_{r}^{2}+i\, \Gamma_{r}\, m_{r}} \right )^{*}\right]
       \frac{1}{{m_Z}^4\,\Delta_{Zi}^2}
     C_{S}^{\chi^0_3 \chi^0_3 r} \nonumber \\
 & & \times \bigg\{
      2\,{m_{\chi^0_3}}\,({m^{2}_{\chi^0_3}}-{m^{2}_Z})\,[-3\,{m^{4}_Z}
       -4\,{m^{3}_{\chi^0_3}}\,{m_{\chi^0_i}}+2\,{m^{2}_Z}\,{m_{\chi^0_3}}\,({m_{\chi^0_3}}+{m_{\chi^0_i}})]
            . \nonumber \\
 & &    + \Delta_{Zi}
      \left[ -4\,{m_{\chi^0_3}}^4\,(2\,{m_{\chi^0_3}}+3\,{m_{\chi^0_3}})
         +2\,{m^{2}_Z}\,{m_{\chi^0_3}}^2\,(5\,{m_{\chi^0_3}}+6\,{m_{\chi^0_i}})    -{m^{4}_Z}\,(5\,{m_{\chi^0_3}}+9\,{m_{\chi^0_i}}) \right] \bigg\}.
         {\hskip -0.4in}\nonumber\\
    \end{eqnarray}
Utilizing results for $\Delta_{Z\ 1,2,3}$ and (\ref{eq:NchichiZ}), value of $m_{H_1} =125 GeV, m_{H_2}\sim {\cal V}^{-\frac{85}{72}} M_P$, $m_{\chi^0_3}\sim {\cal V}^{-\frac{4}{3}}M_P, m_{Z}\sim 90 GeV $
and  $C^{ZZH_i} \sim 10^2,  C^{\chi^0_3 \chi^0_3 Z}\sim  {\cal V}^{-\frac{35}{36}} $ from above, after simplifying, we have:
\begin{eqnarray}
\label{eq:bhchiZZ}
& & \widetilde{b}_{ZZ}^{(h,H-\chi^{0})}\sim  \frac{1}{16\,\pi}\frac{{\cal V}^2 m^3_{\frac{3}{2}}}{m^4_Z}\left (\frac{10^2.{\cal V}^{-\frac{35}{18}}}{4.m_{\chi^0_3}^{2}}+ \frac{10^2. {\cal V}^{-\frac{35}{18}}}{m_{H}^{2}}\right)  \sim \frac{1}{16\,\pi}\frac{ m^3_{\frac{3}{2}}}{m^4_Z}\left (\frac{10^2}{4.m_{\chi^0_3}^{2}}\right )\nonumber\\
& & \sim  {\cal O}(10)^{-10}{GeV}^{-2}~{\rm for}~ {\cal V}\sim {10}^5.
\end{eqnarray}
Utilizing results of equations (\ref{eq:bhZZ}), (\ref{eq:achiZZ}), (\ref{eq:bchiZZ}) and (\ref{eq:bhchiZZ}),
\begin{eqnarray}
\label{eq:sumbZZ}
 \widetilde{b}_{ZZ} &=&
 \widetilde{b}_{ZZ}^{(h,H)}
+ \widetilde{b}_{ZZ}^{(\chi^0)}
+ \widetilde{b}_{ZZ}^{(h,H-\chi^{0})} \sim {\cal O}(10)^{-10}{GeV}^{-2};\nonumber\\
 \widetilde{a}_{ZZ} &=&
 \widetilde{a}_{ZZ}^{(h,H)}
+\widetilde{a}_{ZZ}^{(\chi^{0})}
+\widetilde{a}_{ZZ}^{(h,H-\chi^{0})}\sim  {\cal O}(10)^{-54} {GeV}^{-2}. \end{eqnarray}
 {$ ~~\bf{{\chi}^0_3 {\chi}^0_3\rightarrow f {\bar f}}$}
 
This channel involves annihilation diagrams mediated via the $s$-channel Higgs--boson ($H_1$, $H_2$)
and $Z$ boson exchange and the $t$- and $u$-channel sfermion
($\widetilde{f}_{a}$) exchange. Hence, one needs to evaluate the following partial wave coefficients:
\begin{eqnarray}
 \widetilde{a}_{\bar{f}f} &=&
 \widetilde{a}_{\bar{f}f}^{(h,H)}
+\widetilde{a}_{\bar{f}f}^{(Z)}
+\widetilde{a}_{\bar{f}f}^{(\widetilde{f})}
+\widetilde{a}_{\bar{f}f}^{(h,H-\widetilde{f})}
+\widetilde{a}_{\bar{f}f}^{(Z-\widetilde{f})}, \nonumber \\
 \widetilde{b}_{\bar{f}f} &=&
 \widetilde{b}_{\bar{f}f}^{(h,H)}
+\widetilde{b}_{\bar{f}f}^{(Z)}
+\widetilde{b}_{\bar{f}f}^{(\widetilde{f})}
+\widetilde{b}_{\bar{f}f}^{(h,H-\widetilde{f})}
+\widetilde{b}_{\bar{f}f}^{(Z-\widetilde{f})}.
\end{eqnarray}
 We estimate their values by using analytic expressions of the same as given in \cite{Takeshi_Leszek}.
 
\underline{Higgs--boson $(H_1,H_2)$ exchange:}
   \begin{eqnarray}
 &&  \widetilde{a}_{\bar{f}f}^{(H_1,H_2)}= 0,  \widetilde{b}_{\bar{f}f}^{(H_1,H_2)}= \frac{3}{4\, \pi}
     \left | \sum_{r=H_1,H_2} \frac{C^{ffr}\:
      C^{\chi^0_3 \chi^0_3 r}}{4\, {m^{2}_{\chi^0_3}}-m_{r}^{2}
        +i\, \Gamma_{r}\, m_{r}} \right |^{2} ({m^{2}_{\chi^0_3}}-m^{2}_{f}).
    \end{eqnarray}
    
Again, utilizing the value of masses as mentioned in earlier cases and  $C^{ffh}\sim{\cal O}(1), C^{\chi^0_3 \chi^0_3 h}\sim C^{\chi^0_3 \chi^0_3 H}\sim  {\cal V}^{-\frac{35}{36}} $ from above, after simplifying, we have:
\begin{eqnarray}
\label{eq:bhff}
& & \widetilde{b}_{ff}^{(h,H)}\sim {\cal O}(10)^{-46}GeV^{-2}~{\rm for}~ {\cal V}\sim {10}^5.
\end{eqnarray}
\underline{$Z$--boson exchange:}
  \begin{eqnarray}
      \widetilde{a}_{\bar{f}f}^{(Z)} &=& \frac{1}{2\, \pi}
       \left |  \frac{C^{ffZ}\: C_{A}^{\chi\chi Z}}
         {4 {m^{2}_{\chi^0_3}}-{m^{2}_Z}+i\, \Gamma_{Z}\, {m_Z}} \right |^{2}
       \frac{m^{2}_{f}\,({m^{2}_Z}-4\, {m^{2}_{\chi^0_3}})^{2}}{{m_Z}^{4}}.
       \end{eqnarray}
Utilizing the value of masses as mentioned in earlier cases; $C^{ffZ}\sim {\cal O}(1)$ and $C^{\chi^0_3 \chi^0_3 Z}\sim  {\cal V}^{-\frac{35}{36}} $ from above, after simplifying, we have
\begin{eqnarray}
\label{aZff}
& & \widetilde{a}_{{\bar f}f}^{(h,H)} \sim  O(10)^{-25}{GeV}^{-2}~{\rm for}~ {\cal V}\sim {10}^5;
\end{eqnarray}
\begin{eqnarray}
\label{eq:bZff1}
     \widetilde{b}_{\bar{f}f}^{(Z)} &=& \frac{1}{2\, \pi}
      \Bigl |  \frac{C^{ffZ}\: C^{\chi^0_3 \chi^0_3 Z}}
       {4\, {m^{2}_{\chi^0_3}}-{m^{2}_Z}+i\, \Gamma_{Z}\, {m_Z}} \Bigr |^{2}
        \frac{1}{{m^{2}_Z}\,((4\, {m^{2}_{\chi^0_3}}-{m^{2}_Z})^{2}
          +(\Gamma_{Z}\, {m^{2}_Z}))} \nonumber \\
      & &  \times \bigg[2 \,|C^{ffZ}|^{2}\, \Big\{{m^{2}_Z}\,
             ({m^{2}_{\chi^0_3}}-m^{2}_{f})\,({m^{2}_Z}-4 {m^{2}_{\chi^0_3}})^{2}  +\Gamma_{Z}^{2} [{m^{2}_{\chi^0_3}}\, {m^{4}_Z} + m^{2}_{f}\,
           (24\, {m^{4}_{\chi^0_3}}-\nonumber\\
           &&{\hskip -0.7in} 6\, {m^{2}_Z}\, {m^{2}_{\chi^0_3}}-{m^{4}_Z})] \Big\}
              +{m^{2}_Z}\, |C^{ffZ}|^{2}\,
        \Big\{(2\, {m^{2}_{\chi^0_3}}+m^{2}_{f})\,[(4\, {m^{2}_{\chi^0_3}}-{m^{2}_Z})^{2}
             + {m^{2}_Z}\,\Gamma_{Z}^{2}]\Big\} \bigg]. 
             \end{eqnarray}
As was shown in \cite{gravitino_DM} by incorporating values of relevant masses and interaction vertices $C^{ffZ}$ and $C^{\chi^0_3 \chi^0_3 Z}$, one gets:
             \begin{eqnarray}
             \label{eq:bZff}
  &&  \widetilde{b}_{\bar{f}f}^{(Z)} \sim {\cal O}(10)^{-34}{GeV}^{-2}~{\rm for}~ {\cal V}\sim {10}^5.
           \end{eqnarray}
 \underline{sfermion\ ($\widetilde{f}_{a}$) exchange:}
   \begin{eqnarray}
   \label{eq:atildefff}
    \widetilde{a}_{\bar{f}f}^{(\widetilde{f})} &=& \frac{1}{32\, \pi}
      \sum_{a,b}\frac{(m_{f}\,C_{+}^{a}
           +{m_{\chi^0_3}}\,D_{+}^{a})\,(m_{f}\,C_{+}^{b}
           +{m_{\chi^0_3}}\,D_{+}^{b})}
         {\Delta_{{\widetilde f}_{a}}\, \Delta_{{\widetilde f}_{b}}},
             \end{eqnarray}
a is the index for sfermion mass eigenstates so that $a= 1,..6$ for  squark and charged sleptons and $a=1,..3$ corresponds for sneutrino, and:
\begin{eqnarray}
C_{\pm}^{a} &=& |\Lambda_{fL}^{a}|^{2}\pm|\Lambda_{fR}^{a}|^{2},  D_{\pm}^{a} = \Lambda_{fL}^{a}(\Lambda_{fR}^{a})^{*}
                \pm(\Lambda_{fL}^{a})^{*}\Lambda_{fR}^{a};
\end{eqnarray}
$\Lambda_{fL}^{a}$ corresponds to the Neutralino-fermion-sfermion interactions mediated by L- handed
 squarks/sleptons and $\Lambda_{fR}^{a}$ corresponds to the Neutralino-fermion-fermion interactions
 mediated by R-handed squarks/sleptons. Using results from {\bf section 3}:
 $C^{\chi^0_3 {l_L}\tilde {l_L}}=\tilde{f} {\cal V}^{-\frac{1}{2}}$, $C^{{\chi^{0}_3} {u_L}\tilde {u_L}}\sim \tilde{f}{\cal V}^{-\frac{4}{5}}$.
 With exactly similar procedure, we find: $C^{\chi^0_3 {l_L}\tilde {l_R}}=\tilde{f} {\cal V}^{-\frac{12}{15}}, C^{{\chi^{0}_3} {u_L}\tilde {u_R}}\sim \tilde{f}{\cal V}^{-\frac{25}{36}} $.
Utilizing above
\begin{eqnarray}
\label{CaDa}
\sum_{a} C_{\pm}^{a}& = & Max \left({\left|C^{\chi^0_3 {l_L}\tilde {l_L}}\right|^2  \pm  \left|C^{\chi^0_3 {l_L}\tilde {l_R}} \right|^2}, {\left| C^{{\chi^{0}_3} {u_L}\tilde {u_L}}\right|^2  \pm  \left| C^{{\chi^{0}_3} {u_L}\tilde {u_R}}\right|^2}\right) \sim   \tilde{f^2} {\cal V}^{-1}, \nonumber\\
& & {\hskip -0.8in}\sum_{a}D_{\pm}^{a} = Max\left( C^{\chi^0_3 {l_L}\tilde {l_L}}(C^{\chi^0_3 {l_L}\tilde {l_R}})^{*},C^{{\chi^{0}_3} {u_L}\tilde {u_L}}(C^{{\chi^{0}_3} {u_L}\tilde {u_R}})^{*}\right) \pm  c.c.. \sim    \tilde{f^2}{\cal V}^{-\frac{13}{10}}.
\end{eqnarray}
and $\Delta_{{\widetilde f}_{a}}\equiv
m^{2}_{f}-{m^{2}_{\chi^0_3}}-m^{2}_{\widetilde{f}_{a}} \sim -{m^{2}_{\chi^0_3}}.$
Considering only  first two generations of squarks/sleptons and assuming the universality in scalar masses in both generations, equation (\ref{eq:atildefff}) reduces to the following simplified form:
\begin{eqnarray}
  \widetilde{a}_{\bar{f}f}^{(\widetilde{f})} &=& \frac{1}{ \pi}
      (\frac{m_{f}\, \tilde{f^2} {\cal V}^{-1}
           +{m_{\chi^0_3}}\,\tilde{f^2}{\cal V}^{-\frac{13}{10}})\,(m_{f}\,\tilde{f^2} {\cal V}^{-1}
           +{m_{\chi^0_3}}\,\tilde{f^2}{\cal V}^{-\frac{13}{10}})}
         {m_{\chi^0_3}^{4}} \nonumber\\
        & &  \sim  {10}^{-52}{GeV}^{-2}, {\rm for}~ {m_{\chi^0_3}}\sim {\cal V}^{-\frac{4}{3}}M_P ~{\rm and}~ {\cal V}\sim 10^5.
             \end{eqnarray}
Now, utilizing the numerical estimates of coupling summed over first two generation of squarks/sleptons given above, we will evaluate the value of $\widetilde{b}_{\bar{f}f}^{(\widetilde{f})}$. Using the form of expression from \cite{Takeshi_Leszek}, for ${m_{\chi^0_3}}\sim {\cal V}^{-\frac{4}{3}}M_P$ and $ \Delta_{{\widetilde f}_{a}}
 \sim -{m^{2}_{\chi^0_3}}$, after solving, one gets:
\begin{eqnarray}
\label{eq:btildefff}
 \widetilde{b}_{\bar{f}f}^{(\widetilde{f})} &=& \frac{1}{\pi}\times\frac{{\tilde f}^4 {\cal V}^{-2}}{m^2_{\chi^0_3}} \sim O(10)^{-48} {GeV}^{-2}.
\end{eqnarray}
\underline{Higgs $(H_1,H_2)$--sfermion\ ($\widetilde{f}_{a}$) interference term:}
   \begin{eqnarray}
    && \widetilde{a}_{\bar{f}f}^{(H_1,H_2-\widetilde{f})}= 0, \widetilde{b}_{\bar{f}f}^{(H_1,H_2-\widetilde{f})}= -\frac{1}{8\,\pi}\,
       \sum_{a} Re\,\left[ \sum_{r=H_1,H_2} \frac{C_{S}^{ffr}\:
          C_{S}^{\chi^{0}_3\chi^{0}_3 r}}{4\, {m^{2}_{\chi^0_3}}-m_{r}^{2}
          +i\, \Gamma_{r}\, m_{r}} \right]
            \nonumber \\
      & & \times \frac{({m^{2}_{\chi^0_3}}-m^{2}_{f})}{\Delta_{{\widetilde f}_{a}}^2} \left[C_{+}^{a}\,2\,m_{f}\,{m_{\chi^0_3}}
            + D_{+}^{a}\,(2\,{m^{2}_{\chi^0_3}}+3\,\Delta_{{\widetilde f}_{a}})
                  \right];
                  \end{eqnarray}
Simplifying the above by incorporating values of relevant masses and interaction vertices $C^{f fH_i}$ and $C^{\chi^{0}_3\chi^{0}_3 H_i}$ as given in equation (\ref{Cffhi}) and (\ref{eq:Nchichih}), one gets:
 \begin{eqnarray}
  \label{eq:bhfff}
& &  \widetilde{b}_{\bar{f}f}^{(H_1,H_2-\widetilde{f})} \sim  O(10)^{-47}{GeV}^{-2}~{\rm for}~ {\cal V}\sim {10}^5.
\end{eqnarray}
\underline {$Z$--sfermion\ ($\widetilde{f}_{a}$) interference term:}
    \begin{eqnarray}
     & &   \widetilde{a}_{\bar{f}f}^{(Z-\widetilde{f})}=-\frac{1}{4\, \pi}
        \sum_{a} Re \Bigl[  \frac{C^{ffZ}\:
           C^{\chi^0_3 \chi^0_3 Z}}{4\, {m^{2}_{\chi^0_3}}-{m^{2}_Z}
             +i\, \Gamma_{Z}\, {m_Z}} \Bigr]\frac{m_{f}\,
                 ({m^{2}_Z}-4\,{m^{2}_{\chi^0_3}})}{{m^{2}_Z}} \frac{(m_{f}\,C_{+}^{a}
           +{m_{\chi^0_3}}\,D_{+}^{a})}{\Delta_{{\widetilde f}_{a}}};\nonumber\\
\end{eqnarray}
Utilizing the value of $m_{Z} =90 GeV$, $m_{\chi^0_3}\sim {\cal V}^{\frac{2}{3}}m_{\frac{3}{2}}\sim {\cal V}^{-\frac{4}{3}}M_P$;  $C^{ffZ}\sim {\cal O}(1), C^{\chi^0_3 \chi^0_3 Z}\sim {\tilde f} {\cal V}^{-\frac{11}{18}} $ and equation (\ref{CaDa}), after simplifying, we have
\begin{eqnarray}
\label{eq:afZff}
\label{afZff}
 && \widetilde{a}_{\bar{f}f}^{(Z-\widetilde{f})}=\frac{1}{4\, \pi}
  \Bigl[  \frac{ 
           {\tilde f} {\cal V}^{-\frac{11}{18}} }{4\, {m^{2}_{\chi^0_3}}} \Bigr] \frac{m_{f}\,
                 ({m^{2}_Z}-4\,{m^{2}_{\chi^0_3}})}{{m^{2}_Z}}  \frac{(m_{f}\,C_{+}^{a}
           +{m_{\chi^0_3}}\,D_{+}^{a})}{\Delta_{{\widetilde f}_{a}}} \sim  {\cal O}(10)^{-38} GeV^{-2}.\nonumber\\
           \\
     & &  \widetilde{b}_{\bar{f}f}^{(Z-\widetilde{f})} =-\frac{1}{8\, \pi}
        \sum_{a} Re \biggl[ \left( \frac{ C^{\chi^0_3 \chi^0_3 Z}}
          {(4\, {m^{2}_{\chi^0_3}}-{m^{2}_Z}+i\, \Gamma_{Z}\, {m_Z})^{2}} \right) \,
           \frac{1}{{m^{2}_Z}\, \Delta_{{\widetilde f}_{a}}^{3}} 
             \nonumber \\
             \pagebreak
      & & \times \bigg [C^{ffZ}\,C_{-}^{a}\,
           \Big\{2\,{m^{2}_Z}\,P_Z\,\Delta_{{\widetilde f}_{a}}
            [2\,{m^{2}_{\chi^0_3}}\,({m^{2}_{\chi^0_3}}+\Delta_{{\widetilde f}_{a}})
             +m^{2}_{f}\,(-2\,{m^{2}_{\chi^0_3}}+\Delta_{{\widetilde f}_{a}})]\,\Big\}
          \nonumber \\
      & &  +\,C^{ffZ}\Big\{C_{+}^{a}
         \Big[2\,m^{2}_{f}\,{m^{2}_{\chi^0_3}}\,({m_{\chi^0_3}}^2-m^{2}_{f})\,
            ({m^{2}_Z}-4\,{m^{2}_{\chi^0_3}})\,P_Z \, {m^{2}_{\chi^0_3}} [m^{2}_{f} {m^{2}_Z}
         + \nonumber\\
         &&\,2\, {m^{2}_{\chi^0_3}}\,({m^{2}_Z}-6\,m^{2}_{f})]\,P_Z\,
         \Delta_{{\widetilde f}_{a}} + \,2\,{m_Z}\,\{-{m_Z}\,({m_{\chi^0_3}}^2-m^{2}_{f})\,
         ({m^{2}_Z}-4\,{m^{2}_{\chi^0_3}})  \nonumber\\
         &&+\,i\,\Gamma_{Z}\,[{m^{2}_Z}\,{m^{2}_{\chi^0_3}}
         -m^{2}_{f}({m^{2}_Z}+3\,{m^{2}_{\chi^0_3}})]\}\,
            \Delta_{{\widetilde f}_{a}}^{2}\Big]  + m_{f}\, {m_{\chi^0_3}}\,D_{+}^{a}\Big[4\,{m^{2}_{\chi^0_3}}\,
         ({m^{2}_{\chi^0_3}}-m^{2}_{f})\, \nonumber\\
        &&({m^{2}_Z}-4\,{m^{2}_{\chi^0_3}})\,P_Z+2\,[6\,{m^{2}_Z}\,{m^{2}_{\chi^0_3}}
        -16\ {m^{4}_{\chi^0_3}} -m^{2}_{f}\,(3\,{m^{2}_Z}
        -4\,{m^{2}_{\chi^0_3}})]\,P_Z\,\Delta_{{\widetilde f}_{a}}\nonumber\\
        && -3\,[({m^{2}_Z}-4\,{m^{2}_{\chi^0_3}})^{2}
        -i\, {m_Z}\,\Gamma_{Z}\,({m^{2}_Z}-8\,{m^{2}_{\chi^0_3}})]\,
         \Delta_{{\widetilde f}_{a}}^{2}\Big]\bigg\}\bigg] \biggr];  
\end{eqnarray}
where $P_Z\equiv 4\,{m^{2}_{\chi^0_3}}-{m^{2}_Z}+i\, \Gamma_{Z}\, {m_Z}$.
Again using the numerical values of masses and relevant couplings and assuming $\Gamma_{Z}{m_Z}<< {m^{2}_{\chi^0_3}}$, the above expression reduces to
\begin{eqnarray}
\label{eq:bfZff}
\widetilde{b}_{\bar{f}f}^{(Z-\widetilde{f})} &=& {\cal O}(10)^{-46} {GeV}^{-2}.
\end{eqnarray}
Utilizing the results from eqs. (\ref{eq:bhff}), (\ref{eq:bZff}), (\ref{eq:btildefff}), (\ref{eq:bhfff}), (\ref{eq:bfZff}),  (\ref{aZff}) and (\ref{eq:afZff}):
\begin{eqnarray}
\label{eq:sumbZZ}
 &&\widetilde{b}_{\bar{f}f} =
 \widetilde{b}_{\bar{f}f}^{(h,H)}
+ \widetilde{b}_{\bar{f}f}^{(Z)} +
\widetilde{b}_{\bar{f}f}^{(\widetilde{f})}+
+ \widetilde{b}_{\bar{f}f}^{(h,H-\widetilde{f})}+
\widetilde{b}_{\bar{f}f}^{(Z-\widetilde{f})}
\sim  O(10)^{-34}{GeV}^{-2};\nonumber\\
&& \widetilde{a}_{\bar{f}f} =
 \widetilde{a}_{\bar{f}f}^{(h,H)}+
\widetilde{a}_{\bar{f}f}^{(Z)} +
\widetilde{a}_{\bar{f}f}^{(\widetilde{f})}+
+\widetilde{a}_{\bar{f}f}^{(h,H-\widetilde{f})}+
\widetilde{a}_{\bar{f}f}^{(Z-\widetilde{f})}\sim   {\cal O}(10)^{-25} {GeV}^{-2}. \end{eqnarray}
Relative velocity $v_{f_1 f_2}$ is defined as $
v_{f_1 f_2}\equiv
\Bigl[1-\frac{(m_{f_1}+m_{f_2})^2}{4m^2_{\chi^0_3}}\Bigr]^{1/2}
\Bigl[1-\frac{(m_{f_1}-m_{f_2})^2}{4m^2_{\chi^0_3}}\Bigr]^{1/2}$. For $f_1, f_2 = hh, v_{h h} \equiv 1$; $ f_1, f_2 = ZZ, v_{ZZ} \equiv 1$; $ f_1, f_2 = f{\bar f}, v_{f{\bar f}} \equiv 1$.
Having estimated the partial wave coefficients for each possible annihilation processes, and summing up their contribution as according to (\ref{eq:abf1f2}):
\begin{eqnarray}
a  &=&
\widetilde{a}_{hh}+ \widetilde{a}_{ZZ}+ \widetilde{a}_{f {\bar f}} \equiv  O(10)^{-25} {GeV}^{-2} \nonumber \\
b  &=& \widetilde{b}_{hh}+ \widetilde{b}_{ZZ}+ \widetilde{b}_{f {\bar f}} \equiv  O(10)^{-10} {GeV}^{-2}.
\end{eqnarray}
\begin{equation}
{\rm Now}~~\jxf\equiv \int_0^{x_f}dx \langle\sigma v_{\rm M\o l}\rangle(x)= \int_0^{x_f}dx  (a + b{x_f})= a{x_f} + b\frac{{x^2_f}}{2},
\end{equation}
where
$x=T/m_\chi$. The value of $x_f$ is given by by solving iteratively the equation $
x_f^{-1} = \ln \left( \frac{m_\chi}{2 \pi^3} \sqrt{\frac{45}{2g_* G_N}}
\langle\sigma v_{\rm M\o l}\rangle({x_f})\, x_f^{1/2} \right)$,
where $g_*$ represents the effective
number of degrees of freedom at freeze-out ($\sqrt{g_*}\simeq 9$).
Solving this, $x_f$ comes out to be around $1/33$. The relic abundance  is given as \cite{Jame_D_wells}:
\begin{eqnarray}
\label{eq:relic-density}
\Omega_\chi h^2 & = & \frac{1}{{\mu}^2 \sqrt{g_*}J({x_F})}.
\end{eqnarray}
where ${\mu}= 1.2 \times 10^5 GeV$.
For $J(x_f) \sim  10^{-10}\frac{{x_f}^2}{2}{GeV}^{-2}$, $x_f= \frac{1}{33}$ and $ \sqrt{g_*}=9$,
\begin{eqnarray}
\Omega_{\chi^0_3}h^2\sim\frac{2\cdot(33)^2}{1.44 \times 10^{10}\cdot 9\cdot 10^{-10}}\equiv {\cal O}(10^2).
\end{eqnarray}
For $m_{\frac{3}{2}}\sim {\cal V}^{-2} M_P\sim 10^8 GeV$ and $m_{\chi^0_3}\sim {\cal V}^{\frac{2}{3}}m_{\frac{3}{2}}\sim 10^{11} GeV$ as given in Table~2.1, relic abundance of gravitino is given as :
\begin{eqnarray}
\label{eq:omega1}
\Omega_{\tilde G}h^2& = & \Omega_{\chi^0_3}h^2\times \frac{m_{\frac{3}{2}}}{m_{\chi^0_3}}=\Omega_{\chi^0_3}\times {\cal V}^{-\frac{2}{3}}  \sim 0.1,
\end{eqnarray}
clearly a very desirable value!
\subsection{Slepton Relic Density Calculations}
For the case of slepton NLSPs, the dominant annihilation channel possible in our set up are: ${\slp{a}\slp{b}^*\rightarrow ZZ}$, ${\slp{a}\slp{b}^*\rightarrow Z H_1}$,${\slp{a}\slp{b}^*\rightarrow  H_1 H_1}$, ${\slp{a}\slp{b}^*\rightarrow \gm\gm}$, ${\slp{a}\slp{b}^*\rightarrow \gm  H_1}$, ${\slp{a}\slp{b}^*\rightarrow ll}$. The  analytical expressions for  $\widetilde{w}(s)$  are given  in \cite{takeshi_coannihilation}.
Once again, the approach is to first calculate required vertices in the context of ${\cal N}=1$ gauged supergravity and then utilize the same to calculate partial wave coefficients.

\underline{\bf{Slepton-slepton-Higgs vertex}}

By expanding ${\cal N}=1$ supergravity potential  $V= e^K G^{T_S T_S}|D_{T_S}W|^2$ in the fluctuations around  ${\cal Z}_i \rightarrow {\cal Z}_i +{\cal V}^{\frac{1}{36}}M_P$ , ${\cal A}_1 \rightarrow {\cal A}_1 +{\cal V}^{-\frac{2}{9}}M_P$,  contribution of term quadratic in ${\cal A}_1$ as well as ${\cal Z}_i$ is of the order $ {\cal V}^{\frac{-89}{36}}\langle{\cal Z}_i\rangle $, which after giving VEV to  one of the ${\cal Z}_i$, will be given as:
\begin{equation}
\label{eq:Clalbh}
 C^{\slp{a} \slp{b} H_1}\sim \frac{1}{{\sqrt{(\hat{K}_{{\cal Z}_i{\bar{\cal Z}}_i})^2 (\hat{K}_{{\cal A}_1{\bar{\cal A}}_1})^2}}}\left[{\cal V}^{-\frac{89}{36}}\langle{{\cal Z}^i}\rangle{{\cal \bar {Z}}^i}{ {\cal A}^1}{ {\cal A}^{*1}}\right]\sim O({\cal V}^{-\frac{34}{15}}).
\end{equation}
\begin{equation}
\label{eq:Clalbh1}
 {\rm Similarly},~C^{\slp{a} \slp{b} H_1 H_1 }\sim \frac{1}{{\sqrt{(\hat{K}_{{\cal Z}_i{\bar{\cal Z}}_i})^2 (\hat{K}_{{\cal A}_1{\bar{\cal A}}_1})^2}}}\left[{\cal V}^{-\frac{89}{36}}{{\cal Z}^i}{{\cal \bar{Z}}^j}{ {\cal A}^1}{ {\cal A}^{*1}}\right]\sim O({\cal V}^{-\frac{41}{18}}).
\end{equation}
\underline{\bf{Slepton-slepton-Z boson-Z boson vertex}}

In the context of supergravity action, contribution of required vertex will be given by:
${\bar{\partial}}_{{\bar {\cal A}}_{1}}\partial_{{\cal A}_1}G_{T_B{\bar T}_B} X^{T_B} X^{\bar {T_B}} {A}^{\mu} A_{\nu}$. Using equation (\ref{eq:K}), one gets $
{\bar{\partial}}_{{\bar {\cal A}}_{1}}\partial_{{\cal A}_1}G_{T_B{\bar T}_B}\sim {\cal O}(1) {\bar{\partial}}_{{\bar a}_{\bar 1}}\partial_{a_1}G_{T_B{\bar T}_B}\sim {\cal V}^{-\frac{8}{9}}{\cal A}^{*}_{1}{\cal A}_{1}$.
 Incorporating value of $X^B$ as mentioned earlier in this chapter, the physical slepton-slepton-Z boson- Z boson vertex is proportional to
\begin{equation}
\label{eq:CllZZ}
C^{\slp{a} \slp{b}^{*}ZZ}\sim \frac{{\cal V}^{-\frac{8}{9}}{\tilde f}^2 {\cal V}^{-\frac{4}{3}}}{\sqrt{(K_{A_1 A_1})^2}}\sim \frac{{\tilde f}^2 {\cal V}^{-\frac{20}{9}}}{O(10)^{4}}\sim {\tilde f}^2 {\cal V}^{-3}.
\end{equation}
\underline{\bf{Slepton-slepton-Z boson vertex}}

The gauge kinetic term for slepton-slepton-Z boson vertex will be given by the term: $ \frac{e^K}{\kappa_4^2}G^{T_B\bar{T}_B}\tilde{\bigtriangledown}_\mu T_B\tilde{\bigtriangledown}^\mu {\bar T}_{\bar B}$. This implies that the following term generates the required slepton-slepton-gauge boson vertex i.e:,
\begin{eqnarray}
\label{eq:sq sq gl}
& & C^{\slp{a} \slp{b}^{*}ZZ} \sim {\frac{6i\kappa_4^2\mu_72\pi\alpha^\prime Q_BG^{T_B{\bar T}_B}}{\kappa_4^2{\cal V}^2}\kappa_4^2}A^\mu\partial_\mu\left(\kappa_4^2\mu_7(2\pi\alpha^\prime)^2C_{1{\bar 1}}{\cal A}_1{\bar {\cal A}}_{\bar 1}\right)\nonumber\\
&&\xrightarrow[{\small \kappa_4^2\mu_7(2\pi\alpha^\prime)^2C_{1{\bar 1}}\sim{\cal V}^{\frac{10}{9}}}]{\small G^{T_B{\bar T}_B}\sim{\cal V}^{\frac{7}{3}},}\frac{{\cal V}^{\frac{7}{9}}}{\left(\sqrt{\hat{K}_{{\cal A}_1{\bar {\cal A}}_1}}\right)^2} \sim \tilde{f} {\cal V}, ~{\rm for~{\cal V}\sim {10}^5}.
\end{eqnarray}
${~~\bf {\slp{a}\slp{b}^*\rightarrow ZZ}}$

In this case, we consider annihilation diagrams of leptons mediated via $s$--channel Higgs exchange and t-channel $\slp{c}$ exchange.  Quoting directly the analytical expressions
\begin{eqnarray}
\label{eq:wZZ}
 {\widetilde{w}}_{\slp{a}\slp{b}^*\rightarrow ZZ}=
 {\widetilde{w}}_{ZZ}^{(H_1,H_2,P)} + {\widetilde{w}}_{ZZ}^{(\slp{})} + {\widetilde{w}}_{ZZ}^{(H_1,H_2,P-\slp{})}.
\end{eqnarray}
 \underline{Higgs ($H_1,H_2$) exchange ($+$ Point interaction):}
\begin{eqnarray}
\pagebreak
{\widetilde{w}}_{ZZ}^{(H_1,H_2,P)} & = &
\left|
\sum_{r=H_1,H_2} \frac{C^{ZZr}C^{\slp{b}^*\slp{a}r}}{\prop{H_1}}
-C^{\slp{b}^*\slp{a}ZZ}
\right|^2
\frac{s^2-4 \mz^2 s+12\mz^4}{8\mz^4}\nonumber\\
&& {\hskip -1.3in }\sim
\left|
 \frac{(10^2 V^{-\frac{34}{15}})M_P}{\prop{H_1}}+ \frac{(10^2 V^{-\frac{34}{15}})M_P}{\prop{H_2}}
-{\tilde f}^2 {\cal V}^{-3}
\right|^2
\frac{s^2-4 \mz^2 s+12\mz^4}{8\mz^4}.
\end{eqnarray}
 \underline{slepton\ ($\slp{c}$) exchange:}
\begin{eqnarray}
\label{eq:w_ZZ_sl}
& & {\widetilde{w}}_{ZZ}^{(\slp{})}  =
\frac{1}{\mz^4}\sum_{c,d=1}^{2}
C^{\slp{b}^*\slp{c}Z} C^{\slp{c}^*\slp{a}Z}
C^{\slp{b}^*\slp{d}Z *} C^{\slp{d}^*\slp{a}Z *} \nonumber \\
 & & {\hskip -0.1in} \times  \left[
       \T_4 - 2(\slpsl{2}+2\mz^2) \T_3 \right.  +[ \slpsl{4} +4\slsl{2} + 2\mz^2(\slpsl{2}+3\mz^2) ]\T_2 \nonumber \\
 & & -2[ (\slpsl{2})(\slsl{2}-\mz^4)
         + \mz^2(\slpsl{4} -4\slsl{2} +2\mz^4) ]\T_1 \nonumber \\
 & & {\hskip -0.1in}+(\msli^2-\mz^2)^2(\mslj^2-\mz^2)^2\T_0  -\Y_4
      +[ s(\slpsl{2}-2\mz^2)
               -2(\msli^2-\mz^2)(\mslj^2-\mz^2)  ] \nonumber\\
               & & \hskip -0.1in \Y_2-[ s^2(\msli^2-\mz^2)(\mslj^2-\mz^2)
  +s\{ -\slsl{2}(\slpsl{2})+3\mz^2(\slpsl{4})\nonumber\\
  &&-3\mz^4(\slpsl{2})+2\mz^6 \} \left.
 + (\msli^2-\mz^2)^2(\mslj^2-\mz^2)^2
]\Y_0
\right].
\end{eqnarray}
Considering only first-generation squarks, $
\sum_{c,d=1}^{2}
C^{\slp{b}^*\slp{c}Z} C^{\slp{c}^*\slp{a}Z}
C^{\slp{b}^*\slp{d}Z *} C^{\slp{d}^*\slp{a}Z *}\sim \tilde{f^4}{\cal V}^4$. Hence equation (\ref{eq:w_ZZ_sl}) yields:
\begin{eqnarray}
{\widetilde{w}}_{ZZ}^{(\slp{})} & = &
\frac{{f^4}{\cal V}^4}{\mz^4} \bigg[
       \T_4 - 4 \msli^2\T_3 +  6 \msli^4 \T_2  -4 \msli^6 \T_1 \ + \msli^8 \T_0  -\Y_4
      +[2 s\msli^2 -2 \msli^4 ]\Y_2 -  \nonumber\\
      && [ s^2 \msli^4  - 2s\msli^6 + \msli^8 ]\Y_0 \bigg];
\end{eqnarray}
where $\T_4, \T_3,\Y_0...$ etc are evaluated in \cite{gravitino_DM}.

 \underline{Higgs ($H_1,H_2$) ($+$ Point) -- slepton\ ($\slp{c}$) interference:}
\begin{eqnarray}
&&{\widetilde{w}}_{ZZ}^{(H_1,H_2,P-\slp{})}=
\frac{1}{2\mz^4}\sum_{c=1}^2 \Re
\left[ \left( \sum_{r=H_1,H_2}
\frac{C^{ZZr}C^{\slp{b}^*\slp{a}r}}{\prop{r}}-C^{\slp{b}^*\slp{a}ZZ}
\right)^* C^{\slp{b}^*\slp{c}Z} C^{\slp{c}^*\slp{a}Z} \right]
\nonumber \\
 & &  \times \Bigl[
s^2 + s (\slpsl{2}-2\mslk^2-4\mz^2)
+ 2\mz^2(\slpsl{2}-2\mslk^2+2\mz^2) \nonumber \\
 & &
-2 [ s(\msli^2-\mslk^2-\mz^2)(\mslj^2-\mslk^2-\mz^2) +2\mz^2 \{ \slpsl{4}-\slsl{2}
\nonumber \\
 & & +\mslk^2(\mslk^2-\msli^2-\mslj^2-2\mz^2)  -\mz^2 (\slpsl{2}-\mz^2) \} ]\F \Bigr].
\end{eqnarray}
Using the universality in squark masses, it leads to:
\pagebreak
\begin{eqnarray}
&& {\widetilde{w}}_{ZZ}^{(H_1,H_2,P-\slp{})} \equiv
\frac{1}{2\mz^4} \Re
\left[ \left( \sum_{r=H_1,H_2}
\frac{(10^2 {\cal V}^{-\frac{34}{15}}M_P)}{\prop{r}}-{\tilde f}^2 {\cal V}^{-3}
\right)^* \tilde{f^2}{\cal V}^2 \right]
\nonumber \\
 & & \times \left[
(s^2 -4s \mz^2 + 4 \mz^4) -2 (s\mz^4 -8 \mz^4\msli^2)\F  \right]
\end{eqnarray}
where $ \F\equiv \frac{\log \left(\frac{s-2 {\cal V}^{\frac{4}{5}}-\sqrt{s-4 {\cal V}^{\frac{4}{5}}} \sqrt{s-4 {\cal V}^{\frac{21}{5}}}}{s-2 {\cal V}^{\frac{4}{5}}+\sqrt{s-4 {\cal V}^{\frac{4}{5}}} \sqrt{s-4 {\cal V}^{\frac{21}{5}}}}\right)}{\sqrt{s-4 {\cal V}^{\frac{4}{5}}} \sqrt{s-4
   {\cal V}^{\frac{21}{5}}}}$. Summing up the contribution of ${\widetilde{w}}$ for all s, t and u-channels as according to equation (\ref{eq:wZZ}):
\begin{eqnarray}
\widetilde{w}_{ZZ}|_{s=(4 \msli^2)}\sim {\cal V}^{\frac{34}{5}}, \widetilde{w}^{\prime}_{ZZ}|_{s=(4 \msli^2)}\sim {\cal V}^{\frac{13}{5}}.
\end{eqnarray}
The ``reduced'' coefficients $\widetilde{a}_{ZZ}$ and
$\widetilde{b}_{ZZ}$ will be given by 
\begin{eqnarray}
\label{eq:aZZ}
&& \widetilde{a}_{ZZ}  \sim  \frac{1}{32 \pi \msli^2} {\cal V}^{\frac{34}{5}}\sim 10^{11} {GeV}^{-2};
\widetilde{b}_{ZZ} \sim  \frac{3}{64 \pi \msli^2}{\cal V}^{\frac{13}{5}}\sim  10^{-10} {GeV}^{-2}.
\end{eqnarray}
${~~\bf  {\slp{a}\slp{b}^*\rightarrow Z H_1}}$

The process ${\slp{a}\slp{b}^*\rightarrow Z H_1}$ involves
the $s$--channel Z boson exchange,and the $t$-- and $u$--channel slepton ($\slp{a}$, $a=1,2$) exchange:
\begin{eqnarray}
\label{eq:wzh}
 {\widetilde{w}}_{\slp{a}\slp{b}^*\rightarrow Z H_1}=
  {\widetilde{w}}_{Z H_1}^{(Z)} + {\widetilde{w}}_{Z H_1}^{(\slp{})}
+  {\widetilde{w}}_{Z H_1}^{(Z-\slp{})}.
\end{eqnarray}
 \underline{$Z$ exchange:}
\begin{eqnarray}
& &  {\widetilde{w}}_{Z H_1}^{(Z)}  =
\frac{1}{12\mz^6}
\left| \frac{C^{ZZ H_1}C^{\slp{b}^*\slp{a}Z}}{\prop{Z}} \right|^2\nonumber\\
 & & \times\Bigl[ s^2 \Big\{ 3(\slmsl{2})^2+\mz^4 \Big\} -2s \Big\{ 3(\slmsl{2})^2(\mh^2+2\mz^2)\nonumber\\
 && + \mz^4(\slpsl{2}+\mh^2-5\mz^2) \Big\}  +\mz^4 \Big\{ (\mh^2-\mz^2)^2 +4(\slpsl{2})(\mh^2-5\mz^2) \Big\}
 \nonumber\\
 &&+(\slmsl{2})^2(3\mh^4+6\mz^2\mh^2+19\mz^4) - \frac{2}{s}\mz^2 \Big\{ (\slmsl{2})^2(3\mh^4-2\mz^2\mh^2+\mz^4)
      \nonumber\\
      && +\mz^2(\mh^2-\mz^2)^2(\slpsl{2}) \Big\}    + \frac{4}{s^2}\mz^4(\mh^2-\mz^2)^2(\slmsl{2})^2 \Bigr].
\end{eqnarray}
\pagebreak
For $ C^{ZZ H_1} \sim \frac{m^{2}_{Z}}{v}, C^{\slp{b}^*\slp{a}Z}\sim {\tilde f}{\cal V}$, and assuming universality in slepton masses, we have:
\begin{eqnarray}
{\widetilde{w}}_{Z H_1}^{(Z)}=
\frac{1}{12\mz^6}
\Bigl| \frac{10^{2}.{\tilde f}{\cal V}}{\prop{Z}} \Bigr|^2 \Bigl[ s^2 \mz^4 -4s \msli^2\mz^4 +8 \mz^4 \msli^2 {m^{2}_ {H_1}}- \frac{2}{s}\mz^4 \msli^2 {m^{4}_{H_1}}\Bigr].\nonumber\\
\end{eqnarray}
 \underline{Slepton\ ($\slp{c}$) exchange:}
\begin{eqnarray}
\label{eq:wZhl}
& & {\widetilde{w}}_{Z H_1}^{(\slp{})}  =
\frac{1}{\mz^2}\sum_{c,d=1}^{2}
C^{\slp{c}^*\slp{a}Z} C^{\slp{b}^*\slp{c} H_1}
C^{\slp{d}^*\slp{a}Z *} C^{\slp{b}^*\slp{d} H_1 *}  \left[ \Tt_2 - 2(\msli^2+\mz^2) \Tt_1
                  +(\msli^2-\mz^2)^2 \Tt_0 \right] \nonumber \\
& &  +
\frac{1}{\mz^2}\sum_{c,d=1}^{2}
C^{\slp{c}^*\slp{a}h} C^{\slp{b}^*\slp{c}Z}
C^{\slp{d}^*\slp{a}h *} C^{\slp{b}^*\slp{d}Z *}  \left[ \Tu_2 - 2(\mslj^2+\mz^2) \Tu_1
                  +(\mslj^2-\mz^2)^2 \Tu_0 \right] \nonumber \\
 & &  +
\frac{1}{\mz^2} \Re \sum_{c,d=1}^{2}
C^{\slp{c}^*\slp{a}Z} C^{\slp{b}^*\slp{c} H_1}
C^{\slp{d}^*\slp{a} H_1 *} C^{\slp{b}^*\slp{d}Z *} \Bigl[ -2\Y_2
+\frac{1}{s}(s-\mh^2+\mz^2)(\slmsl{2}) \Y_1 + \nonumber\\
&& \Big\{ s(\slpsl{2}-2\mz^2)-(\slpsl{4}) -2\mz^2\mh^2   +(3\mz^2-\mh^2)(\slpsl{2})
       \nonumber \\
 & &   +\frac{1}{s}(\mh^2-\mz^2)(\slmsl{2})^2   - \frac{1}{2s^2}(\mh^2-\mz^2)^2(\slmsl{2})^2 \Big\} \Y_0 \Bigr].
\end{eqnarray}
Assuming universality in slepton masses of first two generations, we have:\\
$ \sum_{c,d=1}^{2} C^{\slp{c}^*\slp{a} H_1} C^{\slp{b}^*\slp{c}Z}
C^{\slp{d}^*\slp{a} H_1 *} C^{\slp{b}^*\slp{d}Z *}\sim \sum_{c,d=1}^{2}
C^{\slp{c}^*\slp{a} H_1} C^{\slp{b}^*\slp{c}Z}
C^{\slp{d}^*\slp{a} H_1 *} C^{\slp{b}^*\slp{d}Z *}\sim $\\
$ \sum_{c,d=1}^{2}
C^{\slp{c}^*\slp{a}Z} C^{\slp{b}^*\slp{c} H_1}
C^{\slp{d}^*\slp{a} H_1 *} C^{\slp{b}^*\slp{d}Z *} \equiv
{\tilde f}^2 {\cal V}^{-\frac{38}{15}} {M^{2}_p}$ .
Therefore, (\ref{eq:wZhl}) reduces to:
\begin{eqnarray}
\label{eq:wZhfin}
&& {\widetilde{w}}_{Z H_1}^{(\slp{})} \sim
\frac{{\tilde f}^2 {\cal V}^{-\frac{38}{15}} {m^2_{pl}}}{\mz^2}\times \Bigl[(\Tt_2 +\Tu_2) - \msli^2(\Tt_1 +\Tu_1)+ \msli^4 (\Tt_0 +\Tu_0) -2\Y_2 \nonumber\\
&&+ (2s \msli^2-2 \msli^4)\Y_0 \Bigr].  
\end{eqnarray}
 \underline{$Z$ -- slepton\ ($\slp{c}$) interference:}
\begin{eqnarray}
\label{eq:w_Zhslp}
& & {\widetilde{w}}_{Z H_1}^{(Z-\slp{})}  =
\frac{1}{\mz^4}\Re \sum_{c=1}^{2}
\left( \frac{C^{ZZ H_1}C^{\slp{b}^*\slp{a}Z}}{\prop{Z}} \right)^* \Bigl[
C^{\slp{c}^*\slp{a}Z} C^{\slp{b}^*\slp{c} H_1}
\Big\{ -(s - \mh^2)(\slmsl{2}) \nonumber \\
 & &  +2\mz^2(\msli^2-\mslk^2+\mz^2)
     -\frac{1}{s}\mz^2(\mh^2-\mz^2)(\slmsl{2})  + \nonumber\\
     && \Big[ s(\slmsl{2}+\mz^2)(\msli^2-\mslk^2-\mz^2) + \mz^2 \{ -2\mslk^4+\mslk^2(3\msli^2+\mslj^2+3\mz^2)    \nonumber \\
         \pagebreak
 & &-3\msli^4 +3\slsl{2}-2\mslj^4+\mz^2(4\msli^2+\mslj^2)-\mz^4 \} -\mh^2(\slmsl{2}+\mz^2)\nonumber\\
 && (\msli^2-\mslk^2+\mz^2) \Big] \Ft \Big\}  + C^{\slp{c}^*\slp{a}h} C^{\slp{b}^*\slp{c}Z}
\Big\{ (s - \mh^2)(\slmsl{2})  +2\mz^2(\mslj^2-\mslk^2\nonumber\\
     &&+\mz^2)
      +\frac{1}{s}\mz^2(\mh^2-\mz^2)(\slmsl{2}) + \Big[ -s(\slmsl{2}-\mz^2)(\mslj^2-\mslk^2-\mz^2) \nonumber\\
     && +\mz^2 \{ -2\mslk^4+\mslk^2(3\mslj^2+\msli^2+3\mz^2)-3\mslj^4
 +3\slsl{2} -2\msli^4\nonumber\\
 & &+\mz^2(4\mslj^2+\msli^2)-\mz^4 \} +\mh^2(\slmsl{2}-\mz^2)(\mslj^2-\mslk^2+\mz^2) \Big] \Fu \Big\}
 \Bigr].
\end{eqnarray}
For $ C^{ZZ H_1}\sim{\cal O}(10^2), C^{\slp{c}^*\slp{a}Z}\sim {\tilde f}{\cal V}, C^{\slp{b}^*\slp{c} H_1}\sim {\cal V}^{-\frac{34}{15}}$, equation (\ref{eq:w_Zhslp}) reduces to:
\begin{eqnarray}
&& {\widetilde{w}}_{Z H_1}^{(Z-\slp{})} \equiv
\frac{1}{\mz^4}\Re \sum_{c=1}^{2}
\left( \frac{(10^2. {\tilde f}{\cal V})M_P}{\prop{Z}} \right)^*
 C^{\slp{c}^*\slp{a}Z} C^{\slp{b}^*\slp{c}h}\Bigl[
4\mz^4 + (-2s \mz^4 + \msli^2 \mz^4)\nonumber\\
&&(\Ft + \Fu) \Bigr]
\end{eqnarray}
where $ \Ft , \Fu$ are evaluated in \cite{gravitino_DM}. Summing up the contribution of ${\widetilde{w}}$ for all s, t and u-channels as according to equation (\ref{eq:w_Zhslp}),
\begin{eqnarray}
\widetilde{w}_{Z H_1}|_{s=(4 \msli^2)}\sim {\cal V}^{\frac{34}{15}}, \widetilde{w}^{\prime}_{Z H_1}|_{s=(4 \msli^2)}\sim {\cal V}^{\frac{29}{15}},
\end{eqnarray}
and the``reduced'' coefficients $\widetilde{a}_{Z H_1}$ and
$\widetilde{b}_{Z H_1}$ will be given by
\begin{eqnarray}
\label{eq:aZh}
&& \widetilde{a}_{Z H_1}  \sim  \frac{1}{32 \pi \msli^2} {\cal V}^{\frac{34}{15}}\sim 10^{-11} {GeV}^{-2};
\widetilde{b}_{Z H_1} \sim  \frac{3}{64 \pi \msli^2}{\cal V}^{\frac{29}{15}}\sim  10^{-21} {GeV}^{-2}.
\end{eqnarray}
${~~\bf {\slp{a}\slp{b}^*\rightarrow \gm\gm}}$

The annihilation diagrams for this channel include point interaction along with
the $t$-- and $u$--channel slepton ($\slp{a}$, $a=1,2$) exchange:
\begin{eqnarray}
\label{eq:wgmgm}
 {\widetilde{w}}_{\slp{a}\slp{b}^*\rightarrow \gm\gm} = {\widetilde{w}}_{\gm\gm}^{(P)}
             + {\widetilde{w}}_{\gm\gm}^{(\slp{})} + {\widetilde{w}}_{\gm\gm}^{(P-\slp{})}.\nonumber
\end{eqnarray}
\begin{eqnarray}
&& {\rm  \underline{Contact~ interaction}}:  {\widetilde{w}}_{\gm\gm}^{(P)} \sim   (C^{\slp{a} \slp{b}^* \gm\gm})^2 \sim e^4 ({\tilde f}^2 {\cal V}^{-3})^2 \delta_{ab}.\\
&&
{\rm \underline{slepton\ (\slp{a}) exchange}}:\nonumber\\
\pagebreak
 &&{\widetilde{w}}_{\gm\gm}^{(\slp{})}=
 (C^{\slp{a} \slp{b}^* \gm})^4 \delta_{ab} \left[
 4(\T_2 + 2 \msli^2 \T_1 + \msli^4 \T_0) - (s-4\msli^2)^2 \Y_0
\right]\ \sim {\tilde f}^4{\cal V}^4\Bigl[
 4(\T_2 +\nonumber\\
 && 2 \msli^2 \T_1 + \msli^4 \T_0) - (s-4\msli^2)^2 \Y_0
\Bigr].
\end{eqnarray}
where $ \T_0\, \T_1, \T_2, \Y_0$ are evaluated in \cite{gravitino_DM}.

 \underline{Contact -- slepton\ ($\slp{c}$) interference:}
\begin{eqnarray}
&&{\widetilde{w}}_{\gm\gm}^{(P-\slp{})} =
2 C^{\slp{a} \slp{b}^* \gm\gm} (C^{\slp{a} \slp{b}^* \gm})^2 e^4 \delta_{ab}
\Bigl[ - 4 + (s-8\msli^2)\F \Bigr]\ \sim ({\tilde f}^4 {\cal V}^{-1}) e^4 \delta_{ab}
\Bigl[ - 4 \nonumber\\
&&+ (s-8\msli^2)\F \Bigr],
\end{eqnarray}
where $\F$ is evaluated in \cite{gravitino_DM}. Summing up the contribution of ${\widetilde{w}}$ for all channels:  
\begin{eqnarray}
\widetilde{w}_{\gm\gm}|_{s=(4 \msli^2)}\sim {\cal V}^{\frac{21}{5}}, \widetilde{w}^{\prime}_{\gm\gm}|_{s=(4 \msli^2)}\sim {\cal O}(1),
\end{eqnarray}
and the ``reduced'' coefficients $\widetilde{a}_{\gm\gm}$ and
$\widetilde{b}_{\gm\gm}$ will be given by
\begin{eqnarray}
\label{eq:agmgm}
&& \widetilde{a}_{\gm\gm}  \sim  \frac{1}{32 \pi \msli^2} {\cal V}^{\frac{21}{5}}\sim 10^{-2} {GeV}^{-2};
\widetilde{b}_{\gm\gm} \sim  \frac{3}{64 \pi \msli^2}{\cal I}(1)\sim  10^{-23} {GeV}^{-2}.
\end{eqnarray}
${\bf~~{\slp{a}\slp{b}^*\rightarrow \gm  H_1}}$

 The annihilation diagrams for this channel include  
the $t$-- and $u$--channel slepton ($\slp{a}$, $a=1,2$) exchange:

 \underline{Slepton\ ($\slp{c}$) exchange:}
\begin{eqnarray}
&& {\widetilde{w}}_{\slp{a}\slp{b}^*\rightarrow \gm  H_1}={\widetilde{w}}_{\gm  H_1}^{(\slp{})} =
-2 \left| C^{\slp{b}^*\slp{a}\gm} \right|^2\left| C^{\slp{b}^*\slp{a} H_1} \right|^2
\left[ (\Tt_1+\msli^2\Tt_0)+(\Tu_1+\mslj^2\Tu_0) \right. \nonumber \\
 & & \left. + (s+\mh^2-2\msli^2-2\mslj^2)\Y_0 \right]\nonumber\\
 && \sim ({\tilde f} {\cal V}^{-19/15})^2 \left[ (\Tt_1+\msli^2\Tt_0)+(\Tu_1+\mslj^2\Tu_0) + (s+\mh^2-4\msli^2)\Y_0 \right] \nonumber
\end{eqnarray}
where $\Tt_0 , \Tu_0, \Tt_1, \Tu_1, \Y_0$ are evaluated in \cite{gravitino_DM}.  For $s= 4 \msli^2$
\begin{eqnarray}
\widetilde{w}_{\gm  H_1}|_{s=(4 \msli^2)}\sim {\cal V}^{-\frac{61}{5}}, \widetilde{w}^{\prime}_{ZZ}|_{s=(4 \msli^2)}\sim {\cal V}^{-13},
\end{eqnarray}
and the ``reduced'' coefficients $\widetilde{a}_{\gm h}$ and
$\widetilde{b}_{\gm  H_1}$ will be given by
\begin{eqnarray}
\label{eq:agmh}
&& \widetilde{a}_{\gm  H_1}  \sim  \frac{1}{32 \pi\msli^2} {\cal V}^{-\frac{61}{5}}\sim 10^{-83} {GeV}^{-2};
\widetilde{b}_{\gm  H_1} \sim  \frac{3}{64 \pi \msli^2}{\cal V}^{-13}\sim  10^{-87} {GeV}^{-2}.
\end{eqnarray}

${\bf  {\slp{a}\slp{b}^*\rightarrow  H_1 H_1}}$
 
The annihilation diagrams for this channel are mediated via $s$--channel Higgs exchange and t-channel $\slp{c}^*$ exchange. Hence,
\begin{eqnarray}
\label{eq:whh}
 {\widetilde{w}}_{\slp{a}\slp{b}^*\rightarrow  H_1 H_1}=
 {\widetilde{w}}_{ H_1 H_1}^{( H_1, H_2,P)} + {\widetilde{w}}_{ H_1 H_1}^{(\slp{})} + {\widetilde{w}}_{ H_1}^{( H_1, H_2,P-\slp{})}
\end{eqnarray}
 \underline{Higgs ($ H_1, H_2$) exchange ($+$ Point interaction):}
\begin{eqnarray}
{\widetilde{w}}_{ H_1 H_1}^{(h,H,P)} & = &
\left|
\sum_{r= H_1, H_2} \frac{C^{ H_1 H_2r}C^{\slp{b}^*\slp{a}r}}{\prop{r}}
-C^{\slp{b}^*\slp{a}hh}
\right|^2.
\end{eqnarray}
For $m_{H_1} =125 GeV, m_{H_2}\sim {\cal V}^{\frac{59}{72}} m_{\frac{3}{2}}$, $m_{\chi^0_3}\sim {\cal V}^{\frac{2}{3}}m_{\frac{2}{3}}\sim {\cal V}^{-\frac{4}{3}}M_P$,
and  $C^{ H_1 H_1 H_1}\sim \frac{m^{2}_{H_1}}{v}, C^{\slp{a} \slp{b}  H_1}\sim  C^{\slp{a} \slp{b}  H_2}\sim
{\cal V}^{-\frac{34}{15}}, C^{\slp{a} \slp{b}hh}\sim
{\cal V}^{-\frac{41}{18}}$ from above, after simplifying, we have
\begin{eqnarray}
{\widetilde{w}}_{ H_1 H_1}^{( H_1,H_2,P)} & = &
\left|
 \frac{(10^2.{\cal V}^{-\frac{34}{15}}){M_P}}{\prop{h}} + \frac{(10^2.{\cal V}^{-\frac{34}{15}})M_P}{\prop{H}}
-{\cal V}^{-\frac{41}{18}}
\right|^2.
\end{eqnarray}
 \underline{Slepton\ ($\slp{c}$) exchange:}
\begin{eqnarray}
{\widetilde{w}}_{ H_1 H_1}^{(\slp{})} & = &
\sum_{c,d=1}^2
C^{\slp{c}^*\slp{a} H_1} C^{\slp{b}^*\slp{c}H_2}
C^{\slp{d}^*\slp{a} H_1 *} C^{\slp{b}^*\slp{d}H_2 *} \Tt_0
     + \sum_{c,d=1}^2 C^{\slp{c}^*\slp{a} H_2} C^{\slp{b}^*\slp{c} H_1}
C^{\slp{d}^*\slp{a}H_2 *} C^{\slp{b}^*\slp{d} H_1 *} \Tu_0 \nonumber \\
 & & - 2 \Re \sum_{c,d=1}^2
C^{\slp{c}^*\slp{a} H_1} C^{\slp{b}^*\slp{c}H_2}
C^{\slp{d}^*\slp{a} H_1 *} C^{\slp{b}^*\slp{d} H_1 *} \Y_0.
\end{eqnarray}
Strictly speaking we are considering first  generation of squarks which get identified with same modulus ${\cal A}_1$.Therefore, $\sum_{c,d=1}^2
C^{\slp{c}^*\slp{a} H_1} C^{\slp{b}^*\slp{c}H_2}
C^{\slp{d}^*\slp{a} H_1 *} C^{\slp{b}^*\slp{d}H_2 *} \sim ({\cal V}^{-\frac{34}{15}})^4 \sim {\cal V}^{-9}{M_P^4}$.
Hence $
{\widetilde{w}}_{hh}^{(\slp{})}= {\cal V}^{-9}(\Tt_0 +  \Tu_0 -\Y_0)M_P^4$.The contribution of $\Tt_0, \Tu_0, \Y_0$ for  $\slp{a} \slp{b}^*\ra  H_1 H_1$ is of the same order as $\slp{a} \slp{b}^* \ra ZZ $.

 \underline{Higgs ($ H_1, H_2$) ($+$ Point) -- slepton ($\slp{c}$) interference:}
\begin{eqnarray}
& & {\widetilde{w}}_{ H_1 H_1}^{( H_1, H_2,P-\slp{})}  =
2 \Re \sum_{c=1}^2
   \Bigl(
\sum_{r= H_1,H_2} \frac{C^{ H_1 H_2r}C^{\slp{b}^*\slp{a}r}}{\prop{r}}
-C^{\slp{b}^*\slp{a} H_1 H_1}
\Bigr)^*  \Big[
C^{\slp{c}^*\slp{a} H_1} C^{\slp{b}^*\slp{c}H_2} \Ft \nonumber\\
&&
+ C^{\slp{c}^*\slp{a}H_2} C^{\slp{b}^*\slp{c} H_1} \Fu \Big],
\end{eqnarray}
Incorporating values of relevant interaction vertices, and $\Ft,\Fu$ from \cite{gravitino_DM}, by summing up the contribution of ${\widetilde{w}}$ for all channels:
\begin{eqnarray}
\widetilde{w}_{ H_1 H_1}|_{s=(4 \msli^2)}\sim {\cal V}^{-3}, \widetilde{w}^{\prime}_{ H_1 H_1}|_{s=(4 \msli^2)}\sim {\cal V}^{-\frac{36}{5}},
\end{eqnarray}
and the ``reduced'' coefficients $\widetilde{a}_{ H_1 H_1}$ and
$\widetilde{b}_{ H_1 H_1}$ will be given by
\begin{eqnarray}
\label{eq:ahh}
&& \widetilde{a}_{ H_1 H_1}  \sim  \frac{1}{32 \pi \msli^2} {\cal V}^{-3}\sim 10^{-37} {GeV}^{-2};
\widetilde{b}_{ H_1 H_1} \sim  \frac{3}{64 \pi \msli^2}{\cal V}^{-\frac{36}{5}}\sim  10^{-48} {GeV}^{-2}
\end{eqnarray}
{\boldmath  ${\slp{a}\slp{b}\rightarrow \ell\ell}$}

This process proceeds only  $t$-- and $u$--channel neutralino ($\neu{i}$, $i=1,2,3$) exchange. 

\underline{Neutralino\ ($\neu{i}$) exchange:}
{\small
\begin{eqnarray}
& &  {\widetilde{w}}_{\slp{a}\slp{b}\rightarrow \ell\ell}^{(\neu{})}  =
\sum_{i,j=1}^3 \Bigl[
\Dijqp{LLLL} \mnmn (s-2\ml^2)\T_0  -2\ml^2 \Big[ \Dijqp{LLRR}\mnmn\T_0+\Dijqp{LRRL}\T_1 \Big]\nonumber\\
&& + \Dijqp{LRLR}\Big[ -\T_2
  -(s-\msli^2-\mslj^2)\T_1  -(\msli^2-\ml^2)(\mslj^2-\ml^2)\T_0 \Big]   -\ml\mnq \Big[ \Dijqp{LLLR}\nonumber\\
  & & \{ \T_1-(\msli^2-\ml^2)\T_0 \}
     + \Dijqp{LLRL} \{ \T_1-(\mslj^2-\ml^2)\T_0 \} \Big]  -\ml\mnp \Big[ \Dijqp{LRLL} \{ \T_1-(\msli^2-\ml^2) \nonumber\\
   &&\T_0 \}
    + \Dijqp{RLLL} \{ \T_1-(\mslj^2-\ml^2)\T_0 \} \Big] \Bigg]  + \frac{1}{2}\sum_{i,j=1}^3
\Bigg[
-2 \Dijqp{LLLL} \mnmn (s-2\ml^2)\Y_0 \nonumber\\
&&  +4\Dijqp{LLRR}\mnmn\ml^2\Y_0
     -2\Dijqp{LRLR}\ml^2(\slpsl{2}-2\ml^2)\Y_0 + 2 \Dijqp{LRRL}\nonumber\\
     &&\Big[ -\Y_2 -(\slsl{2}-\ml^4)\Y_0 \Big] -\ml\mnq \Big[ \Dijqp{LLLR} \{ \Y_1+(s+\slmsl{2}-4\ml^2)\Y_0 \}
      \nonumber \\
 & & 
     + \Dijqp{LLRL} \{ \Y_1+(s-\msli^2+\mslj^2-4\ml^2) \Y_0 \}
     \Big]
     +\ml\mnp \Big[ \Dijqp{LRLL} \{ \Y_1\nonumber\\
& &   -(s+\slmsl{2}-4\ml^2)\Y_0 \}  + \Dijqp{RLLL} \{ \Y_1- (s-\msli^2+\mslj^2-4\ml^2) \Y_0 \}
\Big] \Bigr],
\end{eqnarray}}
where $\Dijqp{LLLL} ,\Dijqp{LLRR}$ etc. are evaluated in \cite{gravitino_DM}. Given that $m_{\chi^0_3}\sim {\cal V}^{-\frac{4}{3}}M_P, m_{\chi^0_1}=  m_{\chi^0_2}\sim {\cal V}^{-1}M_P$, $ C^{\neu{i}\slp{a}^*\ell *}_R \sim {\tilde f} {\cal V}^{-\frac{1}{2}}, C^{\neu{i}\slp{a}^*\ell *}_R \sim {\tilde f} {\cal V}^{-\frac{12}{15}}$, For $s=4\msli^2$, we have:
\begin{eqnarray}
\widetilde{w}_{\ell\ell}|_{s=(4\msli^2)}\sim {\cal V}^{-\frac{19}{5}}, \widetilde{w}^{\prime}_{\ell\ell}|_{s=(4 \msli^2)}\sim {\cal V}^{-8}.
\end{eqnarray}
The ``reduced'' coefficients $\widetilde{a}_{\ell\ell}$ and
$\widetilde{b}_{\ell\ell}$ will be given by
\begin{eqnarray}
\label{eq:aellell}
&& \widetilde{a}_{\ell\ell}  \sim  \frac{1}{32 \pi \msli^2} {\cal V}^{-\frac{19}{5}}\sim 10^{-41} {GeV}^{-2};
\widetilde{b}_{\ell\ell} \sim  \frac{3}{64 \pi \msli^2}{\cal V}^{-8}\sim  10^{-62} {GeV}^{-2}.
   \end{eqnarray}
 Summing up the values of partial wave coefficients for all annihilation channels,  
\begin{eqnarray}
a_{{\slp{a}\slp{b}^*\rightarrow f_1f_2}}  &=&
\widetilde{a}_{ZZ}+ \widetilde{a}_{ H_1Z}+ \widetilde{a}_{h\gm}+ \widetilde{a}_{\gm\gm}+ \widetilde{a}_{ H_1 H_1}+ \widetilde{a}_{ll} \equiv  O(10)^{11} {GeV}^{-2} \nonumber \\
b_{{\slp{a}\slp{b}^*\rightarrow f_1f_2}}  &=& \widetilde{b}_{ZZ}+ \widetilde{b}_{ H_1Z}+ \widetilde{b}_{ H_1\gm}+ \widetilde{b}_{\gm\gm}+ \widetilde{b}_{ H_1 H_1}+ \widetilde{b}_{ll} \equiv  O(10)^{-10} {GeV}^{-2},
\end{eqnarray}
Now, we have $x_f=T/\msli$, and the relic abundance is given as:
\begin{eqnarray}
\label{eq:relic-density1}
\Omega_{\slp{a}} & = & \frac{1}{{\mu}^2 \sqrt{g_*}J({x_F})},
\end{eqnarray}
where ${\mu}= 1.2 \times 10^5 GeV$. For $J(x_f)  \sim  10^{11} {x_f}~{GeV}^{-2}$, $x_f= \frac{1}{33}$ and $ \sqrt{g_*}=9$, we have
\begin{eqnarray}
\Omega_{\slp{a}}\sim\frac{33}{1.44 \times 10^{10}\cdot 9\cdot 10^{11}}\equiv 10^{-20},
\end{eqnarray}
For $m_{\frac{3}{2}}\sim 10^8 GeV$ and $\msli\sim {\cal V}^{\frac{1}{2}}m_{\frac{3}{2}}$, relic abundance of gravitino will be given as :
\begin{eqnarray}
\label{eq:omega2}
\Omega_{\tilde G}& = & \Omega_{\slp{a}}\times \frac{m_{\frac{3}{2}}}{\msli}=10^{-20} \times {\cal V}^{-\frac{1}{2}} \sim 10^{-22}~{\rm  for}~ {\cal V}\sim 10^5.
\end{eqnarray}
From (\ref{eq:omega1}) and (\ref{eq:omega2}), it appears that relic abundance of gravitino turns out to be extremely suppressed in case of slepton (NLSP) (co-)annihilations as compared to relic abundance of gravitino in case of neutralino (NLSP) aannihilations, for almost similar value of thermal cross-section. 
\section{Results and Discussion}
In this chapter, we have investigated, in detail, the possibility of gravitino as a viable CDM candidate in the context of our local large volume $D3/D7$ $\mu$- split SUSY model. After calculating the masses of various SM and their superpartners, gravitino appeared as the Lightest Supersymmetric particle (LSP) for Calabi-Yau volume ${\cal V} \sim 10^5$ which motivated our query of considering gravitino as a potential CDM candidate in our model.

In big bang cosmology, gravitino population depends on two kinds of mechanisms: thermal as well as non-thermal production. We assumed that reheating temperature will be low enough to produce the effective  relic abundance of gravitino in agreement with experimental observations, therefore almost all of the gravitinos will be produced by electromagnetic as well as hadronic decays of unstable NLSP. The scale of masses of various superpartners suggested that because of an ${\cal O}(1)$ difference between the masses of the sleptons/squarks and the lightest neutralino and given that in our calculations we have not bothered about such ${\cal O}(1)$ factors, both could exist as valid  NLSP candidates if life-times of the same are  such that they do not disturb the bounds given by Big Bang Nucleosynthesis (BBN). Based on this fact, by considering the contributions of all required couplings in ${\cal N}=1$ gauged supergravity action and assuming that similar to the effective Yukawa couplings, there is only an ${\cal O}(1)$ multiplicative change in all three-point interaction vertices in RG-flowing  from the string to the EW scale, we estimated the mean lifetime by calculating two-body and three-body decay widths of gravitino decaying into SM particles, which in both cases turned out to be greater than the age of the universe and hence satisfied the primary requirement of  an appropriate DM candidate. Then we discussed in detail the leading two- and three-body decays of gaugino/neutralino and  determined that life-time of these decays too short to affect the predictions of BBN. Because of considerably high life time of gluino as evaluated in chapter {\bf 2}, it is manifest that the same can not be considered as an appropriate NLSP's because late decays of the gluino into goldstino (longitudinal component of gravitino) can surely elude the constraints coming from BBN. Furthermore, we studied in detail the decay width of three-body decay of neutralino into SM particles, the high life time of which, as compared to life time of neutralino into gravitino, ensured that gravitino abundance does not get diluted. On a similar ground, we also calculated three-body decay width of sleptons which similar to neutralino decays appeared small enough to affect bounds on BBN.  The numerical estimates of various N(LSP)'s are provided in Table~3.1. In short, the explicit calculation of life times of both, LSP and NLSP, confirmed the possibility of gravitino as an appropriate CDM candidate. Finally, by considering non-thermal production mechanism of gravitino, we explicitly estimated values of thermal annihilation cross-section of sleptons and neutralino which ultimately gave gravitino relic abundance from sleptons to be extremely suppressed while the same from neutralino turned out to be $0.1$, almost in agreement with the value of $\Omega_{C}h^2$ suggested by direct and indirect experiments. Thus we conclude that gravitino qualifies as a potential CDM candidate in  local large volume $D3/D7$ $\mu$-split-like  SUSY model.

\chapter{Electric Dipole Moment (EDM) of Electron/Neutron}
\vskip -0.5in
{\hskip1.4in{\it ``Epistemology has always been affected by technologies like the telescope and the microscope, things that have created a radical shift in how we sense physical reality."}}

\hskip4.2in - Ken Goldberg.
\graphicspath{{Chapter4/}{Chapter4/}}
\vskip -0.5in
 \section{Introduction}
The  existence of split SUSY models may not be foreseeable in the near future via precise measurements to be carried at the Large Hadron Collider (LHC) however indirect methods can be made available to test some of the signatures of this scenario. In this context, Electric Dipole Moment (EDM) of electron/neutron serves itself as another testing ground for split SUSY scenario. In Standard Model, the CP-odd phases generated through Cabibo-Kobayashi-Masakawa Matrix (CKM) give a theoretical bound on EDM which is far below the experimental limits. According to the results recently published by particle data group \cite{beringer}, the current experimental limit on the electron EDM is $|d_e/e| < 10.5 \times 10^{-28} cm$, and neutron EDM is $|d_n/e| < 0.29 \times 10^{-25} cm$. In supersymmetric models, new sources of CP violation are introduced from soft SUSY breaking terms and can be used as powerful probes to provide an order-of-magnitude estimate of EDM of the electron/neutron which is within reach of experimental results. The new CP-odd phases are not only sufficient to generate accurate bound on EDM of electron/neutron but also play an essential role in the generation of the baryon asymmetry of the universe \cite{Mayumi_et_al}. However, it is known in literature \cite{Altraev,Smith} that by considering small values of soft SUSY breaking terms and considering an ${\cal O}(1)$ phase, one generates an EDM of electron/neutron which is in excess of experimental limits. The issue can be avoided if one considers either CP violating phase to be of the order {\cal O}(${10}^{-2}-10^{-3}$) or scalar masses to be heavy of the order of few $TeV$ scale. Therefore, it is interesting to get the estimate EDM of electron/neutron in the context of split SUSY scenario. In typical split SUSY models, all possible one-loop contributions to EDM are highly suppressed by the super-heavy scalar mass in the loop and leading contributions to the EDM starts at the two-loop level due to presence of SM particles and EW charginos and neutralinos in the loops (for the analysis of two-loop Barr-Zee diagrams in different models, see \cite{giudice_splitsusy, chang_splitsusy, Pilaftsis_2loop,Fukuyama} and references therein). \emph{However in our model, the gaugino and neutralino/chargino are almost as heavy as neutral scalars except one light Higgs. Based on that, one can not ignore the contribution of one-loop diagrams because of partial compensation of suppression factors appearing from heavy sfermion masses with heavy fermion (neutralino, chargino and gaugino) masses}. 

We provide a complete quantitative analysis of the neutron and electron EDM for all possible one-loop as well as two-loop diagrams in the context of large volume $D3/D7$ $\mu$-split-like  supersymmetry in this chapter which is organized as follows. In section {{\bf 2}}, we explain the origin of non-zero complex phases obtained in the context of ${\cal N}=1$ gauged supergravity limit of our local large volume $D3/D7$ $\mu$-split-like  SUSY model. We also argue that phases of effective Yukawa couplings do not change in the renormalization group flow from string scale down to the EW scale in our model. In section {{\bf 3}}, we turn towards the quantitative discussion of EDM of electron/neutron for various possible one-loop diagrams evaluated in the context of ${\cal N}=1$ gauged supergravity action. The complex phases can be made to appear through the complex off-diagonal components of sfermion/Higgs mass matrix and complex effective Yukawa couplings appearing in all one-loop diagrams. We assume the phase of both off-diagonal component of scalar mass matrix as well as possible Yukawa's to lie in the range $(0, \frac{\pi}{2}]$ in all the calculations. In addition to discussing the one-loop diagrams that exhibit non-zero phases through mixing between sfermions in section {{\bf 3.1}}, we also take into account the loop diagrams in which a unique phase appears through mixing between two Higgs at EW scale in section {{\bf 3.2}}.  Because of the presence of light Higgs existing as a propagator in the loop, one can expect to get reasonable order of magnitude of EDM of electron/neutron from one-loop diagrams involving Higgs and other SM/supersymmetric particles. For two-loop diagrams, in section {\bf 4}, we mainly focus on the Barr-Zee diagrams which involve fermion, sfermions and W boson as part of an internal loop, and are mediated through $h \gamma$ exchange except one R-parity violating diagram which involve fermion in the internal loop and is mediated through ${\nu}_{L}\gamma$ exchange. For the complete analysis, we also calculated the contribution of Rainbow type two-loop diagrams involving R-parity violating as well as R-parity conserving vertices. For all two-loop diagrams discussed in this chapter, the complex effective Yukawa couplings are sufficient to produce non-zero complex phases to generate non-zero EDM.
\vskip -0.5in
 \section{CP Violating Phases}
In this section we explain the possible origin of CP violating phases in the ${\cal N}=1$ gauged supergravity limit of our local large volume $D3/D7$ $\mu$-split-like  SUSY model. The electric dipole moment of a spin-$\frac{1}{2}$ particle is defined by
 the effective CP-violating dimension-5 operator given as:
\beq
{\cal L}_I=-\frac{i}{2} d_f \bar{\psi} \sigma_{\mu\nu} \gamma_5 \psi F^{\mu\nu}.
\eeq
Given that the effective operator is non-renormalizable, the effective lagrangian
can be induced at the loop level provided the theory contains a source of CP violation. In Standard Model, CP violating phases, in general, appear from the complex Kobayashi Maskawa (CKM) phase in the quark mass matrix but the same gets a non-zero contribution only at three-loop level. However, in supersymmetric theories, instead of CKM phase generated in Standard Model, one can consider the new phases appearing from complex parameters of soft SUSY breaking terms, complex Yukawa couplings as well as supersymmetric mass terms.

We consider the existence of non-zero phases appearing from complex effective Yukawa couplings present in ${\cal N}=1$ gauged supergravity action. As discussed in chapter {\bf 2}, position as well as Wilson line moduli identification with SM-like particles  generate  effective Yukawa couplings including R-parity conserving  as well as R-parity violating ones in the context of ${\cal N}=1$  gauged supergravity action \cite{gravitino_DM}, and the solution of RG evolution of effective Yukawa couplings at one-loop level yields:
\beqn
\label{eq:Yhat_sol-2}
\hat{Y}_{\Lambda\Sigma\Delta}(t)\sim\hat{Y}_{\Lambda\Sigma\Delta}(M_s)
\prod_{(a)=1}^3\left(1 + \beta_{(a)}t\right)^{\frac{-2\left(C_{(a)}(\Lambda)
    +C_{(a)}(\Sigma) + C_{(a)}(\Delta)\right) }{b_{(a)}}}.\nonumber
\eeqn
Using the fact that quadratic Casimir invariants as well as beta functions are real, we see that magnitudes of Yukawa couplings $\hat{Y}_{\Lambda\Sigma\Delta}$s change only by ${\cal O}(1)$ while phases of all Yukawas do not change at all as one RG-flows down from the string to the EW scale. Also, given that all four-Wilson line moduli ${\cal A}_I$ as well as position moduli ${\cal Z}_I$ are stabilized at different values; we make an assumption that there will be a distinct phase factor associated with all position as well as Wilson line moduli superfields which produces an overall distinct phase factor for each possible effective Yukawa coupling corresponding to four Wilson line moduli as well as position moduli.   
  
 The other important origin of generation of non-zero phases are given by complex soft SUSY breaking parameters (${m}^{2}_i, {\cal A}_{IJK}, \mu B$)  as well supersymmetric mass term $\mu$. The soft SUSY scalar mass terms can be made real by phase redefinition. However, in addition to the diagonal entries of sfermions corresponding to fermions with L-handed as well as R-handed chirality in the sfermion mass matrix, one gets an off-diagonal contribution because of mixing between L-R sfermion masses  after EW-symmetry breaking. The contribution of the same is governed by complex trilinear couplings as well as supersymmetric mass parameter $\mu$ at EW scale. Therefore, the scalar (sfermion) fields  $\tilde{f}_L$ and $\tilde{f}_R$ have been considered as linear combinations of the mass eigenstates which are obtained by diagonalizing sfermion $(mass)^2$  matrices \cite{ibrahim+nath} i.e.
\beqn
&& \tilde{f}_L=D_{f_{11}} \tilde{f}_1 +D_{f_{12}} \tilde{f}_2, \tilde{f}_R=D_{f_{21}} \tilde{f}_1 +D_{f_{22}} \tilde{f}_2.
\eeqn
where $f$ corresponds to first generation leptons and quarks and
\begin{equation}
D_f=\left(\begin{array}{ccccccc}
\cos \frac{\theta_f}{2} &-\sin \frac{\theta_f}{2} e^{-i\phi_{f}}\\
 \sin \frac{\theta_f}{2} e^{i\phi_{f}} & \cos \frac{\theta_f}{2}
 \end{array} \right),
\end{equation}
and  the mass matrix is given as follows:
\begin{equation}
\label{eq:massmatrix}
M_{\tilde{f}}^2=\left(\begin{array}{ccccccc}
 M_{\tilde{f_L}}^2 & m_u({\cal A}_{f}^{*}-\mu \cot\beta)\\
 m_u({\cal A}_{f}-\mu^{*} \cot\beta) & M_{\tilde{f_R}}^2
 \end{array} \right)_{EW},
\end{equation}
where  ${{\cal A}_{IJK}}$ corresponds to complex trilinear coupling parameter. Diagonalizing the above matrix by performing unitary transformation: $D_{f}^\dagger M_{\tilde{f}}^2 d_f={\rm diag}(M_{\tilde{f}1}^2, M_{\tilde{f}2}^2)$, where $\tan \theta_f=
\frac{2|M_{\tilde{f}21}^2|}{M_{\tilde{f}11}^2-M_{\tilde{f}22}^2}$.
The eigenvalues $M_{\tilde{f}1}^2$ and $M_{\tilde{f}2}^2$ are as follows:
\beq
\label{eq:Me1e2}
M_{\tilde{f}(1)(2)}^2=\frac{1}{2} (M_{\tilde{f}11}^2+M_{\tilde{f}22}^2)
	(+)(-)\frac{1}{2}[(M_{\tilde{f}11}^2-M_{\tilde{f}22}^2)^2 +
		4|M_{\tilde{f}21}^2|^2]^{\frac{1}{2}}.
\eeq
 For $f=e, {\cal A}_{e}^{*}= {{\cal A}_{{\cal Z}_I {\cal A}_1 {\cal A}_3}} $; for  $f=({u,d}), {\cal A}_{u/d}^{*}= {{\cal A}_{{\cal Z}_I {\cal A}_2 {\cal A}_4}} $. In our model as discussed in chapter {{\bf 2}}, we have universality in trilinear couplings w.r.t position as well as Wilson line moduli. Assuming the same to be true at EW scale, the values of trilinear coupling parameters are ${\cal A}_{{\cal Z}_I {\cal A}_1 {\cal A}_3}= {\cal A}_{{\cal Z}_I {\cal A}_2 {\cal A}_4}= {\cal V}^{\frac{37}{36}}m_{\frac{3}{2}}$.
As given in chapter {{\bf 2}}, the value of  supersymmetric mass parameter  $\mu$ at EW scale is ${\cal V}^{\frac{59}{72}}m_{\frac{3}{2}}$. Also we have universality in slepton (squark) masses of first two generations. Therefore, $M_{\tilde{e}11}^2= M_{\tilde{e}22}^2= M_{\tilde{u}11}^2 =M_{\tilde{u}22}^2 \sim  {\cal V} m^{2}_{\frac{3}{2}}$, and
\beqn
\label{me21}
&& |M_{\tilde{e}21}^2|^2= m_e|{\cal A}_{e}^{*}-{\mu}\cot{\beta}| \equiv  ({\cal V}^{\frac{37}{36}})m_{e}m_{\frac{3}{2}} << M_{\tilde{e}11}^2, \nonumber\\
&& |M_{\tilde{u}21}^2|^2= m_u|{\cal A}_{u}^{*}-{\mu}\cot{\beta}| \equiv  ({\cal V}^{\frac{37}{36}})m_{e}m_{\frac{3}{2}} << M_{\tilde{u}11}^2.
\eeqn
Using the above, one can show that eigenvalues of sfermion mass matrix $M_{\tilde{f}(1)(2)}^2 \sim  M_{{\tilde f}^{2}_{L,R}}={\cal V}m^{2}_{\frac{3}{2}}$. The aforementioned mass eigenstates can be utilized to produce non-zero phase responsible to generate finite  EDM of electron as well as neutron in the one-loop diagrams involving sfermions as scalar propagators, and  gaugino and neutralino as fermionic propagators.
\vskip -1.0in
\section{One-Loop Contribution to EDM}
At one-loop level, for a theory of fermion $\psi_f$ interacting with other heavy fermions $\psi_i$'s and heavy scalars $\phi_k$'s
with masses $m_i$, $m_k$ and charges $Q_i $, $Q_k$, the interaction that contains CP violation in
general is given by \cite{ibrahim+nath}:
\beqn
\label{eq:eff1loop}
-{\cal L}_{int}=\sum_{i,k} \bar{\psi}_f
                \left(K_{ik} \PL +L_{ik} \PR\right) \psi_i\phi_k+ h.c
\eeqn
Here  ${\cal L}$ violates CP invariance iff
${\rm Im}(K_{ik} L_{ik}^*)\neq 0$.
and one-loop EDM of the fermion f in this  case is given by
\beq
\sum_{i,k} \frac{m_i }{(4\pi)^2 m_k^2}{\rm Im}(K_{ik} L_{ik}^*)
        \left(Q_i A\left(\frac{m_i^2}{m_k^2}\right)+Q_k    B\left(\frac{m_i^2}{m_k^2}\right)\right),
\eeq
where $A(r)$ and $B(r)$ are defined by
\beq
\label{eq:A}
 A(r)=\frac{1}{2(1-r)^2}\Bigl(3-r+\frac{2lnr}{1-r}\Bigr), 
B(r)=\frac{1}{2(r-1)^2}\Bigl(1+r+\frac{2rlnr}{1-r}\Bigr),
\eeq
where, $Q_k=Q_f-Q_i$. 

We use the above-mentioned analytical result to  get the 
estimate of magnitude of EDM of electron/quark in the context of ${\cal N}=1$ gauged supergravity by inducing all Standard Model as well as supersymmetric particles in the loop diagram. The EDMs of the neutron can be estimated by calculating the contribution of $u$ and $d$ quarks  by using relation $d_n = (4d_{d}-d_{u})/3$. Since in our model, we have identified both up- as well as down-quark with single Wilson line modulus, we will have same contribution to EDM for both up and down-quarks. Hence, the neutron EDM is same as up-quark EDM. Therefore, in the calculations below, we have estimated EDM of electron and up quark only.
\vskip -0.5in
\subsection{One-Loop Diagrams Involving Neutral Sfermions} {\bf{Gaugino Contribution:}}
In this subsection, we estimate the contribution of  electron/neutron EDM at one-loop level due to presence of heavy gaugino nearly isospectral with heavy sfermions. In traditional split SUSY models discussed in literature, the masses of sfermions are very heavy while  gauginos as well as higgsinos are kept light because of the gauge coupling unification. Therefore, one-loop diagrams involving sfermion-gaugino exchange do not give any significant contribution to EDM of fermion. However in the large volume $D3/D7$ model that we have discussed in chapter {\bf 2}, the gaugino as well as higgsino turn out to be very heavy. As it is clear from the equation (\ref{eq:eff1loop}), the order of magnitude of EDM at one-loop level is directly proportional to fermion mass and inversely proportional to sfermion masses circulating in the loop whereas the one-loop function can almost be of the order $ {\cal O}(0.1-1)$ provided either the difference between fermion and sfermion mass is of order ${\cal O}(1)$  or the fermion mass is very light as compared to sfermion mass.  Therefore, naively one would expect an enhancement in the order of magnitude of one-loop EDM due to presence of heavy fermions circulating in a loop. In view of this, we estimate the contribution of one-loop EDM of electron as well as neutron in ${\cal N}=1$ gauged supergravity limit of large volume $D3/D7$ $\mu$-split-like  SUSY model discussed in chapter {{\bf 2}}. The CP violation (imaginary phases) can be induced in the loop diagram by considering diagonalized eigenstates of sfermion mass matrix as propagators in the loop. The loop diagram is given in Figure~4.1.
\begin{figure}
\begin{center}
\begin{picture}(145,147) (130,40)
   \Line(100,50)(330,50)
   \DashCArc(215,50)(80,0,180){4}
   \Photon(222,130)(230,190){4}{4}
   \Text(150,110)[]{${\tilde f}_{i}$}
   \Text(290,100)[]{${\tilde f}_{i}$}
   \Text(90,50)[]{{$f_{L}$}}
   \Text(234,150)[]{{$\gamma$}}
   \Text(340,50)[]{{$f_{R}$}}
   \Text(235,40)[]{{{$\tilde{\lambda}^0$}}}
   \end{picture}
\end{center}
\caption{One-loop diagram involving gaugino.}
 \end{figure}
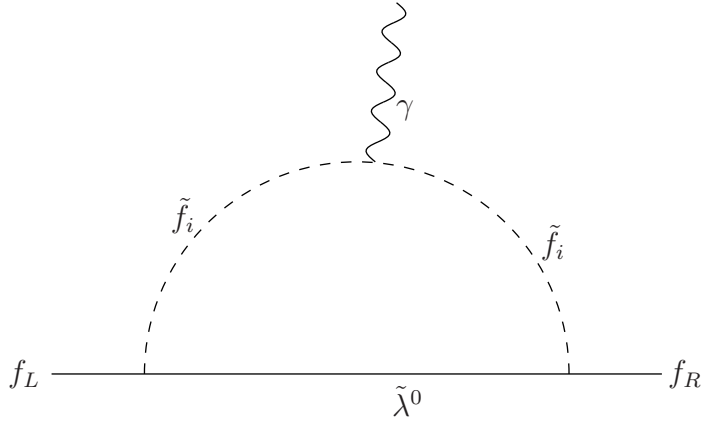
 The effective one-loop operator given in equation (\ref{eq:eff1loop}) can be recasted in the following form:
\beq
\label{eq:eff1loop1}
{\cal L}_{int}=\sum_{i=e,u,d} \bar{\psi}_{f_i}
                \left(K_{i} \PL +L_{i} \PR\right) {\phi_{{\tilde f_i}}}{\tilde \lambda^0}+ h.c
\eeq
For $i=1,2$, above equation can be expanded as:
 \beq
 \label{eq:eff1loop2}
-{\cal L}_{int}= \bar{\psi}_{f}
                \left(K_{1} \PL +L_{1} \PR\right) {\phi_{\tilde f_1}}{\tilde \lambda^0}+\bar{\psi}_{f}
                \left(K_{2} \PL +L_{2} \PR\right) {\phi_{\tilde f_2}}{\tilde \lambda^0}+ h.c.,
\eeq
and one-loop EDM of the fermion $f$ in this  case will be given as:
\beq
\label{eq:EDM}
\frac{m_{\tilde{\lambda^{0}_i}}}{(4\pi)^2 } \left[\frac{1}{m_{{\tilde f_1}^2}}   {\rm Im}(K_{1} L_{1}^*)\left(Q'_{\tilde {f_1}} B\left(\frac{m_{\tilde {\lambda^0}}^2}{m_{{\tilde f_1}^2}}\right)\right)+  \frac{1}{m_{{\tilde f_2}^2}}  {\rm Im}(K_{2} L_{2}^*)
       \left(Q_{\tilde {f_2}}B\left(\frac{m_{\tilde {\lambda^0}}^2}{m_{{\tilde f_2}^2}}\right)\right)\right].
\eeq
where $m_{\tilde {\lambda^0}}$ corresponds to gaugino mass, $m_{\tilde f_1}$ and $m_{\tilde f_2}$ correspond to masses of eigenstates of diagonalized sfermion mass matrix. and $Q'_{\tilde {f_i}}$ corresponds to effective charge defined as: $Q'_{\tilde {f_i}} \sim  Q_{\tilde {f_i}} C_{{\tilde {f_i}}{\tilde {f_i}}{\gamma}}$, where
$C_{{\tilde {f_i}}{\tilde {f_i}}{\gamma}}$ will be volume suppression factor coming from sfermion-photon-sfermion vertex.

To determine  the value of one-loop EDM in this case, we first calculate the contribution of required vertices involved in Figure~4.1. As discussed in earlier chapters, in ${\cal N}=1$ gauged supergravity, lepton(quark)-slepton(squark)-gaugino interaction is governed by  ${\cal L}_{f-\tilde{f}-\tilde{\lambda^{0}}}= g_{YM}g_{{ J}\bar{T}_B}X^{{*}B}{\bar\chi}^{\bar J}\tilde{\lambda^{0}}+ \partial_{J}T_B D^{B}{\bar\chi}^{\bar J}\lambda^{0}$, 
where ${\bar\chi}^{\bar J}$ corresponds to spin $\frac{1}{2}$ fermion,  $X^{{*}B}$ is killing isometry vector and $\tilde{\lambda^{0}}$ corresponds to $SU(2)$-singlet neutral gaugino. Though the gauge coupling $g_{YM}$ is real, the non-zero phase factor is produced from moduli space metric component $g_{{ J}\bar{T}_B}$, and associated with the volume suppression factor arising from the same. Hence, effective gauge coupling interaction vertex generate a particular phase factor which we consider to be ${\cal O}(1)$.

We repeatedly mention that to get the numerical estimate of contribution of aforementioned vertices, we use the identification described in \cite{gravitino_DM} and in chapter {\bf 2} according to which  fermionic superpartners of ${\cal A}_1$ and ${\cal A}_3$ can be identified, respectively with $e_L$ and $e_R$, and the fermionic superpartners of ${\cal A}_2$ and ${\cal A}_4$ can be identified, respectively with the first generation quarks: $u_L/d_L$ and $u_R/d_R$. In principle, incoming left-handed electron (quark) can couple with  scalar superpartners of both left-handed as well as right-handed leptons (quarks).
Therefore, for a  left handed electron $e_L$ interacting with slepton as well as gaugino, interaction vertex will be given as: ${\cal L}_{e_{L}-\tilde{e}-\tilde{\Lambda^{0}}}= g_{YM}g_{{\cal A}_1\bar{T}_B}X^{{*}B}{\bar\chi}^{\bar {\cal A}_1}\tilde{\lambda^{0}}+ \partial_{a_1}T_B D^{B}{\bar\chi}^{\bar a_1}\lambda^{0}$.
To calculate the contribution of $e_{L}-\tilde{e_L}-\tilde{\lambda^{0}}$ vertex, we expand $g_{{\cal A}_1\bar{T}_B}$ in the fluctuations linear in ${\cal A}_1$ around its stabilized VEV. In terms of undiagonalized basis, we have: $ g_{{T_B} {\bar a}_1}\rightarrow-{\cal V}^{-\frac{1}{4}} (a_1-{\cal V}^{-\frac{2}{9}})$. Using $T_{B}=Vol(\sigma_B)- C_{I{\bar J}}a_I{\bar a}_{\bar J} + h.c.$, where the values of intersection matrices $C_{I{\bar J}}$ are given in the appendix of \cite{gravitino_DM}. Utilizing those values,  we get $\partial_{a_1} T_B\rightarrow {\cal V}^{\frac{10}{9}} (a_1-{\cal V}^{-\frac{2}{9}})$. Using the argument that $g_{YM}g_{{T_B} {\bar a}_1} \sim {\cal O}(1)g_{YM}g_{{T_B} {\bar {\cal A}}_1}$ as shown in \cite{gravitino_DM}; incorporating values of  $X^{B}=-6i\kappa_4^2\mu_7Q_{T_B}, \kappa_4^2\mu_7\sim \frac{1}{\cal V}, D^B=\frac{4\pi\alpha^\prime\kappa_4^2\mu_7Q_Bv^B}{\cal V}$ and  $Q_{T_B}\sim{\cal V}^{\frac{1}{3}}(2\pi\alpha^\prime)^2\tilde{f}$, we get the contribution of physical gaugino$(\tilde{\lambda^{0}})$- lepton$(e_{L})$-slepton$(\tilde{e_L})$ interaction vertex as follows :

\begin{equation}
\label{eq:eLeLlambda}
|C_{e_{L}\tilde{e_L}\tilde{\lambda^{0}}}|\equiv \frac{ {\cal V}^{-\frac{2}{9}} {\tilde f}}{{ \sqrt{\hat{K}_{{\cal A}_1{\bar {\cal A}}_1}} }{ \sqrt{\hat{K}_{{\cal A}_1{\bar {\cal A}}_1}} }}{\tilde {\cal A}_1}{\bar\chi}^{\bar {\cal A}_1}\tilde{\lambda^{0}}
 \equiv\tilde{f}\left({\cal V}^{-1}\right),
\end{equation}
where ${\tilde f}$ is dilute flux and value of the same as calculated in \cite{Dhuria+Misra_mu_Split_SUSY} is roughly of  ${\cal O}(10^{-4})$ for Calabi-Yau volume ${\cal V}\sim 10^5$.  

Similarly, contribution of physical gaugino$(\tilde{\lambda^{0}})$- quark$(u_{L})$-squark$(\tilde{u_L})$ interaction vertex will be given by expanding $g_{{\cal A}_2\bar{T}_B}$ in the fluctuations linear in ${\cal A}_2$ around its stabilized VEV. Doing so, one will get: $ g_{{T_B} {\bar a}_2}\rightarrow-{\cal V}^{-\frac{5}{4}} (a_2-{\cal V}^{-\frac{1}{3}}), \partial_{a_2}T_B \rightarrow {\cal V}^{\frac{1}{9}} (a_2-{\cal V}^{-\frac{1}{3}})$,
 and
\begin{equation}
\label{eq:uLuLlambda}
 |C_{u_{L}\tilde{u_L}\tilde{\lambda^{0}}}| \equiv \frac{  {\cal V}^{-\frac{11}{9}} {\tilde f}}{{ \sqrt{\hat{K}_{{\cal A}_2{\bar {\cal A}}_2}} }{ \sqrt{\hat{K}_{{\cal A}_2{\bar {\cal A}}_2}} }}{\tilde {\cal A}_2}{\bar\chi}^{\bar {\cal A}_2}\tilde{\lambda^{0}}
 \equiv\tilde{f}\left({\cal V}^{-\frac{4}{5}}\right).
 \end{equation}
 
The gaugino$(\tilde{\lambda^{0}})$ - fermion($f_L$)-sfermion(${\tilde f}_R$) vertex does not possess $SU(2)$  electroweak symmetry. However, terms in supergravity Lagrangian preserve $SU(2)$ EW symmetry. Therefore, we first generate a term of the type $f_{L} \tilde{f_R} \tilde{\lambda^{0}}H_L$ wherein  $H_L$ is $SU(2)_L$  Higgs doublet.  After spontaneous breaking of the EW symmetry when $H_L$ acquires a non-zero vev $\langle H^0\rangle$, this term generates: $\langle H^0\rangle f_{L} \tilde{f_R} \tilde{\lambda^{0}}$.
For $f_{L,R}= e_{L,R}$, by expanding $g_{{\cal A}_1\bar{T}_B}$ in the fluctuations linear in ${\cal Z}_i$ and then linear in ${\cal A}_3$ around their stabilized value, we have: $ g_{{T_B} {\bar {\cal A}}_1}\rightarrow {\cal V}^{-\frac{13}{36}} \langle {\cal Z}_1 \rangle ({\cal A}_3- {\cal V}^{-\frac{13}{18}})$. The contribution of physical gaugino$(\tilde{\lambda^{0}})$- lepton$(e_{L})$-slepton$(\tilde{e_R})$ interaction vertex will be as follows :
\beq
\label{eq:eLeRlambda}
| C_{e_{L}\tilde{e_{R}}\tilde{\lambda^{0}}}| \equiv \frac{g_{YM}g_{{T_B} {\bar {\cal A}_1}}X^{T_B}\sim {\cal V}^{-\frac{19}{18}} {\tilde f}}{{ \sqrt{\hat{K}_{{\cal Z}_1{\bar {\cal Z}}_1}\hat{K}_{{\cal A}_1{\bar {\cal A}}_1}\hat{K}_{{\cal A}_3{\bar {\cal A}}_3}}}}{\tilde {\cal A}_3 }{\bar\chi}^{\bar {\cal A}_1}\tilde{\lambda^{0}}
 \equiv\tilde{f}\left({\cal V}^{-\frac{15}{9}}\right).
 \eeq
For $f_{L,R}= u_{L,R}$, by expanding $g_{{\cal A}_2\bar{T}_B}$ in the fluctuations linear in ${\cal Z}_i$ and then linear in ${\cal A}_4$ around their stabilized value, we have: $ g_{{T_B} {\bar {\cal A}}_2}\rightarrow {\cal V}^{-\frac{13}{36}} \langle{\cal Z}_1\rangle ({\cal A}_4- {\cal V}^{-\frac{11}{9}})$,
and
\beq
\label{eq:uLeRlambda}
|C_{u_{L}\tilde{u_{R}}\tilde{\lambda^{0}}}|\equiv \frac{g_{YM}g_{{T_B} {\bar {\cal A}_2}}X^{T_B}\sim {\cal V}^{-\frac{19}{18}} {\tilde f}}{{ \sqrt{\hat{K}_{{\cal Z}_1{\bar {\cal Z}}_1}\hat{K}_{{\cal A}_2{\bar {\cal A}}_2}\hat{K}_{{\cal A}_4{\bar {\cal A}}_4}}}}{\tilde {\cal A}_4}{\bar\chi}^{\bar {\cal A}_2}\tilde{\lambda^{0}}
 \equiv\tilde{f}\left({\cal V}^{-\frac{14}{9}}\right).
\eeq
Similarly, outgoing right-handed electron (quark) can couple with both left handed as well as right handed sleptons(squarks) and include the gaugino$(\tilde{\lambda^{0}})$-fermion($f_R$)-sfermion(${\tilde f}_L$) vertex in a loop diagram. The same does not possess $SU(2)$ EW symmetry. For $f_{L,R}= e_{L,R}$, by expanding $g_{{\cal A}_3\bar{T}_B}$ first in the fluctuations linear in ${\cal Z}_1$ and then linear in ${\cal A}_1$ around their stabilized VEV's, we have: $ g_{{T_B} {\bar {\cal A}}_3}\rightarrow-{\cal V}^{-\frac{13}{36}} \langle {\cal Z}_1\rangle ({\cal A}_1- {\cal V}^{-\frac{2}{9}})$. The physical gaugino$(\tilde{\lambda^{0}})$- lepton$(e_{R})$-slepton$(\tilde{e_L})$ interaction vertex will be given as :
\beq
\label{eq:eReLlambda}
|C_{e_{R}\tilde{e_{L}}\tilde{\lambda^{0}}}|\equiv \frac{g_{YM}g_{{T_B} {\bar {\cal A}_3}}X^{T_B}\sim {\cal V}^{-\frac{19}{18}} {\tilde f}}{{ \sqrt{\hat{K}_{{\cal Z}_1{\bar {\cal Z}}_1}\hat{K}_{{\cal A}_1{\bar {\cal A}}_1}\hat{K}_{{\cal A}_3{\bar {\cal A}}_3}}}}{\tilde {\cal A}_1}{\bar\chi}^{\bar {\cal A}_3}\tilde{\lambda^{0}}
 \equiv\tilde{f}\left({\cal V}^{-\frac{15}{9}}\right).
\eeq
For $f_{L,R}= u_{L,R}$, one gets: $ g_{{T_B} {\bar {\cal A}}_4}\rightarrow-{\cal V}^{-\frac{13}{36}} \langle {\cal Z}_1\rangle ({\cal A}_2- {\cal V}^{-\frac{1}{3}})$, and
\beq
\label{eq:uRuLlambda}
|C_{u_{R}\tilde{u_{L}}\tilde{\lambda^{0}}}|\equiv \frac{g_{YM}g_{{T_B} {\bar {\cal A}_2}}X^{T_B}\sim {\cal V}^{-\frac{19}{18}} {\tilde f}}{{ \sqrt{\hat{K}_{{\cal Z}_1{\bar {\cal Z}}_1}\hat{K}_{{\cal A}_2{\bar {\cal A}}_2}\hat{K}_{{\cal A}_4{\bar {\cal A}}_4}}}}{\tilde {\cal A}_2}{\bar\chi}^{\bar {\cal A}_4}\tilde{\lambda^{0}}
 \equiv\tilde{f}\left({\cal V}^{-\frac{14}{9}}\right).
\eeq
 To calculate the contribution of $e_{R}-\tilde{e_R}-\tilde{\lambda^{0}}$ vertex, we expand  $g_{{\cal A}_3\bar{T}_B}$ in the fluctuations linear in ${\cal A}_3$, and obtain: $ g_{{T_B} {\bar {\cal A}}_3}\rightarrow-{\cal V}^{\frac{7}{9}} ({\cal A}_3-{\cal V}^{\frac{13}{18}}),  \partial_{{\cal A}_3}T_B \rightarrow {\cal V}^{\frac{19}{9}} ({\cal A}_3-{\cal V}^{-\frac{13}{18}})$. Utilizing this, the physical gaugino$(\tilde{\lambda^{0}})$- lepton$(e_{R})$-slepton$(\tilde{e_R})$ interaction vertex will be given as :
\beq
\label{eq:eReRlambda}
|C_{e_{R}\tilde{e_{R}}\tilde{\lambda^{0}}}|\equiv \frac{{\cal V}^{ \frac{7}{9}} {\tilde f}}{{\sqrt{\hat{K}_{{\cal A}_3{\bar {\cal A}}_3} \hat{K}_{{\cal A}_3{\bar {\cal A}}_3}}} }{\tilde {\cal A}_3}{\bar\chi}^{\bar {\cal A}_3}\tilde{\lambda^{0}}
 \equiv\tilde{f}\left({\cal V}^{-\frac{3}{5}}\right).
\eeq
Similarly, by expanding $g_{{\cal A}_4 \bar{T}_B}$ in the fluctuations linear in ${\cal A}_4$, we will have
$ g_{{T_B} {\bar {\cal A}}_4}\rightarrow-{\cal V}^{\frac{16}{9}} ({\cal A}_4-{\cal V}^{\frac{11}{9}}), \partial_{{\cal A}_4}T_B \rightarrow {\cal V}^{\frac{28}{9}} ({\cal A}_4-{\cal V}^{-\frac{11}{9}})$, and
\beq
\label{eq:uRuRlambda}
|C_{u_{R}\tilde{u_{R}}\tilde{\lambda^{0}}}|\equiv \frac{{\cal V}^{ \frac{16}{9}} {\tilde f}}{{\sqrt{\hat{K}_{{\cal A}_4{\bar {\cal A}}_4}{\hat{K}_{{\cal A}_4{\bar {\cal A}}_4}} }} }{\tilde {\cal A}_4}{\bar\chi}^{\bar {\cal A}_4}\tilde{\lambda^{0}}
 \equiv\tilde{f}\left({\cal V}^{-\frac{3}{5}}\right).
\eeq
 To determine the contribution of effective charge $Q'_i$, we need to evaluate the contribution of sfermion(${\tilde f_i}$)-photon($\gamma$)-sfermion({$\tilde f_i$}) vertices which are expressed  in terms of ${\tilde f_{L/R}}$ basis as below:
\beqn
\label{eq:Ce1e1gamma}
&& C_{{\tilde {f_1}}{\tilde {f_1}}{\gamma}}\sim D_{f_{11}} D_{f_{11}}^*C_{{\tilde {f_L}}{\tilde {f_L}^*}{\gamma}}+ (D_{f_{11}} D_{f_{12}}^*+ D_{f_{12}} D_{f_{12}}^*) C_{{\tilde {f_L}}{\tilde {f_R}^*}{\gamma}}+ D_{f_{12}} D_{f_{12}}^*C_{{\tilde {f_R}}{\tilde {f_R}^*}{\gamma}}, \nonumber\\
&& C_{{\tilde {f_2}}{\tilde {f_2}}{\gamma}}\sim D_{f_{21}} D_{f_{21}}^*C_{{\tilde {f_L}}{\tilde {f_L}^*}{\gamma}}+ ( D_{f_{21}} D_{f_{22}}^*+ D_{f_{22}} D_{f_{21}}^*)C_{{\tilde {f_L}}{\tilde {f_R}^*}{\gamma}}+ D_{f_{22}} D_{f_{22}}^*C_{{\tilde {f_R}}{\tilde {f_R}^*}{\gamma}}.\nonumber\\
\eeqn
The sfermion-sfermion-photon vertex can be evaluated from the bulk kinetic term in the ${\cal N}=1$ gauged supergravity action as given below:
 \beqn
 {\cal L}= {\frac{1}{\kappa_4^2{\cal V}^2}}G^{T_B{\bar T}_B}\tilde{\bigtriangledown}_\mu T_B\tilde{\bigtriangledown}^\mu {\bar T}_{\bar B},
  \eeqn
\begin{eqnarray}
\label{eq:sigmas_TB}
&&{\rm where}~~ \tilde{\bigtriangledown}_\mu T_B = \partial_\mu T_B + 6i \kappa_4^2\mu_7 lQ_{T_B}A_{\mu}; \nonumber\\
& & T_B \sim  \sigma_B + \left( i\kappa_{B bc}c^b{\cal B}^c + \kappa_B + \frac{i}{(\tau - {\bar\tau})}\kappa_{B bc}{\cal G}^b({\cal G}^c
- {\bar {\cal G}}^c) i\delta^B_B\kappa_4^2\mu_7l^2C_B^{I{\bar J}}a_I{\bar a_{\bar J}} + \frac{3i}{4}\delta^B_B\tau Q_{\tilde{f}}\right.\nonumber\\
 & & \left.  + i\mu_3l^2(\omega_B)_{i{\bar j}} z^i\left({\bar z}^{\bar j}-\frac{i}{2}{\bar z}^{\tilde{a}}({\bar{\cal P}}_{\tilde{a}})^{\bar j}_lz^l\right)\right).\nonumber\\
\eeqn
The form of expression that eventually leads to give the contribution of required sfermion-sfermion-photon vertex is as given below:
\begin{eqnarray}
\label{eq:sq sq gl}
& &  C_{f_{L/R} f^{*}_{L/R} \gamma} \sim {\frac{6i\kappa_4^2\mu_72\pi\alpha^\prime Q_BG^{T_B{\bar T}_B}}{\kappa_4^2{\cal V}^2}}A^\mu\partial_\mu\left(\kappa_4^2\mu_7(2\pi\alpha^\prime)^2C_{i{\bar j}}{\cal A}_i{\bar {\cal A}}_{\bar j}\right) 
\end{eqnarray}
Using $G^{T_B{\bar T}_B}(EW)\sim {\cal V}^{\frac{7}{3}}$(the large value is justified by obtaining ${\cal O}(1)$ SM fermion-fermion-photon coupling vertex in ${\cal N}=1$ gauged supergravity action; see details therein), $Q_B\sim {\cal V}^{\frac{1}{3}}{\tilde f}, \kappa_4^2\mu_7\sim \frac{1}{\cal V}$, the above expression reduces to $ |C_{f_{L/R} f^{*}_{L/R} \gamma}|\equiv  {\cal V}^{\frac{1}{3}}A^\mu\partial_\mu\left(\kappa_4^2\mu_7(2\pi\alpha^\prime)^2C_{i{\bar j}}{\tilde {\cal A}_i} {\tilde {\cal A}}_{\bar j}\right)$.
For $i=j=1$,  $\kappa_4^2\mu_7(2\pi\alpha^\prime)^2C_{1{\bar 1}}\sim {\cal V}^{\frac{10}{9}} $ as given in appendix of \cite{gravitino_DM}. Using the same
\beq
\label{eq:Celelgamma}
|C_{{\tilde {e_L}}{\tilde {e_L}^*}{\gamma}}|\equiv \frac{{\cal V}^{\frac{16}{9}} {\tilde f}} { \sqrt{\hat{K}_{{\cal A}_1{\bar {\cal A}}_1}\hat{K}_{{\cal A}_1{\bar {\cal A}}_1}} }\equiv ({\cal V}^{\frac{44}{45}}{\tilde f}){\tilde {\cal A}_1}  A^\mu\partial_\mu {\tilde {\cal A}_{1}}.
\eeq
For $i=1,j=3$;  $\kappa_4^2\mu_7(2\pi\alpha^\prime)^2C_{1{\bar 3}}\sim {\cal V}^{\frac{29}{18}}$, we have
\beq
\label{eq:CeleRgamma}
|C_{{\tilde {e_L}}{\tilde {e_R}^*}{\gamma}}|\equiv \frac{{\cal V}^{\frac{41}{18}}{\tilde f}} { \sqrt{\hat{K}_{{\cal A}_1{\bar {\cal A}}_1}\hat{K}_{{\cal A}_3{\bar {\cal A}}_3}} }\equiv ({\cal V}^{\frac{53}{45}}{\tilde f}){\tilde {\cal A}_1} A^\mu\partial_\mu {\tilde {\cal A}}_{3}.
\eeq
For $i=j=3$; $\kappa_4^2\mu_7(2\pi\alpha^\prime)^2C_{3{\bar 3}}\sim {\cal V}^{\frac{19}{9}}$ and
\beq
\label{eq:CeReRgamma}
|C_{{\tilde {e_R}}{\tilde {e_R}^*}{\gamma}}|\equiv \frac{{\cal V}^{\frac{25}{9}}{\tilde f}} { \sqrt{\hat{K}_{{\cal A}_3{\bar {\cal A}}_3}\hat{K}_{{\cal A}_3{\bar {\cal A}}_3}}}\equiv({\tilde f}{\cal V}^{\frac{62}{45}}){\tilde {\cal A}_3} A^\mu\partial_\mu {\tilde {\cal A}}_{3}.
\eeq
For i=j=2, $\kappa_4^2\mu_7(2\pi\alpha^\prime)^2C_{2{\bar 2}}\sim {\cal V}^{\frac{1}{9}}$ and
\beq
\label{eq:Cululgamma}
|C_{{\tilde {u_L}}{\tilde {u_L}^*}{\gamma}}|\equiv \frac{{\cal V}^{\frac{7}{9}}{\tilde f}}  { \sqrt{\hat{K}_{{\cal A}_2{\bar {\cal A}}_2}\hat{K}_{{\cal A}_2{\bar {\cal A}}_2}} }\equiv ({\tilde f}{\cal V}^{\frac{53}{45}}){\tilde {\cal A}_2} A^\mu\partial_\mu {\tilde {\cal A}}_{2}.
\eeq
For $i=2,j=4$,  $\kappa_4^2\mu_7(2\pi\alpha^\prime)^2C_{2{\bar 4}}\sim {\cal V}^{\frac{29}{18}}$ and
\beq
\label{eq:CuluRgamma}
|C_{{\tilde {u_L}}{\tilde {u_R}^*}{\gamma}}|\equiv \frac{{\cal V}^{\frac{41}{18}}{\tilde f}} { \sqrt{\hat{K}_{{\cal A}_2{\bar {\cal A}}_2}\hat{K}_{{\cal A}_4{\bar {\cal A}}_4}} }\equiv ({\tilde f}{\cal V}^{\frac{23}{18}}){\tilde {\cal A}_2} A^\mu\partial_\mu {\tilde {\cal A}}_{4}.
\eeq
For $i=j=4$, $\kappa_4^2\mu_7(2\pi\alpha^\prime)^2C_{4{\bar 4}}\sim {\cal V}^{\frac{28}{9}}$ and
\beq
\label{eq:CuRuRgamma}
|C_{{\tilde {u_R}}{\tilde {u_R}^*}{\gamma}}|\equiv \frac{ {\cal V}^{\frac{34}{9}}{\tilde f}} { \sqrt{\hat{K}_{{\cal A}_4{\bar {\cal A}}_4}\hat{K}_{{\cal A}_4{\bar {\cal A}}_4}}}\equiv({\tilde f}{\cal V}^{\frac{62}{45}}){\tilde {\cal A}_4} A^\mu\partial_\mu {\tilde {\cal A}}_{4}.
\eeq
Substituting the results given in eqs.~(\ref{eq:Celelgamma})-(\ref{eq:CuRuRgamma}) in equation~(\ref{eq:Ce1e1gamma}), the volume suppression factors corresponding to scalar-scalar-photon vertices are given as follows:
\beqn
\label{eq:Ce1e1gamma1}
&& C_{{\tilde {e_1}}{\tilde {e_1}}{\gamma}}\equiv {\tilde f}  \left( {\cal V}^{\frac{44}{45}}\cos^2  {\theta_e}  -  {\cal V}^{\frac{53}{45}} \cos {\theta_e} \sin {\theta_e}  (e^{i\phi_{e}} + e^{-i\phi_{e}})e^{i\phi_{g_e}} +  {\cal V}^{\frac{62}{45}}\sin^2  {\theta_e} \right) ,\nonumber\\
&&C_{{\tilde {e_2}}{\tilde {e_2}}{\gamma}}\equiv {\tilde f}  \left(  {\cal V}^{\frac{44}{45}} \sin^2  {\theta_e}  + {\cal V}^{\frac{53}{45}} \cos  {\theta_e} \sin {\theta_e}  (e^{i\phi_{e}} + e^{-i\phi_{e}}))e^{-i\phi_{g_e}}+ {\cal V}^{\frac{62}{45}} \cos^2 {\theta_e}\right), \nonumber\\
&& C_{{\tilde {u_1}}{\tilde {u_1}}{\gamma}}\equiv {\tilde f}  \left( {\cal V}^{\frac{53}{45}}\cos^2  {\theta_u}  -  {\cal V}^{\frac{23}{18}} \cos {\theta_u} \sin {\theta_e}  (e^{i\phi_{u}} + e^{-i\phi_{u}}))e^{i\phi_{g_u}}+  {\cal V}^{\frac{62}{45}}\sin^2  {\theta_u} \right),\nonumber\\
&&C_{{\tilde {u_2}}{\tilde {u_2}}{\gamma}}\equiv {\tilde f}  \left(  {\cal V}^{\frac{53}{45}} \sin^2  {\theta_u}  + {\cal V}^{\frac{23}{18}} \cos  {\theta_u} \sin {\theta_u}  (e^{i\phi_{u}} + e^{-i\phi_{u}}))e^{-i\phi_{g_u}} + {\cal V}^{\frac{62}{45}} \cos^2 {\theta_u}\right),
\eeqn
where $\phi_{g_e}$ and $\phi_{g_u}$ are phase factors associated with $C_{{\tilde {e_L}}{\tilde {e^{*}_R}}{\gamma}}$ and $C_{{\tilde {u_L}}{\tilde {u^{*}_R}}{\gamma}}$- we consider the same to be ${\cal O}(1).$
Now,  the Lagrangian relevant to couplings involved in one-loop diagram shown in Figure~4.1 is given as:
\begin{eqnarray}
\label{eq:LagLR}
{\cal L}= C_{f_L {\tilde f}^{*}_L {\tilde \lambda}^{0}_i} f_{L}\tilde{f_{L}}\tilde{\lambda^{0}}+C_{f_L {\tilde f}^{*}_R {\tilde \lambda}^{0}_i} f_{L}\tilde{f_{R}}\tilde{\lambda^{0}}+ C_{f^{*}_R {\tilde f}_L {\tilde \lambda}^{0}_i} f_{R}\tilde{f_{L}}\tilde{\lambda^{0}}+  C_{f^{*}_R {\tilde f}_R {\tilde \lambda}^{0}_i} f_{R}\tilde{f_{R}}\tilde{\lambda^{0}}.
\end{eqnarray}
where, from  eqs.~(\ref{eq:eLeLlambda})-(\ref{eq:uRuRlambda}), we have:
\beqn
\label{verticesgaugino}
&& | C_{e_L {\tilde e}^{*}_L {\tilde \lambda}^{0}_i}|\equiv{\tilde f} {\cal V}^{-1}, |C_{e_R {\tilde e}^{*}_R {\tilde \lambda}^{0}_i}|\equiv{\tilde f}{\cal V}^{-\frac{3}{5}}, |C_{e^{*}_R {\tilde e}_L {\tilde \lambda}^{0}_i}| \equiv |C_{e^{*}_L {\tilde e}_R {\tilde \lambda}^{0}_i}|\equiv{\tilde f}{\cal V}^{-\frac{15}{9}} \nonumber\\
&& |C_{u_L {\tilde u}^{*}_L {\tilde \lambda}^{0}_i}|\equiv{\tilde f} {\cal V}^{-\frac{4}{5}}, |C_{u_R {\tilde u}^{*}_R {\tilde \lambda}^{0}_i}|\equiv{\tilde f}{\cal V}^{-\frac{3}{5}}, |C_{u^{*}_R {\tilde u}_L {\tilde \lambda}^{0}_i}| \equiv |C_{u^{*}_L {\tilde u}_R {\tilde \lambda}^{0}_i}|\equiv{\tilde f}{\cal V}^{-\frac{14}{9}}.
\eeqn
Writing $\tilde{f}_L$ as well as $\tilde{f}_R$ given in equation (\ref{eq:LagLR}) in terms of diagonalized basis $\tilde{f}_1$ and $\tilde{f}_2$, the equation takes the form as of equation (\ref{eq:eff1loop2}):
\beqn
\label{eq:eff1loopg}
&& {\cal L}_{int}= \bar{\chi}_{f}
                \Bigl(( C_{{\lambda}^0_i {f_L}\tilde {f_L}} D_{f_{11}}+ C_{{\lambda}^0_i {f_L}\tilde {f_R}}  D_{f_{21}}) \PR +(C_{{\lambda}^0_i {f_R}\tilde {f_L}}  D_{f_{11}}+ C_{{\lambda}^0_i {f_R}\tilde {f_R}} D_{f_{21}}) \PL\Bigr) \nonumber\\
                &&{\phi_{f_1}}{\tilde \lambda^0}+  \bar{\chi}_{f}
                \Bigl((C_{{\lambda}^0_i {f_L}\tilde {f_L}}  D_{f_{12}}+ C^{\chi^0_i {f_L}\tilde {f_R}} D_{f_{22}}) \PR +( C_{{\lambda}^0_i {f_R}\tilde {f_L}} D_{f_{12}}+ C_{{\lambda}^0_i {f_R}\tilde {f_R}}D_{f_{22}}) \PL \Bigr) \nonumber\\
                &&{\phi_{f_2}}{\tilde \lambda^0}+ h.c..  
\eeqn
Using equation (\ref{eq:EDM}), the EDM expression will take the form:
\beqn
&& \frac{d_f}{e}|_{\lambda^{0}_i}=  \frac{m_{\tilde{\lambda}_i^0}}{{(4\pi)^2 }} \Bigl[ \frac{1}{m^{2}_{\tilde {f_1}}} {\rm Im}\left(C_{{\lambda}^0_i {f_L}\tilde {f_L}}C_{{\lambda}^0_i {f_R}\tilde {f_R}} D_{f_{11}} D_{f_{21}}^*+ C_{{\lambda}^0_i {f_L}\tilde {f_R}}C_{\chi^0_i {f_R}\tilde {f_L}} D_{f_{21}}D_{f_{11}}^*\right)\nonumber\\
&& Q'_{\tilde {f_1}} B\Bigl(\frac{m_{\tilde{\lambda}_i^0}^2}{m_{{\tilde f_1}^2}}\Bigr)+  {\frac{1}{m^{2}_{\tilde {f_2}}}} {\rm Im}\left(C_{{\lambda}^0_i {f_L}\tilde {f_L}}C_{\chi^0_i {f_R}\tilde {f_R}}  D_{f_{12}}D_{f_{22}}^*+ C_{{\lambda}^0_i {f_L}\tilde {f_R}}C_{{\lambda}^0_i {f_R}\tilde {f_L}} D_{f_{22}}D_{f_{12}}^*\right)\nonumber\\
&& Q'_{\tilde {f_2}} B\Bigl(\frac{m_{\tilde{\lambda}_i^0}^2}{m_{{\tilde f_2}^2}}\Bigr)\Bigr].
\eeqn
Considering $f_{L,R}=e_{L,R}$, incorporating results of interaction vertices as given in equation (\ref{verticesgaugino}) and using the assumption that phase factors associated with effective gauge couplings are ${\cal O}(1)$, the dominant contribution of electron EDM is given as:
\beqn
&& {\hskip -0.2in} \frac{d_e}{e}|_{\lambda^{0}_i}\equiv \frac{m_{\tilde \lambda^0} ({\tilde f}^2 {\cal V}^{-\frac{8}{5}}sin {\theta_e} \sin  {\phi_e})}{(4\pi)^2} \left[\frac{C_{{\tilde {e_2}}{\tilde {e_2}^*}{\gamma}}}{m_{\tilde {e_2}}^2}B\left(\frac{m_{\tilde {\lambda^0}}^2}{m_{{\tilde e_2}^2}}\right)-\frac{C_{{\tilde {e_1}}{\tilde {e_1}^*}{\gamma}}}{m_{\tilde {e_1}}^2}B\left(\frac{m_{\tilde {\lambda^0}}^2}{m_{{\tilde e_1}^2}}\right) \right].
\eeqn
For $f_{L,R}=u_{L,R}$, quark EDM will be given as:
\beqn
&& {\hskip -0.2in}  \frac{d_u}{e}|_{\lambda^{0}_i}\equiv \frac{m_{\tilde \lambda^0} ({\tilde f}^2 {\cal V}^{-\frac{7}{5}}sin {\theta_u} \sin  {\phi_u})}{(4\pi)^2} \left[\frac{C_{{\tilde {u_2}}{\tilde {u_2}^*}{\gamma}}}{m_{\tilde {u_2}}^2}B\left(\frac{m_{\tilde {\lambda^0}}^2}{m_{{\tilde u_2}^2}}\right)-\frac{C_{{\tilde {u_1}}{\tilde {u_1}^*}{\gamma}}}{m_{\tilde {u_1}}^2}B\left(\frac{m_{\tilde {\lambda^0}}^2}{m_{{\tilde u_1}^2}}\right) \right].
\eeqn
Putting the values \footnote{We only incorporate the volume suppression coming from $ C_{{\tilde {e_i}}{\tilde {e_i}^*}{\gamma}}$ and $C_{{\tilde {u_i}}{\tilde {u_i}^*}{\gamma}}$. The momentum dependence of both vertices have already been included in the one-loop functions $A(r)$ and $B(r)$.}  of $C_{{\tilde {e_i}}{\tilde {e_i}^*}{\gamma}}$  and $C_{{\tilde {u_i}}{\tilde {u_i}^*}{\gamma}}$ as given in equation no (\ref{eq:Ce1e1gamma1}), we get
\beqn
\label{eq:delambda}
&&  {\hskip -0.25in}\frac{d_e}{e}|_{\lambda^0}\equiv \frac{m_{\tilde \lambda^0}({\tilde f}^2 {\cal V}^{-\frac{8}{5}}sin {\theta_e} \sin  {\phi_e}) }{(4\pi)^2 }{\cal V}^{\frac{62}{45}}{\tilde f} \left[\frac{\cos^2 {\theta_e}}{m_{\tilde {e_2}}^2}B\left(\frac{m_{\tilde {\lambda^0}}^2}{m_{{\tilde e_2}^2}}\right)-\frac{\sin^2 {\theta_e}}{m_{\tilde {e_1}}^2}B\left(\frac{m_{\tilde {\lambda^0}}^2}{m_{{\tilde e_1}^2}}\right) \right], \nonumber\\
&&  {\hskip -0.25in}{\rm and}\nonumber\\
&&  {\hskip -0.25in}\frac{d_u}{e}|_{\lambda^0}\equiv \frac{m_{\tilde \lambda^0}({\tilde f}^2 {\cal V}^{-\frac{7}{5}} sin {\theta_u} \sin  {\phi_u}) }{(4\pi)^2 }{\cal V}^{\frac{62}{45}}{\tilde f} \left[\frac{\cos^2 {\theta_u}}{m_{\tilde {u_2}}^2}B\left(\frac{m_{\tilde {\lambda^0}}^2}{m_{{\tilde u_2}^2}}\right)-\frac{\sin^2 {\theta_u}}{m_{\tilde {u_1}}^2}B\left(\frac{m_{\tilde {\lambda^0}}^2}{m_{{\tilde u_1}^2}}\right) \right].
\eeqn
 Here, $\sin \theta_f=
\frac{2|M_{\tilde{f_{21}^2}}|}{\sqrt{\left(M_{\tilde{f_L}}^2-M_{\tilde{f_R}}^2\right)^2+4 M_{\tilde{f}21}^4}}$.  As discussed in chapter {\bf 2}, in our model, we have $M_{\tilde{e_L}}^2= M_{\tilde{e_R}}^2= M_{\tilde{u_L}}^2= M_{\tilde{u_R}}^2 \sim {m^{2}_0}$. Using the same, we get $\sin \theta_e=\sin \theta_u=1$.  Also, we assume $\phi_{e,u}= (0,\frac{\pi}{2}]$. As explained in equation (\ref{eq:mass_Zi}),  ${m^{2}_{\tilde {f_1}}}={m^{2}_{\tilde {f_2}}}= m_{\tilde{f_L}/\tilde{f_R}}^2 ={\cal V} m^{2}_{\frac{3}{2}}$. Utilizing the same and the value of gaugino mass $m_{\tilde {\lambda^0}}^2= {\cal V}^{\frac{4}{3}}m^{2}_{\frac{2}{3}}$, we get:
\beqn
 {\hskip -0.2in}B\Bigl(\frac{m_{\tilde {\lambda^0}}^2}{m_{{\tilde f_i}^2}}\Bigr)=\frac{1}{2\Bigl(\frac{m_{\tilde {\lambda^0}}^2}{m_{{\tilde f_i}^2}}-1\Bigr)^2}\Bigl(1+\frac{m_{\tilde {\lambda^0}}^2}{m_{{\tilde f_i}^2}}+\frac{2\frac{m_{\tilde {\lambda^0}}^2}{m_{{\tilde f_i}^2}}ln\Bigl(\frac{m_{\tilde {\lambda^0}}^2}{m_{{\tilde f_i}^2}}\Bigr)}{1-\frac{m_{\tilde {\lambda^0}}^2}{m_{{\tilde f_i}^2}}}\Bigr) \sim \frac{m_{{\tilde f_i}^2}}{m_{\tilde {\lambda^0}}^2}\sim~{\cal V}^{-\frac{1}{3}},
\eeqn
where for $i=1,2$, $f_i=({e_1},{e_2}),({u_1},{u_2})$. Incorporating the value of masses in equation (\ref{eq:delambda}), using ${\tilde f}\sim {\cal V}^{-\frac{23}{30}}$ as obtained in chapter {\bf 2}, and Calabi-Yau volume ${\cal V}\sim 10^5$, the dominant contribution of EDM of electron will be given as:
\beqn
{\hskip -0.2in} \frac{d_e}{e}|_{\lambda^0}\equiv \frac{{\cal V}^{\frac{2}{3}}m_{\frac{3}{2}}\Bigl({\tilde f}^2 {\cal V}^{-\frac{8}{5}} \Bigr)}{(4\pi)^2 } \times{\tilde f} {\cal V}^{\frac{62}{45}}\Bigl( \frac{ {\cal V}^{-\frac{1}{3}}}{{\cal V} m^{2}_{\frac{3}{2}}} \Bigr) \equiv \frac{{\tilde f}^3 {\cal V}^{\frac{2}{3}+ \frac{62}{45}-\frac{8}{5}-\frac{1}{3}-1}}{(4\pi)^2 m_{\frac{3}{2}}} \equiv 10^{-39} cm,
\eeqn
and the dominant contribution of EDM of neutron/quark will be given as:
\beqn
{\hskip -0.2in} \frac{d_n}{e}|_{\lambda^0}\equiv \frac{{\cal V}^{\frac{2}{3}}m_{\frac{3}{2}}\left({\tilde f}^2 {\cal V}^{-\frac{7}{5}} \right)}{(4\pi)^2 } \times{\tilde f} {\cal V}^{\frac{62}{45}}\left( \frac{ {\cal V}^{-\frac{1}{3}}}{{\cal V} m^{2}_{\frac{3}{2}}} \right) \equiv \frac{{\tilde f}^3 {\cal V}^{\frac{2}{3}+ \frac{62}{45}-\frac{7}{5}-\frac{1}{3}-1}}{(4\pi)^2 m_{\frac{3}{2}}} \equiv 10^{-38} cm.
\eeqn
\\
{\bf{Neutralino Contribution:}}   The physical eigenstates of neutralino mass matrix, as discussed in section {\bf 5} of chapter {\bf 2}, are given as:
 \begin{eqnarray}
\label{eq:neutralinos_I}
& & \tilde{\chi}_1^0\sim\frac{-\tilde{H}^{0}_u+\tilde{H}^{0}_d}{\sqrt{2}};\  m_{{\chi}_1^0} \sim{\cal V}^{\frac{59}{72}} m_{\frac{3}{2}},\nonumber\\
& & \tilde{\chi}_2^0\sim \left(\frac{v}{M_P}{\tilde f}{\cal V}^{\frac{5}{6}}\right)\lambda^0+\frac{\tilde{H}^{0}_u+\tilde{H}^{0}_d}{\sqrt{2}};\ m_{{\chi}_2^0}\sim{\cal V}^{\frac{59}{72}} m_{\frac{3}{2}}, \nonumber\\
&&\tilde{\chi}_3^0\sim - \lambda^0+\left(\frac{v}{M_P}{\tilde f}{\cal V}^{\frac{5}{6}}\right) \left(\tilde{H}^{0}_u+\tilde{H}^{0}_d\right); \ m_{{\chi}_3^0} \sim
{\cal V}^{\frac{2}{3}} m_{\frac{3}{2}}.
\end{eqnarray}
where v is value of Higgs VEV at electroweak scale. $\tilde{H}^{0}_u$ and $\tilde{H}^{0}_d$ correspond to SU(2)-doublet higgsino. $\tilde{\chi}_1^0$ is purely a higgsino and $\tilde{\chi}_2^0$ ($\tilde{\chi}_3^0$) are formed by linear combination of gaugino (higgsino) with a very small admixture of higgsino (gaugino). Since  neutralinos are also very heavy, we evaluate the contribution of the same to one-loop electron/neutron EDM involving heavy sfermions. Though the neutralino ($\chi^{0}_{1,2}$)-fermion-sfermion couplings are complex in this case, the phase disappears due to presence of both complex coupling as well as its conjugate in the EDM expression. Therefore, the non-zero EDM arises due to CP violating phases appearing from mass eignstates of sfermion mass matrix only.  The one-loop diagram is given in Figure~4.2.

We have already calculated the contribution of gaugino-lepton(quark)-slepton (squark) vertices in the above calculations. Now we estimate coefficients of vertices corresponding to  higgsino-lepton(quark)-slepton(squark) interaction vertices.
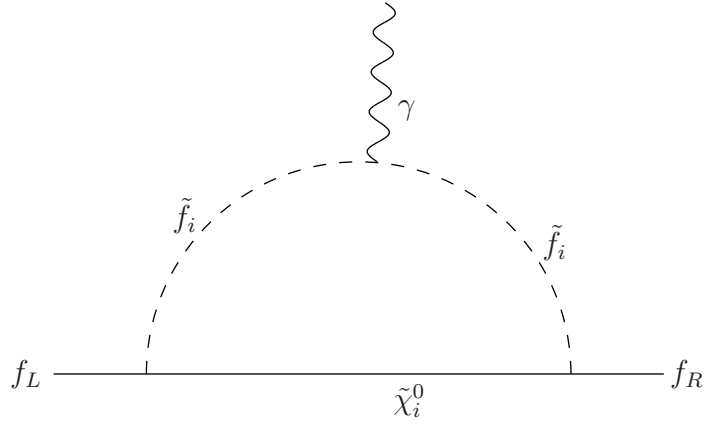
\begin{figure}
\begin{center}
\begin{picture}(145,147) (130,40)
   \Line(100,50)(330,50)
   \DashCArc(215,50)(80,0,180){5}
   \Photon(222,130)(225,190){4}{4}
    \Text(234,150)[]{{$\gamma$}}
   \Text(150,110)[]{${\tilde f}_{i}$}
   \Text(290,100)[]{${\tilde f}_{i}$}
   \Text(90,50)[]{{$f_{L}$}}
   \Text(340,50)[]{{$f_{R}$}}
   \Text(235,40)[]{{{$\tilde{\chi}_i^0$}}}
   \end{picture}
\end{center}
\caption{One-loop diagram involving neutralinos.}
 \end{figure}
In ${\cal N}=1$ gauged supergravity, higgsino-fermion-sfermion interaction is governed by \cite{Wess_Bagger}:
$ {\cal L}_{f \tilde{f} \tilde{H^{0}}}= \frac{e^{\frac{K}{2}}}{2}\left({\cal D}_iD_{\bar J}W\right)\chi^i_{L}{\chi^{j}_L}^{c}+  ig_{i{\bar J}}{\bar\chi}^{\bar J}\Bigl[{\bar\sigma}\cdot\partial\chi^i+\Gamma^i_{Lk}{\bar\sigma}\cdot\partial a^L\chi^k
+\frac{1}{4}\left(\partial_{a_L}K{\bar\sigma}\cdot a_L - {\rm c.c.}\right)\chi^i\Bigr]$. Using this, we evaluate the coefficients of higgsino-lepton(quark)-slepton(squark) interaction vertices. For an incoming electron ($e^{-}_L$) interacting with slepton as well as neutralino, the contribution of higgsino-lepton ($e^{-}_L$)-slepton (${\tilde e_L}$) vertex in gauged supergravity action of Wess and Bagger \cite{Wess_Bagger}, is given by: $\frac{e^{\frac{K}{2}}}{2}\left({\cal D}_{{\cal Z}_1}D_{\bar {\cal A}_1}W\right)\chi^{{\cal A}_i}{\chi^{c^{{\cal Z}_I}}} +ig_{{\bar I}{ {{\cal A}_1}}}{\bar\chi}^{Z_i}\left[{\bar\sigma}\cdot\partial{{\chi^{{\cal A}_1}}}+\Gamma^{{\cal A}_1}_{{\cal A}_1{\bar A_1}}{\bar\sigma}\cdot\partial {\cal A}_1{\chi^{{\cal A}_1}}
+\frac{1}{4}\left(\partial_{{\cal A}_3}K{\bar\sigma}\cdot {\cal A}_1 - {\rm c.c.}\right){\chi^{{\cal {A}}_1}}\right]$, $\chi^{\cal {Z}}$ and $\chi^{c^{{\cal Z}_1}}$ correspond to $SU(2)_L$ higgsino and its charge conjugate, $\chi^{{\cal A}_1}$ corresponds to $SU(2)_L$ electron and ${\tilde {\cal A}_1}$ corresponds to left-handed slepton. In diagonalized set of basis, $g_{ I{\bar {\cal A}_1}}=0$. Since  $SU(2)$ EW symmetry is not conserved for higgsino-lepton-slepton vertex, therefore to calculate the contribution of same, we  generate a term of the type $e_L \tilde{e}_L{\tilde H^{c}}_L H_L$ wherein $e_L$ and $H_L$ are respectively the $SU(2)_L$ electron and Higgs doublets, $\tilde{l}_L$ is also an $SU(2)_L$ doublet and ${\tilde H^{c}}_L$ is $SU(2)_L$  higgsino doublet. After giving VEV to one of the Higgs doublet $H_L$, one gets required vertex. By considering $a_1\rightarrow a_1+{\cal V}^{-\frac{2}{9}}{M_P}$, and further picking up the component of $ {\cal D}_i D_{{\bar a}_1}W$ linear in $z_i$ as well as linear in fluctuation $ (a_1-{\cal V}^{-\frac{2}{9}}{M_P})$, we see that $e^{\frac{K}{2}}{\cal D}_i D_{{\bar a}_1}W \sim  {\cal V}^{-\frac{31}{18}} z_i (a_1-{\cal V}^{-\frac{2}{9}}{M_P})$. As was shown in chapter {\bf 2}, $ e^{\frac{K}{2}}{\cal D}_I D_{\bar {{\cal A}_1}}W \sim {\cal O}(1) e^{\frac{K}{2}}{\cal D}_i D_{{\bar a}_1}W$. Utilizing the same, magnitude of the physical  higgsino($\tilde {H^{c}_L}$)-lepton (${e_L}$)-slepton(${\tilde {e_L}}$) vertex after giving  VEV to ${\cal Z}_I$  will be given as :
\begin{eqnarray}
\label{eq:elelZi}
& &  |C_{{\tilde H^{c}}_L {e_L} \tilde {e_L}}|\equiv\frac{{\cal V}^{-\frac{31}{18}}\langle{\cal Z}_I\rangle}{{\sqrt{\hat{K}^{2}_{{\cal Z}_1{\bar{\cal Z}}_1}{\hat{K}_{{\cal A}_1{\bar {\cal A}}_1}}{\hat{K}_{{\cal A}_1{\bar {\cal A}}_1}}}}}{\tilde {\cal A}_1}\chi^{c^{{\cal Z}_I}}{\chi^{{\cal A}_1}} \equiv {\cal V}^{-\frac{3}{2}}.
\end{eqnarray}
The coefficient of  higgsino($\tilde {H^{c}_L}$)-lepton(${u_L}$)-slepton(${\tilde {u_L}}$) vertex can be determined by expanding ${\cal D}_I D_{\bar {{\cal A}_2}}W$ linear in ${\cal Z}_I$ as well as linear in fluctuation $ ({\cal A}_2-{\cal V}^{-\frac{1}{3}}{M_P})$. The magnitude of the same has already been calculated in chapter {\bf 2} and given as:
\begin{eqnarray}
\label{eq:ululZi}
& &  |C_{{\tilde H^{c}}_L {u_L} \tilde {u_L}}|\equiv\frac{{\cal V}^{-\frac{20}{9}}\langle{\cal Z}_i\rangle}{{\sqrt{\hat{K}^{2}_{{\cal Z}_1{\bar{\cal Z}}_1}{\hat{K}_{{\cal A}_2{\bar {\cal A}}_2}}{\hat{K}_{{\cal A}_2{\bar {\cal A}}_2}}}}}{\tilde {\cal A}_2}\chi^{c^{{\cal Z}_I}}{\chi^{{\cal A}_2}} \equiv {\cal V}^{-\frac{4}{5}}.
\end{eqnarray}
To determine the contribution of higgsino-lepton($e^{-}_L$)-slepton(${\tilde e_R}$) vertex, one needs to expand  $\frac{e^{\frac{K}{2}}}{2}\left({\cal D}_{{\cal Z}_I}D_{\bar {\cal A}_1}W\right) $ in the fluctuations linear in ${\cal A}_3$ about its stabilized value. Considering  $a_3\rightarrow a_3+{\cal V}^{-\frac{13}{18}}{M_P}$ and picking up the component of ${\cal D}_I D_{\bar {a_1}}W$ linear in $a_3$, we have: $e^{\frac{K}{2}}{\cal D}_i D_{{\cal A}_1}W\equiv e^{\frac{K}{2}}{\cal D}_I D_{\bar {a_1}}W\sim {\cal V}^{-\frac{43}{36}} (a_3-{\cal V}^{-\frac{13}{18}}{M_P})$ the contribution of physical higgsino($\tilde H^{c}_L$)-lepton(${e_L}$)-slepton(${\tilde {e_R}}$) vertex  will be given as :
\begin{eqnarray}
\label{eq:eleRZi}
& &  |C_{{\tilde H^{c}}_L {e_L} \tilde {e_R}}|\equiv\frac{{\cal V}^{-\frac{43}{36}} }{{\sqrt{{\hat{K}_{{\cal Z}_1{\bar {\cal Z}}_1}}{\hat{K}_{{\cal A}_1{\bar {\cal A}}_1}}{\hat{K}_{{\cal A}_3{\bar {\cal A}}_3}}}}}{\tilde {\cal A}_3}\chi^{Z_i}{\chi^{c^{{\cal A}_1}}}\equiv{\cal V}^{-\frac{9}{5}}.
\end{eqnarray}
Similarly, one can calculate higgsino($\tilde H^{c}_L$)-lepton (${u_L}$)-slepton(${\tilde {u_R}}$) vertex by expanding $\frac{e^{\frac{K}{2}}}{2}\left({\cal D}_{{\cal Z}_I}D_{\bar {\cal A}_2}W\right)$ in the fluctuations linear in ${\cal A}_4$ about its stabilized value. Considering  $a_4\rightarrow a_4+{\cal V}^{-\frac{11}{9}}{M_P}$ and picking up the component of ${\cal D}_i D_{\bar {a_1}}W$ linear in $a_4$, we have:  $\frac{e^{\frac{K}{2}}}{2}\left({\cal D}_{{\cal Z}_I}D_{\bar {\cal A}_2}W\right)\equiv e^{\frac{K}{2}}{\cal D}_i D_{\bar {a_1}}W\sim {\cal V}^{-\frac{43}{36}} (a_4-{\cal V}^{-\frac{11}{9}}{M_P})$, the coefficient of higgsino($\tilde H^{c}_L$)-lepton(${u_L}$)-slepton(${\tilde {u_R}}$) vertex will be given as:
\begin{eqnarray}
\label{eq:uluRZi}
& &  |C_{{\tilde H^{c}}_L {u_L} \tilde {u_R}}|\equiv\frac{{\cal V}^{-\frac{43}{36}} }{{\sqrt{{\hat{K}_{{\cal Z}_1{\bar {\cal Z}}_1}}{\hat{K}_{{\cal A}_2{\bar {\cal A}}_2}}{\hat{K}_{{\cal A}_4{\bar {\cal A}}_4}}}}}{\tilde {\cal A}_4}\chi^{Z_i}{\chi^{c^{{\cal A}_2}}} \equiv{\cal V}^{-\frac{5}{3}}.\end{eqnarray}
For an outgoing electron $e^{-}_R$ interacting with slepton as well as neutralino, the  contribution of  higgsino-lepton ($e^{-}_R$)-slepton(${\tilde e_L}$) vertex is given by expanding  $\frac{e^{\frac{K}{2}}}{2}\left({\cal D}_{{\cal Z}_1}D_{{\cal A}_3}W\right) $ linear in ${\cal A}_1$. Considering  $a_1\rightarrow a_1+{\cal V}^{-\frac{2}{9}}{M_P}$ and picking up the component of ${\cal D}_i D_{a_3}W$ linear in $a_1$, we have: $e^{\frac{K}{2}}{\cal D}_I D_{{\cal A}_3}W\equiv e^{\frac{K}{2}}{\cal D}_i D_{a_3}W \sim {\cal V}^{-\frac{43}{36}} (a_1-{\cal V}^{-\frac{2}{9}}{M_P})$. The contribution of physical  higgsino($\tilde H^{c}_L$)-lepton (${e_R}$)-slepton(${\tilde {e_L}}$) vertex  will be given as :
\begin{eqnarray}
\label{eq:eRelZi}
& &  |C_{{\tilde H^{c}}_L {e_R} \tilde {e_L}}|\equiv\frac{{\cal V}^{-\frac{37}{72}}}{{\sqrt{\hat{K}_{{\cal Z}_1{\bar{\cal Z}}_1}{\hat{K}_{{\cal A}_1{\bar {\cal A}}_1}}{\hat{K}_{{\cal A}_3{\bar {\cal A}}_3}}}}}{\tilde {\cal A}_1}\chi^{Z_i}{\chi^{c^{{\cal A}_3}}} \equiv {\cal V}^{-\frac{9}{5}}.\end{eqnarray}
Similarly, considering  $a_4\rightarrow a_4+{\cal V}^{-\frac{11}{9}}{M_P}$ and picking up the component of above term linear in $a_1$, we have, $e^{\frac{K}{2}}{\cal D}_I D_{{\cal A}_4}W\equiv e^{\frac{K}{2}}{\cal D}_i D_{a_4}W\sim {\cal V}^{-\frac{43}{36}} (a_2-{\cal V}^{-\frac{1}{3}}{M_P})$, the contribution of physical  higgsino($\tilde H_L$)-quark (${u_R}$)-squark(${\tilde {u_L}}$) vertex is given as :
\begin{eqnarray}
\label{eq:uRulZi}
& &  |C_{{\tilde H}_L {u_R} \tilde {u_L}}|\equiv\frac{{\cal V}^{-\frac{43}{36}}}{{\sqrt{\hat{K}_{{\cal Z}_1{\bar{\cal Z}}_1}{\hat{K}_{{\cal A}_2{\bar {\cal A}}_2}}{\hat{K}_{{\cal A}_4{\bar {\cal A}}_4}}}}}{\tilde A_2}\chi^{Z_i}{\chi^{c^{{\cal A}_4}}} \equiv {\cal V}^{-\frac{5}{3}}.
\end{eqnarray}
The higgsino-lepton($e^{-}_R$)-slepton(${\tilde e_R}$) vertex also does not possess
  $SU(2)$ EW symmetry. Therefore, to calculate the contribution of same, we  generate a term of the type $e_R \tilde{e}_R{\tilde H}_L H_L$, where $H_L$ is one of the $SU(2)_L$ Higgs doublets. Thereafter, we expand  $\frac{e^{\frac{K}{2}}}{2}\left({\cal D}_{{\cal Z}_1}D_{\bar {\cal A}_3}W\right)\chi^{{\cal Z}_i}{\chi^{ {{\cal A}}_3}}$ linear in ${\cal Z}_1$ and then linear in ${\cal A}_3$ about their stabilized VEV's. Considering  $a_3\rightarrow a_3+{\cal V}^{-\frac{13}{18}}{M_P}$ and further picking up the component linear in $z_i$ as well as linear in fluctuation $(a_3-{\cal V}^{-\frac{2}{9}}{M_P})$, we get: $ e^{\frac{K}{2}}{\cal D}_i D_{{\cal A}_3}W\equiv e^{\frac{K}{2}}{\cal D}_i D_{{\bar a}_3}W\sim {\cal V}^{-\frac{13}{18}} \langle z_i \rangle (a_3-{\cal V}^{-\frac{13}{18}}{M_P})$. The magnitude of physical higgsino($\tilde H_L$)-lepton (${e_R}$)-slepton(${\tilde {e_R}}$) vertex after giving VEV to ${\cal Z}_I$  is given as :
\begin{eqnarray}
\label{eq:eReRZi}
& &  |C_{{\tilde H}_L {e_R} \tilde {e_R}}| \equiv\frac{{\cal V}^{-\frac{13}{18}}\langle{\cal Z}_i\rangle}{{\sqrt{\hat{K}^{2}_{{\cal Z}_1{\bar{\cal Z}}_1}{\hat{K}_{{\cal A}_3{\bar {\cal A}}_3}}{\hat{K}_{{\cal A}_3{\bar {\cal A}}_3}}}}}{\tilde {\cal A}_3}\chi^{Z_I}{\chi^{ {\cal A}_3}} \equiv {\cal V}^{-\frac{10}{9}}.
\end{eqnarray}
 The contribution of higgsino-quark($u_R$)-squark(${\tilde u_R}$) vertex has already been evaluated in chapter {\bf 3} by expanding  $\frac{e^{\frac{K}{2}}}{2}\left({\cal D}_{{\cal Z}_I}D_{{\cal A}_4}W\right)\chi^{{\cal Z}_I}{\chi^{ {{\cal A}}_4}}$ in the fluctuations linear in ${\cal Z}_I$ as well as ${\cal A}_4$ about their stabilized VEV's. The magnitude of the same is given as:
 \begin{eqnarray}
\label{eq:uRuRZi}
& &  |C_{{\tilde H}_L {u_R} \tilde {u_R}}|\equiv\frac{{\cal V}^{\frac{5}{18}}\langle{\cal Z}_i\rangle}{{\sqrt{\hat{K}^{2}_{{\cal Z}_1{\bar{\cal Z}}_1}{\hat{K}_{{\cal A}_4{\bar {\cal A}}_4}}{\hat{K}_{{\cal A}_4{\bar {\cal A}}_4}}}}}{\tilde {\cal A}_4}\chi^{Z_I}{\chi^{ {\cal A}_4}} \equiv {\cal V}^{-\frac{10}{9}}.
\end{eqnarray}
 The results of coefficients of both slepton(squark)-lepton(quark)-higgsino as  given in set of eqs.~(\ref{eq:elelZi})-(\ref{eq:uRuRZi}) are as follows:
\beqn
\label{eq:Higgs+gaug}
&& |C_{{\tilde H^{c}}_L {e_L} \tilde {e_L}}|\equiv {\cal V}^{-\frac{3}{2}},|C_{{\tilde H^{c}}_L {e_L} \tilde {e_R}}| \equiv |C_{{\tilde H}_L {e_R} \tilde {e_L}}|\equiv {\cal V}^{-\frac{9}{5}}, |C_{{\tilde H}_L {e_R} \tilde {e_R}}|\equiv {\cal V}^{-\frac{10}{9}},\nonumber\\
&& |C_{{\tilde H^{c}}_L {u_L} \tilde {u_L}}|\equiv {\cal V}^{-\frac{4}{5}}, |C_{{\tilde H^{c}}_L {u_L} \tilde {u_R}}| \equiv |C_{{\tilde H}_L {u_R} \tilde {u_L}}|\equiv{\cal V}^{-\frac{5}{3}}, |C_{{\tilde H}_L {u_R}|\tilde {u_R}}|\equiv {\cal V}^{-\frac{10}{9}}.
\eeqn
Utilizing the aforementioned results and the results of various gaugino-fermion-sfermion vertices as given in equation (\ref{verticesgaugino}), and  by adding the contribution of same as according to equation (\ref{eq:neutralinos_I}), the volume suppression factors coming from the neutralino-lepton-slepton vertices are given as:
\beqn
\label{eq:chiLR}
&& |C_{\chi^0_1 {e_L}\tilde {e_L}}|=|C_{\chi^0_2 {e_L}\tilde {e_L}}|\equiv {\cal V}^{-\frac{3}{2}}, |C_{\chi^0_3 {e_L}\tilde {e_L}}|\equiv {\tilde f}{\cal V}^{-1}, |C_{\chi^0_1 {e_L}\tilde {e_R}}|=|C_{\chi^0_2 {e_L}\tilde {e_R}}|\equiv {\cal V}^{-\frac{9}{5}},  \nonumber\\
&&|C_{\chi^0_3 {e_L}\tilde {e_R}}|\equiv  {\tilde f} {\cal V}^{-\frac{15}{9}}, | C_{\chi^0_1 {e_R}\tilde {e_L}}|=|C_{\chi^0_2 {e_R}\tilde {e_L}}|\equiv {\cal V}^{-\frac{9}{5}}, |C_{\chi^0_3 {e_R}\tilde {e_L}}|\equiv{\tilde f} {\cal V}^{-\frac{15}{9}}, \nonumber\\
&& | C_{\chi^0_1 {e_R}\tilde {e_R}}|=|C_{\chi^0_2 {e_R}\tilde {e_R}}|\equiv {\cal V}^{-\frac{10}{9}}, |C_{\chi^0_3 {e_R}\tilde {e_R}}|\equiv {\tilde f} {\cal V}^{-\frac{3}{5}}.
\eeqn
The volume suppression factors coming from the neutralino-quark-squark vertices are given as:
\beqn
\label{eq:chiLRu}
&& |C_{\chi^0_1 {u_L}\tilde {u_L}}|=|C_{\chi^0_2 {u_L}\tilde {u_L}}|\equiv {\cal V}^{-\frac{4}{5}}, |C_{\chi^0_3 {u_L}\tilde {u_L}}|\equiv {\tilde f}{\cal V}^{-\frac{4}{5}}, |C_{\chi^0_1 {u_L}\tilde {u_R}}|=|C_{\chi^0_2 {u_L}\tilde {u_R}}|\equiv{\cal V}^{-\frac{5}{3}}, \nonumber\\
&&|C_{\chi^0_3 {u_L}\tilde {u_R}}|\equiv  {\tilde f} {\cal V}^{-\frac{14}{9}}  |C_{\chi^0_1 {u_R}\tilde {u_L}}|=|C_{\chi^0_2 {u_R}\tilde {u_L}}|\equiv {\cal V}^{-\frac{5}{3}}, |C_{\chi^0_3 {u_R}\tilde {u_L}}|\equiv {\tilde f} {\cal V}^{-\frac{14}{9}}, \nonumber\\
&&| C_{\chi^0_1 {u_R}\tilde {u_R}}|=|C_{\chi^0_2 {u_R}\tilde {u_R}}|\equiv {\cal V}^{-\frac{10}{9}}, |C_{\chi^0_3 {u_R}\tilde {u_R}}|\equiv {\tilde f} {\cal V}^{-\frac{3}{5}}.
\eeqn
The interaction Lagrangian governing the neutralino-slepton(squark)-lepton(quark) interaction can be written as:
\begin{eqnarray}
\label{eq:LagLR1}
{\cal L}= \sum_{i=1,3} C_{\chi^0_i {f_L}\tilde {f_L}} f_{L}\tilde{f_{L}}\chi^0_i + C_{\chi^0_i {f_L}\tilde {f_R}} f_{L}\tilde{f_{R}}\chi^0_i +C_{\chi^0_i {f_R}\tilde {f_L}} f_{R}\tilde{f_{L}}\chi^0_i +C_{\chi^0_i {f_R}\tilde {f_R}}f_{ R}\tilde{f_{R}}\chi^0_i,\nonumber\\
\end{eqnarray}
 where f=(e,u). Rewriting $f_L$ as well as  $f_R$ in term of diagonalized basis states  $f_1$ and $f_2$,  the equation takes the form as of equation (\ref{eq:eff1loop2}):
\beqn
\label{eq:eff1loop1}
&& {\cal L}_{int}= \bar{\chi}_{f}
                \left(( C_{\chi^0_i {f_L}\tilde {f_L}} D_{f_{11}}+ C_{\chi^0_i {f_L}\tilde {f_R}}  D_{f_{21}}) \PR +(C_{\chi^0_i {f_R}\tilde {f_L}}  D_{f_{11}}+ C_{\chi^0_i {f_R}\tilde {f_R}} D_{f_{21}}) \PL\right)  \nonumber\\
                && {\phi_{f_1}}\chi^0_i+\bar{\chi}_{f}
                \left((C_{\chi^0_i {f_L}\tilde {f_L}}  D_{f_{12}}+ C_{\chi^0_i {f_L}\tilde {f_R}} D_{f_{22}}) \PR +( C_{\chi^0_i {f_R}\tilde {f_L}} D_{f_{12}}+ C_{\chi^0_i {f_R}\tilde {f_R}}D_{f_{22}}) \PL \right)  \nonumber\\
                && {\phi_{f_2}}\chi^0_i.  \eeqn
Using equation (\ref{eq:EDM}), dipole moment contribution will follow:
\beqn
&&\frac{d_f}{e}|_{\chi_i}=   \sum_{i=1,3}\frac{m_{\tilde{\chi}_i^0}}{{(4\pi)^2 }} \Bigl[ \frac{1}{m^{2}_{\tilde {f_1}}} {\rm Im}\left(C_{\chi^0_i {f_L}\tilde {f_L}}C_{\chi^0_i {f_R}\tilde {f_R}} D_{f_{11}} D_{f_{21}}^*+ C_{\chi^0_i {f_L}\tilde {f_R}}C_{\chi^0_i {f_R}\tilde {f_L}} D_{f_{21}}D_{f_{11}}^*\right) \nonumber\\
&& Q'_{\tilde {f_1}} B\Bigl(\frac{m_{\tilde{\chi}_i^0}^2}{m_{{\tilde f_1}^2}}\Bigr)+ {\frac{1}{m^{2}_{\tilde {f_2}}}} {\rm Im}\left(C_{\chi^0_i {f_L}\tilde {f_L}}C_{\chi^0_i {f_R}\tilde {f_R}}  D_{f_{12}}D_{f_{22}}^*+ C_{\chi^0_i {f_L}\tilde {f_R}}C_{\chi^0_i {f_R}\tilde {f_L}} D_{f_{22}}D_{f_{12}}^*\right) \nonumber\\
&&  Q'_{\tilde {f_2}} B\Bigl(\frac{m_{\tilde{\chi}_i^0}^2}{m_{{\tilde f_2}^2}}\Bigr)\Bigr].
\eeqn
 Using the  values of first generation scalar/slepton mass $m_{{\tilde f_1}} ={\cal V}^{\frac{1}{2}}m_{\frac{3}{2}}$ and $m_{\tilde{\chi}_1^0}=m_{\tilde{\chi}_2^0}={\cal V}^{\frac{59}{72}}m_{\frac{3}{2}}$, $m_{\tilde{\chi}_3^0}={\cal V}^{\frac{2}{3}}m_{\frac{3}{2}}$; one gets:
 \beqn
&& B\Bigl(\frac{m_{\tilde{\chi}_2^0}^2}{m_{{\tilde f_i}^2}}\Bigr)= B\Bigl(\frac{m_{\tilde{\chi}_1^0}^2}{m_{{\tilde f_i}^2}}\Bigr)=\frac{1}{2\Bigl(\frac{m_{\tilde{\chi}_1^0}^2}{m_{{\tilde f_i}^2}}-1\Bigr)^2}\Bigl(1+\frac{m_{\tilde{\chi}_1^0}^2}{m_{{\tilde f_i}^2}}+\frac{m_{\tilde{\chi}_1^0}^2}{m_{{\tilde f_i}^2}}ln \Bigl(\frac{m_{\tilde{\chi}_1^0}^2}{m_{{\tilde f_i}^2}}\Bigr)\Bigr)\Bigl({1-\frac{m_{\tilde{\chi}_1^0}^2}{m_{{\tilde f_i}^2}}}\Bigr) \nonumber\\
&& \sim \frac{m_{{\tilde f_i}^2}}{m_{\tilde{\chi}_1^0}^2} \sim\frac{1}{{\cal V}^{\frac{23}{36}}}, \nonumber\\
&& B\Bigl(\frac{m_{\tilde{\chi}_3^0}^2}{m_{{\tilde f_i}^2}}\Bigr)=\frac{1}{2\Bigl(\frac{m_{\tilde{\chi}_3^0}^2}{m_{{\tilde f_i}^2}}-1\Bigr)^2}\Bigl(1+\frac{m_{\tilde{\chi}_3^0}^2}{m_{{\tilde f_i}^2}}+\frac{m_{\tilde{\chi}_3^0}^2}{m_{{\tilde f_i}^2}}ln \Bigl(\frac{m_{\tilde{\chi}_3^0}^2}{m_{{\tilde f_i}^2}}\Bigr)\Bigr)\Bigl({1-\frac{m_{\tilde{\chi}_3^0}^2}{m_{{\tilde f_i}^2}}}\Bigr)  \nonumber\\
&& \sim \frac{m_{{\tilde f_i}^2}}{m_{\tilde{\chi}_3^0}^2}={\cal V}^{-\frac{1}{3}}.
\eeqn
Utilizing above and the results of   $C_{\chi^0_i {e_{L/R}}\tilde {e_{L/R}}}$ as given in  equation (\ref{eq:chiLR}), and further simplifying,  dominant contribution of EDM of electron will be given as\footnote{We use the assumption that the complex phases appearing in effective Yukawa couplings are of O(1).}:
\beqn
&& \frac{d_e}{e}|_{\chi_i}\equiv \frac{{\cal V}^{\frac{59}{72}}m_{\frac{3}{2}}\left(  {\cal V}^{-\frac{8}{3}} \sin {\theta_e} \sin {\phi_e} \right)}{(4\pi)^2 {\cal V}^{\frac{23}{36}} } \left[\frac{C_{{\tilde {e_2}}{\tilde {e_2}^*}{\gamma}}}{m_{\tilde {e_2}}^2} -\frac{C_{{\tilde {e_1}}{\tilde {e_1}^*}{\gamma}}}{m_{\tilde {e_1}}^2} \right]. 
\eeqn
Similarly, using results of   $C^{\chi^0_i {u_{L/R}}\tilde {u_{L/R}}}$ as given in  equation (\ref{eq:chiLRu}), the  dominant contribution of EDM of quark will be given as:
\beqn
&& \frac{d_u}{e}|_{\chi_i}\equiv \frac{{\cal V}^{\frac{59}{72}}m_{\frac{3}{2}}\left( {\cal V}^{-\frac{17}{9}} \sin {\theta_e} \sin {\phi_e} \right)}{(4\pi)^2  {\cal V}^{\frac{23}{36}}} \left[\frac{C_{{\tilde {u_2}}{\tilde {u_2}^*}{\gamma}}}{m_{\tilde {u_2}}^2} -\frac{C_{{\tilde {u_1}}{\tilde {u_1}^*}{\gamma}}}{m_{\tilde {u_1}}^2}  \right].  
\eeqn

Incorporating the value of $ C_{{\tilde {e_i}}{\tilde {e_i}}{\gamma}}$ from equation no (\ref{eq:Ce1e1gamma1}) , one gets
\beqn
&&\frac{d_e}{e}|_{\chi_i}\equiv \frac{{\cal V}^{\frac{59}{72}}m_{\frac{3}{2}}\left({\cal V}^{-\frac{8}{3}} \sin {\theta_e} \sin {\phi_e}\right)}{(4\pi)^2  {\cal V}^{\frac{23}{36}}}{\cal V}^{\frac{62}{45}}{\tilde f}\left[  \frac{\cos^2 {\theta_e}}{m_{\tilde{e_2}}^2} -\frac{\sin^2  {\theta_e} }{m_{\tilde {e_1}}^2} \right],\nonumber\\
&& {\hskip -0.4in}{\rm and}~~\frac{d_u}{e}|_{\chi_i}\equiv \frac{{\cal V}^{\frac{59}{72}}m_{\frac{3}{2}}\left({\cal V}^{-\frac{17}{9}} \sin {\theta_u} \sin {\phi_u}\right)}{(4\pi)^2  {\cal V}^{\frac{23}{36}} }{\cal V}^{\frac{62}{45}}{\tilde f}\left[  \frac{\cos^2 {\theta_u}}{m_{\tilde{u_2}}^2} -\frac{\sin^2  {\theta_u} }{m_{\tilde {u_1}}^2} \right]. 
\eeqn
Incorporating value of $\sin {\theta_{e}}=\sin {\theta_{u}}= 1$, $\sin{\phi_e}=\sin{\phi_u}= (0,1]$, ${\tilde f}\sim {\cal V}^{-\frac{23}{30}}$, and value of scalar masses $m_{{\tilde e_i}}=m_{{\tilde u_i}}= {\cal V}^{\frac{1}{2}}m_{\frac{3}{2}}$, the numerical value of EDM of electron for this case will be:
 \beqn
\hskip -0.2in\frac{d_e}{e}|_{\chi_i}\equiv \frac{{\cal V}^{\frac{59}{72}}m_{\frac{3}{2}}}{(4\pi)^2{\cal V}^{\frac{23}{36}}  }({\tilde f}{\cal V}^{\frac{62}{45}})\times{\cal V}^{-\frac{8}{3}}\left( \frac{1}{{\cal V} m^{2}_{\frac{3}{2}}}\right) \equiv \frac{{\tilde f} {\cal V}^{\frac{59}{72}+ \frac{62}{45}-\frac{8}{3}-\frac{23}{36}-1}}{(4\pi)^2 m_{\frac{3}{2}}} \equiv 10^{-37} cm. \nonumber\\
\eeqn
and the numerical value of EDM of neutron/quark will be:
 \beqn
\hskip -0.2in \frac{d_n}{e}|_{\chi_i}\equiv \frac{{\cal V}^{\frac{59}{72}}m_{\frac{3}{2}}}{(4\pi)^2{\cal V}^{\frac{23}{36}}  }({\tilde f} {\cal V}^{\frac{62}{45}})\times{\cal V}^{-\frac{17}{9}}\left( \frac{1}{{\cal V} m^{2}_{\frac{3}{2}}}\right) \sim \frac{{\tilde f} {\cal V}^{\frac{59}{72}+ \frac{62}{45}-\frac{17}{9}-\frac{23}{36}-1}}{(4\pi)^2 m_{\frac{3}{2}}} \equiv 10^{-34} cm.\nonumber\\
\eeqn

{{\bf R-parity violating vertices contribution:}}
 \begin{figure}
\begin{center}
\begin{picture}(250,270)(70,10)
    \Line(10,50)(25,50)
   \DashLine(25,50)(135,50){5}
    \Line(135,50)(150,50)
   \CArc(80,50)(50,0,180)
   \Photon(80,100)(80,140){4}{4}
   \Text(80,50)[]{$\times$}
   \Text(40,100)[]{$u^{c}$}
   \Text(120,100)[]{$u^{c}$}
   \Text(0,40)[]{{$e_{L}$}}
   \Text(160,40)[]{{$e_{R}$}}
   \Text(80,40)[]{{{${\tilde d}$}}}
   \Text(80,20)[]{$(c)$}
     \Line(250,50)(265,50)
   \DashLine(265,50)(375,50){5}
    \Line(375,50)(390,50)
   \CArc(320,50)(50,0,180)
   \Photon(320,100)(320,140){4}{4}
   \Text(320,50)[]{$\times$}
    \Text(280,100)[]{${\tilde d}$}
   \Text(360,100)[]{${\tilde d} $}
   \Text(240,40)[]{{$e_{L}$}}
   \Text(400,40)[]{{$e_{R}$}}
   \Text(320,40)[]{{{$u^c$}}}
    \Text(320,20)[]{$(d)$}
    \Line(10,200)(150,200)
   \DashCArc(80,200)(50,0,180){5}
   \Photon(80,250)(80,290){4}{4}
   \Text(100,247)[]{$\times$}
   \Text(40,250)[]{${\tilde u}^{c} $}
   \Text(120,250)[]{${\tilde u}^{c} $}
   \Text(0,190)[]{{$e_{L}$}}
   \Text(160,190)[]{{$e_{R}$}}
   \Text(80,190)[]{{{$d$}}}
   \Text(80,170)[]{$(a)$}
   \Line(250,200)(390,200)
    \DashCArc(320,200)(50,0,180){5}
   \Photon(320,250)(320,290){4}{4}
   \Text(340,247)[]{$\times$}
    \Text(280,250)[]{${\tilde d}$}
   \Text(360,250)[]{${\tilde d} $}
   \Text(240,190)[]{{$e_{L}$}}
   \Text(400,190)[]{{$e_{R}$}}
   \Text(320,190)[]{{{$u^c$}}}
    \Text(320,170)[]{$(b)$}
   \end{picture}
\end{center}
\caption{One-loop diagrams involving R-parity violating couplings.}
 \end{figure}
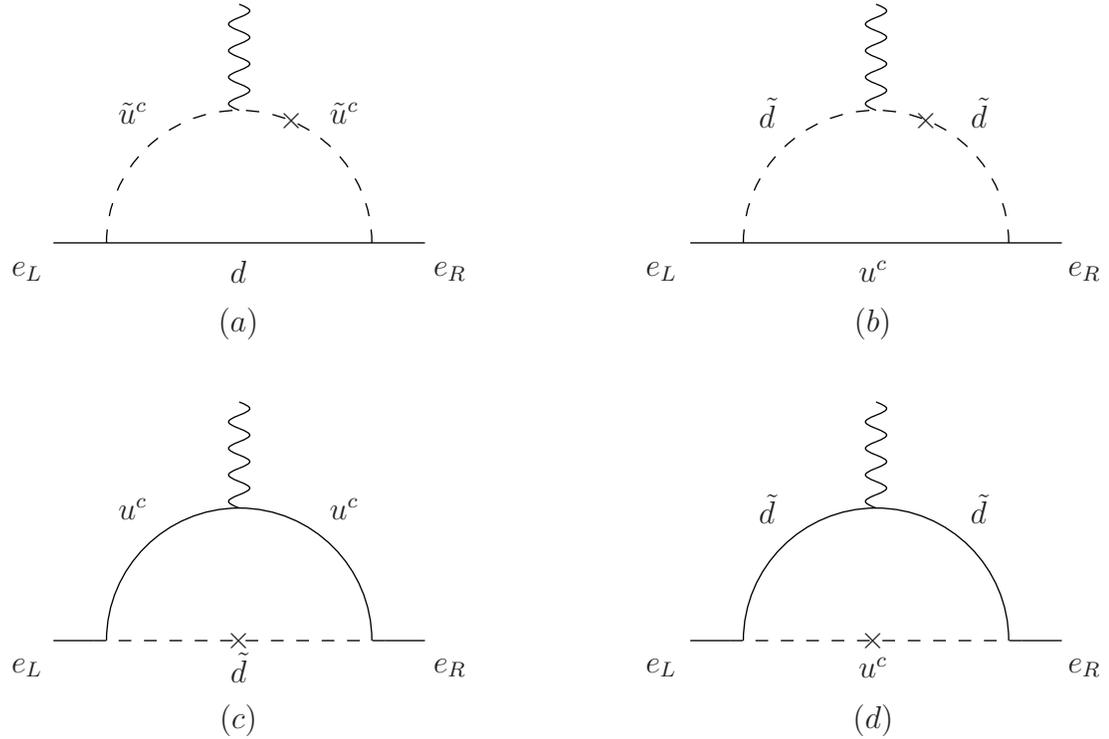
 We have explicitly taken into account the contribution of R-parity violating couplings in the context of ${\cal N}=1$ gauged supergravity limit of $\mu$-split-like  SUSY set-up discussed in \cite{gravitino_DM}. Though one would certainly expect a very suppressed value of EDM because of presence of heavy sfermions as well as vanishing contribution of R-parity violating vertices, we discuss the effect of the same on EDM of electron/neutron  just to compare the order of magnitude  of EDM with respect to R-parity conserving loop diagrams. Though the R-parity violating interaction vertices are complex but due to presence of both R-parity violating vertex as well as its conjugate in the one-loop diagrams as given in Figure~4.3,  the complex phase  disappears and therefore, contribution of the same to EDM will vanish. However, as similar to neutralino and gaugino one-loop diagrams, the non-zero phase corresponding to CP-violating effect can appear only by considering the chirality flip between slepton(squark) fields appearing as a propagators in the one-loop diagram. Due to chirality flip, the matrix amplitude depends on the off-diagonal component of slepton(squark) mass matrix, the contribution of which further depends on complex trilinear coupling ${\cal A}_{IJK}$ as well as supersymmetric mass parameter $\mu$.

The one-loop Feynman diagrams for electron EDM mediated by R-parity violating interaction vertices are given in Figure~4.3. Using the analytical results as given in \cite{frank_hamidian} to get the numerical estimate of EDM of electron, we have
\beqn
\label{EDMRPV}
&& {\hskip -0.4in} \frac{d_e}{e}|_{RPV}=-|C_{e_{L}{\tilde u}^{c}_R d_L}|^2 C_{{\tilde u}^{c}_{R} {\tilde u^{c *}_{R}} \gamma}\ \frac{2 e}{3}\ |{{\cal A}_{u_j}}| \frac{m_{d_{k}}}{m^{3}_{\tilde u}}\ \sin {\theta_u}\ \sin {\phi_{A_u}}B(r_{d_k})\nonumber\\
&&  -|C_{e_{L}{\tilde d_L} u^{c}_R }|^2 C_{{\tilde d_L}^{*} \tilde d_L \gamma} \ \frac{ e}{3}\ |{{\cal A}_{d_j}}| \frac{m_{u_{k}}}{m^{3}_{\tilde d}}\ \sin {\theta_d}\ \sin {\phi_{A_d}}B(r_{u_k})\nonumber\\
&& -|C_{e_{L}{\tilde u}^{c}_R d_L}|^2 C_{{ u}^{c}_R {u^{c *}_R} \gamma}\ \frac{2 e}{3}\ |{{\cal A}_{d_j}}| \frac{m_{u_{j}}}{m^{3}_{\tilde d}}\ \sin {\theta_d}\ \sin {\phi_{A_d}}A(r_{u_k}) \nonumber\\
&& -|C_{e_{L}{\tilde d_L} u^{c}_R }|^2 C_{{d}^{*}_L d_L \gamma} \ \frac{e}{3}\ |{{\cal A}_{u_j}}| \frac{m_{d_{j}}}{m^{3}_{\tilde u}}\ \sin {\theta_u}\ \sin {\phi_{A_u}}A(r_{d_k}).
\eeqn
where $r_{(u_k/d_k)}=  {m^{2}_{(u_k/d_k)}}/{m^{2}_{\tilde f}}$ and form of one-loop functions $A(r)$ and $B(r)$ are already defined in (\ref{eq:A}).

One can draw the similar R-parity violating one-loop diagram to calculate quark EDM ($u$) by replacing $(e_L,e_R) \leftrightarrow(u_L,u_R)$ and ${\tilde u}^{c}\leftrightarrow {\tilde e}^{c}$. The analytical expression in case of quark EDM will  be of the following form:
\beqn
\label{EDMRPV}
&& {\hskip -0.4in} \frac{d_u}{e}|_{RPV}=-|C_{u_{L}{\tilde e}^{c}_R d_L}|^2 C_{{\tilde e}^{c}_{R} {\tilde e^{c *}_{R}} \gamma}\ e \ |{{\cal A}_{e_j}}| \frac{m_{d_{k}}}{m^{3}_{\tilde e}}\ \sin {\theta_e}\ \sin {\phi_{A_e}}B(r_{d_k})\nonumber\\
&&  -|C_{u_{L}{\tilde d_L} e^{c}_R }|^2 C_{{\tilde d}^{*}_L \tilde d_L \gamma} \ \frac{ e}{3}\ |{{\cal A}_{d_j}}| \frac{m_{e_{k}}}{m^{3}_{\tilde d}}\ \sin {\theta_d}\ \sin {\phi_{A_d}}B(r_{e_k})\nonumber\\
&& -|C_{u_{L}{\tilde e}^{c}_R d_L}|^2 C_{{ e}^{c}_{R} {e^{c *}_{R}} \gamma}\ \frac{ e}{3}\ |{{\cal A}_{d_j}}| \frac{m_{e_{j}}}{m^{3}_{\tilde d}}\ \sin {\theta_d}\ \sin {\phi_{A_d}}A(r_{e_k}) \nonumber\\
&& -|C_{u_{L}{\tilde d_L} e^{c}_R }|^2 C_{{d}^{*}_L d_L \gamma} \ e \ |{{\cal A}_{e_j}}| \frac{m_{d_{j}}}{m^{3}_{\tilde e}}\ \sin {\theta_e}\ \sin {\phi_{A_u}}A(r_{d_k}).
\eeqn
The magnitude of the coefficient of interaction vertices $C_{e_{L}{\tilde u}^c d}$, $C_{e_{L}{\tilde d} u^c }$, $C_{u_{L}{\tilde e}^{c}_R d_L}$ and $C_{u_{L}{\tilde d_L} e^{c}_R }$  have already been obtained in chapter {\bf 3} and given as:
\beqn
C_{e_{L}{\tilde u}^{c}_R d_L} = C_{e_{L}{\tilde d}u^{c}_R }=C_{u_{L}{\tilde e}^{c}_R d_L}=C_{u_{L}{\tilde d_L} e^{c}_R } \equiv {\cal V}^{\frac{5}{3}}e^{i \phi_{y_\alpha}},
\eeqn
where $\phi_{y_\alpha}$ is the phase factor associated with complex R-parity violating interaction vertices.
 
The volume suppression factors coming from $C_{{\tilde u}^{c}_R {\tilde u^{c *}_R} \gamma}$, $C_{{\tilde e}^{c}_R {\tilde e^{c *}_R} \gamma}$ and  $C_{{\tilde d_L} {\tilde d_L}^{*}\gamma}$ vertices have already been obtained in the case of gaugino one-loop diagrams and given as:
\beqn
C_{{\tilde u}^{c}_R {\tilde u^{c *}_R} \gamma}= C_{{\tilde e}^{c}_R {\tilde e^{c *}_R} \gamma}\equiv {\cal V}^{\frac{62}{45}}{\tilde f}, C_{{\tilde d_L}{\tilde d_L}^{*}\gamma} \equiv {\cal V}^{\frac{53}{45}}{\tilde f}.
\eeqn
We set $C_{f f^{*}\gamma}|_{\rm EW}$ to be the charge of the quark $d_L$. The reason for the same is as follows. Consider the following kinetic-term-like term contributing to the quark-quark-photon vertex in ${\cal N}=1$ gauged supergravity action of Wess and Bagger: $g_{YM} g_{a_2 {\bar a}_2}\frac{{\bar\chi}^{{\bar a}_2}_L\Gamma^{a_2}_{T_B a_2}X^B\slashed{A}\chi^{a_2}}{\left(\sqrt{K_{{a_2}{\bar a}_2}}\right)^2}\in
\frac{g_{a_2{\bar a}_2}{\bar\chi}^{{\bar a}_2}\slashed{D}\chi^{a_2}}{\left(\sqrt{K_{a_2{\bar a}_2}}\right)^2}$. For the purpose of demonstrating the possibility of obtaining a SM-like quark-quark-photon coupling at the EW scale, let us assume that all moduli save $T_B, a_2$ have been stabilized at values indicated earlier and $\left(n^0_\beta\right)_{\rm max}\sim{\cal V}$ and consequently we take the K\"{a}hler potential to be:
\begin{eqnarray}
\label{K_TBa2}
K&\sim&-2 ln \Bigl[\left(T_B + {\bar T}_B - a_2\left\{C_{2{\bar 2}} {\bar a}_2 + C_{2{\bar 1}}\langle {\bar a}_1\rangle + C_{2{\bar 3}}\langle{\bar a}_3\rangle + C_{2{\bar 4}}\langle{\bar a}_4\rangle\right\} +  c.c. + {\cal V}^{\frac{2}{3}}\right)^{\frac{3}{2}} + {\cal V}\biggr]\nonumber\\
& \equiv & - 2 ln\biggl[\left(T_B + {\bar T}_B - C_{2{\bar 2}}|a_2|^2 - a_2{\bar \Sigma}_2 + h.c. + {\cal V}^{\frac{2}{3}}\right)^{\frac{3}{2}} + {\cal V}\Bigr].
\end{eqnarray}
Consider having frozen all moduli save $T_B$ and $a_2$. Then from:
$$\left(\begin{array}{cc}g_{T_B{\bar T}_B} & g_{T_B{\bar a}_2} \\ g_{a_2{\bar T}_B} & g_{a_2{\bar a}_2}\end{array}\right)^{-1}=\frac{1}{g_{T_B{\bar T}_B} g_{a_2{\bar a}_2} - |g_{T_B{\bar a}_2}|^2}\left(\begin{array}{cc} g_{a_2{\bar a}_2} & - g_{T_B{\bar a}_2} \\
- g_{a_2{\bar T_B}} & g_{T_B{\bar T}_B}\end{array}\right),$$ if $g_{T_B{\bar a}_2}|_{EW}$ is small such that
\begin{equation}
\label{metric_componets_small_large}
|g_{T_B{\bar a}_2}|^2_{EW} > g_{T_B{\bar T}_B} g_{a_2{\bar a}_2}|_{EW}
\end{equation}
 then:
\begin{eqnarray}
\label{inverse_EW}
& & g^{T_B{\bar T}_B}|_{EW}\sim \frac{g_{a_2{\bar a}_2}}{|g_{T_B{\bar a}_2}|^2},\
g^{a_2{\bar a}_2}|_{EW}\sim \frac{g_{T_B{\bar T}_B}}{|g_{T_B{\bar a}_2}|^2}|_{EW},\nonumber\\
& &  g^{T_B{\bar a}_2}|_{EW}\sim \frac{1}{g_{T_B{\bar a}_2}}|_{EW}\equiv{\rm large}.
\end{eqnarray}
From (\ref{K_TBa2}), we see that:
{\footnotesize
\begin{eqnarray}
\label{qqgamma_i}
& & \partial_{a_2}K\sim - \frac{3\left(C_{2{\bar 2}}{\bar a}_2 + {\bar\Sigma}_2 \right)\sqrt{T_B + {\bar T}_B - C_{2{\bar 2}}|a_2|^2 - a_2{\bar \Sigma}_2 + h.c. + {\cal V}^{\frac{2}{3}}}}{\left(T_B + {\bar T}_B - C_{2{\bar 2}}|a_2|^2 - a_2{\bar \Sigma}_2 + h.c. + {\cal V}^{\frac{2}{3}}\right)^{\frac{3}{2}} + {\cal V}},\nonumber\\
& & {\bar\partial}_{{\bar T}_B}\partial_{T_B}K=\nonumber\\
&& -\frac{3}{\sqrt{T_B + {\bar T}_B - C_{2{\bar 2}}|a_2|^2 - a_2{\bar \Sigma}_2 + h.c. + {\cal V}^{\frac{2}{3}}}\biggl[{\left(T_B + {\bar T}_B - C_{2{\bar 2}}|a_2|^2 - a_2{\bar \Sigma}_2 + h.c. + {\cal V}^{\frac{2}{3}}\right)^{\frac{3}{2}} + {\cal V}}\biggr]}\nonumber\\
& & + \frac{9\left(T_B + {\bar T}_B - C_{2{\bar 2}}|a_2|^2 - a_2{\bar \Sigma}_2 + h.c. + {\cal V}^{\frac{2}{3}}\right)}{\biggl[{\left(T_B + {\bar T}_B - C_{2{\bar 2}}|a_2|^2 - a_2{\bar \Sigma}_2 + h.c. + {\cal V}^{\frac{2}{3}}\right)^{\frac{3}{2}} + {\cal V}}\biggr]^2}\nonumber\\
&  &  {\bar\partial}_{{\bar a}_2}\partial_{a_2}K\sim -\frac{3 C_{2{\bar 2}}\sqrt{T_B + {\bar T}_B - C_{2{\bar 2}}|a_2|^2 - a_2{\bar \Sigma}_2 + h.c. + {\cal V}^{\frac{2}{3}}}}{\left(T_B + {\bar T}_B - C_{2{\bar 2}}|a_2|^2 - a_2{\bar \Sigma}_2 + h.c. + {\cal V}^{\frac{2}{3}}\right)^{\frac{3}{2}} + {\cal V}} \nonumber\\ -
& & \frac{3|C_{2{\bar 2}}a_2 + \Sigma_2|^2}{2\sqrt{T_B + {\bar T}_B - C_{2{\bar 2}}|a_2|^2 - a_2{\bar \Sigma}_2 + h.c. + {\cal V}^{\frac{2}{3}}}\biggl[\left(T_B + {\bar T}_B - C_{2{\bar 2}}|a_2|^2 - a_2{\bar \Sigma}_2 + h.c. + {\cal V}^{\frac{2}{3}}\right)^{\frac{3}{2}} + {\cal V}\biggr]}\nonumber\\
& &  + \frac{9|C_{2{\bar 2}}a_2 + \Sigma_2|^2\left(T_B + {\bar T}_B - C_{2{\bar 2}}|a_2|^2 - a_2{\bar \Sigma}_2 + h.c. + {\cal V}^{\frac{2}{3}}\right)}{2\biggl[\left(T_B + {\bar T}_B - C_{2{\bar 2}}|a_2|^2 - a_2{\bar \Sigma}_2 + h.c. + {\cal V}^{\frac{2}{3}}\right)^{\frac{3}{2}} + {\cal V}\biggr]^2}\nonumber\\
& &  {\bar\partial}_{{\bar T}_B}\partial_{a_2}K\sim\nonumber\\
&& -\frac{3\left(C_{2{\bar 2}}{\bar a}_2 + {\bar \Sigma}_2\right)}{2\sqrt{T_B + {\bar T}_B - C_{2{\bar 2}}|a_2|^2 - a_2{\bar \Sigma}_2 + h.c. + {\cal V}^{\frac{2}{3}}}\biggl[\left(T_B + {\bar T}_B - C_{2{\bar 2}}|a_2|^2 - a_2{\bar \Sigma}_2 + h.c. + {\cal V}^{\frac{2}{3}}\right)^{\frac{3}{2}} + {\cal V}\biggr]} \nonumber\\
 & & + \frac{9(C_{2{\bar 2}} {\bar a}_2 + {\bar\Sigma}_2)}{2\biggl[\left(T_B + {\bar T}_B - C_{2{\bar 2}}|a_2|^2 - a_2{\bar \Sigma}_2 + h.c. + {\cal V}^{\frac{2}{3}}\right)^{\frac{3}{2}} + {\cal V}\biggr]^2},\nonumber\\
 & &  \partial_{T_B}{\bar\partial}_{{\bar T}_B}\partial_{a_2}K\sim \nonumber\\
 && \frac{3\left(C_{2{\bar 2}} {\bar a}_2 + {\bar\Sigma}_2\right)}{4\left({T_B + {\bar T}_B - C_{2{\bar 2}}|a_2|^2 - a_2{\bar \Sigma}_2 + h.c. + {\cal V}^{\frac{2}{3}}}\right)^{\frac{3}{2}}\left[{\left(T_B + {\bar T}_B - C_{2{\bar 2}}|a_2|^2 - a_2{\bar \Sigma}_2 + h.c. + {\cal V}^{\frac{2}{3}}\right)^{\frac{3}{2}} + {\cal V}}\right]}+\nonumber\\
  & &  \frac{9\left(C_{2{\bar 2}} {\bar a}_2 + {\bar\Sigma}_2\right)}{4\left[{\left(T_B + {\bar T}_B - C_{2{\bar 2}}|a_2|^2 - a_2{\bar \Sigma}_2 + h.c. + {\cal V}^{\frac{2}{3}}\right)^{\frac{3}{2}} + {\cal V}}\right]^2} \nonumber\\
& &   - \frac{27\left(C_{2{\bar 2}} {\bar a}_2 + {\bar\Sigma}_2\right)\left({T_B + {\bar T}_B - C_{2{\bar 2}}|a_2|^2 - a_2{\bar \Sigma}_2 + h.c. + {\cal V}^{\frac{2}{3}}}\right)^{\frac{3}{2}}}{2\left[{\left(T_B + {\bar T}_B - C_{2{\bar 2}}|a_2|^2 - a_2{\bar \Sigma}_2 + h.c. + {\cal V}^{\frac{2}{3}}\right)^{\frac{3}{2}} + {\cal V}}\right]^3}.
\end{eqnarray}}
Hence, if ${\bar\partial}_{{\bar T}_B}\partial_{a_2}K|_{EW}\sim\delta<<1$ such that (\ref{metric_componets_small_large}) is satisfied then:
{ 
\begin{eqnarray}
\label{smallBbara2_EW}
&& 3\langle\left(T_B + {\bar T}_B - C_{2{\bar 2}}|a_2|^2 - a_2{\bar \Sigma}_2 + h.c. + {\cal V}^{\frac{2}{3}}\right)\rangle_{EW}^{\frac{3}{2}} \nonumber\\
&& \sim \langle\biggl[\left(T_B + {\bar T}_B - C_{2{\bar 2}}|a_2|^2 - a_2{\bar \Sigma}_2 + h.c. + {\cal V}^{\frac{2}{3}}\right)^{\frac{3}{2}} + {\cal V}\biggr]\rangle_{EW}.
\end{eqnarray}}
Using (\ref{smallBbara2_EW}) in the fourth equation in (\ref{qqgamma_i}), one sees that:
\begin{equation}
\label{dTBga2Tbbar}
\partial_{T_B}g_{a_2{\bar T}_B}|_{EW\ {\rm near}\ (\ref{smallBbara2_EW})}\sim\frac{9\left(C_{2{\bar 2}}{\bar a}_2 + {\bar\Sigma}_2\right)}{\biggl[\left(T_B + {\bar T}_B - C_{2{\bar 2}}|a_2|^2 - a_2{\bar \Sigma}_2 + h.c. + {\cal V}^{\frac{2}{3}}\right)^{\frac{3}{2}} + {\cal V}\biggr]^2}\sim {\cal V}^{-\frac{29}{18}},
\end{equation}
assuming $\langle a_{1,2,3,4}\rangle|_{EW}\sim{\cal O}(1)\langle a_{1,2,3,4}\rangle|_{M_s}$. If $g_{a_2{\bar a}_2}|_{M_s}\sim g_{a_2{\bar a}_2}|_{EW}\sim10^{-2}$, then from (\ref{metric_componets_small_large}), one sees:
\begin{equation}
\label{gBBbar_EW}
g_{T_B{\bar T_B}}|_{EW}\sim\delta^\prime<10^2\delta^2.
\end{equation}
Noting that:
\begin{equation}
\label{aff-con}
\Gamma^{a_2}_{T_ba_2}=\frac{g^{a_2{\bar T}_B}}{2}\left(\partial_{T_B}g_{a_2{\bar T}_B} +
\partial_{a_2}g_{T_B{\bar T}_B}\right) + \frac{g^{a_2{\bar a}_2}}{2}\left(\partial_{T_B}g_{a_2{\bar a}_2} + \partial_{a_2}g_{T_B{\bar a}_2}\right),
\end{equation}
we see that one can get a large contribution to (\ref{aff-con}) from $g^{a_2{\bar T}_B}\partial_{T_B}g_{a_2{\bar T}_B}|_{EW}$ given by:
\begin{equation}
\label{ginva2TbdTbga2Tb}
g^{a_2{\bar T}_B}\partial_{T_B}g_{a_2{\bar T}_B}|_{EW,\ {\cal V}\sim10^4}\sim\frac{10^{-6.5}}{\delta}.
\end{equation}
Let us look at implementation of (\ref{gBBbar_EW}) and its consequences. From (\ref{qqgamma_i}), one notes that
(\ref{gBBbar_EW}) is identically satisfied if (\ref{smallBbara2_EW}) is satisfied. Consider working with
$\tau_{S,B},z^i,a_I,...$ instead of $T_{S,B},z^i,a_I,...$ having frozen $G^a$ and other open-string moduli. Noting then that:
\begin{equation}
K^{\alpha{\bar\beta}}=\frac{1}{K_{\tau_S{\bar\tau}_S}K_{\tau_B{\bar\tau}_B} - |K_{\tau_S{\bar\tau}_B}|^2}\left(\begin{array}{cc}K_{\tau_B{\bar\tau}_B} & - K_{\tau_B{\bar\tau}_S} \\
- K_{\tau_S{\bar\tau}_B} & K_{\tau_S{\bar\tau}_S}\end{array}
\right),
\end{equation}
and assuming $K_{\tau_B{\bar\tau}_B}|_{EW}\sim\delta^\prime<<1, K_{\tau_S{\bar\tau}_S}|_{M_s}\sim K_{\tau_S{\bar\tau}_S}|_{EW}\sim{\cal V}^{-1}, K_{\tau_S{\bar\tau}_B}|_{M_s}\sim
K_{\tau_S{\bar\tau}_B}|_{EW}\sim{\cal V}^{-\frac{5}{3}}$ implying $|K_{\tau_S{\bar\tau}_B}|^2>K_{\tau_S{\bar\tau}_S}K_{\tau_B{\bar\tau}_B}$, one obtains:
\begin{eqnarray}
\label{metric_bulk_EW}
& & K^{\tau_S{\bar\tau}_S}|_{EW}\sim\frac{K_{\tau_B{\bar\tau}_B}|_{EW}}{|K_{\tau_S{\bar\tau}_B}|^2_{EW}}\sim\delta^\prime {\cal V}^{\frac{10}{3}},  K^{\tau_B{\bar\tau}_B}|_{EW}\sim\frac{K_{\tau_S{\bar\tau}_S}|_{EW}}{|K_{\tau_S{\bar\tau}_B}|^2_{EW}}\sim{\cal V}^{\frac{7}{3}},\nonumber\\
& & K^{\tau_S{\bar\tau}_B}|_{EW}\sim\frac{1}{K_{\tau_S{\bar\tau}_B}|_{EW}}\sim{\cal V}^{\frac{5}{3}}.
\end{eqnarray}
Equation (\ref{metric_bulk_EW}) implies:
\begin{eqnarray}
\label{F-nonrenormalization}
& & {\bar F}^{\tau_S}|_{M_s}\sim K^{\tau_S{\bar\tau}_S}D_{{\tau}_S}{W}\nonumber\\
& & \sim {\bar F}^{\tau_S}|_{EW}=e^{\frac{K}{2}}\left(K^{\tau_S{\bar\tau}_S}D_{\tau_S}{ W} + K^{\tau_B{\bar\tau}_S}D_{{\tau}_B}{W}\right)\sim\left(\delta^\prime {\cal V}^{\frac{10}{3}} + {\cal V}\right)m_{3/2}\sim\frac{1}{\cal V}.
\end{eqnarray}
So, the $F^{\tau_S}$-term (potential $||F^{\tau_S}||^2$) is 1-loop RG-invariant! Further, the complete $F$-term potential:
\begin{eqnarray}
\label{F_term_1L_RG_inv}
& &  V|_{M_s}\sim e^K K^{\tau_S{\bar\tau}_S}|D_{\tau_S}W|^2\sim {\cal V}m_{3/2}^2\nonumber\\
& &   \sim e^K\left(K^{\tau_S{\bar\tau}_S}|D_{\tau_S}W|^2 + K^{\tau_B{\bar\tau}_B}|D_{\tau_B}W|^2 + K^{\tau_S{\bar\tau}_B}D_{\tau_S}D_{{\bar\tau}_B}{\bar W} + {\rm h.c.}\right)_{EW}\nonumber\\
&& \sim\left(\delta^\prime {\cal V}^{\frac{10}{3}} + {\cal V}\right)m_{3/2}^2\sim{\cal V}m_{3/2}^2, 
 \end{eqnarray}
is also 1-loop RG-invariant. So, the quark-quark-photon vertex  can be made to be of ${\cal O}(1)$ for $\delta\sim10^{-13}$, i.e., one can hope that the coupling $C_{ff^{*}\gamma}\sim{\cal O}(1)$ for $f({\rm fermion})\equiv e, u$.
 
For $r_{(u_k/d_k/e_k)}=  {m^{2}_{(u_k/d_k/e_k)}}/{m^{2}_{\tilde f_i}}$, $A(r_{u_k/d_k/e_k})= B(r_{u_k/d_k/e_k})=1$.
As mentioned in equation (\ref{me21}), $|{\cal A}'_{e}|= |{\cal A}_{e}^{*}-{\mu}\cot{\beta}| \equiv {\cal V}m_{\frac{3}{2}}$, $|{\cal A}'_{u}|= |{\cal A}_{u}^{*}-{\mu}\cot{\beta}| \equiv {\cal V}m_{\frac{3}{2}}$. Using the these results and results of coefficient of interaction vertices as given above; considering $\sin{\phi_u}=\sin{\phi_d}= (0,1]$, $\sin{\theta_e}=\sin{\theta_u}=1$, the magnitude of dominant contribution of EDM of electron will be given as:
\beqn
\frac{d_e}{e}|_{\rm RPV}\sim  \frac{2}{3} \frac{{\cal V}^{-\frac{10}{3}+1 +\frac{62}{45}}}{{\cal V}^{\frac{3}{2}} m^{2}_{\frac{3}{2}}}m_{u_k} \equiv 10^{-31}GeV^{-1} \equiv 10^{-45}cm,
\eeqn
and the magnitude of dominant contribution of EDM of neutron/quark will be given as follows:
\beqn
\frac{d_n}{e}|_{\rm RPV}\sim  \frac{{\cal V}^{-\frac{10}{3}+1 +\frac{62}{45}}}{{\cal V}^{\frac{3}{2}} m^{2}_{\frac{3}{2}}}m_{e_k} \equiv 10^{-31}GeV^{-1} \equiv 10^{-45}cm.
\eeqn
\subsection{One-loop Diagrams Involving Neutral Higgs}
In this subsection, we explicitly evaluate the contribution of one-loop diagrams involving fermions and Higgs as propagators to electric dipole moment of fermion. The fine-tuning argument given by N.~Arkani-Hamed and S.~Dimopoulos in \cite{HamidSplitSUSY} is not just able to provide a light Higgs by diagonalizing the Higgs mass matrix, it is important to give a reasonable order of magnitude of EDM by considering diagonalized Higgs mass eigenstates (light Higgs as one of the eigenstate of Higgs mass matrix) as scalar propagators in the loop.  In the discussions so far, we have argued that CP-violating phases in the one-loop diagrams contributing to EDM of electron/neutron are accomplished by considering an off-diagonal contribution of sfermion mass matrix at electroweak scale. In this subsection, we will discuss the one-loop diagrams in which non-zero CP violating phases appears through mixing between Higgs doublets in the Higgs mass matrix. Using the same approach, we have already got a light as well as heavy Higgs formed by linear combination of two Higgs doublets (for detailed explanation, see section {\bf 4} of chapter {\bf 2}). Now, we implement this approach to evaluate the contribution of non-zero EDM of electron/neutron by considering eigenstates of Higgs mass matrix as propagators in the one-loop diagrams shown in Figure~4.4 and Figure~4.5.

{{\bf SM-like Yukawa coupling contribution}}: The one-loop diagram mediated by SM-like Yukawa coupling is given in Figure~4.4. The effective one-loop operator given in equation (\ref{eq:eff1loop}) can be recasted in the following form:
\beqn
\label{eq:eff1loop1H}
{\cal L}_{int}=\sum_{i} \bar{\chi}_{f}
                (C_{f^{*}_{L}f_{R}{H_i}} \PL+ C_{f^{*}_{L}f_{R}{H_i}} \PR) {\phi_{H_i}}{ \chi_f}+h.c. 
\eeqn
For $i=1,2$, above equation can be expanded as:
 \beqn
\label{eq:eff1loop2H}
&&-{\cal L}_{int}= \bar{\chi}_{f}
                \Bigl(C_{f^{*}_{L}f_{R}{H_1}} \PL +C_{f^{*}_{L}f_{R}{H_1}}  \PR\Bigr){\phi_{H_1}}{ \chi_f}+\bar{\chi}_{f}
                 \Bigl(C_{f^{*}_{L}f_{R}{H_2}}  \PL \nonumber\\
                 && +C_{f^{*}_{L}f_{R}{H_2}}  \PR \Bigr){\phi_{H_2}}{ \chi_f}+ h.c. 
\eeqn
where ${\phi_{H_1}}$ and ${\phi_{H_2}}$ correspond to eigenstates of mass matrix of Higgs doublets and ${\chi_f}$ corresponding to fermion. Using equation (\ref{massmat}), the aforementioned vertices can be expressed in terms of undiagonalized ($H_u,H_d$) basis as follows:
\begin{eqnarray}
\label{Celerh1}
&& C_{f^{*}_{L}f_{R}{H_1}}= D_{h_{11}}C_{f^{*}_{L}f_{R}{H_u}}+D_{h_{12}}C_{f^{*}_{L}f_{R}{H_d}} \nonumber\\
&& C_{f^{*}_{L}f_{R}{H_2}}= D_{h_{21}}C_{f^{*}_{L}f_{R}{H_u}}+D_{h_{22}}C_{f^{*}_{L}f_{R}{H_d}}.
\end{eqnarray}
\begin{figure}
\begin{center}
\begin{picture}(145,115) (130,20)
   \ArrowLine(100,50)(135,50)
    \ArrowLine(215,50)(135,50)
    \Text(215,50)[]{$\times$}
    \ArrowLine(215,50)(295,50)
     \ArrowLine(330,50)(295,50)
     \DashCArc(215,50)(80,0,180){5}
   \Photon(185,50)(240,20){4}{4}
   \Text(150,110)[]{$H^{0}_{i}$}
   \Text(290,100)[]{$H^{0}_{i}$}
   \Text(90,50)[]{{$f_{L}$}}
   \Text(340,50)[]{{$f^{c}_{R}$}}
   \Text(170,40)[]{{{$f^{c}_{R}$}}}
   \Text(250,40)[]{{{$f_{L}$}}}  \end{picture}
\end{center}
\vskip -0.1in
\caption{One-loop diagram involving Higgs and SM-like fermions.}
 \end{figure}
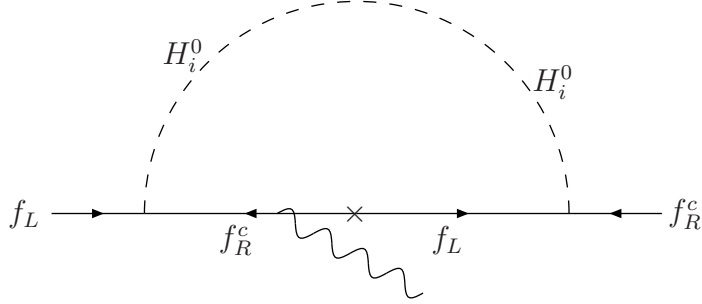
In ${\cal N}=1$ gauged supergravity, the interaction vertices $C_{e^{*}_{L}e_{R}{H_u}/{H_d}}$ and $C_{u^{*}_{L}u_{R}{H_u}/{H_d}}$ will be given by expanding $e^{\frac{K}{2}}{\cal D}_{{\cal A}_1}D_{{\cal A}_3}W$ and $e^{\frac{K}{2}}{\cal D}_{{\cal A}_2}D_{{\cal A}_4}W$ respectively in the fluctuations linear in ${\cal Z}_{i}$ about its stabilized VEV. The values of the same, as already given in chapter {\bf 2}, are as follows:
\beqn
\label{eq:Yhatz1a1a3}
&&{\rm for}~f=e,~ C_{e^{*}_{L}e_{R}{H_u}/{H_d}}= \hat{Y}^{\rm eff}_{{\cal Z}_I{\cal A}_1{\cal A}_3} =\frac{{\cal O}({\cal Z}_I-{\cal V}^{\frac{1}{36}})\ {\rm term\ in}\
e^{\frac{K}{2}}{\cal D}_{{\cal A}_1}D_{{\cal A}_3}W}{\sqrt{K_{{\cal Z}_i\bar{\cal Z}_i}K_{{\cal A}_1\bar{\cal A}_1}K_{{\cal A}_3\bar{\cal A}_3}}}\equiv   {\cal V}^{-\frac{47}{45}} e^{i \phi_{{\cal Y}_e}}\nonumber\\
&&
{\rm for}~f=u,~ C_{u^{*}_{L}u_{R}{H_u}/{H_d}}= \hat{Y}^{\rm eff}_{{\cal Z}_I{\cal A}_2{\cal A}_4} =\frac{{\cal O}({\cal Z}_I-{\cal V}^{\frac{1}{36}})\ {\rm term\ in}\
e^{\frac{K}{2}}{\cal D}_{{\cal A}_2}D_{{\cal A}_4}W}{\sqrt{K_{{\cal Z}_1\bar{\cal Z}_1}K_{{\cal A}_1\bar{\cal A}_1}K_{{\cal A}_3\bar{\cal A}_3}}}\equiv {\cal V}^{-\frac{19}{18}} e^{i \phi_{{\cal Y}_u}}.\nonumber
\eeqn
where $e^{i \phi_{{\cal Y}_e}}$ and $e^{i \phi_{{\cal Y}_e}}$ are the phase factors associated with complex 
effective Yukawa

{\hskip -0.25in}couplings. Going back to equation (\ref{Celerh1}),
\begin{eqnarray}
\label{Celerh2}
&& C_{e^{*}_{L}e_{R}{H_1}}\equiv  {\cal V}^{-\frac{47}{45}}e^{i \phi_{{\cal Y}_e}}(D_{h_{11}} +D_{h_{21}}), C_{e^{*}_{L}e_{R}{H_2}}\equiv{\cal V}^{-\frac{47}{45}}e^{i \phi_{{\cal Y}_e}}( D_{h_{12}} +D_{h_{22}}),\nonumber\\
&&  C_{u^{*}_{L}u_{R}{H_1}}\equiv{\cal V}^{-\frac{19}{18}}e^{i \phi_{{\cal Y}_u}}(D_{h_{11}} +D_{h_{21}}), C_{u^{*}_{L}u_{R}{H_2}}\equiv{\cal V}^{-\frac{19}{18}}e^{i \phi_{{\cal Y}_u}}( D_{h_{12}} +D_{h_{22}}).
\end{eqnarray}
Now, the one-loop EDM of the electron (quark) in this  case will be given as \cite{keum+kong}:
\beqn
\label{eq:EDMh}
&&{\frac{d}{e}}|_{H_{1,2}}=\frac{m_f Q_f}{(4\pi)^2} \left( \frac{1}{m^{2}_{H_1}}{\rm Im}(C_{f^{*}_{L}f_{R}{H_1}}C_{f^{*}_{L}f_{R}{H_1}}^*) A\left(\frac{m^{2}_{f}}{m^{2}_{H_1}}\right) + \frac{1}{m^{2}_{H_2}} {\rm Im}(C_{f^{*}_{L}f_{R}{H_2}} C_{f^{*}_{L}f_{R}{H_2}}^*)\right.\nonumber\\
&& {\hskip 0.5in} \left. A\left(\frac{m^{2}_{f}}{m^{2}_{H_2}}\right)\right). 
\eeqn
where $m_{f}$ corresponds to fermion mass and $m_{H_{1,2}}$ correspond to eigenstates of Higgs mass matrix.
Since  we are considering only first generation fermions in our $D3/D7$ $\mu$-split-like  SUSY set up, physical mass eigenstate of fermion is same as usual Dirac mass term corresponding to first generation lepton/quark only. Using the fact that phase factors associated with Wilson line modulus ${\cal A}_{1/2}$ (identified with first generation L-handed lepton/quark), Wilson line modulus ${\cal A}_{3/4} $(identified with first generation R-handed lepton/quark) and position modulus (identified with Higgs doublet) are distinct; and the coefficient of Yukawa coupling coupling (evaluated dynamically by expanding  $e^{\frac{K}{2}}{\cal D}_{{\cal A}_1}D_{{\cal A}_3}W$ term in the gauged supergravity action in the fluctuations linear in ${\cal Z}_1$ and then giving it VEV) also produces a non-zero phase factor, the mass of fermion can be complex. Therefore, we assume that overall phase formed by adding all phase factors associated with fields as all coefficient of Yukawa coupling add up in such a way that  overall phase vanishes and fermion mass is real. Using (\ref{Celerh2}),
\beqn
&& {\rm Im}(C_{e^{*}_{L}e_{R}{H_2}}C_{e^{*}_{L}e_{R}{H_2}}^*) = - {\rm Im}(C_{e^{*}_{L}e_{R}{H_1}}C_{e^{*}_{L}e_{R}{H_1}}^*) \equiv \frac{1}{2}{\cal V}^{-\frac{94}{45}}\sin{\theta_h}\sin{\phi_h}\nonumber\\
&& {\rm Im}(C_{u^{*}_{L}u_{R}{H_2}}C_{u^{*}_{L}u_{R}{H_2}}^*) = - {\rm Im}(C_{u^{*}_{L}u_{R}{H_1}}C_{u^{*}_{L}u_{R}{H_1}}^*) \equiv \frac{1}{2}{\cal V}^{-\frac{19}{18}}\sin{\theta_h}\sin{\phi_h}.  \eeqn
 Given that $\sin \theta_h=
\frac{2|{\hat \mu} {\cal B}|}{\sqrt{\left(M_{H_u}^2-M_{H_d}^2\right)^2+4 ( {{\hat \mu} {\cal B}})^2}}$. Using the values given above, $\sin{\theta_h} \in[0,1]$. We also make an assumption that $\phi_h~ \exists ~(0,\frac{\pi}{2}]$. Using equation (\ref{eq:A}), and value of $m_{e}= 0.5 MeV$, $m_{H_1}\sim 125 GeV$ and $m_{H_2}\sim {\cal V}^{\frac{59}{72}}m_{\frac{3}{2}}$;
$A\left(\frac{m^{2}_{e}}{m^{2}_{H_1}}\right)=  A\left(\frac{m^{2}_{e}}{m^{2}_{H_2}}\right)\equiv 1$.
Using (\ref{A}), the dominant contribution of electron EDM in this case will be given as
\beq
\label{eq:EDMh1}
{\frac{d_e}{e}}|_{H_{1,2}}=\frac{10^{-3}}{4(4\pi)^2} {\cal V}^{-\frac{94}{45}} \left(\frac{1}{m^{2}_{H_1}}-\frac{1}{m^{2}_{H_2}}\right)\equiv 10^{-20} GeV^{-1}\equiv{\cal O}(10^{-34}) cm.
\eeq
The  numerical estimate of neutron/quark EDM will be given as:
\beq
\label{eq:EDMh2}
{\frac{d_n}{e}}|_{H_{1,2}}=\frac{10^{-3}}{2 (4\pi)^2} {\cal V}^{-\frac{19}{9}} \left(\frac{1}{m^{2}_{H_1}}-\frac{1}{m^{2}_{H_2}}\right)\equiv10^{-29} GeV^{-1}\equiv{\cal O}(10^{-33}) cm.
\eeq
{{\bf Chargino contribution: }} The one-loop diagram corresponding to electron EDM mediated via Higgs and chargino exchange is given in Figure~4.5. Due to presence of heavy fermions and  light as well as heavy scalars existing as propagators in the loop, using analytical expression of one-loop EDM as given in equation (\ref{eq:EDM}), one can expect an enhancement in the order of magnitude of EDM. We explicitly analysis the contribution of this loop diagram to EDM at one loop in the context of ${\cal N}=1$ gauged SUGRA. One can not have similar diagram for quark because of violation of charge conservation. So we use the loop diagram given in Figure~4.5 to get the analysis of EDM of electron only. The effective one-loop operator will be of the following form:
\beqn
\label{eq:eff1loop1}
{\cal L}_{int}=\sum_{i,j} \bar{\chi}_{f}
                (C_{f^{*}_{L}\chi^{+}_{j}{H_i^{0}}} \PL){\phi_{H_i^{0}}}{\tilde \chi}^{+}_{j}+\bar{\chi}_{f}( C_{q^{*}_{R}\chi^{-}_{j}{H_i^{0}}} \PR) {\phi_{H_i^{0}}}{\tilde \chi}^{-}_{j}+h.c., \ i,j=1,2. \nonumber
\eeqn
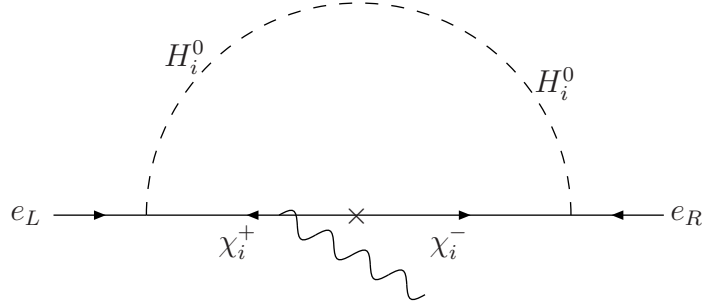
\begin{figure}
\begin{center}
\begin{picture}(145,110) (130,20)
   \ArrowLine(100,50)(135,50)
    \ArrowLine(215,50)(135,50)
    \Text(215,50)[]{$\times$}
    \ArrowLine(215,50)(295,50)
     \ArrowLine(330,50)(295,50)
     \DashCArc(215,50)(80,0,180){5}
   \Photon(185,50)(240,20){4}{4}
   \Text(150,110)[]{$H^{0}_{i}$}
   \Text(290,100)[]{$H^{0}_{i}$}
   \Text(90,50)[]{{$e_{L}$}}
   \Text(340,50)[]{{$e_{R}$}}
   \Text(170,40)[]{{{$\chi^{+}_{i}$}}}
   \Text(250,40)[]{{{$\chi^{-}_{i}$}}}  \end{picture}
\end{center}
\caption{One-loop diagram involving Higgs and charginos.}
 \end{figure}
Using equation (\ref{massmat}), one can represent coefficient of interaction vertices in terms of undiagonalized basis of Higgs mass matrix as follows:
\begin{eqnarray}
\label{Celerchi}
&&C_{e^{*}_{L}\chi^{+}_{j}{H_1^{0}}} = D_{h_{11}}C_{e^{*}_{L}\chi^{+}_{j}{H_u^{0}}} +D_{h_{12}}C_{e^{*}_{L}\chi^{+}_{j}{H_d^{0}}}, C_{e^{*}_{L}\chi^{+}_{j}{H_2^{0}}}= D_{h_{21}}C_{e^{*}_{L}\chi^{+}_{j}{H_u^{0}}}+D_{h_{22}}C_{e^{*}_{L}\chi^{+}_{j}{H_u^{0}}}, \nonumber\\
&&C_{e^{*}_{R}\chi^{-}_{j}{H_1^{0}}} = D_{h_{11}}C_{e^{*}_{R}\chi^{-}_{j}{H_u^{0}}} +D_{h_{12}}C_{e^{*}_{R}\chi^{-}_{j}{H_d^{0}}}, C_{e^{*}_{R}\chi^{-}_{j}{H_2^{0}}}= D_{h_{21}}C_{e^{*}_{R}\chi^{-}_{j}{H_u^{0}}}+D_{h_{22}}C_{e^{*}_{R}\chi^{-}_{j}{H_u^{0}}}\nonumber\\
\end{eqnarray}
As given in equation (\ref{eq:charginoI}),
\beqn
&& {\tilde \chi}^{+}_{1}= -{\tilde H}^{+}_{u}+\left(\frac{v}{M_P}{\tilde f}{\cal V}^{\frac{5}{6}}\right){\tilde \lambda}^{+}_{i},{\tilde \chi}^{-}_{1}=  -{\tilde H}^{-}_{d}+\left(\frac{v}{M_P}{\tilde f}{\cal V}^{\frac{5}{6}}\right){\tilde \lambda}^{-}_{i},~m_{{\tilde \chi}^{\pm}_{1}}\sim {\cal V}^{\frac{59}{72}}m_{\frac{3}{2}}\nonumber\\
&&  {\tilde \chi}^{+}_{2}= {\tilde \lambda}^{+}_{i} +\left(\frac{v}{M_P}{\tilde f}{\cal V}^{\frac{5}{6}}\right){\tilde H}^{+}_{u},{\tilde \chi}^{-}_{2}= {\tilde \lambda}^{-}_{i} +\left(\frac{v}{M_P}{\tilde f}{\cal V}^{\frac{5}{6}}\right) {\tilde H}^{-}_{d},~m_{{\tilde \chi}^{\pm}_{2}}\sim {\cal V}^{\frac{2}{3}}m_{\frac{3}{2}}.
\eeqn
Using the above,
\beqn
\label{eq:Ceechi}
&&C_{e^{*}_{L}\chi^{+}_{1}{H_u^{0}}/{H_d^{0}}}= -C_{e^{*}_{L}{\tilde H^{+}_u}{H_u^{0}}/H_d^{0}}+ \left(\frac{v}{M_P}{\tilde f}{\cal V}^{\frac{5}{6}}\right)C_{e^{*}_{L}{\tilde {\tilde \lambda}^{+}_i}{H_u^{0}}/H_d^{0}}\nonumber\\
&& C_{e^{*}_{L}\chi^{+}_{2}{H_u^{0}}/{H_d^{0}}}= C_{e^{*}_{L}{\tilde {\tilde \lambda}^{+}_i}{H_u^{0}}/H_d^{0}}+\left(\frac{v}{M_P}{\tilde f}{\cal V}^{\frac{5}{6}}\right) C_{e^{*}_{L}{\tilde H^{+}_u}{H_u^{0}}/H_d^{0}}\nonumber\\
&& C_{e^{*}_{R}\chi^{-}_{1}{H_u^{0}}/{H_d^{0}}}= -C_{e^{*}_{R}{\tilde H^{-}_d}{H_u^{0}}/H_d^{0}}+ \left(\frac{v}{M_P}{\tilde f}{\cal V}^{\frac{5}{6}}\right)C_{e^{*}_{R}{\tilde {\tilde \lambda}^{+}_i}{H_u^{0}}/H_d^{0}}\nonumber\\
&&C_{e^{*}_{R}\chi^{-}_{2}{H_u^{0}}/{H_d^{0}}}= C_{e^{*}_{R}{\tilde {\tilde \lambda}^{+}_i}{H_u^{0}}/H_d^{0}}+\left(\frac{v}{M_P}{\tilde f}{\cal V}^{\frac{5}{6}}\right) C_{e^{*}_{R}{\tilde H^{-}_d}{H_u^{0}}/H_d^{0}}.
\eeqn
 The interaction vertices $C_{e^{*}_{L}{\tilde H^{+}_u}{H_u^{0}/H_d^{0}}}$ and $ C_{e^{*}_{R}{\tilde H^{-}_d}{H_u^{0}}/H_d^{0}}$ corresponding to Figure 5. will be given by expanding the $e^{\frac{K}{2}}{\cal D}_{{\cal Z}_1}D_{{\cal A}_1}W$  and $e^{\frac{K}{2}}{\cal D}_{{\cal Z}_1}D_{{\cal A}_3}W$ in the fluctuations linear in ${\cal Z}_{i}$ about its stabilized VEV. The contributions of $e^{\frac{K}{2}}{\cal D}_{z_1}D_{a_1}W$  as well as $e^{\frac{K}{2}}{\cal D}_{z_1}D_{a_3}W$ have been given  in terms of undiagonalized $(z_i,a_i)$  basis in chapter {\bf 2}. We assume that $ e^{\frac{K}{2}}{\cal D}_i D_{\bar {{\cal A}_1}}W \sim {\cal O}(1) e^{\frac{K}{2}}{\cal D}_i D_{{\bar a}_1}W$. Since the EW symmetry gets broken for the higgsino(${\tilde H^{+}_u}$)-lepton (${e_L}$)-Higgs(${H^{0}_{u}/H^{0}_d}$) vertex, we evaluate the contribution of the same by expanding $e^{\frac{K}{2}}{\cal D}_{z_1}D_{a_1}W$ in the fluctuations linear in $z_1$ as well as $(z_i- {\cal V}^{\frac{1}{36}})$, and then giving  VEV to $z_i$. Doing so, the magnitude of coefficient of this vertex  will be given as :
\beqn
\label{eq:Yhatz1z1a1}
|C_{e^{*}_{L}{\tilde H^{+}_u}{H_u^{0}/H_d^{0}}}|\sim  \frac{\langle{\cal Z}_i\rangle {\cal O}({\cal Z}_i-{\cal V}^{\frac{1}{36}})\ {\rm term\ in}\
e^{\frac{K}{2}}{\cal D}_{{\cal Z}_1}D_{{\cal A}_1}W}{\sqrt{(K_{{\cal Z}_1\bar{\cal Z}_1})^3 K_{{\cal A}_1\bar{\cal A}_1}}}\equiv {\cal V}^{-\frac{1}{10}},~{\rm for}~{\cal V}=10^5.\nonumber\\
\eeqn
Similarly, the contribution of physical  higgsino(${\tilde H^{-}_d}$)-lepton(${e_R}$)-Higgs(${H^{0}_{u}/H^{0}_d}$) vertex  will be given as
\beqn
\label{eq:Yhatz1z1a3}
|C_{e^{*}_{R}{\tilde H^{-}_d}{H_u^{0}/H_d^{0}}}| \sim \frac{{\cal O}({\cal Z}_i-{\cal V}^{\frac{1}{36}})\ {\rm term\ in}\
e^{\frac{K}{2}}{\cal D}_{{\cal Z}_1}D_{{\cal A}_3}W}{\sqrt{K_{{\cal Z}_1\bar{\cal Z}_1}K_{{\cal Z}_1\bar{\cal Z}_1}K_{{\cal A}_3\bar{\cal A}_3}}}\equiv   {\cal V}^{\frac{1}{10}},~{\rm for}~{\cal V}=10^5. 
\eeqn
 The coefficient of interaction vertex $e^{-}_{L}-H^{0}_{u}- {\tilde \lambda^{+}_i}$ corresponding to Figure 4.5 will be given by $
{\cal L}_{e^{-}_{L}-H^{0}_{u}-{\tilde \lambda^{+}_i}}= g_{YM}g_{ {{\cal A}_1}\bar{T}_B}X^{{*}B}{\bar\chi}^{\bar {\cal A}_1}\tilde{\lambda^{+}_i}+\partial_{{\cal A}_1}T_B D^{B} {\bar\chi}^{\bar {\cal A}_1}\tilde{\lambda^{+}_i}$.
Since $\partial_{{\cal A}_1}T_B$ does not give any term which is linear in ${\cal Z}_i$, so the second term contributes zero to the given vertex. By expanding $g_{{\cal A}_1\bar{T}_B}$ in the fluctuation linear in ${\cal Z}_1$ around its stabilized VEV, in terms of undiagonalized basis, we have: $ g_{{T_B} {\bar a}_1}\rightarrow-{\cal V}^{-\frac{13}{12}} (z_1-{\cal V}^{-\frac{1}{36}}),~{\rm and}~g_{YM}\sim{\cal V}^{-\frac{1}{36}}$.
 Considering $g_{YM}g_{{T_B} {\bar a}_1} \sim {\cal O}(1)g_{YM}g_{{T_B} {\bar {\cal A}}_1}$ as shown in \cite{gravitino_DM}; incorporating values of  $X^{B}=-6i\kappa_4^2\mu_7Q_{T_B}$, $\kappa_4^2\mu_7\sim \frac{1}{\cal V}$ and  $Q_{T_B}\sim{\cal V}^{\frac{1}{3}} (2\pi\alpha^\prime)^2\tilde{f}$, we get the contribution of physical gaugino(${\tilde \lambda^{+}_i}$)-lepton$(e_{L})$-Higgs($H^{0}_{u}$) interaction vertex given as follows:
\begin{equation}
\label{eq:eLHambda}
|C_{e_{L}{\tilde \lambda^{+}_i}{H_u^{0}}/H_d^{0}}|\equiv \frac{g_{YM}g_{{T_B} {\bar {\cal A}_1}}X^{T_B}\sim {\cal V}^{-\frac{47}{36}} {\tilde f}}{{ \sqrt{\hat{K}_{{\cal A}_1{\bar {\cal A}}_1}\hat{K}_{{\cal Z}_1{\bar {\cal Z}}_1}} }}{\cal Z}_1{\bar\chi}^{\bar {\cal A}_1}\tilde{\lambda^{0}}
 \equiv\tilde{f}\left({\cal V}^{-\frac{3}{2}}\right).
\end{equation}
To calculate the coefficient of  interaction vertex $e^{*}_{R}-H^{0}_{u}-\tilde{\lambda^{-}_i}$, we need to expand $g_{{\cal A}_3\bar{T}_B}$ in the fluctuation quadratic in ${\cal Z}_1$ to first conserve $SU(2)_L$ symmetry and after giving  VEV to one of the ${\cal Z}_i$, we get the required contribution
\begin{equation}
\label{eq:eRHlambda}
|C_{e^{*}_{R}{\tilde \lambda^{+}_i}{H_u^{0}}/H_d^{0}}|\equiv \frac{g_{YM}g_{{T_B} {\bar {\cal A}_3}}X^{T_B}\sim {\cal V}^{-\frac{16}{9}}\langle{\cal Z}\rangle {\tilde f}}{{ \sqrt{\hat{K}_{{\cal A}_3{\bar {\cal A}}_3}\hat{K}^{2}_{{\cal Z}_1{\bar {\cal Z}}_1}} }}{\cal Z}_1{\bar\chi}^{\bar {\cal A}_3}\tilde{\lambda^{0}}
 \equiv\tilde{f}\left({\cal V}^{-\frac{15}{9}}\frac{\langle{\cal Z}_i\rangle}{M_P}\right).
\end{equation}
Incorporating the results given in eqs.~(\ref{eq:Yhatz1z1a1})-(\ref{eq:eRHlambda}) in equation (\ref{eq:Ceechi}),
\pagebreak
we have,
\beqn
\label{eq:results}
 && |C_{e^{*}_{L}\chi^{+}_{1}{H_u^{0}}/{H_d^{0}}}|\equiv{\cal V}^{-\frac{1}{10}}, |C_{e^{*}_{L}\chi^{+}_{2}{H_u^{0}}/{H_d^{0}}}|\equiv{\cal V}^{\frac{1}{10}}, \nonumber\\
 && |C_{e^{*}_{R}\chi^{-}_{1}{H_u^{0}}/{H_d^{0}}}|\equiv\tilde{f}{\cal V}^{-\frac{3}{2}},
 |C_{e^{*}_{R}\chi^{-}_{2}{H_u^{0}}/{H_d^{0}}}|\equiv \tilde{f}{\cal V}^{-\frac{15}{9}}\frac{\langle{\cal Z}_i\rangle}{M_P}.
 \eeqn
Now, the one-loop EDM of the electron in this  case will be given as \cite{keum+kong}:
\beqn
\label{eq:EDMchi}
{\frac{d}{e}}|_{\chi^{\pm}_i}=\sum_{i}\frac{m_{\chi^{\pm}_j} Q'_{{e}(i)}}{(4\pi)^2} \left[ \frac{1}{m^{2}_{H^{0}_i}}{\rm Im}\left(\left(C_{e^{*}_{L}\chi^{+}_{i}{H_i^{0}}}C_{e^{*}_{R}\chi^{-}_{j}{H_i^{0}}}^*\right) A\left(\frac{m^{2}_{\chi^{\pm}_i}}{m^{2}_{H^{0}_i}}\right) \right)\right]. 
\eeqn
where $m_{\chi^{\pm}_j}$ and $m^{2}_{H^{0}_i}$ corresponds to masses eigenstates of chargino and Higgs mass matrix. The effective charge for this loop diagram will be $Q'_{{e}(i)}= Q_e C_{{\chi^{+}_i}{\chi^{-}_i}{\gamma}}$ where $C_{{\chi^{+}_1}{\chi^{-}_1}{\gamma}}=C_{{{\tilde H}^{+}_i}{{\tilde H}^{-}_i}{\gamma}},C_{{\chi^{+}_2}{\chi^{-}_2}{\gamma}}=C_{{{\tilde \lambda}^{+}_{i}{{\tilde  \lambda}^{-}_i}{\gamma}}}$.
The contributions of both higgsino-higgsino-gauge boson vertex and gaugino-gaugino-gauge boson have already obtained in the context of ${\cal N}=1$ gauged  supergravity in chapter {\bf 3}. Using the same,
\begin{equation}
C_{{\chi^{+}_1}{\chi^{-}_1}{\gamma}}\equiv {\tilde f}{\cal V}^{-\frac{5}{18}}, C_{{\chi^{+}_2}{\chi^{-}_2}{\gamma}}\equiv {\tilde f}{\cal V}^{-\frac{11}{18}}.
\end{equation}
Utilizing the results of $C_{e^{*}_{L/R}\chi^{\pm}_{i}{H_i^{0}}}$ vertices given in (\ref{eq:results}) and the assumption that value of phase factor associated with these couplings are of ${\cal O}(1)$; $m_{\chi^{\pm}_1}=m_{H_2}= {\cal V}^{\frac{59}{72}}m_{\frac{3}{2}}$,$m_{\chi^{\pm}_2} = {\cal V}^{\frac{2}{3}}m_{\frac{3}{2}}$ and $m_{H_1}\sim 125 GeV$ as given in chapter {\bf 2}, $\sin \theta_h=
 (0,1]$, $\phi_e= (0,\frac{\pi}{2}]$, and
$A(\frac{m^{2}_{\chi^{\pm}_i}}{m^{2}_{H^{0}_i}})\equiv \frac{m^{2}_{H^{0}_i}}{m^{2}_{\chi^{\pm}_i}}$ by using (\ref{eq:A}),  we have:
\beq
\label{eq:EDMchi1}
{\frac{d}{e}}|_{\chi^{\pm}_i}\equiv \frac{1}{\sqrt{2}(4\pi)^2}({\cal V}^{-\frac{1}{10}+\frac{1}{10}})\times\frac{{\tilde f}{\cal V}^{-\frac{5}{18}}} {{\cal V}^{\frac{59}{72}}m_{\frac{3}{2}}} \equiv {\cal O}(10^{-32}) cm,~ {\rm for} ~{\cal V}={\cal O}(1)\times 10^4.
\eeq
\subsection{One-Loop Diagrams Involving Gravitino and Sgoldstino}
{{\bf Gravitino contribution:}} In this subsection, we have estimated the EDM of electron (quark) by considering the gravitino as a propagator in one-loop diagrams despite the fact that these are logarithmically divergent. The loop diagrams are given in Figure~4.6. To get the numerical estimate of EDM corresponding to these diagrams, we first need to determine the contribution of relevant vertices in ${\cal N}=1$ gauged supergravity. 
\begin{figure}
\begin{center}
\begin{picture}(100,137) (280,40)
   \Line(120,50)(130,50)
   \Line(130,50)(210,50)
 \ZigZag(130,50)(210,50){3}{8}
   \Line(210,50)(220,50)
   \DashLine(130,50)(170,120){5}
   \Photon(170,120)(170,170){5}{4}
   \DashLine(170,120)(210,50){5}
   \Text(185,160)[]{$\gamma$}
   \Text(140,100)[]{${\tilde f}_{i}$}
   \Text(200,100)[]{${\tilde f}_{i}$}
   \Text(110,50)[]{{$f_{L}$}}
   \Text(230,50)[]{{$f_{R}$}}
   \Text(170,40)[]{{{${\tilde G}$}}}
   \Text(170,20)[]{{{$(a)$}}}
      \Line(280,50)(290,50)
      \DashLine(290,50)(370,50){5}
      \Line(370,50)(380,50)
   \Line(290,50)(330,120)
   \Photon(290,50)(330,120){4}{4}
   \Photon(330,120)(330,170){5}{4}
   \Line(330,120)(370,50)
   \ZigZag(330,120)(370,50){3}{8}
   \Text(345,160)[]{$\gamma$}
   \Text(300,100)[]{${\tilde \gamma}$}
   \Text(360,100)[]{${\tilde G}$}
   \Text(270,50)[]{{$f_{L}$}}
   \Text(390,50)[]{{$f_{R}$}}
   \Text(330,40)[]{{{${\tilde f}$}}}
   \Text(330,20)[]{{{$(b)$}}}
   \Line(450,50)(490,120)
   \Photon(490,120)(490,170){5}{4}
   \DashCArc(500,102.5)(20,300,118){5}
   \ZigZag(490,120)(510,85){3}{8}
   \Line(490,120)(510,85)
   \Line(510,85)(530,50)
   \Text(505,160)[]{$\gamma$}
   \Text(490,100)[]{${\tilde G}$}
   \Text(530,110)[]{${\tilde f}_{i}$}
   \Text(450,40)[]{{$f_{L}$}}
   \Text(530,40)[]{{$f_{R}$}}
   \Text(490,20)[]{{$(c)$}}
   \end{picture}
\end{center}
\vskip 0.2in
\caption{One-loop diagram involving gravitino.}
 \end{figure}
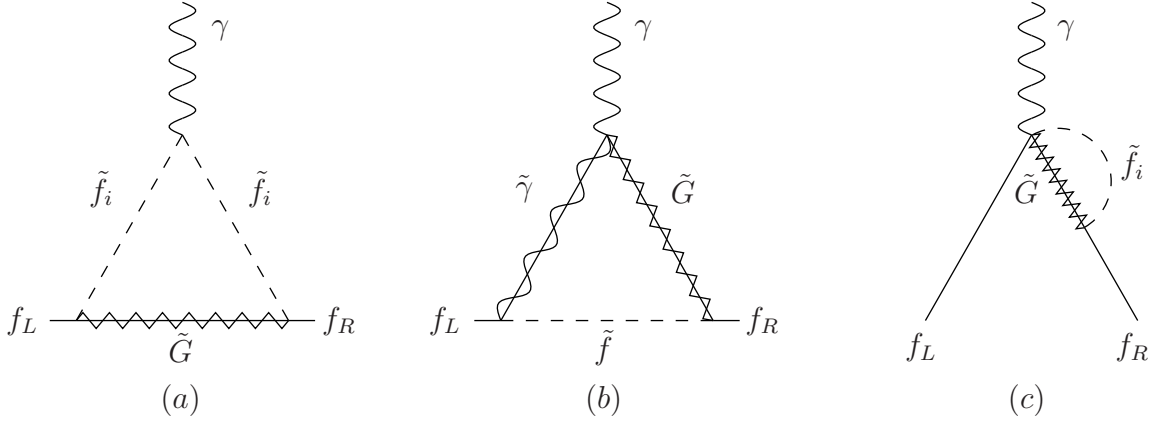
  In ${\cal N}=1$ gauged supergravity, the gravitino-fermion-sfermion vertex will be given as: $
 {\cal L}_{\tilde G-f-{\tilde f}}=-\frac{1}{2}\sqrt{2}eg_{ij}\partial_{\mu}{\phi}^i \chi^{j}{\gamma}^{\mu}{\gamma}^{\nu}\psi_{\mu}$.  The physical ${\psi_{\mu}}$- lepton(quark)-slepton(squark) vertex will be given as:
 \beqn
 \label{Cpsilltilde}
&& |C_{\tilde G\ e_L \ {\tilde e_L}}| \equiv \frac{g_{{\cal A}_1 {\bar {\cal A}}_1}}{\sqrt{{\cal K}_{{\cal A}_1 {\bar {\cal A}_1}}{\cal K}_{{\cal A}_1 {\bar {\cal A}_1}}}}\partial_{\mu}{\cal A}_1 \chi^{{\cal A}_1}{\gamma}^{\mu}{\gamma}^{\nu}\psi_{\mu}\equiv\partial_{\mu}{\cal A}_1 \chi^{{\cal A}_1}{\gamma}^{\mu}{\gamma}^{\nu}\psi_{\mu},\nonumber\\
 && |C_{\tilde G\ u_L \ {\tilde u_L}}| \equiv \frac{g_{{\cal A}_2 {\bar {\cal A}}_2}}{\sqrt{{\cal K}_{{\cal A}_2 {\bar {\cal A}_2}}{\cal K}_{{\cal A}_2 {\bar {\cal A}_2}}}}\partial_{\mu}{\cal A}_2 \chi^{{\cal A}_2}{\gamma}^{\mu}{\gamma}^{\nu}\psi_{\mu}\equiv \partial_{\mu}{\cal A}_2 \chi^{{\cal A}_2}{\gamma}^{\mu}{\gamma}^{\nu}\psi_{\mu}.
 \eeqn
 The contribution of physical sfermion -sfermion -photon vertices have already been obtained in subsection {\bf 3.1} and the values of the same are given as:
 \beq
 |C_{\tilde e_L \tilde e_L \gamma}|\equiv {\cal V}^{\frac{44}{45}}{\tilde {\cal A}}_1 \partial_{\mu}{\tilde {\cal A}}_1 A^{\mu},  |C_{\tilde u_L \tilde u_L \psi_{\mu}}| \equiv {\cal V}^{\frac{53}{45}}{\tilde {\cal A}}_2 \partial_{\mu}{\tilde {\cal A}}_2 A^{\mu}.
 \eeq
 The contribution of the fermion-sfermion-photino($\gamma$) vertex in the context of ${\cal N}=1 $ gauged supergravity action will given by $
{\cal L}_{f {\tilde f} {\tilde \gamma}}= g_{YM}g_{{\cal A}_I \bar{T}_B}X^{{*}B}{\bar\chi}^{\bar {\cal A}_I}\tilde{\gamma}+\partial_{{\cal A}_1}T_B D^{B} {\bar\chi}^{\bar {\cal A}_I}\tilde{\gamma}$.
 For $f={e}$,  by expanding $g_{a_1 \bar{T}_B}$ in the fluctuations linear in $a_1$ around its stabilized VEV, we have: $ g_{\bar {a_1}T_B}= {\cal V}^{-\frac{2}{9}}(a_1- {\cal V}^{-\frac{2}{9}}), ~{\rm and}~ \partial_{{\cal A}_1}T_B \rightarrow {\cal V}^{\frac{10}{9}}({\cal A}_1- {\cal V}^{-\frac{2}{9}})$.
Assuming that $g_{\bar { {\cal A}_1}T_B}={\cal O}(1)g_{\bar {a_1} T_B}$ and using  $X^{{*}B}={\kappa}^{2}_4 {\mu}_7 Q_B, D^B=\frac{4\pi\alpha^\prime\kappa_4^2\mu_7Q_Bv^B}{\cal V}$ where $ Q_B \sim {\cal V}^{\frac{1}{3}}{\tilde f}$ and ${\kappa}^{2}_4 {\mu}_7\sim \frac{1}{\cal V}$, we get:
\beqn
  |C_{e_L {\tilde e_L} {\tilde \gamma}}|\sim  \frac{{\cal V}^{-\frac{2}{9}}{\tilde f}}{\sqrt{{\cal K}_{{\cal A}_1 {\bar {\cal A}_1}}{\cal K}_{{\cal A}_1 {\bar {\cal A}_1}}}}{\tilde {\cal A}}_1 {\bar\chi}^{\bar {\cal A}_1}\tilde{\gamma}\equiv {\tilde f}{\cal V}^{-1}{\tilde {\cal A}}_1 {\bar\chi}^{\bar {\cal A}_1}\tilde{\gamma}.\nonumber
  \eeqn
For $f={u}$,  using  $ g_{\bar {{\cal A}_2}T_B} \sim {\cal O}(1) g_{\bar {a_2}T_B}= {\cal V}^{-\frac{5}{4}}(a_2- {\cal V}^{-\frac{1}{3}})$ and $\partial_{{\cal A}_2}T_B \rightarrow {\cal V}^{\frac{1}{9}}({\cal A}_2- {\cal V}^{-\frac{1}{3}})$, we have:
   \beqn
  |C_{u_L {\tilde u_L} {\tilde \gamma}}|\sim  \frac{{\cal V}^{-\frac{11}{9}}{\tilde f}}{\sqrt{{\cal K}_{{\cal A}_2 {\bar {\cal A}_2}}{\cal K}_{{\cal A}_2 {\bar {\cal A}_2}}}}{\tilde {\cal A}}_2 {\bar\chi}^{\bar {\cal A}_2}\tilde{\gamma}\equiv {\tilde f}{\cal V}^{-\frac{4}{5}}{\tilde {\cal A}}_2 {\bar\chi}^{\bar {\cal A}_2}\tilde{\gamma}.
  \eeqn
 The contribution of the gravitino-fermion-sfermion-photon vertex in the context of ${\cal N}=1 $ gauged supergravity action will be given as: $
  {\cal L}=-\frac{1}{2}{\sqrt 2}e g_{{\cal A}_I T^{*}_B}X^{B}A_{\mu}{\bar\chi}^{{\cal A}_I}{\gamma}^{\mu}{\gamma}^{\nu}{\psi_{\mu}}$.  Using the above-mentioned value of $g_{{\cal A}_1 T^{*}_B}$, $g_{{\cal A}_2 T^{*}_B}$  and $X^{B}$, the coefficient of physical  gravitino-lepton(quark)-slepton(squark)-photon vertex will be given as:
  \beqn
  \label{Gelel}
  && | C_{\tilde G e_L {\tilde e_L}\gamma}|\sim \frac{{\cal V}^{-\frac{8}{9}}{\tilde f}}{\sqrt{{\cal K}_{{\cal A}_1 {\bar {\cal A}_1}}{\cal K}_{{\cal A}_1 {\bar {\cal A}_1}}}}A_{\mu}{\bar\chi}^{{\cal A}_1}{\gamma}^{\mu}{\gamma}^{\nu}{\psi_{\mu}}\equiv {\tilde f}{\cal V}^{-\frac{5}{3}} A_{\mu}{\bar\chi}^{{\cal A}_1}{\gamma}^{\mu}{\gamma}^{\nu}{\psi_{\mu}},\nonumber\\
  &&| C_{\tilde G u_L {\tilde u_L}\gamma}|\sim \frac{{\cal V}^{-\frac{35}{18}}{\tilde f}}{\sqrt{{\cal K}_{{\cal A}_2 {\bar {\cal A}_2}}{\cal K}_{{\cal A}_2 {\bar {\cal A}_2}}}}A_{\mu}{\bar\chi}^{{\cal A}_2}{\gamma}^{\mu}{\gamma}^{\nu}{\psi_{\mu}}\equiv {\tilde f}{\cal V}^{-\frac{5}{3}} A_{\mu}{\bar\chi}^{{\cal A}_2}{\gamma}^{\mu}{\gamma}^{\nu}{\psi_{\mu}}.
  \eeqn
The contribution of photon ($\gamma$)-photino (${\tilde \gamma}$)-gravitino($\gamma$) vertex will be given $
{\cal L}= \frac{i}{4}e\bar {\gamma}^{\mu}{\lambda}[\slashed{\partial},\slashed{\cal A}]{\psi}_{\mu}$. We notice that there is no moduli space-dependent factor coming from this vertex.
 
The above Feynman diagrams involving gravitino in a loop have been explicitly worked out in \cite{mandez+orteu} to calculate the magnetic moment of muon in the context of spontaneously broken minimal ${\cal N}=1 $ gauged supergravity. We explicitly utilize their results in a modified form to get the estimate of EDM of electron/quark in the ${\cal N}=1$ gauged supergravity. The modified  results of magnetic moment of electron after multiplying with volume suppression factors coming from relevant vertices as calculated in equations (\ref{Cpsilltilde})-(\ref{Gelel}) are as follows:

For Figure~4.6(a):
\pagebreak
 \beqn
&& {a}^{\rm div}_{f}|_{4.6(a)}\equiv{\tilde f}{\cal V}^{a}(G_N m^{2}_{f}/{\pi})\sum_{j=1,2}\Bigl[\Gamma(\epsilon-1)[-\frac{1}{90}{\mu}^2+\frac{1}{18}{\mu}^{2}_{j}]+
 \Gamma(\epsilon)[\frac{2}{45}{\mu}^2+\frac{2}{9}]\nonumber\\
 && +(-1)^j \sin{\theta} \Gamma(\epsilon-1)[-{\mu}^{2}_{j}/3{\mu}]\Bigr],
 \eeqn
 where ${\cal V}^{a}$ is the Calabi-Yau volume-suppression factor.
 Here $\mu={m_f}/m_{\frac{3}{2}}$ and $\mu_j=m_{{\tilde f}_j}/m_{3/2}$, $j=1,2$; m is lepton mass, $m_{\frac{3}{2}}$ is gravitino mass. $m_{{\tilde f}_1}$ and $m_{{\tilde f}_2}$ are eigenvalues of diagonalized slepton(squark) mass matrix. In our set-up $\sin{\theta}=1$. Using $m_{{\tilde f}_1}=m_{{\tilde f}_2}={\cal V}^{\frac{1}{2}}m_{\frac{3}{2}}$, $m_{\frac{3}{2}}={\cal V}^{-2}M_P$  and $m_e={\cal O}(1) MeV$, we have
 $\mu_1=\mu_2= \frac{1}{\cal V}~~{\rm and}~~\mu= 10^{-11}~{\rm for}~{\cal V}=10^5$.
 
 For $f=e$, incorporating these values, dominant contribution will be of the form:
 \beqn
&&  {a}^{\rm div}_{e}|_{4.6(a)}\equiv {\tilde f}{\cal V}^{\frac{44}{45}}(G_N m^{2}_{e}/{\pi}) \Bigl[\frac{1}{18 {\cal V}^2}\Gamma(\epsilon-1) + \frac{2}{9}  \Gamma(\epsilon)\Bigr]  \equiv {\tilde f}{\cal V}^{\frac{44}{45}}(G_N m^{2}_{e}/{\pi}) \nonumber\\
&&{\hskip 0.8in} \Bigl[\frac{1}{18 {\cal V}^2}\Gamma(\epsilon-1)+\frac{1}{18 {\cal V}^2}\Gamma(\epsilon) + a'\Bigr].
 \eeqn
 where $ a'=(\frac{2}{9} -\frac{1}{18 {\cal V}^2}) \Gamma(\epsilon)$ is divergent piece.
 Using $-\Gamma(\epsilon-1)=\Gamma(\epsilon)(1+\epsilon)$, the finite contribution will be given as:
 ${a}^{\rm finite}_{e}|_{4.6(a)}\equiv \frac{1}{18}{\tilde f}{\cal V}^{-\frac{46}{45}}(G_N m^{2}_{e}/{\pi})$. Similarly, using the volume suppression factor coming from quark-quark photon vertex, we get:
${a}^{\rm finite}_{u}|_{4.6(a)}\equiv \frac{1}{18}{\tilde f}{\cal V}^{-\frac{37}{45}}(G_N m^{2}_{u}/{\pi})$. Now we use the relation between anomalous magnetic moment and electric dipole moment to get the numerical estimate of EDM of electron in this case. As given in \cite{Graesser+thomas}, 
\begin{equation}
\label{eq:afdf}
a_f= \frac{2 |m_f|}{e Q_f}|d_f|\cos{\phi},
\end{equation}
 where $m_f$ and $Q_f$ correspond to mass and charge of fermion; $d_f$ is electric dipole moment of fermion and $\phi$ is defined as $
\phi\equiv {\rm Arg}(d_f m^{*}_f)$. We consider that in the loop diagrams involving sfermion as propagators, the non-trivial phase responsible to generate EDM  appears from eigenstates of sfermion mass matrix (off-diagonal component of slepton mass matrix) and we assume the value of same as ${\phi_{d_f}}~ \exists~ (0,\frac{\pi}{2}]$. The first generation electron/quark mass has been calculated from complex effective Yukawa coupling(${\cal Y}^{\rm eff}_{{\cal Z}_I {\cal A}_{1/3}{\cal A}_{2/4}}$) in ${\cal N}=1$ gauged supergravity and there is a distinct phase factor ${\phi_{y_e/y_u}}$ associated with the same. Using the fact that ${\phi_{d_f}}\neq {\phi_{y_e/y_u}}$  the relative phase between two will be in the interval $\phi~ \exists~(0,\frac{\pi}{2})\sim {\cal O}(1)$. Hence,
\beqn
&& \frac{d_e}{e}|_{4.6(a)}= 2{|m_e|}\ {a}^{\rm finite}_{e}|_{4.6(a)}\equiv  \frac{1}{18}{\tilde f}{\cal V}^{-\frac{46}{45}}(G_N m_{e}/{\pi})\equiv 10^{-67}cm, \nonumber\\
&& \frac{d_u}{e}|_{4.6(a)}= 2{|m_u|}\ {a}^{\rm finite}_{u}|_{4.6(a)}\equiv  \frac{1}{18}{\tilde f}{\cal V}^{-\frac{37}{45}}(G_N m_{u}/{\pi})\equiv 10^{-67}cm.
\eeqn
For Figure~4.6(b):
 \beqn
 && {a}^{\rm div}_{f}|_{4.6(b)}\equiv{\tilde f}{\cal V}^{-a}(G_N m^{2}_{f}/{\pi})\sum_{j=1,2}\Bigl[\Gamma(\epsilon-1)[\frac{1}{20}{\mu}^2-\frac{1}{6}{\mu}^{2}_{j}]+
 \Gamma(\epsilon)[-\frac{7}{60}{\mu}^{2}]\nonumber\\
 &&{\hskip 0.7in} +(-1)^j \sin{2\alpha}\Gamma(\epsilon-1)[{\mu}^{2}_{j}/{\mu}]\Bigr]; ~{\rm where}~ {f=e,u}. \nonumber
 \eeqn
 For $f=e$, incorporating the values of masses and simplifying, now we will have
  \beqn
 &&   {a}^{\rm div}_{e}|_{4.6(b)}\equiv{\tilde f}{\cal V}^{-1}(G_N m^{2}_{e}/{\pi})\Bigl[-\frac{1}{6 {\cal V}^2}\Gamma(\epsilon-1) -\frac{7}{60}{\mu}^{2}  \Gamma(\epsilon)\Bigr]\equiv {\tilde f}{\cal V}^{-1}(G_N m^{2}_{f}/{\pi})\nonumber\\
 && {\hskip 0.7in}\Bigl[ -\frac{7}{60}{\mu}^{2}  \Gamma(\epsilon-1)-\frac{7}{60}{\mu}^{2}  \Gamma(\epsilon)+a'\Bigr]. 
 \eeqn
 where $a'=(-\frac{1}{6 {\cal V}^2}+\frac{7}{60}{\mu}^{2})\Gamma(\epsilon-1)$ is divergent piece.
 Picking up the finite contribution, we get $  {a}^{\rm finite}_{e}|_{4.6(b)}\equiv  \frac{7}{60}{\tilde f}{\cal V}^{-1}(G_N m^{2}_{f}/{\pi}){\mu}^{2}$, and therefore,
\beq
\frac{d_e}{e}|_{4.6(b)} \equiv2 |m_e| \ {{a}^{finite}_{e}|_{4.6(b)}}\equiv 10^{-65}GeV^{-1} \equiv 10^{-79}cm.
\eeq
Similarly, using volume suppression factor coming from quark-quark photon vertex,
 \beqn
 \frac{d_u}{e}|_{4.6(b)}\equiv 10^{-64}GeV^{-1} \equiv 10^{-78}cm. 
 \eeqn
For Figure 4.6(c):
 \beqn
 && {a}^{\rm div}_{f}|_{6(c)}\equiv {\tilde f}{\cal V}^{-\frac{5}{3}}(G_N m^{2}_{f}/{\pi})\sum_{j=1,2}\Bigl[\Gamma(\epsilon-1)[-\frac{1}{90}{\mu}^2+\frac{1}{9}{\mu}^{2}_{j}]+
 \Gamma(\epsilon)[\frac{1}{10}{\mu}^2-\frac{2}{9}]\nonumber\\
 && +(-1)^j \sin{2\alpha}\Gamma(\epsilon-1)[-2{\mu}^{2}_{j}/3{\mu}]\Bigr]. \nonumber
  \eeqn
As Similar to the above, incorporating the value of masses and further simplifying, dominant contribution is given by:
 \beqn
 &&   {a}^{\rm div}_{f}|_{6(c)}\equiv{\tilde f}{\cal V}^{-\frac{5}{3}}(G_N m^{2}_{f}/{\pi}) \Bigl[\frac{1}{9 {\cal V}^2}\Gamma(\epsilon-1)-\frac{2}{9}
 \Gamma(\epsilon)\Bigr]\equiv {\tilde f}{\cal V}^{-\frac{5}{3}}(G_N m^{2}_{f}/{\pi})\nonumber\\
 && {\hskip 0.7in} \Bigl[\frac{1}{9 {\cal V}^2} \Gamma(\epsilon-1)+\frac{1}{9 {\cal V}^2} \Gamma(\epsilon)+a'\Bigr].
  \eeqn
  where $a'= (-\frac{2}{9}-\frac{1}{9 {\cal V}^2} )\Gamma(\epsilon)$ is divergent piece.
  Considering the finite piece,  ${a}^{\rm finite}_{q}|_{4.6(c)}= \frac{1}{9}{\tilde f}{\cal V}^{-\frac{11}{3}}(G_N m^{2}_{f}/{\pi})$. Again using $\frac{d_f}{e}|_{4.6(c)}= 2 |m_f| \ {{a}^{\rm finite}_{e}|_{4.6(c)}}$, we get
 \beq
\frac{d_e}{e}|_{4.6(c)}= \frac{d_u}{e}|_{4.6(c)}\equiv 10^{-66} GeV^{-1} \equiv 10^{-80} cm.
\eeq
Hence, the overall contribution of EDM of electron as well as neutron/quark in case of one-loop Feynman diagrams involving gravitino is:
\beq
\frac{d_{e}}{e}|_{\tilde G}=\frac{d_{n}}{e}|_{\tilde G}=\frac{d_{e/u}}{e}|_{4.6(a)}+\frac{d_{e/u}}{e}|_{4.6(b)}+\frac{d_{e/u}}{e}|_{4.6(c)}\equiv 10^{-67}cm.
\eeq

{{\bf Sgoldstino contribution:}} In supersymmetric models, the sgoldstino is the bosonic component of the superfield corresponding to which there is an $F$-term (D-term) supersymmetry breaking. In our model, supersymmetry is broken in the bulk sector and the scale of the same is governed by $F$-term (assuming that in dilute flux approximation $V_{D} << V_{F}$) corresponding to bulk fields ($F^{\tau_{S}},F^{\tau_{B}},{\cal G}^a$) where $\tau_{S}$ and ${\tau_{B}}$ correspond to `small' and `big' divisor volume moduli and ${\cal G}^a$ correspond to complexified NS-NS and RR axions. It was shown in \cite{gravitino_DM}, at $M_s$,  $|F^{\tau_{S}}|>|F^{{\cal G}^a}|, |F^{\tau_B}|$. From section {\bf 3.1}, the requirement of the quark-quark-photon coupling to be the SM at the EW scale, we see that $|F^{\tau_B}|$ is the most dominant $F$-term at the EW scale. To obtain an estimate of the off-shell goldstino multiplet, we consider the same to be: $(\tau_B, \chi_{B},F^B)$, where $\tau_B$ is a complex scalar field. Here, we identify $\sigma_B$ with scalar (sgoldstino) field and $\rho_B$ with pseudo-scalar (sgoldstino) field.

{\emph {Mass of sgoldstino}}: The dominant contribution to $F$-term potential, at the string scale $M_s$, is given by $V = ||F^{\tau_S}||^2$, where $F^{{\bar\tau}_S}= e^{K/2} {\bar\partial}^{{\bar\tau}_S}\partial^{\beta}K D_{\beta}W$\footnote{We note that $e^{K({\tau}_{S,B},G^a,z^i, a_I,...)}
{\bar\partial}^{\bar{\cal I}}\partial^{\cal J}K({\tau}_{S,B},G^a,z^i, a_I,...) D_{\cal J}W D_{\bar{\cal I}}{\bar W}({\cal I}\equiv T_{s,b},G^a,z^i,a_I,...)=e^{K(\tau_{z,b},G^a,z^i,a_I,...)}{\bar\partial}^{\bar{\alpha}}\partial^\beta K(\tau_{S,B},G^a,z^i,a_I,...) D_\beta W D_{\bar\alpha}{\bar W}(\beta=\tau_{S,B},G^a,z^i,a_I)$; however $G_{{\cal I}{\bar{\cal J}}}=\partial_{\cal I}{\bar\partial}_{\bar{\cal J}} K(T_{S,B},G^a,z^i, a_I,...),
G_{\alpha{\bar\beta}}\neq\partial_\alpha{\bar\partial}_{\bar{\beta}}K(\tau_{S,B},G^a,z^i,a_I,...)$ as $\tau_{S,B}$ is not an ${\cal N}=1$ chiral coordinate.}. At the EW scale the $F$-term potential receives the dominant contribution
from the $||D_{\tau_B}W||^2$ term and is estimated to be: $V(n_s=2)|_{EW}\sim e^KK^{\tau_S{\bar\tau}_B}D_{\tau_S}W D_{{\bar\tau}_B}{\bar W} + e^KK^{\tau_B{\bar\tau}_B}|D_{\tau_B}W|^2$, near $\langle\sigma_S\rangle\sim\frac{ln {\cal V}}{({\cal O}(1))_{\sigma_S}^4}, \langle\sigma_B\rangle\sim \frac{{\cal V}^{\frac{2}{3}}}{({\cal O}(1)_{\sigma_B})^4}$ yields $$\frac{\partial^2V}{\partial\sigma_B^2}\biggr|_{EW}\sim {\cal V}^{-\frac{1}{3}}m_{3/2}^2({\cal O}(1)_{\sigma_B})^2\left(({\cal O}(1)_{\sigma_S})^2 + \frac{({\cal O}(1)_{\sigma_S})^6}{ln {\cal V}}\right).$$ For the aforementioned ${\cal O}(1)_{\sigma_B}=\frac{{\cal O}(1)_{\sigma_S}}{2}\sim3.5$ for ${\cal V}\sim10^4,
 \left(\frac{\partial^2V}{\partial\sigma_B^2}\right)_{EW}\sim{\cal V}^{\frac{4}{3}}$ and the canonically normalized coefficient quadratic in the fluctuations, yields the sgoldstino mass estimate:
$$m_{\tau_B} \sim \sqrt{\frac{\left(\frac{\partial^2V}{\partial\sigma_B^2}\right)_{EW}}{\kappa^2_4\mu_7 K^{\tau_B{\bar\tau}_B}_{EW}}}\sim {\cal O}(1)m_{3/2}.$$
It will be interesting to get the contribution of the same to electron/neutron EDM.
\begin{figure}
\begin{center}
\begin{picture}(100,137) (200,20)
   \Line(100,50)(110,50)
   \DashLine(110,50)(190,50){5}
   \Line(190,50)(200,50)
   \Line(110,50)(150,120)
   \Photon(150,120)(150,170){5}{4}
   \Line(150,120)(190,50)
   \Text(165,160)[]{$\gamma$}
   \Text(125,100)[]{${f}_{i}$}
   \Text(175,100)[]{${f}_{i}$}
   \Text(90,50)[]{{$f_{L}$}}
   \Text(210,50)[]{{$f_{R}$}}
   \Text(150,40)[]{{{${{\tau}_B}$}}}
   \Text(150,20)[]{{{$(a)$}}}
      \Line(280,50)(290,50)
      \Line(290,50)(370,50)
      \Line(370,50)(380,50)
   \Photon(290,50)(330,120){5}{4}
   \Photon(330,120)(330,170){5}{4}
   \DashLine(330,120)(370,50){5}
   \Text(345,160)[]{{$\gamma$}}
   \Text(305,100)[]{${\tilde \gamma}$}
   \Text(355,100)[]{${\tau}_B$}
   \Text(270,50)[]{{$f_{L}$}}
   \Text(390,50)[]{{$f_{R}$}}
   \Text(330,40)[]{{{${f_L}$}}}
   \Text(330,20)[]{{{$(b)$}}}
   \end{picture}
\end{center}
\caption{One-loop diagram involving sgoldstino.}
 \end{figure}
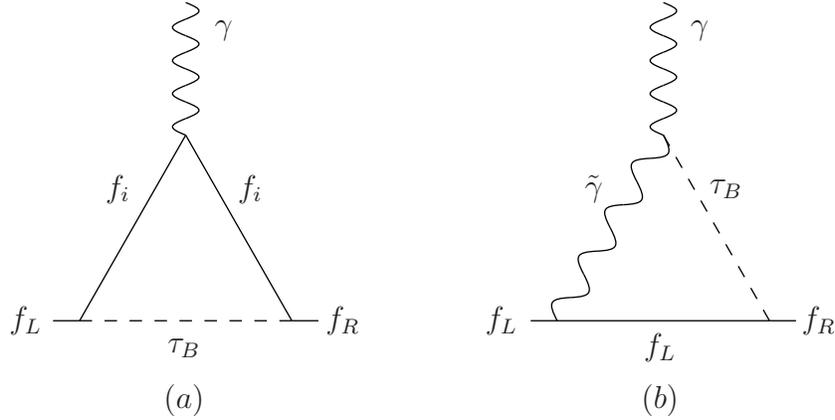
 To get the analysis of one-loop diagrams involving sgoldstino,  we consider only scalar sgoldstino field ($\sigma_s$) and first calculate the contribution of vertices involving sgoldstino in the context of ${\cal N}=1$ gauged supergravity. The coefficient of lepton($e_L$)-scalar(sgoldstino($\sigma_B$))-lepton($e_R$) vertex has been calculated by expanding $\frac{e^{\frac{K}{2}}}{2}\left({\cal D}_{{\cal A}_1} D_{{\cal A}_3}W\right){\bar \chi}^{{\cal A}_1}\chi^{{\cal A}_3}$ in the fluctuations linear in $\sigma_B$ in ${\cal N}=1$ gauged supergravity. By expanding above in the fluctuations linear in $\sigma_B\rightarrow \sigma_B+{\cal V}^{\frac{2}{3}}{M_P}$ , on simplifying,
we have: $\frac{e^{\frac{K}{2}}}{2} {\cal D}_{{\cal A}_1}D_{{\cal A}_3}W\chi^{{\cal A}_1}{\bar\chi}^{{\cal A}_3} \sim {\cal V}^{-\frac{13}{3}}\delta{\sigma_B}\chi^{{\cal A}_1}{\bar\chi}^{{\cal A}_3}$. The physical lepton($e_L$)-sgoldstino($\sigma_B$)-lepton($e_R$) vertex will be given as
\begin{eqnarray}
\label{eq:2a2}
|C_{{\delta \sigma_B} {e_{L}} {e^{c}_{R}}}| \equiv \frac{{\cal V}^{-\frac{13}{3}}}{{\sqrt{k^{2}_4 \mu_7 G^{\tau_B \bar {\tau_B}}\hat{K}_{{\cal A}_1{\bar{\cal A}}_1}{\hat{K}_{{\cal A}_3{\bar{\cal A}}_3}}}}}\equiv {\cal V}^{-\frac{92}{15}}, {\rm for}~{\cal V}\sim {10^5}.
\end{eqnarray}
Similarly,  the coefficient of quark($u_L$)-scalar(sgoldstino($\sigma_B$))-lepton ($u_R$) vertex can be calculated by expanding $\frac{e^{\frac{K}{2}}}{2}\left({\cal D}_{{\cal A}_2} D_{{\cal A}_4}W\right){\bar \chi}^{{\cal A}_2}\chi^{{\cal A}_4}$ in the fluctuations linear in $\sigma_B$ in $N=1$ gauged supergravity. Using the similar procedure, we get: $\frac{e^{\frac{K}{2}}}{2} {\cal D}_{{\cal A}_2}D_{{\cal A}_4}W\chi^{{\cal A}_2}{\bar\chi}^{{\cal A}_4}\\
\sim{\cal V}^{-4}\delta{\sigma_B}\chi^{{\cal A}_2}{\bar\chi}^{{\cal A}_4}$.
Therefore, the physical quark($u_L$)-sgoldstino ($\sigma_B$)-quark($u_R$)vertex will be given as
\begin{eqnarray}
\label{eq:2a2u}
|C_{{\delta \sigma_B} {u_{L}} {u^{c}_{R}}}|\sim \frac{{\cal V}^{-5}}{{\sqrt{k^{2}_4 \mu_7 G^{\tau_B \bar {\tau_B}}\hat{K}_{{\cal A}_2{\bar{\cal A}}_2}{\hat{K}_{{\cal A}_4{\bar{\cal A}}_4}}}}}\equiv {\cal V}^{-\frac{33}{5}}, {\rm for}~{\cal V}\sim {10^5}.
\end{eqnarray}
In ${\cal N}=1$ supergravity, the contribution of photon-sgoldstino(scalar)-photon will be accommodated by gauge kinetic term ${\cal L}= Re(T_B) F\wedge *_{4}F$,
where $Re(T_B)=  \sigma_B - C_{i \bar j}a_i a_{\bar j}$.  Considering $\sigma_B \rightarrow \langle \sigma_B \rangle+\delta \sigma_B $, coefficient of the physical vertex will be given as:
\beqn
|C_{\gamma\gamma \delta\sigma_B}| \equiv \frac{1/M_{p}}{\sqrt{k^{2}_4 \mu_7 G^{\tau_B \bar {\tau_B}}}}\sim \frac{{\cal V}^{-\frac{2}{3}}}{M_{p}}
\eeqn
 The possibility of getting fermion-fermion-photon vertex $C_{f f^{*} \gamma}\equiv {\cal O}(1)$ has been shown in subsection {\bf 3.1}.
 
Now we use the values of coefficients of relevant vertices to evaluate the estimate of EDM for loop diagrams given in Figure~4.7(a) and 4.7(b).The diagrams have been evaluated in \cite{brignole_et_al} to determine the estimate of magnetic moment of muon in ${\cal N}=1$  global supersymmetry. Utilizing their results in a modified form in the context of ${\cal N}=1$ gauged SUGRA and the relation between magnetic moment and EDM as given in equation (\ref{eq:afdf}), 
\\
 For Figure~4.7(a), the magnitude of electric dipole moment will be
\beqn
|\frac{d_f}{e}|_{4.7(a)}=\frac{m_f}{16 \pi^2}\cos{\phi}\Bigl[ (C_{\delta\sigma_B f_L f^{c}_R})^2 \int^{1}_{0} dx \frac{x^{2}(2-x)}{m^{2}_{\sigma_B}(1-x) + m^{2}_{f}x^2} \Bigr].
\eeqn
Putting the value of $|C_{\delta\sigma_B e_L e^{c}_R}|\equiv {\cal V}^{-\frac{92}{15}}, |C_{\delta\sigma_B u_L u^{c}_R}|\equiv {\cal V}^{-\frac{33}{6}}$, and the value of masses $m_{\sigma_B}= m_{\frac{3}{2}}, m_{e}=0.5 MeV$, we get
\beqn
|\frac{d_e}{e}|_{4.7(a)}\equiv 10^{-95}cm,~ {\rm and}~|\frac{d_n}{e}|_{4.7(a)}\equiv 10^{-89}cm.
\eeqn
For Figure~4.7(b):
\beqn
&& {\hskip -0.25in} |\frac{d_f}{e}|_{4.7(b)}= \frac{C_{\delta\sigma_B f_L f^{c}_R}C_{\gamma\gamma \delta\sigma_B}}{8 \pi^2} \Bigl[ \Delta_{UV} - \frac{1}{2}-  \int^{1}_{0}dx \int^{1-x}_{0} dy  \log \Bigl[\frac {m^{2}_{\sigma_B}y + m^{2}_{f} x^2}{\mu^2} \Bigr] \Bigr].\nonumber\\
\eeqn
where $\Delta_{UV}= \log[\frac{\Delta^{2}_{\rm UV}}{\mu^2}]-1$.
Incorporating values of relevant inputs and considering the finite piece,
\beqn
&& |\frac{d_e}{e}|_{4.7(b)}\equiv 10^{-72}cm,  |\frac{d_n}{e}|_{4.7(b)}\equiv 10^{-68}cm.
\eeqn
Hence, the overall contribution of sgoldstino to EDM of electron/neutron is
\beqn
&&|\frac{d_{e}}{e}|_{sgoldstino} = |\frac{d_{e}}{e}|_{4.7(a)}+ |\frac{d_{e}}{e}|_{4.7(b)}\equiv10^{-72}cm,\\
&& \hskip -0.5in{\rm and}\nonumber\\
&&
|\frac{d_{n}}{e}|_{sgoldstino} = |\frac{d_{n}}{e}|_{4.7(a)}+ |\frac{d_{n}}{e}|_{4.7(b)}\equiv10^{-68}cm.
\eeqn
The results of all possible one-loop diagrams contributing to EDM of electron/neutron  are summarized in a table given below:
\begin{table}[h]
\label{table:decay_lifetime}  
\begin{tabular}{l c c rrrrrrr}  
\hline\hline                       
 One-loop particle exchange  & Origin of ${\mathbb {C}}$ phase & $d_{e}$(e cm)  & $d_n$(e cm)
\\ [3.0ex]
\hline
{$\lambda^{0}{\tilde f}$} &  sfermion mass eigenstates & $10^{-39}$& $10^{-38}$    \\[1ex]
{$\chi^{0}_i \tilde f$} & ''& $10^{-37}$ & $ 10^{-34} $ & \\[1ex]
{$f \tilde f$}& ''& $10^{-45}$& $ 10^{-45}  $   \\[1ex]
{$f H^{0}_i$} &  Higgs mass eigenstates  & $10^{-34}$ & $ 10^{-33}  $   \\[1ex]
{$\chi^{\pm} H^{0}_i$}  &  ''  & $10^{-32}$ &  $-$   \\[1ex]
{${\rm gravitino}(\tilde G)\ {\tilde f}$}    & sfermion mass eigenstates & $10^{-67}$ & $10^{-67}$   \\[1ex]
{${\rm sgoldstino}\ {\tilde f}$}   & sfermion mass eigenstates  & $10^{-72}$ & $10^{-68}$   \\[1ex]
 \hline                          
\end{tabular}
\caption{Results of EDM of electron/neutron for all possible one-loop diagrams.}
\label{tab:PPer}
\end{table}
\section{Two-loop Level Barr-Zee Type Contribution to EDM}
In the two-loop diagrams discussed in this section, the CP-violating effects are mainly demonstarted by complex effective Yukawa couplings which include R-parity violating couplings, SM-like Yukawa couplings as well as couplings involving higgsino, and complex scalar trilinear couplings in the context of ${\cal N}=1$ gauged supergravity. In the subsection given below, we present the contribution of individual Barr-Zee type diagrams formed by including an internal fermion loop generated by R-parity violating interactions, SM-like Yukawa interactions and gaugino(gaugino)-higgsino(higgsino)-Higgs couplings. The two-loop diagrams are shown in Figure~4.8.
\subsection{Two-Loop Level Barr-Zee Feynman Diagrams Involving Internal Fermion loop}
\begin{figure}
\begin{center}
\begin{picture}(100,200) (290,-35)
   \ArrowLine(130,60)(280,60)
   \DashLine(140,60)(175,110){4}
   \Text(185,112)[]{{$P_L$}}
   \CArc(200,120)(25,0,180)
   \ArrowArc(200,120)(25,180,0)
   \Photon(225,110)(270,60){3}{7}
   \Photon(200,145)(200,180){2}{4}
    \Text(150,100)[]{$H^{0}_i$}
   \Text(215,165)[]{$\gamma^0$}
    \Text(165,130)[]{$u(e)$}
   \Text(115,53)[]{{$e_{L}(u_L)$}}
   \Text(295,53)[]{{$e_{R}(u_R)$}}
   \Text(260,100)[]{{$\gamma^0$}}
   \Text(200,40)[]{{$(a)$}}
 \ArrowLine(380,60)(530,60)
  \DashLine(390,60)(425,110){4}
  \CArc(450,120)(25,0,180)
   \ArrowArc(450,120)(25,180,0)
    \Photon(475,110)(520,60){3}{7}
   \Photon(450,145)(450,180){2}{4}
    \Text(400,100)[]{$H^{0}_i$}
   \Text(465,165)[]{$\gamma^0$}
    \Text(420,130)[]{$\chi^{\pm}_i$}
   \Text(365,53)[]{{$e_{L}(u_L)$}}
   \Text(545,53)[]{{$e_{R}(u_R)$}}
   \Text(510,100)[]{{$\gamma^0$}}
   \Text(450,40)[]{{$(b)$}}
    \Line(245,-100)(415,-100)
    \DashLine(255,-100)(300,-40){4}
     \Text(310,-40)[]{{$P_L$}}
     \CArc(325,-40)(25,0,180)
     \ArrowArc(325,-40)(25,180,0)
    \Photon(350,-40)(405,-100){3}{7}
    \Photon(325,-15)(325,20){2}{4}
    \Text(265,-70)[]{${\tilde \nu}_{iL}$}
   \Text(340,-5)[]{$\gamma^0$}
    \Text(290,-30)[]{$u(e)$}
   \Text(235,-107)[]{{$e_{L}(u_L)$}}
   \Text(425,-107)[]{{$e_{R}(u_R)$}}
   \Text(390,-60)[]{{$\gamma^0$}}
   \Text(325,-120)[]{{$(c)$}}
\end{picture}
\end{center}
\vskip 1.02in
\caption{Two-loop diagram involving fermions in the internal loop.}
 \end{figure}
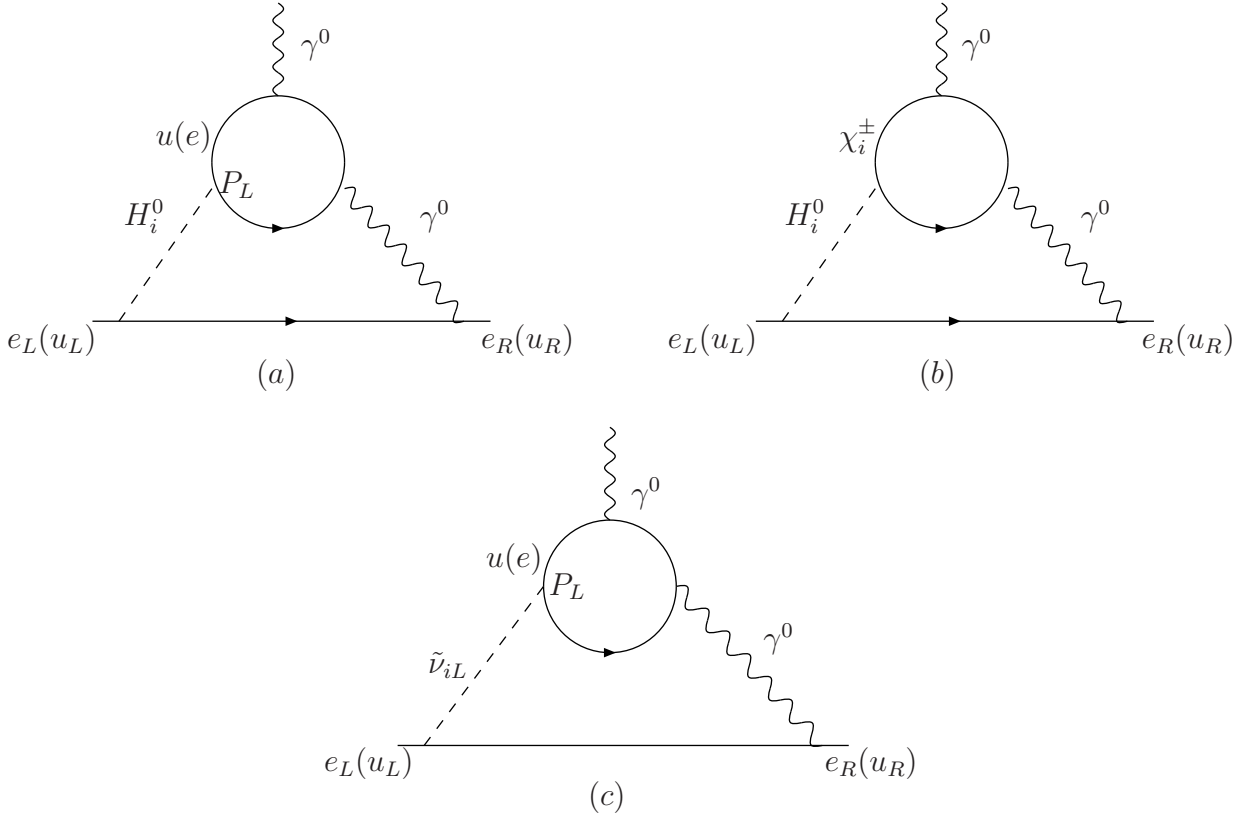
 {{\bf Higgs contribution:}}
For a two-loop diagram given in Figure~4.8(a), the interaction Lagrangian is governed by Yukawa couplings given as:
\beqn
{\cal L} \supset  {\hat Y}_{H^{0}_i u_{L} u^{c}_{R}}{H}^{0}_{i} u_{kL} u^{c}_{kR}+ {\hat Y}^{*}_{H^{0}_i e_{L} e^{c}_{R}} H^{0}_{i} e_{jL} e^{c}_{jR}+h.c..
\eeqn
We have already given the estimate of effective Yukawa couplings for first generation of leptons and quarks in chapter {\bf 2} in the context of ${\cal N}=1$ gauged supergravity. Using those results, we have
\beqn
 {\hat Y}_{H^{0}_i e_{L} e^{c}_{R}}\sim {\hat Y}^{\rm eff}_{{\cal Z}_i {\cal A}_1 {\cal A}_3}\equiv   {\cal V}^{-\frac{47}{45}}e^{i \phi_{y_e}}, {\hat Y}_{H^{0}_i u_{L} u^{c}_{R}}\sim {\hat Y}^{\rm eff}_{{\cal Z}_i {\cal A}_2 {\cal A}_4}\equiv  {\cal V}^{-\frac{17}{18}}e^{i \phi_{y_u}}; {\rm for~ {\cal V}=10^5}\nonumber\\
\eeqn
where $e^{i \phi_{y_e}}$ and $e^{i \phi_{y_u}}$ are non-zero phases of aforementioned Yukawa couplings.

For a two-loop Barr-Zee diagram involving internal fermion loop and taking into account the chirality flip between internal loop and external line, the analytical expression has been derived in \cite{Pilaftsis, Yamanaka_fermion}. Using the same, the electric dipole moment of electron for a loop diagram given in Figure 4.8(b) will be\footnote{We consider $Q^{'}_{e}= C_{e e^{*}\gamma} Q_{e} \sim Q_e$ because $ C_{e e^{*}\gamma}\sim {\cal O}(1)$ as shown in {{\bf subsection 4.3}}. Similarly $Q^{'}_{u}\sim Q_{u}$.}:
\pagebreak
\beqn
\label{eq:2loopdH1}
\frac{d}{e}|_{H}= \sum_{i=1,2} Im \left({\hat Y}_{H^{0}_i e_{L} e^{c}_{R}}{\hat Y}_{H^{0}_i u_{L} u^{c}_{R}} \right)\frac{\alpha_{em}Q^{2}_{u}Q_{e}}{16 \pi^3 m_{e j}} \left(f(z_1)- g(z_1)\right).
\eeqn
and EDM of neutron will be given as:
\beqn
\label{eq:2loopdH1n}
\frac{d}{n}|_{H}= \sum_{i=1,2} Im \left({\hat Y}_{H^{0}_i e_{L} e^{c}_{R}}{\hat Y}_{H^{0}_i u_{L} u^{c}_{R}} \right)\frac{\alpha_{em}Q^{2}_{e}Q_{u}}{16 \pi^3 m_{e j}} \left(f(z_2)- g(z_2)\right).
\eeqn
where $z_1= \frac{m^{2}_{e}}{m^{2}_{H^{0}_i}}; z_2= \frac{m^{2}_{u}}{m^{2}_{H^{0}_i}}$ and
\beqn
&& f(z)=\frac{z}{2}\int^{1}_{0} dx \frac{1- 2x(1-x)}{x(1-x)-z}{\rm ln} \left(\frac{x(1-x)}{z} \right)\nonumber\\
&& g(z)=\frac{z}{2}\int^{1}_{0} dx \frac{1}{x(1-x)-z}{\rm ln} \left(\frac{x(1-x)}{z} \right).
\eeqn
Using the value of masses $m_{H^{0}_1}=125 GeV$,  $m_{H^{0}_2}={\cal V}^{\frac{59}{72}}m_{\frac{3}{2}}$ and $m_{e}=0.5GeV$, $
f\left( \frac{m^{2}_{e}/m^{2}_{u}}{m^{2}_{H^{0}_1}}\right)= g\left( \frac{m^{2}_{e}/m^{2}_{u}}{m^{2}_{H^{0}_1}}\right)= 10^{-10}; f\left( \frac{m^{2}_{e}/m^{2}_{u}}{m^{2}_{H^{0}_2}}\right)= g\left( \frac{m^{2}_{e}/m^{2}_{u}}{m^{2}_{H^{0}_2}}\right)= 10^{-23}$, 
Utilizing the same and assuming $e^{i (\phi_{y_e}-\phi_{y_u})}=(0,1]$, equation (\ref{eq:2loopdH1}) and (\ref{eq:2loopdH1n}) reduces to give EDM result as follows:
\beqn
\label{eq:2loopdH12}
\frac{d}{e}|_{H}=\frac{d}{n}|_{H}\sim {\cal V}^{-2} \times 10^{-2} \times 10^{-10}= 10^{-22}GeV^{-1} \equiv 10^{-36} cm.
\eeqn
{{\bf Chargino contribution:}} In a loop diagram 4.8(b), the general Lagrangian governing the interaction of chargino's will be as:
\beqn
{\cal L} \supset  {C}_{ikk}{H}^{0}_{i} {\chi}^{+}_{kL} {\chi}^{-}_{kR}+ {\hat Y}^{*}_{H^{0}_i{e_L}{e^{c}_R}}H^{0}_{i} e_{jL} e^{c}_{jR}+ {\hat Y}^{*}_{H^{0}_i{u_L}{u^{c}_R}}H^{0}_{i} u_{jL} u^{c}_{jR}+ h.c. 
\eeqn
We evaluate the the contribution of chargino(${\chi}^{\pm}_i$)-Higgs-chargino (${\chi}^{\pm}_1$)  vertex in ${\cal N}=1 $ gauged supergravity. As described in section {\bf 5} of chapter 2, ${\chi}^{\pm}_1$ and ${\chi}^{\pm}_2$  correspond to a higgsino (${\tilde H}^{\pm}_{i}$) with a very small admixture of gaugino (${\lambda}^{\pm}_{i}$) and vice versa. So $C_{\chi^{+}_i \chi^{-}_1 H^{0}_i}\equiv C_{{\tilde H^{+}_{i}{\tilde H^{-}_{i} H^{0}_i}}}$; and $C_{\chi^{+}_2 \chi^{-}_2 H^{0}_i} \equiv C_{{{\tilde \lambda}^{+}_i {\lambda}^{-}_{i}H^{0}_i}}$.
 
  \underline{Higgsino($\chi^{-}_{kL}$)-Higgs-higgsino($\chi^{+}_{kR}$) vertex}:
Given that higgsino is a majorana particle, therefore $\chi^{+}_{kR}= (\chi^{-}_{kL})^c$. In our model,  higgsino has been identified with position moduli ${\cal Z}_i$, the contribution of this vertex in ${\cal N}=1$ gauged sypergravity will be given by expanding $e^{\frac{K}{2}} {\cal D}_{{\cal Z}_i} {\cal D}_{\bar {\cal Z}_i}W $ in the fluctuations linear in ${\cal Z}_i$ about its stabilized VEV. Since $SU(2)_L$ symmetry is not conserved for this vertex. We will expand the above in the fluctuations quadratic in ${\cal Z}_i$; giving  VEV to one of the ${\cal Z}_i$.  Considering ${z_i} \rightarrow {\cal V}^{\frac{1}{18}}+ \delta {z_i}$; we have
${\cal D}_{z_i} {\cal D}_{\bar z_i}W= {\cal V}^{-\frac{16}{9}}z_i \langle z_i\rangle$. Using an argument that ${\cal D}_{{\cal Z}_i} {\cal D}_{\bar {\cal Z}_i}W = {\cal D}_{ \bar {z_1}}D_{z_i}W$, the physical vertex will be given as:
\beqn
C_{\chi^{+}_i \chi^{-}_1 H^{0}_i} \equiv C_{{\tilde H^{+}_{i}{\tilde H^{-}_{i} H^{0}_i}}} =\frac{{\cal V}^{-\frac{7}{4}}}{(\sqrt{\hat{K}_{{\cal Z}_1{\bar{\cal Z}}_1}})^4}= {\cal V}^{\frac{1}{4}}e^{i{ \phi_{\chi_1}}}.
\eeqn
where ${{ \phi_{\chi_1}}}$ correspond to non-zero phase associated with the aforementioned coupling.
 
\underline{Gaugino($\lambda^{+}_{kR}$)-Higgs- gaugino ($\lambda^{+}_{kL}$) vertex}: The coefficient of this vertex will be given from the kinetic term of gaugino. The interaction term corresponding to this coupling will be given by considering term $ {\cal L}= i{\bar {\lambda_L}}{\gamma}^m {\frac{1}{4}(K_{{\cal Z}_i}{\partial}_m {{\cal Z}_i}- c.c.)}{\lambda}_{L}+(\partial_{{\cal Z}_i} T_B){\bar {\lambda_L}}{\gamma}^m {\frac{1}{4}(K_{{\cal Z}_i}{\partial}_m {{\cal Z}_i}\\-
 c.c.)}{\lambda}_{L}$, where  ${\lambda}_{L}$ corresponds to gaugino. Given that charged(gaugino's) are either $SU(2)_L$ singlets or triplets, the aforementioned vertex  does not preserve $SU(2)_L$ symmetry - one has to obtain the term bilinear in ${\cal Z}_i$ such that we give  VEV to one of the ${\cal Z}_i$. Since $(\partial_{{\cal Z}_i} T_B)$ does not contain terms bilinear in ${\cal Z}_i$ which is needed to ensure $SU(2)_L$ symmetry, second term contributes zero to the given vertex. In terms of undiagonalized basis,
$\partial_{z_i}K\sim{\cal V}^{-\frac{2}{3}}\langle z_i\rangle$, and using
$\partial_{{\cal Z}_i}K\sim  {\cal O}(1) \partial_{z_i}K$, we have:
$\partial_{{\cal Z}_i}K\sim {\cal V}^{-\frac{2}{3}}\langle{\cal Z}_i\rangle$, incorporating the same
\begin{eqnarray}
\label{eq:Cggh}
 & & {\cal L}=  \frac{{\cal V}^{-\frac{2}{3}}\langle{\cal Z}_i\rangle{\bar {\lambda_L}}\frac{\slashed{\partial}{{\cal Z}_i}}{M_P}{\lambda_L}}{{\sqrt{(\hat{K}_{{\cal Z}_1{\bar {\cal Z}}_1}})^2}}\sim   {\cal V}^{\frac{13}{36}}h {\bar {\lambda_L}}\frac{\slashed{p}_h}{M_P}{\lambda_L}
 \sim {\cal V}^{\frac{13}{36}}{h}{\bar {\lambda_L}}\frac{{\gamma}\cdot({p_{e_L} + p_{e_R}})}{M_P}{\lambda}_{L}  
 \end{eqnarray}
Therefore,
 \beqn
 C_{\chi^{+}_2 \chi^{-}_2 H^{0}_i} \equiv C_{H^{0}_i{{\lambda^{-}_{R}}}{\lambda}^{+}_{L}}\sim {\cal V}^{\frac{13}{36}}\frac{m_{e}}{M_P}e^{i{{ \phi_{{\tilde\lambda}^{0}_1}}}}.
 \eeqn
where ${{ \phi_{{\tilde\lambda}^{0}_1}}}$ correspond to non-zero phase associated with the aforementioned coupling.

 The contribution of gaugino-gaugino-gauge boson as well as higgsino-higgsino-gauge boson have been already evaluated in the context of ${\cal N}=1$ gauged supergravity. The volume suppression factors corresponding to these vertices are as follows:
\beqn
\label{eq:vertices}
&& |C_{\chi^{+}_{1}\chi^{-}_{1}\gamma}|\equiv |C_{{\tilde H}^{+}_{u}{\tilde H}^{-}_{d}\gamma}|\equiv {\tilde f} {\cal V}^{-\frac{5}{18}};  |C_{\chi^{+}_{2}\chi^{-}_{2}\gamma}|\equiv  |C_{{\tilde \lambda}^{+}_{i}{\tilde \lambda}^{-}_{i}\gamma}|\equiv {\tilde f} {\cal V}^{-\frac{11}{18}}
\eeqn
 
Now, EDM of electron for a loop diagram given in Figure 4.8(b) will be given as:
\beqn
\label{eq:2loopdchi}
\frac{d_e}{e}|_{\chi^{\pm}_k}= \sum_{i= {1,2}}\sum_{k=1,2} Im \left( {\hat Y}_{H^{0}_i{e_L}{e^{c}_{R}}}{C}_{\chi^{+}_{k}\chi^{-}_{k}h} \right)\left({C}_{\chi^{+}_{k}\chi^{-}_{k}\gamma} \right)^2\frac{\alpha_{em}Q^{2}_{\chi_i}Q_{e}}{16 \pi^3 m_{\chi^{\pm}_k}} \left(f(z)- g(z)\right)\nonumber
\eeqn
and EDM of neutron for a loop diagram given in Figure 4.8(b) will be given as:
\beqn
\label{eq:2loopdchin}
\frac{d_{u/n}}{e}|_{\chi^{\pm}_k}= \sum_{i= {1,2}}\sum_{k=1,2} Im \left( {\hat Y}_{H^{0}_i{e_L}{e^{c}_{R}}}{C}_{\chi^{+}_{k}\chi^{-}_{k}h} \right)\left({C}_{\chi^{+}_{k}\chi^{-}_{k}\gamma} \right)^2\frac{\alpha_{em}Q^{2}_{\chi_i}Q_{u}}{16 \pi^3 m_{\chi^{\pm}_k}} \left(f(z)- g(z)\right)\nonumber
\eeqn
where $z= \frac{m^{2}_{\chi^{\pm}_k}}{m^{2}_{H^{0}_i}}; f\Bigl( \frac{m^{2}_{\chi^{\pm}_1}}{m^{2}_{H^{0}_1}}\Bigr)- g\Bigl( \frac{m^{2}_{\chi^{\pm}_1}}{m^{2}_{H^{0}_1}}\Bigr)= 10; f\Bigl( \frac{m^{2}_{\chi^{\pm}_1}}{m^{2}_{H^{0}_2}}\Bigr)- g\Bigl( \frac{m^{2}_{\chi^{\pm}_1}}{m^{2}_{H^{0}_2}}\Bigr)= 1, f\Bigl( \frac{m^{2}_{\chi^{\pm}_2}}{m^{2}_{H^{0}_1}}\Bigr)- g\Bigl( \frac{m^{2}_{\chi^{\pm}_2}}{m^{2}_{H^{0}_1}}\Bigr)= 10 ; f\Bigl( \frac{m^{2}_{\chi^{\pm}_2}}{m^{2}_{H^{0}_2}}\Bigr)- g\Bigl( \frac{m^{2}_{\chi^{\pm}_2}}{m^{2}_{H^{0}_2}}\Bigr)= 0.1$.
Considering  $({ \phi_{\chi_i}}-{\phi_{y_e}})= ({ \phi_{\chi_i}}-{\phi_{{\tilde\lambda}^{0}_1}})\sim (0,\frac{\pi}{2}]$; for $m_{\chi^{\pm}_1}= {\cal V}^{\frac{59}{72}}m_{\frac{3}{2}}$, $m_{\chi ^{\pm}_2}= {\cal V}^{\frac{2}{3}}m_{\frac{3}{2}}$, ${\hat Y}_{H^{0}_i{e_L}{e^{c}_{R}}}={\cal V}^{-\frac{47}{45}}$, ${\hat Y}_{H^{0}_i{e_L}{e^{c}_{R}}}={\cal V}^{-\frac{17}{18}}$; the value of EDM of electron and neutron will be given as:
\beqn
\label{eq:2loopdchi2}
\frac{d_e}{e}|_{\chi_i}=\frac{d_n}{e}|_{\chi_i}\sim  \frac{ {\tilde f}^2{\cal V}^{-\frac{8}{3}}}{m_{\frac{3}{2}}} \times 10^{-5} \times 10^{1}\equiv10^{-33}GeV^{-1} \equiv 10^{-47} cm.
\eeqn
{{\bf R Parity violating contribution:}}
For a loop diagram given in Figure 4.8(c), the Lagrangian governing the interaction of neutrino will correspond to R-parity violating interactions given as:
\beqn
{\cal L} \supset  {\tilde \lambda }_{{\tilde \nu}_L u_L u^{c}_R}{\nu}_{iL} u_{kL} u^{c}_{kR}+ {\tilde \lambda }_{{\tilde \nu}_L e_L e^{c}_R}{\nu}_{i L} e_{j L} e^{c}_{j R}+h.c.. 
\eeqn
The contribution of R-parity violating interaction terms ${\hat \lambda}_{ikk}$ and ${\hat \lambda}^{*}_{ijj}$
are given by expanding ${\cal D}_{{\cal A}_1}{\cal D}_{{\cal A}_3} W$ and ${\cal D}_{{\cal A}_2}{\cal D}_{{\cal A}_4} W$ in the fluctuations linear in ${\cal A}_{1}$  around its stabilized VEV. The values of the same have already been calculated in the context of ${\cal N}=1$ gauged supergravity action and given as follows:
\beqn
&& {\tilde \lambda }_{{\tilde \nu}_L e_L e^{c}_R} \equiv{\cal V}^{-\frac{5}{3}}e^{i \phi_{\lambda_e}}, {\tilde \lambda }_{{\tilde \nu}_L u_L u^{c}_R} \equiv{\cal V}^{-\frac{5}{3}}e^{i \phi_{\lambda_u}},
 \eeqn
 where $e^{i \phi_{\lambda_e}}$ and $e^{i \phi_{\lambda_u}}$ are non-zero phases corresponding to above-mentioned complex R-parity violating couplings. The EDM of electron in this case will be:
\beqn
\label{eq:2loopdnu}
\frac{d}{e}|_{RPV}= Im \left( {\tilde \lambda }_{{\tilde \nu}_L e_L e^{c}_R}{\tilde \lambda }_{{\tilde \nu}_L u_L u^{c}_R} \right)\frac{\alpha_{em}Q^{2}_{u}Q_{e}}{16 \pi^3 m_{e j}} \left(f(z_1)- g(z_1)\right),
\eeqn
and EDM of neutron will be given as:
\beqn
\label{eq:2loopdnun}
\frac{d}{u/n}|_{RPV}= Im \left( {\tilde \lambda }_{{\tilde \nu}_L e_L e^{c}_R}{\tilde \lambda }_{{\tilde \nu}_L u_L u^{c}_R} \right)\frac{\alpha_{em}Q^{2}_{e}Q_{u}}{16 \pi^3 m_{e j}} \left(f(z_2)- g(z_2)\right).
\eeqn
where $z_1= \frac{m^{2}_{e}}{m^{2}_{\nu_{iL}}}; z_2= \frac{m^{2}_{u}}{m^{2}_{\nu_{iL}}}$. Using the value of masses $m_{{\nu}_{iL}}= {\cal V}^{\frac{1}{2}}m_{\frac{3}{2}}$, $m_{e}=0.5GeV$ and $m_{u}={\cal O}(1)$, $f\left( \frac{m^{2}_{e}/m^{2}_{u}}{m^{2}_{{\nu}_{iL}}}\right)= g\left( \frac{m^{2}_{e}/m^{2}_{u}}{m^{2}_{{\nu}_{iL}}}\right)= 10^{-27}$,  and assuming $( \phi_{\lambda_e}-\phi_{\lambda_u})= (0,\frac{\pi}{2}]$; equation (\ref{eq:2loopdnu}) and (\ref{eq:2loopdnun}) reduce to give EDM result as follows:
\beqn
\frac{d}{e}|_{RPV}= \frac{d}{n}|_{RPV} \sim {\cal V}^{-\frac{10}{3}} \times 10^{-2} \times 10^{-27}\equiv 10^{-55}GeV^{-1} \equiv 10^{-70} cm.
\eeqn
\subsection{Two-loop Level Barr-Zee Feynman Diagrams Involving Internal Sfermion Loop}
In this subsection, we evaluate the contribution of heavy sfermion loop generated by trilinear scalar interactions including Higgs. The loop diagrams are mediated by $\gamma h$ exchange. Unlike one-loop diagrams, here we do not have to consider the mixing sleptons(squarks) because of the fact that non-zero phase associated with complex scalar trilinear interaction is sufficient to generate non-zero EDM of electron/neutron. We first evaluate the contribution of relevant vertices in the context of ${\cal N}=1$ gauged supergravity for two-loop diagrams shown in Figure 4.9.
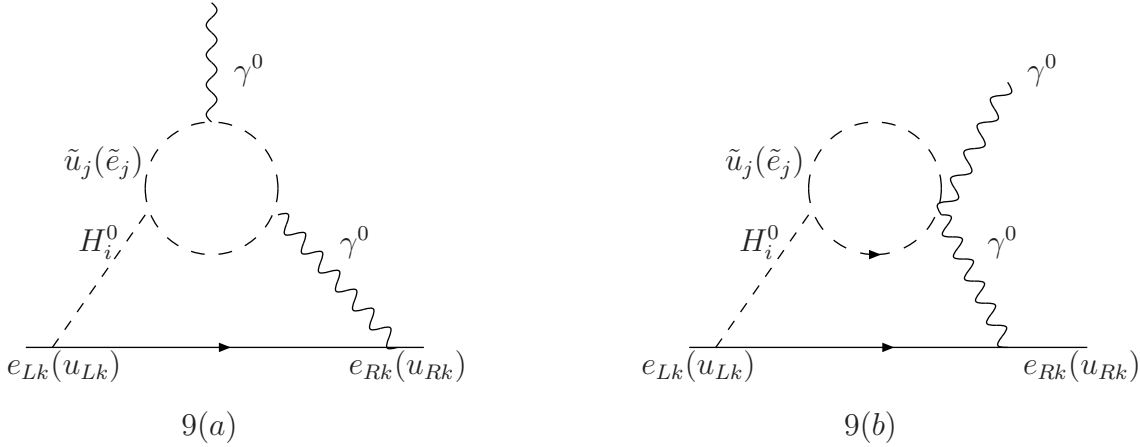
\begin{figure}
\begin{center}
\begin{picture}(100,200) (290,-10)
   \ArrowLine(130,60)(280,60)
   \DashLine(140,60)(175,110){4}
   \DashCArc(200,120)(25,0,180){5}
   \DashCArc(200,120)(25,180,0){5}
   \Photon(225,110)(270,60){3}{7}
   \Photon(200,145)(200,190){2}{4}
    \Text(158,100)[]{$H^{0}_i$}
   \Text(215,165)[]{$\gamma^0$}
    \Text(160,130)[]{${\tilde u_j}({\tilde e_j})$}
   \Text(145,53)[]{{$e_{L k}(u_{L k})$}}
   \Text(275,53)[]{{$e_{R k}(u_{R k})$}}
   \Text(255,100)[]{{$\gamma^0$}}
   \Text(200,30)[]{{$9(a)$}}
 \ArrowLine(380,60)(530,60)
  \DashLine(390,60)(425,110){4}
  \DashCArc(450,120)(25,0,180){5}
   \DashArrowArc(450,120)(25,180,0){5}
    \Photon(475,110)(500,60){3}{6}
   \Photon(475,110)(500,160){3}{5}
    \Text(408,100)[]{$H^{0}_i$}
   \Text(515,165)[]{$\gamma^0$}
    \Text(410,130)[]{${\tilde u_j}({\tilde e_j})$}
   \Text(385,53)[]{{$e_{L k}(u_{L k})$}}
   \Text(530,53)[]{{$e_{R k}(u_{R k})$}}
   \Text(500,100)[]{{$\gamma^0$}}
   \Text(450,30)[]{{$9(b)$}}
\end{picture}
\end{center}
\vskip -0.6in
\caption{Two-loop diagram involving sfermions in the internal loop.}
\end{figure}

\underline{Slepton(${\tilde e}_{jR}$)-slepton(${\tilde e}_{j R}$)-Higgs vertex:} by expanding effective supergravity potential
$V|_{EW}\sim e^KK^{\tau_S{\bar\tau}_B}D_{\tau_S}W D_{{\bar\tau}_B}{\bar W} + e^KK^{\tau_B{\bar\tau}_B}|D_{\tau_B}W|^2$ in the fluctuations linear in ${\cal Z}_i \rightarrow {\cal Z}_i +{\cal V}^{\frac{1}{36}}M_P$ , ${\cal A}_3 \rightarrow {\cal A}_3 +{\cal V}^{-\frac{13}{18}}M_P$,  contribution of the term quadratic in ${\cal A}_3$ as well as ${\cal Z}_i$ is of the order $ {\cal V}^{-\frac{59}{36}}\langle{\cal Z}_i\rangle $, which after giving VEV to  one of the ${\cal Z}_i$, will be given as:

\begin{equation}
\label{eq:CeReRh}
 C_{{\tilde e_R}{\tilde e_R}^{*} H^{0}_i}\equiv \frac{1}{{\sqrt{(\hat{K}_{{\cal Z}_i{\bar{\cal Z}}_i})^2 (\hat{K}_{{\cal A}_3{\bar{\cal A}}_3})^2}}}\left[{\cal V}^{-\frac{59}{36}}\langle {\cal Z}_i\rangle\right]\equiv ( {\cal V}^{-2}M_P) e^{i\phi_{{\tilde e_R}}}.
\end{equation}
where $\phi_{{\tilde e_R}}$ is non-zero phase corresponding to aforementioned complex scalar 3-point interaction vertex. Using the similar procedure, the  of coefficient of slepton(${\tilde e}_{jL}$)-slepton(${\tilde e}_{j L}$)-Higgs vertex will be given as:
    \beqn
    C_{{\tilde e_L}{\tilde e_L}^{*} H^{0}_i}\equiv   \frac{1}{{\sqrt{(\hat{K}_{{\cal Z}_i{\bar{\cal Z}}_i})^2 (\hat{K}_{{\cal A}_1{\bar{\cal A}}_1})^2}}}\left[{\cal V}^{-\frac{95}{36}}\langle {\cal Z}_i\rangle\right]\equiv( {\cal V}^{-\frac{12}{5}}M_P) e^{i\phi_{{\tilde e_L}}}.
    \eeqn
$\phi_{{\tilde e_L}}$ is non-zero phase corresponding to this particular complex scalar 3-point interaction vertex.
 
 \underline{Squark(${\tilde u}_{jR}$)-squark(${\tilde u}_{j R}$)-Higgs vertex:} By expanding $V|_{EW}$ in the fluctuations around  ${\cal Z}_i \rightarrow {\cal Z}_i +{\cal V}^{\frac{1}{36}}M_P$ , ${\cal A}_4 \rightarrow {\cal A}_4 +{\cal V}^{-\frac{11}{9}}M_P$,  contribution of term quadratic in ${\cal A}_4$ as well as ${\cal Z}_i$ is of the order $ {\cal V}^{-\frac{23}{36}}\langle{\cal Z}_i\rangle $, which after giving VEV to  one of the ${\cal Z}_i$, will be given as:
\begin{equation}
\label{eq:CuRuRh}
 C_{{\tilde u_R}{\tilde u_R}^{*} H^{0}_i}\equiv \frac{1}{{\sqrt{(\hat{K}_{{\cal Z}_i{\bar{\cal Z}}_i})^2 (\hat{K}_{{\cal A}_4{\bar{\cal A}}_4})^2}}}\left[{\cal V}^{-\frac{23}{36}}\langle {\cal Z}_i\rangle\right]\equiv( {\cal V}^{-2}M_P) e^{i\phi_{{\tilde u_R}}}.
\end{equation}
where $\phi_{{\tilde u_R}}$ is non-zero phase corresponding to aforementioned complex scalar 3-point interaction vertex.
 
\underline{Squark(${\tilde u}_{jL}$)-squark(${\tilde u}_{j L}$)-Higgs vertex:} By expanding $V|_{EW}$ in the fluctuations around  ${\cal Z}_i \rightarrow {\cal Z}_i +{\cal V}^{\frac{1}{36}}M_P$ , ${\cal A}_2 \rightarrow {\cal A}_2 +{\cal V}^{-\frac{1}{3}}M_P$,  contribution of term quadratic in ${\cal A}_2$ as well as ${\cal Z}_i$ is of the order $ {\cal V}^{-\frac{131}{36}}\langle{\cal Z}_i\rangle $, which after giving VEV to  one of the ${\cal Z}_i$, will be given as:
\begin{equation}
\label{eq:CuLuLh}
 C_{{\tilde u_L}{\tilde u_L}^{*} H^{0}_i}\equiv \frac{1}{{\sqrt{(\hat{K}_{{\cal Z}_i{\bar{\cal Z}}_i})^2 (\hat{K}_{{\cal A}_2{\bar{\cal A}}_2})^2}}}\left[{\cal V}^{-\frac{131}{36}}\langle {\cal Z}_i\rangle\right]\equiv ( {\cal V}^{-\frac{20}{9}}M_P) e^{i\phi_{{\tilde u_R}}}.
\end{equation}
where $\phi_{{\tilde u_L}}$ is non-zero phase corresponding to aforementioned complex scalar 3-point interaction vertex.

The contribution of  slepton (${\tilde e}_{j R}$)-slepton (${\tilde e}_{j R}$)-photon($\gamma$)- photon($\gamma$) vertex  will be given by:
${\bar{\partial}}_{{\bar {\cal A}}_{3}}\partial_{{\cal A}_3}G_{T_B{\bar T}_B} X^{T_B} X^{\bar {T_B}} {A}^{\mu} A_{\nu}$. On solving: $
{\bar{\partial}}_{{\bar {\cal A}}_{3}}\partial_{{\cal A}_3}G_{T_B{\bar T}_B} \sim {\cal V}^{\frac{1}{9}}{\cal A}^{*}_{1}{\cal A}_{1}$, Incorporating values of $X^B$ as mentioned earlier, the real physical slepton(${\tilde e}_{jR}$)-slepton(${\tilde e}_{j R}$)-photon($\gamma$)-photon($\gamma$) vertex is proportional to
\begin{equation}
\label{eq:CllZZ}
C_{{\tilde e}_R {\tilde e}^{*}_R \gamma \gamma}\equiv \frac{{\cal V}^{\frac{1}{9}}{\tilde f}^2 {\cal V}^{-\frac{4}{3}}}{\sqrt{(K_{{\cal A}_3 {\cal A}_3})^2}}\equiv  {\tilde f}^2 {\cal V}^{-\frac{13}{5}}.
\end{equation}
   The coefficient of  real physical ${\tilde e}_{j L}$ - ${\tilde e}_{j L}$ $\gamma$- $\gamma$ vertex has been obtained in chapter {\bf 3}. The value of the same is given by $C_{{\tilde e}_L {\tilde e}^{*}_L \gamma \gamma}\equiv {\tilde f}^2 {\cal V}^{-3}$.
 Similarly, the coefficient of real  physical ${\tilde u}_{j R}-{\tilde u}_{j R}-\gamma-\gamma$ vertex  will be given by:
${\bar{\partial}}_{{\bar {\cal A}}_{4}}\partial_{{\cal A}_4}G_{T_B{\bar T}_B} X^{T_B} X^{\bar {T_B}} {A}^{\mu} A_{\nu}$. On solving, the volume suppression factor corresponding to this  vertex will be given as:
\begin{equation}
C_{{\tilde u}_R {\tilde u}^{*}_R \gamma \gamma}\sim \frac{{\rm coefficient~of}~{\bar{\partial}}_{{\bar {\cal A}}_{4}}\partial_{{\cal A}_4}G_{T_B{\bar T}_B}}{\sqrt{(K_{{\cal A}_4 {\cal A}_4})^2}}\equiv {\tilde f}^2 {\cal V}^{-\frac{118}{45}}.
\end{equation}
The coefficient of  real physical (${\tilde u}_{jL}-{\tilde u}_{j L}-\gamma-\gamma$) vertex  will be given by:
${\bar{\partial}}_{{\bar {\cal A}}_{2}}\partial_{{\cal A}_2}G_{T_B{\bar T}_B} \\
X^{T_B} X^{\bar {T_B}} {A}^{\mu} A_{\nu}$. On solving, the volume suppression factor corresponding to this  vertex will be given as:
\begin{equation}
C_{{\tilde u}_L {\tilde u}^{*}_L \gamma \gamma}\sim \frac{{\rm coefficient~of}~{\bar{\partial}}_{{\bar {\cal A}}_{2}}\partial_{{\cal A}_2}G_{T_B{\bar T}_B}}{\sqrt{(K_{{\cal A}_2 {\cal A}_2})^2}}\equiv {\tilde f}^2{\cal V}^{-\frac{127}{45}}.
\end{equation}
The contribution of  real scalar-scalar-photon vertices have already been obtained in section {\bf 2}, and given as:
\beqn
&& C_{{\tilde e}_L {\tilde e}^{*}_L \gamma}\equiv ({\tilde f} {\cal V}^{\frac{44}{45}}){\tilde {\cal A}_1} A^\mu\partial_\mu {\tilde {\cal A}}_{1}, C_{{\tilde e}_R {\tilde e}^{*}_R \gamma} \equiv ({\tilde f}{\cal V}^{\frac{53}{45}}){\cal A}_3 A^\mu\partial_\mu {\bar {\cal A}}_{3}, \nonumber\\
&& C_{{\tilde u}_L {\tilde u}^{*}_L \gamma}\equiv({\tilde f}{\cal V}^{\frac{53}{45}}){\tilde {\cal A}_2} A^\mu\partial_\mu {\tilde {\cal A}}_{2}, C_{{\tilde u}_R {\tilde u}^{*}_R \gamma}\equiv ({\tilde f}{\cal V}^{\frac{62}{45}}){\tilde {\cal A}_4} A^\mu\partial_\mu {\tilde {\cal A}}_{4}.
\eeqn
The analytical expression for the EDM involving sfermion/scalar in an internal loop has been provided in \cite{Yamanaka_sfermion}. Using the same, for Figure~4.9(a), EDM of electron will be given as:
\beqn
\label{eq:deksfermion}
{\hskip -0.1in} \frac{d_{e}}{e}|^{\rm sfermion}_{4.9(a)}= \sum_{i=1,2} \ \sum_{{j={{\tilde u}_L, {\tilde u}_R}}} Im ({\hat Y}_{H^{0}_ie_L e^{c}_R}{C_{H^{0}_i j j^{*}}})(C_{j j^{*}\gamma})^2\times \frac{\alpha_{em} \eta_c Q_{e_j}q^{2}_{j}}{32 \pi^3 m^{2}_{H^{0}_i}}F({\tilde z}),
\eeqn
and EDM of neutron/quark will be given as:
\beqn
\label{eq:deksfermionn}
{\hskip -0.1in}  \frac{d_{n}}{e}|^{\rm sfermion}_{4.9(a)}= \sum_{i=1,2} \ \sum_{{j={{\tilde e}_L, {\tilde e}_R}}} Im ({\hat Y}_{H^{0}_iu_L u^{c}_R}{C_{H^{0}_i j j^{*}}})(C_{j j^{*}\gamma})^2\times \frac{\alpha_{em} \eta_c Q_{u_j}q^{2}_{j}}{32 \pi^3 m^{2}_{H^{0}_i}}F({\tilde z}).
\eeqn
where $z=\frac{m^{2}_{j}}{m^{2}_{H^{0}_i}}; 
F(z)=-\int^{1}_{0} dx \frac{x(1-x)}{x(1-x)-z}{\rm ln} \left(\frac{x(1-x)}{z} \right)$. 
Considering $(\phi_{{\tilde u_{L/R}}}-\Phi_{y_e})=(\phi_{{\tilde e_{L/R}}}-\Phi_{y_u})=(0,\frac{\pi}{2}]$; $|{\hat Y}_{H^{0}_ie_L e^{c}_R}|\equiv {\cal V}^{-\frac{47}{45}}, |{\hat Y}_{H^{0}_iu_L u^{c}_R}|\equiv {\cal V}^{-\frac{19}{18}}$ and using the value of masses, $m_{{\tilde e}_L}=m_{{\tilde e}_R}= m_{{\tilde u}_L}=m_{{\tilde u}_R}={\cal V}^{\frac{1}{2}}m_{\frac{3}{2}}$, $m_{H^{0}_1}=125 GeV$ and $m_{H^{0}_2}={\cal V}^{\frac{59}{72}}m_{\frac{3}{2}}$ , we have $F\Bigl(\frac{m^{2}_{j}}{m^{2}_{H^{0}_1}}\Bigr)= 10^{-17} , F\Bigl(\frac{m^{2}_{j}}{m^{2}_{H^{0}_2}}\Bigr)= 1$. Incorporating the value of interaction vertices, equation (\ref{eq:deksfermion}) and (\ref{eq:deksfermionn}) reduce to give EDM results as follows:
\beqn
\label{eq:deksfermion1}
&&\frac{d_{e}}{e}|^{\rm sfermion}_{4.9(a)}= 10^{-8}\times {\cal V}^{-\frac{4}{15}}{\tilde f}^2 \equiv10^{-15} GeV^{-1}\equiv 10^{-29}cm;~{\rm for~{\cal V}=10^4},\nonumber\\
&&
\frac{d_{n}}{e}|^{\rm sfermion}_{4.9(a)}= 10^{-8}\times {\cal V}^{-\frac{3}{10}}{\tilde f}^2\equiv 10^{-15} GeV^{-1}\equiv 10^{-29}cm;~{\rm for~{\cal V}= 10^4}.
\eeqn
For a loop diagram given in Figure 4.9(b), EDM of electron will be given as:
\beqn
\frac{d_{e}}{e}|^{\rm sfermion}_{4.9(b)}= \sum_{i=1,2} \ \sum_{{j={{\tilde u}_L, {\tilde u}_R}}} Im ({\hat Y}_{H^{0}_ie_L e^{c}_R}{C_{H^{0}_i j j^{*}}})(C_{j j^{*}\gamma \gamma})\times \frac{\alpha_{em}Q_{e_j}q^{2}_{j}}{32 \pi^3 m^{2}_{H^{0}_i}}F({\tilde z}),
\eeqn
and EDM of neutron will be given as:
\beqn
{\hskip -0.1in} \frac{d_{n/u}}{e}|^{\rm sfermion}_{4.9(b)}= \sum_{i=1,2} \ \sum_{{j={{\tilde e}_L, {\tilde e}_R}}} Im ({\hat Y}_{H^{0}_iu_L u^{c}_R}{C_{H^{0}_i j j^{*}}})(C_{j j^{*}\gamma \gamma})\times \frac{\alpha_{em}Q_{u_j}q^{2}_{j}}{32 \pi^3 m^{2}_{H^{0}_i}}F({\tilde z}).
\eeqn
where $F\left(\frac{m^{2}_{j}}{m^{2}_{H^{0}_1}}\right)= 10^{-17}, F\left(\frac{m^{2}_{j}}{m^{2}_{H^{0}_1}}\right)= 1 $.
Incorporating the value of masses and estimate of relevant coupling veretx, EDM of electron will be
 \beqn
\frac{d_{e}}{e}|^{\rm sfermion}_{4.9(b)}= 10^{-9} \times {\cal V}^{-\frac{17}{3}}{\tilde f}^2 \equiv 10^{-43} GeV^{-1}\equiv 10^{-57}cm.
\eeqn
The EDM of neutron in this case will be given as:
 \beqn
\frac{d_{n}}{e}|^{\rm sfermion}_{4.9(b)}= 10^{-9} \times {\cal V}^{-\frac{91}{18}}{\tilde f}^2 \equiv 10^{-40} GeV^{-1}\equiv 10^{-54}cm.
\eeqn
The overall contribution of EDM of electron as well as neutron corresponding to 2-loop diagram involving sfermions is:
\beqn
\frac{d_{e/n}}{e}|^{\rm sfermion}= \frac{d_{e/n}}{e}|^{\rm sfermion}_{4.9(a)} +\frac{d_{e/n}}{e}|^{\rm sfermion}_{4.9(b)}\equiv10^{-29}cm.
\eeqn
\subsection{Two-loop Level Barr-Zee Feynman Diagram Involving $W^{\pm}$ Boson in the Internal Loop}

In this subsection, we discuss the important contribution of Barr-Zee diagram involving W boson as an internal loop. In the one-loop as well as two-loop diagrams discussed so far, we have discussed the contribution mediated by Higgs exchange. The non-zero phases in one-loop diagram are affected by considering a mixing between Higgs doublet in $\mu$-split-like  SUSY model while in two-loop diagrams, the phases are affected through complex effective Yukawa couplings.  It has been found in \cite{Leigh_et_al} that two-loop graphs involving W boson loop can induce electric dipole moment of $d_e$ of the order of the experimental bound ($10^{-27} cm$) in the multi-Higgs models provided there is an exchange of Higgs  in the Higgs propagator and the CP violation in the neutral Higgs sector is fairly maximal. The approach was given by S. Weinberg in \cite{higgs_1,weinberg_higgs_2}. In this work, he pointed out that dimension-six purely gluonic operator gives a large value for the EDM of the neutron, which is just below the
present experimental bound if one considers CP violation through the exchange of
Higgs particles, whose interactions involve one or more complex phases. The approach was extended by Barr and Zee who have found that Higgs exchange can also give an electric dipole moment to the electron of the order of experimental limits by considering an EDM operator involving a top quark also.  In this spirit, we present an analysis of EDM of electron/neutron involving W boson loop in the context of $\mu$ split-SUSY model which, as already discussed involves a light Higgs and a heavy Higgs doublet. As already described in chapter {\bf 2}, the neutral Higgs are defined as:
\beqn
\label{massmat14}
&& h_{1}=D_{h_{11}} h_{u} +D_{h_{12}}h_{d}, h_{2}=D_{h_{21}}h_{u} +D_{h_{22}} h_{d}.
\eeqn
where $
D_h=\Bigl(\begin{array}{ccccccc}
\cos \frac{\theta_h}{2} &-\sin \frac{\theta_h}{2} e^{-i\phi_{h}}\\
 \sin \frac{\theta_h}{2} e^{i\phi_{h}} & \cos \frac{\theta_h}{2}
 \end{array} \Bigr)$. 
 
 In the notations of Weinberg, the CP-violating phase can appear from the neutral Higgs-boson exchange through imaginary terms in the amplitude and Higgs propagators are represented as: $
A(q^2)= \sqrt{2}G_f \sum_{n} \frac{Z_n}{q^2+ m^{2}_{H_n}}$,
where $Z_n$ is non-zero phase appearing due to exchange between Higgs doublet in the propagator.
We address this argument of generation of non-zero phase in the ${\cal N}=1$ gauged supergravity action. 
We first provide the analysis of  required SM-like coupling involved in Figure~4.10 in the context of ${\cal N}=1$  gauged supergravity action.
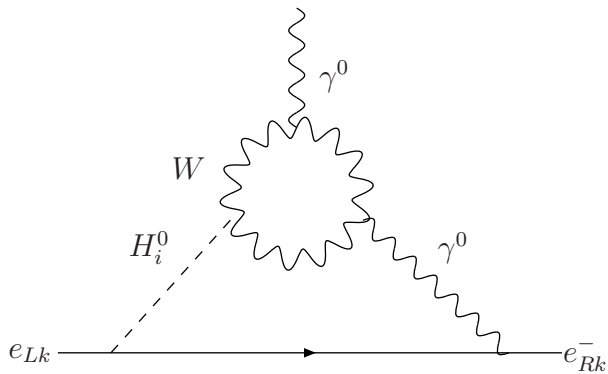
\begin{figure}
\begin{center}
\begin{picture}(100,200) (170,0)
   \ArrowLine(110,60)(300,60)
   \DashLine(130,60)(175,110){4}
   \PhotonArc(200,120)(25,0,180){4}{7}
   \PhotonArc(200,120)(25,180,0){4}{7}
   \Photon(225,110)(280,60){3}{7}
   \Photon(200,145)(200,190){3}{4}
    \Text(145,100)[]{$H^{0}_i$}
   \Text(215,165)[]{$\gamma^0$}
    \Text(160,130)[]{$W$}
   \Text(100,60)[]{{$e_{L k}$}}
   \Text(310,60)[]{{$e^{-}_{R k}$}}
   \Text(260,100)[]{{$\gamma^0$}}
  \end{picture}
\end{center}
\vskip -0.8in
\caption{Two-loop diagram involving W boson in the internal loop.}
\end{figure}

The contribution of $W^{+}$-photon-$W^{-}$ vertex is evaluated by CP-even interaction term given as \cite{tripleboson}$:
{\cal L}= - Re(f)A^{\mu}W^{-}_{\mu\nu}W^{+\nu}+ Re(f)W^{+\mu}W^{-\nu}F_{\mu\nu}$, where $W^{-}_{\mu\nu}=\partial_{\mu}W^{-}_{\nu}-\partial_{\nu}W^{-}_{\mu}$ and $F_{\mu\nu}=\partial_{\mu}A_{\nu}-\partial_{\nu}A_{\mu}$ and $ Re(f)$ is gauge kinetic function, which in our set-up is given by 'big' divisor volume modulus $Re(T_B)\sim{\cal V}^{\frac{1}{18}}\equiv {\cal O}(1)$ for Calabi-Yau ${\cal V}=10^5$.
Therefore, the volume suppression corresponding to this interaction vertex is:
\beqn
C_{W^{+}W^{-}\gamma}\equiv {\cal V}^{\frac{1}{18}}\equiv {\cal O}(1).
\eeqn
 
The effective $W^{+}$-Higgs-$W^{-}$ vertex can be evaluated in the effective supergravity action as follows.
 Consider the gauge kinetic term: $Re(T) F^2$ and then choose the term $C_{1{\bar3}} a_1 a_{\bar3}$ in $Re(T_B)$ with the understanding that one first gives  VEV to the predominantly $SU(2)_L$-doublet valued $a_1$, then one picks out the ${\cal Z}$-dependent contribution in $a_3$ and also use the value of the intersection component $C_{1{\bar 3}}$. One will therefore consider:
$C_{1{\bar 3}} \langle a_1\rangle{\cal V}^{-\frac{7}{5}}{\cal Z}
\frac{p_1\cdot p_2}{\sqrt{K_{{\cal A}_1{\bar{\cal A}}_1}}\sqrt{K_{{\cal Z}{\bar{\cal Z}}}}\left(\sqrt{Re(T)}\right)^2}$; $K_{{\cal Z}{\bar{\cal Z}}}|_{M_s}\sim10^{-5}, K_{{\cal A}_1{\bar{\cal A}}_1}|_{M_s}\sim10^4$ which at the EW scale we will assume to be $\frac{10^{-5}}{({\cal O}(1))^2}$ and $\frac{10^4}{({\cal O}(1))^2}$. For non-relativistic gauge bosons, $p_1\cdot p_2\sim M_{W/Z}^2, Re(T)|_{EW} \sim {\cal O}(1)M_P\sim v {\cal V}^3\ {\rm GeV}, C_{1{\bar 3}}\sim {\cal V}^{\frac{29}{18}}, \langle a_1\rangle|_{EW}\sim{\cal O}(1){\cal V}^{-\frac{2}{9}}M_P$ (related to the requirement of obtaining ${\cal O}(10^2)$ GeV $W/Z$-boson mass at the EW scale- see \cite{gravitino_DM}). We thus obtain the following:
${\cal V}^{\frac{29}{18}}\times({\cal O}(1))^2 \times {\cal O}(1) \times {\cal V}^{-\frac{2}{9}} \times {\cal V}^{-\frac{7}{5}} M_{W/Z}^2 \times \frac{\sqrt{10}}{({\cal O}(1) \times v \times {\cal V}^3 )}  \sim
({\cal O}(1))^2 \times \sqrt{10}{\cal V}^{-3} \frac{M_{W/Z\ {\rm in GeV}}^2}{v ({\rm GeV})}$.
Now, in the superspace notation, the kinetic terms for the gauge field are generically written as:
$\int d^2 \theta f_{ab}(\Phi) W^a W^b$ where $W^a$ is the gauge-invariant super-field strength and $W=W^a T^a$ for a non-abelian group - as $f_{ab}$ is an apriori arbitrary holomorphic function of $\Phi$. Consider hence $\Phi=T, f\sim e^T$ and look at $\int d^2\theta (T)^{2m+1}_{\theta,{\bar\theta}=0} W^2$ which will consist of $({\cal O}(1)^2 \times C_{1{\bar 3}} \langle a_1 {\bar a}_3\rangle )^{2m}  \times \sqrt{10} \times {\cal V}^{-3}
\frac{M_{W/Z\ {\rm in\ GeV}}^2}{GeV}$, which, e.g., for $m=2$ yields
$({\cal O}(1))^2 \times \sqrt{10} \times  {\cal V}^{\frac{8}{3}-3} \times \frac{M_{W/Z\ {\rm in\ GeV}}^2}{v ({\rm GeV})}$ or for ${\cal V}\sim10^4$, one obtains ${\cal O}(1)\frac{M^2_{W/Z ({\rm in\ GeV})}}{v ({\rm in\ GeV})}$.
Utilizing this, at EW scale,
\begin{equation}
 C_{W^{+}H^{0}_{i}W^{-}}\equiv  \frac{M^{2}_{W}}{v}e^{i \phi_{W}}.
 \end{equation}
 The value of complex Yukawa coupling to be used to evaluate EDM corresponding to Figure 4.10 have already been obtained in chapter {\bf 2} and given as: ${\hat Y}_{H^{0}_i e_{L}e^{c}_{R}} \sim {\cal V}^{-\frac{47}{45}}e^{i\phi_{y_e}}$ and of ${\hat Y}_{H^{0}_i u_{L}u^{c}_{R}} \sim {\cal V}^{-\frac{19}{18}}e^{i\phi_{y_u}}$. The matrix amplitude as well as analytical expression for W boson related loop diagrams has been worked out in \cite{Leigh_et_al}. We utilize the same in a modified form to get the numerical estimate of EDM corresponding to a loop diagram given in Figure~4.10.
\beqn
&& \frac{d}{e}|_W = \frac{\alpha}{(4 \pi)^3 M^{2}_W}C_{W^{+}W^{-}\gamma}\sum_{i} Im ({\hat Y}_{H^{0}_i e_{L}e^{c}_{R}}  C_{W^{+}H^{0}_{i}W^{-}} )[5g(z^{W}_{i})+3f(z^{W}_{i})\nonumber\\
&& +\frac{3}{4}(g(z^{W}_{i})+h(z^{W}_{i}))],
\eeqn
where f(z) and g(z) are already defined in subsection {\bf 4.1} and
\beqn
h(z)=\frac{z}{2}\int^{1}_{0}dx \frac{1}{x(1-x)-z}\left(\frac{z}{x(1-x)-z}\ln\left(\frac{x(1-x)-z}{x}\right)-1\right);
\eeqn
where $z^{W}_{i}=\frac{m^{2}_{H^{0}_i}}{m^{2}_W}$. Considering $(\phi_{W}-\phi_{y_e})= (0,\frac{\pi}{2}]$;  using the values $m_{H^{0}_{1}}=125 GeV$ and $m_{H^{0}_{2}}={\cal V}^{\frac{59}{72}}m_{\frac{3}{2}}$, we get: 
$f\Bigl(\frac{m^{2}_{H^{0}_i}}{m^{2}_W}\Bigr)= g\left(\frac{m^{2}_{H^{0}_i}}{m^{2}_W}\right)= h\Bigl(\frac{m^{2}_{H^{0}_i}}{m^{2}_W}\Bigr)={\cal O}(1)$,
and the EDM result for electron will be given as:
\beqn
\frac{d_e}{e}|_W \sim \frac{\alpha}{(4 \pi)^3} \frac{1}{v} \times {\cal V}^{-\frac{47}{45}}\equiv10^{13} GeV^{-1}\equiv 10^{-27} cm.
\eeqn
Similarly, by considering $(\phi_{W}-\phi_{y_u})= (0,\frac{\pi}{2}]$;  EDM of neutron will be given as:
\beqn
\frac{d_n}{e}|_W \sim \frac{\alpha}{(4 \pi)^3} \frac{1}{v} \times {\cal V}^{-\frac{19}{18}}\equiv10^{13} GeV^{-1}\equiv 10^{-27} cm.
\eeqn
\subsection{Two-loop Level Rainbow Type Contribution to EDM}
The two-loop level analysis of the supersymmetric effects to the fermion electric dipole moment has been extended by considering rainbow diagrams in addition to famous Barr-Zee diagrams with the expectation that they might give a significant contribution to fermionic EDM. The importance of these diagrams is discussed in detail in \cite{Pilaftsis}. In this subsection, we estimate the contribution of two-loop rainbow type of diagrams involving R-parity conserving supersymmetric interaction vertices and R-parity violating vertices. The CP violating phases  appear from the diagonalized eigenstates in the inner loop as well as from complex effective Yukawa couplings in the higgsino sector. The Feynman diagrams have been classified based on different types of inner one-loop insertions. One corresponds to one-loop effective higgsino-gaugino-gauge boson vertex and the other corresponds to one-loop effective higgsino-gaugino transition. The matrix amplitudes as well analytic expressions to estimate the EDM for above rainbow diagrams are calculated in \cite{Yamanaka_rainbow} to the first order in the external momentum carried by the gauge boson. We utilize their expressions to get the order of magnitude of EDM of electron as well as neutron in our case.
\begin{figure}
\begin{center}
\begin{picture}(345,100) (80,-20)
   \ArrowLine(95,50)(115,50)
   \Line(115,50)(135,50)
   \ArrowLine(135,50)(175,50)
   \Line(175,50)(195,50)
   \ArrowLine(195,50)(215,50)
   \DashArrowArc(155,50)(40,180,0){4}
     \DashArrowArcn(155,50)(20,0,180){4}
   \Photon(165,50)(170,80){4}{4}
   \Text(125,58)[]{${\tilde h}^{0}$}
      \Text(150,58)[]{$f_i$}
   \Text(184,58)[]{${\tilde \lambda}^{0}_{i}$}
   \Text(95,42)[]{{$f_{L}$}}
   \Text(177,82)[]{{$\gamma$}}
   \Text(215,42)[]{{$f_{R}$}}
   \Text(155,23)[]{{${\tilde f_i}$}}
   \Text(155,2)[]{{${\tilde f_R}$}}
   \Text(155,-20)[]{{$(a)$}}
   \ArrowLine(255,50)(275,50)
   \Line(275,50)(295,50)
   \ArrowLine(335,50)(295,50)
   \Line(335,50)(355,50)
   \ArrowLine(355,50)(375,50)
   \DashArrowArc(315,50)(40,180,0){4}
     \DashArrowArc(315,50)(20,180,0){4}
   \Photon(325,50)(330,80){4}{4}
   \Text(285,58)[]{${\tilde h}^{0}$}
    \Text(310,58)[]{$f_i$}
   \Text(345,58)[]{${\tilde \lambda}^{0}_{i}$}
   \Text(255,42)[]{{$f_{L}$}}
   \Text(340,83)[]{{$\gamma$}}
   \Text(375,42)[]{{$f_{R}$}}
     \Text(320,23)[]{{${\tilde f_i}$}}
      \Text(320,2)[]{{${\tilde f_R}$}}
   \Text(320,-20)[]{{$(b)$}}
   \ArrowLine(95,-100)(115,-100)
   \Line(115,-100)(135,-100)
   \DashArrowLine(135,-100)(175,-100){4}
   \Line(175,-100)(195,-100)
   \ArrowLine(195,-100)(215,-100)
   \DashArrowArc(155,-100)(40,180,0){4}
     \ArrowArcn(155,-100)(20,0,180)
   \Photon(165,-100)(170,-70){4}{4}
   \Text(125,-92)[]{${\tilde h}^{0}$}
      \Text(150,-92)[]{${\tilde f_i}$}
   \Text(184,-92)[]{${\tilde \lambda}^{0}_{i}$}
   \Text(95,-108)[]{{$f_{L}$}}
   \Text(178,-70)[]{{$\gamma$}}
   \Text(215,-108)[]{{$f_{R}$}}
   \Text(150,-128)[]{{${ f_i}$}}
   \Text(150,-148)[]{{${\tilde f_R}$}}
   \Text(150,-170)[]{{$(c)$}}
   \ArrowLine(255,-100)(275,-100)
   \Line(275,-100)(295,-100)
   \DashArrowLine(335,-100)(295,-100){4}
   \Line(335,-100)(355,-100)
   \ArrowLine(355,-100)(375,-100)
   \DashArrowArc(315,-100)(40,180,0){4}
     \ArrowArc(315,-100)(20,180,0)
   \Photon(325,-100)(330,-70){4}{4}
   \Text(285,-92)[]{${\tilde h}^{0}$}
      \Text(310,-92)[]{${\tilde f_i}$}
   \Text(345,-92)[]{${\tilde \lambda}^{0}_{i}$}
   \Text(255,-108)[]{{$f_{L}$}}
   \Text(338,-70)[]{{$\gamma$}}
   \Text(375,-108)[]{{$f_{R}$}}
     \Text(315,-129)[]{{${ f_i}$}}
      \Text(315,-148)[]{{${\tilde f_R}$}}
   \Text(315,-170)[]{{$(d)$}}
   \end{picture}
  \end{center}
  \vskip 2.0in
\caption{Two-loop level Rainbow type diagrams involving higgsino-gaugino-gauge boson vertex.}
 \end{figure}
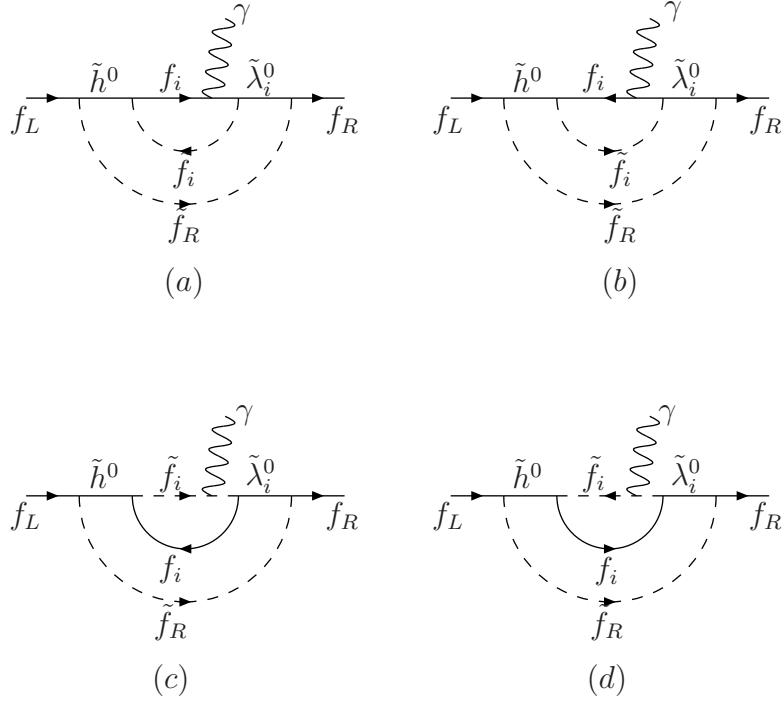
  
{{\bf R-parity conserving rainbow type contribution:}} for the loop diagrams given in Figure~4.11 and Figure~4.12, the result of EDM will given by the following formulae respectively:
 \begin{eqnarray}
d_f^1
&\approx &
\sum
\frac{n_c (Q_f+ Q'_f) C_{{\tilde H^{0}}f_L {\tilde f}^{*}_R}  C_{{\tilde H^{0}}f_i f^{*}_i}}{64\pi^3} \sum_{n=1,2} |m_{\lambda^{0}_n}| \sin (\delta_f -\theta_n ) \times \frac{e ({g'}^{(n)}_{\tilde f_L} -{g'}^{(n)}_{\tilde f_R})  }{4\pi} 
\nonumber\\
&& 
{\hskip -0.25in} \sin \theta_f \cos \theta_f  \sum_{\tilde f= \tilde f_L,\tilde f_R} \hspace{-1em} s \, {g'}^{(n)}_{\tilde f}
\Bigl[ 
F' (|m_{\lambda_n}|^2 , |\mu|^2 , m_{\tilde f_{L/R}}^2 , m_{\tilde f_{2}}^2) -F' (|m_{\lambda^{0}_n}|^2 , |\mu|^2 , m_{\tilde f_{L/R}}^2, m_{\tilde f_{1}}^2)
\Bigr]
,
\label{eq:edmhgg}
\nonumber\\
d_f^2
&=&
\sum_f
\frac{n_c Q'_{f_R} C_{{\tilde H^{0}}f_L{\tilde f}^{*}_R}  C_{{\tilde H^{0}}f_i f^{*}_i }}{64\pi^3}  \sum_{n=1,2} |m_{\lambda^{0}_n}| \sin (\delta_f -\theta_n) 
 \times \frac{ e ({g'}^{(n)}_{\tilde f_L} +{g'}^{(n)}_{\tilde f_R})}{4\pi} \sin \theta_f \cos \theta_f 
\nonumber\\
&& 
\times \sum_{\tilde f= \tilde f_L,\tilde f_R} \hspace{-1em}  {g'}^{(n)}_{\tilde f} m_{\tilde F}^2 
\Bigl[ 
F''(|m_{\lambda_n}|^2 , |\mu|^2 ,m_{\tilde f_{L/R}}^2, , m_{\tilde f_{1}}^2 )
-F''(|m_{\lambda_n}|^2 , |\mu|^2 , m_{\tilde f_{L/R}}^2, , m_{\tilde f_{2}}^2 )
\Bigr],\nonumber\\ 
\label{eq:edmhg}
\end{eqnarray}
where $n_c=3$ for the inner quark-squark loop and $n_c=1$ for the inner lepton-slepton loop. The fields $\tilde f_1$ and $\tilde f_2$ correspond to mass eigenstates of the sfermion $\tilde f$. The value of constant $s$ is $+1$ for left-handed sfermion $\tilde f_L$ and $-1$ for right-handed sfermion $\tilde f_R$.  
The effective electric charges are given by $Q'_{f}= C_{ f_i f^{*}_i \gamma}Q_{f}$ and $Q'_{f_R}= C_{ f_R f^{*}_R\gamma}$.  The interaction vertices $C_{{\tilde H^{0}}f_L f^{*}_R} $ and $C_{{\tilde H^{0}}f f^{*} }$ correspond to effective Yukawa couplings. ${g'}^{(n)}_{\tilde f_L}$ and ${g'}^{(n)}_{\tilde f_R}$ denote effective gauge couplings corresponding to supersymmetric sfermions. 
The functions $F'$ and $F''$ as defined in \cite{Yamanaka_rainbow} are given below: 
\begin{eqnarray}
F' (a,b,c,f)
&\equiv &
\frac{a \ln a - f \, {\rm Li}_2 \left( 1 - \frac{a}{f} \right)}{(a-b)(a-c)}
+\frac{b \ln b - f \, {\rm Li}_2 \left( 1 - \frac{b}{f} \right)}{(b-a)(b-c)}
+\frac{c \ln c - f \, {\rm Li}_2 \left( 1 - \frac{c}{f} \right)}{(c-a)(c-b)}
\, ,
\nonumber\\
&& \hskip -1.1inF''(a,b,c,f) \equiv \nonumber\\
&&
 {\hskip -1.0in} a^2 R_{13}(a,b,c) J_1(a,f)
+b^2 R_{13}(b,a,c) J_1(b,f)
+(a^2 b^2 - 3abc^2 +ac^3 +bc^3 ) R_{33} (c,a,b) 
\nonumber\\
&& {\hskip -0.8in}
J_1 (c,f)+c(ac+bc-2ab) R_{22} (c,a,b) J_2 (c,f) 
+c^2 R_{11} (c,a,b ) J_3 (c,f)
\nonumber\\
&&{\hskip -0.8in} -f (1+2\ln f) 
\Bigl[
 a\ln a R_{13} (a,b,c)
+b\ln b R_{13} (b,a,c)
+(a^2 b +ab^2 -3abc +c^3) \ln c 
\nonumber\\
&&  {\hskip -0.8in}
R_{33} (c,a,b) +\frac{1}{c}(ab-c^2) R_{22} (c,a,b) 
-\frac{1}{2c} R_{11} (c,a,b)
\Bigr]
\nonumber\\
&&{\hskip -0.8in}
-2f \left[
 a R_{13} (a,b,c) J_1 (a,f)
+b R_{13} (b,a,c) J_1 (b,f)
+(a^2b+ab^2 -3abc +c^3 )
\right.
\nonumber\\
&&  {\hskip -0.8in}
\left.
 R_{33} (c,a,b) J_1 (c,f) +(c^2-ab) R_{22}(c,a,b) J_2 (c,f)
+c R_{11}(c,a,b) J_3 (c,f)
\right]
\nonumber\\
&& {\hskip -0.8in} +f^2 \ln f 
\Bigl[
 \ln a R_{13} (a,b,c)
+\ln b R_{13} (b,a,c)
+\ln c (a^2 +b^2 +ab -3ac -3bc +3c^2 ) \nonumber\\
&& {\hskip -0.5in}
R_{33} (c,a,b)
+\frac{1}{c} (a+b-2c) R_{22} (c,a,b)
-\frac{1}{2c^2} R_{11} (c,a,b)
\Bigr]
\nonumber\\
&& {\hskip -0.8in}
+f^2 \left[
 R_{13} (a,b,c) J_1 (a,f)
+R_{13} (b,a,c) J_1 (b,f)
+(a^2 +b^2 +ab -3ac -3bc +3c^2 )
\right.
\nonumber\\
&& {\hskip -0.8in} \left.
  R_{33} (c,a,b) J_1 (c,f) -(a+b-2c) R_{22} (c,a,b) J_2 (c,f)
+R_{11} (c,a,b) J_3 (c,f)
\right].
\end{eqnarray}
where $R_{nm} (a,b,c) \equiv \frac{1}{(a-b)^n (a-c)^m}$, $J_1 (a,f) \equiv {\rm Li}_2 (1-a/f ) +\frac{1}{2} \ln^2 f -\ln a \ln f$, $J_2 (c,f) \equiv \frac{c \ln c -f \ln f}{c(c-f)}$, $J_3 (c,f) \equiv \frac{c^2 \ln c - c(c-f) -f (2c-f) \ln f}{2c^2 (c-f)^2}$ and  ${\rm Li}_2 (x)$ denotes dilogarithm function.

The effective Yukawas as well as gauge interaction vertices are already calculated in section {\bf 3}. The magnitude of the values of the same are:
\begin{eqnarray}
&& |C^{{\tilde H^{0}}e_L {\tilde e}^{*}_R}| \equiv  {\cal V}^{-\frac{9}{5}}, |C^{{\tilde H^{0}}e_i {{\tilde e}_i}^{*} }| \equiv {\cal V}^{-\frac{10}{9}}, |C^{\tilde e_i \tilde e_i \gamma}|\equiv {\tilde f}{\cal V}^{\frac{53}{45}}, |C^{{\tilde H^{0}}u_L {\tilde u}^{*}_R}| \equiv  {\cal V}^{-\frac{5}{3}},\nonumber\\
&& |C^{{\tilde H^{0}}u_i {{\tilde u}_i}^{*} }|\equiv {\cal V}^{-\frac{10}{9}},  |C^{\tilde u_i \tilde e_i \gamma}|\equiv {\tilde f}{\cal V}^{\frac{53}{45}},   {g'}^{(n)}_{\tilde e_R} \equiv |C^{e_R {\tilde e}^{*}_R {\tilde \lambda}^{0}_i}| \equiv  {\tilde f} {\cal V}^{-\frac{3}{5}}, {g'}^{(n)}_{\tilde e}\equiv  |C^{e_i {\tilde e}^{*}_i {\tilde \lambda}^{0}_i}| \equiv {\tilde f}{\cal V}^{-\frac{3}{5}},\nonumber\\
&& 
  {g'}^{(n)}_{\tilde e_R} \equiv |C^{u_R {\tilde u}^{*}_R {\tilde \lambda}^{0}_i}| \equiv  {\tilde f} {\cal V}^{-\frac{3}{5}}, {g'}^{(n)}_{\tilde u}\equiv  |C^{u_i {\tilde u}^{*}_i {\tilde \lambda}^{0}_i}| \equiv {\tilde f}{\cal V}^{-\frac{3}{5}}; ~~i=1,2.
\end{eqnarray}
Using the value of $ m_{{\tilde\lambda}^{0}_1}= m_{\lambda^{0}_2}= {\cal V}^{\frac{2}{3}}m_{\frac{3}{2}}$, $m^{2}_{\tilde e_{1}}=m^{2}_{\tilde u_{1}}=  {\cal V}m^{2}_{\frac{3}{2}} + m^{2}_{{\tilde e}_{12}}$, $m^{2}_{\tilde e_{2}}=m^{2}_{\tilde u_{2}}=  {\cal V}m^{2}_{\frac{3}{2}} - m^{2}_{{\tilde e}_{12}}$,  $m_{\tilde H^{0}}\sim {\cal V}^{\frac{59}{72}}m_{\frac{3}{2}}$, $ F' (| m_{\lambda_n}|^2 , |\mu|^2 , m_{\tilde f_{L/R}}^2, m_{\tilde f_{2}}^2) - F' (|m_{\lambda_n}|^2 , |\mu|^2 , m_{\tilde f_{L/R}}^2 , m_{\tilde f_{1}}^2)\\
\sim 10^{-24}, F' (| m_{\lambda_n}|^2 , |\mu|^2 , m_{\tilde f_{L/R}}^2, m_{\tilde f_{2}}^2) - F' (|m_{\lambda_n}|^2 , |\mu|^2 , m_{\tilde f_{L/R}}^2 , m_{\tilde f_{1}}^2)\sim 10^{-43}$.
 Incorporating the above results in the analytical expression as given in (\ref{eq:edmhgg}) and (\ref{eq:edmhg})  (with the assumption that value of phase factor associated with all effective R-parity conserving Yukawa couplings are of ${\cal O}(1)$), 
  \begin{eqnarray}
d_e^1/e
&\equiv &
{\tilde f}^{3} {\cal V}^{-\frac{18}{5}} \times  {\cal V}^{\frac{2}{3}}m_{\frac{3}{2}} \times (10^{-24}) GeV^{-2} \equiv {10}^{-57} cm.
\nonumber\\
d_e^2/e
&\equiv &
{\tilde f}^{3} {\cal V}^{-\frac{18}{5}}\times  {\cal V}^{\frac{5}{3}}m^{3}_{\frac{3}{2}} \times (10^{-43} )GeV^{-4} \equiv {10}^{-55} cm.\nonumber\\
&& {\rm \hskip -0.5in Similarly} \nonumber\\
d_u^1/e
&\equiv &
{\tilde f}^{3} {\cal V}^{-\frac{10}{3}} \times  {\cal V}^{\frac{2}{3}}m_{\frac{3}{2}} \times (10^{-24}) GeV^{-2} \equiv {10}^{-56} cm.
\nonumber\\
d_u^2/e 
&\equiv&
{\tilde f}^{3} {\cal V}^{-\frac{10}{3}}\times  {\cal V}^{\frac{5}{3}}m^{3}_{\frac{3}{2}} \times (10^{-43}) GeV^{-4} \equiv {10}^{-54} cm.
\label{eq:edmhg2}
\end{eqnarray}
So, the final EDM of electron as well as quark or neutron in case of R-parity conserving supersymmetric Feynman diagrams are given as:
 \begin{equation}
{d_e}/{e} = d^{1}_e/{e} + {d^{2}_e}/{e} \equiv {10}^{-55} cm, {d_u}/{e} = d^{1}_u/{e} + {d^{2}_u}/{e} \equiv {10}^{-54} cm.
\end{equation}
 \begin{figure}
\begin{center}
\begin{picture}(345,110) (80,-30)
   \ArrowLine(95,50)(115,50)
   \Line(115,50)(135,50)
   \ArrowLine(175,50)(135,50)
   \Line(175,50)(195,50)
   \ArrowLine(195,50)(215,50)
   \DashArrowArc(155,50)(40,180,0){4}
     \DashArrowArc(155,50)(20,180,0){4}
   \Photon(165,11)(172,-18){4}{4}
   \Text(125,58)[]{${\tilde h}^{0}$}
    \Text(150,58)[]{$f_i$}
   \Text(184,58)[]{${\tilde \lambda}^{0}_{i}$}
   \Text(95,42)[]{{$f_{L}$}}
   \Text(178,-18)[]{{$\gamma$}}
   \Text(215,42)[]{{$f_{R}$}}
   \Text(155,21)[]{{${\tilde f_i}$}}
   \Text(155,2)[]{{${\tilde f_R}$}}
   \Text(155,-20)[]{{$(a)$}}
   \ArrowLine(255,50)(275,50)
   \Line(275,50)(295,50)
   \ArrowLine(295,50)(335,50)
   \Line(335,50)(355,50)
   \ArrowLine(355,50)(375,50)
   \DashArrowArc(315,50)(40,180,0){4}
     \DashArrowArcn(315,50)(20,0,180){4}
   \Photon(327,11)(340,-18){4}{4}
   \Text(285,58)[]{${\tilde h}^{0}$}
    \Text(310,58)[]{$f_i$}
   \Text(345,58)[]{${\tilde \lambda}^{0}_{i}$}
   \Text(255,42)[]{{$f_{L}$}}
   \Text(347,-20)[]{{$\gamma$}}
   \Text(375,42)[]{{$f_{R}$}}
     \Text(320,21)[]{{${\tilde f_i}$}}
      \Text(320,2)[]{{${\tilde f_R}$}}
   \Text(320,-20)[]{{$(b)$}}
      \end{picture}
  \end{center}
  \vskip 0.2in
\caption{Two-loop level rainbow type diagrams involving higgsino-gaugino transition.}
 \end{figure}
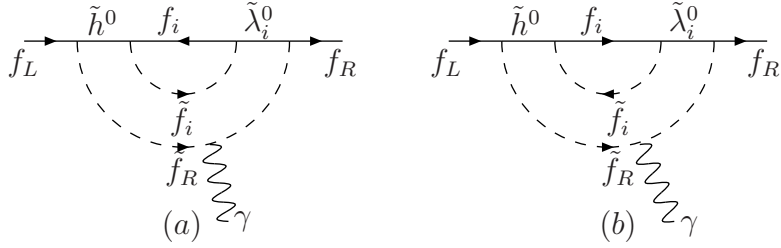
{{\bf R-Parity violating rainbow type contribution:}}
The similar kind of Feynman diagrams can be drawn by replacing the neutral higgsino component with Dirac massless neutrino in Figure~4.11 and Figure~4.12. 
The formulae of the EDM of the fermion f for two types of Feynman diagrams as defined in \cite{Yamanaka_Rparityvi} are given below:
\begin{eqnarray}
d^1_{f}
&=&
\sum_{n=1,2} 
{\rm Im} ( C_{{\tilde H^{0}}f_L f^{*}_R}  C_{{\tilde H^{0}}f_i f^{*}_i } e^{i(\theta_n -\delta_{f_j})})
\frac{(Q_f+ Q'_f)n_c }{64 \pi^3} |m_{\lambda^{0}_n }| \frac{e (g^{(n)}_{\tilde f_L} -g^{(n)}_{\tilde f_R}) }{4\pi} \sin \theta_{f_j} \cos \theta_{f_j} 
\nonumber\\
&& \times
\sum_{\tilde f= \tilde f_L , \tilde f_R} s\, g^{(n)}_{\tilde f} 
\left[ f'(|m_{\lambda^{0}_n}|^2, 0 , m_{\tilde f_{L/R}}^2, m_{\tilde f_{1j}}^2 \, )
-f'(|m_{\lambda_n}|^2, 0 , m_{\tilde f_{L/R}}^2, m_{\tilde f_{2j}}^2 \, ) \right],
\nonumber\\
d^2_{f}
&=&
-\sum_{n=1,2} 
{\rm Im} ( C_{{\tilde H^{0}}f_L f^{*}_R}  C_{{\tilde H^{0}}f_i f^{*}_i }e^{i(\theta_n -\delta_{f_j})} )
\frac{Q'_{f_R}  n_c }{64 \pi^3} |m_{\lambda^{0}_n }| \frac{e (g^{(n)}_{\tilde f_L} +g^{(n)}_{\tilde f_R}) }{4\pi} \sin \theta_{f_j} \cos \theta_{f_j} 
\nonumber\\
&&\hspace{-1em} \times
\sum_{\tilde f= \tilde f_L , \tilde f_R} g^{(n)}_{\tilde f} m_{\tilde f_{k}}^2
\left[ f''(|m_{\lambda^{0}_n}|^2, 0 ,m_{\tilde f_{L/R}}^2, m_{\tilde f_{1}}^2 \, )
-f''(|m_{\lambda_n}|^2, 0 , m_{\tilde f_{L/R}}^2, m_{\tilde f_{2}}^2 \, ) \right].\nonumber\\
\label{eq:edmnug}
\end{eqnarray}
The interaction vertices $C_{\nu^{0} f_L {\tilde f}^{*}_R} $ and $C_{\nu^0 f_i f^{*}_i }$ correspond to effective R-parity violating couplings. ${g'}^{(n)}_{\tilde f_L}$ and ${g'}^{(n)}_{\tilde f_R}$ denote effective gauge couplings corresponding to supersymmetric sfermions.

The functions $F'$ and $F''$ as defined in  \cite{Yamanaka_Rparityvi} are quoted below:
\begin{eqnarray}
F' (a,0,c,f)
&\equiv &
\frac{1}{a-c} \left[ \ln \frac{a}{c} + \frac{f}{c} {\rm Li}_2 \left( 1 - \frac{c}{f} \right) - \frac{f}{a} {\rm Li}_2 \left( 1 - \frac{a}{f} \right) \right]
-\frac{f}{ac} \frac{\pi^2}{6}
\, ,
\label{eq:F'}
\\
\nonumber\\
F''(a,0,c,f)
&\equiv &
\frac{(a-f)^2 }{a(a-c)^3} \left[ {\rm Li}_2 (1-a/f ) -{\rm Li}_2 (1-c/f ) \right] 
- \frac{f^2}{ac^3} \left[ {\rm Li}_2 (1-c/f ) - \frac{\pi^2}{6} \right]
\nonumber\\
&&
{\hskip -1.0in} +\frac{1}{a(c-f)} \left[ \frac{(a-f)^2}{(a-c)^2} -\frac{f^2}{c^2} \right] \ln \frac{c}{f}
- \frac{c \ln c -  c+2f }{2c^2 (a-c)} 
+ \frac{f+a \ln f}{(a-c)^2} \left[ \frac{1}{c} -\frac{1}{a-c} \ln \frac{a}{c} \right]
 \label{eq:F''}\nonumber\\
\end{eqnarray}
with ${\rm Li}_2 (x)$ denoting the dilogarithm function.\\
{\hskip 0.2in} Using the value of $ m_{{\tilde\lambda}^{0}_1}= m_{\lambda^{0}_2}= {\cal V}^{\frac{2}{3}}m_{\frac{3}{2}}$, $m^{2}_{\tilde e_{1}}=m^{2}_{\tilde u_{1}}=  {\cal V}m^{2}_{\frac{3}{2}} + m^{2}_{{\tilde e}_{12}}$, $m^{2}_{\tilde e_{2}}=m^{2}_{\tilde u_{2}}=  {\cal V}m^{2}_{\frac{3}{2}} - m^{2}_{{\tilde e}_{12}}$, we get: $F' (| m_{\lambda^{0}_n}|^2 ,0 , m_{\tilde f_{L/R}}^2, m_{\tilde f_{2}}^2) - F' (|m_{\lambda^{0}_n}|^2 , |\mu|^2 , m_{\tilde f_{L/R}}^2 , m_{\tilde f_{1}}^2)\sim 10^{-22},  F' (| m_{\lambda^{0}_n}|^2 ,0 , m_{\tilde f_{L/R}}^2, m_{\tilde f_{2}}^2) - F' (|m_{\lambda^{0}_n}|^2 , |\mu|^2 , m_{\tilde f_{L/R}}^2 , m_{\tilde f_{1}}^2)\sim 10^{-42}$. The contribution of R-parity violating vertices are already calculated in chapter {\bf 3}  in the context of ${\cal N}=1$ gauged supergravity action. The values of the same are as follows:
\begin{eqnarray}
&& |C^{\nu^{0} e_L {\tilde e}^{*}_R}| \equiv |C^{\nu^{0} e_i {{\tilde e}_i}^{*} }| \equiv {\cal V}^{-\frac{5}{3}}, |C^{\nu^{0} u_L {\tilde u}^{*}_R}| \equiv |C^{\nu^{0} u_i {{\tilde u}_i}^{*} }| \equiv {\cal V}^{-\frac{5}{3}}  ; i=1,2 .
 \end{eqnarray}
 Incorporating values of above-mentioned R-parity violating interaction vertices and the values of effective gauge couplings in the analytic expressions given in equation (\ref{eq:edmnug})(with the assumption that value of phase factor associated with all effective R-parity violating Yukawa couplings are of ${\cal O}(1)$), 
, 
\begin{eqnarray}
d_e^1/e= d_u^1/e
&\equiv &
{\tilde f}^{3} {\cal V}^{-\frac{10}{3}} \times  {\cal V}^{\frac{2}{3}}m_{\frac{3}{2}} \times (10^{-22}) GeV^{-2} \equiv {10}^{-53} cm.
\nonumber\\
d_e^1/e=d_u^2/e
&\equiv&
{\tilde f}^{3} {\cal V}^{-\frac{10}{3}}\times  {\cal V}^{\frac{5}{3}}m^{3}_{\frac{3}{2}} \times (10^{-42}) GeV^{-4} \equiv {10}^{-52} cm.
\label{eq:edmhg2}
\end{eqnarray}
So, the final EDM of electron as well as quark or neutron in case of R-parity violating Feynman diagrams are given as:
 \begin{equation}
{d_n}/{e} = {d_e}/{e} \equiv {10}^{-52} cm.
\end{equation}
The results of all two-loop diagrams contributing to EDM of electron/neutron are summarized in a table given below:
 \begin{table}[h]
\label{table:decay_lifetime}
\begin{tabular}{l c c rrrrrrr}  
\hline\hline                       
 Two-loop particle exchange&  Origin of $\mathbb C$ phase & $d_{e}$(e cm)  & $d_n$(e cm)
\\ [3.0ex]
\hline
{$ H^{0}_i \gamma f$} &  ${ \hat{Y}}_{\rm eff}\in\mathbb C$ &  $10^{-36}$& $10^{-36}$    \\[1ex]
{$H^{0}_i \gamma \chi^{\pm}_{i}$} & ''& $10^{-47}$ & $ 10^{-47} $ & \\[1ex]
{${\tilde f} f \gamma$}&''  & $10^{-70}$& $ 10^{-70} $   \\[1ex]
{${\tilde f} ^{0}_i H^{0}_i \gamma$} &''& $10^{-29}$ & $ 10^{-29}  $   \\[1ex]
{$\gamma W^{\pm} H^{0}_i$}&Higgs exchange & $10^{-27}$ & $ 10^{-27} $   \\[1ex]
{${\tilde H^{0}} {\tilde f}{\lambda^{0}_i}$}(Rainbow type)& Diagonalized sfermion mass  & $10^{-55} $ & $10^{-54}$ \\[1ex]
&   eigenstates and effective Yukawa's & &  \\[1ex]
{$\nu^0{\tilde f}{\lambda^{0}_i}$} (Rainbow type)& '' & $10^{-52} $ & $10^{-52}$ \\[1ex]
  \hline                          
\end{tabular}
\label{tab:PPer1}
\caption{Results of EDM of electron/neutron for all possible two-loop diagrams.}  

\end{table}
\section{Results and Discussion}
To summarize, we have performed a quantitative analysis of EDM of electron and neutron in a phenomenological model which provides a local realization of large volume $D3/D7$ $\mu$-split-like  supersymmetry. Having considered the ${\cal N}=1$ gauged supergravity limit of the aforementioned model, we have estimated all possible one-loop as well as two-loop diagrams. The non-zero CP-violating phase corresponding to dimension-five non-renormalizable EDM operator   appeared at one-loop and two-loop level from  off-diagonal component of scalar mass matrix and complex effective Yukawa couplings respectively. We have considered the order of phases to exist in (0,$\frac{\pi}{2}$]. In the one-loop graphs involving sfermions, the neutralino-mediated loop diagrams gave the dominant contributions to the electron(neutron) EDM values as compared to gaugino-mediated one-loop diagrams and the diagrams involving R-parity violating vertices because in ${\cal N}=1$ gauged supergravity gaugino interaction vertices are dependent on suppressed dilute non-abelian fluxes and contribution of R-parity violating vertices are generally suppressed. However, all of the three-loop diagrams gave a very suppressed contribution to electron and neutron EDM. Then we considered one-loop diagrams involving Higgs and other supersymmetric/SM particles. By considering SM-like fermions with Higgs in a loop, we got the electron EDM estimate ($d_{e}/e\equiv 10^{-34}$ cm) and neutron EDM estimate ($d_{n}/e\equiv10^{-33}$ cm), considerably higher than the value predicted by SM. Interestingly, by considering one-loop diagrams involving chargino and Higgs, the  electron EDM value turned out to be ($d_e/e\equiv 10^{-32}cm$) i.e  a healthy EDM of electron even in the presence of super-heavy chargino in the loop.   All of the above one-loop diagrams involved MSSM-like superfields and CP-violating phases appeared from visible sector fields only. For a full-fledged analysis, we took into account goldstino supermultiplet also as the physical degrees of freedom in one-loop diagrams. As sgoldstino corresponds to the bosonic component of the superfield corresponding to which there is a supersymmetry breaking and same occurs maximally in our large volume $D3/D7$ model via complex `big' divisor volume modulus ($\tau_B$), we have identified the sgoldstino field with complex $\tau_B$ field.  Since, the fermionic component goldstino gets absorbed into the gravitino and becomes a longitudinal component of the massive gravitino, we basically considered one-loop diagrams involving gravitino and sgoldstino in the loop. In such kind of loop diagrams, CP-violating phases appeared from hidden sector fields. However, by evaluating the matrix amplitudes of these loop diagrams, we got a very suppressed contribution of electron and neutron EDM. The results of all one-loop diagrams are summarized in Table~4.1. In case of two-loop diagrams, we have evaluated contribution of Barr-Zee diagrams involving sfermions/fermions in an internal loop and mediated via $\gamma h$ exchange, and R-parity violating diagram involving fermions but mediated via ${\tilde f} h$ exchange. Here, the two-loop Barr-Zee diagrams involving heavy sfermions and a light Higgs  gave a most dominant contribution of EDM (${d_{(e/n)}}/{e}\equiv 10^{-29} cm$) as compared to two-loop diagrams involving only SM-like particles. With substantial fine tuning in Calabi-Yau volume, one could hope to produce EDM results same as experimental limits. Next, inspired by the approach given in \cite{Leigh_et_al, Barr_Zee,higgs_1, weinberg_higgs_2} to obtain large EDM value (almost same as an experimental bound) from Barr-Zee diagrams involving top quarks and W bosons loop in multi-Higgs models, we  provided an estimate of the same by using complex eigenstates of two Higgs doublets  as discussed in chapter {\bf 2}. By showing the possibility of obtaining the numerical estimate of all SM-like vertices relevant for these diagrams  to be same as their standard values in the context of ${\cal N}=1$ gauged supergravity model, we  produced EDM ($d_{(e,n)}/e\sim 10^{-27}$) cm in case of  Barr-Zee diagram involving $W$ boson.  By evaluating explicitly,  we also showed that two-loop rainbow diagrams give a very suppressed contribution as compared to Barr-Zee diagrams. The results of all two-loop diagrams are summarized in Table~4.2. Thus, we close the chapter by concluding that in our large volume $D3/D7$ $\mu$-split-like  SUSY model, despite the presence of very heavy supersymmetric scalars/fermions in the loops, we are able to produce a contribution to electric dipole moment of  both electron as well as neutron close to experimental bound at two-loop level and a sizable contribution even at one-loop level.

\chapter{Thermo/hydrodynamical aspects of M-Theory Uplifts}
\vskip -0.3in
{\hskip0.5in{\it ``Where there is matter, there is geometry."}}

\hskip4.2in - Johannes Kepler.
\graphicspath{{Chapter5/}{Chapter5/}}
\vskip -0.5in
\section{Introduction}
During the past few years there has been a lot of progress toward constructing the string theoretic dual descriptions of large-N gauge theories. The experimental data obtained at the Relativistic Heavy Ion Collider (RHIC) has provided  valuable information that shows characteristics of non-perturbative gauge theories obtained in the context of Quark Gluon Plasma (QGP) etc. The lack of reliable computing methods to studying non-perturbative gauge theories in various theoretical models motivates study of non-compact manifolds relevant to obtaining gravity duals used to studying various aspects of strongly coupled (IR) behavior of gauge theory. The AdS/CFT correspondence has been widely used in establishing, in particular, the connection between strongly coupled ${\cal N} = 4$ supersymmetric Yang-Mills (SYM) theory in large N limit and classical ten-dimensional supergravity. However, the analysis based on that relies on AdS spaces whose dual is a conformal field theory (CFT) with no running couplings \cite{maldacena, witten}. Though AdS background suffices to give complete picture of IR behavior, yet it does not provide UV completion of gauge theories. These limitations were overcome in models of gauge-gravity duality that are not supersymmetric, and are non-conformal. One of the popular non-AdS backgrounds used is to study D3-branes is a conifold, which is inspired from studies of branes at conical singularities, resulting in ${\cal N}=1$ superconformal gauge theory \cite{klebanov_witten, kacru+silverstein,lawrence_et_al}. The non-conformal or the confining behavior of gauge theories have been well understood in `Klebanov-Strassler' background \cite{Klebanov+Strassler} (as well in \cite{Klebanov+Tseytlin}) formed by placing $M$ fractional $D5$-branes wrapping over a vanishing  $S^2$ of $T^{1,1}$ base of conifold in addition to  $D3$-branes placed at the tip of the conifold. Though in this model, quarks transform  in the bi-fundamental representations of the two possible UV gauge groups, but eventually cascade away in the far IR which is not relevant to studying thermal aspects. Therefore, in order to allow the presence of fundamental quarks at finite temperature to make the discussion more appropriate to study of the deconfined phase of strongly coupled QCD i.e ``Quark Gluon Plasma", one has to incorporate co-incident $D7$-branes in `Klebanov-Strassler' construction of warped deformed conifold. The effect of $D7$-branes in KS background has been computed locally in \cite{metrics} just by introducing back-reaction due to presence of $D7$-branes with the assumption that axio-dilation vanishes locally and therefore, is not justified once one approaches the full global solution. The other subtle issues related to the same are discussed in \cite{metrics}. The very well known background given in the presence of non-zero axio-dilaton has been formed by introducing coincident $D7$-branes in `Klebanov-Tseytlin' background/in the large-$r$ limit of `Klebanov-Strassler' background via Ouyang embedding as explained in \cite{ouyang}. However, in the presence of non-zero temperature, Ouyang geometry also gets modified and has been given in the context of type IIB string theory in \cite{metrics} by inserting a black hole in a warped resolved conifold, named as modified Ouyang-Klebanov-Strassler (OKS)-Black Hole (BH) background. In that background, backreaction due to presence of black hole as well as D7-branes is included in 10-D warp factor. 

In this chapter, by extensively using modified OKS-BH background  including deformations,  we discuss the investigation of the possibility of getting thermodynamically stable local 11-dimensional uplift of warped deformed conifolds relevant to study of particular phase of  ${\cal N}=1$ strongly coupled QCD known as ``Quark Gluon Plasma". Since deformed conifolds do not possess global isometries along three directions, it is not possible to achieve M-theory uplift globally. However as given in \cite{SYZ3_Ts}, by applying suitable co-ordinate transformation on complex structure of  base of $T^3$ fiber, one can obtain an isometry along the third direction locally. Therefore, the main aim is to first satisfy the requirements of implementing mirror symmetry (obtained by applying three successive T-dualities on a supersymmetric $T^3$ fiber inside the Calabi-Yau) given by Strominger-Yau-Zaslow to transform the type IIB metric to its type IIA mirror so that one can obtain a local M-theory uplift of the latter. In this work, we will consider the following two limits:
\begin{eqnarray}
\label{limits_Dasguptaetal-i}
& (i)   & ~ {\rm weak}(g_s)~{\rm coupling-large\ t'Hooft\ coupling\ limit}:\nonumber\\
& &  g_s<<1, g_sN_f<<1, \frac{g_sM^2}{N}<<1, g_sM>>1, g_sN>>1\nonumber\\
& & {\rm effected\ by}: g_s\sim\epsilon^{d}, M\sim\epsilon^{-\frac{3d}{2}}, N\sim\epsilon^{-19d}, \epsilon\leq{\cal O}(10^{-2}), d>0
 \end{eqnarray}
(the limit in the first line of (\ref{limits_Dasguptaetal-i}) though not its realization in the second line, considered in \cite{metrics});
\begin{eqnarray}
\label{limits_Dasguptaetal-ii}
& (i i) &~ {\rm \bf MQGP\ limit}: \frac{g_sM^2}{N}<<1, g_sN>>1, {\rm finite}\
 g_s, M\ \nonumber\\
& &{\rm effected\ by}:  g_s\sim\epsilon^d, M\sim\epsilon^{-\frac{3d}{2}}, N\sim\epsilon^{-39d}, \epsilon\lesssim 1, d>0.
\end{eqnarray}
To begin with, in section {\bf 2}, we outline the basics of (resolved) warped deformed conifolds and mention suitable co-ordinate transformations to generate, locally, a `third' isometry and to produce a large volume of the base of a local $T^3$ fibration to permit application of mirror symmetry a al SYZ in type IIB background. In section {\bf  3}, we calculate the chemical potential arising from a $U(1)$ gauge field arising from the $U(N_f)\sim SU(N_f)\times U(1)$ living on the $D7$-brane world volume as a function of temperature using the type IIB background of \cite{metrics}. We also examine the behavior of same as a function of increasing number of quarks i.e $N_f$ at constant temperature to investigate thermodynamical stability of system in type IIB background. In section  {\bf 4}, using the 10-dimensional metric of (resolved) warped deformed conifolds \cite{metrics} and analytic
expressions given in \cite{SYZ3_Ts} to mirror transform, locally, the components of type IIB metric as well as NS-NS B-field components, we first obtain the form of the metric of the type IIA mirror manifold. In the process, at  $\theta_i= 0,\pi$ and $r\sim \sqrt{3}a$, we show how to construct a large base of a local $T^3$ fibration so that mirror symmetry between resolved and deformed conifolds can be established locally; in fact if one assumes that the local uplift is valid globally, we show that one can (more readily) ensure a large-base $T^3$-fibration near $\theta_{1,2}=\frac{\pi}{2},
r\sim\sqrt{3}a$ as well. Next, by applying T-duality rules given in \cite{T-dual_Hassan} to T-dualize RR odd-form field strengths and performing successive three T-dualities along three, locally toroidal
 isometry directions, we get various non-zero components of type IIA  2-form field strength using which we obtain one-form gauge field potential which eventually leads to a local M-theory metric. We then consider two limits - one which involves taking the weak coupling
limit of M-theory with a small string coupling and the other, which we call as the `MQGP' limit involves considering a finite string coupling
( $ g_s \lesssim 1$ and hence relevant to strongly coupled QGP).  We argue that the M-theory uplift for both limits
yields an $M3$-brane with $\frac{1}{8}$ near-horizon supersymmetry near $\theta_{1,2}=0,\pi$. Interested in studying hydro/thermodynamical aspects of the local M-theory uplift, in {\bf 5.1} and 
{\bf 5.2}, assuming the formula for $\frac{\eta}{s}$ and the diffusion coefficient $D$  of \cite{KovtunSonStarinets} obtained for only
 radial-coordinate-dependent
metric (i.e. $\mu_C=0$) to also be valid for the type IIA mirror and its  local M-theory uplift having frozen the angular dependence (tunably
small $\mu_C$ of section {\bf 3}),  we  calculate shear viscosity-entropy density ratio $\frac{\eta}{s}$ in both limits  and show that one gets exactly
$\frac{\eta}{s}=\frac{1}{4\pi}, D\sim\frac{1}{T}$; $G^{IIA}_{tt,rr}, G^{\cal M}_{tt,rr}$ having no angular dependence and  the angular
dependences entering via the metric determinants,  cancel out neatly. After that, again using the same limits, in {\bf 5.3}  we get simplified expressions of
11-dimensional Euclideanized space-time action which includes contribution from 11-dimensional bulk and  flux terms, 10-dimensional Gibbons-Hawking
term as well as contributions arising from ${\cal O}(R^4)$-terms given that the horizon radius turns out to be of the order of the string
scale making higher derivative contributions important; despite the fact that we end up
with a string-scale-sized horizon, the finite parts of the ${\cal O}(R^4)$-terms vanish in the aforementioned limits.
 We show
that in the limits (\ref{limits_Dasguptaetal-i}) and (\ref{limits_Dasguptaetal-ii}), the divergent portion of, e.g., $\int_{r=r_\Lambda}\sqrt{h}R$
(and/or $\int_{r=r_\Lambda}\sqrt{h}$ and/or $\int_{r=r_\Lambda}\sqrt{h}|G_4|^2$),
can act as an appropriate counter-term cancelling the
divergent portions ($r_\Lambda\rightarrow\infty$) of the Einstein-Hilbert + Gibbons-Hawking-York + Flux + ${\cal O}(R^4)$ terms of the action; in
limit (\ref{limits_Dasguptaetal-i}), we show that one can provide an asymptotically-linear-dilaton-type interpretation to the counter term(s). In the aforementioned calculations,
we introduce  cut-offs $\epsilon_{\theta}$  near $\theta_{1,2}=0,\pi$, which after
identification with $\epsilon^{\gamma_{(i),(ii)}}$ for appropriate $\gamma_{(i),(ii)}$
for the two limits (\ref{limits_Dasguptaetal-i}) and (\ref{limits_Dasguptaetal-ii}) with the $\epsilon$ also the same as the ones appearing in the limits
(\ref{limits_Dasguptaetal-i}) and (\ref{limits_Dasguptaetal-ii}), ensure that  the
finite part of the action - generated entirely by the  Gibbons-Hawking-York surface term - is independent of the cut-offs $\epsilon_{\theta}$. After evaluating the 11-D action/partition function in the aforementioned limit, we
calculate various thermodynamical quantities, e.g, entropy and specific heat in 11-dimensional M-theory background the sign of which has been
considered to be a criterion to check the thermodynamical stability of the $M$-theory uplift. Finally, we summarize our results in section {\bf 6}.
 \section{Singular, Resolved and Deformed Conifolds}
We briefly discuss the generic expressions of singular, resolved as well as deformed conifold metrics relevant to discussions in this chapter. Analogous to a two-dimensional cone embedded in $\mathbb{R}^3: x^2 + y^2 - u^2=0$, a real six-dimensional conifold can be represented as a quadric: $\sum_{i=1}^4 z_i^2=0$ in $\mathbb{C}^4$, which is smooth everywhere except at the apex/node: $\vec{z}=0$ which is a double-point. Writing $z_i=x_i+\imath y_i$, the base of the cone is $S^2(y_i)$ fibered over $S^3(x_i)$ and as all fibrations are trivial, the base is $S^2\times S^3$. The conifolds are cones over Sasaki-Einstein spaces $T^{p,q}, p$ and $q$ being relatively prime integers, which means \cite{Candela+delaOssa}:
 \begin{eqnarray}
&& ds^2=dr^2 + r^2 g_{T^{p,q}}, g_{T^{p,q}}=g_{ab}dx^a dx^b
=\lambda^2(d\psi + p cos\theta_1 d\phi_1 + q cos\theta_2 d\phi_2)^2 + \nonumber\\
&& \Lambda_1^{-1}(d\theta_1^2 + sin^2\theta_1 d\phi_1^2) + \Lambda_2^{-1}(d\theta_2^2 + sin^2\theta_2 d\phi_2^2),
\end{eqnarray}
$\psi=\psi_1+\psi_2$, where the Euler angles for the two SU(2)s are ($\phi_{1,2}\in[0,2\pi],\theta_{1,2}\in[0,\pi],\psi_{1,2}\in[0,4\pi]$).
An $n$-conifold is Ricci-flat if the base is Einsteinian: $R_{ab}(g)=(n-2)g_{ab}$. For $n=6$, one obtains: $4=\frac{\lambda^2}{2}\left\{(p\Lambda_1)^2 + (q\Lambda_2)^2\right\}=\Lambda_1-\frac{1}{2}(\lambda p \Lambda_1)^2=\Lambda_2 - \frac{1}{2}(\lambda q \Lambda_2)^2$. For $p=1,q=1$ one obtains:   \begin{equation}
  g_{T^{1,1}}=\frac{1}{9}(d\psi +  cos\theta_1 d\phi_1 +  cos\theta_2 d\phi_2)^2 + \frac{1}{6}(d\theta_1^2 + sin^2\theta_1 d\phi_1^2) + \frac{1}{6}(d\theta_2^2 + sin^2\theta_2 d\phi_2^2). 
  \end{equation}
The conifold base can be viewed as a coset space $\frac{SU(2)\times SU(2)}{U(1)}$. To see this, convenient to introduce $W  = \frac{1}{\sqrt{2}} \left ( \begin{array}{cc} z_3 + \imath z_4 & z_1 - \imath z_2 \\ z_1 + \imath z_2 & - z_3 + \imath z_4  \end{array} \right ).$  By a rescaling $ Z = \frac{W}{r}$, the defining equation for the conifold  and the base $\det Z  =  0, \mathrm{tr} Z^\dagger Z  =  1$. The metric is K\"{a}hler iff $g_{i{\bar j}}=\partial_i{\bar\partial}_{\bar j}K(r^2)\Rightarrow ds^2 = K'(r^2) tr(dW^\dagger dW) + K''(r^2)|tr(W^\dagger dW)|^2$, $r^2=Tr(W^\dagger W)$. It is shown in \cite{Candela+delaOssa} that $T^{p,q}$  corresponds to K\"{a}hler Ricci-flat metric only for choice of $p=q=1$.
 
 The singular conifold can be smoothed by deforming as: $\sum_{i=1}^4z_i^2=\epsilon^2$.  For $r\neq\epsilon$ one gets back $T^{1,1}$ and for $r=\epsilon$(`origin of coordinates'), one gets an $S^3(\vec{x})$.
 The deformed conifold metric as given in \cite{Minasian+tsimpis} is as follows:
 \begin{eqnarray}
& &  ds^2_{\rm def} = |tr(W_\epsilon^\dagger d W_\epsilon)|^2K''(r^2) + tr(dW_\epsilon^\dagger dW_\epsilon) K'(r^2)
\nonumber\\
& & = \Bigl[\left(r^2 \hat{\gamma}'-\hat{\gamma}\right)
\,\left(1-\frac{\epsilon^4}{r^4}\right)+\hat{\gamma}\Bigr]  \times\Bigl(\frac{dr^2}{r^2(1-\epsilon^4/r^4)}
+ \frac{1}{4}\,(d\psi+\cos\theta_1\,d\phi_1+\cos\theta_2\,d\phi_2)^2\Bigr)\nonumber\\
&&  +\, \frac{\hat{\gamma}}{4}\,\left[(\sin\theta_1^2\,d\phi_1^2+d\theta_1^2)
+(\sin\theta_2^2\,d\phi_2^2+d\theta_2^2)\right] +\frac{\hat{\gamma}\epsilon^2}{2 r^2}\,\left[\cos\psi(d\theta_1
d\theta_2-\sin\theta_1\sin\theta_2 d\phi_1
d\phi_2) \right.\nonumber\\
& & \left.  +\sin\psi(\sin\theta_1 d\phi_1 d\theta_2+\sin\theta_2 d\phi_2
d\theta_1)\right],
\end{eqnarray}
$\psi\equiv\psi_1+\psi_2, \hat{\gamma}\equiv r^2 K$ determined by the solution of the Ricci-flatness condition: $r^2(r^4-\epsilon^4)(\hat{\gamma}^3)'+3\epsilon^4\hat{\gamma}^3\,=\,2 r^8\,$; as $r\rightarrow\infty$, $\hat{\gamma}\rightarrow r^{\frac{4}{3}}$ (see \cite{Candela+delaOssa}).

The resolved conifold is obtained by resolving the singularity at the apex of the cone by non-trivial $S^2$, i.e., replace
Replace $\sum_{i=1}^4 z_i^2=0$ by $\begin{pmatrix} x & u \\ v & y\end{pmatrix}
\begin{pmatrix} \lambda_1 \\ \lambda_2 \end{pmatrix}
\,=\, 0, (\lambda_1,\lambda_2)\neq(0,0)$. So, for $(u,v,x,y)\ne 0$ (away from the tip),  one gets back the conifold. Due to the overall scaling freedom $(\lambda_1,\lambda_2)\sim (\lambda\lambda_1,\lambda\lambda_2)$, $(\lambda_1,\lambda_2)$ actually describe a $\mathbb{CP}^1\sim S^2$ at the tip of the cone. The most general K\"{a}hler potential is given by $K=F(r^2) + 4 a^2 ln (1 + |\lambda|^2)$ implying the resolved conifold metric \cite{zayas+tseytlin} is given by: \begin{eqnarray}
& & ds^2_{\rm res}  = F'(r^2) tr(dW^\dagger dW) + F''(r^2)|tr(W^\dagger dW)|^2 + 4a^2\frac{|d\lambda|^2}{(1 + |\lambda|^2)^2} \nonumber\\
& & = \widetilde{\gamma}'\,dr^2 + \frac{\widetilde{\gamma}'}{4}\,r^2\big(d\psi+\cos\theta_1\,d\phi_1
+\cos\theta_2\,d\phi_2\big)^2\nonumber\\
& &+ \frac{\widetilde{\gamma}}{4}\,\big(d\theta_1^2+\sin^2\theta_1\,d\phi_1^2\big)  +
\frac{\widetilde{\gamma}+4a^2}{4}\,\big(d\theta_2^2+\sin^2\theta_2\,d\phi_2^2\big); \widetilde{\gamma}\stackrel{r\rightarrow\infty}{\longrightarrow}r^{\frac{4}{3}}
\end{eqnarray}
  Defining $\rho^2=3/2\,\widetilde{\gamma}$,
\begin{eqnarray}
& & ds^2_{\rm res}  =  \frac{\kappa(\rho)}{9}\,\rho^2\big(d\psi+\cos\theta_1\,d\phi_1
+\cos\theta_2\,d\phi_2\big)^2 \nonumber \\
& &+ \frac{\rho^3}{6}\,\big(d\theta_1^2+\sin^2\theta_1\,d\phi_1^2\big)+
\frac{\rho^2+6a^2}{6}\,\big(d\theta_2^2+\sin^2\theta_2\,d\phi_2^2\big) +  \kappa(\rho)^{-1}\,d\rho^2\,,
\end{eqnarray}
with $\kappa(\rho)=(\rho^2+9a^2)/(\rho^2+6a^2)$.We will always work in the limit wherein $\rho$, relabelled as $r$ in subsequent sections, is much larger than $a$.
\section{ Type IIB (Resolved) Warped Deformed Conifold Background}
The basic motivation in using non-AdS background to explain the properties of QCD was to provide a geometric background that helps to explain not just the physics of QCD in the IR region but is also  sufficient to unravel the various key points needed to explain UV completion of the theory. The knowledge of UV completion is important to handle the issues related to finiteness of the  solution at short distances as well as to capture certain aspects of large-N thermal QCD. The different models were proposed to incorporate the effect of renormalization group in the dual background by  connecting conformal fixed points at IR as well as UV \cite{Freedman_et_al, Freedman_et_al1}. As discussed in section {\bf 1}, the first successful attempt to explain the RG flow (without any fixed point/surface) in the dual background was made by Klebanov and Strassler in \cite{Klebanov+Strassler} by embedding D-branes in a conifold background which was further extended to OKS background in the presence of fundamental quarks and finally, followed by modified OKS-BH background in the presence of black-hole.  Before going into details of (resolved) warped deformed conifold relevant to study of QGP, let us first review the modified OKS-BH geometry in the presence of the black hole given in \cite{metrics}.

Starting from  Klebanov-Strassler model \cite{Klebanov+Strassler}, in this background, $N\ D3$-branes are placed at the tip of six-dimensional conifold whereas  $M\ D5$-branes are wrapped over the vanishing two cycle $S^2$ of conifold base. Introducing $M\ D5$-branes/fractional D3-branes thus produces $SU(N +M)\times SU(N)$ supersymmetric gauge theory such that the matter fields transforming as bi-fundamental representation of UV gauge group $SU(N + M) \times SU(N)$ and under renormalization group flow, eventually cascade down to $SU(M)$ in the far IR region. However, the behavior of gauge theories so-obtained is not quite simple under RG flow. The subtle issues are explained in \cite{Klebanov+Strassler} and elaborated  in more detail in \cite{metrics} also. Due to the presence of complicated gauge theory, there are numerous points where  beta functions $\beta_{g_1}, \beta_{g_2}$ corresponding to two gauge coupling constants $(g_1, g_2)$ are very small, the RG-fixed surfaces involve infinite number of choppy Seiberg dualities and hence RG flow is not smooth as one goes from one surface to another surface at different energy scales and therefore, is not able to provide the dual gravity background.  However, it has been discussed in \cite{metrics}, that inspite of the choppy nature of RG flow at the boundary, there is a smooth RG flow if one hovers towards center of two-dimensional surface which eventually leads to give weakly-coupled gravity description. Therefore, there is a very small regime of smooth RG flow in the gauge theory side that can be captured by weakly coupled supergravity description. This means the KS picture is able to explain the UV completion of the so-formed QCD description. However, the KS description does not include fundamental quarks. To discover the background relevant to study of certain aspects of gauge theory at finite temperature, one has to introduce fundamental quarks at high temperature. This is done by insert co-incident $D7$-branes in Klebanov-Strassler warped deformed construction. There are various subtle issues explained in \cite {metrics} as related to the validity of the same once one goes to explain the full Global scenario. Therefore, $D7$ branes are embedded in large r regime of KS geometry via Ouyang embedding \cite{ouyang} (that takes into account the effect of the axio-dilaton field on the metric) given as:
\begin{equation}
r^{\frac{3}{2}}e^{\frac{i}{2}(\psi-\phi_1-\phi_2)}\sin\frac{\theta_1}{2} \sin\frac{\theta_2}{2}=\mu
\end{equation}
where $\mu$ is embedding parameter. This discussion so far is valid at $T=0$. The extended background in the presence of non-zero temperature has been discussed in \cite{metrics} by inserting a black-hole to the OKS background which results in both resolution as well as deformation of the two and three cycles of the conifold respectively at $r=0$. In the presence of a black-hole in a warped deformed conifold, the metric is given by \cite{metrics},\cite{finite_mu_Dasguptaetal}:
\begin{equation}
\label{metric}
ds^2 = {1\over \sqrt{h}}
\left(-g_1 dt^2+dx_1^2+dx_2^2+dx_3^2\right)+\sqrt{h}\Big[g_2^{-1}dr^2+r^2 d{\cal M}_5^2\Big].
\end{equation}
 where $h_5~(\rm deformation~parameter)^2<<1$. The $g_i$'s demonstrate the presence of black hole in modified OKS-BH background and given as follows:
\begin{eqnarray}
& & g_1(r,\theta_1,\theta_2)= 1-\frac{r_h^4}{r^4} + {\cal O}\left(\frac{g^2_sM}{N}\right),
~~~~ g_2(r,\theta_1,\theta_2) = 1-\frac{r_h^4}{r^4} + {\cal O}\left(\frac{g^2_sM}{N}\right).
\end{eqnarray}
where $r_h$ is the horizon, and the ($\theta_1, \theta_2$) dependence come from the
${\cal O}\left(\frac{g_sM^2}{N}\right)$ corrections (We note that the `black hole' factors $g_i$ are
stated to receive ${\cal O}(g_s^2 M N_f)$ corrections in \cite{metrics}, but as shown in \cite{finite_mu_Dasguptaetal},
the six-dimensional warp factors $h_i$ are expected to receive corrections of
${\cal O}\left(\frac{g_sM^2}{N}\right)$ - we assume the same to also be true of the `black hole
functions' $g_{1,2}$)\footnote{This will have extremely non-trivial consequence that one can use
the same choice of $h_i$ and $g_i$ in the weak($g_s$)coupling-large t'Hooft couplings as well as the
`MQGP' limits, later in the chapter.} and
\begin{eqnarray}
d{\cal M}_5^2 = && h_1 (d\psi + {\rm cos}~\theta_1~d\phi_1 + {\rm cos}~\theta_2~d\phi_2)^2 +
h_2 (d\theta_1^2 + {\rm sin}^2 \theta_1 ~d\phi_1^2) +   \nonumber\\
&& + h_4 (h_3 d\theta_2^2 + {\rm sin}^2 \theta_2 ~d\phi_2^2) + h_5~{\rm cos}~\psi \left(d\theta_1 d\theta_2 -
{\rm sin}~\theta_1 {\rm sin}~\theta_2 d\phi_1 d\phi_2\right) + \nonumber\\
&&  + h_5 ~{\rm sin}~\psi \left({\rm sin}~\theta_1~d\theta_2 d\phi_1 +
{\rm sin}~\theta_2~d\theta_1 d\phi_2\right).
\end{eqnarray}
Due to presence of Black-hole, $h_i$ appearing in internal metric as well as $M, N_f$ are not constant and upto linear order depend on $g_s, M, N_f$ as given below:
\begin{eqnarray}
& & h_1 = {1\over 9} + {\cal O}(\frac{g_sM^2}{N}), ~~~~~ h_2 = h_4 = {1\over 6} + {\cal O}(\frac{g_sM^2}{N}), ~~~~~ h_3 = 1 + {\cal O}(\frac{g_sM^2}{N}),\nonumber\\
&& L=\left(4\pi g_s N\right)^{\frac{1}{4}}, M_{\rm eff} = M + \sum_{m\ge n} a_{mn} (g_sN_f)^m (g_sM)^n, \nonumber\\
&&
N_{f}^{\rm eff} = N_f + \sum_{m \ge n} b_{mn} (g_sN_f)^m (g_sM)^n. 
\end{eqnarray}
The warp factor that includes the back-reaction due to fluxes as well as black-hole is given as:
\begin{eqnarray}
\label{eq:h}
&& h =\frac{L^4}{r^4}\Bigg[1+\frac{3g_sM_{\rm eff}^2}{2\pi N}{\rm log}r\left\{1+\frac{3g_sN^{\rm eff}_f}{2\pi}\left({\rm
log}r+\frac{1}{2}\right)+\frac{g_sN^{\rm eff}_f}{4\pi}{\rm log}\left({\rm sin}\frac{\theta_1}{2}
{\rm sin}\frac{\theta_2}{2}\right)\right\}\Bigg].\nonumber\\
\end{eqnarray}
Though the results are calculated to order ${\cal O}(g_s N_f)$ at small r,  the conjectural solution is given at large r by embedding the model in F-theory set-up so that one is able to avoid the issue of singularities appearing in the background at large r and therefore, explain the holographic renormalisibility of theory. This is connected with the fact that at large r, dilaton, warp factor as well as fluxes should not blow up as r approaches infinity. As realized explicitly in \cite{metrics}, the analytical expression of warp factor given in equation (\ref{eq:h}) does not show up singularity at large r if parameters $N, M, g_s$ are tuned in such a way that following limits are satisfied.
\begin{equation}
\label{limits_Dasguptaetal}
g_s\rightarrow 0, g_sN_f\rightarrow 0,\frac{g_sM^2}{N}\rightarrow 0,g_sM\rightarrow\infty, g_sN\rightarrow\infty.
\end{equation}
Therefore,  to get the results using OKS-BH background, one should strictly satisfy aforementioned constraints on $(g_s, M, N)$.

We utilize the above background to study the thermodynamic stability of solution in a ``(resolved) warped deformed conifold background" obtained by considering limits
\begin{equation}
h_5 \not = 0, h_3 = 1, h_4 - h_2 = a, g_i = 1-\frac{r_h^4}{r^4} + {\cal O}\left(\frac{g_sM^2}{ N}\right).
\end{equation}
 i.e by considering deformation parameter $h_5$ to be very small but not equal to zero, and $r>>a$.

The three-form fluxes by including black hole factors are given by \cite{metrics}:
\begin{eqnarray}
\label{three-form fluxes}
{\widetilde F}_3 & = & 2M { A_1} \left(1 + {3g_sN_f\over 2\pi}~{\rm log}~r\right) ~e_\psi \wedge
\frac{1}{2}\left({\rm sin}~\theta_1~ d\theta_1 \wedge d\phi_1-{ B_1}~{\rm sin}~\theta_2~ d\theta_2 \wedge
d\phi_2\right)\nonumber\\
&& -{3g_s MN_f\over 4\pi} { A_2}~{dr\over r}\wedge e_\psi \wedge \left({\rm cot}~{\theta_2 \over 2}~{\rm sin}~\theta_2 ~d\phi_2
- { B_2}~ {\rm cot}~{\theta_1 \over 2}~{\rm sin}~\theta_1 ~d\phi_1\right)\nonumber \\
&& -{3g_s MN_f\over 8\pi}{ A_3} ~{\rm sin}~\theta_1 ~{\rm sin}~\theta_2 \left({\rm cot}~{\theta_2 \over 2}~d\theta_1 +
{ B_3}~ {\rm cot}~{\theta_1 \over 2}~d\theta_2\right)\wedge d\phi_1 \wedge d\phi_2\label{brend}, \nonumber\\
H_3 &=&  {6g_s { A_4} M}\Bigr(1+\frac{9g_s N_f}{4\pi}~{\rm log}~r+\frac{g_s N_f}{2\pi}
~{\rm log}~{\rm sin}\frac{\theta_1}{2}~
{\rm sin}\frac{\theta_2}{2}\Bigr)\frac{dr}{r}\nonumber \\
&& \wedge \frac{1}{2}\Bigr({\rm sin}~\theta_1~ d\theta_1 \wedge d\phi_1
- { B_4}~{\rm sin}~\theta_2~ d\theta_2 \wedge d\phi_2\Bigr)
+ \frac{3g^2_s M N_f}{8\pi} { A_5} \Bigr(\frac{dr}{r}\wedge e_\psi -\frac{1}{2}de_\psi \Bigr)\nonumber  \\
&& \hspace*{1.5cm} \wedge \Bigr({\rm cot}~\frac{\theta_2}{2}~d\theta_2
-{ B_5}~{\rm cot}~\frac{\theta_1}{2} ~d\theta_1\Bigr),\nonumber\\
& & {\rm implying}\nonumber\\
 B_2 & = & 6 g_s { A_4} M\left( ln r + \frac{9 g_s N_f}{4\pi} (ln r)^2 + \frac{g_s N_f}{4\pi}(1 + 2 ln r) ln \left(sin\frac{\theta_1}{2}sin\frac{\theta_2}{2}\right)\right)\nonumber\\
& & \times \frac{1}{2}\Bigr({\rm sin}~\theta_1~ d\theta_1 \wedge d\phi_1
- { B_4}~{\rm sin}~\theta_2~ d\theta_2 \wedge d\phi_2\Bigr) \nonumber\\
& &  + \frac{3g_s^2MN_f}{8\pi} { A_5} ln r e_\psi\wedge \left(cot\frac{\theta_2}{2} d\theta_2 - { B_5} cot\frac{\theta_1}{2} d\theta_1\right),\nonumber\\
& & e^{-\Phi} = {1\over g_s} -\frac{N_f}{8\pi} ~{\rm log} \left(r^6 + 9a^2 r^4\right) -
\frac{N_f}{2\pi} {\rm log} \left({\rm sin}~{\theta_1\over 2} ~ {\rm sin}~{\theta_2\over 2}\right).
\end{eqnarray}
where the asymmetry factors ${ A_i}, { B_i}$ encode all the information of the black hole etc in our background. To order ${\cal O}(g_sN_f)$, the same are given by:
\begin{eqnarray}
\label{asymmetry}
&& { A_1} ~=~ 1 + {9g_s N_f \over 4\pi} \cdot {a^2\over r^2}\cdot (2 - 3~{\rm log}~r) + {\cal O}(a^2 g_s^2 N_f^2) \nonumber\\
&& { B_2} ~=~ 1 + {36 a^2~{\rm log}~r \over r^3 + 18 a^2 r ~{\rm log}~r} + {\cal O}(a^2 g_s^2 N_f^2)\nonumber\\
&& { A_2} ~= ~1 + {18 a^2 \over r^2} \cdot {\rm log}~r + {\cal O}(a^2 g_s^2 N_f^2) \nonumber\\
&& { B_1} ~=~ 1 + {81\over 2} \cdot
{g_s N_f a^2 {\rm log}~r \over 4\pi r^2 + 9 g_s N_f a^2 (2 - 3~{\rm log}~r)} + {\cal O}(a^2 g_s^2 N_f^2)\nonumber\\
&& { A_3} ~=~ 1 - {18 a^2 \over r^2}\cdot {\rm log}~r +  {\cal O}(a^2 g_s^2 N_f^2)\nonumber\\
&& { B_3} ~ = ~ 1 + {36 a^2 {\rm log}~r \over r^2 - 18 a^2 {\rm log}~r} + {\cal O}(a^2 g_s^2 N_f^2)\nonumber\\
&& { A_4} ~ = ~ 1 - {3a^2 \over r^2} + {\cal O}(a^2 g_s^2 N_f^2), ~~~~~ { B_4} ~ = ~ 1 + {3g_s a^2 \over r^2 - 3 a^2} + {\cal O}(a^2 g_s^2 N_f^2)\nonumber\\
&& { A_5} ~ = ~ 1 + {36 a^2 {\rm log}~r \over r} +  {\cal O}(a^2 g_s^2 N_f^2), ~~~~
{ B_5} ~ = ~ 1 + {72 a^2 {\rm log}~r \over r + 36 a^2 {\rm log}~r} + {\cal O}(a^2 g_s^2 N_f^2).\nonumber\\
\end{eqnarray}
The ${\cal O}(a^2/r^2)$ corrections included in asymmetry factors correspond to modified Ouyang background  in the presence of black hole. The values for the axion $C_0$ and the five form $F_5$ are given by \cite{metrics}:
\begin{eqnarray}
\label{axfive}
&&C_0 ~ = ~ {N_f \over 4\pi} (\psi - \phi_1 - \phi_2),\nonumber\\
&& F_5 ~ = ~ {1\over g_s} \left[ d^4 x \wedge d h^{-1} + \ast(d^4 x \wedge dh^{-1})\right].
\end{eqnarray}
with the dilaton to be taken as approximately a constant near the D7-brane.

Working in a local limit around: $r\approx\langle r\rangle, \theta_{1,2}\approx\langle\theta_{1,2}\rangle,\psi\approx\langle\psi\rangle$ equivalent to replacing the ${\bf CP}^1(\theta_1,\phi_1), {\bf CP}^2(\theta_2,\phi_2)$ locally by $T^2(\theta_1,x), T^2(\theta_2,y)$. Define T-duality coordinates, $(\phi_1,\phi_2,\psi)\rightarrow(x,y,z)$ (\cite{SYZ3_Ts}):
\begin{equation}
\label{xyz defs}
x = \sqrt{h_2}h^{\frac{1}{4}}sin\langle\theta_1\rangle\langle r\rangle \phi_1,\ y = \sqrt{h_4}h^{\frac{1}{4}}sin\langle\theta_2\rangle\langle r\rangle \phi_2,\ z=\sqrt{h_1}\langle r\rangle h^{\frac{1}{4}}\psi.
\end{equation}
Interestingly, around $\psi=\langle\psi\rangle$, under the coordinate transformation (\cite{SYZ3_Ts}):
\begin{equation}
\label{transformation_psi}
\left(\begin{array}{c} sin\theta_2 d\phi_2 \\ d\theta_2\end{array} \right)\rightarrow \left(\begin{array}{cc} cos\langle\psi\rangle & sin\langle\psi\rangle \\
- sin\langle\psi\rangle & cos\langle\psi\rangle
\end{array}\right)\left(\begin{array}{c}
sin\theta_2 d\phi_2\\
d\theta_2
\end{array}
\right),
\end{equation}
the $h_5$ term becomes $h_5\left[d\theta_1 d\theta_2 - sin\theta_1 sin\theta_2 d\phi_1d\phi_2\right]$. Further, $e_\psi\rightarrow d(\psi + sin\langle\psi\rangle  ln \\ sin\theta_2) + cos\theta_1 d\phi_1 + cos\theta_2 d\phi_2$, which under $\psi\rightarrow\psi - sin\langle\psi\rangle ln sin\theta_2$, implies:
$e_\psi\rightarrow e_\psi$. Locally, thus one introduces an isometry along $\psi$ in addition to the isometries along $\phi_{1,2}$. This clearly is not valid globally - the deformed conifold does not possess a third global isometry. We will be taking $\langle\psi\rangle\sim0,2\pi,4\pi$ so that the
one-forms $sin\theta_2 d\phi_2$ and $ d\theta_2$ do not change appreciably.

To enable use of SYZ-mirror duality via three T dualities, one needs to ensure a large base (implying large complex structures of the aforementioned two two-tori) of the $T^3(x,y,z)$ fibration. This is effected via \cite{large_base}:
\begin{eqnarray}
\label{SYZ-large base}
& & d\psi\rightarrow d\psi + f_1(\theta_1)cos\theta_1 d\theta_1 + f_2(\theta_2)d\theta_2, d\phi_{1,2}\rightarrow d\phi_{1,2} - f_{1,2}(\theta_{1,2})d\theta_{1,2},
\end{eqnarray}
for appropriately chosen large values of $f_{1,2}(\theta_{1,2})$. The three-form fluxes of (\ref{three-form fluxes}) remain invariant under (\ref{SYZ-large base}). The fact that one can choose such large values of $f_{1,2}(\theta_{1,2})$, is justified in section {\bf 4}.

\section{Chemical Potential in the Type IIB Background}
The chemical potential in gravity dual is generated via D7-brane gauge fields in a certain geometric background. The temporal component of bulk $U(1)$ field on the D7-brane worldvolume is related to chemical potential  since it is the conjugate field to the electric charge. So, it is defined in a
gauge-invariant manner as follows: $\mu_C=\int_{r_h}^\infty dr F_{rt}$.
$F_{rt}$ can be evaluated by solving Lagrange equation of motion for DBI Action. Since the background given in \cite{metrics} has been chosen to give non-conformal gauge theories, it might be the case that warp factors $h_i$ appearing in the metric vary while RG flow from IR scale to UV scale. Therefore, as similar to \cite{nonextremel_dasgpta}, the analysis given below is based on the assumption that $h_i$  is constant at a fixed energy scale.

Assuming $\mu(\neq0)\in{\bf R}$ in Ouyang's embedding (\cite{real_mu_Ouyang_et_al}): \\
$r^{\frac{3}{2}}e^{\frac{i}{2}(\psi-\phi_1-\phi_2)}\sin\frac{\theta_1}{2} \sin\frac{\theta_2}{2}=\mu$, implies that $r^{\frac{3}{2}}\sin\left(\frac{\psi-\phi_1-\phi_2}{2}\right)\sin\frac{\theta_1}{2} \sin\frac{\theta_2}{2}=0$ which could be satisfied for $\psi=\phi_1+\phi_2$ and $r^{\frac{3}{2}}\sin\frac{\theta_1}{2} \sin\frac{\theta_2}{2}=\mu$. Using the same, one obtains the following metric for a space-time-filling wrapped $D7$-brane embedded in the warped deformed conifold:
{\small
\beqn
&&ds^2 = {1\over \sqrt{h\left(r,\theta_2,\theta_1(r,\theta_2)\right)}}
\left(-g_1(r) dt^2+dx^2+dy^2+dz^2\right)+\sqrt{h\left(r,\theta_2,\theta_1(r,\theta_2)\right)}\Big[\frac{dr^2}{g_2(r)}\nonumber\\
&&+r^2 d{\cal M}_3^2\Big],\nonumber\\
& & {\rm where,}~~~d{\cal M}_3^2 = {h_1} \left({d\phi_2} (\cos ({\theta_2})+1)+{d\phi_1} \left(2-\frac{2 \mu ^2 \csc
   ^2\left(\frac{{\theta_2}}{2}\right)}{r^3}\right)\right)^2+ \nonumber\\
   & & {h_2} \left(\left(1-\left(1-\frac{2 \mu ^2
   \csc ^2\left(\frac{{\theta_2}}{2}\right)}{r^3}\right)^2\right) {d\phi_1}^2+\frac{\mu ^2 \left(\frac{3
   {dr}}{r}+{d\theta_2} \cot \left(\frac{{\theta_2}}{2}\right)\right)^2}{r^3 \left(\sin
   ^2\left(\frac{{\theta_2}}{2}\right)-\frac{\mu ^2}{r^3}\right)}\right) +{h_5} \cos
   ({\phi_1}+{\phi_2})\nonumber\\
   & & \left(\frac{{-d\theta_2} \mu  \left(\frac{3 {dr}}{r}+{d\theta_2} \cot
   \left(\frac{{\theta_2}}{2}\right)\right)}{r^{3/2} \sqrt{\sin
   ^2\left(\frac{{\theta_2}}{2}\right)-\frac{\mu ^2}{r^3}}}-{d\phi_1} {d\phi_2} \sqrt{1-\left(1-\frac{2
   \mu ^2 \csc ^2\left(\frac{{\theta_2}}{2}\right)}{r^3}\right)^2} \sin ({\theta_2})\right)+{h_5} \nonumber\\
   & & \sin
   ({\phi_1}+{\phi_2}) \left(\frac{-\mu  \left(\frac{3 {dr}}{r}+{d\theta_2} \cot
   \left(\frac{{\theta_2}}{2}\right)\right) \sin ({\theta_2}) {d\phi_2}}{r^{3/2} \sqrt{\sin
   ^2\left(\frac{{\theta_2}}{2}\right)-\frac{\mu ^2}{r^3}}}+{d\phi_1}{d\phi_2} \sqrt{1-\left(1-\frac{2 \mu ^2 \csc
   ^2\left(\frac{{\theta_2}}{2}\right)}{r^3}\right)^2} \right)\nonumber\\
   & & +{h_4} \left({h_3}
   {d\theta_2}^2+{d\phi_2}^2 \sin ^2({\theta_2})\right).
\end{eqnarray}}
To start with, we will neglect $B_2$ of (\ref{three-form fluxes}) in the DBI action and include a $U(1)$ field strength $F=\partial_r A_t dr\wedge dt$ which would give the following:
\begin{eqnarray}
\label{eq:DBI-i}
& &  \sqrt{det\left(i^*g + F\right)} = \frac{\sqrt{\cal G}}{2 \sqrt{2} \pi ^{3/4}},
\end{eqnarray}
{\footnotesize
\begin{eqnarray}
\label{eq:DBI-i_i}
\pagebreak
& & {\rm where,}~~{\cal G}\equiv \frac{1}{{g_s} N \sqrt{\frac{{g_s} N}{r^4}}}\Biggl[r^4 \Biggl\{2 \sqrt{\pi } \sqrt{\frac{{g_s} N}{r^4}} \left(\frac{\mu ^2 \cot ^2\left(\frac{\theta_2}{2}\right)}{6 r^3 \left(\sin
   ^2\left(\frac{\theta_2}{2}\right)-\frac{\mu ^2}{r^3}\right)}-\frac{{h_5} \mu  \cos (\phi_1+\phi_2) \cot
   \left(\frac{\theta_2}{2}\right)}{r^{3/2} \sqrt{\sin ^2\left(\frac{\theta_2}{2}\right)-\frac{\mu ^2}{r^3}}}+\frac{1}{6}\right)\nonumber\\
   & & \times \Biggl\{2 \sqrt{\pi }
   \sqrt{\frac{{g_s} N}{r^4}} r^2 \left(-\frac{2 \mu ^4 \csc ^4\left(\frac{\theta_2}{2}\right)}{9 r^6}-\frac{2 \mu ^2 \csc
   ^2\left(\frac{\theta_2}{2}\right)}{9 r^3}+\frac{4}{9}\right)\nonumber\\
    & & \times\Biggl(2 \sqrt{\pi } \sqrt{\frac{{g_s} N}{r^4}} r^2
   \Biggl({F_{rt}}^2+\frac{\left(\frac{{r_h}^4}{r^4}-1\right) \left(\frac{3 \mu ^2}{2 r^3 \left(\sin ^2\left(\frac{\theta_2}{2}\right)-\frac{\mu
   ^2}{r^3}\right)}+\frac{2 \sqrt{\pi } \sqrt{\frac{{g_s} N}{r^4}}}{1-\frac{{r_h}^4}{r^4}}\right)}{2 \sqrt{\pi } \sqrt{\frac{{g_s} N}{r^4}}}\Biggr)\nonumber\\
& & \times   \left(\frac{\cos ^2(\theta_2)}{9}+\frac{2 \cos (\theta_2)}{9}+\frac{\sin ^2(\theta_2)}{6}+\frac{1}{9}\right)-\frac{18 {h_5}^2 \mu ^2 \sqrt{\pi
   } \sqrt{\frac{{g_s} N}{r^4}} \left(\frac{{r_h}^4}{r^4}-1\right) \sin ^2(\phi_1+\phi_2) \sin ^2(\theta_2)}{r \left(\sin
   ^2\left(\frac{\theta_2}{2}\right)-\frac{\mu ^2}{r^3}\right)}\Biggr)\nonumber\\
   & & -4 {g_s} N \pi  \Biggl({F_{rt}}^2+\frac{\left(\frac{{r_h}^4}{r^4}-1\right)
   \left(\frac{3 \mu ^2}{2 r^3 \left(\sin ^2\left(\frac{\theta_2}{2}\right)-\frac{\mu ^2}{r^3}\right)}+\frac{2 \sqrt{\pi } \sqrt{\frac{{g_s}
   N}{r^4}}}{1-\frac{{r_h}^4}{r^4}}\right)}{2 \sqrt{\pi } \sqrt{\frac{{g_s} N}{r^4}}}\Biggr)\nonumber\\
   & & \times \biggl[\frac{2}{9} (\cos (\theta_2)+1) \left(2-\frac{2 \mu
   ^2 \csc ^2\left(\frac{\theta_2}{2}\right)}{r^3}\right)+{h_5} \sqrt{1-\left(1-\frac{2 \mu ^2 \csc ^2\left(\frac{\theta_2}{2}\right)}{r^3}\right)^2}
   \sin (\phi_1+\phi_2)\nonumber\\
   & & -{h_5} \cos (\phi_1+\phi_2) \sqrt{1-\left(1-\frac{2 \mu ^2 \csc
   ^2\left(\frac{\theta_2}{2}\right)}{r^3}\right)^2} \sin (\theta_2)\biggr]^2\Biggr\} r^2\nonumber\\
   & & +\frac{1}{\sqrt{\sin
   ^2\left(\frac{\theta_2}{2}\right)-\frac{\mu ^2}{r^3}}}\biggl\{2 {h_5} \mu  \sqrt{\pi } \sqrt{\frac{{g_s} N}{r^4}}
   \cot \left(\frac{\theta_2}{2}\right) \sin (\phi_1+\phi_2) \sin (\theta_2) \times \nonumber\\
    & & \Biggl\{\frac{4 {g_s} {h_5}\mu N}{r^{3/2} \sqrt{\sin ^2\left(\frac{\theta_2}{2}\right)-\frac{\mu
   ^2}{r^3}}}\Biggl\{   \pi  \cot
   \left(\frac{\theta_2}{2}\right) \sin (\phi_1+\phi_2) \Biggl({F_{rt}}^2+\frac{\left(\frac{{r_h}^4}{r^4}-1\right) \left(\frac{3 \mu ^2}{2 r^3
   \left(\sin ^2\left(\frac{\theta_2}{2}\right)-\frac{\mu ^2}{r^3}\right)}+\frac{2 \sqrt{\pi } \sqrt{\frac{{g_s}
   N}{r^4}}}{1-\frac{{r_h}^4}{r^4}}\right)}{2 \sqrt{\pi } \sqrt{\frac{{g_s} N}{r^4}}}\Biggr)\nonumber\\
   & & \times\sin (\theta_2) \biggl(\frac{2}{9} (\cos (\theta_2)+1)
   \left(2-\frac{2 \mu ^2 \csc ^2\left(\frac{\theta_2}{2}\right)}{r^3}\right)+{h_5} \sqrt{1-\left(1-\frac{2 \mu ^2 \csc
   ^2\left(\frac{\theta_2}{2}\right)}{r^3}\right)^2} \sin (\phi_1+\phi_2)\nonumber\\
   & & -{h_5} \cos (\phi_1+\phi_2) \sqrt{1-\left(1-\frac{2 \mu ^2
   \csc ^2\left(\frac{\theta_2}{2}\right)}{r^3}\right)^2} \sin (\theta_2)\biggr)\Biggr\}\nonumber\\
   & & -\frac{1}{r^{3/2} \sqrt{\sin ^2\left(\frac{\theta_2}{2}\right)-\frac{\mu ^2}{r^3}}}\Biggl\{4 {g_s} {h_5} \mu  N \pi  \cot \left(\frac{\theta_2}{2}\right) \left(-\frac{2 \mu ^4 \csc ^4\left(\frac{\theta_2}{2}\right)}{9
   r^6}-\frac{2 \mu ^2 \csc ^2\left(\frac{\theta_2}{2}\right)}{9 r^3}+\frac{4}{9}\right)\nonumber\\
    & & \times\sin (\phi_1+\phi_2)
   \Biggl({F_{rt}}^2+\frac{\left(\frac{{r_h}^4}{r^4}-1\right) \left(\frac{3 \mu ^2}{2 r^3 \left(\sin ^2\left(\frac{\theta_2}{2}\right)-\frac{\mu
   ^2}{r^3}\right)}+\frac{2 \sqrt{\pi } \sqrt{\frac{{g_s} N}{r^4}}}{1-\frac{{r_h}^4}{r^4}}\right)}{2 \sqrt{\pi } \sqrt{\frac{{g_s} N}{r^4}}}\Biggr) \sin
   (\theta_2)\Biggr\}\Biggr\} \sqrt{r}\biggr\}\Biggr\}\Biggr].
\end{eqnarray}}
One sees that:
\begin{equation}
\label{mu0}
\int_0^{2\pi}\int_0^{2\pi}d\phi_1d\phi_2\int_0^\pi d\theta_2\left(\sqrt{det(i^*(g) + F)}\right)^{\mu^0}\sim r^3\sqrt{1 - F_{rt}^2},
\end{equation}
{\small
\begin{eqnarray}
\label{mu1}
& &{\rm and}~~ \left(\sqrt{det(i^*(g) + F)}\right)^{\mu}\nonumber\\
& & = \frac{1}{3 \sqrt{2}
   \sqrt{\left(1-F_{rt}^2\right) r^6 \cos ^2\left(\frac{\theta_2}{2}\right) (3 \cos (\theta_2)+1)}}\Biggl[\left(F_{rt}^2-1\right) h_5 r^{9/2} \Biggl(\sqrt{2} \cot ^3\left(\frac{\theta_2}{2}\right) \cos (\phi_1+\phi_2)\nonumber\\
   & & \times \Biggl(2 r^{3/2} \cos
   (2 \theta_2) \sqrt{-\frac{1}{r^3 (\cos (\theta_2)-1)}}+6 r^{3/2} \sqrt{-\frac{1}{r^3 (\cos (\theta_2)-1)}}\nonumber\\
   & & +\cos (\theta_2) \left(3
   \sqrt{1-\cos (\theta_2)}-8 r^{3/2} \sqrt{-\frac{1}{r^3 (\cos (\theta_2)-1)}}\right) +\sqrt{1-\cos (\theta_2)}\Biggr)\nonumber\\
   & & -8 r^{3/2} \cos
   ^2\left(\frac{\theta_2}{2}\right) \sin (\phi_1+\phi_2) \sqrt{\frac{\csc ^2\left(\frac{\theta_2}{2}\right)}{r^3}}\Biggr)\Biggr];\nonumber\\
& &  \int_0^{2\pi}\int_0^{2\pi}d\phi_1d\phi_2 \left(\sqrt{det(i^*(g) + F)}\right)^{\mu}=0.
\end{eqnarray}}
The finite part of the DBI action for $D7$ for the Ouyang embedding is hence:
\begin{equation}
\label{DBI_finite}
\int dr e^{-\phi(r)} r^3\sqrt{1 - F_{rt}^2},
\end{equation}
where $e^{-\phi(r)}=\frac{1}{g_s} - \frac{N_f}{8\pi}ln\left(r^6 + a^2 r^4\right) - \frac{Nf}{2\pi} ln\left(\mu r^{-\frac{3}{2}}\right)\stackrel{r>>a}{\longrightarrow}\frac{1}{g_s} - \frac{N_f}{2\pi}ln\mu$. We will be assuming that the embedding parameter is (real and) less than unity; if  $\mu\sim0$ then one assumes that $\mu\sim\epsilon^\alpha,\alpha>0$ so that using (\ref{limits_Dasguptaetal}) $g_s N_f ln\mu\rightarrow0$. The Euler-Lagrange eom corresponding to (\ref{DBI_finite}) is:
\begin{equation}
\label{DBI-i}
\partial_r\left(\frac{\left[\frac{1}{g_s} - \frac{N_f}{2\pi}ln\mu\right]r^3\partial_rA_t}{\sqrt{1- \left(\partial_rA_t\right)^2}}\right)=0.
\end{equation}
We therefore obtain:
\begin{equation}
\label{DBI-ii}
\partial_r A_t(r)=\frac{C e^{\phi(r)}}{\sqrt{C^2 e^{2\phi(r)} + r^6}},
\end{equation}
for large $r$ implying
\begin{equation}
\label{DBI-iii}
\mu_C=\int_{r_h}^\infty \frac{C e^{\phi(r)}}{\sqrt{C^2 e^{2\phi(r)} + r^6}}=\frac{C g_s\pi}{2\pi - g_s N_f ln\mu}\frac{\ _2F_1\left(\frac{1}{3},\frac{1}{2};\frac{4}{3};-\frac{C^2}{\left(\frac{1}{g_s} - \frac{N_f}{2\pi}ln\mu\right)^2r_h^6}\right)}{r_h^2},
\end{equation}
where $r_h=\pi \sqrt{4\pi g_s N} T$. From (\ref{DBI-iii}) one sees that:
\begin{equation}
\label{dmuoverdT-i}
\frac{\partial\mu_C}{\partial T}\Bigg|_{N_f; C=\left(\frac{1}{g_s} - \frac{N_f}{2\pi} ln\mu\right)\pi^3\left(4\pi g_s N\right)^{\frac{3}{2}}} \sim - \frac{1}{\sqrt{1 + T^6}}<0.
\end{equation}
In order to study the thermodynamical stability of the type IIB solution, we need to consider:
\begin{equation}
\label{eq:e_-phi_improved}
 e^{-\phi(r)}\approx \frac{1}{g_s} - \frac{N_f ln\mu}{2\pi} - \frac{N_f a^2}{8\pi r^2},
\end{equation}
along the Ouyang embedding. This implies that:
\begin{eqnarray}
\label{eq:mu_C_improved}
& & \mu_C = C \int_{r_h}^\infty dr\frac{1}{\left(\frac{1}{g_s} - \frac{N_f ln\mu}{2\pi} - \frac{N_f a^2}{8\pi r^2}\right)\sqrt{r^6 + \frac{C^2}{\left(\frac{1}{g_s} - \frac{N_f ln\mu}{2\pi} - \frac{N_f a^2}{8\pi r^2}\right)^2}}}\nonumber\\
& & \approx C\int_{r_h}^\infty dr\frac{g_s}{\sqrt{C^2g_s^2 + r^6}}
+ g_s N_f\int_{r_h}^\infty dr\frac {r^4\left(a^2 C g_s + 4 C g_s r^2 ln\mu\right)}{8\pi\left(C^2 g_s^2 + r^6\right)^{\frac{3}{2}}}\nonumber\\
& &  = C g_s\left[\frac{\, _2F_1\left(\frac{1}{3},\frac{1}{2};\frac{4}{3};-\frac{C^2 g_s^2}{{r_h}^6}\right)}{2 {r_h}^2}\right] + \nonumber\\
 && g_s N_f\left[ \frac{C g_s \left(\, _2F_1\left(\frac{2}{3},\frac{3}{2};\frac{5}{3};-\frac{C^2 g_s^2}{{r_h}^6}\right) a^2+8
   {r_h}^2 \, _2F_1\left(\frac{1}{3},\frac{3}{2};\frac{4}{3};-\frac{C^2 g_s^2}{r_h^6}\right) ln (\mu
   )\right)}{32 \pi  r_h^4}\right]. 
\end{eqnarray}
From (\ref{eq:mu_C_improved}), one sees:
\begin{equation}
\label{dmuoverdT-ii}
\frac{\partial\mu_C}{\partial T}\Biggr|_{N_f,a = f r_h:f<<<1; C=\frac{\pi^3\left(4\pi g_s N\right)^{\frac{3}{2}}}{g_s}}\sim - \frac{1}{8\pi}\frac{T^6(f^2 + 4 ln\mu)}{\left(1 + T^6\right)^{\frac{3}{2}}}
+ f^2 g_s N_f\frac{\, _2F_1\left(\frac{2}{3},\frac{3}{2};\frac{5}{3}; - \frac{1}{T^6}\right)}{16\pi T^3}<0.
\end{equation}
So, from (\ref{dmuoverdT-i}) and (\ref{dmuoverdT-ii}) it is clear that  $\frac{\partial\mu_C}{\partial T}\Biggr|_{N_f} = - \frac{\partial S}{\partial N_f}\Biggr|_T < 0$.
Apart from $C_v>0$, thermodynamic stability requires: $\frac{\partial\mu_C}{\partial N_f}\left.\right|_{T}>0$, which using (\ref{eq:mu_C_improved}) for $C>0, \mu = \lim_{\epsilon\rightarrow0^+}1 - \epsilon$ is satisfied!

Let us now analyse the inclusion of the  NS-NS  $B_2$ of (\ref{three-form fluxes}) in the DBI action. From (\ref{three-form fluxes}), using the Ouyang embedding (implying for a real $\mu$: $d\psi=d\phi_1 + d\phi_2, d\theta_1 = - \tan\left(\frac{\theta_1}{2}\right)\left(3 \frac{dr}{r} + \cot\left(\frac{\theta_2}{2}\right)d\theta_2\right)$):
\begin{eqnarray}
\label{B_Ouyang}
& & B_2 = - \frac{3}{r}\tan\frac{\theta_1}{2}\left(B_{\theta_1\phi_1} + B_{\theta_1\psi}\right)dr\wedge d\phi_1 +
 \left[B_{\theta_2\phi_1}- \tan\frac{\theta_1}{2}\cot\frac{\theta_2}{2}\left(B_{\theta_1\phi_1} + B_{\theta_1\psi}\right)\right] \nonumber\\
 & &d\theta_2\wedge d\phi_1
 - \frac{3}{r}\tan\frac{\theta_1}{2}\left(B_{\theta_1\phi_2} + B_{\theta_1\psi}\right)dr\wedge d\phi_2
+\left[ B_{\theta_2\phi_2} - \tan\frac{\theta_1}{2}\cot\frac{\theta_2}{2}\left(B_{\theta_1\phi_2} + B_{\theta_1\psi}\right)\right]\nonumber\\
&& d\theta_2\wedge d\phi_2.\end{eqnarray}
Working in the $g_sM\rightarrow \infty, g_sM\rightarrow \infty, g_s\rightarrow0, \frac{g_sM^2}{N}\rightarrow0$ limit of \cite{metrics}, one obtains:
{\small
\begin{eqnarray}
\label{DBI_B-i}
& & \left(\sqrt{det\left(i^*(g + B) + F\right)}\right)^{\mu^0}=\nonumber\\
& &  \frac{\sqrt{\frac{\left({F_{rt}}^2-1\right) r^2 \left(\sin ^2(\theta_2) \left(81(g_s M)^2 \left(r^2-3 a^2\right)^2 ln ^2(r)+\pi  r^4 {g_s N}\right)-2 \pi  r^4 {g_s N} \cos ^2(\theta_2)-4 \pi  r^4 {g_s N} \cos (\theta_2)-2 \pi  r^4
   {g_s N}\right)}{{g_s N}}}}{9 \sqrt{\pi }};\nonumber\\
&&{\rm and}~~ \int_0^{2\pi}\int_0^{2\pi}d\phi_1d\phi_2\int_0^\pi d\theta_2\left(\sqrt{det\left(i^*(g + B) + F\right)}\right)^{\mu^0}
    \sim r^3\sqrt{1 - F_{rt}^2} + {\cal O}\left(\sqrt{\frac{g_sM^2}{N}}\right); 
    \nonumber\\
    \\
   & &  \left(\sqrt{det\left(i^*(g + B) + F\right)}\right)^{\mu}=\nonumber\\
    & & \frac{\left(F_{rt}^2-1\right) {h_5} \sqrt{r}}{12 \sqrt{\frac{{g_sN}}{r^4}}
   (9\sqrt{\pi}(\sqrt{det(i^*(g+B) + F)})^{\mu^0})} \times \nonumber\\
   & &  \Biggl\{\frac{\sqrt{\pi } {g_sN} \sqrt{\sin ^2\left(\frac{\theta_2}{2}\right)} \left(10 \cos
   \left(\frac{\theta_2}{2}\right)+3 \left(\cos \left(\frac{3 \theta_2}{2}\right)+\cos \left(\frac{5 \theta_2}{2}\right)\right)\right) \csc
   ^3\left(\frac{\theta_2}{2}\right) \cos (\phi_1+\phi_2)}{\sqrt{\frac{{g_sN}}{r^4}}} \nonumber\\
   & &   -2 \Biggl[2 \sqrt{2} \cos \left(\frac{\theta_2}{2}\right)
   \cot \left(\frac{\theta_2}{2}\right) \sin (\phi_1+\phi_2) \Biggl(9 {g_sM} \left(3 a^2-r^2\right) ln (r) \sqrt{1-\cos (\theta_2)}\cos
   \left(\frac{3 \theta_2}{2}\right)\nonumber\\
   & & \left(9 a^4 {g_s}-3 a^2 {g_s} r^2-1\right) +27 {g_sM} \left(3 a^2-r^2\right) ln (r) \cos
   \left(\frac{\theta_2}{2}\right) \sqrt{1-\cos (\theta_2)}   \nonumber\\
   & &   +8 \sqrt{\pi } r^{11/2}
   \sqrt{\frac{{g_sN}}{r^4}} \sin \left(\frac{\theta_2}{2}\right) \sqrt{-\frac{1}{r^3 (\cos (\theta_2)-1)}}\Biggr)-\nonumber\\
   & &  \frac{\sqrt{\pi } {g_sN} \sin (2
   (\phi_1+\phi_2)) \csc (\phi_1+\phi_2) \Bigl(2   \sin ^3(\theta_2)  \csc
   ^2\left(\frac{\theta_2}{3}\right)  +  \sin (2 \theta_2) \csc
   ^3\left(\frac{\theta_2}{2}\right)\Bigr)}{\sqrt{\frac{{g_sN}}{r^4}}}\Biggr]\Biggr\};\nonumber\\
   & & {\rm and}~~\int_0^{2\pi}\int_0^{2\pi}d\phi_1d\phi_2\left(\sqrt{det\left(i^*(g + B) + F\right)}\right)^{\mu}=0. \end{eqnarray}
\begin{eqnarray}
\label{eq:DBI_B-iii}
&&{\rm Hence,}~~\int_0^{\pi}d\theta_2\int_0^{2\pi}\int_0^{2\pi}d\phi_1d\phi_2\left(\sqrt{det\left(i^*(g + B) + F\right)}\right)\sim \left(r^3\sqrt{1 - F_{rt}^2}+ {\cal O}\left(\sqrt{\frac{g_sM^2}{N}}\right)\right) \nonumber\\
&& + {\cal O}(\mu^2).
\end{eqnarray}}
such that in the $g_sM\rightarrow\infty,g_sN\rightarrow\infty$ limit, the ${\cal O}(\mu^0)$ results
and consequences thereafter, for $B=0$ and $B\neq0$, match.

For {\bf 5.1} and {\bf 5.2} one would need to have a very small $\mu_C$, which we see from
(\ref{DBI-iii}) and (\ref{eq:mu_C_improved}), can be arranged by tuning the $C$ to an appropriately small
value, e.g., for $C=0.001$, $\mu_C=0.05, 0.0006$ respectively for the two scalings of $g_s, M, N$
referred to in the abstract and use in and after Sec. {\bf 4}. The choice of $C$ in (\ref{dmuoverdT-i})
and (\ref{dmuoverdT-ii}) was only for convenience.
\vskip -0.5in
\section{The Local type IIA SYZ Dual and Local M-Theory Uplift}
The starting metric in the type IIB theory has the following components
\begin{eqnarray}
&& ds^2 =
g_{\mu \nu}dx^\mu ~dx^\nu + g_{x\mu} dx ~dx^\mu +  g_{y \mu} dy~
dx^\mu +  g_{z\mu} dz ~ dx^\mu +  g_{xy} dx ~dy
  + g_{xz} dx ~dz   \nonumber\\
 &&+  g_{zy} dz ~ dy+  g_{xx} dx^2  +  g_{yy}dy^2
 +  g_{zz}~dz^2,
  \end{eqnarray}
where $\mu, \nu \neq x, y, z$.
As shown in \cite{SYZ3_Ts}, the form of the metric of the mirror manifold after performing three T-dualities, first along $x$, then along $y$ and finally along $z$:
\begin{eqnarray}
\label{mirror_metric}
& & ds^2 =
\left( G_{\mu\nu} - {G_{z\mu}G_{z\nu} - {\cal B}_{z\mu} {\cal
B}_{z\nu} \over G_{zz}} \right) dx^\mu~dx^\nu +2 \left( G_{x\nu} -
{G_{zx}G_{z\nu} - {\cal B}_{zx} {\cal B}_{z\nu}
 \over G_{zz}} \right) dx~dx^\mu  \nonumber\\
& &  + 2\left( G_{y\nu} - {G_{zy}G_{z\nu} - {\cal B}_{zy} {\cal B}_{z\nu}
 \over G_{zz}}\right) dy~dx^\nu +
2\left( G_{xy} - {G_{zx}G_{zy} - {\cal B}_{zx} {\cal B}_{zy} \over
G_{zz}}\right) dx~dy  \nonumber\\
& &  + {dz^2\over G_{zz}} + 2{{\cal
B}_{\mu z} \over G_{zz}} dx^\mu~dz + 2{{\cal B}_{xz} \over G_{zz}}
dx~dz + 2{{\cal B}_{yz} \over G_{zz}} dy~dz \nonumber\\
& & +  \left( G_{xx}
- {G^2_{zx} - {\cal B}^2_{zx} \over G_{zz}} \right) dx^2 + \left(
G_{yy} - {G^2_{zy} - {\cal B}^2_{zy} \over G_{zz}} \right)
dy^2.
\end{eqnarray}
To implement mirror symmetry a al SYZ prescription, one needs to ensure that the base of the local $T^3$-
fibration is large. Near $\theta_1=\theta_2=0$ we will show that it is possible to obtain a large base for which $f_{1,2}(\theta_{1,2})>>a$ (small resolution factor, i.e. $a<<1$). The guiding principle is that one requires that the metric obtained after SYZ-mirror transformation applied to the resolved warped deformed conifold is like a warped resolved conifold at least locally, then $G^{IIA}_{\theta_1\theta_2}$ needs to vanish. We will {\bf first} implement $\frac{g_s^2M}{N_f}\rightarrow0$
(common to both limits (\ref{limits_Dasguptaetal-i}) and (\ref{limits_Dasguptaetal-ii})) for simplifying the type IIA metric components which means that we will set $h(r,\theta_1,\theta_2)=\frac{4\pi g_s N}{r^4}$; this yields the type IIA metric components enumerated in (\ref{metric-mirror}). Interestingly, if one assumes that the local uplift is also
valid globally, then we show that the same guiding principle is also applicable at  $\theta_1=\theta_2=\frac{\pi}{2}$, in fact more readily.
Assuming one has found appropriate $f_i(\theta_i)$, after T-dualizing along $x, y, z$, equation (\ref{mirror_metric}) yields (\ref{G_munu}) - (\ref{Gzx}), which yields the components of the type IIA mirror metric as given in (\ref{metric-mirror}).
\\Along $r=a\sqrt{3},\theta_1\rightarrow0,\theta_2\rightarrow m\theta;\theta\rightarrow0$ where $m\sim{\cal O}(1)$, (\ref{metric-mirror}) yields:
{\small
\begin{eqnarray}
\label{metric-mirror-iii}
& & G^{IIA}_{\theta_1\theta_1},G^{IIA}_{\theta_2\theta_2}\sim\sqrt{g_sN},\nonumber\\
& & G^{IIA}_{\theta_1\theta_2}\sim \lim_{g_sN\to\infty,\theta\to0}h_5\sqrt{g_sN}\theta^2\left[f_1(0)f_2(0)
- \frac{2 }{m\theta^2}\right],\nonumber\\
& & {\rm which\ vanishes\ for}\ f_1(0)\sim f_2(0)\sim\lim_{\theta\to0}\frac{\sqrt{2}}{\sqrt{m}\theta}>>a\ {\rm for\ small\ unconstrained}\ h_5;\nonumber\\
& & G^{IIA}_{\phi_1\theta_1}\sim\lim_{\theta\to0}\frac{g_s^2MN_f  ln a}{h_5^2(g_sN)^{\frac{1}{4}}\theta^4}\sim0, G^{IIA}_{\phi_1\theta_2}=0,  G^{IIA}_{\psi\theta_1}\sim \lim_{\theta\to0}\frac{g_s^2MN_f  ln a}{h_5^2(g_sN)^{\frac{1}{4}}\theta^3}\sim0,  G^{IIA}_{\psi\theta_2}=0,\nonumber\\
& & G^{IIA}_{\phi_2\theta_1}\sim\lim_{\theta\to0}\frac{g_s^2MN_f  ln a}{(g_sN)^{\frac{1}{4}}\theta^2}\sim0,  G^{IIA}_{\phi_2\theta_2}\sim(g_sN)^{\frac{1}{4}}\theta f_2(0)\sim(g_sN)^{\frac{1}{4}}>>1; G^{IIA}_{\phi_1\phi_1}\sim G^{IIA}_{\phi_2\phi_2}\sim\theta^2;\nonumber\\
& & G^{IIA}_{\psi\psi}\sim{\cal O}(1); G^{IIA}_{\phi_2\psi}\sim{\cal O}(1);\nonumber\\
& & G^{IIA}_{\phi_1\psi}\sim\frac{-243 \sqrt{6} h_5^2 m^2 \theta ^5+162 h_5^2 m^2 \theta ^4+18 \sqrt{6} m (9 h_5+m) \theta ^3-72 h_5
   m \theta ^2-12 \sqrt{6} \theta +8}{1458 h_5^2 m^2 \theta ^3},\nonumber\\
& & G^{IIA}_{\phi_1\phi_2}\sim\frac{1}{27 \left(9 h_5^2-1\right) m^2 \theta ^2 \left(3
   \theta ^2+2\right) \left(-27 h_5^2 \theta ^2 m^2+\left(3 \theta ^2+2\right) m^2+12 h_5 m+2\right)}\nonumber\\
 & &\times\Bigl[-243 h_5^3 m^3 \left(-27 \theta ^3+3 \sqrt{6} \theta ^2-36 \theta +2 \sqrt{6}\right) \theta ^4-27 h_5^2 m^2
   \Bigl(9 \sqrt{6} \theta ^4+54 \theta ^3-6 \sqrt{6} \theta ^2
   \nonumber\\
   && +144 \theta -8 \sqrt{6}\Bigr)\theta ^2   -12 \sqrt{6} \theta ^2+72
   \theta -3 h_5 m \Bigl(243 m^2 \theta ^7+162 \left(2 m^2+1\right) \theta ^5-36 \sqrt{6} \theta ^4\nonumber\\
   && +108 \left(m^2+3\right)
   \theta ^3-12 \sqrt{6} \theta ^2-72 \theta +8 \sqrt{6}\Bigr) -8 \sqrt{6}\Bigr].
\end{eqnarray}}
If $\theta\rightarrow0,h_5\rightarrow0, m\sim{\cal O}(1)$ such that $-72 h_5
   m \theta ^2-12 \sqrt{6} \theta +8=0$, then $\theta\rightarrow\frac{-\sqrt{6} + \sqrt{2}\sqrt{3 + 8 h_5 m}}{12 h_5 m}\\
   =\frac{1}{3}\sqrt{\frac{2}{3}} - \frac{2}{9}\sqrt{\frac{2}{3}} h_5 m + {\cal O}
   \left((h_5 m)^2\right)$. So, if $h_5m<<1$ then $\theta\sim0.3<1$. The understanding now is that $\theta\rightarrow0$ by dividing out the same by a large number. So, the numerator of $G^{IIA}_{\phi_1\psi}$ will go as ${\cal O}(\theta^3)$ implying that $G^{IIA}_{\phi_1\psi}$ will be finite as $\theta_{1,2}\rightarrow0$. The terms up to ${\cal O}(\theta=\frac{-\sqrt{6} + \sqrt{2}\sqrt{3 + 8 h_5 m}}{12 h_5 m})$ in the numerator of $G^{IIA}_{\phi_1\phi_2}$ will be given by $16\sqrt{\frac{2}{3}} h_5 m
   + {\cal O}\left((h_5 m)^2\right)$, which for $h_5 m<<1$ is negligible. So, $G^{IIA}_{\phi_1\phi_2}$ is finite as $\theta_1\rightarrow0,\theta_2\rightarrow0$. Near $r=a \sqrt{3}, \theta_1=\theta_2=\frac{\pi}{2}$, using (\ref{metric-mirror}):
{\small
\begin{eqnarray}
\label{metric-mirror-ii}
& & G^{IIA}_{\theta_1\theta_1},G^{IIA}_{\theta_2\theta_2}\sim\sqrt{g_sN}, G^{IIA}_{\phi_1\phi_1/\phi_2\phi_2}=\frac{1}{1 - 9 h_5^2}\nonumber\\
& & G^{IIA}_{\theta_1\theta_2}\sim h_5\sqrt{g_sN}\left(f_1\left(\frac{\pi}{2}\right)f_2\left(\frac{\pi}{2}\right) - 2\right)=0\ {\rm for}\ f_1\left(\frac{\pi}{2}\right)=f_2\left(\frac{\pi}{2}\right)=\sqrt{2}>>a,\nonumber\\
& & G^{IIA}_{\phi_1\theta_1}\sim\frac{h_5^2g_s^2MN_f}{(g_sN)^{\frac{1}{4}}}\sim0,  G^{IIA}_{\phi_1\theta_2}=0,  G^{IIA}_{\psi\theta_1}\sim\frac{g_s^2MN_f}{(g_sN)^{\frac{1}{4}}}\sim0,  G^{IIA}_{\psi\theta_2}=0, G^{IIA}_{\phi_2\theta_1}=0,\nonumber\\
& & G^{IIA}_{\phi_2\theta_2}=0,  G^{IIA}_{\psi\psi}=\frac{1}{9}, G^{IIA}_{\phi_1\psi}\sim h_5^2, G^{IIA}_{\phi_2\psi}=0, G^{IIA}_{\phi_1\phi_2}\sim h_5\sim0, G^{IIA}_{\phi_1\phi_2}\sim h_5\sim0.
\end{eqnarray}
}
We see from (\ref{metric-mirror-ii}) that the metric locally will look like a resolved conifold metric along $\theta_1=\theta_2=\frac{\pi}{2}$ and $r=a\sqrt{3}$.

We will now discuss how one obtains, locally, a one-form type IIA potential from the triple T-dual (along $x, y, z$) of the type IIB $F_{1,3,5}$. From (\ref{three-form fluxes}), one sees that the following are the non-zero components with respect to the T-duality coordinates $(x,y,z)$ of $F_3$: $ F_{z\theta_i x}, F_{z\theta_i y}, F_{x\theta_i y}, F_{rzy}, F_{rzx}, F_{rxy}, F_{\theta_i xy}$.

Using the T-duality rules for RR field strengths of \cite{T-dual_Hassan}, one sees that:
{\small \begin{eqnarray}
\label{Ftildetildetildeythetai}
& & \tilde{\tilde{\tilde{F}}}_{y\theta_i/r}=\Biggl[- 3 F_{z\theta_i/rx} - 6\biggl( \biggl\{F_z - 2 g^{-1}_{xx}\left(g_{xz}F_z - g_{xx}F_z\right)\biggr\}b_{x\theta_i/r}\biggr)\nonumber\\
& & - 2 g^{-1}_{yy}\Biggl\{\tilde{g}_{yz}\biggl[F_{y\theta_i/rx} - 2 \biggl[F_y - 2 g^{-1}_{xx}\left(g_{xy}F_x - g_{xx}F_y\right)\biggr]b_{x\theta_i/r}\biggr]\Biggr\}\nonumber\\
& & -2\Biggl(- \tilde{\tilde{B}}_{zy}\Biggl[F_{y\theta_i/rx} + 2\left(F_y - 2 g^{-1}_{xx}\left(g_{xy} F_x - g_{xx}F_y\right)\right)b_{x\theta_i/r}\Biggr]\Biggr)\Biggr],
\end{eqnarray}
where $\tilde{\tilde{B}}_{zy}=\tilde{g}^{-1}_{yy}\tilde{g}_{yz}=\left(g_{yy} - g^{-1}_{xx}g^2_{xy}\right)^{-1}
\left(g_{yz} - g^{-1}_{xx}g_{xy}g_{xz}\right)$. }The components of (\ref{Ftildetildetildeythetai}) are explicitly given in an appendix of \cite{MQGP}.Now,
\pagebreak
\begin{eqnarray}
\label{Ftildetildetildezthetair}
& & \tilde{\tilde{\tilde{F}}}_{z\theta_i/r} = -\Biggl[F_{y\theta_i/rx} + 2\left(F_y - 2g^{-1}_{xx}\left(g_{xy}F_x - g_{xx}F_y\right)b_{x\theta_i/r}\right)\Biggr],
\end{eqnarray}
Similarly,
{\small \begin{eqnarray}
\label{eq:Ftildetildetildexthetair}
& & \tilde{\tilde{\tilde{F}}}_{x\theta_i/r}=-\left(F_{yz\theta_i/r} - 3 g^{-1}_{xx}\left\{g_{xy}F_{xz\theta_i/r} + g_{xx}F_{zy\theta_i/r} + g_{xz}F_{yx\theta_i/r}\right\}\right)\nonumber\\
& & \hskip-0.4in- 3 \Biggl[-\tilde{B}_{yx}\Biggl\{-F_{z\theta_i/rx} - 2\left(F_z - 2g^{-1}_{xx}\left(g_{xz}F_x - g_{xx}F_z\right)\right)b_{x\theta_i/r} - 2\tilde{g}_{yy}^{-1}\Biggl(\tilde{g}_{yz}\Bigl\{F_{xy\theta_i/r} + 2 b_{x\theta_i/r}\nonumber\\
& & \left[F_y - g^{-1}_{xx} g_{yy}F_x\right]\Bigr\} + \tilde{g}_{yy}\left\{F_{z\theta_i/rx} - 2 \left[F_z - 2 g^{-1}_{xx}\left(g_{xz}F_x - g_{xx}F_z\right)\right]b_{x\theta_i/r}\right\} \Biggr)\Biggr\}\Biggr]
\end{eqnarray}}
Using the above, the exact expressions are given in \cite{MQGP}.
We therefore can construct the following gauge field one-form in the local limit:
\begin{eqnarray}
\label{A}
& & A^{F_3} = \Biggl[\tilde{\tilde{\tilde{F}}}_{xr}x dr + \tilde{\tilde{\tilde{F}}}_{x\theta_1}x d\theta_1 + \tilde{\tilde{\tilde{F}}}_{x\theta_2}x d\theta_2 + \tilde{\tilde{\tilde{F}}}_{y\theta_1}y d\theta_1 + \tilde{\tilde{\tilde{F}}}_{y\theta_2}y d\theta_2 +  \tilde{\tilde{\tilde{F}}}_{z\theta_2}z d\theta_2\nonumber\\
 & &   +  \tilde{\tilde{\tilde{F}}}_{z\theta_1}z d\theta_1 +   \tilde{\tilde{\tilde{F}}}_{zr}z dr
 + \tilde{\tilde{\tilde{F}}}_{yr}y dr\Biggr]\left(\theta_{1,2}\rightarrow\langle\theta_{1,2}\rangle,\phi_{1,2}\rightarrow\langle \phi_{1,2}\rangle,\psi\rightarrow\right\langle\psi\rangle,r\rightarrow\langle r\rangle).
\end{eqnarray}
The two-form field strength obtained from three T-dualities to $F_1$ can be obtained via application of T-duality rules of \cite{T-dual_Hassan}:
\begin{eqnarray}
\label{eq:Fitldetildetildebeta1beta2}
& & \tilde{\tilde{\tilde{F}}}_{zx} = - 2 \Biggl[F_y - g^{-1}_{xx}\left\{g_{xy} F_x - g_{xx}F_y\right\}\Biggr];\nonumber\\
& & \tilde{\tilde{\tilde{F}}}_{yx} = - 3 \left[F_z - g^{-1}_{xx}\left(g_{xz}F_x - g_{xx}F_z\right)\right]
-2\tilde{g}^{-1}_{yy}\left[- \tilde{g}_{yz}\left\{F_y - g^{-1}_{xx}\left(g_{xy}F_x - g_{xx}F_y\right)\right\}\right]\nonumber\\
& & - 4 \tilde{\tilde{b}}_{zy}\Biggl[F_y - g^{-1}_{xx}\left(g_{xy}F_x - g_{xx}F_y\right)\Biggr];  \tilde{\tilde{\tilde{F}}}_{yz}=0.
\end{eqnarray}
where $\tilde{\tilde{b}}_{yz}= - \frac{g_{yz}}{g_{yy} - \frac{g_{xy}^2}{g_{xx}}}$.  The exact expressions of the same are given in \cite{MQGP}.
One therefore obtains the following two-form field strength in the mirror type IIA:
\begin{eqnarray}
\label{F2_IIA}
& & \tilde{\tilde{\tilde{F}}}_2 = \tilde{\tilde{\tilde{F}}}_{xr}dx\wedge dr + \tilde{\tilde{\tilde{F}}}_{x\theta_1}dx\wedge d\theta_1 + \tilde{\tilde{\tilde{F}}}_{x\theta_2}dx\wedge d\theta_2 + \tilde{\tilde{\tilde{F}}}_{y\theta_1}dy\wedge d\theta_1 + \tilde{\tilde{\tilde{F}}}_{y\theta_2}dy\wedge d\theta_2\nonumber\\
 & &  +  \tilde{\tilde{\tilde{F}}}_{z\theta_2}dz\wedge d\theta_2 +  \tilde{\tilde{\tilde{F}}}_{z\theta_1}dz\wedge d\theta_1 +   \tilde{\tilde{\tilde{F}}}_{zr}dz\wedge dr
 + \tilde{\tilde{\tilde{F}}}_{yr}dy\wedge dr.
\end{eqnarray}
Hence,
 {\small  \begin{equation}  
  A^{F_1}=\left(\tilde{\tilde{\tilde {F}}}_{1yx} y dx + \tilde{\tilde{\tilde{F}}}_{1zx} z dx\right)\left(\theta_{1,2}\rightarrow\langle\theta_{1,2}\rangle,\phi_{1,2}\rightarrow\langle \phi_{1,2}\rangle,\psi\rightarrow\right\langle\psi\rangle,r\rightarrow\langle r\rangle).
  \end{equation}}
 The two-form field strength components obtained from three T-dualities applied to the self-dual five-form field strength are, using \cite{T-dual_Hassan}, given via:
  {\small \begin{eqnarray}
   \label{eq:HodgedualFtildetildetildebeta1beta2}
  & & \widetilde{\widetilde{\widetilde{(*F_5)}}}_{\beta_1\beta_2} = - \Biggl[ - (*F_5)_{xyz\beta_1\beta_2} - 4 \Biggl(- b_{x\beta_1}\left\{F_{yz\beta_2} - 2 g_{xx}^{-1}\left[g_{xy}F_{xz\beta_2} - g_{xx}F_{z\beta_2y} + g_{xz}F_{\beta_2yx}\right]\right\}\nonumber\\
   & & - b_{x\beta_2}\left\{F_{yz\beta_1} - 2 g_{xx}^{-1}\biggl[g_{xy}F_{xz\beta_1} - g_{xx}F_{z\beta_1y} + g_{xz}F_{\beta_1yx}\biggr]\right\}\Biggr)\nonumber\\
  & & - 3 \Biggl(\tilde{b}_{y\beta_1}\left[- 3 F_{x\beta_2z} + 6 b_{x\beta_2}\left\{F_z - g^{-1}_{xx}\left(g_{xz}F_x - g_{xx}F_z\right)\right\} - 2 \tilde{g}_{yy}^{-1}\left(\tilde{g}_{yz}g^{-1}_{xx}g_{x\beta_2}F_x\right)\right]\nonumber\\
  & & - \tilde{b}_{y\beta_2}\Biggl[- 3 F_{x\beta_1z} + 6 b_{x\beta_1}\left\{F_z - g^{-1}_{xx}(g_{xz}F_x - g_{xx}F_z)\right\} - 2 \tilde{g}^{-1}_{yy}\left(\tilde{g}_{yz}g^{-1}_{xx}g_{x\beta_2}F_x\right)\Biggr]\Biggr)\Biggr]\nonumber\\
  & & - 2 \left(- \tilde{\tilde{b}}_{z\beta_1}\left[- 2 g^{-1}_{xx}g_{x\beta_2}F_x\right] - b_{z\beta_2}\left[- 2 g^{-1}_{xx}g_{x\beta_1}F_x\right]\right),
   \end{eqnarray}
 where $*F_5=F_5, \beta_{1/2}\equiv r,\theta_{1/2}$, and $
 \tilde{b}_{y\beta_i} = b_{y\beta_i} - \frac{g_{xy}b_{x\beta_i}}{g_{xx}}$.}
 
Therefore using (\ref{eq:HodgedualFtildetildetildebeta1beta2}),  we obtain the form of  triple T-dual of self-dual five-form field in \cite{MQGP}. So,
{\small\begin{eqnarray}
& &  A^{F_5}=\left( \tilde{\tilde{\tilde{F}}}_{5r\theta_1}r d\theta_1 + \tilde{\tilde{\tilde{F}}}_{5\theta_1\theta_2}\theta_1 d\theta_2 + \tilde{\tilde{\tilde{F}}}_{5r\theta_2}r d\theta_2\right)\left(\theta_{1,2}\rightarrow\langle\theta_{1,2}\rangle,\phi_{1,2}\rightarrow\langle \phi_{1,2}\rangle,\psi\rightarrow\langle\psi\rangle,r\rightarrow\langle r\rangle\right);\nonumber\\
& &\tilde{\tilde{\tilde{\phi}}}=\frac{1}{2}\left[2\phi - ln\left(g_{xx}\left\{g_{yy} - \frac{g_{xy}^2}{g_{xx}}\right\}\left\{g_{zz}-\frac{g_{xz}^2}{g_{xx}} - \frac{\left(g_{yz} - \frac{g_{xy}g_{xz}}{g_{xx}}\right)^2}{g_{yy} - \frac{g_{xy}^2}{g_{xx}}}\right\}\right)\right].
\end{eqnarray}}
Therefore, the uplifted $M-$theory metric is given by:
\begin{eqnarray}
\label{Mtheory met}
& &  ds^2_{11} = e^{-\frac{2\tilde{\tilde{\tilde{\phi}}}}{3}} \Biggl[
\frac{1}{\sqrt{h\left(r,\theta_1,\theta_2\right)}}\Bigl(-g_1 dt^2+dx_1^2+dx_2^2+dx_3^2\Bigr)+\sqrt{h\left(r,\theta_1,\theta_2\right)}\Bigl(g_2^{-1}dr^2\Bigr)\nonumber\\
&& +  ds^2_{{\rm equation}\ (\ref{mirror_metric})}\Biggr]\  + e^{\frac{4\tilde{\tilde{\tilde{\phi}}}}{3}}\Bigl(dx_{11} + A^{F_1}+A^{F_3}+A^{F_5}\Bigr)^2.
\end{eqnarray}
The horizon area, using (\ref{metric}) or the black $M3$-brane metric (\ref{Mtheory met}), assuming as in \cite{Gubser+Klebanov+Peet} that the world-volume coordinates $x_{1,2,3}$ are wrapped around a $T^3$ of a very large radius, will be proportional to:
\begin{equation}
\label{horizon_area}
{\rm Horizon\ area}\sim \left(\frac{1}{h^{\frac{1}{4}}(r_h)}\right)^3\left(h^{\frac{1}{4}}(r_h)\right)^5r_h^5\ \stackrel{h\sim\frac{4\pi g_sN}{r^4}}{\longrightarrow\ }\ r_h^3.
\end{equation}
In order to check for the supersymmetry of the $M3$-brane solution of (\ref{Mtheory met}) and to get the the explicit dependence of 11-dimensional action on parameters ${g_s},M/M_{\rm eff}, N/N_{\rm eff}$, we have first simplified metric components in the (i) weak($g_s$) coupling - large t'Hooft couplings limit: (\ref{limits_Dasguptaetal-i})
as well  as the (ii) `MQGP limit': (\ref{limits_Dasguptaetal-ii})
.  The simplified expressions for all non-zero 11-dimensional  metric components  in
either limit using (\ref{metric-mirror}) (which too is valid in either limit)
$\forall\theta_{1,2}\in[\epsilon_{\theta_{1,2}},\pi-\epsilon_{\theta_{1,2}}]$, assuming that globally one can replace $x,y,z$ respectively by\\ $\sqrt{h_2}\left(4\pi g_sN\right)^{\frac{1}{4}}sin\theta_1 sin\phi_1, \sqrt{h_4}\left(4\pi g_sN\right)^{\frac{1}{4}}sin\theta_2 sin\phi_2,2\sqrt{h_1}\left(4\pi g_s N\right)^{\frac{1}{4}}sin\frac{\psi}{2}$,  are as follows\footnote{Remark: For a chosen scaling, ${r_{\Lambda}\rightarrow \epsilon^{-a}}$, $M/M_{\rm eff}\sim {\epsilon^{-\frac{3}{2}}d}$, $N/N_{\rm eff}\sim{\epsilon^{-19d}} $, $g_s \sim {\epsilon^{d}} $, we observe that metric components ${G^{\cal M}_{00},G^{\cal M}_{11},G^{\cal M}_{22},G^{\cal M}_{33}}$ asymptotically approach the flat metric on ${\bf R}^{1,3}$ as $r_{\Lambda}\rightarrow \infty$ for $ a={\frac{25}{6}} $.}:
{\footnotesize
\begin{eqnarray}
\label{eq:simplifiedmetriccomponents}
&(i)& G^{\cal M}_{0 0}\sim  -\frac{3^{2/3} r^2 \left(1-\frac{r_h^4}{r^4}\right)}{2 {g_s}^{7/6} \sqrt{N} \sqrt{\pi }},~
(ii)~
G^{\cal M}_{1 1}\sim \frac{3^{2/3} r^2}{2 {g_s}^{7/6} \sqrt{N} \sqrt{\pi }},~(iii)~
G^{\cal M}_{2 2}\sim \frac{3^{2/3} r^2}{2 {g_s}^{7/6} \sqrt{N} \sqrt{\pi }},\nonumber\\
&(iv)&
G^{\cal M}_{3 3}\sim \frac{3^{2/3} r^2}{2 {g_s}^{7/6} \sqrt{N} \sqrt{\pi }},~(v)~
G^{\cal M}_{r r}\sim \frac{2~3^{2/3} \left(1-\frac{r_h^4}{r^4}\right)^{-1} \sqrt{N} \sqrt{\pi }}{\sqrt[6]{{g_s}} r^2},\nonumber\\
&(vi)&
G^{\cal M}_{x r}\sim  \frac{1}{\sqrt[3]{3} \pi ^2 (\cos (2 {\theta_1})-5)^2}\nonumber\\
& & \times\Bigl(4 {g_s}^{4/3} {N_f}^2 {\sin (\phi_1)} \sin (\frac{\psi}{2})  \sin ({\theta_1}) \left(9 \sin ^2({\theta_1})+6 \cos ^2({\theta_1})+4 \cos ({\theta_1})\right)
   \Bigl(9 {h_5} \sin ({\theta_1}) \nonumber\\
   && +4 \cos ^2({\theta_1}) \csc ({\theta_2})-2 \cos ({\theta_1}) \cot ({\theta_2})+6 \sin ^2({\theta_1}) \csc
   ({\theta_2})\Bigr)\Bigr) \nonumber\\
&(vii)& G^{\cal M}_{ r_{\theta_1}}\sim \frac{3\ 3^{2/3} {g_s}^{4/3} {N^{2}_f}  {\sin^2 (\phi_1)}  \sin ^2({\theta_1}) (8 \cos ({\theta_1})-3 \cos (2 {\theta_1})+15)^2}{2 \pi ^2 (\cos (2
   {\theta_1})-5)^2} \nonumber\\
&(viii)& G^{\cal M}_{ r_{\theta_2}}\sim \frac{3\ 3^{2/3} {g_s}^{4/3} {N^{2}_f} {\sin^2 (\phi_1)} \sin ^2({\theta_1}) (8 \cos ({\theta_1})-3 \cos (2 {\theta_1})+15)^2}{2 \pi ^2 (\cos (2
   {\theta_1})-5)^2} \nonumber\\
&(ix)& G^{\cal M}_{10 r}\sim -\frac{\sqrt[6]{3} {g_s}^{4/3} {N_f} {\sin (\phi_1)} \sin ({\theta_1}) (8 \cos ({\theta_1})-3 \cos (2 {\theta_1})+15)}{\sqrt{2} \pi  (\cos (2
   {\theta_1})-5)}\nonumber\\
           \pagebreak
&(x)&
G^{\cal M}_{\theta_1 \theta_1}\sim \frac{10^{-4}  \left(1-\frac{r_h^4}{r^4}\right)^{-2}  {g_s}^{23/6} {M_{\rm eff}}^4 {N_f}^2 \cot ^2\left(\frac{{\theta_2}}{2}\right) ln ^2(r)}{ \left(9 {g_s}^2 {N_f}
   ln ^2(r) {M^{2}_{\rm eff}}+6 {g_s} \pi  ln (r) {M_{\rm eff}}^2+4 N \pi ^2\right)^{3/2} \left(\left(\sin ^2({\theta_1})-\sin ^2({\theta_2})\right)
   {f_2}({\theta_2})^2+1\right)} \nonumber\\
   && + \frac{1}{2 \sqrt[3]{3} \sqrt[6]{g_s} \sqrt{N} \pi }\Bigr( 2 N \pi  \left({f_1}({\theta_1})^2 \sin ^2({\theta_1})+1\right)- \nonumber\\
   && \frac{324 {g_s} M^2 \left(r^2-3 a^2\right)^2 ln ^2(r) \sin ^2({\theta_1}) (2 \cos ({\theta_1}) \cos ({\theta_2}))^2}{r^4 (\cos (2 {\theta_1})-5) \left(2 \cos ^2({\theta_2}) \sin ^2({\theta_1})+2 \cos ^2({\theta_1}) \sin ^2({\theta_2})+3
   \left(\sin ^2({\theta_1}) \sin ^2({\theta_2})\right)\right)}\Bigr)\nonumber\\
&(xi)&  G^{\cal M}_{\theta_1\theta_2}\sim \frac{{g_s}^{13/3}}{r^4} \Bigr(\frac{3^{2/3} {g_s} \left(1-\frac{r_h^4}{r^4}\right)^{-2} {M_{\rm eff}}^4 {N_f}^2 \cot \left(\frac{{\theta_1}}{2}\right) \cot \left(\frac{{\theta_2}}{2}\right) ln ^2(r)}{204800  \pi
   ^{7/2} \left(\frac{{g_s} N}{r^4}\right)^{3/2} r^2} + \nonumber\\
     && \frac{1}{2 \sqrt{\pi}{g_s}^{9/2} N^{1/2}}\Bigr(\frac{{h_5} N \pi  r^4 (4 (\cos (2 {\theta_1})-5)+{f_1}({\theta_1}) {f_2}({\theta_2}) (13 \sin ({\theta_1})+\sin (3 {\theta_1})) \sin
   ({\theta_2}))}{\cos (2 {\theta_1})-5}+ \nonumber\\
   && \frac{54 {g_s}  M^2 \left(r^2-3 a^2\right)^2 ln ^2(r) \sin ({\theta_1}) \sin ({\theta_2}) (9 {h_5} \sin
   ({\theta_1}) \sin ({\theta_2})-2 \cos ({\theta_1}) \cos ({\theta_2}))}{2 \cos ^2({\theta_2}) \sin ^2({\theta_1})+3 \sin ^2({\theta_2}) \sin
   ^2({\theta_1})+2 \cos ^2({\theta_1}) \sin ^2({\theta_2})+3 {h_5} \sin (2 {\theta_1}) \sin (2 {\theta_2})}\Bigr)\Bigr)\nonumber\\
 &(xii)& G^{\cal M}_{x \theta_1}\sim -\frac{  {g_s} ^{4/3} }{\pi ^2 (\cos (2 {\theta_1})-5)^2 \left(3
   \sin ^2({\theta_1}) \sin ^2({\theta_2})+2 \sin ^2({\theta_1}) \cos ^2({\theta_2})+2 \cos ^2({\theta_1}) \sin ^2({\theta_2})\right)}\times \nonumber\\
   && \hskip -0.1in \Bigl(3^{2/3} {N^{2}_f} \sin{\phi_1} \sin{\phi_2} \sin ({\theta_1}) \sin ^2({\theta_2}) \left(9 \sin ^2({\theta_1})+6 \cos ^2({\theta_1})+4 \cos
   ({\theta_1})\right) \Bigl(-4 \cos ^4({\theta_1}) (4 \cot ({\theta_2})\nonumber\\
   &&-9 \sin ({\theta_2}))  +2 \cos ^3({\theta_1}) (9 \sin ({\theta_2})+4 \cos
   ({\theta_2}) \cot ({\theta_2}))-6 \sin ({\theta_1}) \cos ^2({\theta_1}) (8 \sin ({\theta_1}) \cot ({\theta_2})\nonumber\\
   && -6 \sin ({\theta_1}) \cos
   ({\theta_2}) \cot ({\theta_2})) +3 \sin ^2({\theta_1}) \cos ({\theta_1}) (9 \sin ({\theta_2})+10 \cos ({\theta_2}) \cot ({\theta_2}))+9
   \Bigl(3 \bigl(3 \sin ^4({\theta_1})\nonumber\\
   &&+\sin ^2(2 {\theta_1})\bigr) \sin ({\theta_2})   -4 \sin ^4({\theta_1}) \cot ({\theta_2})+ 6 \sin
   ^4({\theta_1}) \cos ({\theta_2}) \cot ({\theta_2})\Bigr)\Bigr)\Bigr) +\nonumber\\
   && \frac{1}{16 \pi ^{5/4}\sqrt[4]{N} r}\Bigr( {g_s}^{13/12} M  {N_f} \cot \left(\frac{{\theta_1}}{2}\right) \left(9 {h_5}+\left(3 \sqrt{6}-2 \cot ({\theta_1})\right)   \cot
   ({\theta_2})\right) \csc ({\theta_1}) \csc ({\theta_2}) ln (r)\nonumber\\
 &&  \left(108 ln (r) a^2+r\right) (2 \cos ({\theta_1}) \cos ({\theta_2})-9
   {h_5} \sin ({\theta_1}) \sin ({\theta_2}))\Bigr) \nonumber\\
   &(xiii)&   G^{\cal M}_{y \theta_1}\sim \Bigr(11 {g_s}^{1/12} M \csc ({\theta_1}) ln (r) \Bigr(-3 {g_s} {h_5} {N_f} r \cot \left(\frac{{\theta_1}}{2}\right)
   \left(108 ln (r) a^2+r\right) \cos ^3({\theta_1})+ 8 \pi r^2 \nonumber\\
   &&     \cot ({\theta_2}) \sin ^2({\theta_1}) \cos
   ({\theta_1})-\Bigr(\frac{1}{2} {g_s} {N_f} r (4-2 \cos (2 {\theta_2})) \cot \left(\frac{{\theta_1}}{2}\right) \cot ({\theta_2}) \csc
   ^2({\theta_2}) \left(108 ln (r) a^2+r\right) \Bigr) \nonumber\\
   &&  \sin ^3({\theta_1})\Bigr) \sin ^2({\theta_2})\Bigr)/\Bigr(4\pi
   ^{5/4}\sqrt[4]{N}  r^2 \Bigr(2 \cos ^2({\theta_2}) \sin ^2({\theta_1})+2 \cos ^2({\theta_1}) \sin ^2({\theta_2}) + \nonumber\\
   && 3 \left(\sin ^2({\theta_1}) \sin
   ^2({\theta_2})+{h_5} \sin (2 {\theta_1}) \sin (2 {\theta_2})\right)\Bigr)\Bigr) \nonumber\\
  &(xiv)& G^{\cal M}_{z \theta_1}\sim \frac{1}{8 \sqrt{2} \pi ^{5/4}\sqrt[4]{N} r}\Bigr(3^{2/3} {g_s}^{13/12} M  {N_f} \cot \left(\frac{{\theta_1}}{2}\right) \csc ^2({\theta_1}) ln (r) \left(108 ln (r) a^2+r\right) \nonumber\\
  &&  \left(2
   \cos ^2({\theta_1})+\left(2 \cot ^2({\theta_2})+3\right) \sin ^2({\theta_1})+6 {h_5} \cot ({\theta_2}) \sin (2 {\theta_1})\right)\Bigr)\nonumber\\
   &(xv)& G^{\cal M}_{10 \theta_1}\sim \frac{{g_s}^{11/6}   {N_f} r ln (r)}{652 \sqrt{N} \pi ^{7/4}} \Bigr( \frac{  \left(1-\frac{r_h^4}{r^4}\right)^{-1}{g_s}^{3/4} {M_{\rm eff}}^2 \cot \left(\frac{{\theta_2}}{2}\right)}{  \sqrt[4]{N} r \sqrt{\left(\sin ^2({\theta_1})-\sin ^2({\theta_2})\right)
   {f_2}({\theta_2})^2+1}}- \nonumber\\
   && \frac{17280 M \sqrt[4]{{g_s} N} \sqrt{\pi } \left(r^2-3 a^2\right) \cos ({\theta_2}) {f_1}({\theta_1}) \sin ^3({\theta_1}) (2 \cos ({\theta_1})
   \cos ({\theta_2}) )}{r^2 (\cos (2 {\theta_1})-5) \left(2 \cos ^2({\theta_2}) \sin
   ^2({\theta_1})+2 \cos ^2({\theta_1}) \sin ^2({\theta_2})+3 \sin ^2({\theta_1}) \sin ^2({\theta_2})\right)}\Bigr)\nonumber\\
   & & -\frac{\sqrt[6]{3} {g_s}^{4/3} {N_f} {\sin (\phi_1)} \sin ({\theta_1}) (8 \cos ({\theta_1})-3 \cos (2 {\theta_1})+15)}{\sqrt{2} \pi  (\cos (2
   {\theta_1})-5)}\nonumber\\
   &(xvi)& G^{\cal M}_{\theta_2 \theta_2}\sim  -\frac{3^{2/3} \left(1-\frac{r_h^4}{r^4}\right)^{-2}{g_s}^{23/6} {M^{4}_{\rm eff}} {N^{2}_f} \cot ^2\left(\frac{{\theta_1}}{2}\right) \left((\cos (2 {\theta_1})-\cos (2 {\theta_2}))
   {f^{2}_2}({\theta_2})-2\right) ln ^2(r)}{51200 g^2 \sqrt{\pi } \left(9 {g_s}^2 {N_f} ln ^2(r) {M_{\rm eff}}^2+6 {g_s} \pi  ln (r)
   {M_{\rm eff}}^2+4 N \pi ^2\right)^{3/2}}  \nonumber\\
   && + \frac{1}{32 \sqrt[3]{3} \pi ^{5/2}\sqrt{g_s N} r^4 (\cos (2 {\theta_1})-5)} \Bigr(-64 \sqrt[3]{{g_s}} N \pi ^3 r^4 {f_2}({\theta_2})^2 \sin ^3({\theta_1}) (2 \cos ({\theta_1}) \cot ({\theta_2}) )) -\nonumber\\
      \pagebreak
   && \frac{1296 {g_s}^{4/3} M^2 \pi ^2 \left(r^2-3 a^2\right)^2 \left(-9 {g_s} a^4+3 {g_s} r^2 a^2+1\right)^2 (\cos (2 {\theta_1})-5)^2 ln ^2(r) \sin
   ^2({\theta_2})}{2 \cos ^2({\theta_2}) \sin ^2({\theta_1})+2 \cos ^2({\theta_1}) \sin ^2({\theta_2})+3 \left(\sin ^2({\theta_1}) \sin
   ^2({\theta_2})+{h_5} \sin (2 {\theta_1}) \sin (2 {\theta_2})\right)}\Bigr)\nonumber\\
   &(xvii)& G^{\cal M}_{x \theta_2}\sim -\frac{  {g_s} ^{4/3} }{\pi ^2 (\cos (2 {\theta_1})-5)^2 \left(3
   \sin ^2({\theta_1}) \sin ^2({\theta_2})+2 \sin ^2({\theta_1}) \cos ^2({\theta_2})+2 \cos ^2({\theta_1}) \sin ^2({\theta_2})\right)}\times \nonumber\\
   && \hskip -0.1in \Bigl(3^{2/3} {N^{2}_f} \sin{\phi_1} \sin{\phi_2} \sin ({\theta_1}) \sin ^2({\theta_2}) \left(9 \sin ^2({\theta_1})+6 \cos ^2({\theta_1})+4 \cos
   ({\theta_1})\right) \Bigl(-4 \cos ^4({\theta_1}) (4 \cot ({\theta_2})\nonumber\\
   &&-9 \sin ({\theta_2}))  +2 \cos ^3({\theta_1}) (9 \sin ({\theta_2})+4 \cos
   ({\theta_2}) \cot ({\theta_2}))\nonumber\\
   && -6 \sin ({\theta_1}) \cos ^2({\theta_1}) (8 \sin ({\theta_1}) \cot ({\theta_2})-6 \sin ({\theta_1}) \cos
   ({\theta_2}) \cot ({\theta_2})) +3 \sin ^2({\theta_1}) \cos ({\theta_1}) (9 \sin ({\theta_2})\nonumber\\
   &&+10 \cos ({\theta_2}) \cot ({\theta_2}))-9
   \Bigl(3 \left(-3 \sin ^4({\theta_1})-\sin ^2(2 {\theta_1})\right) \sin ({\theta_2}) \nonumber\\
   && +4 \sin ^4({\theta_1}) \cot ({\theta_2})-6 \sin
   ^4({\theta_1}) \cos ({\theta_2}) \cot ({\theta_2})\Bigr)\Bigr)\Bigr) +\nonumber\\
   &&   \frac{72 \sqrt[6]{3} M  {g_s}^{1/12}  \left(3 a^2-r^2\right) \left(9 {g_s} a^4-3 {g_s} r^2 a^2-1\right) \cos
   ^2({\theta_1}) \cot ({\theta_1}) \cot ({\theta_2}) ln (r)}{r^2 \sqrt[4]{\pi } \sqrt[4]{N}(\cos (2 {\theta_1})-5) \left(2 \cot ^2({\theta_1})+2 \cot
   ^2({\theta_2})+3\right)}\nonumber\\
   &(xviii)& G^{\cal M}_{y \theta_2}\sim \frac{ \sqrt[4]{N} {g_s}^{-5/12}  \pi ^{1/4} (\cos (2 {\theta_1})-5) \cos ({\theta_2}) {f_2}({\theta_2}) \sin ^2({\theta_2})}{3 \sqrt{2}
   \sqrt[3]{3}  \left(2 \cos ^2({\theta_2}) \sin ^2({\theta_1})+2 \cos ^2({\theta_1}) \sin ^2({\theta_2})+3\sin ^2({\theta_1})
   \sin ^2({\theta_2}) \right)}\nonumber\\
   &(xix)& G^{\cal M}_{z \theta_2}\sim \frac{3^{2/3} {g_s}^{13/12} M  {N_f}}{256 \sqrt{2} \pi ^{5/4} \sqrt[4]{N} r (\cos (2 {\theta_1})-5)} \times \nonumber\\
    && \Bigr((6 \cos (2 {\theta_1})+(  \cos (2 ({\theta_1}-{\theta_2}))+6 \cos (2 {\theta_2})+12 {h_5} \cos (2 ({\theta_1}+{\theta_2}))+\cos
   (2 ({\theta_1}+{\theta_2}))-14) \nonumber\\
   && \csc ^2({\theta_1}) \csc ^3\left(\frac{{\theta_2}}{2}\right) ln (r) \left(36 ln (r) a^2+r\right) \sec
   \left(\frac{{\theta_2}}{2}\right) \left(\cos ^2({\theta_1})+4 \cos ({\theta_2}) \cos ({\theta_1})-\sin ^2({\theta_1})-5\right)\Bigr) \nonumber\\
   &(xx)& G^{\cal M}_{10 \theta_2}\sim  \frac{{g_s}^{11/6} {N_f} r ln (r)}{640 \sqrt{2} \sqrt[3]{3} \sqrt{N} \pi ^{7/4}} \times \nonumber\\
   && \Bigr(\frac{2   \left(1-\frac{r_h^4}{r^4}\right)^{-1} {g_s}^{3/4} {M_{\rm eff}}^2 \cot \left(\frac{{\theta_1}}{2}\right) \sqrt{\left(\sin ^2({\theta_1})-\sin ^2({\theta_2})\right)
   {f_2}({\theta_2})^2+1}}{\sqrt[4]{N} r}- \nonumber\\
   && \frac{51840 {h_5} M \sqrt[4]{{g_s} N} \sqrt{\pi } \left(r^2-3 a^2\right) \left(-9 {g_s} a^4+3 {g_s} r^2 a^2+1\right) \cos ({\theta_2})
   {f_2}({\theta_2}) \sin ({\theta_1}) \sin ^2({\theta_2})}{r^2 \left(2 \cos ^2({\theta_2}) \sin ^2({\theta_1})+2 \cos ^2({\theta_1}) \sin
   ^2({\theta_2})+3 \left(\sin ^2({\theta_1}) \sin ^2({\theta_2})+{h_5} \sin (2 {\theta_1}) \sin (2 {\theta_2})\right)\right)}\Bigr)\nonumber\\
   & &  -\frac{\sqrt[6]{3} {g_s}^{4/3} {N_f} {\sin (\phi_1)} \sin ({\theta_1}) (8 \cos ({\theta_1})-3 \cos (2 {\theta_1})+15)}{\sqrt{2} \pi  (\cos (2
   {\theta_1})-5)}\nonumber\\
   &(xxi)& G^{\cal M}_{x x}\sim -\frac{3^{2/3} \  (\cos (2 {\theta_2})-5) \sin ^2({\theta_1})}{2 {g_s}^{2/3} \left(2 \cos ^2({\theta_2}) \sin ^2({\theta_1})+2
   \cos ^2({\theta_1}) \sin ^2({\theta_2})\right)}\nonumber\\
   &(xxii)& G^{\cal M}_{x y}\sim -\frac{ \cos ^2({\theta_1}) (\cos (2 {\theta_1})-5) \cos ^3({\theta_2}) \sin ({\theta_2})}{3 \sqrt{2} 3^{5/6} {g_s}^{2/3}
   \left(\cos ^2({\theta_2}) \sin ^2({\theta_1})+\cos ^2({\theta_1}) \sin ^2({\theta_2})\right)^2}\nonumber\\
   &(xxiii)& G^{\cal M}_{x z}\sim \frac{2 \csc ({\theta_1}) \csc ({\theta_2})}{27 \sqrt[3]{3}{g_s}^{2/3} (\cos (2 {\theta_1})-5)} \times \nonumber\\
   && \left[2 \left(\cos ({\theta_2}) \left(6 \sqrt{6} \cot ({\theta_2})-4 \cot ({\theta_1}) \cot ({\theta_2})\right)-9 \sqrt{6} \sin ({\theta_2})\right) \cos
   ^3({\theta_1})+12 \cos ({\theta_2}) \cot ({\theta_2}) \sin ({\theta_1}) \right. \nonumber\\
   && \left.\cos ^2({\theta_1})- 27 \sin ^2({\theta_1}) \left(4 {h_5} \cos
   ({\theta_2})-\sqrt{6} \sin ({\theta_2})\right) \cos ({\theta_1})+81 )\sqrt{6}{h_5} \sin ^3({\theta_1} \cos ({\theta_2}) \right] \nonumber\\
   &(xxiv)& G^{\cal M}_{yy}\sim \frac{  3^{2/3} }{{g_s}^{2/3}}, ~(xxv)~G^{\cal M}_{y z}\sim  -\frac{  \sqrt{2  } \csc ({\theta_1}) (3 {h_5} \cos ({\theta_1})+\cot ({\theta_2}) \sin ({\theta_1}))}{3^{5/6} {g_s}^{2/3}},~(xxvi)~G^{\cal M}_{10\ 10}\sim \frac{{g_s}^{4/3}}{3 \sqrt[3]{3}} \nonumber\\
   &(xxvii)& G^{\cal M}_{zz }\sim -\frac{  \csc ^2({\theta_1}) \left(-2 \cos ^2({\theta_1})+\left(-2 \cot ^2({\theta_2})-3\right) \sin ^2({\theta_1})-6 {h_5}
   \cot ({\theta_2}) \sin (2 {\theta_1})\right)}{9 \sqrt[3]{3} {g_s}^{2/3}} 
      \end{eqnarray}}
This is similar in spirit to the M-theory uplift of the warped resolved
conifold in \cite{large_base}. For (\ref{limits_Dasguptaetal-ii}), we also need to keep in mind the following. For
finite $g_s, N_f$, one sees that $- ln\left[\frac{1}{g_s} - \frac{N_f}{8\pi} ln(9a^2r^2+r^6)\right.$
$\left. - \frac{N_f}{2\pi}
ln\left(\sin\frac{\theta_1}{2}\sin\frac{\theta_2}{2}\right)\right]\in\tilde{\tilde{\tilde{\phi}}}$ is expected to
receive the most dominant contribution near $\theta_{1,2}=0,\pi, r=r_\Lambda\rightarrow\infty$. We will assume that
$\theta_{1,2}\rightarrow0$ as $\epsilon^\gamma$  and $r_\Lambda\sim\epsilon^{-\beta},\epsilon\rightarrow0,\gamma,\beta>0$ 
such that: $\gamma\sim\frac{3}{4}\beta$ implying thereby that in both limits (\ref{limits_Dasguptaetal-i}) as well 
as (\ref{limits_Dasguptaetal-ii}), the aforementioned log is $ln g_s$. One hence obtains 
{\footnotesize
\begin{eqnarray}
\label{ds11squared}
& & ds_{11}^2= \frac{-\left(1 - \frac{r_h^4}{r^4}\right)r^2}{g_s^{\frac{7}{6}}\sqrt{N}} dt^2 + \frac{r^2}{g_s^{\frac{2}{3}}\sqrt{g_sN}}\left(ds^2_{{\bf R}^3}\right) + \sqrt{\frac{N}{g_s^{\frac{1}{3}}}}\frac{dr^2}{\left(1 - \frac{r_h^4}{r^4}\right)} \nonumber\\
& & + g_s^{\frac{4}{3}}\left(g_s N\right)^{\frac{1}{4}}\sin\phi_1 sin\frac{\psi}{2}\Upsilon_1(\theta_1,\theta_2) d\phi_1 dr + g_s^{\frac{4}{3}}sin^2\phi_1\Upsilon_2(\theta_1) dr d\theta_1
 + g_s^{\frac{4}{3}}sin^2\phi_1\Upsilon_3(\theta_1) dr d\theta_2\nonumber\\
& & + \left[\frac{g_s^{\frac{23}{6}}M^4N_f^2\Gamma_1(\theta_{1,2})\left(ln r\right)^2}{N^{\frac{3}{2}}\left(1 - \frac{r_h^4}{r^4}\right)^2} + \frac{\sqrt{N}\Gamma_1^\prime(\theta_{1,2})}{g_s^{\frac{1}{6}}} + \frac{g_sM^2\left(r^2 - 3 a^2\right)^2\left(ln r\right)^2\Gamma_1^{\prime\prime}(\theta_{1,2})}{r^4g_s^{\frac{1}{6}}\sqrt{N}}\right]d\theta_1^2\nonumber\\
& & +\frac{g_s^{\frac{13}{3}}}{r^4}\left[\frac{M^4N_f^2g_sr^4\left(ln r\right)^2\Gamma_2(\theta_{1,2})}{\left(1 - \frac{r_h^4}{r^4}\right)^2\left(g_sN\right)^{\frac{3}{2}}} + \frac{1}{\sqrt{N}g_s^{\frac{9}{2}}}\left\{h_5Nr^4\Gamma_3(\theta_{1,2}) + g_sM^2\left(ln r\right)^2(r^2 - 3 a^2)\Gamma_4(\theta_{1,2})\right\}\right]d\theta_1d\theta_2\nonumber\\
& & + \left[ - \frac{g_s^{\frac{23}{6}}M^4N_f^2\left(ln r\right)^2\Gamma_5(\theta_{1,2})}{N^{\frac{3}{2}}\left(1 - \frac{r_h^4}{r^4}\right)^2} + \frac{1}{\sqrt{g_sN}r^4}
\left\{ - g_s^{\frac{1}{3}}Nr^4\Gamma_2(\theta_{1,2}) - g_s^{\frac{4}{3}}M^2(r^2 - 3 a^2)^2\Gamma_7(\theta_{1,2})\left(ln r\right)^2\right\}\right]d\theta_2^2\nonumber\\
& & +(g_sN)^{\frac{1}{4}}\left\{\frac{g_s^{\frac{13}{12}}MN_f ln r \Gamma_8(\theta_{1,2})(108 a^2 ln r + r)}{N^{\frac{1}{4}}r} + g_s^{\frac{4}{3}}N_f^2\sin\phi_1\sin\phi_2\Gamma_8^\prime(\theta_{1,2})\right\}d\phi_1d\theta_1\nonumber\\
& &  + \left\{g_s^{\frac{4}{3}}N_f^2\sin\phi_1\sin\phi_2\Gamma_8^\prime(\theta_{1,2}) + ln r\frac{g_s^{\frac{1}{12}}M (3 a^2 - r^2)(9g_sa^4 - 3 g_sa^2r^2 -1)
\Gamma_8^{\prime\prime}(\theta_{1,2})}{N^{\frac{1}{4}}r^2}\right\}d\phi_1d\theta_2\nonumber\\
& & + \frac{g_s^{\frac{1}{12}}M(g_sN)^{\frac{1}{4}}}{N^{\frac{1}{4}}r^2}\Upsilon(r,\theta_{1,2})d\phi_2d\theta_1
+ \frac{N^{\frac{1}{4}}(g_sN)^{\frac{1}{4}}\Gamma_9(\theta_{1,2}}{g_s^{\frac{5}{12}}}d\phi_2d\theta_2 \nonumber\\
& &
+ \frac{(g_sN)^{\frac{1}{4}}g_s^{\frac{13}{12}}MN_f ln r (108 a^2 ln r + r)\Gamma_{10}(\theta_{1,2})}{rN^{\frac{1}{4}}}d\psi d\theta_1+ \frac{(g_sN)^{\frac{1}{4}}g_s^{\frac{13}{12}}MN_f ln r \Gamma_{11}(\theta_{1,2})}{rN^{\frac{1}{4}}}d\psi d\theta_1 +  \nonumber\\
& & \frac{\sqrt{g_sN}}{g_s^{\frac{2}{3}}}\left(\Gamma_{12}(\theta_{1,2})d\phi_1^2 + \Gamma_{13}(\theta_{1,2})d\phi_2^2 + \Gamma_{14}(\theta_{1,2})d\psi^2\right. + \Gamma_{15}(\theta_{1,2})d\phi_1d\phi_2 +  \Gamma_{16}(\theta_{1,2})d\phi_1d\psi + \nonumber\\
  && \left. \Gamma_{17}(\theta_{1,2})d\phi_2d\psi\right) + g_s^{\frac{4}{3}}dx_{10}^2 +
 - g_s^{\frac{4}{3}}sin\phi_1\Upsilon_4(\theta_1) dr dx_{10}
  \nonumber\\
   & & + \frac{g_s^{\frac{11}{6}}N_f r ln r}{\sqrt{N}}\left\{\frac{g_s^{\frac{3}{4}}M^2\Gamma_{18}(\theta_{1,2})}{N^{\frac{1}{4}}\left(1 - \frac{r_h^4}{r^4}\right)} - \frac{(g_sN)^{\frac{1}{4}}(r^2 - 3 a^2)\Gamma_{19}(\theta_{1,2})}{r^2} - g_s^{\frac{4}{3}}sin\phi_1\Upsilon_5(\theta_1)\right\}dx_{10}d\theta_1\nonumber\\
& &  + \frac{g_s^{\frac{11}{6}}N_f r ln r}{\sqrt{N}}\left\{\frac{g_s^{\frac{3}{4}}M^2\Gamma_{20}(\theta_{1,2})}{N^{\frac{1}{4}}r\left(1 - \frac{r_h^4}{r^4}\right)} - \frac{h_5M(g_sN)^{\frac{1}{4}}(r^2 - 3 a^2)\Gamma_{21}(\theta_{1,2})}{r^2} - g_s^{\frac{4}{3}}sin\phi_1\Upsilon_5(\theta_1)\right\}dx_{10}d\theta_2,\nonumber
\end{eqnarray}}
\noindent [the angular parts $\Gamma_i(\theta_{1,2}), \Upsilon_j(\theta_{1,2})$, etc. can be read off from (\ref{eq:simplifiedmetriccomponents})] which, similar to \cite{Cvetic_et_al_M3-i}, is an $M3$-brane solution.

Let us look at the near horizon limit of the $G^{\cal M}_{tt}$ and $G^{\cal M}_{rr}$ of (\ref{ds11squared}). Before doing so, these components in the limits (\ref{limits_Dasguptaetal-i})  takes the form:
\begin{equation}
\label{near_horison_i}
- \epsilon^{\frac{25 d}{3}}\left(1 - \frac{r_h^4}{r^4}\right)r^2 dt^2 + \frac{dr^2}{\epsilon^{\frac{29 d}{3}}\left(1 - \frac{r_h^4}{r^4}\right)}.
\end{equation}
In the near-horizon limit $r = r_h + \epsilon^\prime\chi$, implying $1 - \frac{r_h^4}{r^4} = \frac{4 \epsilon^\prime\chi}{r_h} + {\cal O}\left(\epsilon^\prime\ ^2\right)$,
\begin{equation}
\label{near_horizon_ii}
\frac{dr^2}{\epsilon^{\frac{29 d}{3}}\left(1 - \frac{r_h^4}{r^4}\right)} \sim \xi \frac{(d\chi)^2}{\chi},\ {\rm where}\ \xi\equiv\frac{\epsilon^\prime r_h}{\epsilon^{\frac{29 d}{3}}}.
\end{equation}
Writing $\xi \frac{(d\chi)^2}{\chi} = du^2$ or $\chi=\frac{u^2}{4\xi}$, one obtains: $
- 4\pi^2 u^2 T^2 dt^2 + du^2$, 
where $T^2(r_h\approx1\ [{\rm See\ {\bf 5}}])\sim\frac{1}{\left(\sqrt{g_sN}\right)^2}$, in conformity with \cite{metrics} as well as (\ref{T}). Similarly, in the MQGP limit (\ref{limits_Dasguptaetal-ii}), we obtains:
$G^{\cal M}_{00}\sim\epsilon^{\frac{55d}{3}}r^2\left(1 - \frac{r_h^4}{r^4}\right),
G^{\cal M}_{rr}\sim\frac{\epsilon^{-\frac{59d}{3}}}{r^2
\left(1 - \frac{r_h^4}{r^4}\right)}$, one can rewrite $G_{rr}dr^2
=\xi^\prime\frac{d\omega^2}{\omega}$ where $\xi^\prime\sim
\frac{r_h\epsilon^\prime}{\epsilon^{\frac{59d}{3}}}$. Once again writing
$\xi^\prime \frac{d\omega^2}{\omega}=dv^2$ or $\omega=\frac{v^2}{4\xi^\prime}$,
one sees that near $r=r_h\sim1, G^{\cal M}_{tt}dt^2\sim\epsilon^{38d}u^2dt^2$ implying
again $T^2\sim\frac{1}{\left(\sqrt{g_sN}\right)^2}$ in conformity with
\cite{metrics} and (\ref{T}).

As we will see in {\bf 5.3}, the action using (\ref{eq:simplifiedmetriccomponents}) is singular at $\theta_{1,2}=0,\pi$; we regulate these
pole-singularities by introducing a small angle cut-off $\epsilon_\theta$: $\theta_{1,2}\in[\epsilon_\theta,\pi-\epsilon_\theta]$. We then show
that the finite part of the action turns out to be independent of this $\epsilon_\theta$ if one identifies $\epsilon_\theta=\epsilon^\gamma$,
for an appropriate $\gamma$. We will henceforth follow this.

We discuss the supersymmetry of the M-theory uplift in the two limits. We give detailed arguments for (\ref{limits_Dasguptaetal-i}) and sketch out the arguments for (\ref{limits_Dasguptaetal-ii}).

In the limit (\ref{limits_Dasguptaetal-i}), the near-horizon (with $r_h\sim1$) limit of (\ref{ds11squared}) near $\theta_{1,2}=0,\pi$,
after appropriate rescaling of ${\bf R}^3$-variables and using $G^M_{\bullet r}=\epsilon^{\frac{29d}{3}} u G^M_{\bullet u},
G_{\phi_{1,2}\bullet}\sim\left(g_s N\right)^{\frac{1}{4}}\sin\theta_{1,2}G_{x/y\bullet}, G_{\psi\bullet}\sim\left(g_s N\right)^{\frac{1}{4}}G_{z\bullet}$, reduces to:
{\small
\begin{eqnarray}
\label{metric_simp}
& & ds_{11}^2 = - 4\pi^2 u^2 T^2 dt^2 + du^2 +  ds^2_{{\bf R}^{3}} + \epsilon^{12d}u sin^2\phi_1 du (d\theta_1 + d\theta_2)
 + \frac{d\theta_1^2}{\epsilon^{\frac{29d}{3}}} + \left(\frac{h_5}{\epsilon^{\frac{29d}{3}}} \right)d\theta_1d\theta_2  \nonumber\\
 \pagebreak
& &  + \frac{d\theta_2^2}{\epsilon^{\frac{29d}{3}}} +
\frac{d\phi_1^2}{\epsilon^{\frac{26d}{3}}} + \frac{d\phi_2^2}{\epsilon^{\frac{26}{3}}} + \frac{d\psi^2}{\epsilon^{\frac{32d}{3}}} + \frac{d\phi_1d\phi_2}{\epsilon^{\frac{61d}{6}}} + \frac{d\phi_1d\psi}{\epsilon^{\frac{43d}{6}}}
+ \frac{d\phi_2d\psi}{\epsilon^{\frac{29d}{3}}} + \epsilon^{\frac{23d}{2}} u sin\phi_1 du dx_{10}  \nonumber\\
 & &+
\epsilon^{7d} u sin\phi_1 sin\frac{\psi}{2} du d\phi_1 + \epsilon^{-\frac{8d}{3}}\sin\phi_1\sin\phi_2d\phi_1(d\theta_1 + d\theta_2) + \epsilon^{\frac{4d}{3}}dx_{10}^2 + d\theta_1dx_{10} \epsilon^{\frac{11d}{6}}sin\phi_1\nonumber\\
&& + \epsilon^{\frac{11d}{6}} sin\phi_1d\theta_2dx_{10}. 
 \end{eqnarray}}
The terms relevant to $G^{\cal M}_{\phi_1\bullet}, G^{\cal M}_{u\bullet}, G^{\cal M}_{\theta_{1,2}\bullet}$ in (\ref{metric_simp}) are:
\begin{eqnarray}
\label{dphi1_du}
& & \frac{d\phi_1}{\epsilon^{\frac{61d}{6}}}\left(\epsilon^{\frac{3d}{2}}d\phi_1 + d\phi_2 + \epsilon^{\frac{9d}{3}}d\psi
+ u \epsilon^{\frac{103}{6}}\sin\phi_1\sin\frac{\psi}{2}d\psi + \epsilon^{\frac{45d}{6}}\sin\phi_1\sin\phi_2[d\theta_1 + d\theta_2]
\right)\nonumber\\
& & + du\left(du + u \epsilon^{12d}\sin^2\phi_1[d\theta_1 + d\theta_2] + u \epsilon^{\frac{23d}{2}}\sin\phi_1dx_{10}\right) + \frac{d\theta_1^2}{\epsilon^{\frac{29d}{3}}} + \left(\frac{h_5}{\epsilon^{\frac{29d}{3}}} \right)d\theta_1d\theta_2 + \frac{d\theta_2^2}{\epsilon^{\frac{29d}{3}}}\nonumber\\
& & + d\theta_1dx_{10} \epsilon^{\frac{11d}{6}}sin\phi_1 + \epsilon^{\frac{11d}{6}} sin\phi_1d\theta_2dx_{10},
\end{eqnarray}
which for $h_5<<1$ will be approximated by:
\begin{equation}
\label{dphi1_du_simp}
 \frac{d\phi_1}{\epsilon^{\frac{61d}{6}}}\left(\epsilon^{\frac{3d}{2}}d\phi_1 + d\phi_2 + \epsilon^{\frac{9d}{3}}d\psi\right)+ du^2 + \frac{d\theta_1^2}{\epsilon^{\frac{29d}{3}}} + \frac{d\theta_2^2}{\epsilon^{\frac{29d}{3}}} .
\end{equation}
The metric for $d=1$ restricted to the fiber $T^3(\phi_1,\phi_2,\psi): \frac{d\phi_1}{\epsilon^{\frac{61}{6}}}\left(\epsilon^{\frac{3}{2}} d\phi_1 + d\phi_2 + \epsilon^3 d\psi\right) + \frac{d\phi_2}{\epsilon^{\frac{26}{3}}}\left(d\phi_2 + \frac{d\psi}{\epsilon}\right) + \frac{d\psi^2}{\epsilon^{\frac{32}{3}}}$ can be diagonalised and the metric of (\ref{metric_simp}) implies the following elfbeins for the $D=11$ space-time which locally is  $\mathbb{R}^4(t,x^{1,2,3})\times M_7(u,\theta_{1,2},\phi_{1,2},\psi,x_{10})$:
\begin{eqnarray}
\label{elfbeins}
& & e^0 = 2\pi T u dt,  e^{1,2,3} = dx^{1,2,3},  e^4 = du,  e^{5,6}=\frac{d\theta_{1,2}}{\epsilon^{\frac{29}{6}}},e^{7} = \frac{\epsilon^{-\frac{61}{12}}}{2^{\frac{1}{4}}}\left(d\phi_1-d\phi_2\right),\nonumber\\
& & e^{8}=\frac{\epsilon^{-\frac{61}{12}}}{2^{\frac{1}{4}}}\left(-d\phi_1-d\phi_2\right),  e^{9}=\epsilon^{-\frac{16}{3}}d\psi, e^{10} = \epsilon^{\frac{2}{3}}dx_{10}.
\end{eqnarray}
From (\ref{B-3tduals}), the non-zero components of the four-form field strength $G_4$  in the limit (\ref{limits_Dasguptaetal-i}) or (\ref{limits_Dasguptaetal-ii}),  are:
\begin{eqnarray}
\label{non_zero_G4_components}
& & G_{\theta_1\theta_2 \phi_1\phi_2}, G_{r \theta_1\theta_2 \phi_2},  G_{r \theta_1 \theta_2 \phi_1}, G_{r \theta_2 \phi_1 \psi},G_{r \theta_1 \phi_1 \psi}, G_{r \theta_2 \phi_1 \phi_2}, G_{r \theta_1 \theta_2 \psi},G_{r \theta_1 \phi_2 \psi}, G_{r \theta_2 \phi_2 \psi}, G_{\theta_1 \theta_2 \phi_2 \psi}, \nonumber\\
&& G_{\theta_1 \phi_1 \phi_2 \psi}, G_{\theta_1 \theta_2 \phi_1\psi}, ({\rm from}\ H_3\wedge A_1);\nonumber\\
 & & G_{ r \theta_1 \phi_110}, G_{r \theta_2 \phi_1 10}, G_{r \theta_1 \phi_2 10}, G_{r \theta_2 \phi_2 10}, G_{r \theta_2 \psi 10}, G_{r \theta_1 \psi 10}, G_{ r \phi_1 \psi 10}, G_{ r \phi_2 \psi 10}, G_{ r \theta_1 \theta_2}, G_{\theta_1 \phi_2 \psi 10}, \nonumber\\
 \pagebreak
&& G_{\theta_2 \psi_2 \psi 10}, G_{\theta_1 \phi_1 \psi 10}, G_{\theta_2 \phi_1 \psi 10}, G_{\theta_1 \theta_2 \psi 10}, G_{\theta_1 \theta_2 \phi_1 10}, G_{\theta_1 \theta_2 \phi_2 10}, ({\rm from}\ H_3\wedge dx_{10}).
\end{eqnarray}
The dominant contribution appear from following components:
{\footnotesize
\begin{eqnarray}
\label{G4s}
&& G_{\theta_1\theta_2 \phi_1\phi_2}= \Bigl({N_f} {f_ 2}({\theta_2}) \sin ({\theta_1}) \cos
   ({\theta_1}) \sin (2 {\theta_2}) \cos ({\theta_2}) \Bigl(4 \cos
   ^3({\theta_1}) \sin ^2({\theta_2}) \cos ({\theta_2}) (6 {\phi_2}
   \cos ({\theta_1})-\psi )\nonumber\\
   && +\cos ^2({\theta_2}) \left(-12 {\phi_2} \cos
   ^3({\theta_1}) \sin ^2({\theta_2})+2 \psi  \sin ^4({\theta_1})-\psi
   \sin ^2({\theta_1}) (\cos (2 {\theta_1})-9)\right) \nonumber\\
   && -4 \psi  \sin
   ^2({\theta_1}) \cos ({\theta_1}) \cos ^3({\theta_2})-2 \psi  (\cos
   (2 {\theta_1})-5) \cos ^2({\theta_1}) \sin
   ^2({\theta_2})\Bigr)\Bigr)/\Bigl(\sqrt{6} \pi  (\cos (2 {\theta_1})-5) \nonumber\\
   && \left(\sin
   ^2({\theta_1}) \cos ^2({\theta_2})+\cos ^2({\theta_1}) \sin
   ^2({\theta_2})\right)^3\Bigr)\sim\epsilon^{-\frac{3}{2}}_{(\ref{limits_Dasguptaetal-i})}/
   \epsilon^{-\frac{9}{2}}_{(\ref{limits_Dasguptaetal-ii})}\nonumber\\
   && G_{r \theta_1\theta_2 \phi_2}= \frac{1}{160 \sqrt{2} \pi ^{7/4}}\Biggl( \Bigl(\sqrt[4]{{g_s}}
   {M_{eff}}^2 {\theta_1} {f_ 1}({\theta_1}) \csc ({\theta_2})
   (3 {h_ 5} \cot ({\theta_1})+\cot ({\theta_2}))  \Bigl(12 {g_s}
   {N_f} \log (r) \nonumber\\
  && +{g_s} {N_f} \log \Bigl(\sin
   \left(\frac{{\theta_1}}{2}\right) \sin
   \left(\frac{{\theta_2}}{2}\right)\ \Bigr)+3 {g_s} {N_f}+4 \pi
   \Bigr)\Bigr)/\Bigl(N^{3/4} r \left(2 \cot ^2({\theta_1})+2 \cot
   ^2({\theta_2})+3\right)\Bigr)+\nonumber\\
   &&  \frac{30 \sqrt{\pi } {N_f}
   {f_ 2}({\theta_2}) \sin (2{\theta_1}) \sin^2 (2 {\theta_2})   \left(27 {g_s} M
   {\phi_2} \cot \left(\frac{{\theta_1}}{2}\right) \sin ({\theta_2})+4
   \sqrt{3} \sqrt[4]{\pi } {\phi_1} r \sqrt[4]{{g_s} N} \sin
   ({\theta_1})\right)}{\sqrt[4]{{g_s} N}r \left(\sin ^2({\theta_1}) \cos
   ^2({\theta_2})+\cos ^2({\theta_1}) \sin
   ^2({\theta_2})\right)^2} \Biggr)\sim
   \epsilon^{-\frac{1}{2}}_{(\ref{limits_Dasguptaetal-i})/(\ref{limits_Dasguptaetal-ii})}
    \nonumber\\
   &&  G_{r \theta_1 \theta_2 \phi_1} = \Bigl(32 \sqrt{6} {g_s} M {N_f} {\phi_2} {f_ 1}({\theta_1})
   \cos ^4({\theta_1}) \sin ^2({\theta_2}) \cos ^2({\theta_2}) (2 \cos
   ({\theta_1})-\cos ({\theta_2}))\Bigr)/\Bigl(2\pi  r (\cos (2 {\theta_2})\nonumber\\
   && cos (2 {\theta_1}) +\cos (2 ({\theta_1}+{\theta_2}))+6 \cos
   (2 {\theta_1})+6 \cos (2 {\theta_2})-14) \left(\sin ^2({\theta_1})
   \cos ^2({\theta_2})+\cos ^2({\theta_1}) \sin
   ^2({\theta_2})\right)\Bigr)-\nonumber\\
   && \frac{\sqrt[4]{{g_s}} {M^{2}_{eff}}
   {\theta_1} {f_ 1}({\theta_1}) \left(12 {g_s} {N_f} \log
   (r)+{g_s} {N_f} \log \left(\sin \left(\frac{{\theta_1}}{2}\right)
   \sin \left(\frac{{\theta_2}}{2}\right)\right)+3 {g_s} {N_f}+4 \pi
   \right)}{320 \sqrt{2} \pi ^{7/4} N^{3/4} r}\sim\frac{1}{\epsilon r}_{(\ref{limits_Dasguptaetal-i})}/\frac{1}{\epsilon^2 r}_{(\ref{limits_Dasguptaetal-ii})}\nonumber\\
   && G_{\theta_1\theta_2 \phi_2 \ 10}=\frac{{f_ 2}({\theta_2}) \sin ({\theta_1}) \cos ({\theta_1}) \sin
   ({\theta_2}) \sin (2 {\theta_2}) \cos ({\theta_2})}{\left(\sin
   ^2({\theta_1}) \cos ^2({\theta_2})+\cos ^2({\theta_1}) \sin
   ^2({\theta_2})\right)^2}\sim\epsilon^{-1}_{(\ref{limits_Dasguptaetal-i})}/\epsilon^{-3}_{(\ref{limits_Dasguptaetal-ii})}
   \nonumber\\
   &&  G_{\theta_1\theta_2\phi_1 \ 10}=-\frac{1}{\left(\left(2 \cot
   ^2({\theta_1})+3\right) \sin ^2({\theta_2})+2 \cos
   ^2({\theta_2})\right)^2}\nonumber\\
   && \times \Bigl(2 {f_ 2}({\theta_2}) \csc ({\theta_1}) \sin ^2({\theta_2})
   \cos ({\theta_2}) \Bigl(-24 {h_ 5} \cot ^3({\theta_1}) \sin
   ({\theta_2}) \cos ({\theta_2})+2 \cot ^2({\theta_1}) \cos
   ^2({\theta_2})\nonumber\\
   && +\left(-2 \cot ^4({\theta_1})+\cot
   ^2({\theta_1})+3\right) \sin ^2({\theta_2})\Bigr)\Bigr)\sim\epsilon^{-1}_{(\ref{limits_Dasguptaetal-i})}/\epsilon^{-3}_{(\ref{limits_Dasguptaetal-ii})}
   \nonumber\\
   && G_{r \theta_1 \theta_2 \ 10}=\frac{8 {g_s} M {f_ 1}({\theta_1}) (9 \cos ({\theta_1})-\cos (3
   {\theta_1})) \sin ({\theta_2}) \cos ({\theta_2})}{r (\cos (2
   ({\theta_1}-{\theta_2}))+\cos (2 ({\theta_1}+{\theta_2}))+6 \cos
   (2 {\theta_1})+6 \cos (2 {\theta_2})-14)}\sim\frac{1}{r\sqrt{\epsilon}}_{(\ref{limits_Dasguptaetal-i})/(\ref{limits_Dasguptaetal-ii})}\nonumber\\
   \end{eqnarray}}
The amount of near-horizon supersymmetry will be determined by solving for the killing spinor $\epsilon$ by the vanishing superysmmetric
variation of the gravitino in $D=11$ supergravity, which is given by:
\begin{equation}
\label{gravitino_susy}
\delta\psi_M = D_M\varepsilon - \frac{1}{288}\Gamma_M G^{M_1M_2M_3M_4}\Gamma_{M_1M_2M_3M_4}\varepsilon + \frac{1}{36}G_M^{\ M_1M_2M_3}\Gamma_{M_1M_2M_3}\varepsilon = 0,
\end{equation}
where $D_M\varepsilon = \partial_M\varepsilon + \frac{1}{4}\omega_M^{AB}\Gamma_{AB}\varepsilon$ where $M, M_.$ and $A,B$ are respectively the curved space and tangent space indices.  
Also,
\begin{eqnarray}
\Gamma_t &=& e^0_t\Gamma_0; 
\Gamma_{x_i} = e^i_{x_i}\Gamma_i,\ i=1, 2, 3; 
\Gamma_r = e^4_r\Gamma_4; 
\Gamma_{\theta_{1,2}} = e^{5,6}_{\theta_{1,2}}\Gamma_{5,6};\nonumber\\
\Gamma_{\phi_{1,2},\psi} &=&  e^7_{\phi_{1,2},\psi}\Gamma_7
+ e^8_{\phi_{1,2},\psi}\Gamma_8 + e^9_{\phi_{1,2},\psi}\Gamma_9.
\end{eqnarray}
Similarly, $G_u^{\ M_1M_2M_3} = G^M_{uu}G^{uM_1M_2M_3}(M_{1,2,3}\neq u)$ and $G_{\theta_1}^{\ M_1M_2M_3} = G^M_{\theta_1\theta_1}G^{\theta_1M_1M_2M_3}\\(M_{1,2,3} \neq \theta_1)$. 
The elfbeins near $r=r_h$ and $\theta_{1,2}=0,\pi$ are, in the limit (\ref{limits_Dasguptaetal-i}) and (\ref{limits_Dasguptaetal-ii}), constants implying the vanishing of the spin connection.

In the limit (\ref{limits_Dasguptaetal-ii}), from (\ref{three-form fluxes}), we see that that from $H_3\wedge A, H_3\wedge dx_{10}$, we end up with two types of fluxes: $G_{u M_1 M_2 M_3}(M_{1,2,3}\equiv
\theta_{1,2},\phi_{1,2},x_{10})$ and $G_{\theta_1 m_1 m_2 m_3}(m_{1,2,3}\equiv\theta_2,\phi_{1,2},x_{10})$. The former, using $G_{u\bullet\bullet\bullet} = u \epsilon^{\frac{59d}{3}}G_{r\bullet\bullet\bullet}$, is sub-dominant as compared to the latter and therefore will be dropped in the subsequent analysis. Hence only $G_{\theta_1\theta_2\phi_1\phi_2}, G_{\theta_1\theta_2\phi_1\ 10}, G_{\theta_1\theta_2\phi_2\ 10}$ survive.

Now, $\delta\psi_{u,t,x^{1,2,3}}\sim G^{M_1M_2M_3M_4}\Gamma_{t,u,x^{1,2,3}}\Gamma_{M_1M_2M_3M_4}\varepsilon $ (because
$G_{u,t,x^{1,2,3}}^{\ \ \ \ \ M_1M_2M_3}=0$), which given that in (\ref{limits_Dasguptaetal-i}) or (\ref{limits_Dasguptaetal-ii}) and near $\theta_{1,2}=0,\pi$ is approximately diagonal in $u,\theta_{1,2},x_{10}$, is proportional to $G^{\theta_1\theta_2M_1M_2}(\epsilon)\Gamma_{t,u,x^{1,2,3}}\Gamma_{\theta_1\theta_2M_1M_2}\varepsilon
\sim G^{\theta_1\theta_2M_1M_2}(\epsilon)\Gamma_{t,u,x^{1,2,3}}\\\Gamma_{\theta_1\theta_2}\Gamma_{M_1M_2}\varepsilon,  (M_1,M_2,M_3)=(\phi_{1,2},x_{10})$. So, these variations will vanish upon imposing:
\begin{equation}
\label{G4.Gamma4ii}
\sum_{(M_1,M_2)=(\phi_{1,2},x_{10})}G^{\theta_1\theta_2M_1M_2}(\epsilon)\Gamma_{M_1M_2}\varepsilon = 0.
\end{equation}
From (\ref{G4.Gamma4ii}), we see that that $\delta\psi_{\theta_{1,2}}=0$ is automatically satisfied.

Finally, $\delta\psi_{\phi_{1,2},\psi,x_{10}}=0$ will be considered. We will work in the basis $(u,\theta_{1,2},
\phi_{1,2}\equiv \pm\phi_1 - \phi_2,\psi,x_{10})$ with corresponding tangent space indices given by $(4;5,6;7,8;9;10)$. So, we end up with the following set of equations:
\begin{eqnarray}
\label{killing-components_a}
& & \frac{\partial\varepsilon}{\partial\Phi_1} + \beta'G_{\Phi_1}^{\ \  abc}\Gamma_{abc}\varepsilon = 0, \frac{\partial\varepsilon}{\partial\Phi_2} + \beta'G_{\Phi_2}^{\ \ abc}\Gamma_{abc}\varepsilon = 0,\nonumber\\
& & \frac{\partial\varepsilon}{\partial\psi} + \beta'G_{\psi}^{\ \ abc}\Gamma_{abc}\varepsilon = 0,  \frac{\partial\varepsilon}{\partial x_{10}} + \beta'G_{x_{10}}^{\ \ \ \ abc}\Gamma_{abc}\varepsilon = 0,
\end{eqnarray}
which utilizing the facts that the most dominant contributions of the $G_4$ flux components of the type $G_{\psi\bullet\bullet\bullet}$ are $G_{\psi\Phi_{1,2}\theta_1\theta_2}$ and of the type $G_{x_{10}\bullet\bullet\bullet}$ are $G_{x_{10}\theta_1\theta_2\Phi_{1,2}}$, are respectively equivalent to the following set of equations:
{\small \begin{eqnarray}
\label{killing-components_b}
& & \frac{\partial\varepsilon}{\partial\Phi_1} +\beta g_{\Phi_1\Phi_1}E^{\Phi_1}_{\ 7}G^{7568}\Gamma_{568}\varepsilon=\frac{\partial\varepsilon}{\partial\Phi_1} +\beta e^{-\frac{61}{12}}G^{7568}\Gamma_{568}\varepsilon=0\nonumber\\
& &  \frac{\partial\varepsilon}{\partial\Phi_2} + \beta g_{\Phi_2\Phi_2}E^{\Phi_2}_{\ 8}G^{8567}\Gamma_{567}\varepsilon=\frac{\partial\varepsilon}{\partial\Phi_1} + \beta e^{-\frac{61}{12}}G^{8567}\Gamma_{567}\varepsilon=0\nonumber\\
& & \frac{\partial\varepsilon}{\partial\psi} + \beta g_{\psi\psi_1}E^{\psi}_{\ 9}\left(G^{9567}\Gamma_{567} + G^{9568}\Gamma_{568}\right)\varepsilon=\frac{\partial\varepsilon}{\partial\Phi_1} + \beta e^{-\frac{16}{3}}\left(G^{9567}\Gamma_{567} + G^{9568}\Gamma_{568}\right)\varepsilon=0,\nonumber\\
& & \frac{\partial\varepsilon}{\partial x_{10}} + \beta g_{x_{10}x_{10}}E^{x_{10}}_{ \overline{10}}\left(G^{\overline{10}567}\Gamma_{567} + G^{\overline{10}568}\Gamma_{568}\right)\varepsilon=\frac{\partial\varepsilon}{\partial x_{10}} + \beta\epsilon^{\frac{2}{3}}\left(G^{\overline{10}567}\Gamma_{567} + G^{\overline{10}568}\Gamma_{568}\right)\varepsilon\nonumber\\
&& =0.
\end{eqnarray}}
 In addition to (\ref{G4.Gamma4ii}), one imposes: $
\Gamma_7\varepsilon=\pm\varepsilon;\ \Gamma_8\varepsilon=\pm\varepsilon$.
This implies that (\ref{G4.Gamma4ii}) becomes:
\begin{equation}
\label{G4.Gamma4iii}
\left(G^{5678}(\epsilon)  \pm G^{567\overline{10}}(\epsilon)\Gamma_{\overline{10}} \pm G^{568\overline{10}}\Gamma_{\overline{10}}\right)\varepsilon=0,
\end{equation}
and the following solution of the killing spinor equation is obtained:
\begin{eqnarray}
\label{killing-spinor-solutionii}
& & \varepsilon(\theta_{1,2},\Phi_{1,2},\psi,x_{10})= e^{\mp\beta\Phi_1\epsilon^{-\frac{61}{12}}G^{5678}\Gamma_{56}}.e^{\mp\beta\Phi_2\epsilon^{-\frac{61}{12}}G^{5678}\Gamma_{56}}.e^{-\beta\psi\epsilon^{-\frac{16}{3}}(\mp G^{5679}\Gamma_{56} \mp G^{5689}\Gamma_{56})}\nonumber\\
& &.e^{x_{10}\epsilon^{\frac{2}{3}}(G^{567\overline{10}}\Gamma_{56} \pm G^{568\overline{10}}\Gamma_{56})}\varepsilon_0.
\end{eqnarray}
So, we obtain, once again, a near-horizon $\frac{1}{8}$-supersymmetric $M3$-brane solution near $\theta_{1,2}=0,\pi$.

For values of $\theta_{1,2}$ away from 0, $\pi$ and
$r>r_h$, we expect a reduced amount of supersymmetry. In other words we expect a near-horizon
enhancement of supersymmetry (See \cite{enhancement-SUSY-horizon}). We hope to get back to this issue
in a subsequent work.
\section{Hydrodynamics and  Thermodynamics of M-Theory Uplift}
In this section, we calculate hydrodynamical as well as thermodynamical quantities of local M-theory uplift obtained in {\bf Section 4} in
the limits (\ref{limits_Dasguptaetal-i}) and (\ref{limits_Dasguptaetal-ii}) of  M-theory. From \cite{KovtunSonStarinets}, for black-brane solutions
of the type:
\begin{equation}
\label{KSS-a}
ds^2 = G_{00}(r)dt^2 + G_{rr}dr^2 + G_{xx}(r)\sum_{i=1}^p \left(dx^i\right)^2 + Z(r)K_{mn}(y)dy^i dy^j,
\end{equation}
where in the vicinity of the horizon $r=r_0$, $G_{00}$ vanishes, $G_{rr}$ diverges and $G_{xx}(r)$ and $Z(r)$ remain finite. Then demanding the absence of a conical singularity at the origin, the Hawking temperature associated with (\ref{KSS-a}) is given by (\cite{schmude}):
\begin{equation}
\label{KSS-b}
T=\frac{\partial_r G_{00}}{4\pi \sqrt{G_{00}G_{rr}}}.
\end{equation}
Now, in both limits (\ref{limits_Dasguptaetal-i}) and (\ref{limits_Dasguptaetal-ii}), $G_{00}^{\cal M}, G_{rr}^{\cal M}$ have no angular dependence and hence
the temperature (\ref{KSS-b})  of the black $M3$-brane (\ref{Mtheory met}) then turns out to be given by:
\begin{eqnarray}
\label{T}
& &\hskip-0.2in T =\frac{\sqrt{2}}{{r_h} \sqrt{\pi} \sqrt{\frac{{g_s} \left(18 {g_s}^2 {N_f} ln ^2({r_h}) {M_{\rm eff}}^2+3 {g_s} (4 \pi -{g_s} {N_f}
   (-3+ln (2))) ln ({r_h}) {M_{\rm eff}}^2+8 N \pi ^2\right)}{{r_h}^4}}}\stackrel{(\ref{limits_Dasguptaetal})}{\longrightarrow}\frac{r_h}{\pi L^2}. \nonumber\\
   \end{eqnarray}
To get a numerical estimate for $r_h$,  we see that equating $T$ to $\frac{r_h}{\pi L^2}$, (\ref{T}) is solved, in
both limits (\ref{limits_Dasguptaetal-i}) and the MQGP limit  (\ref{limits_Dasguptaetal-ii}) by $r_h=1+\epsilon$, where $0<\epsilon<1$.
\subsection{ Shear-Viscosity-to-Entropy-Density Ratio}
Now, the shear-viscosity-to-entropy-density ratio in the hydrodynamical gravity dual of \cite{KovtunSonStarinets}  is given by:
\begin{equation}
\label{etaovers-i}
\frac{\eta}{s}= T \frac{\sqrt{|G |}}{\sqrt{|G _{00}G _{rr}|}}\Biggr|_{r=r_h}\int_{r_h}^\infty dr \frac{|G_{00}G_{rr}|}{G_{{\bf R}^p}\sqrt{|G |}}.
\end{equation}
We first check the estimate evaluate $\eta/s$ for type IIA mirror metric of (\ref{metric-mirror}).
The simplified expressions of relevant metric components and its determinant in the limits (\ref{limits_Dasguptaetal-i}) as well as (\ref{limits_Dasguptaetal-ii}), freezing the angular dependence
on $\theta_{1,2}$ (there being no dependence on $\phi_{1,2},\psi,x_{10}$ in
 (\ref{limits_Dasguptaetal-i}) or (\ref{limits_Dasguptaetal-ii})), are:
{\small  \begin{eqnarray}
  \label{simGIIA}
&&G^{IIA}_{00}\sim \frac{r^2 \left(\frac{{r_h}^4}{r^4}-1\right)}{2 \sqrt{\pi } \sqrt{{g_s} N}}, G^{IIA}_{{\bf R}^3}\sim \frac{r^2}{2 \sqrt{\pi } \sqrt{{g_s} N}}, G^{IIA}_{rr} \sim \frac{2 \sqrt{\pi } \sqrt{{g_s} N}}{r^2 \left(1-\frac{{r_h}^4}{r^4}\right)} ,\nonumber\\
& & \sqrt{|G^{IIA}|}\sim \nonumber\\
&& \left\{\frac{r^3 {f_2}({\theta_2}) (\cos (2 {\theta_1})-5)^3 \left(3 \sqrt{6}-2 \cot ({\theta_1})\right) \cot ({\theta_1})  
   \sin ^2({\theta_2}) \cos ({\theta_2}) \sqrt{{f_1}({\theta_1})^2 \sin ^2({\theta_1})+1}}{11664 \sqrt{3} \sqrt[4]{\pi } \sqrt[4]{{g_s}}
   \sqrt[4]{N} \left(3 \sin ^2({\theta_1}) \sin ^2({\theta_2})+2 \sin ^2({\theta_1}) \cos ^2({\theta_2})+2 \cos ^2({\theta_1}) \sin
   ^2({\theta_2})\right)^2}\right\} .\nonumber\\
\end{eqnarray}}
Again we note that in (\ref{limits_Dasguptaetal-i}) or (\ref{limits_Dasguptaetal-ii}),
$G_{00,rr,{\bf R}^3}^{IIA}$ are independent of the angular coordinates;
additionally it was possible to tune the chemical potential $\mu_C$ to a small value - as shown in Sec. {\bf 4}. This permits use of
(\ref{etaovers-i}). Utilizing above,
{\small
\begin{eqnarray}
\label{int}
&& \int_{r_h}^\infty dr\frac{|G^{IIA}_{00}G^{IIA}_{rr}|}{G^{IIA}_{{\bf R}^3}\sqrt{|G^{IIA}|}} \sim \nonumber\\
&&  \frac{23328 \sqrt{3} \pi ^{3/4} {g_s}^{3/4} N^{3/4}}{4\ {r_h}^4} \times \nonumber\\
&& \left\{\frac{ \sin ^2({\theta_1}) \tan ({\theta_1}) \csc ^3({\theta_2})   \left(3 \sin
   ^2({\theta_1}) \sin ^2({\theta_2})+2 \sin ^2({\theta_1}) \cos ^2({\theta_2})+2 \cos ^2({\theta_1}) \sin ^2({\theta_2})\right)^2}{
   {f_2}({\theta_2}) (\cos (2 {\theta_1})-5)^3 \left(3 \sqrt{6}-2 \cot ({\theta_1})\right) \sqrt{{f_1}({\theta_1})^2 \sin
   ^2({\theta_1})+1}}\right\}. \nonumber\\
&& {\rm and} \nonumber\\
&& \lim_{r\rightarrow r_h+\epsilon}\frac{\sqrt{|G^{IIA}|}}{\sqrt{|G^{IIA}_{tt}G^{IIA}_{rr}|}}\sim \frac{{r_h}^3}{11664 \sqrt{3} \sqrt[4]{\pi } \sqrt[4]{{g_s}}
   \sqrt[4]{N} } \times \nonumber\\
   && \left\{\frac{  {f_2}({\theta_2}) (\cos (2 {\theta_1})-5)^3 \left(3 \sqrt{6}-2 \cot ({\theta_1})\right) \cos ({\theta_1}) 
   \sin ^4({\theta_2}) \cos ({\theta_2}) \sqrt{{f_1}({\theta_1})^2 \sin ^2({\theta_1})+1}}{\sin ^3({\theta_1})\left(3 \sin ^2({\theta_1}) \sin ^2({\theta_2})+2 \sin ^2({\theta_1}) \cos ^2({\theta_2})+2 \cos ^2({\theta_1}) \sin
   ^2({\theta_2})\right)^2}\right\}.\nonumber\\
\end{eqnarray}}
 Multiplying above expressions as according to equation (\ref{etaovers-i}) and  putting value of $T \sim \frac{1}{\pi \sqrt{4 \pi g_s N}}$ for $r_h \sim 1$, we get
\begin{eqnarray}
\label{eq:etaoversIIA}
\frac{\eta}{s}=\frac{r_h}{\pi \sqrt{4 \pi g_s N}}\times \frac{1}{2}\sqrt{{g_s} N}\sqrt{\pi}=\frac{1}{4 \pi}.
\end{eqnarray}
The simplified expressions of relevant metric components as well as determinant of 11D metric corresponding to most dominant contribution in the limits (\ref{limits_Dasguptaetal-i})
or (\ref{limits_Dasguptaetal-ii}):
\pagebreak
{\small \begin{eqnarray}
&& G^{\cal M}_{00}\sim -\frac{3^{2/3} r^2 \left(1-\frac{r_h^4}{r^4}\right)}{2 \sqrt{\pi } {g_s}^{2/3} \sqrt{ {g_s} N}},  G^{\cal M}_{{\bf R}^3}\sim\frac{3^{2/3}r^2}{2 \sqrt{\pi } {g_s}^{2/3} \sqrt{ {g_s} N}},  G^{\cal M}_{rr} \sim \frac{2\ 3^{2/3} \sqrt{\pi } \sqrt{ {g_s} N}}{ {g_s}^{2/3} r^2\left(1-\frac{r_h^4}{r^4}\right)},\nonumber\\
& & \sqrt{|G^{\cal M}|}\sim \frac{4 r^3 {f_2}({\theta_2}) \cos ^2({\theta_1}) \cot ^2({\theta_1}) \cos ^3({\theta_2})}{27\ 3^{5/6} \sqrt[4]{\pi } {g_s}^{8/3}
   \sqrt[4]{{g_s} N} \left(2 \sin ^2({\theta_1}) \cos ^2({\theta_2})+2 \cos ^2({\theta_1}) \sin ^2({\theta_2})\right)}.
\end{eqnarray}}
Utilizing above,
{\small \begin{eqnarray}
\label{int}
&& \int_{r_h}^\infty dr \frac{|G^{\cal M}_{00}G^{\cal M}_{rr}|}{G^{\cal M}_{{\bf R}^3}\sqrt{|G^{\cal M}|}} \sim \nonumber\\
&&  \int_{1+\epsilon}^\infty \frac{81 \sqrt{3} \pi ^{3/4} {g_s}^2 \sqrt[4]{{g_s} N} \sin ^2({\theta_1})  \sec ^3({\theta_2}) \sqrt{ {g_s} N}
   \left( \sin ^2({\theta_1}) \cos ^2({\theta_2})+ \cos ^2({\theta_1}) \sin ^2({\theta_2})\right)}{2 r^5 {f_2}({\theta_2})\cos ^4({\theta_1})} \,dr,\   \nonumber\\
   && \sim  \frac{81 \sqrt{3} \pi ^{3/4} {g_s}^2 \sqrt[4]{{g_s} N} \sin ^2({\theta_1})  \sec ^3({\theta_2}) \sqrt{ {g_s} N}
   \left(\sin ^2({\theta_1}) \cos ^2({\theta_2})+ \cos ^2({\theta_1}) \sin ^2({\theta_2})\right)}{4 {f_2}({\theta_2})\cos ^4({\theta_1})}. \nonumber\\
&& {\rm and} \nonumber\\
&& \lim_{r\rightarrow r_h+\epsilon}\frac{\sqrt{|G^{\cal M}|}}{\sqrt{|G^{\cal M}_{00}G^{\cal M}_{rr}|}}\sim  \frac{4 {f_2}({\theta_2}) \cos ^2({\theta_2}) \cot ^2({\theta_2}) \cos ^3({\theta_2})}{81 \sqrt{3} \sqrt[4]{\pi } \sqrt[4]{{g_s} N} \left(2 \sin
   ^2({\theta_2}) \cos ^2({\theta_2})+2 \cos ^2({\theta_2}) \sin ^2({\theta_2})\right)}.
\end{eqnarray}}
 Multiplying above expressions as according to equation (\ref{etaovers-i}) and  putting value of $T \sim \frac{1}{\sqrt{4 \pi g_s N}}$ for $r_h \sim 1$, we get
\begin{eqnarray}
\frac{\eta}{s}=\frac{r_h}{\pi \sqrt{4 \pi g_s N}}\times \frac{1}{2}\sqrt{{g_s} N}\sqrt{\pi}=\frac{1}{4 \pi}.
\end{eqnarray}
Needless to say, as mentioned earlier, given that $G_{00,rr,{\bf R}^3}^{IIA, {\cal M}}$ in both limits (\ref{limits_Dasguptaetal-i}) and (\ref{limits_Dasguptaetal-ii})
are independent of the angular coordinates and that one integrates only w.r.t. $r$ in (\ref{etaovers-i}), the angular portions of the metric's
determinants is bound to (and is explicitly verified above) to cancel out.
\subsection{Diffusion Coefficient }
The general expression of Diffusion coefficient in the hydrodynamical gravity dual of \cite{KovtunSonStarinets} is given by:
\begin{equation}
\label{D}
D= \frac{\sqrt{|G|}Z(r)}{G\sqrt{|G_{00}G_{rr}|}}\Biggr|_{r=r_h}\int_{r_h}^\infty dr
\frac{|G_{00}G_{rr}|}{\sqrt{|G|}Z(r)}.
\end{equation}
where $Z(r) =\sqrt{h}r^2$ and $\sqrt{h}$ is warp factor in type IIB/IIA background. Justifications
similar to the ones given in {\bf 5.1} also permit the use of (\ref{D}).

Here, we first check the estimate of diffusion coefficient in type IIB background using type IIB metric given in (\ref{metric}).
The simplified expressions of relevant metric components and determinant of  using the limits of (\ref{limits_Dasguptaetal}) are:
{\small \begin{eqnarray}
&& g_{00}^{IIB}\sim \frac{r^2 \left(-1+\frac{{r_h}^4}{r^4}\right)}{2 \sqrt{\pi } \sqrt{{g_s} N}},  g_{rr}^{IIB} \sim \frac{2 \sqrt{\pi } \sqrt{{g_s} N}}{r^2 \left(1-\frac{{r_h}^4}{r^4}\right)} ,  \sqrt{|g^{IIB}|}\sim \frac{r^6 \left(1-{f_2}({\theta_2})^2 \left(\sin ^2({\theta_2})-\sin ^2({\theta_1})\right)\right)}{8 \sqrt{\pi } \sqrt{{g_s} N}}.\nonumber\\
\end{eqnarray}}
Incorporating above-mentioned results in equation (\ref{D}) and further simplifying, 
\begin{equation}
\label{D1}
D= \frac{\sqrt{2} \pi ^{3/4} {r_h} ({g_s} N)^{3/4}}{ {\cal A}(\theta_1,\theta_2)}\int_{r_h}^\infty {\cal A}(\theta_1,\theta_2) \frac{\sqrt{2}}{\sqrt[4]{\pi } r^3 \sqrt[4]{{g_s} N}} dr;
\end{equation} $
{\cal A}(\theta_1,\theta_2)\sim \sqrt{1-{f_2}({\theta_2})^2 \left(\sin ^2({\theta_2})-\sin ^2({\theta_1})\right)}$.
On solving equation (\ref{D1}), we get:
\begin{equation}
D = \frac{\sqrt{ {g_s} N} \sqrt{\pi }}{r_h}=\frac{1}{2 \pi T}.
\end{equation}

Similarly, for type IIA background formed by applying transformation rules  on (\ref{metric}), the simplified expressions of relevant metric components in the limits of (\ref{limits_Dasguptaetal}) are already given in (\ref{simGIIA}). Incorporating the same in eq.~(\ref{D}) and further simplifying,
\begin{equation}
\label{D2}
D= \frac{5832  \pi ^{3/4} {r_h} ({g_s} N)^{3/4}}{ {\cal A^{\prime}}(\theta_1,\theta_2)}\int_{r_h}^\infty {\cal A^{\prime}}(\theta_1,\theta_2) \frac{1}{2916 \sqrt[4]{\pi } r^3 \sqrt[4]{{g_s} N}} dr;
\end{equation}
{\small \begin{eqnarray}
\hskip -0.2in {\cal A^{\prime}}(\theta_1,\theta_2)\sim \left\{\frac{\sin ^3({\theta_1})  \csc ^4({\theta_2})  \left(3 \sin ^2({\theta_1}) \sin ^2({\theta_2})+2 \sin
   ^2({\theta_1}) \cos ^2({\theta_2})+2 \cos ^2({\theta_1}) \sin ^2({\theta_2})\right)^2}{{f_2}({\theta_2}) \cos ({\theta_1}) \cos ({\theta_2})(\cos (2 {\theta_1})-5)^3 \left(3
   \sqrt{6}-2 \cot ({\theta_1})\right) \sqrt{{f_1}({\theta_1})^2 \sin ^2({\theta_1})+1}}\right\}.\nonumber
\end{eqnarray}}
After solving equation (\ref{D2}), one gets:
\begin{equation}
D = \frac{\sqrt{ {g_s} N} \sqrt{\pi }}{r_h}=\frac{1}{2 \pi T}.
\end{equation}
\subsection{Partition Function}
To study, further, the thermodynamic properties of the solution (\ref{Mtheory met}), we need to
evaluate the partition
function ${\cal Z} = e^{- \cal{S}_E}$, where keeping in mind  $l_s \sim r_h$,
 higher order $\alpha^{\prime}$ corrections become important, the action we will consider
will include ${\cal O}(R^4)$-terms (See \cite{d_11_R_fourth}):
\begin{eqnarray}
  \label{eq:GravitationalAction}
  {\cal S_{\cal E}} & = & \frac{1}{16\pi}\int_{\cal{M}} d^{11}\!x \sqrt{G^{\cal M}} R^{\cal M}
   + \frac{1}{8\pi}\int_{\partial M} d^{10}\!x K^{\cal M} \sqrt{\hat{h}}-\frac{1}{4}\int_{M} \Bigl( |G_4|^2-{C_3\wedge G_4\wedge G_4}\bigr) \nonumber\\
& & \frac{T_2}{{2\pi}^4. 3^2.2^{13}}\int_{\cal{M}} d^{11}\!x \sqrt{G^{\cal M}}(J-\frac{1}{2}E_8)+T_2 \int C_3 \wedge X_8 - {\cal S}^{\rm ct},
\end{eqnarray}
where   $(J,E_8,X_8)$ are quartic polynomials in 11-dimensional space and defined as:
\begin{equation}
\label{J}
J = 3.2^8 \Bigl(R^{mijn}R_{pijq} R_{m}^{\ rsp}R^{q}_{\ rsn}+\frac{1}{2}R^{mnij}R_{pqij} R_{m}^{\ rsp}R^{q}_{\ rsn}\Bigr),
\end{equation}
\begin{equation}
\label{E_8}
E_8= \epsilon^{abc m_1 n_1...m_4 n_4} \epsilon_{abc m^{\prime}_1 n^{\prime}_1...m^{\prime}_4 n^{\prime}_4}R^{m^{\prime}_1 n^{\prime}_1}_{\ \ \ \ \ \ m_1 n_1}...R^{m^{\prime}_4 n^{\prime}_4}_{\ \ \ \ \ \ m_4 n_4},
\end{equation}
   \begin{equation}
 X_8 \sim \frac{1}{192 \cdot (2 \pi^2)^4}\Bigl[tr(R^4)- (tr{R^2})^2\Bigr],
 \end{equation}
for Euclideanised space-time where $\cal{M}$ is a volume of spacetime defined by $r <r_\Lambda$, where the counter-term ${\cal S}^{\rm ct}$ is added such that the Euclidean action ${\cal S}_{\cal E}$ is finite \cite{S_ct-Perry_et_al},\cite{Mann+Mcnees}.  The action (\ref{eq:GravitationalAction}), apart from being divergent (as $r\rightarrow\infty$) also possesses pole-singularities near $\theta_{1,2}=0,\pi$. We will regulate the second divergence by taking a
small $\theta_{1,2}$-cutoff $\epsilon_\theta$, $\theta_{1,2}\in[\epsilon_\theta,\pi-\epsilon_\theta]$, and demanding $\epsilon_\theta
\sim\epsilon^\gamma$, for an appropriate $\gamma$.  We will then explicitly check that the finite part
of (\ref{eq:GravitationalAction}) turns out to be independent of this cut-off $\epsilon/\epsilon_\theta$.

For M-theory thermodynamical calculations, we provide a slightly more detailed explanation of
the limits (\ref{limits_Dasguptaetal-i}) and (\ref{limits_Dasguptaetal-ii}):
 
    In weak coupling  - large t'Hooft coupling(s) limit of M-theory, we consider the limits:
$$g_s<<1, g_s M>>1, g_s N>>, \frac{g_s M^2}{N}<<1, {g_s}^2 M N_f<<1$$  similar to \cite{metrics}. Based on that, we assume that these parameters scale with $\epsilon$ as
  $g_s \sim {\epsilon^{d}} $, $N/N_{\rm eff}\sim  {\epsilon^{-a}} $, $M/M_{\rm eff}\sim{\epsilon^{-b}}$
 such that  ${2b-a-d}<0, {2d-b}<0, {b-d}>0,{a-d}>0$, $\forall$ $ \epsilon\rightarrow 0 $, $a,b,d > 0 $.  Further, for
(\ref{limits_Dasguptaetal-i}),  we choose $a=19 d, {b}={\frac{3}{2}}d, \gamma={\frac{1}{2}}d $. For the purpose of obtaining
the simplified type IIA mirror components (\ref{metric-mirror}) and its M-theory uplift simplified metric components of
(\ref{eq:simplifiedmetriccomponents}), we had set $\epsilon\leq0.01$.  However, it is understood
that in the identification $\epsilon_\theta=\epsilon^\gamma$, one will eventually have to take $\epsilon_\theta$ and
thus $\epsilon$ to be very small.
 
In MQGP limit:
$$g_s, g_s M, {g_s}^2 M N_f\equiv{\rm finite}, g_s N>>1, \frac{g_s M^2}{N}<<1.$$
 Now we assume the scaling of these parameters with $\epsilon$ as
 $g_s \sim {\alpha_1} {\epsilon^{d}} $, $N/N_{\rm eff}\sim\alpha_2 {\epsilon^{-a}} $, $M/M_{\rm eff}\sim\alpha_3{\epsilon^{-b}}$,  such that  ${2b-a-d}<0, {2d-b}<0, {b-d}>0,{a-d}>0$, $ a,b,d > 0 $. To obey these constraints,  we define $a=39 d, {b}={\frac{3}{2}}d, \gamma={\frac{3}{2}}d $. For the purpose of obtaining
the simplified type IIA mirror components (\ref{metric-mirror}) and its M-theory uplift simplified metric components of
(\ref{eq:simplifiedmetriccomponents}), we had taken $\alpha_{1,2,3}\sim{\cal O}(1), \epsilon\lesssim 1$. However, it is understood
that in the identification $\epsilon_{\theta_{1,2}}\sim\epsilon^\gamma$ given that one will eventually have to take $\epsilon_\theta$ and
thus $\epsilon$ to be very small, this would numerically imply taking large $\alpha_1$ and small $\alpha_3$ such that $g_s, M$ are
individually finite in the MQGP limit. Further, writing $\epsilon_{\theta_1}=\alpha_4\epsilon^\gamma, \epsilon_{\theta_2}=\alpha_5\epsilon^\gamma$, one will see in {\bf 6.3.1 - 6.3.4} that the Gibbons-Hawking-York surface term will be the only term containing an finite contribution which would be proportional to $\frac{1}{\alpha_1^{\frac{9}{4}}\alpha_2^{\frac{1}{4}}\alpha_4^4\alpha_5}$. In the
$\epsilon_{\theta_{1,2}}\rightarrow0$-limit one will take $\alpha_{1,3,4,5}:\alpha_1^{\frac{9}{4}}\alpha_4^4\alpha_5\equiv$finite; $\alpha_2\sim{\cal O}(1)$ always and the simplified (\ref{metric-mirror})
and (\ref{eq:simplifiedmetriccomponents}) continue to be valid.

\subsubsection{Einstein-Hilbert Action}

We now evaluate the contribution of Einstein-Hilbert action in both limits
(\ref{limits_Dasguptaetal-i}) and (\ref{limits_Dasguptaetal-ii}) of M-theory. In either limit,
the simplified expression  of determinant of 11 dimensional M-theory uplift is  given as:
\pagebreak
{\small
\begin{eqnarray}
\label{eq:g}
&& {G^{\cal M}}\sim \Bigl(r^6 \cos ^2({\theta_2}) \csc ^2({\theta_1}) {f_2}({\theta_2})^2 \left({f_1}({\theta_1})^2 \sin ^2({\theta_1})+1\right) \sin
   ^2({\theta_2}) \Bigl(\Bigl(4 \cos ({\theta_2}) \left(2 \cot ({\theta_1})-3 \sqrt{6}\right)\nonumber\\
   && \cot ({\theta_2}) +18 \sqrt{6} \sin ({\theta_2})\Bigr)
   \cos ^3({\theta_1})+12 \cos ({\theta_2}) \cot ({\theta_2}) \sin ({\theta_1}) \cos ^2({\theta_1})+27 \sin ^2({\theta_1}) \Bigl(\sqrt{6} \sin
   ({\theta_2})\nonumber\\
   && -4 {h_5} \cos ({\theta_2})\Bigr)\cos ({\theta_1})  +81 {h_5} \sin ^3({\theta_1}) \left(\sqrt{6} \cos ({\theta_2})+3 {h_5}
   \sin ({\theta_2})\right)\Bigr)^2\Bigr)/\Bigl(  {g_s}^{16/3} \sqrt{{g_s} N}\sqrt{\pi } \nonumber\\
   &&   \Bigl(2 \cos ^2({\theta_2}) \sin ^2({\theta_1}) +3 \sin
   ^2({\theta_2}) \sin ^2({\theta_1})+2 \cos ^2({\theta_1}) \sin ^2({\theta_2})+3 {h_5} \sin (2 {\theta_1}) \sin (2 {\theta_2})\Bigr)^2\Bigr).
\end{eqnarray}}
Ricci scalar is given as: $
R= G^{MN} G^{PQ} R_{M  P N Q}\ {\rm where}\ M, N, P, Q = 0 ,..10.$
For particular choice's of scaling parameters, we see that the contribution of Ricci scalar is dominated by $G^{\cal M}\ ^{\theta_1 \theta_1} G^{\cal M}\ ^{x z} R^{\cal M}_{ x \theta_1 z \theta_1}$ component. On simplifying the same using equation (\ref{eq:simplifiedmetriccomponents}), we have: 
{\small
\begin{eqnarray}
\label{eq:invYtheta2}
&&   G^{\cal M}\ ^{x z}\sim \Biggl(27 \sqrt[3]{3} {g_s}^{2/3} (\cos (2 {\theta_1})-5) \sin ({\theta_1}) \sin ({\theta_2})\Biggr)/\Biggl(2 \Bigl(\Bigl(4 \cos ({\theta_2}) \left(2 \cot
   ({\theta_1})-3 \sqrt{6}\right)  \cot ({\theta_2})\nonumber\\
   &&+18 \sqrt{6} \sin ({\theta_2})\Bigr)  \cos ^3({\theta_1})+12 \cos ({\theta_2}) \cot
   ({\theta_2}) \sin ({\theta_1}) \cos ^2({\theta_1})+27 \sin ^2({\theta_1}) \Bigl(\sqrt{6} \sin ({\theta_2})\nonumber\\
   &&-4 {h_5} \cos ({\theta_2})\Bigr)
   \cos ({\theta_1})  +81 {h_5} \sin ^3({\theta_1}) \left(\sqrt{6} \cos ({\theta_2})+3 {h_5} \sin ({\theta_2})\right)\Bigr)\Biggr),
\end{eqnarray}
\begin{eqnarray}
\label{eq:invtheta1theta1}
&&  G^{\cal M}\ ^{\theta_1 \theta_1}\sim \frac{\sqrt[3]{3} \sqrt[6]{{g_s}}}{\sqrt{N} \sqrt{\pi } \left({f_1}({\theta_1})^2 \sin ^2({\theta_1})+1\right)},
\end{eqnarray}}
and
{\small
\begin{eqnarray}
\label{eq:R7696}
 &&  R_{x \theta_1 z \theta_1}^{\cal M} \sim -\biggl(2 \csc ({\theta_1}) \csc ({\theta_2}) \Bigl(12 \cos ({\theta_2}) \cot ({\theta_2}) \cos ^3({\theta_1})-36 \cot ({\theta_2}) \sin
   ({\theta_1}) \Bigl(\sqrt{6} \cos ({\theta_2}) \nonumber\\
    &&-6 {h_5} \sin ({\theta_2})\Bigr) \cos ^2({\theta_1})+\Bigl(\cos ({\theta_2}) \left(8 \cot
   ({\theta_2}) \cot ^2({\theta_1})+3 \left(8 \cot ({\theta_2})-81 \sqrt{6} {h_5}\right) \sin ^2({\theta_1})\right)\nonumber\\
   &&-729 {h_5}^2 \sin
   ^2({\theta_1}) \sin ({\theta_2})\Bigr)   \cos ({\theta_1})+27 \sin ^3({\theta_1}) \left(\sqrt{6} \sin ({\theta_2})-4 {h_5} \cos
   ({\theta_2})\right)\Bigr)^2\biggr)/\biggl(27 \sqrt[3]{3} {g_s}^{2/3}\nonumber\\
   &&  (\cos (2 {\theta_1})-5) \Bigl(2 \Bigl(2 \cos ({\theta_2})  \left(3 \sqrt{6}-2 \cot
   ({\theta_1})\right) \cot ({\theta_2})-9 \sqrt{6} \sin ({\theta_2})\Bigr) \cos ^3({\theta_1})-12 \cos ({\theta_2})\nonumber\\
   && \cot ({\theta_2}) \sin
   ({\theta_1}) \cos ^2({\theta_1})  -27 \sin ^2({\theta_1}) \left(\sqrt{6} \sin ({\theta_2})-4 {h_5} \cos ({\theta_2})\right) \cos
   ({\theta_1})-81 {h_5} \sin ^3({\theta_1}) \nonumber\\
   &&\left(\sqrt{6} \cos ({\theta_2})+3 {h_5} \sin ({\theta_2})\right)\Bigr)\biggr).
\end{eqnarray}}
Using set of equations (\ref{eq:g})- (\ref{eq:R7696}), the most dominant contribution of 11-Dimensional Bulk term is given by analytical expression given as:
\pagebreak
{\small
\begin{eqnarray}
&&   \sqrt{G^{\cal M}} R^{\cal M}\sim \biggl(r^3 \cos ({\theta_2}) \csc ({\theta_1}) {f_2}({\theta_2}) \sin ({\theta_2}) \Bigl(12 \cos ({\theta_2}) \cot ({\theta_2}) \cos
   ^3({\theta_1})-36 \cot ({\theta_2}) \sin ({\theta_1}) \nonumber\\
   &&\Bigl(\sqrt{6} \cos ({\theta_2}) -6 {h_5} \sin ({\theta_2})\Bigr) \cos
   ^2({\theta_1})+\Bigl(\cos ({\theta_2}) \Bigl(8 \cot ({\theta_2}) \cot ^2({\theta_1})+3 \left(8 \cot ({\theta_2})-81 \sqrt{6} {h_5}\right)\nonumber\\
   && \sin
   ^2({\theta_1})\Bigr) -729 {h_5}^2 \sin ^2({\theta_1}) \sin ({\theta_2})\Bigr) \cos ({\theta_1})+27 \sin ^3({\theta_1}) \left(\sqrt{6} \sin
   ({\theta_2})-4 {h_5} \cos ({\theta_2})\right)\Bigr)^2\biggr)/\nonumber\\
   && \biggl(54 {g_s}^{11/4} N^{3/4} \pi ^{3/4} \sqrt{3 {f_1}({\theta_1})^2 \sin
   ^2({\theta_1})+3} \Bigl(\Bigl(4 \cos ({\theta_2}) \left(2 \cot ({\theta_1})-3 \sqrt{6}\right) \cot ({\theta_2})\nonumber\\
    &&+18 \sqrt{6} \sin
   ({\theta_2})\Bigr) \cos ^3({\theta_1})+12 \cos ({\theta_2}) \cot ({\theta_2}) \sin ({\theta_1}) \cos ^2({\theta_1})+27 \sin ^2({\theta_1})
   \bigl(\sqrt{6} \sin ({\theta_2})\nonumber\\
   &&-4 {h_5} \cos ({\theta_2})\bigr) \cos ({\theta_1})+81 {h_5} \sin ^3({\theta_1})   \Bigl(\sqrt{6} \cos
   ({\theta_2})  +3 {h_5} \sin ({\theta_2})\Bigr)\Bigr) \Bigl(2 \cos ^2({\theta_2}) \sin ^2({\theta_1})\nonumber\\
   &&+3 \sin ^2({\theta_2}) \sin
   ^2({\theta_1})+2 \cos ^2({\theta_1}) \sin ^2({\theta_2}) +3 {h_5} \sin (2 {\theta_1}) \sin (2 {\theta_2})\Bigr)\biggr).
   \end{eqnarray}}
The Einstein-Hilbert action receiving the most dominant contribution near $\theta_{1,2}=0,\pi$, we simplify the above near the same and obtain:
{\small
 \begin{eqnarray}
   && \sqrt{G^{\cal M}} R^{\cal M}\sim \frac{125  r^3 \cos ({\theta_2}) \cot ^2({\theta_2}) \csc ^4({\theta_1}) {f_2}({\theta_2})}{864 \sqrt{3} {g_s}^{11/4} N^{3/4} \pi ^{3/4}}.
 \end{eqnarray}}
We assume that result of
integration with respect to ${\theta_{1,2}}$ variables, is simply given by summing up the contribution of integrand near ${\theta_{1,2}}=\epsilon_{\theta_{1,2}}$ and  ${\theta_{1,2}}= \pi-\epsilon_{\theta_{1,2}}$.
Integrating other angular as well as radial variables, we have
{\small
\begin{eqnarray}
\label{eq:intSbulk}
&&  \frac{1}{16 \pi} \int_{x_{10}\in[0,{2\pi}], r\in[r_h,r_\Lambda],\theta_{1,2}\in[\epsilon_{\theta_{1,2}},\pi-\epsilon_{\theta_{1,2}}],\phi_{1,2}\in[0,2\pi], \psi\in[0,4\pi]}\sqrt{G^{\cal M}} R^{\cal M} \nonumber\\
&\sim&    \frac{125 {\pi}^{\frac{9}{4}} r^{4}_{\Lambda} \cos ({\theta_2}) \cot ^2({\theta_2}) \csc ^4({\theta_1}) {f_2}({\theta_2})}{1728 \sqrt{3} {g_s}^{11/4} {N}^{3/4}}\Biggr|_{\theta_{1,2}=\epsilon_{\theta_{1,2}} + \theta_{1,2}=\pi-\epsilon_{\theta_{1,2}}}.
\end{eqnarray}}
\underline{Limit (\ref{limits_Dasguptaetal-i})}:
Incorporating   $g_s \sim \ {\epsilon}$, $N/N_{\rm eff}\sim \ {\epsilon^{-19}} $, $M/M_{\rm eff}\sim {\epsilon^{-\frac{3}{2}}}$
$\theta_{1,2}\sim   {\epsilon^{\frac{1}{2}}}$, ${f_2}({{\theta_{1,2}}})\sim { \frac{1}{{\theta_{1,2}}}}$, we see that
\begin{eqnarray}
\label{eq:intS_EH1}
  &&  { {\cal S}_{\rm EH}}\sim \frac{1}{16 \pi} \int_{x^0
\in \left[0,\frac {\pi \sqrt{4\pi g_s N}}{r_h}\right],x_{10}\in[0,{2\pi}], r\in[r_h,r_\Lambda],\theta_{1,2}\in[\epsilon_{\theta_{1,2}},\pi-\epsilon_{\theta_{1,2}}],\phi_{1,2}\in[0,2\pi], \psi\in[0,4\pi]} \sqrt{G^{\cal M}} R^{\cal M} \nonumber\\
 &&=a_{\rm EH}{\frac{r^{4}_{\Lambda}}{\epsilon r_h}}.
  \end{eqnarray}
 \underline{MQGP limit (\ref{limits_Dasguptaetal-ii})}:
  Incorporating  $g_s={\alpha_1}\epsilon $, $N/N_{\rm eff}=\alpha_2  {\epsilon^{-39}} $, $M/M_{\rm eff}=\alpha_3{\epsilon^{-\frac{3}{2}}}$,  
   $\theta_{1,2}=\alpha_{4,5} {\epsilon^{\frac{3}{2}}}$, ${f_2}(\epsilon_{\theta_{1,2}})\sim  \frac{1}{\epsilon_{\theta_{1,2}}}$,
\begin{eqnarray}
\label{eq:intS_EH2}
  && {\hskip -0.6in} { {\cal S}_{\rm EH}}\sim \frac{1}{16 \pi} \int_{x^0
\in \left[0,\frac {\pi \sqrt{4\pi g_s N}}{r_h}\right],x_{10}\in[0,{2\pi}], r\in[r_h,r_\Lambda],\theta_{1,2}\in[\epsilon_{\theta_{1,2}},\pi-\epsilon_{\theta_{1,2}}],\phi_{1,2}\in[0,2\pi], \psi\in[0,4\pi]}\sqrt{G^{\cal M}} R^{\cal M}\nonumber\\
 &=& a_{EH}(\alpha_{1,3,4,5})\frac{r^{4}_{\Lambda}}{\epsilon^3 r_h}.
  \end{eqnarray}
Therefore, the contribution corresponding to ${S_{\rm EH}}$ is divergent as $r_\Lambda$ becomes large and $\epsilon$ becomes small.
\subsubsection{Gibbons-Hawking-York Surface Action}
Similarly, using equations (\ref{Mtheory met}) and (\ref{eq:simplifiedmetriccomponents}), simplified form of the Gibbons-Hawking-York surface action will be given as under:
{\small
\begin{eqnarray}
\label{eq:Gibbons_Hawking_i}
& &  K^{\cal M} \sqrt{\hat{h}}\sim \Bigl(4 r^4 \left(1-\frac{r_h^4}{r^4}\right)\cos ({\theta_2}) \csc ({\theta_1}) {f_2}({\theta_2}) \sqrt{{f_1}({\theta_1})^2 \sin ^2({\theta_1})+1} \sin ({\theta_2})
   \Bigl(\Bigl(4 \cos ({\theta_2}) \nonumber\\
   &&\bigl(2 \cot ({\theta_1}) -3 \sqrt{6}\bigr) \cot ({\theta_2}) +18 \sqrt{6} \sin ({\theta_2})\Bigr)   \cos
   ^3({\theta_1})+12 \cos ({\theta_2}) \cot ({\theta_2}) \sin ({\theta_1}) \cos ^2({\theta_1})\nonumber\\
   &&+27 \sin ^2({\theta_1}) \left(\sqrt{6} \sin
   ({\theta_2})-4 {h_5} \cos ({\theta_2})\right) \cos ({\theta_1})  +81 {h_5} \sin ^3({\theta_1}) \left(\sqrt{6} \cos ({\theta_2})+3 {h_5}
   \sin ({\theta_2})\right)\Bigr)\Bigr)/\nonumber\\
   &&\Bigl(324 \sqrt{3} {g_s}^{11/4} N^{3/4} \pi ^{3/4} \Bigl(2 \cos ^2({\theta_2}) \sin ^2({\theta_1})   +3 \sin
   ^2({\theta_2}) \sin ^2({\theta_1})+2 \cos ^2({\theta_1}) \sin ^2({\theta_2})\nonumber\\
   &&+3 {h_5} \sin (2 {\theta_1}) \sin (2 {\theta_2})\Bigr)\Bigr),
\end{eqnarray}}
where
   $K^{\cal M}\equiv\frac{1}{2}\sqrt{G^{\cal M}\ ^{rr}}\frac{\partial_r {\rm det} h_{ab}}{{\rm det} h_{ab}}\Biggr|_{r=r_\Lambda} $, $h_{ab}$ is the pull-back of $G^{\cal M}_{MN}$ on to $r=r_\Lambda$ where $M=(r,a)$.
   Further simplifying above near $\theta_{1,2}=0,\pi$, the analytical expression reduces to
   {\small
   \begin{eqnarray}
   \label{eq:Gibbons_Hawking ii}
 &&  K^{\cal M} \sqrt{\hat{h}} \sim \frac{ 4 \left(1-\frac{r_h^4}{r^4_{\Lambda}}\right) r^4 \cos ^2({\theta_1}) \cos ^3({\theta_2}) \cot ^2({\theta_1}) {f_2}({\theta_2})}{81 \sqrt{3} {g_s}^{11/4} N^{3/4} \pi ^{3/4} \left(
   \cos ^2({\theta_2}) \sin ^2({\theta_1})+  \cos ^2({\theta_1}) \sin ^2({\theta_2})\right)}.
      \end{eqnarray}}
      \pagebreak
Utilizing the same approach as used in equation (\ref{eq:intSbulk}) and integrating, we have,
 {\small \begin{eqnarray}
  & &  \frac{1}{8 \pi} \int_{x_{10}\in[0,{2\pi}], \theta_{1,2}\in[0,\pi],\phi_{1,2}\in[0,2\pi], \psi\in[0,4\pi]} K^{\cal M}\sqrt{\hat h} \Biggr|_{r=r_\Lambda}  \nonumber\\
   & \sim & \frac{16 \pi ^{9/4} \left(1-\frac{r_h^4}{r^4_{\Lambda}}\right) {r^4_{\Lambda}} \cos ^2({\theta_1}) \cos ^3({\theta_2}) \cot ^2({\theta_1}) {f_2}({\theta_2})}{81\sqrt{3} {g_s}^{11/4} N^{3/4} \left(2
   \cos ^2({\theta_2}) \sin ^2({\theta_1})+2 \cos ^2({\theta_1}) \sin ^2({\theta_2})\right)}\Biggr|_{\theta_{1,2}=\epsilon_{\theta_{1,2}} + \theta_{1,2}=\pi-\epsilon_{\theta_{1,2}}}.
  \end{eqnarray}}
   \underline{Limit (\ref{limits_Dasguptaetal-i})}:
Incorporating   $g_s \sim  {\epsilon^{1}} $, $N/N_{\rm eff}\sim  {\epsilon^{-19}} $, $M/M_{\rm eff}\sim  {\epsilon^{-\frac{3}{2}}}$ $\theta_{1,2 }\sim  {\epsilon^{\frac{1}{2}}}$,  ${f_2}({{\theta_{1,2}}})\sim { \frac{1}{{\theta_{1,2}}}}$, we see that:
\begin{eqnarray}
\label{eq:Gibbons-Hawking_Hawking iii}
{\cal S}_{\rm GHY-boundary}&\sim& \frac{1}{8 \pi} \int_{x^0
\in \left[0,\frac {\pi \sqrt{4\pi g_s N}}{r_h}\right],x_{10}\in[0,{2\pi}], ,\theta_{1,2}\in[\epsilon_{\theta_{1,2}},\pi-\epsilon_{\theta_{1,2}}],\phi_{1,2}\in[0,2\pi], \psi\in[0,4\pi]}  K^{\cal M} \sqrt{\hat{h}}  \Biggr|_{r=r_\Lambda}\nonumber\\
 &\sim&(+{\rm ive})\bigl(\frac{ {r^{4}_\Lambda}-  {r_h}^4}{r_h}\bigr).
  \end{eqnarray}
So,
  ${\cal S}_{\rm GHY-boundary}^{\rm finite}\sim -{r_h}^3$ and ${\cal S}_{\rm GHY-boundary}^{\rm Infinite}=a_{{\rm GHY-boundary}}\frac{{r^{4}_\Lambda}}{r_h}$.

\underline{MQGP limit (\ref{limits_Dasguptaetal-ii})}:
  Incorporating  $g_s \sim \ {\epsilon} $, $N/N_{\rm eff}\sim   {\epsilon^{-39}} $, $M/M_{\rm eff}\sim  {\epsilon^{-\frac{3}{2}}}$,   $\epsilon_{\theta_{1,2}}\sim\epsilon^{\frac{3}{2}}$, ${f_2}({{\theta_{1,2}}})\sim { \frac{1}{{\theta_{1,2}}}}$,
  \begin{eqnarray}
\label{eq:Gibbons_Hawking iv}
{\cal S}_{\rm GHY-boundary}&\sim & \frac{1}{8 \pi} \int_{x^0
\in \left[0,\frac {\pi \sqrt{4\pi g_s N}}{r_h}\right],x_{10}\in[0,{2\pi}], ,\theta_{1,2}\in[\epsilon_{\theta_{1,2}},\pi-\epsilon_{\theta_{1,2}}],\phi_{1,2}\in[0,2\pi], \psi\in[0,4\pi]}  K^{\cal M} \sqrt{\hat{h}}  \Biggr|_{r=r_\Lambda}\nonumber\\
&& \sim (+{\rm ive})  \left( \frac{r^{4}_\Lambda-  {r_h}^4}{r_h}\right).
  \end{eqnarray}
  Further, for $\epsilon_\theta<<1$, writing $g_s=\alpha_1\epsilon, N=\alpha_2\epsilon^{-39}, M=\alpha_3\epsilon^{-\frac{3}{2}}, \epsilon_{\theta_1}=\alpha_4\epsilon^\gamma, \epsilon_{\theta_2}=\alpha_5\epsilon^\gamma$, one can show that
   (\ref{eq:Gibbons_Hawking iv}) will be proportional to $\frac{1}{\alpha_1^{\frac{9}{4}}\alpha_2^{\frac{1}{4}}\alpha_4^4\alpha_5}$.  We will take the large $\alpha_1$ and the small $\alpha_{4,5} :\alpha_1^{\frac{9}{4}}\alpha_4^4\alpha_5\equiv$finite; $\alpha_2$ is always finite.
Therefore, ${\cal S}_{\rm GHY-boundary}^{\rm Infinite}\sim a_{{\rm GHY-boundary}}(\alpha_{1,2,4,5})\frac{{r^{4}_\Lambda}}{r_h}$,  and ${\cal S}_{\rm GHY-boundary}^{\rm finite}\sim - r_h^3$, i.e., {\it independent of the cut-off $\epsilon_{\theta_{1,2}}/\epsilon$}.

\subsubsection{Flux Action}

Now,  $G_4=d C_3 + A_1 \wedge d B_2 + dx_{10}\wedge dB_2$, and $C_{\mu \nu 10}^M = B_{\mu \nu}^{IIA}, C_{\mu \nu \rho}^M = C_{\mu \nu \rho}^{IIA}$. Now, $F_4^{IIA}$ will be obtained via a triple
T-dual of type IIB $F_{1,3,5}$ where $F_1\sim F_{x/y/z}, F_3\sim F_{xy r/\theta_1/\theta_2}, F_{xz r/\theta_1/\theta_2},
 F_{yz r/\theta_1/\theta_2}$ and $F_5\sim F_{xyz \beta_1\beta_2}$ where $\beta_i=r/\theta_i$.

 Consider $T_x$ followed by $T_y$ followed by $T_z$ where $T_i$ means T-dualizing along i-th direction. As an example, $T_x F_x^{IIB}\rightarrow
{\rm non-dynamical\ 0-form\ field\ strength}^{IIA}$\cite{kiritsis-book}, $ T_y T_x F_x^{IIB} 
\rightarrow F_y^{IIB}$, $T_z F_y^{IIB} \rightarrow F_{yz}^{IIA}$ implying one can never generate
$F_4^{IIA}$ from $F_1^{IIB}$. As also an example consider $T_x F_{xy\beta_i}^{IIB}\rightarrow F_{y \beta_i}^{IIA}, T_yF_{y \beta_i}^{IIA}\rightarrow F_{\beta_i}^{IIB}, T_z F_{\beta_i}^{IIB}
\rightarrow F_{\beta_i z}^{IIA}$ again not generating $F_4^{IIA}$;
   $ T_x F_{xyz \beta_1 \beta_2}^{IIB}\rightarrow F_{yz \beta_1 \beta_2}^{IIA}$,
    $T_y F_{yz \beta_1 \beta_2}^{IIA}\rightarrow F_{z \beta_1 \beta_2}^{IIB},
T_z F_{z \beta_1 \beta_2}^{IIB}\rightarrow F_{\beta_1 \beta_2}^{IIB}$; thus one can not generate $F_4^{IIA}$.
Thus, the four-form flux $G_4=d\left(C_{\mu\nu10}dx^\mu\wedge dx^\nu \right.\\
\left.\wedge dx_{10}\right)
 + \left(A^{F_1}_1 + A^{F_3}_1 + A^{F_5}_1\right)\wedge H_3=H_3\wedge dx_{10} + A\wedge H_3$, where $C_{\mu\nu10}\equiv B_{\mu\nu}$ implying that the flux-dependent D=11 action is given by the following two terms:
{\small \begin{eqnarray}
\label{flux_action_D=11-i}
& & {\hskip -0.3in} \int C_3\wedge G_4\wedge G_4 = \int B\wedge dx_{10} \wedge \left(H\wedge dx_{10} + A\wedge H\right)\wedge (H\wedge dx_{10} + A\wedge H)=0,
\end{eqnarray}}
{\small \begin{eqnarray}
\label{flux_action_D=11-ii}
& & {\hskip -0.6in}{\rm and}~~\int G_4\wedge *_{11}G_4 = \int \left(H_3\wedge dx_{10} + A\wedge H_3\right)\wedge *_{11}\left(H_3\wedge dx_{10} + A\wedge H_3\right).
\end{eqnarray}}
Now, $H_3\wedge dx_{10}\wedge *_{11}\left(H_3\wedge A\right)=0$ as neither $H_3$ nor $A$ has support along $x_{10}$. Hence:
{\small \begin{eqnarray}
\label{flux_action_D=11-iii}
& & H_3\wedge dx_{10}\wedge *_{11}\left(H_3\wedge dx_{10}\right)=\sqrt{G}H_{\mu\nu\rho10}G^{\mu\mu_1}G^{\nu\nu_1}G^{\rho\rho_1}G^{10\lambda_1}H_{\mu_1\nu_1\rho_1\lambda_1}
dt\wedge...dx_{10}\nonumber\\
& & = \sqrt{G} H_{\mu\nu\rho10}\left(-G^{\mu10}G^{\nu\nu_1}G^{\rho\rho_1}G^{10\lambda_1}H_{\nu_1\rho_1\lambda_1} + G^{\mu\mu_1}G^{\nu10}G^{\rho\rho_1}G^{10\lambda_1}H_{\mu_1\rho_1\lambda_1} \right.\nonumber\\
& & \left.- G^{\mu\mu_1}G^{\nu\nu_1}G^{\rho10}G^{10\lambda_1}H_{\mu_1\nu_1\lambda_1} + G^{\mu\mu_1}G^{\nu\nu_1}G^{\rho\rho_1}G^{10\ 10}H_{\mu_1\nu_1\rho_1}\right)dt\wedge...dx_{10},
\end{eqnarray}}
where $H_{\mu\nu\rho10}=H_{\mu\nu\rho}$, and
\begin{eqnarray}
\label{flux_action_D=11-iv}
& & \left(H\wedge A\right)\wedge *_{11}\left(H\wedge A\right)=\sqrt{G} H_{[\mu\nu\rho}A_{\lambda]}G^{\mu\mu_1}G^{\nu\nu_1}G^{\lambda\lambda_1}H_{[\mu_1\nu_1\rho_1}A_{\lambda_1]},
\end{eqnarray}
where
$H_{[\mu_1\mu_2\mu_3}A_{\mu_4]}\equiv H_{\mu_1\mu_2\mu_3}A_{\mu_4} - \left(H_{\mu_2\mu_3\mu_4}A_{\mu_1} - H_{\mu_3\mu_4\mu_1}A_{\mu_2} + H_{\mu_4\mu_1\mu_2}A_{\mu_3}\right)$.
 
Considering the same scaling behavior as used to calculate the contribution of Einstein-Hilbert Action as well as Gibbons-Hawking-York surface action terms,  we see that for both limits (\ref{limits_Dasguptaetal-i}) and (\ref{limits_Dasguptaetal-ii}), in equation (\ref{flux_action_D=11-iii}), contribution of $ H_3\wedge dx^{11}\wedge *_{11}(H_3\wedge dx^{11})$ is always dominated by  $ \sqrt{G} H^{2}_{\theta_1 \theta_2 y} G^{\cal M}\ ^{\theta_1\theta_1}G^{\cal M}\ ^{\theta_2 \theta_2}G^{\cal M}\ ^{yy}G^{\cal M}\ ^{10\ 10}$ term  and in equation (\ref{flux_action_D=11-iv}), contribution of $(H\wedge A)\wedge *_{11}(H\wedge A)$ is dominated by $ \sqrt{G} H^{2}_{ \theta_1 \theta_2 y} {A^2_{y}} G^{\cal M}\ ^{\theta_1\theta_1}G^{\cal M}\ ^{\theta_1\theta_1} G^{\cal M}\ ^{\theta_2\theta_2}G^{yy}  $ term. Therefore, for simplicity in calculations, we assume that leading contribution in equations ($\ref{flux_action_D=11-iii}$) and ($\ref{flux_action_D=11-iv}$) are governed by aforementioned terms. The
 relevant inverses of the 11-dimensional metric components  of (\ref{metric-mirror}), in the limit (\ref{limits_Dasguptaetal}), simplify to the following expressions:
 {\small
\begin{eqnarray}
\label{eq:Gtheta2theta2etal}
&\bullet& G^{\cal M}\ ^{\theta_2 \theta_2}\sim \biggl(216 \sqrt[3]{3} \sqrt[6]{{g_s}} \cos ^6({\theta_1}) \cot ^4({\theta_2}) \Bigl(2 \cos ^2({\theta_2}) \sin ^2({\theta_1}) + 3 \sin^2({\theta_2}) \sin ^2({\theta_1})+2 \cos ^2({\theta_1})\nonumber\\
&&  \sin ^2({\theta_2}) +3 {h_5} \sin (2 {\theta_1}) \sin (2
   {\theta_2})\Bigr)^2\biggr)/\biggl(\sqrt{N} \sqrt{\pi } {f_2}({\theta_2})^2 \left({f_1}({\theta_1})^2 \sin ^2({\theta_1})+1\right) \Bigl(\cos
   ^2({\theta_2}) \sin ^2({\theta_1})\nonumber\\
&&+\cos ^2({\theta_1}) \sin ^2({\theta_2})\Bigr)^2  \Bigl(\left(4 \cos ({\theta_2}) \left(2 \cot ({\theta_1})-3
   \sqrt{6}\right) \cot ({\theta_2})+18 \sqrt{6} \sin ({\theta_2})\right) \cos ^3({\theta_1})\nonumber\\
   && +12 \cos ({\theta_2}) \cot ({\theta_2}) \sin
   ({\theta_1}) \cos ^2({\theta_1})  +27 \sin ^2({\theta_1}) \left(\sqrt{6} \sin ({\theta_2})-4 {h_5} \cos ({\theta_2})\right) \cos
   ({\theta_1})\nonumber\\
   &&+81 {h_5} \sin ^3({\theta_1}) \left(\sqrt{6} \cos ({\theta_2})+3 {h_5} \sin ({\theta_2})\right)\Bigr)^2\biggr)\nonumber\\
  &\bullet& G^{\cal M}\ ^{yy} \sim \biggl(36 \sqrt[3]{3} {g_s}^{2/3} \csc ^4({\theta_2}) \sec ^2({\theta_2}) \sin ^3({\theta_1}) (9 {h_5} \sin ({\theta_1})-2 \cos ({\theta_1})
   \cot ({\theta_2})) \Bigl(2 \cos ^2({\theta_2}) \nonumber\\
   &&  \sin ^2({\theta_1})+3 \sin ^2({\theta_2}) \sin ^2({\theta_1})+2 \cos ^2({\theta_1}) \sin
   ^2({\theta_2})+3 {h_5} \sin (2 {\theta_1}) \sin (2 {\theta_2})\Bigr)^2\biggr)/\bigl(\cos (2 {\theta_1})-5)^3\bigr)\nonumber\\
   &\bullet& G^{\cal M}\ ^{r r}\sim \frac{r^2 \sqrt[6]{{g_s}} \left(1-\frac{r_h^4}{r^4}\right) }{2~3^{2/3} \sqrt{\pi } \sqrt{N}  }, G^{\cal M}\ ^{10\ 10} \sim  3 \sqrt[3]{3} \left(\frac{1}{{g_s}}\right)^{4/3}\nonumber\\
 &\bullet&H_{r \theta_1 y}\sim \frac {{f_ 2} ({\theta_ 2}) \sin ({\theta_ 1}) \cos ({\theta_ 1}) \sin ({\theta_ 2}) \sin (2 {\theta_ 2}) \cos
      ({\theta_ 2})} {\left (\sin ^2 ({\theta_ 1}) \cos ^2 ({\theta_ 2}) + \cos ^2 ({\theta_ 1}) \sin ^2 ({\theta_ 2}) \right)^2}
   \nonumber\\
   &\bullet& A_{\theta_2}\sim \frac{\sqrt {\frac {3} {2}} {N_f} {\phi_ 2} \sin ^2 ({\theta_ 2}) \left (8 \cos ^3 ({\theta_ 1}) \cos ({\theta_ 2}) \cot ({\theta_ 2}) - 16
           \cos ^4 ({\theta_ 1}) \cot ({\theta_ 2}) \right)} {\pi  (\cos (2 {\theta_ 1}) -
       5) \left (2 \sin ^2 ({\theta_ 1}) \cos ^2 ({\theta_ 2}) + 2 \cos
            ^2 ({\theta_ 1}) \sin ^2 ({\theta_ 2}) \right)}-\nonumber\\
            && \frac{\sqrt {\frac {2} {3}} {N_f}  \psi  \csc ({\theta_ 2}) \left (-2 \cos ({\theta_ 1})
           \cos ({\theta_ 2}) + \sin ^2 ({\theta_ 1}) - \cos ^2 ({\theta_ 1}) + 5 \right)} {\pi  (\cos (2 {\theta_ 1}) - 5)} .
   \end{eqnarray}}
On simplifying equations (\ref{flux_action_D=11-iii}) and (\ref{flux_action_D=11-iv}) with the help of equations (\ref{eq:Gtheta2theta2etal}), the most dominant contribution near $\theta_{1,2}=0,\pi$ will be given by the following analytical expression:
{\small
\begin{eqnarray}
\label{flux_action_D=11-v}
& &  \hskip -0.2in H_3\wedge dx^{11}\wedge *_{11}\left(H_3\wedge dx^{11}\right) \sim  \frac {1134 \sqrt {3}  r^3 {f_ 2} ({\theta_ 2}) (\sin ^5 ({\theta_ 1}) \cos^5 ({\theta_ 1}) \cos ^3 ({\theta_ 2}) \cot ^3 ({\theta_ 2})} {\pi ^{5/4} {g_s}^{13/ 4} N^{5/4} \left (\sin ^2 ({\theta_ 1}) \cos^2 ({\theta_ 2}) + \cos ^2 ({\theta_ 1}) \sin ^2 ({\theta_ 2}) \right)^3} dx^{1} \wedge...dx^{11}
    \nonumber\\
    \pagebreak
   && \hskip 1.7in \sim \frac{1134 \sqrt{3}  r^3  {f_2}( {\theta_2}) \cos^4( {\theta_1}) \cot ( {\theta_1}) \csc ^3( {\theta_2})}{\pi ^{5/4} {g_s}^{13/ 4} N^{5/4}}dx^{1} \wedge...dx^{11}.\\
& & \hskip -0.6in \left(H\wedge A\right)\wedge *_{11}\left(H\wedge A\right)\sim \frac {243 \sqrt {3} {N_f}^2 r^3 {f_ 2} ({\theta_ 2}) \sin ^7 ({\theta_ 1}) \cos ^3 ({\theta_ 1}) \sin ^4 (2 {\theta_ 2}) \csc({\theta_ 2}) (\psi -  6 {\phi_ 2} \cos ({\theta_ 1}))^2} {2 \pi ^{13/4} ({g_s} N)^{5/4} (\cos (2 {\theta_ 1}) - 5)^2 \left (\sin^2 ({\theta_ 1}) \cos ^2 ({\theta_ 2}) + \cos ^2 ({\theta_ 1}) \sin ^2 ({\theta_ 2}) \right)^5} dx^{1} \wedge...dx^{11} \nonumber\\
   &&{\hskip 0.8in}  \sim  \frac {243 \sqrt {3} {N_f}^2 r^3 {f_ 2} ({\theta_ 2}) \cot ^3 ({\theta_ 1}) \tan ^3 ({\theta_ 2}) \sec ^3 ({\theta_ 2}) (\psi - 6 {\phi_ 2} \cos ({\theta_ 1}))^2} {2 \pi ^{13/4} ({g_s} N)^{5/4}}dx^{1} \wedge...dx^{11}.
\end{eqnarray}}
Integrating above:
{\small
\begin{eqnarray}
\label{eq:intG4wedG4i}
&&  \int_{x_{10}\in[0,{2\pi}], r\in[r_h,r_\Lambda],\theta_{1,2}\in[\epsilon_{\theta_{1,2}},\pi-\epsilon_{\theta_{1,2}}],\phi_{1,2}\in[0,2\pi], \psi\in[0,4\pi]} G_4\wedge *_{11}G_4   \sim  \nonumber\\
  && \frac{r^3  {f_2}( {\theta_2}) \cos^4( {\theta_1}) \cot ( {\theta_1}) \csc ^3( {\theta_2})}{{g_s}^{13/ 4} N^{5/4}}+ \frac {{N^{2}_f} r^3 {f_ 2} ({\theta_ 2}) \cos ^5 ({\theta_ 1}) \sin ^3 ({\theta_ 2})   } {\sin ^3 ({\theta_ 1}) \cos ^6 ({\theta_ 2})({g_s} N)^{5/4}}
   \Biggr|_{\theta_{1,2}=\epsilon_{\theta_{1,2}} + \theta_{1,2}=\pi-\epsilon_{\theta_{1,2}}}.
\end{eqnarray}}
\underline{Limit (\ref{limits_Dasguptaetal-i})} :
Incorporating   $g_s \sim  {\epsilon} $, $N/N_{\rm eff}\sim  {\epsilon^{-19}} $, $M/M_{\rm eff}\sim  {\epsilon^{-\frac{3}{2}}}$ $\theta_{1 } {\epsilon^{\frac{1}{2}}}$, $\theta_{2}\sim   {\epsilon^{\frac{1}{2}}}$, ${f_2}({{\theta_{1,2}}})\sim { \frac{1}{{\theta_{1,2}}}}$, we see that:
\begin{eqnarray}
\label{eq:intG4wedG4ii}
&& \int_{x^0
\in \left[0,\frac {\pi \sqrt{4\pi g_s N}}{r_h}\right],x_{10}\in[0,{2\pi}], r\in[r_h,r_\Lambda],\theta_{1,2}\in[\epsilon_{\theta_{1,2}},\pi-\epsilon_{\theta_{1,2}}],\phi_{1,2}\in[0,2\pi], \psi\in[0,4\pi]} G_4\wedge *_{11}G_4\nonumber\\
& &   \sim - {\epsilon}^{9}  \Bigl(\frac{r^{4}_{\Lambda}}{r_h} \Bigr).
\end{eqnarray}
 Therefore, ${\cal S}_{\rm flux}^{\rm Infinite}\sim   - {\epsilon}^{9}   \frac{r^{4}_{\Lambda}}{r_h}$.
\\
\underline{MQGP limit (\ref{limits_Dasguptaetal-ii})}:
  Incorporating  $g_s \sim \ {\alpha_1} \,{\epsilon} $, $N/N_{\rm eff}\sim \alpha_2  {\epsilon^{-39}} $, $M/M_{\rm eff}\sim  \alpha_3{\epsilon^{-\frac{3}{2}}}$,   $\theta_{1,2 }\sim \alpha_{4,5} {\epsilon^{\frac{3}{2}}}$, ${f_2}({{\theta_{1,2}}})\sim { \frac{1}{{\theta_{1,2}}}}$, we see that
\begin{eqnarray}
\label{eq:intG4wedG4iii}
&& \int_{x^0
\in \left[0,\frac {\pi \sqrt{4\pi g_s N}}{r_h}\right],x_{10}\in[0,{2\pi}], r\in[r_h,r_\Lambda],\theta_{1,2}\in[0,\pi],\phi_{1,2}\in[0,2\pi], \psi\in[0,4\pi]} G_4\wedge *_{11}G_4\nonumber\\
& &   \sim - a_{G_4}(\alpha_{1,2,3,4,5}){\epsilon}^{19}   \Bigl(\frac{r^{4}_{\Lambda}}{r_h}+ \frac{{r_h}^3} {ln(r_{\Lambda})}\Bigr).
\end{eqnarray}
Therefore,  ${\cal S}_{\rm flux}^{\rm Infinite}\sim {\epsilon}^{19}a_{G_4}(\alpha_{1,2,3,4,5}) \frac{r^{4}_{\Lambda}}{r_h}$.
\subsubsection{${\cal O}(R^4)$ Action Terms}
In either limits (\ref{limits_Dasguptaetal-i}) and (\ref{limits_Dasguptaetal-ii}), the dominant contribution to J is given by
{\small \begin{equation}
\label{eq:J}
 J \sim  \bigl(G^{\cal M}\ ^{\theta_1 \theta_1}\bigr)^3  G^{\cal M}\ ^{\theta_1 \theta_2}  G^{\cal M}\ ^{\theta_2 \theta_2}G^{\cal M}\ ^{\theta_1 y}G^{\cal M}\ ^{\theta_2 y} G^{\cal M}\ ^{y y} \bigl(R_{\theta_2 \theta_1 y \theta_1}\bigr)^4 +\frac{1}{2}\bigl(R_{\theta_2 \theta_1 \theta_1 y }\bigr)^2 \bigl(R_{\theta_2 \theta_1 y \theta_1}\bigr)^2,
\end{equation}}
and the dominant contribution to $E_8$ is given by:
{\small \begin{eqnarray}
\label{eq:E8}
&& E_8 \sim  G^{\cal M}\ ^{0 0}G^{\cal M}\ ^{1 1}G^{\cal M}\ ^{2 2} G^{\cal M}\ ^{3 3}\bigl( G^{\cal M}\ ^{r r} G^{\cal M}\ ^{\theta_1 \theta_1}G^{\cal M}\ ^{\theta_2 \theta_2}G^{\cal M}\ ^{x x} G^{\cal M}\ ^{y y}G^{\cal M}\ ^{z z}G^{\cal M}\ ^{10 10}\bigr)^2 \nonumber\\
 &&   R_{3 r 3 r}R_{\theta_1 \theta_2 \theta_1 \theta_2} R_{ x y xy}R_{z 10 z 10}.
\end{eqnarray}}
The simplified form of analytic expressions of $G^{\cal M}\ ^{\theta_1 \theta_1}$, $ G^{\cal M}\ ^{\theta_2 \theta_2}$ and $G^{\cal M}\ ^{r r}$ are given in equation no (\ref{eq:invtheta1theta1}) and (\ref{eq:Gtheta2theta2etal}). The simplified expressions of other inverse components as well as covariant 4-rank tensor relevant to get the estimate of J and $E_8$  are as follows:
{\small
\begin{eqnarray}
&\bullet& G^{\cal M}\ ^{\theta_1 \theta_2}=-\frac{1}{64 \sqrt{\pi } \sqrt{N} {f_2}({\theta_2})^2}\Bigl(81 \sqrt[3]{3} \sqrt[6]{{g_s}} {h_5} \csc ^4({\theta_2}) \sec ^2({\theta_2}) \bigl(2 \sin ^2({\theta_1}) \cos ^2({\theta_2})\nonumber\\
&&+2 \cos
   ^2({\theta_1}) \sin ^2({\theta_2})\bigr)^2  ({f_1}({\theta_1}) {f_2}({\theta_2}) (13 \sin ({\theta_1})+\sin (3 {\theta_1})) \sin
   ({\theta_2})-16)\bigr) \nonumber\\
&\bullet& G^{\cal M}\ ^{\theta_1 y}=-\frac{1}{\sqrt[4]{\pi } \sqrt[4]{N} {f_2}({\theta_2})}\Bigl({h_5} ({g_s} N)^{5/12} \csc ^2({\theta_2}) \sec ({\theta_2}) \Bigl(2 \sin ^2({\theta_1}) \cos ^2({\theta_2})\nonumber\\
&&+2 \cos ^2({\theta_1})
   \sin ^2({\theta_2})\Bigr)({f_1}({\theta_1}) {f_2}({\theta_2}) (13 \sin ({\theta_1})+\sin (3 {\theta_1})) \sin
   ({\theta_2})-16)\Bigr)\nonumber\\
&\bullet&G^{\cal M}\ ^{\theta_2 y}= -\frac{{g_s}^{5/12} \csc ^2({\theta_2}) \sec ({\theta_2}) \left(2 \sin ^2({\theta_1}) \cos ^2({\theta_2})+2 \cos ^2({\theta_1}) \sin
   ^2({\theta_2})\right)}{\sqrt[4]{\pi } \sqrt[4]{N} {f_2}({\theta_2})}\nonumber\\
&\bullet& G^{\cal M}\ ^{y y}=
{g_s}^{2/3} \sin ^3({\theta_1}) \cos ({\theta_1}) \csc ^5({\theta_2}) \sec ({\theta_2}) \left(2 \sin ^2({\theta_1}) \cos ^2({\theta_2})+2 \cos
   ^2({\theta_1}) \sin ^2({\theta_2})\right)^2\nonumber\\
&\bullet&
G^{\cal M}\ ^{x x}=-\frac{81}{2} \sqrt[3]{3} {g_s}^{2/3} \tan ^2({\theta_1}) \sec ^6({\theta_1}) \tan ^4({\theta_2})\nonumber\\
&\bullet& G^{\cal M}\ ^{z z}=-\frac{243 \sqrt[3]{3} {g_s}^{2/3} \sin ^4({\theta_1}) \tan ^3({\theta_1}) \sin ({\theta_2}) \cos ({\theta_2})}{2 \left(\sin ^2({\theta_1}) \cos
   ^2({\theta_2})+\cos ^2({\theta_1}) \sin ^2({\theta_2})\right)^2}\nonumber\\
&\bullet& G^{\cal M}\ ^{0 0}\sim  -\frac{2 {g_s}^{7/6} \sqrt{N} \sqrt{\pi }}{3^{2/3} r^2 \left(1-\frac{r_h^4}{r^4}\right)}, G^{\cal M}\ ^{1 1}\sim \frac{2 {g_s}^{7/6} \sqrt{N} \sqrt{\pi }}{3^{2/3} r^2}, G^{\cal M}\ ^{2 2}\sim \frac{2 {g_s}^{7/6} \sqrt{N} \sqrt{\pi }}{3^{2/3} r^2}\nonumber\\
&\bullet& G^{\cal M}\ ^{3 3}\sim \frac{2 {g_s}^{7/6} \sqrt{N} \sqrt{\pi }}{3^{2/3} r^2}\nonumber\\
\pagebreak
&\bullet& R_{\theta_2 \theta_1 y \theta_1}\sim \frac{40 \sqrt{2} \sqrt[4]{\pi } {f_2}({\theta_2}) \sin ^2({\theta_1}) \cos ^2({\theta_1}) \sin ^2({\theta_2}) \cos ^5({\theta_2})}{3 \sqrt[3]{3}
   {g_s}^{5/12} \sqrt[4]{N} \left(2 \sin ^2({\theta_1}) \cos ^2({\theta_2})+2 \cos ^2({\theta_1}) \sin ^2({\theta_2})\right)^3}\nonumber\\
    &\bullet& R_{ 3 r 3 r}\sim \frac{{N_f} \sin (2 {\theta_1}) \cot \left(\frac{{\theta_1}}{2}\right)}{2 \sqrt[3]{3} \pi ^{3/2} \sqrt[6]{{g_s}} \sqrt{N}
   \left(1-\frac{{r_h}^4}{r^4}\right)}\nonumber\\
   &\bullet& R_{\theta_1 \theta_2 \theta_1 \theta_2}\sim \frac{2\ 3^{2/3} \sqrt{\pi } \sqrt{N}-\frac{2}{3} {f_2}({\theta_2})^2 \sin ^3({\theta_1}) \cos ^3({\theta_1}) \cos ^2({\theta_2}) \cot
   ({\theta_2})}{2 \sqrt[6]{{g_s}} \sin ^2({\theta_1}) \cos ^2({\theta_2})+2 \cos ^2({\theta_1}) \sin ^2({\theta_2})}\nonumber\\
   &\bullet& R_{x y x x y}\sim -\frac{128 \sin ^2({\theta_1}) \cos ^6({\theta_1}) \sin ^2({\theta_2}) \cos ^{10}({\theta_2})}{27 \sqrt[3]{3} \sqrt{\pi } {g_s}^{7/6} \sqrt{N}
   \left(\sin ^2({\theta_1}) \cos ^2({\theta_2})+\cos ^2({\theta_1}) \sin ^2({\theta_2})\right)^6}\nonumber\\
   &\bullet& R_{z 10 z 10}\sim \frac{4 {g_s}^{11/6} {N_f} \cot \left(\frac{{\theta_1}}{2}\right) \cot ^3({\theta_1})}{81 \sqrt[3]{3} \pi ^{3/2} \sqrt{N}}.
   \end{eqnarray}}
 Utilizing above and simplified form of $ G^{\cal M} $ as given in equation (\ref{eq:g}), we have
 {\small
 \begin{eqnarray}
&&   \frac{T_2}{{2\pi}^4. 3^2.2^{13}}\int_{\cal{M}} d^{11}\!x \sqrt{G^{\cal M}}(J-\frac{1}{2}E_8)\sim \nonumber\\
 && \frac{(2 \pi^2)^{1/3}}{{2\pi}^4. 3^2.2^{13}} \Bigg( \frac{10^5 {h_5}^3 r^{4}_{\Lambda} \sin ^9({\theta_1}) \cos ^{13}({\theta_1})   \cos ^{15}({\theta_2}) ({f_1}({\theta_1})
   {f_2}({\theta_2}) \sin ({\theta_1}) \sin ({\theta_2})-1)^3}{{g_s}^2 N^2 {f_2}({\theta_2})^2  \sin^{11}({\theta_2})  \left(2 \sin ^2({\theta_1}) \cos
   ^2({\theta_2})+2 \cos ^2({\theta_1}) \sin ^2({\theta_2})\right)^4}+ \nonumber\\
   && \frac{10^8 {g_s}^{67/12} {N_f}^2 \sin ^{20}({\theta_1}) \sin (2 {\theta_1}) \tan ^8({\theta_1}) \cot ^2\left(\frac{{\theta_1}}{2}\right)
   \cos ^5({\theta_2}) \cot ^3({\theta_2})}{N^{7/4} r^{2}_{\Lambda} {f_2}({\theta_2}) \left(\sin ^2({\theta_1}) \cos ^2({\theta_2})+\cos ^2({\theta_1})
   \sin ^2({\theta_2})\right)^8}\Bigg)\Biggr|_{\theta_{1,2}=\epsilon_{\theta_{1,2}} + \theta_{1,2}=\pi-\epsilon_{\theta_{1,2}}}.
   \end{eqnarray}}
\underline{ Limit (\ref{limits_Dasguptaetal-i})}:
Incorporating   $g_s \sim \ {\epsilon} $, $N/N_{\rm eff}\sim {\epsilon^{-19}} $, $M/M_{\rm eff}\sim  {\epsilon^{-\frac{3}{2}}},\theta_{1,2}\sim  {\epsilon^{\frac{1}{2}}}, {f_2}({{\theta_{1,2}}})\sim { \frac{1}{{\theta_{1,2}}}}$, 

we see that
{\small \begin{eqnarray}
\label{eq:higherderv1}
&&  \hskip -0.2in \frac{T_2}{{2\pi}^4. 3^2.2^{13}} \int_{x^0
\in \left[0,\frac {\pi \sqrt{4\pi g_s N}}{r_h}\right],x_{10}\in[0,{2\pi}], r\in[r_h,r_\Lambda],\theta_{1,2}\in[\epsilon_{\theta_{1,2}},\pi-\epsilon_{\theta_{1,2}}],\phi_{1,2}\in[0,2\pi], \psi\in[0,4\pi]} \sqrt{G^{\cal M}}(J-\frac{1}{2}E_8)  \nonumber\\
 && \sim  \frac{{\cal O}(10^{-14})}{{\epsilon}^{-23} } \frac{r^{4}_{\Lambda}}{r_h} + \frac{{\cal O}(1 0)}{{\epsilon}^{-26} } \Bigl(\frac{1}{ r^{2}_{\Lambda} r_h}\Bigr).
\end{eqnarray}}
Thus, $ {\cal S}_{\rm {\cal O}(R^4)}^{\rm finite}\sim   {{\cal O}(10 )}{{\epsilon}^{26} } \Bigl(\frac{1}{ r^{2}_{\Lambda} r_h}\Bigr)\rightarrow 0, r_{\Lambda}\rightarrow \infty~ {\rm and}~ {\epsilon}\rightarrow0$;  ${\cal S}_{\rm {\cal O}(R^4)}^{\rm Infinite}\sim   {{\cal O}(10^{-14})}{{\epsilon}^{23} } \frac{r^{4}_{\Lambda}}{r_h}$.

\underline{ MQGP limit (\ref{limits_Dasguptaetal-ii})}:
 Incorporating  $g_s \sim \ {\alpha_1} \,{\epsilon} $, $N/N_{\rm eff}\sim \alpha_2  {\epsilon^{-39}} $, $M/M_{\rm eff}\sim \alpha_3 {\epsilon^{-\frac{3}{2}}}$,   $\theta_{1,2}\sim \alpha_{4,5}  {\epsilon^{\frac{3}{2}}}$, ${f_{1,2}}({{\theta_{1,2}}})\sim { \frac{1}{{\theta_{1,2}}}}$,
{\small\begin{eqnarray}
\label{eq:higherderv2}
&&  \hskip -0.2in \frac{T_2}{{2\pi}^4. 3^2.2^{13}} \int_{x^0
\in \left[0,\frac {\pi \sqrt{4\pi g_s N}}{r_h}\right],x_{10}\in[0,{2\pi}], r\in[r_h,r_\Lambda],\theta_{1,2}\in[\epsilon_{\theta_{1,2}},\pi-\epsilon_{\theta_{1,2}}],\phi_{1,2}\in[0,2\pi], \psi\in[0,4\pi]} \sqrt{G^{\cal M}}(J-\frac{1}{2}E_8) \nonumber\\
 && \sim  a_{R^4}(\alpha_{1,2,4,5}){{\cal O}(10^{-15})}{{\epsilon}^{45} } \frac{r^{4}_{\Lambda}}{r_h} +  b_{R^4}(\alpha_{1,2,4,5}){{\cal O}(10^6 )}{{\epsilon}^{68} } \Bigl(\frac{1}{ r^{2}_{\Lambda} r_h}\Bigr).
\end{eqnarray}}
Therefore, $ {\cal S}_{\rm {\cal O}(R^4)}^{\rm finite}\sim  b_{R^4}(\alpha_{1,2,4,5}){\cal O}(10 ^6){{\epsilon}^{68} } \Bigl(\frac{1}{ r^{2}_{\Lambda} r_h}\Bigr) \rightarrow 0, r_{\Lambda}\rightarrow \infty$;  ${\cal S}_{\rm {\cal O}(R^4)}^{\rm Infinite}\sim  a_{R^4}(\alpha_{1,2,4,5}){{\epsilon}^{45} } \frac{r^{4}_{\Lambda}}{r_h}$.
 
The other quartic term in D=11 Supergravity Action is
   $$T_2 \int C_3 \wedge X_8, $$ where $C_3$ is 3-form flux in D=11 Supergravity  defined as $C_3 =B_{\mu \nu} dx^{\mu}\wedge dx^{\nu} \wedge dx_{10}$. Now,
   $R_{\mu\nu}^{\ \ \ ab} =  R_{\mu\nu}^{\ \ \ \rho\lambda}e^{\ a}_{\rho} e^{\ b}_{\lambda}$ where $e^a_{\ \mu}$ are frames in components. Now: $tr(R^2) =
R_{\mu_1\nu_1}^{\ \ \ \ ab}R_{\mu_2\nu_2 b a}  dx^{\mu_1}\wedge dx^{\nu_1}\wedge dx^{\mu_2}\wedge dx^{\nu_2}
= e^{\ a}_{\rho_1} e^{\ b}_{\lambda_1} E_b^{\ \rho_2}E_a^{\ \lambda_2} R_{\mu_1\nu_1}^{\rho_1\lambda_1} R_{\mu_2\nu_2\rho_2\lambda_2}
 dx^{\mu_1}\wedge dx^{\nu_1}\wedge dx^{\mu_2}\wedge dx^{\nu_2}$,
where $E_a^{\ \mu}$ are the inverse frames in components. Using:
$e^{\ a}_{\rho_1}E_a^{\ \lambda_2}=\delta^{\lambda_2}_{\rho_1}$, etc. the above gives:
$$tr(R^2)=R_{\mu_1\nu_1}^{\ \ \ \rho_1\lambda_1} R_{\mu_2\nu_2\lambda_1\rho_1}
 dx^{\mu_1}\wedge dx^{\nu_1}\wedge dx^{\mu_2}\wedge dx^{\nu_2}.$$  Writing $tr(R^4)$ and $(tr(R^2))^2$ in terms of purely covariant curvature tensor,  one similarly has:
{\small  \begin{eqnarray}
 \label{eq:trR4}
 && tr(R^4)= G^{\cal M}\ ^{\lambda_1 \lambda^{\prime}_1} G^{\cal M}\ ^{\rho_1 \rho^{\prime}_1}G^{\cal M}\ ^{\rho_2 \rho^{\prime}_2}G^{\cal M}\ ^{\lambda_2 \lambda^{\prime}_2} R_{\rho^{\prime}_1 \lambda^{\prime}_1  \mu_1 \nu_1}R_{\mu_2 \nu_2 \lambda_1 \rho_2}  R_{\rho^{\prime}_2 \lambda^{\prime}_2 \mu_3 \nu_3  } R_{\mu_4 \nu_4 \lambda_2 \rho_1}dx^{\mu_1}\wedge dx^{\nu_1}\nonumber\\
 &&\wedge dx^{\mu_2}\wedge dx^{\nu_2}\wedge  dx^{\mu_3}\wedge dx^{\nu_3}\wedge dx^{\mu_4}\wedge dx^{\nu_4},\\
  \label{eq:trR2}
 &&{\rm and}~~ (tr(R^2))^2 = G^{\cal M}\ ^{\rho_1 \rho^{\prime}_1} G^{\cal M}\ ^{\lambda_1 \lambda^{\prime}_1}    G^{\cal M}\ ^{\rho_2 \rho^{\prime}_2} G^{\cal M}\ ^{\lambda_2 \lambda^{\prime}_2}R_{ \rho^{\prime}_1 \lambda^{\prime}_1 \mu_1 \nu_1 }\nonumber\\
 &&R_{\lambda_1\rho_1\mu_2 \nu_2 }R_{ \rho^{\prime}_2 \lambda^{\prime}_2 \mu_3 \nu_3 }R_{\lambda_2 \rho_2\mu_4 \nu_4 } dx^{\mu_1}\wedge dx^{\nu_1}\wedge dx^{\mu_2}\wedge dx^{\nu_2}\wedge  dx^{\mu_3}\wedge dx^{\nu_3}\wedge dx^{\mu_4}\wedge dx^{\nu_4}.
 \end{eqnarray}}
From equation (\ref{three-form fluxes}), the non-zero components of 2-form flux $B_2$ include: $B_{\theta_1 x},B_{\theta_2 y},\\B_{\theta_1 z}, B_{\theta_2 z}$
and therefore, the non-zero components of $C_3$ are $C_{\theta_1 x 10}, C_{\theta_2 y 10}, C_{\theta_1 z 10}$ and $ C_{\theta_2 z 10}$.
Since non-zero three form flux components do not include space-time indices, the same must be included in $X_8$ to have non vanishing $ \int C_3 \wedge X_8$. Effectively, one needs to calculate $X_8$ for a Euclideanized eight-fold $M_8$ that is locally $S^1(x^0)\times{\bf R}^{3}_{\rm conf}\times M_4(r,\beta_i,x_{a_1},x_{a_2})$ where
$\beta_i\equiv\theta_1\ {\rm or}\ \theta_2$, $x_{a_{1,2}}\equiv (y,z)\ {\rm or}\ (x,z)\ {\rm or}\ (x,y)$ and ${\bf R}^3_{\rm conf}$ implies conformally Euclidean. Now, in Dasgupta et al's limit (\ref{limits_Dasguptaetal}), one makes the following observations:
\vskip 0.1in
(i)~ Let $a$, etc. index the $S^1(x^0)\times{\bf R}^3_{\rm conf}(x^{1,2,3})$ coordinates, $m$ the $M_4(r,\beta_i,x_{a_1},x_{a_2})$ coordinates and let $\alpha=(a,m)$. From $R^{ab}_{\ \ \alpha\beta}=g^{bb}(\partial_{[\alpha}\Gamma^a_{b|\beta]} + \Gamma^\gamma_{b[\beta}\Gamma^a_{\gamma|\alpha]})$, one sees that only $R^{ab}_{\ \ ab} = - \frac{g^{bb}g^{rr}g^{aa}}{4}(\partial_rg_{bb})(\partial_rg_{aa})\neq0, a\neq b$.
\vskip 0.1in
(ii)~ $R^{ab}{\ \ m c}=0$.
\vskip 0.1in
(iii)~ Using $\Gamma^{\alpha^\prime}_{\beta^\prime a}=\frac{g^{aa}}{2}\partial_rg_{aa}\delta^{\alpha^\prime}_a\delta^r_{\beta^\prime} - \frac{g^{rr}}{2}\partial_rg_{aa}\delta^a_{\beta^\prime}\delta^{\alpha^\prime}_r$, one obtains:
 $R^{\alpha\beta}_{\ \ ma} = g^{rr}\partial_m\Gamma^a_{ra}\delta^\alpha_a\delta^\beta_r - g^{aa}\partial_m\Gamma^r_{aa}\delta^\alpha_r\delta^\beta_a + g^{rr}\Gamma^a_{ra}\Gamma^a_{am}\delta^\alpha_a\delta^\beta_r - g^{aa}\Gamma^r_{aa}\Gamma^n_{rm}\delta^\alpha_n\delta^\beta_a - g^{\beta\beta^\prime}\Gamma^r_{\beta^\prime m}\Gamma^a_{ra}\delta^\alpha_a + g^{aa}\Gamma^a_{am}\Gamma^r_{aa}\delta^\beta_a\delta^\alpha_r$. Therefore, e.g., $R^{ab}_{\ \ ab}\neq0,\ R^{ar}_{\ \ ar}\neq0$.

 Using these, and noting that $tr(R^4)$ will involve terms of the following three types:
{\small  \begin{eqnarray}
 \label{trRfourth_types}
& & R^{\rho_1\lambda_1}_{\ \ \ \ a n_1}R_{\lambda_1\rho_2a_2n_2}R^{\rho_2\lambda_3}_{\ \ \ \ a_3n_3}R_{\lambda_3\rho_1a_4n_4}\prod_{i=1}^4dx^{a_i}\wedge dx^{n_i}\nonumber\\
& & R^{\rho_1\lambda_1}_{\ \ \ \ a_1a_2}R_{\lambda_1\rho_2m_2n_2}R^{\rho_2\lambda_3}_{\ \ \ \ a_3n_3}R_{\lambda_3\rho_1a_4n_4}dx^{a_1}\wedge dx^{a_2}\wedge dx^{m_2}\wedge dx^{n_2}\wedge dx^{a_3}\wedge dx^{n_3}\wedge dx^{a_4}\wedge dx^{n_4}\nonumber\\
& & R^{\rho_1\lambda_1}_{\ \ \ \ a_1a_2}R_{\lambda_1\rho_2a_3a_4}R^{\rho_2\lambda_3}_{\ \ \ \ m_3n_3}R_{\lambda_3\rho_1m_4n_4}\prod dx^{a_i}\wedge \prod_{i=1}^2dx^{m_i}\wedge dx^{n_i},
\end{eqnarray}}
let us look at the three types of terms in (\ref{trRfourth_types}) individually below. The first possibility will hence consist of the following set of terms:
{\small \begin{eqnarray}
\label{trRfourth-typei}
& & R^{a_1r}_{\ \ \ a_1n_1}R_{r\rho_2a_2n_2}R^{\rho_2\lambda_3}_{\ \ \ \ a_3n_3}R_{\lambda_3a_1a_4n_4}\prod_{i=1}^4dx^{a_i}\wedge dx^{n_i},\nonumber\\
& & R^{a_1r}_{\ \ \ a_1n_1}R_{a_1\rho_2a_2n_2}R^{\rho_2\lambda_3}_{\ \ \ \ a_3n_3}R_{\lambda_3ra_4n_4}\prod_{i=1}^4dx^{a_i}\wedge dx^{n_i},\nonumber\\
& & R^{na_1}_{\ \ \ a_1n_1}R_{a_1\rho_2a_2n_2}R^{\rho_2\lambda_3}_{\ \ \ \ a_3n_3}R_{\lambda_3na_4n_4}\prod_{i=1}^4dx^{a_i}\wedge dx^{n_i},\nonumber\\
& & R^{a_1\lambda_1}_{\ \ \ a_1n_1}R_{\lambda_1\rho_2a_2n_2}R^{\rho_2\lambda_3}_{\ \ \ \ a_3n_3}R_{\lambda_3a_1a_4n_4}\prod_{i=1}^4dx^{a_i}\wedge dx^{n_i}.
\end{eqnarray}}
Based on the three observations made above, each of the four terms in (\ref{trRfourth-typei})vanishes for reasons similar to:
{\small \begin{eqnarray}
& & R_{\lambda_3a_1a_4n_4}\prod_{i=1}^4dx^{a_i}\wedge dx^{n_i}\sim\delta^{a_4}_{\lambda_3}{\bf \delta^{a_1}_{n_4}}\prod_{i=1}^4dx^{a_i}\wedge dx^{n_i}=0\ {\rm or}\
\delta^r_{\lambda_3}\delta^r_{n_4}{\bf\delta_{a_1a_4}}\prod_{i=1}^4dx^{a_i}\wedge dx^{n_i}=0.\nonumber\\
\end{eqnarray}}
 The second possibility vanishes because:
{\small \begin{eqnarray}
\label{trRfourth-typeiii}
& & R_{a_2\rho_2a_3a_4}\prod dx^{a_i}\wedge \prod_{i=1}^2dx^{m_i}\wedge dx^{n_i} \sim\left({\bf \delta^{a_2}_{a_3}}\delta^{a_4}_{\rho_2}\ {\rm or}\ {\bf\delta^{a_2}_{a_4}}\delta^{a_3}_{\rho_2}\right)\prod dx^{a_i}\wedge \prod_{i=1}^2dx^{m_i}\wedge dx^{n_i}=0.\nonumber\\
\end{eqnarray}}
 The third possibility vanishes because:
{\small \begin{eqnarray}
\label{trRfourth-typeii}
& &  R_{\lambda_3a_1a_4n_4}dx^{a_1}\wedge dx^{a_2}\wedge dx^{m_2}\wedge dx^{n_2}\wedge dx^{a_3}\wedge dx^{n_3}\wedge dx^{a_4}\wedge dx^{n_4}\sim\nonumber\\
& & \hskip -0.8in \left(\delta^{a_4}_{\lambda_3}{\bf\delta^{a_1}_{n_4}}\ {\rm or}\ \delta^r_{\lambda_3}\delta^r_{n_4}{\bf \delta^{a_1a_4}}\right)dx^{a_1}\wedge dx^{a_2}\wedge dx^{m_2}\wedge dx^{n_2}\wedge dx^{a_3}\wedge dx^{n_3}\wedge dx^{a_4}\wedge dx^{n_4}=0.
\end{eqnarray}}
 Similarly, for $tr(R^2)^2=R^{\rho_1\lambda_1}_{\ \ \ \ \mu_1\nu_1}R_{\lambda_1\rho_1\mu_2\nu_2}R^{\rho_2\lambda_2}_{\ \ \ \ \mu_3\nu_3}R_{\lambda_2\rho_2\mu_4\nu_4}\prod_{i=1}^4dx^{\mu_i}\wedge dx^{\nu_i}$ there are the following types of terms:
{\small \begin{eqnarray}
\label{trRsqsq-typei}
& & R^{\rho_1\lambda_1}_{\ \ \ \ a_1n_1}R_{\lambda_1\rho_1a_2n_2}R^{\rho_2\lambda_2}_{\ \ \ \ a_3n_3}R_{\lambda_2\rho_2a_4n_4}\prod_{i=1}^4dx^{a_i}\wedge dx^{n_i},
\end{eqnarray}}
which vanishes because of reasons similar to:
{\small \begin{eqnarray}
\label{type-i_poss}
& & R^{a_1r}_{\ \ \ a_1n_1}R_{ra_1a_2n_2}\prod_{i=1}^4dx^{a_i}\wedge dx^{n_i}\sim\delta^r_{n_2}{\bf\delta_{a_1a_2}}\prod_{i=1}^4dx^{a_i}\wedge dx^{n_i}=0.\\
\label{trsqsq-typeii}
& & R^{\rho_1\lambda_1}_{a_1a_2}R_{\ \ \ \lambda_1\rho_1m_2n_2}R^{\rho_2\lambda_2}_{\ \ \ \ a_3n_3}R_{\lambda_2\rho_2a_4n_4}dx^{a_1}\wedge dx^{a_2}\wedge dx^{m_2}\wedge dx^{n_2}\wedge dx^{a_3}\wedge dx^{n_3}\wedge dx^{a_4}\wedge dx^{n_4}\nonumber\\
& & \sim R^{a_1a_2}_{\ \ \ a_1a_2}{\bf R_{a_2a_1a_3n_3}}dx^{a_1}\wedge dx^{a_2}\wedge dx^{m_2}\wedge dx^{n_2}\wedge dx^{a_3}\wedge dx^{n_3}\wedge dx^{a_4}\wedge dx^{n_4}=0.\\
\label{trsqsq-typeiii}
& & R^{\rho_1\lambda_1}_{\ \ \ \ a_1a_2}R_{\lambda_1\rho_1a_3a_4}R^{\rho_2\lambda_2}_{\ \ \ \ m_3n_3}R_{\lambda_2\rho_2m_4n_4}\prod_{i=1}^4dx^{a_i}\wedge\prod_{j=3}^4dx^{m_j}\wedge dx^{n_j}\nonumber\\
& & \sim R^{a_1a_2}_{\ \ \ a_1a_2}R_{a_2a_1a_3a_4}\prod_{i=1}^4dx^{a_i}\wedge\prod_{j=3}^4dx^{m_j}\wedge dx^{n_j}\nonumber\\
& & \left({\bf\delta_{a_2a_3}\delta_{a_1a_4}}\ {\rm or}\ {\bf\delta_{a_2a_4}\delta_{a_1a_3}}\right)\prod_{i=1}^4dx^{a_i}\wedge\prod_{j=3}^4dx^{m_j}\wedge dx^{n_j}=0.
\end{eqnarray}}
 One sees, therefore,  the first and second Pontryagin classes of $TM_8$ satisfy: $p_1^2(TM_8)=p_2(TM_8)=0$ implying $X_8(TM_8) = 4 p_2 - p_1^2=0,$ and hence $$C_3\wedge X_8=0.$$

{{ \bf Counter-term evaluation :}} To summarize, in the limit (\ref{limits_Dasguptaetal-i}),
from (\ref{eq:intS_EH1}), (\ref{eq:Gibbons-Hawking_Hawking iii}), (\ref{eq:intG4wedG4ii}), (\ref{eq:higherderv1}), one sees that the divergent part of the action is given by:
\begin{equation}
\label{IR_i}
\frac{r_\Lambda^4}{r_h}\left(\frac{a_{EH}}{\epsilon } + a_{\rm GHY-boundary} + \epsilon^{9} a_{G_4} +
a_{R^4}\epsilon^{23} \right).
\end{equation}
The divergent part of action can be compensated by adding an appropriate counter-term corresponding to intrinsic boundary geometry so that the overall 11-dimensional Action renders finite contribution. Using (\ref{limits_Dasguptaetal-i}) and equation (\ref{eq:simplifiedmetriccomponents}) and further simplifying near $\theta_{1,2}=0,\pi$, we have
{\small
\begin{eqnarray}
 && \sqrt{h^{\cal M}} \sim \left\{\frac{\sqrt{2} \sqrt{1 - \frac{r_h^4}{r_\lambda^4}} r^4 {f_2}({\theta_2}) \cot ^4({\theta_1}) \cos ({\theta_2})}{81 \sqrt[6]{3} \sqrt{\pi } {g_s}^{7/3} \sqrt{{g_s}
   N}}\right\},\\
   &&{\hskip -0.4in} {\rm and}~~ \sqrt{h^{\cal M}} R^{\cal M}\sim\left\{\frac{125 r^4 {f_2}({\theta_2})\sqrt{1-\frac{r^{4}_h}{r^4}} \cot ^2({\theta_1}) \csc ^4({\theta_1}) \cos ({\theta_2})}{864 \sqrt{2} 3^{5/6} \pi  {g_s}^{8/3} N}\right\}.
 \end{eqnarray} }
 resulting in
{\small
\begin{eqnarray}
\label{eq:intSct}
&& {\hskip -0.4in} \frac{1}{8 \pi} \int_{x_{10}\in[0,{2\pi}], r\in[r_h,r_\Lambda],\theta_{1,2}\in[\epsilon_{\theta_{1,2}},\pi-\epsilon_{\theta_{1,2}}],\phi_{1,2}\in[0,2\pi], \psi\in[0,4\pi]}\sqrt{h^{\cal M}} R^{\cal M}|_{r=r_{\Lambda}}    \nonumber\\
&& {\hskip -0.4in} \sim \frac{125 r^{4}_{\Lambda} \pi^2 {f_2(\theta_2)} \sqrt{1-\frac{r^{4}_h}{ r^{4}_{\Lambda}}} \cot ^2({\theta_1}) \csc ^4({\theta_1}) \cos ({\theta_2})}{216 \sqrt{2} 3^{5/6} {g_s}^{8/3} N}  \Biggr|_{\theta_{1,2}=\epsilon_{\theta_{1,2}} +
\theta_{1,2}=\pi-\epsilon_{\theta_{1,2}}}.  
\end{eqnarray}
}
We hence see that
\begin{eqnarray}
\label{eq:ct1}
  && {\hskip -0.6in}  \frac{1}{8 \pi} \int_{x^0
\in \left[0,\frac {\pi \sqrt{4\pi g_s N}}{r_h}\right],x_{10}\in[0,{2\pi}], \theta_{1,2}\in[\epsilon_{\theta_{1,2}},\pi - \epsilon_{\theta_{1,2}}],\phi_{1,2}\in[0,2\pi], \psi\in[0,4\pi]} \sqrt{h^{\cal M}} R^{\cal M}|_{r=r_{\Lambda}} \nonumber\\
 &&\sim \epsilon^{\frac{77}{6}} \frac{r^4_{\Lambda}}{r_h}\sqrt{1 - \frac{r_h^4}{r_\Lambda^4}}\stackrel{r_\Lambda\rightarrow\infty}{\longrightarrow}\epsilon^{\frac{77}{6}}\frac{r_\Lambda^4}{r_h}.
  \end{eqnarray}
So, from (\ref{IR_i}) and (\ref{eq:ct1}), one sees that an appropriate counter term will be:
\begin{equation}
\label{ct-i}
-\epsilon^{-\frac{77}{6}}\left(\frac{a_{EH}}{\epsilon } + a_{\rm GHY-boundary} + \epsilon^{9} a_{G_4} +
a_{R^4}\epsilon^{23} \right)\int_{r=r_\Lambda}\sqrt{h}R^{\cal M}.
\end{equation}
One can show that:
\begin{eqnarray}
\label{sqrth+sqrthG4sq}
& & \int_{r=r_\Lambda}\sqrt{h}\sim\epsilon^{\kappa^{(i)}_{\rm cosmo}}\frac{r_\Lambda^4}{r_h},  \int_{r=r_\Lambda}\sqrt{h}|G_4|^2\sim\epsilon^{\kappa^{(i)}_{\rm flux}}\frac{r_\Lambda^4}{r_h},
\end{eqnarray}
for appropriate $\kappa^{(i)}_{\rm cosmo/flux}$  using which one sees that one can also construct the following as the appropriate counter-terms:
\begin{eqnarray}
\label{ct-i2_3}
& & - \epsilon^{-\kappa^{(i)}_{\rm cosmo}}\left(\frac{a_{EH}}{\epsilon } + a_{\rm GHY-boundary} + \epsilon^{9} a_{G_4} +
a_{R^4}\epsilon^{23} \right)\int_{r=r_\Lambda}\sqrt{h},\nonumber\\
& & - \epsilon^{-\kappa^{(i)}_{\rm flux}}\left(\frac{a_{EH}}{\epsilon } + a_{\rm GHY-boundary} + \epsilon^{9} a_{G_4} +
a_{R^4}\epsilon^{23} \right)\int_{r=r_\Lambda}\sqrt{h}|G_4|^2.
\end{eqnarray}
Interestingly in the limit (\ref{limits_Dasguptaetal-i}), we argue below that one can give an asymptotically-linear-dilaton counter-term interpration to (\ref{ct-i}) and (\ref{ct-i2_3}). The behavior of surface counter-terms  in the asymptotic linear dilaton background,
 is discussed in  \cite{Mann+Mcnees}  by defining a set of linear dilaton(ADL) boundary conditions. According to the ADL boundary conditions given in  \cite{Mann+Mcnees}, the  metric is expanded as :
\begin{equation}
\label{eq:ALD1}
ds^2= d{r}^2 +r^2( h^{(0)}_{mn}+r^{2-d}( h^{(1)}_{mn}+r^{1-d}( h^{(0)}_{mn}+...)dy^m dy^n.
\end{equation}
The scalar field can be expanded as:
\begin{equation}
\label{eq:ALD2}
 \phi= {\bar \phi}ln{r} +\phi^{0} +r^{2-d} \phi^{1} + r^{1-d} \phi^{2}+....
 \end{equation}
where ${\bar \phi}$ and $\phi^{0}$ are constants, and remaining $ \phi^{i}$ are smooth functions. Similarly, the p-form field strength is expanded as
follows:
\begin{equation}
\label{eq:ALD3}
 F_{a_1....a_p}=  F^{(0)}_{a_1....a_p}+ r^{2-d}F^{(1)}_{a_1....a_p}+ r^{1-d}F^{(2)}_{a_1....a_p}+ ...
\end{equation}
Again as explained in \cite{Mann+Mcnees}, incorporating the boundary conditions (\ref{eq:ALD1})- (\ref{eq:ALD3}) in EOM corresponding to bulk integral
gives a condition
 \begin{equation}
 \label{eq:ALD4}
 \phi_0 + {\bar \phi}ln{r} = \frac{p-1}{\alpha}ln({\beta r}),
 \end{equation}
where $\beta= \frac{8(p-1)}{\alpha^2 Q^2}$ ($Q$ being related to p-form fluxes' components' magnitute - See \cite{Mann+Mcnees} for more details).
The counter-terms consistent with $\delta S=0$, are given to be:
\begin{equation}
\label{CT_ALD}
{\cal S}_{ct}= \int d^{d} x \,\sqrt{h}\left(c_0 e^{-\frac{\alpha}{p-1}\phi}+ c_1 e^{ \frac{\alpha}{p-1}\phi} R + c_2e^{ \frac{2 p-1}{p-1}\phi} F_p^2 \right)
\end{equation}
 where for the special case wherein $\alpha=\pm\left(\frac{p-1}{d-1}\right)$,
  $c_0 = 2 \beta -(d-1)^2 \beta^2 c_1$ and $c_2= -\frac{1}{2 p!}c_1$ for ALD boundary data and $c_1$ is arbitrary.

Now, the scalar field  corresponds to  geometric $x_{10}$ size modulus (when one takes the weak-coupling limit of $M$ theory compactifying it on a circle
of very small radius)  which is given by the applying triple T-duality on type IIB dilaton to yield the type IIA dilaton, i.e., $ \phi^{IIA}= \tilde {\tilde{\tilde{\phi}}}$. At least in the small $g_s$  and $g_s N_f$ limit (weakly coupled M-theory description mentioned in (\ref{limits_Dasguptaetal-i})), one sees that for $r=r_\Lambda>>1$:
$\phi^{IIA}\sim  ln g_s - ln[1 - (3 g_s N_f/4\pi) ln r]$ and therefore, can be expressed as
\begin{equation}
\label{eq:ALD5}
\phi^{IIA}= \phi^{0} + \ \bar\phi \,ln{r_{\Lambda}} + {\cal O}((g_s N_f)^2)
\end{equation}
From (\ref{eq:ALD4}) and (\ref{eq:ALD5}), we get
$\phi^{IIA}=\frac{p-1}{\alpha}ln({\beta r_{\Lambda}})$. Substituting the same in (\ref{CT_ALD}) will include the following set of terms ($p=4$):
\begin{equation}
\label{CT_ALD2}
c_1\left(-\frac{(d-1)^2\beta^2}{\beta r_\Lambda}\int_{r=r_\Lambda}\sqrt{h} + \beta r_\Lambda\int_{r=r_\Lambda}\sqrt{h}R - \frac{(\beta r_\Lambda)^7}{48}
\int_{r=r_\Lambda}\sqrt{h}|G_4|^2\right).
\end{equation}
The three terms in (\ref{CT_ALD2}) are divergent and using an obvious notation, (\ref{CT_ALD2}) is given by:
\begin{equation}
\label{CT_ALD3}
c_1\frac{r_\Lambda^4}{r_h}\left(-\epsilon^{\kappa^{(i)}_{\rm cosmo}}a^{(i)}_{\rm cosmo}\frac{(d-1)^2\beta^2}{\beta r_\Lambda} + \epsilon^{\kappa^{(i)}_{\rm EH-boundary}}
a^{(i)}_{\rm EH-boundary}\beta r_\Lambda
 -\epsilon^{\kappa^{(i)}_{\rm flux}}a^{(i)}_{\rm flux} \frac{(\beta r_\Lambda)^7}{48} \right).
\end{equation}
Since the choice of $c_1$ is arbitrary, considering
\begin{equation}
\label{c_1}
c_1 =- \left(\frac{\frac{a_{EH}}{\epsilon } + a_{\rm GHY-boundary} + \epsilon^{9} a_{G_4} +
a_{R^4}\epsilon^{23} }{-\epsilon^{\kappa^{(i)}_{\rm cosmo}}a^{(i)}_{\rm cosmo}\frac{(d-1)^2\beta^2}{\beta r_\Lambda} + \epsilon^{\kappa^{(i)}_{\rm EH-boundary}}
a^{(i)}_{\rm EH-boundary}\beta r_\Lambda
 -\epsilon^{\kappa^{(i)}_{\rm flux}}a^{(i)}_{\rm flux} \frac{(\beta r_\Lambda)^7}{48}}\right)
\end{equation}
which cancels off the divergences coming from Einstein- Hilbert Action, Gibbons-Hawking-York surface term, flux term as well as higher order ${\cal O}(R^4)$ as given in set of equations ((\ref{eq:intS_EH1}), (\ref{eq:Gibbons-Hawking_Hawking iii}), (\ref{eq:intG4wedG4ii}), (\ref{eq:higherderv1})) in the weakly coupled description of M-theory.

In limit (\ref{limits_Dasguptaetal-ii}), to summarize, from (\ref{eq:intS_EH2}), (\ref{eq:Gibbons_Hawking iv}), (\ref{eq:intG4wedG4iii}) and (\ref{eq:higherderv2}), we see that the divergent contribution is given by:
\begin{equation}
\label{IR_ii}
\frac{r_\Lambda^4}{r_h}\left(\frac{a_{EH}(\alpha_{1,3,4,5})}{\epsilon^3} + a_{\rm GHY-boundary}(\alpha_{1,2,4,5}) + \epsilon^{19}\alpha_{G_4}(\alpha_{1,2,3,4,5}) + a_{R^4}(\alpha_{1,2,4,5})\epsilon^{45} \right).
\end{equation}
So, by arguments similar to the ones given to yield (\ref{ct-i}) as the counter term in limit (\ref{limits_Dasguptaetal-i}), the required counter term in limit (\ref{limits_Dasguptaetal-ii}) is:
{\small \begin{eqnarray}
\label{ct-ii}
&&-\epsilon^{\frac{155}{6}}\left(\frac{a_{EH}(\alpha_{1,3,4,5})}{\epsilon^3} + a_{\rm GHY-boundary}(\alpha_{1,2,4,5}) +\right. \nonumber\\
&& \left. \epsilon^{19}\alpha_{G_4}(\alpha_{1,2,3,4,5}) + a_{R^4}(\alpha_{1,2,4,5})\epsilon^{45} \right)\int_{r=r_\Lambda}\sqrt{h}R^{\cal M}.
\end{eqnarray}}
Similar to (\ref{ct-i2_3}), one can also have the following counter terms:
{\small
\begin{eqnarray}
\label{ct_ii2_3}
& & -\left(\epsilon^{-\kappa^{(ii)}_{\rm cosmo}}\int_{r=r_\Lambda}\sqrt{h},\epsilon^{-\kappa^{(ii)}_{\rm flux}}\int_{r=r_\Lambda}\sqrt{h}|G_4|^2\right)
\left(\frac{a_{EH}(\alpha_{1,3,4,5})}{\epsilon^3} + a_{\rm GHY-boundary}(\alpha_{1,2,4,5}) +\right. \nonumber\\
&& \left. \epsilon^{19}\alpha_{G_4}(\alpha_{1,2,3,4,5}) + a_{R^4}(\alpha_{1,2,4,5})\epsilon^{45} \right).
\end{eqnarray}}
To investigate the thermodynamic stability of the uplift, let us calculate the specific heat corresponding to the classical partition function/action calculated above wherein the finite part of the action is coming entirely from Gibbons-Hawking-York surface term, and ${\cal S}_{\cal E}^{\rm finite}\sim -r_h^3$ in the limits of (\ref{limits_Dasguptaetal-i}) and (\ref{limits_Dasguptaetal-ii}). Based on the argument given in \cite{S_ct-Perry_et_al}, the negative sign in the action does not represent physical instability of the solution. Now, the average energy which is $\langle E\rangle =
 - T^2\frac{\partial {\cal S}_{\cal E}^{\rm finite}}{\partial T}
= - \frac{T^2(r_h)}{\frac{\partial T}{\partial r_h}}\frac{\partial {\cal S}_{\cal E}^{\rm finite}}{\partial r_h}$. In both limits, one sees from (exact or approximate form of) (\ref{T}) that $\frac{\partial T}{\partial r_h}>0$ implying $\langle E\rangle>0$. Now, the entropy $S = - \frac{1}{T(r_h)}\frac{T^2(r_h)}{\frac{\partial T}{\partial r_h}}
\frac{\partial {\cal S}_{\cal E}^{\rm finite}}{\partial r_h} - {\cal S}_{\cal E}^{\rm finite}$. Therefore the specific heat is given by: $C= \frac{T(r_h)}{\frac{\partial T}{\partial r_h}}\frac{\partial S}{\partial r_h}$. Again the aforementioned limit, one can show:
$\frac{T(r_h)}{\frac{\partial T}{\partial r_h}}\sim {r_h}$, which implies that entropy is positive and one can approximate the same
as $S\sim  r_h^3$. A quick check on this result arises from the fact that the entropy is expected to be proportional to the horizon area - from (\ref{horizon_area}) we see that as expected the entropy should scale like $r_h^3$. Using the same, therefore, $C\sim\frac{T(r_h)}{\frac{\partial T}{\partial r_h}}\frac{\partial \left(  r_h^3\right)}{\partial r_h}\sim  {r_h}^3 >0$ - implying a stable uplift!

\section{Results and Discussion}

We have  constructed a local M-theory uplift of a (resolved) warped deformed conifold using modified OKS-BH background given in \cite{metrics} in the context of type IIB string background relevant to study of thermal QCD with fundamental flavor thermal quarks. Following \cite{SYZ3_Ts}, we defined T-duality coordinates to have a local isometry along third direction $\psi$ in addtition to the global isometries along $\phi_1,\phi_2)$ and then applied suitable coordinate transformations on some of the angular coordinates to ensure the base of the local $T^3$-fibration is large  so that  mirror symmetry a la SYZ, could be applied (locally). Having done so, we first obtained type IIA metric formed by using analytic expressions given in \cite{SYZ3_Ts} to mirror transform the type IIB metric components in the presence of a black hole in a warped deformed conifold and then obtained type IIA RR one-form gauge fields by applying T-duality rules on type IIB RR odd-form field strengths. Using T-dualized metric components, we calculated the contribution of the type IIA metric  along $r= \sqrt{3a}$ and $\theta_i\rightarrow \pi/2$ as well as $\theta_1\rightarrow 0~{\rm and}~\theta_2\rightarrow m \theta_1$ where $m \sim {\cal O}(1)$, and showed that $G^{IIA}_{\theta_1\theta_2}$ will vanish if complex structure base/$f_i(\theta_i)$ along chosen values of $\theta_i$ will be very large - one will get a warped resolved conifold. In other words, the argument that mirror of warped deformed conifold should be warped resolved conifold at least locally, automatically satisfied the condition of having large base required to implement mirror symmetry conditions proposed by Strominger-Yau-Zaslow. All of the above eventually lead to a local 11-dimensional M-theory uplift.

The hydrodynamical as well as thermodynamical properties of strongly coupled (i.e. large t'Hooft coupling) gauge theories at finite temperature are governed by the presence of a black hole in the dual description, and therefore depends on horizon radius. We basically set up an approach to study the behavior of hydrodynamical as well as thermodynamical quantities in both weak coupling but large t'Hooft coupling regime of M-theory accomplished by ($g_s, g_sN_f,
\frac{g_sM^2}{N}<<1,g_sM, g_sN>>1$ (c.f. \cite{metrics})) as well as MQGP limit accomplished by letting $g_s\stackrel{<}{\sim}1$  and $
\frac{g_sM^2}{N}<<1, g_sN>>1, {\rm finite}\ g_s, M, N_f$. The idea of discussing thermodynamical properties was two fold: the first was to check whether the solution  possesses the thermodynamical stability both in type IIB as well as local M-theory uplift and the other was to explicitly verify whether gravity dual so obtained is able to show certain aspects of strongly coupled Plasma i.e QGP. The thermodynamical stability conditions are basically governed by inequalities imposed on certain thermodynamic quantities such as $\Delta S<0, \Delta E>0 ~{\rm and} ~\Delta H>0$ (deviations from
equilibrium values implied). In particular, considering the condition that $\Delta E(S,V,N)$ as well $\partial^2 E(S,V,N) >0$  and expanding $\partial^2 E(S,V,N)$ around equilibrium values of $(S_0, V_0, N_0)$ lead to satisfy three conditions  $C_v>0, \kappa >0 ~{\rm and}~ \frac{\partial\mu}{\partial N_f}\left.\right|_{T}>0$ for the system to be in stable thermodynamic equilibrium at constant value of S, V and N \cite{Bruno}. Keeping this in mind, we  calculated the chemical potential arising from $U(1)$ gauge field living on $N_f$ $D7$-branes wrapping a four-cycle of the warped deformed conifold as a function of temperature by considering Ouyang embedding in type IIB background and then studied the behavior of the same as a function of $N_f$(number of bi-fundamental quark flavors) and showed that $ \frac{\partial{\mu_C}}{\partial N_f}|_{T}>0$  upto linear order in embedding parameter $\mu$, thus obeying one of the conditions to achieve thermodynamic stability in type IIB background. Going ahead, by obtaining the local M-theory uplift, we gave a way to implement  both limits of M-theory in thermodynamical calculations by choosing the scaling of relevant parameters $(g_s,M,N)$ scale $\epsilon$ as ($g_s \sim {\epsilon^{d}} $, $M\sim {\epsilon^{-\frac{3d}{2} }}$, $N \sim {\epsilon^{-19 d}}~ {\rm for}~  \epsilon \leq 0.01, d>0 $) and  ($g_s \sim  {\epsilon^{d}} $, $M\sim {\epsilon^{-\frac{3 d}{2}}}$, $N \sim {\epsilon^{-39 d}}~ {\rm for}~  \epsilon \lesssim 1, d>0$) consistent with
weak-string-coupling - large-t'Hooft-coupling  limit of M-theory and the  `MQGP limit'
corresponding to finite-string-coupling - large-t'Hooft-coupling limit of M-theory, respectively. Interestingly, we realized that in both limits, the uplift produced
a black $M3$-brane whose near-horizon geometry near the $\theta_{1,2}=0,\pi$ branches, preserved $\frac{1}{8}$ supersymmetry. The other very important hydrodynamical quantities that one can obtain using dual description of 11-D supergravity background to verify the aspects of strongly coupled Q(uark) G(luon) P(lasma) is the shear viscosity-to-entropy ratio $(\eta/s)$ and diffusion coefficient D in type IIA and local M-theory uplift. Considering the limits (\ref{limits_Dasguptaetal-i}) and (\ref{limits_Dasguptaetal-ii}), by first calculating the horizon radius and thereafter using the standard expressions
given in \cite{Kovtun} (the absence of angular dependence in $G^{IIA,{\cal M}}_{tt,rr,{\bf R}^3}$ and a tunably small chemical potential permitting the use of same), we showed that shear viscosity-to-entropy ratio naturally come out to be $\frac{1}{4 \pi}$ in type IIA and the limits (\ref{limits_Dasguptaetal-i}) and (\ref{limits_Dasguptaetal-ii}) of M-theory and Diffusion coefficient D turned out to be $\frac{1}{T}$ in both type IIB and type IIA background. The results are consistent with the values that theorists expect for any quantum field theory description which has a gravity dual in (non)-extremal case \cite{buchel+liu} - the ${\cal O}(R^4)$-terms do not modify the value of $\frac{\eta}{s}$ because in both limits
(\ref{limits_Dasguptaetal-i}) and (\ref{limits_Dasguptaetal-ii}), (the finite part of) the same vanishes. We next evaluated the $D=11$ Euclideanized supergravity action (Einstein-Hilbert + Hawking surface + Flux terms+${\cal O}(R^4)$terms) in the two limits. This action was expected to receive the maximum
contribution near $\theta=0,\pi$  as it possessed pole singularities near the
same  which could be regularized by considering a small ${\theta_{1,2}}$ cut-off i.e ${\theta_{1,2}} \in [\epsilon_{\theta},\pi-\epsilon_{\theta}]$ with ${\theta_{1,2}}\sim {\epsilon}^{\gamma_{(i),(ii)}}$ for $\gamma_{(i),(ii)}$ appropriate to the limits
(\ref{limits_Dasguptaetal-i}) and (\ref{limits_Dasguptaetal-ii}) in such a way
that the finite contribution (coming solely from the Gibbons-Hawking-York surface
term)  is independent of the cut-off  ${\epsilon_{\theta}}$. We showed that the
IR divergence could be removed by adding appropriate surface counter terms:
 $\int_{r=r_{\Lambda}}\sqrt{h}\left(1,{R},|G_4|^2\right)$; these counter terms in
the weak-string-coupling - large-t'Hooft-coupling
limit (\ref{limits_Dasguptaetal}) could be understood in terms of asymptotically-
linear-dilaton-gravity type surface counter terms. The finite piece of the
 Gibbons-Hawking surface term turned out to be proportional to $- {r^{3}_h}$.  We also verified entropy as well as specific heat obtained using the partition function turns out to be positive, thus obeying one of the conditions for thermodynamical stability of 11-dimensional M-theory solution.  Thus, in this chapter, we conclude that local 11-Dimensional M-theory uplift so-obtained is able to provide some theoretical insight into the physics of  strongly coupled Quark Gluon Plasma.

\chapter{Summary and Future Directions}
\vskip -0.6in
{\hskip1.4in{\it ``... things are the way they are in our universe because if they Weren't, we would not be here to notice."}}

\hskip4.2in - Brian Greene.
\graphicspath{{Chapter6/}{Chapter6/}}
\vskip 0.2in

In this review article, we have investigated two distinct approaches to  supergravity phenomenology inspired by compactifications
of Type IIB string theory and M-theory. The first part of review article has been concerned with the investigation of important issues related to particle phenomenology as well as cosmology by using a phenomenological model locally realizable as large volume limit of a Type IIB Swiss-Cheese Calabi-Yau orientifold involving a mobile space-time filling D3-brane restricted to a nearly special Lagrangian three-cycle embedded in the ``big" divisor  and multiple fluxed stacks of  space-time filling D7-branes wrapping the same ``big" divisor, as discussed in detail in chapter {\bf 2}. The second (smaller) part has dealt with some important hydro/thermodynamical aspects related to strongly-coupled/thermal gauge theories by constructing local eleven-dimensional uplifts of  non-compact (resolved) warped deformed conifold of Type IIB string theory involving space time filling D3-branes placed at the conifold's apex, $D5$-branes wrapping a two-cycle and D7-branes wrapped around non-compact four-cycle, as discussed in chapter {\bf 5}. In the following paragraphs we will outline
the key results of this work and discuss interesting prospects for future work.

After reviewing the motivation of (split) supersymmetry and the use of string theory as an important framework for BSM physics in chapter {{\bf 1}}, we provided details of a phenomenological model and its possible local string-theory based realization, in chapter {\bf 2}. In this context, we have described  the framework of Type IIB compactified on the orientifold of a ``Swiss-Cheese Calabi-Yau" in the large volume limit including (non-)perturbative $\alpha^{\prime}$ corrections in the K\"{a}hler potential and non-perturbative instanton-corrections in the superpotential, in the presence of $D3/D7$-branes as mentioned above. In a self-consistent manner, we show the possibility of the existence of a metastable $dS$-minimum for non-zero vevs of the mobile $D3$-brane position moduli, $D7$-branes valued Wilson line moduli (apart from bulk moduli) in a region of the moduli space 
where the potential is positive definite.  Next, by turning on different but small two-form fluxes on the different two-cycles homologously non-trivial from the point of view of the big divisor's, we provide explicit representations for $SU(3)_c \times SU(2)_L$ bifundamental first-generation quarks, their $SU(3)_c × U(1)_R$ bifundamental cousins, $SU(2)_L \times U(1)_L$ bifundamental first-generation leptons and Higgs, as well as the $U(1)_L \times U(1)_R$ bifundamental leptonic cousins. 
To evaluate the Wilson line moduli contrbution in one of the ${\cal N}=1$ chiral coordinates $T_B$, due to inclusion of four Wilson line moduli on the world volume of space-time filling D7-branes wrapped around the ``big divisor" restricted to (nearly) a special Lagrangian sub-manifold, we have constructed the distribution harmonic one-forms localized along the mobile space-time filling $D3$-brane (restricted to the 3-cycle)  which, interestingly,  supports the possibility of generating ${\cal O}(1)g_{YM}$  at string scale by wrapping D7-branes around the big divisor due to partial cancellation between geometric big divisor volume modulus and quadratic wilson line moduli term with substantial fine-tuning. Using the modified ${\cal N} = 1$ chiral coordinates, we constructed the form of K\"{a}hler potential and superpotential which has been  used throughout to  evaluate the moduli-space dependent factors in the ${\cal N}=1$ gauged supergravity action and hence the phenomenological results. Foremost, by calculating effective Yukawa couplings in the context of ${\cal N}=1$ gauged supergravity, we showed that Dirac masses corresponding to particular effective Yukawa coupling ${\cal Y}_{{\cal Z}_{I}{\cal A}_{1}{\cal A}_{3}}$ matched with the mass of electron, and  ${\cal Y}_{{\cal Z}_{I}{\cal A}_{1}{\cal A}_{3}}$ with Dirac mass of first generation quarks. Therefore, we were able to make the identification of right-set of four Wilson line moduli('s super-partners) with SM-like particles, and position moduli with Higgs-doublets.  Followed by this, we evaluated soft SUSY parameters  which turned out to be very high, thus indicating one of the signatures of split SUSY kind of scenario. To realize it more explicitly, within chapter {\bf 2}, we have discussed the evaluation of Higgs mass matrix which at EW scale provided one light Higgs of order 125 GeV with some fine-tuning, one heavy Higgs, and a very heavy higgsino mass parameter- hence ``$\mu$-split-like  SUSY" scenario. As another concrete signature of $\mu$-split-like  SUSY, we estimated life time of gluino using ${\cal N}=1$ gauged supergravity action, to be reasonably high.
In chapter 3, we turned our attention toward addressing particle-cosmology-related issues and demonstrated gravitino - the LSP in our model  - as a viable CDM candidate respecting BBN constraints as well as reproducing a relic abundance of around 0.1 in the context of ${\cal N}=1$ gauged supergravity limit of $D3/D7$ $\mu$-split-like  SUSY model. To justify the above, we  explicitly showed that the life time in case of all (co)-NLSP decay channels was smaller than $10^2$ sec (onset of BBN era) and life time of LSP decay to ordinary SM particles was more than age of universe. As another phenomenological consequence of split SUSY scenario's, finally, in chapter {\bf 4}, we evaluated electric dipole moment (EDM) of electron/neutron up to two loops, again in the context of using ${\cal N}=1$ gauged SUGRA. The sizable contribution of (e/n) EDM from one-loop diagram in our model is reasoned out in detail in chapter {\bf 4}. By assuming distinct ${\cal O}(1)$ CP violating phases corresponding to four Wilson line moduli as well as position moduli, we have obtained dominant contribution of the electron around ${\cal O}(10^{-32})$ esu-cm from one-loop diagram involving heavy chargino and a light Higgs as propagators in the loop. The neutron
EDM got a dominant contribution of the order $d_e/e \sim {\cal O}(10^{-33})$cm from one-loop diagram involving SM-like quarks and Higgs.  At two-loop level, we have obtained dominant contribution of  around ${\cal O}(10^{-29})$ esu-cm from diagram involving heavy sfermions and a light Higgs,  By conjecturing that the CP-violating phase could appear from the Higgs mass matrix obtained in the context of $\mu$-split-like  SUSY, we have shown the possibility of getting EDM of e/n around $ {\cal O}(10^{-27})$ esu-cm in case of two-loop diagram involving $W^{±}$ bosons.

Further, we sketch-out some arguments about the dimension-five neutrino mass and the stability of proton in our model. Using the RG-flow arguments of \cite{ferm_masses_MS}, one can show that the  Weinberg-type dimension-five Majorana-mass generating operator: ${\cal O}(\langle z_i\rangle^2)$ coefficient in $
\frac{e^{\frac{K}{2}}
\frac{\partial^2W}{\partial{\cal A}_1^2}}{\sqrt{K_{{\cal Z}_i\bar{\cal Z}_i}^2K_{{\cal A}_1\bar{\cal A}_1}^2}}\left(\bar{\chi}_L^{{\cal A}_1}{\cal Z}_i\right)^2$ or in fact $\frac{e^{\frac{K}{2}}{\cal D}_{\bar{{\cal A}}_1}D_{{\cal A}_1}\bar{W}}{\sqrt{K_{{\cal Z}_i\bar{\cal Z}_i}^2K_{{\cal A}_1\bar{\cal A}_1}^2}}\left(\bar{\chi}_L^{{\cal A}_1}{\cal Z}_i\right)^2$ produces the correct first-generation neutrino mass scale of slightly less than $1eV$ for $\langle z_i\rangle\sim{\cal O}(1){\cal V}^{\frac{1}{36}}$. Secondly,  for trilinear R-parity violating couplings $\lambda'_{ijk}\sim {\cal V}^{-\frac{5}{3}}$(see equation~(\ref{eq:2A2})-(\ref{eq:2C2})) and $\lambda''_{ijk} \sim {\cal V}^{-\frac{43}{30}}$ (see equation~(\ref{eq:4a2})-(\ref{eq:4b2})), using the analytic expression given in \cite{martin}, the rough estimate of proton decay width turns out to be
\begin{equation}
\Gamma_{p\rightarrow e^{+}{\pi}^0}\sim {m^{5}_{proton}}\frac{\left|\lambda'_{ijk}\lambda''_{ijk}\right|^2}{m^4_{{\tilde q}_I}}\sim \frac{{\cal V}^{-6.1}}{m^4_{{\tilde q}_I}}\sim 10^{-73} GeV,
\end{equation} thus giving life time of about $10^{42}$ years, which explicitly governs the stability of proton in $\mu$-split-like  SUSY scenario.

We would mention that results obtained in chapters {\bf 2-4} were simply dependent on dilute  $\tilde f$-fluxes and Calabi-Yau volume ${\cal V}$ which we  fixed to be around $10^{5}$. The lower bound on $\tilde f\sim 10^{-4}$ appeared by imposing that the flux-dependent $D$-term potential is sub-dominant as compared to the $F$-term potential in the dilute flux approximation (see chapter {\bf 2}). Though one could have chosen a viable range of $\tilde f$s satisfying the lower bound of $\tilde f\sim 10^{-4}$, the justification behind choosing a particular value of $\tilde f \sim 10^{-4}$ was because of getting the right amount of relic abundance of Gravitino. The justification behind constraining  a value of Calabi-Yau ${\cal V}$ to be of the order $10^5$ was based on the right identification of Wilson line moduli and position moduli with SM particle spectrum. More importantly, all of the results have been obtained in the context of the model which can be constructed locally near a particular nearly special Lagrangian three-cycle of a Swiss-Cheese Calabi-Yau 3-fold. It would be interesting to determine the global embedding of our model. Further, in the $D3/D7$ set-up described above, we have shown the possibility of identification of  Wilson line moduli only with first or second generation quarks and leptons. By extending the set-up to include Wilson line moduli identifiable with second and third generation quarks, one could hope to obtain other Wilson line moduli existing as effective NLSP, contributing significantly to gravitino abundance and the value of electron/neutron EDM very close to experimental bound for a given choice of the internal complex three-fold volume. The highlights of future goals in this direction are:
\vskip 0.1in
{\hskip -0.3in} $(i)$ The cosmologically inspired moduli  of maases typically greater than that of super-massive WIMP DM candidates, can directly decay into DM particles with branching ratio ${\cal O}(0.01-1)$.  The yield of DM particles can be quite large, since moduli tend to couple to the matter sector universally with gravitational strength couplings (1/$M_{p}$). The large yield might overclose the universe known as `cosmological gravitino problem'. In \cite{gravitino_DM}, we have estimated relic density of gravitino in the context of ${\cal N}=1$ gauged supergravity action by considering annihilation channels involving supersymmetric particles (identified with Wilson line moduli and position moduli)  and showed that relic density does not overclose the universe. The gravitino can also be produced directly from decay of bulk/geometric moduli (``big" as well as ``small" divisor modulus and closed string moduli) and may have branching ratio ${\cal O}(0.01-1)$. Therefore, we plan to evaluate decay modes of bulk moduli to kinematically allowed SM/supersymmetric particles as well as gravitino using ${\cal N}=1$ gauged supergravity limit of large volume $D3/D7$ model given in \cite{gravitino_DM} and calculate the branching ratio of gravitino.
\vskip 0.1in
{\hskip -0.3in} $(ii)$
By calculating reheating temperature corresponding to aforementioned decay modes, we plan to calculate number density of gravitino produced via bulk moduli decays and show that cosmological gravitino problem does not exist.
\vskip 0.1in
{\hskip -0.3in} $(iii)$
We intend to study the possibility to produce baryon asymmetry by direct decay of moduli through CP violating as well as baryon number violating couplings to baryons. The high moduli masses produce very low reheating temperature and it is very difficult to produce appreciable number of baryon asymmetry via standard baryogenesis mechanism such as Affleck-Dine mechanism, EW baryogenesis/leptogenesis mechanism. In \cite{dutta_et_al}, the possibility of  baryogenesis  at very low reheating temperature is discussed by considering extra TeV-scale coloured fields in addition to MSSM fields and showing that moduli decay to coloured fields produce right amount of baryon asymmetry. In this work, the asymmetry parameter is calculated by considering various tree level as well as one-loop decay  diagrams involving moduli fields as well as coloured scalar fields.  In \cite{gravitino_DM},  the colored fields have been identified as the Wilson line moduli. The non-zero interaction terms involving `bulk moduli fields' as well as `coloured fields' (identified with Wilson-line moduli) can be obtained in ${\cal N}=1$ gauged supergravity action. The non-zero CP violating phases can be made to appear because of distinct phases associated with each moduli field (as discussed in \cite{dhuria+misra_EDM}) and baryon number violation can be justified by allowing R-parity violating couplings into consideration (see \cite{gravitino_DM}).  By considering same tree level as well as one-loop diagrams as discussed in \cite{dutta_et_al} and evaluating relevant vertices, we plan to calculate the order of magnitude of baryon asymmetry in our ``local large volume $D3/D7$ $\mu$-split-like  SUSY" model.

As a second part of the review article, in chapter {\bf 5}, we have constructed local M-theory uplift of (resolved) warped deformed conifold using modified OKS-BH background given in \cite{metrics} in the context of Type IIB string theory with $N_f$ D7-branes wrapped around a 4-cycle in the resolved warped deformed conifold with $(M)N$ (fractional) D3-branes. The same is relevant to the study of aspects of strongly coupled QCD at finite temperature. By satisfying the requirements of implementing SYZ mirror symmetry locally (as the resolved warped deformed conifold does not possess a `third' global killing isometry along  the `original' angular variable $\psi\in[0,4\pi]$ for implementing SYZ mirror symmetry) to obtain Type $IIA$ background (near $(\theta_{1,2},\psi)=(\langle\theta_{1,2}\rangle,\left\{0,2\pi,4\pi
\right\})$), we oxidized the so-obtained Type $IIA$ background to $M$ theory  and then argued that there exists a new `MQGP limit' ($\frac{g_sM^2}{N}<<1,g_sN>>1$ for finite $(M,)g_s$ more relevant to strongly-coupled QGP) not considered in \cite{metrics}. After obtaining so, we set up an approach to study the behaviour of hydrodynamical as well as thermodynamical quantities in both weak coupling but large t'Hooft coupling regime of M-theory ($g_s<<1$) as well as MQGP limit  accomplished by letting ($g_s\stackrel{<}{\sim}1$). For both limits:
\vskip 0.1in
  {\hskip -0.3in} $(i)$
We obtained a black M3-brane solution whose near-horizon geometry near the $\theta_{1,2} = 0$, branches, preserved $\frac{1}{8}$ supersymmetry;
 \vskip 0.1in
  {\hskip -0.3in} $(ii)$
Investigated thermodynamical stability of solution in Type IIB background as well as its M-theory uplift. The thermodynamical stability with Ouyang embedded D7-branes in Type IIB background of \cite{metrics} was verified by studying the behaviour of the chemical potential as a function of $N_f$(number of bi-fundamental quark flavors) and showing that $ \frac{\partial{\mu_C}}{\partial N_f}|_{T}>0$. The thermodynamical stability of the M-theory uplift was discussed by first evaluating finite 11D supergravity action and thereafter showing positivity of specific heat;
\vskip 0.1in
  {\hskip -0.3in} $(iii)$
Shear viscosity-to-entropy density ratio naturally turned out to be $\frac{1}{4 \pi}$ in Type IIA and its local M-theory uplift. Diffusion constant for both, Types IIB/IIA backgrounds also come out to be the reciprocal of the temperature;
\vskip 0.1in
  {\hskip -0.3in} $(iv)$
It was verified that the black M3-brane entropy went like the cube of the horizon radius from M-theoretic thermodynamical methods as well as the horizon area calculated from the starting Type IIB, mirrory Type IIA and the black M3-brane solutions.
 
The potentially interesting extensions to this model have been outlined as future goals below:
\vskip 0.1in
 {\hskip -0.3in} $(i)$
We plan to calculate number density of baryons in the context of modified OKS-BH background. We have already investigated the form of finite part of DBI action ($I_{D7}$) using modified OKS-BH background given in \cite{metrics} in the context of Type IIB string background. Now we will extend the line of investigation by studying the asymptotic form of temporal component of gauge field and calculate the number density of quarks in the above-mentioned background using the technique given in  \cite{robert_myers1, robert_myers2}.
\vskip 0.1in
 {\hskip -0.3in} $(ii)$
We propose to obtain the analytic form of number density  ($n_B$) as well as chemical potential $\mu_f$ as a function of temperature in $T\rightarrow 0$ limit by studying complete phase diagram of $U(1)$ gauge field/chemical potential as a function of temperature using modified OKS-BH background of \cite{metrics,MQGP} in the context of Type IIB string theory. Doing so, we can directly incorporate the results of  ($n_B$)  as well as $\mu_f$ to obtain pressure, energy density as well as charge density of neutron star by using analytic expressions given in \cite{Glendenning}.  The numerical values of baryon density can be calculated by choosing specific values of parameters, $g_s$ as well as number of D3-branes $(N/N_{eff})$ so that limits of \cite{metrics} could be satisfied. In this way, stringy background can be used to study astrophysical applications also.
\vskip 0.1in
{\hskip -0.3in} $(iii)$
By introducing D7-brane probes into the black D3-brane background, the resulting theory also contains spectrum of quark-antiquark bound states or mesons. As these mesons are dual to strings with both ends on D7-branes, spectrum of the same can be studied  computing the spectrum of fluctuations of the worldvolume fields on the D7-branes. The transition between Minkowski and black hole embedding can also be characterized by studying the meson spectrum formed by combining quark-aniquark pair.
 At zero temperature, the spectrum of mesons has a mass gap and is discrete while at finite temperature, the mesonic excitations are unstable and are characterized by a discrete spectrum of quasinormal modes. The meson spectrum as well as mass gap has been studied in zero temperature limit using O(uyang) K(lebanov) S(trassler) background in \cite{ouyang_meson}. We plan to study expected continuos meson spectra obtained by introducing black hole in the OKS background background in `(resolved) warped deformed conifold' by considering fluctuations of scalar fields on the world volume of D7-branes.          

\

\renewcommand{\chaptermark}[1]{         
\markboth{ \thechapter.\ #1}{}} %

\appendix
\chapter{Appendix}

\graphicspath{{AppendixA/}{AppendixA/}}

\setcounter{equation}{0} \seceqaa
\vskip -0.5in
\section{Details of Local type IIA SYZ Mirror}
In this appendix, after applying T-duality, locally, along $x, y, z$ and in this order, we give explicitly  (i) the type IIA components obtained, and (ii) the components of type IIA two-form fluxes obtained from type IIB one-form,  

\subsection{Type IIA Metric Components}

 The various components of the metric after three successive T-dualities along $x, y$ and $z$ respectively, can be written as \cite{SYZ3_Ts}:
 {\small
\begin{eqnarray}
\label{G_munu}
& & G_{\mu\nu} = {g_{\mu\nu}g_{xx} -
g_{x\mu}g_{x\nu} + b_{x\mu}b_{x\nu} \over g_{xx}} -
{(g_{y\mu}g_{xx} - g_{xy} g_{x \mu} + b_{xy} b_{x\mu})
(g_{y\nu}g_{xx}
 - g_{xy} g_{x \nu} + b_{xy} b_{x\nu}) \over
g_{xx}(g_{yy}g_{xx}- g_{xy}^2 + b_{xy}^2)} \nonumber\\
&& + {(b_{y\mu}g_{xx} - g_{xy} b_{x \mu} + b_{xy}
g_{x\mu})(b_{y\nu}g_{xx}
 - g_{xy} b_{x \nu} + b_{xy} g_{x\nu})\over
g_{xx}(g_{yy}g_{xx}- g_{xy}^2 + b_{xy}^2)},\\
\label{Gmuz}
& & G_{\mu z} = {g_{\mu z}g_{xx} -
g_{x\mu}g_{xz} + b_{x \mu}b_{xz} \over g_{xx}} - {(g_{y\mu}g_{xx}
- g_{xy} g_{x \mu} + b_{xy} b_{x\mu}) (g_{yz}g_{xx} - g_{xy} g_{x
z} + b_{xy} b_{xz}) \over g_{xx}(g_{yy}g_{xx}- g_{xy}^2 +
b_{xy}^2)} \nonumber\\
& &  + {(b_{y\mu}g_{xx} - g_{xy} b_{x \mu} +
b_{xy} g_{x\mu})(b_{yz}g_{xx} - g_{xy} b_{x z} + b_{xy}
g_{xz})\over g_{xx}(g_{yy}g_{xx}- g_{xy}^2 + b_{xy}^2)},\\
\label{Gzz}
\pagebreak
& & G_{zz} =  {g_{zz}g_{xx} - j^2_{xz} +
b^2_{xz}\over g_{xx}} - {(g_{yz}g_{xx} - g_{xy} g_{xz} + b_{xy}
b_{xz})^2 \over g_{xx}(g_{yy}g_{xx}- g_{xy}^2 + b_{xy}^2)} + {(b_{yz}g_{xx} - g_{xy} b_{x z} + b_{xy} g_{xz})^2 \over
g_{xx}(g_{yy}g_{xx}- g_{xy}^2 + b_{xy}^2)},\\
\label{Gymu}
& & G_{y \mu} = -{b_{y \mu} g_{xx} - b_{x \mu} g_{xy} + b_{xy}
 g_{\mu x} \over g_{yy}g_{xx}- g_{xy}^2 + b_{xy}^2},
~ G_{y z} = -{b_{y z} g_{xx} - b_{x z} g_{xy} + b_{xy} g_{z x}
\over g_{yy}g_{xx}- g_{xy}^2 + b_{xy}^2},\\
\label{Gyy}
& & G_{yy} = {g_{xx} \over g_{yy}g_{xx}- g_{xy}^2 +
b_{xy}^2},~ G_{xx} = {g_{yy} \over g_{yy}g_{xx}- g_{xy}^2 +
b_{xy}^2}, ~G_{xy} = {-g_{xy} \over g_{yy}g_{xx}- g_{xy}^2 +
b_{xy}^2},\\
\label{Gmux}
& & G_{\mu x} = {b_{\mu x} \over g_{xx}} + {(g_{\mu y} g_{xx} -
 g_{xy} g_{x \mu} + b_{xy} b_{x \mu}) b_{xy} \over
g_{xx}(g_{yy}g_{xx}- g_{xy}^2 + b_{xy}^2)}
+ {(b_{y \mu} g_{xx} - g_{xy} b_{x \mu} + b_{xy} g_{x \mu}) g_{xy}
 \over g_{xx}(g_{yy}g_{xx}- g_{xy}^2 + b_{xy}^2)},\\
\label{Gzx}
& & G_{z x} = {b_{z x} \over g_{xx}} + {(g_{z y} g_{xx} -
g_{xy} g_{x z} + b_{xy} b_{x z}) b_{xy} \over
g_{xx}(g_{yy}g_{xx}- g_{xy}^2 + b_{xy}^2)}
 + {(b_{y z} g_{xx} - g_{xy} b_{x z} + b_{xy} g_{xz}) g_{xy}
  \over g_{xx}(g_{yy}g_{xx}- g_{xy}^2 + b_{xy}^2)}.
\end{eqnarray}}
In the above formulae we have denoted the type IIB
$B$ fields as $b_{mn}$.  For the generic case we will switch on all the
components of the $B$ field
{\small
\begin{eqnarray}
b & = &
b_{\mu\nu} ~ dx^\mu \wedge dx^\nu + b_{x \mu} dx \wedge dx^\mu +  b_{y \mu}
~ dy~\wedge dx^\mu + b_{z \mu} ~ dz \wedge dx^\mu \nonumber\\
& & + ~ b_{xy}
~ dx \wedge dy +
 b_{xz} ~ dx  \wedge dz +  b_{zy}~ dz  \wedge dy.
 \end{eqnarray}}
 After applying again the T-dualities, the NS component
of the $B$ field in the mirror set-up will take the form
{\small
\begin{eqnarray}
\label{B}
{\tilde B} & = & \left( {\cal B}_{\mu\nu}
+ {2 {\cal B}_{z[\mu} G_{\nu]z} \over G_{zz}} \right) dx^\mu
\wedge dx^\nu + \left( {\cal B}_{\mu x} + {2 {\cal B}_{z[\mu}
G_{x]z} \over G_{zz}}\right)
 dx^\mu \wedge dx  \nonumber\\
& & \left( {\cal B}_{\mu y} + {2 {\cal B}_{z[\mu} G_{y]z} \over G_{zz}}
 \right) dx^\mu \wedge dy
+ \left( {\cal B}_{xy} + {2 {\cal B}_{z[x} G_{y]z} \over G_{zz}}
\right) dx \wedge dy \nonumber\\
& &  + {G_{z \mu} \over G_{zz}} dx^\mu
\wedge dz + {G_{z x} \over G_{zz}} dx \wedge dz + {G_{z y} \over
G_{zz}} dy \wedge dz.
\end{eqnarray}}
 Here the $G_{mn}$ components have been
given above, and the generic expressions of ${\cal B}$  components are provided in \cite{MQGP}. Using the same, the analytic expressions of non-zero type IIA ${\cal B}$ components are:
{\footnotesize
\begin{eqnarray}
\label{B-3tduals}
& (i) & B^{IIA}_{xz}:=
 \Bigl(-54 \sin ^3({\theta_1}) \cos ({\theta_1}) \left(\sqrt{6} \left(9 h_ 5^2-1\right)+4 h_ 5 \cot
   ({\theta_2})\right)+81 h_ 5 \bigl(2 \sin ^4({\theta_1}) \left(3 h_ 5+\sqrt{6} \cot
   ({\theta_2})\right) \nonumber\\
   && +h_5 \sin ^2(2 {\theta_1})\bigr)-12 \sin ({\theta_1}) \cos ^3({\theta_1}) \left(12
   h_ 5 \cot ({\theta_2})+2 \sqrt{6} \cot ^2({\theta_2})-3 \sqrt{6}\right)+12 \sin ^2({\theta_1}) \cos
   ^2({\theta_1}) \nonumber\\
   && \cot ({\theta_2}) \left(27 \sqrt{6} h_ 5+2 \cot ({\theta_2})\right)+16 \cos ^4({\theta_1})
   \cot ^2({\theta_2}) \Bigr)/\Bigl(3 (\cos (2 {\theta_1})-5) \Bigl(\sin ^2({\theta_1}) \bigl(-27 h_ 5^2+2 \cot
   ^2({\theta_2}) \nonumber\\
   &&+3\bigr) +6 h_ 5 \sin (2 {\theta_1}) \cot ({\theta_2})+2 \cos ^2({\theta_1})\Bigr)\Bigr), \nonumber\\
  & (ii) &B^{IIA}_{yz}:= -\frac{3 \sqrt{6} (3 h_ 5 \cot ({\theta_1})+\cot ({\theta_2}))}{-27 h_ 5^2+12 h_ 5 \cot ({\theta_1})
   \cot ({\theta_2})+2 \cot ^2({\theta_1})+2 \cot ^2({\theta_2})+3} \nonumber\\
      \pagebreak
   & (iii) & B^{IIA}_{\theta1 x}:= \Bigl({f_ 1}({\theta_1}) \sin ^2({\theta_1}) \left(\sin ({\theta_1}) \left(27 h_ 5^2-2 \cot
   ^2({\theta_2})-3\right)-6 h_ 5 \cos ({\theta_1}) \cot ({\theta_2})\right)\nonumber\\
   && \Bigl({g_s} \Bigl(18
   {g_s}^2 {M_{eff}}^2 {N_f} \log ^2(r)+3 {g_s} {M_{eff}}^2 \log (r) \left({g_s} {N_f} \log \left(\sin
   \left(\frac{{\theta_1}}{2}\right) \sin \left(\frac{{\theta_2}}{2}\right)\right)+3 {g_s} {N_f}+4 \pi \right) \nonumber\\
   && +8
   \pi ^2 N\Bigr)\Bigr)^{\frac{1}{4}}\Bigr)/\Bigl(2^{3/4} \sqrt{3} \sqrt[4]{\pi } \left(\sin ^2({\theta_1}) \left(-27 h_ 5^2+2 \cot
   ^2({\theta_2})+3\right)+6 h_ 5 \sin (2 {\theta_1}) \cot ({\theta_2})+2 \cos ^2({\theta_1})\right)\Bigr) \nonumber\\
& (iv) & B^{IIA}_{\theta2 x}:= \Bigl(\sqrt[4]{\frac{2}{\pi }} r {f_ 2}({\theta_2}) \sin ({\theta_1}) \sin ({\theta_2}) \cos ({\theta_2}) (3
   h_ 5 \sin ({\theta_1}) \cos ({\theta_2})+\cos ({\theta_1}) \sin ({\theta_2}))\nonumber\\
   &&  \Bigl({g_s}
   \Bigl(18 {g_s}^2 {M_{eff}}^2 {N_f} \log ^2(r)+3 {g_s} {M_{eff}}^2 \log (r) \left({g_s} {N_f} \log
   \left(\sin \left(\frac{{\theta_1}}{2}\right) \sin \left(\frac{{\theta_2}}{2}\right)\right)+3 {g_s} {N_f}+4 \pi
   \right)\nonumber\\
   && +8 \pi ^2 N\Bigr)\Bigr)^{\frac{1}{4}}\Bigr)/\Bigl(\sqrt{3} \Bigl(3 \left(-9 h_ 5^2 \sin ^2({\theta_1}) \sin
   ^2({\theta_2})+h_ 5 \sin (2 {\theta_1}) \sin (2 {\theta_2})+\sin ^2({\theta_1}) \sin
   ^2({\theta_2})\right)\nonumber\\
   && +2 \sin ^2({\theta_1}) \cos ^2({\theta_2})+2 \cos ^2({\theta_1}) \sin
   ^2({\theta_2})\Bigr)\Bigr)\nonumber\\
& (v) & B^{IIA}_{\theta1y}:=
\Bigl(\sqrt[4]{\frac{2}{\pi }}  {f_ 1}({\theta_1}) \cos ({\theta_1}) (3 h_ 5 \cot ({\theta_1})+\cot
   ({\theta_2})) \Bigl({g_s} \Bigl(18 {g_s}^2 {M_{eff}}^2 {N_f} \log ^2(r)+ \nonumber\\
   &&3 {g_s} {M_{eff}}^2
   \log (r) \left({g_s} {N_f} \log \left(\sin \left(\frac{{\theta_1}}{2}\right) \sin
   \left(\frac{{\theta_2}}{2}\right)\right)+3 {g_s} {N_f}+4 \pi \right)+8 \pi ^2 N\Bigr)\Big)^{\frac{1}{4}}\Bigr)/\Bigl(\sqrt{3} \Bigl(-27
   h_ 5^2 \nonumber\\
   && +12 h_ 5 \cot ({\theta_1}) \cot ({\theta_2})+2 \cot ^2({\theta_1})+2 \cot
   ^2({\theta_2})+3\Bigr)\Bigr)\nonumber\\
& (vi) & B^{IIA}_{\theta_2 y}:=
\Bigl(\sqrt[4]{\frac{2}{\pi }}  {f_ 1}({\theta_1}) \cos ({\theta_1}) (3 h_ 5 \cot ({\theta_1})+\cot
   ({\theta_2})) \Bigl({g_s} \Bigl(18 {g_s}^2 {M_{eff}}^2 {N_f} \log ^2(r)\nonumber\\
   && +3 {g_s} {M_{eff}}^2
   \log (r) \left({g_s} {N_f} \log \left(\sin \left(\frac{{\theta_1}}{2}\right) \sin
   \left(\frac{{\theta_2}}{2}\right)\right)+3 {g_s} {N_f}+4 \pi \right)+8 \pi ^2 N\Bigr)\Bigr)/\Bigl(\sqrt{3} \Bigl(-27
   h_ 5^2\nonumber\\
   && +12 h_ 5 \cot ({\theta_1}) \cot ({\theta_2})+2 \cot ^2({\theta_1})+2 \cot
   ^2({\theta_2})+3\Bigr)\Bigr)\nonumber\\
& (vii) & B^{IIA}_{\theta_1 \theta_2}:=
\frac{1}{4 \pi  r^2}\Bigl({g_s} M \Bigl(-\Bigl(9 \sin ({\theta_1}) \Bigl(\frac{1}{2} {g_s} {N_f} r {f_ 1}({\theta_1}) \log (r)
   \sin ({\theta_1})   \cos \left(\frac{3 {\theta_2}}{2}\right) \csc
   \left(\frac{{\theta_2}}{2}\right)\nonumber\\
   &&  \left(36 a^2 \log (r)+r\right)+3 h_ 5 {f_ 2}({\theta_2}) \sin
   ({\theta_2}) \Bigl({g_s} {N_f} r \log (r) \cos ({\theta_1}) \cot \left(\frac{{\theta_1}}{2}\right)
   \left(108 a^2 \log (r)+r\right)\nonumber\\
   && -2 \left(3 a^2-r^2\right) \sin ({\theta_1}) \Bigl(2 \log (r) \left({g_s} {N_f}
   \log \left(\sin \left(\frac{{\theta_1}}{2}\right) \sin \left(\frac{{\theta_2}}{2}\right)\right)+2 \pi \right)+9
   {g_s} {N_f} \log ^2(r)\nonumber\\
   && +{g_s} {N_f} \log \left(\sin \left(\frac{{\theta_1}}{2}\right) \sin
   \left(\frac{{\theta_2}}{2}\right)\right)\Bigr)\Bigr)\Bigr)\Bigr)/\Bigl(\cos (2 {\theta_1})-5\Bigr)-\nonumber\\
   && \Bigl({f_ 2}({\theta_2})
   \sin ({\theta_2}) \left(\sin ({\theta_2}) \left(27 h_ 5^2-2 \cot ^2({\theta_1})-3\right)-6 h_ 5 \cot
   ({\theta_1}) \cos ({\theta_2})\right) \Bigl(\frac{9}{2} h_ 5 \sin ({\theta_2}) \nonumber\\
   && \biggl({g_s} {N_f} r
   \log (r) \cos ({\theta_1}) \csc ^2\left(\frac{{\theta_1}}{2}\right) \left(108 a^2 \log (r)+r\right)-4 \left(3
   a^2-r^2\right) \nonumber\\
   && \left(2 \log (r) \left({g_s} {N_f} \log \left(\sin \left(\frac{{\theta_1}}{2}\right) \sin
   \left(\frac{{\theta_2}}{2}\right)\right)+2 \pi \right)+9 {g_s} {N_f} \log ^2(r) \right)\biggr)+\cos
   ({\theta_2}) \nonumber\\
   && \Bigl(4 \left(3 a^2-r^2\right) \cot ({\theta_1}) \Bigl(2 \log (r) \left({g_s} {N_f} \log
   \left(\sin \left(\frac{{\theta_1}}{2}\right) \sin \left(\frac{{\theta_2}}{2}\right)\right)+2 \pi \right)+9 {g_s}
   {N_f} \log ^2(r)\nonumber\\
   && +{g_s} {N_f} \log \left(\sin \left(\frac{{\theta_1}}{2}\right) \sin
   \left(\frac{{\theta_2}}{2}\right)\right)\Bigr)+3 {g_s} {N_f} r \log (r) \cot
   \left(\frac{{\theta_1}}{2}\right) \left(108 a^2 \log (r)+r\right)\Bigr)\Bigr)\Bigr)/\Bigl(\sin ^2({\theta_2}) \nonumber\\
   && \left(-27
   h_ 5^2+2 \cot ^2({\theta_1})+3\right)+6 h_ 5 \cot ({\theta_1}) \sin (2 {\theta_2})+2 \cos
   ^2({\theta_2})\Bigr)-\Bigl(2 {f_ 1}({\theta_1}) \cos ({\theta_1}) \sin ({\theta_2})\nonumber\\
   &&  (3 h_ 5 \cot
   ({\theta_1})+\cot ({\theta_2})) \Bigl(\left(r^2-3 a^2\right) (\cos (2 {\theta_1})-5) \csc ^2({\theta_1}) \csc
   ({\theta_2}) \left(9 a^4 {g_s}-3 a^2 {g_s} r^2-1\right) \nonumber\\
   && \left(2 \log (r) \left({g_s} {N_f} \log
   \left(\sin \left(\frac{{\theta_1}}{2}\right) \sin \left(\frac{{\theta_2}}{2}\right)\right)+2 \pi \right)+9 {g_s}
   {N_f} \log ^2(r) \right)\nonumber\\
   && - \frac{1}{8} {g_s} {N_f} r \log (r) \cos ({\theta_2}) \csc
   ^3\left(\frac{{\theta_2}}{2}\right) \sec \left(\frac{{\theta_2}}{2}\right) \left(36 a^2 \log (r)+r\right) \Bigl(-18
   h_ 5 \csc ({\theta_1}) \sin ({\theta_2})\nonumber\\
   && +4 \cot ({\theta_1}) \csc ({\theta_1}) \cos ({\theta_2})+\cot
   ^2({\theta_1})-5 \csc ^2({\theta_1})-1\Bigr)\Bigr)\Bigr)/\Bigl(-27 h_ 5^2+12 h_ 5 \cot ({\theta_1}) \cot
   ({\theta_2})\nonumber\\
   && +2 \cot ^2({\theta_1})+2 \cot ^2({\theta_2})+3\Bigr)\Bigr)\Bigr).
\end{eqnarray}}
The T-dualized NS-NS components and metric components after the aforementioned triple T-dualities  can be easily evaluated, using which, one obtains the following type $IIA$ metric components. The simplified expressions of the same in  the (i) weak($g_s$) coupling - large t'Hooft couplings limit:
$g_s\sim\epsilon^d, M\sim\epsilon^{-\frac{3d}{2}}, N\sim\epsilon^{-19d},\epsilon\leq0.01$, as well
as the (ii) `MQGP limit':$g_s\sim\epsilon^d, M\sim\epsilon^{-\frac{3d}{2}}, N\sim\epsilon^{-39d}, \epsilon\lesssim 1$, 
{\footnotesize
\begin{eqnarray}
\label{metric-mirror}
& (i) & G^{IIA}_{\theta_1\theta_1}\sim \frac{2 \pi \sqrt[]{{g_s} N} \left({f_1}({\theta_1})^2 \sin ^2({\theta_1})+1\right)}{\sqrt[3]{3} }, (ii)~G^{IIA}_{\theta_2\theta_2}\sim \frac{2 \pi  \sqrt[]{{g_s} N} \cos ({\theta_1}) \cos ({\theta_2}) \left({f_2}({\theta_2})^2 \sin ^2({\theta_1})+1\right)}{9 \sqrt[3]{3}
   \left(\frac{\sin ^2({\theta_1})}{6}+\frac{\cos ^2({\theta_1})}{9}\right)} \nonumber\\
   & (iii)& G^{IIA}_{\theta_1\theta_2}\sim -\frac{2 \pi \sqrt[]{{g_s} N} {f_1}({\theta_1}) {f_2}({\theta_2}) \sin ^2({\theta_1})(\cos (2 {\theta_1})-5)^{-1} (\cos (3 {\theta_1})-9 \cos ({\theta_1})) \sin ^2({\theta_2}) \cos
   ({\theta_2})}{\sqrt[3]{3}   \left(3 \left( \sin
   ^2({\theta_1}) \sin ^2({\theta_2})\right)+2 \sin ^2({\theta_1}) \cos ^2({\theta_2})+2 \cos ^2({\theta_1}) \sin ^2({\theta_2})\right)}  \nonumber\\
   & (iv) &  G^{IIA}_{x\theta_1}=\frac{{g_s}^{7/4}}{4 \sqrt{2} \pi ^{3/4}\sqrt[4]{N} r}\Bigl(\sqrt[3]{3}   M {N_f} ln (r) \cot \left(\frac{{\theta_1}}{2}\right) \csc ({\theta_1}) \csc
   ({\theta_2}) \left(108 a^2 ln (r)+r\right)  \nonumber\\
   && \left(9 {h_5}+\left(3 \sqrt{6}-2 \cot ({\theta_1})\right) \cot
   ({\theta_2})\right) (2 \cos ({\theta_1}) \cos ({\theta_2})-9 {h_5} \sin ({\theta_1}) \sin ({\theta_2}))\Bigr)\nonumber\\
   & (v) & G^{IIA}_{x\theta_2}=\frac{1}{6 \sqrt{2}
   \pi ^{5/4} r (\cos (2 {\theta_1})-5)^2 \sqrt[4]{{g_s} N}}\Bigl({g_s}^2 M {N_f} ln (r) \sin ({\theta_1}) \cot \left(\frac{{\theta_2}}{2}\right) \csc ({\theta_2}) \left(36 a^2 ln (r)+r\right)\nonumber\\
   &&{\hskip -0.1in} \Bigl(27
   \sqrt{6} \sin ^2({\theta_1}) \cos ({\theta_1}) \sin ({\theta_2})-2 \cos ^3({\theta_1}) \left(\cos ({\theta_2}) \Bigl(6 \sqrt{6} \cot
   ({\theta_2})-4 \cot ({\theta_1}) \cot ({\theta_2})\Bigr)-9 \sqrt{6} \sin ({\theta_2})\right)\nonumber\\
   && +12 \sin ({\theta_1}) \cos ^2({\theta_1}) \cos
   ({\theta_2}) \cot ({\theta_2})\Bigr) \left(-2 \cot ({\theta_1}) \csc ({\theta_1}) \cos ({\theta_2})+2 \cot ^2({\theta_1})+3\right)\Bigr)\nonumber\\
    & (vi) & G^{IIA}_{y\theta_1}= \frac{9\ 3^{5/6} (g_s)^{3/4}  M ln (r) \csc ({\theta_1}) \sin ^4({\theta_2})  \left(-8 \pi  \left(3 a^2-r^2\right)  \cos ({\theta_1}) \cot ({\theta_2})\right)}{2 N^{1/4}
   \pi ^{3/4} r^2 \left(3 \left( \sin ^2({\theta_1}) \sin ^2({\theta_2})\right)+2 \sin ^2({\theta_1})
   \cos ^2({\theta_2})+2 \cos ^2({\theta_1}) \sin ^2({\theta_2})\right)}
    \nonumber\\
   & (vii) & G^{IIA}_{y\theta_2}=\frac{\sqrt{2} \pi ^{3/4}  (g_s N)^{1/4} {f_2}({\theta_2}) (\cos (2 {\theta_1})-5) \sin ^2({\theta_2}) \cos ({\theta_2})
    }{  \left(3 \left({h_5} \sin (2 {\theta_1}) \sin (2 {\theta_2})+\sin ^2({\theta_1}) \sin
   ^2({\theta_2})\right)+2 \sin ^2({\theta_1}) \cos ^2({\theta_2})+2 \cos ^2({\theta_1}) \sin ^2({\theta_2})\right)}\nonumber\\
    & (viii) & G^{IIA}_{z\theta_1 }=\frac{{g_s}^2 M {N_f} ln (r) \cot \left(\frac{{\theta_1}}{2}\right) \csc ^2({\theta_1}) \left(108 a^2 ln (r)+r\right) \left(\sin ^2({\theta_1})
   \left(2 \cot ^2({\theta_2})+3\right)+2 \cos ^2({\theta_1})\right)}{8 \sqrt{2} \pi ^{5/4} r \sqrt[4]{{g_s} N}}\nonumber\\
   & (ix) &  G^{IIA}_{z\theta_2 }=\frac{1}{4 \sqrt{2} \pi ^{5/4} r \sqrt[4]{{g_s} N} (\cos (2 {\theta_1})-5)}\Bigl({g_s}^2 M {N_f} ln (r) \cot \left(\frac{{\theta_2}}{2}\right) \left(36 a^2 ln (r)+r\right) \nonumber\\
   && \left(-2 \cot ({\theta_1}) \csc ({\theta_1})
   \cos ({\theta_2})+2 \cot ^2({\theta_1})+3\right) \left( \sin ^2({\theta_1}) \left(2 \cot
   ^2({\theta_2})+3\right)+2 \cos ^2({\theta_1})\right)\Bigr)\nonumber\\
   & (x) & G^{IIA}_{yy}=\frac{1}{1-9 {h_5}^2}, (xi)~~ G^{IIA}_{zz}=\frac{1}{27} \left(12 {h_5} \cot ({\theta_1}) \cot ({\theta_2})+2 \cot ^2({\theta_1})+2 \cot ^2({\theta_2})+3\right) \nonumber\\
  & (xii) & G^{IIA}_{xz}= \frac{2 \left(2 \cos ^2({\theta_1}) \left(4 \cot ^2({\theta_1}) \cot ^2({\theta_2})+\cot ({\theta_1}) \left(9 \sqrt{6}-6 \sqrt{6} \cot
   ^2({\theta_2})\right)\right)+27 \sqrt{6} \sin ({\theta_1}) \cos ({\theta_1})\right)}{81 (\cos (2 {\theta_1})-5)}\nonumber\\
  & (xiii) & G^{IIA}_{yz}=-\frac{1}{3} \sqrt{\frac{2}{3}} (3 {h_5} \cot ({\theta_1})+\cot ({\theta_2}))\nonumber\\
   & (xiv) & G^{IIA}_{xy}=-\frac{2 \sin ({\theta_2})\cos ^2({\theta_1}) }{27
   (\cos (2 {\theta_1})-5) \left(3 \sin ^2({\theta_1}) \sin ^2({\theta_2})+2 \sin ^2({\theta_1}) \cos ^2({\theta_2})+2 \cos ^2({\theta_1}) \sin
   ^2({\theta_2})\right)}\times\nonumber\\
    &&{\hskip -0.2in} \left(8 \sqrt{6} \cos ^2({\theta_1}) \cos ({\theta_2}) \cot ^2({\theta_2})-72 \sin ({\theta_1}) \cos ({\theta_1})
   \cos ({\theta_2}) \cot ^2({\theta_2})+12 \sqrt{6} \sin ^2({\theta_1}) \cos ({\theta_2}) \cot ^2({\theta_2})\right).\nonumber\\
\end{eqnarray}}

 
 \clearpage
\begin{spacing}{1.34}
\addcontentsline{toc}{chapter}{Bibliography}
\end{spacing}

\graphicspath{{Bibliography/Bibliography}}



\end{document}